\documentclass[bibliography=totoc]{scrbook}
\usepackage[bmargin=3.4cm, outer=2cm, inner=3cm]{geometry}

\usepackage[german,english]{babel}
\usepackage[T1]{fontenc}
\usepackage{tocloft}

\usepackage[toc,page]{appendix}
\usepackage[font=small,labelfont=bf]{caption}
\usepackage{emptypage}

\usepackage{colortbl}

\usepackage{pdfpages}

\usepackage{rotating}

\usepackage[numbers,sort&compress]{natbib}
\usepackage{scrpage2} 
\ifoot[]{} 
\ofoot[\pagemark]{\pagemark} 

\makeatletter
\renewcommand*{\raggedchapterentry}{\setlength\@tempdima{2.6em}}
\renewcommand*\l@section{\bprot@dottedtocline{1}{0em}{2.6em}}
\renewcommand*\l@subsection{\bprot@dottedtocline{1}{2.7em}{2.5em}}
\renewcommand*\l@subsubsection{\bprot@dottedtocline{1}{5.2em}{3.3em}}
\makeatother

\newcolumntype{x}[1]{!{\centering\arraybackslash\vrule width #1}}

\makeatletter
\newcommand{\thickhline}{%
    \noalign {\ifnum 0=`}\fi \hrule height 1pt
    \futurelet \reserved@a \@xhline
}
\makeatother

\usepackage{hhline}
\usepackage{charter}
\usepackage{wrapfig}
\usepackage{arydshln}
\renewcommand{\arraystretch}{1.3}
\usepackage{mathrsfs} 
\usepackage{amssymb,amsbsy,amsmath,amsfonts}

\definecolor{airforceblue}{rgb}{0.36, 0.54, 0.66}
\definecolor{bluegray}{rgb}{0.4, 0.6, 0.8}
\definecolor{blue(ncs)}{rgb}{0.0, 0.53, 0.74}
\definecolor{antiquefuchsia}{rgb}{0.57, 0.36, 0.51}
\definecolor{byzantium}{rgb}{0.44, 0.16, 0.39}

\usepackage[mathscr]{euscript}
\DeclareMathAlphabet{\mathpzc}{T1}{pzc}{m}{it}

\usepackage[Glenn]{fncychap}
\usepackage[svgnames]{xcolor}

\usepackage{ae}
\usepackage{graphicx}
\usepackage{caption}
\usepackage{upgreek}
\usepackage{enumitem}
\usepackage[normalem]{ulem}
\usepackage{cancel} 
\usepackage{nicefrac}
\usepackage{float}
\usepackage{dsfont}
\usepackage{upgreek}
\usepackage{booktabs}
\usepackage{color}
\numberwithin{equation}{chapter}
\numberwithin{table}{chapter}
\numberwithin{figure}{chapter}
\usepackage{esint}
\usepackage{slashed}
\usepackage{multirow}
\usepackage[customcolors]{hf-tikz}
\ChNameVar{\bfseries\Large\sf}
\ChNumVar{\Huge}
\ChTitleVar{\bfseries\Large\rm}
\usepackage{tikzsymbols} 

\allowdisplaybreaks

\let\oldbibliography\thebibliography
\renewcommand{\thebibliography}[1]{\oldbibliography{#1}
\setlength{\itemsep}{0pt}}

\def\beq{\begin{equation}}
\def\eeq{\end{equation}}
\def\bea{\begin{eqnarray}}
\def\eea{\end{eqnarray}} 
\def\beqa{\begin{equation}\begin{array}{l}}
\def\eeqa{\end{array}\end{equation}}

\def\eqlab#1{\label{eq:#1}}
\def\figlab#1{\label{fig:#1}}

\def\seclab#1{\label{sec:#1}}
\def\chaplab#1{\label{chap:#1}}

\def\eref#1{(\ref{eq:#1})}
\def\eqref#1{eq.~(\ref{eq:#1})}
\def\Eqref#1{Eq.~(\ref{eq:#1})}

\def\Figref#1{Fig.~\ref{fig:#1}}

\def\secref#1#2{Section~\ref{chap:#1}\textcolor{cyan}{.}\ref{sec:#2}}
\def\chapref#1{Chapter~\ref{chap:#1}}
\def\appref#1#2{Appendix~\ref{chap:#1}\textcolor{cyan}{.}\ref{sec:#2}}

\def\half{\mbox{\small{$\frac{1}{2}\;$}}}

\def\third{\mbox{\small{$\frac{1}{3}$}}}
\def\sixth{\mbox{\small{$\frac{1}{6}$}}}

\def\barr{\left(\begin{array}{c}}
\def\earr{\end{array}\right)}
\def\bmat{\left(\begin{array}{cc}}
\def\emat{\end{array}\right)}
\def\al{\alpha}
\def\be{\beta}
\def\ga{\gamma} \def\Ga{{\it\Gamma}}
\def\de{\delta} \def\De{\Delta}
\def\veps{\varepsilon}  \def\eps{\epsilon}

\def\la{\lambda} \def\La{{\it\Lambda}}

\def\si{\sigma} 
\def\th{\theta}  
\def\w{\omega} 

\def\vfi{\varphi}

\def\dd{{\rm d}}
\def\pa{\partial}

\def\La{{\it\Lambda}}

\def\pa{\partial}

\def\nn{\nonumber}

\def\XXint#1#2#3{{
\setbox0=\hbox{$#1{#2#3}{\int}$}
\vcenter{\hbox{$#2#3$}}\kern-.5\wd0}}

\newcommand*\xbar[1]{%
  \hbox{%
    \vbox{%
      \hrule height 0.2pt 
      \kern0.5ex
      \hbox{%
        \kern-0.2em
        \ensuremath{#1}%
        \kern-0.05em
      }%
    }%
  }%
}

\makeatletter
\newcommand{\lambdabar}{{\mathchoice
  {\smash@bar\textfont\displaystyle{0.25}{1.2}\lambda}
  {\smash@bar\textfont\textstyle{0.25}{1.2}\lambda}
  {\smash@bar\scriptfont\scriptstyle{0.25}{1.2}\lambda}
  {\smash@bar\scriptscriptfont\scriptscriptstyle{0.25}{1.2}\lambda}
}}
\newcommand{\smash@bar}[4]{%
  \smash{\rlap{\raisebox{-#3\fontdimen5#10}{$\m@th#2\mkern#4mu\mathchar'26$}}}%
}
\makeatother

\def\nn{\nonumber}
\def\dd{\mathrm{d}}
\def\cO{\mathcal{O}}

\def\lag{{\mathcal L}}

\def\xx{\mathscr{x}}

\def\scA{\mathscr{A}}
\def\cA{\mathcal{A}}
\def\scO{\mathscr{O}}
\def\MM{\mathcal{M}}
\def\cA{{\mathcal A}}

\def\cO{\mathcal{O}}

\def\lag{{\mathscr L}}

\def\bS{{\bf S}}

\def\br{\boldsymbol{r}}

\def\bp{\boldsymbol{p}}

\def\bq{\boldsymbol{q}}
\def\bs{\boldsymbol{s}}
\def\bS{\boldsymbol{S}}
\def\bL{\boldsymbol{l}}
\def\bJ{\boldsymbol{j}}

\def\bQ{\boldsymbol{Q}}

\def\bgamma{\boldsymbol{\gamma}}

\def\3d{3-D}

\def\ol#1{\overline{#1}}

\DeclareMathOperator\arccosh{arccosh}
\DeclareMathOperator\arctanh{arctanh}
\DeclareMathOperator\im{Im}
\DeclareMathOperator\re{Re}
\DeclareMathAlphabet{\mathantt}{OML}{antt}{l}{it}
\usepackage{datetime}
\newdateformat{mydate}{\THEDAY. M\"arz \THEYEAR}
\newdateformat{MyDate}{\THEDAY.03.\THEYEAR}

\DeclareOldFontCommand{\rm}{\normalfont\rmfamily}{\mathrm}
\DeclareOldFontCommand{\sf}{\normalfont\sffamily}{\mathsf}
\DeclareOldFontCommand{\tt}{\normalfont\ttfamily}{\mathtt}
\DeclareOldFontCommand{\bf}{\normalfont\bfseries}{\mathbf}
\DeclareOldFontCommand{\it}{\normalfont\itshape}{\mathit}
\DeclareOldFontCommand{\sl}{\normalfont\slshape}{\@nomath\sl}
\DeclareOldFontCommand{\sc}{\normalfont\scshape}{\@nomath\sc}
\DeclareRobustCommand*\cal{\@fontswitch\relax\mathcal}
\DeclareRobustCommand*\mit{\@fontswitch\relax\mathnormal}

\newcommand{\mychapter}[2]{
    \setcounter{chapter}{#1}
    \setcounter{section}{0}
    \chapter*{#2}
    \addcontentsline{toc}{chapter}{#2}
}

\setheadsepline{0.4pt}
\title{Exciting Nucleon in Compton Scattering and Hydrogen-Like Atoms}
\author{Franziska Hagelstein}
\date{\today}

\begin{document}
\includepdf[pages=-]{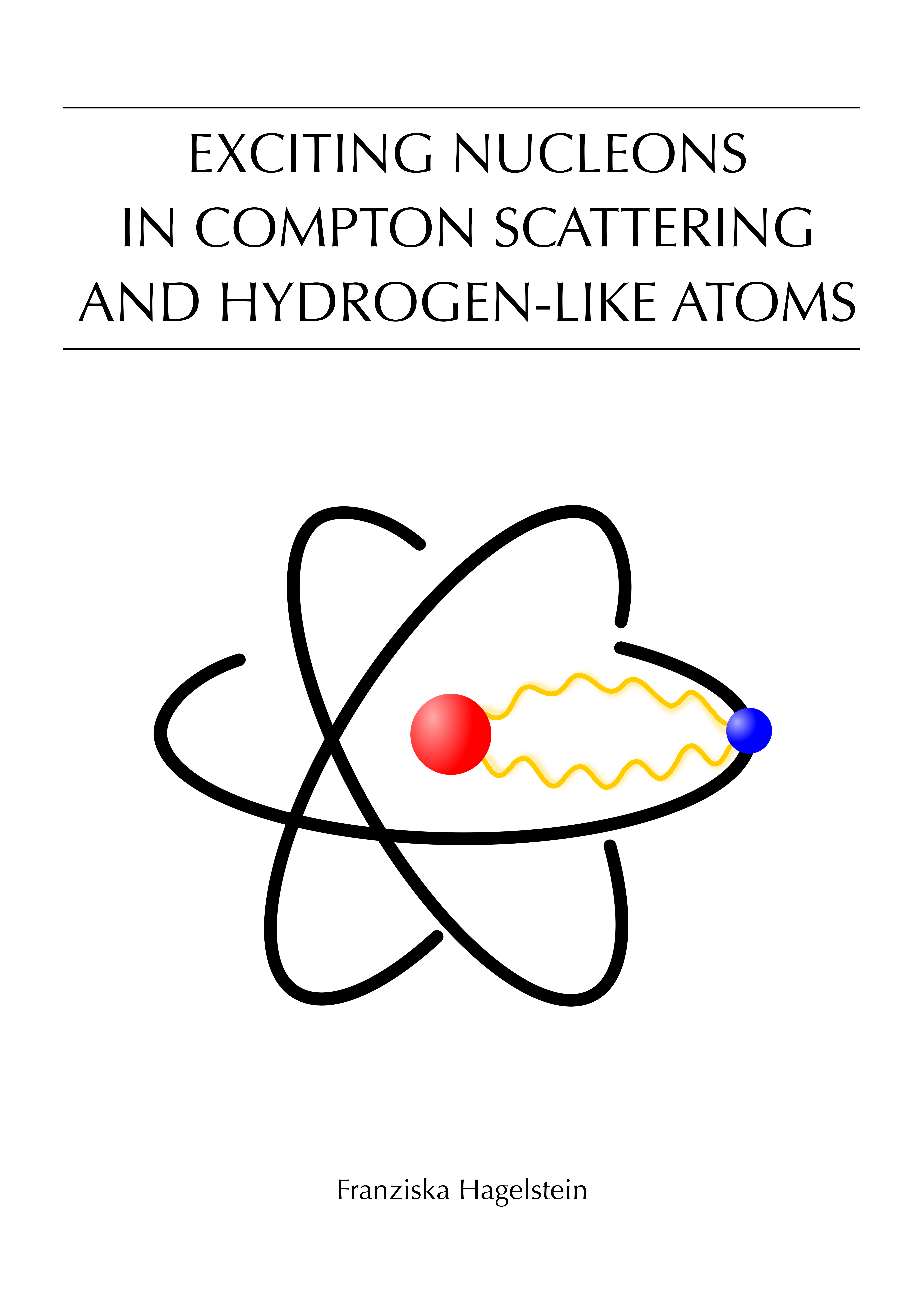} 
\clearpage

\thispagestyle{empty}
\noindent   {
    \Large Datum der mündlichen Prüfung: 25. Juli\ 2017}
 
\vfill
\noindent 
Franziska Hagelstein\\
Institut f\"ur Kernphysik\\
Staudingerweg 7\\
Johannes Gutenberg-Universit\"at Mainz\\
55128 Mainz\\
{\tt \href{mailto:hagelste@uni-mainz.de}{hagelste@uni-mainz.de}}\\[1cm]
 {\bfseries Betreuer (Scientific Advisor):} \\
        Dr.\ Vladimir Pascalutsa \\
  Institut f\"{u}r Kernphysik  \\
  PRISMA Cluster of Excellence\\
  Johannes Gutenberg-Universit\"{a}t Mainz 
 \\
 55128 Mainz \\

\begin{titlepage}
\setlength{\oddsidemargin}{.6cm}
\vspace{30mm}
    \begin{center}
\noindent \makebox[\linewidth]{\rule{16cm}{2pt}}
     \textbf
     {\fontsize{23}{32}\selectfont{\\[0.4cm] EXCITING NUCLEONS\\
      IN      COMPTON SCATTERING \\
     AND 
      HYDROGEN-LIKE ATOMS\\[0.4cm]}}
\noindent \makebox[\linewidth]{\rule{16cm}{2pt}}  
     \\[3cm]
     {\Large Dissertation \\
zur Erlangung des Grades \\
      ``Doktor der Naturwissenschaften''
 \/\\[1.25cm]
        am Fachbereich Physik, Mathematik und Informatik (FB 08) \/\\
        der Johannes Gutenberg-Universit\"at Mainz  \\[4.1cm]}
     {\large vorgelegt von}\\[0.3cm]
     {\LARGE {\bf Franziska Elfriede Hildegard Hagelstein} \\[0.3cm]}
     {\Large geboren in Oberwesel}
     \\[1.75cm]
     {\Large Mainz, März 2017}
   \end{center}
\end{titlepage}
\thispagestyle{empty}

\newpage
\thispagestyle{empty}

\noindent{\LARGE{List of publications}}\\[0.5cm]

\noindent The results presented in this thesis have partially appeared in the following publications.\\[0.1cm]

\noindent {\Large{Articles published in refereed journals}}
\begin{enumerate}
\item F.\ Hagelstein and Vladimir Pascalutsa,\\
{\it Breakdown of the Expansion of Finite-Size Corrections to
                        the Hydrogen Lamb Shift in Moments of Charge
                        Distribution,}\\
Phys.\ Rev.\ {\bf A 91} (2015) 040502(R) [hep-ph/1502.03721].
\item F.\ Hagelstein and Vladimir Pascalutsa,\\
 {\it Reply to ``Comment on `Breakdown of the Expansion of Finite-Size Corrections to
                        the Hydrogen Lamb Shift in Moments of Charge
                        Distribution'",}\\
Phys.\ Rev.\ {\bf A 93} (2016) 026502 [hep-ph/1602.01978].
\item
Oleksii Gryniuk, F.\ Hagelstein and Vladimir Pascalutsa,\\
{\it Evaluation of the Forward Compton Scattering off Protons: I. Spin-Independent Amplitude},\\
Phys.\ Rev.\ {\bf D 92} (2015) 074031 [nucl-th/1508.07952].
\item
Oleksii Gryniuk, F.\ Hagelstein and Vladimir Pascalutsa,\\
{\it Evaluation of the Forward Compton Scattering off Protons: II. Spin-Dependent Amplitude and Observables},\\
Phys.\ Rev.\ {\bf D 94} (2016) 034043 [nucl-th/1604.00789v1].\\[-0.1cm]
\end{enumerate}

\noindent {\Large{Conference proceedings}}

\begin{enumerate}
\addtocounter{enumi}{4}
\item F.\ Hagelstein and Vladimir Pascalutsa, \\
{\it Proton Structure in the Hyperfine Splitting of Muonic Hydrogen},\\
8\textsuperscript{th} International Workshop on Chiral Dynamics, Pisa, June 29 -- July 3, 2015,\\ 
PoS(CD15)077 [nucl-th/1511.04301].
\end{enumerate}
\noindent {\Large{Commissioned review article}}
\begin{enumerate}
\addtocounter{enumi}{5}
\item F.\ Hagelstein, Rory Miskimen and Vladimir Pascalutsa,\\
{\it Nucleon Polarizabilities: from Compton Scattering to Hydrogen Atom,}\\
Prog.\ Part.\ Nucl.\ Phys.\ {\bf 88} (2016) 29-97 [nucl-th/1512.03765].
\end{enumerate}

\newpage
\thispagestyle{empty}
\mbox{ }

\noindent {\LARGE{Abstract}}\\[0.5cm]
This PhD thesis is devoted to the low-energy structure of the nucleon (proton and neutron) as seen through electromagnetic probes, e.g., electron and Compton scattering. The research presented here is based primarily on dispersion theory and 
chiral effective-field theory. The main motivation is the recent \textit{proton radius puzzle}, which is the 
discrepancy between the classic proton charge radius determinations (based on electron-proton scattering and normal hydrogen spectroscopy) and the highly precise extraction based on first muonic-hydrogen experiments by the CREMA
Collaboration.  
The precision of muonic-hydrogen experiments is presently limited by  the knowledge of proton structure effects beyond
the charge radius. A major part of this thesis  is devoted to calculating these effects using everything we know 
about the nucleon electromagnetic structure from both theory and experiment.

The thesis consists of eight chapters. The first and last are, respectively, 
the introduction and conclusion. The remainder of this thesis can roughly be divided into the following three topics:
finite-size effects in hydrogen-like atoms, real and virtual Compton scattering, and two-photon-exchange effects.

The first of these topics is of direct relevance to the proton charge radius extraction from hydrogen and muonic hydrogen. 
We derive the  finite-size effects using a dispersive representation of the proton electromagnetic
form factors. As result, we reveal some limitations in the usual accounting of finite-size effects in terms of 
the expansion in charge and magnetization radii.  We can easily construct a model of nucleon form factors which 
exploits these limitations such as to resolve the proton radius puzzle.

%

The second topic --- Compton scattering --- is important for understanding  the two-photon-exchange
effects. We review the concept of dispersion relations and Compton scattering sum rules, which are based on the general principles of unitarity, causality and analyticity. A new set of sum rules for the elastic-channel contribution to the quasi-static polarizabilities is derived and verified within quantum electrodynamics. We also perform the next-to-next-to-leading order 
calculation of Compton scattering using the SU(2) baryon chiral perturbation theory with $\Delta(1232)$-isobar degrees of freedom.

In the last topic, we use the doubly-virtual Compton scattering off the nucleus to evaluate the two-photon-exchange effects in lepton-nucleus bound states. We focus on the leading and subleading, i.e., order $(Z\al)^5$ and $(Z\al)^6$, polarizability contributions to the spectra of muonic hydrogen, deuterium  and helium. We present the next-to-leading order baryon chiral perturbation theory prediction for the proton-polarizability effect in the Lamb shift and hyperfine splitting of muonic hydrogen
 and a first model-independent prediction of the neutron-polarizability effect in light muonic atoms. Motivated by the large-$N_c$ limit of quantum chromodynamics, we consider the effect of the $\Delta(1232)$-excitation in the hyperfine splitting of muonic hydrogen. We study the neutral-pion exchange and an equivalent to the Coulomb-distortion contribution, both belonging to the class of off-forward two-photon-exchange effects. To allow for a detailed comparison with empirical information, we expand the contribution of non-Born two-photon exchange to the hyperfine splitting in terms of individual spin polarizabilities. 
We conclude by evaluating the impact of our model-independent predictions of polarizability effects on the extractions of proton charge and Zemach radii from muonic-hydrogen spectroscopy.

\newpage
\thispagestyle{empty}
\mbox{}
\newpage
\thispagestyle{empty}

\noindent {\LARGE{Zusammenfassung}}\\[0.5cm]
{\small Die vorliegende Doktorarbeit beschäftigt sich mit der Niederenergiestruktur von Nukleonen (Protonen und Neutronen), wie sie durch elektromagnetische Sonden, beispielsweise Elektronen- und Comptonstreuung, beobachtet wird. Die hier präsentierte Forschung  beruht hauptsächlich auf Dispersionstheorie und chiraler effektiver Feldtheorie. Als Hauptmotivation dient das aktuelle \textit{Protonenradiusproblem}, d.h.\ die Diskrepanz zwischen den klassischen Ergebnissen (der Elektron-Proton-Streuung oder der Spektroskopie von normalem Wasserstoff) für den Protonenradius und der hochpräzisen Bestimmung durch die Experimente der CREMA Kollaboration an myonischem Wasserstoff. Die Genauigkeit der Experimente mit myonischem Wasserstoff ist derzeit durch das Wissen über Effekte der Protonenstruktur beschränkt, welche über den Ladungsradius hinausgehen. Ein gro\ss er Teil dieser Arbeit widmet sich der Berechnung ebendieser Effekte, wobei auf alles zurückgegriffen wird, was über die elektromagnetische Struktur der Nukleonen aus Theorie und Experiment bekannt ist.

Die Arbeit besteht aus acht Kapiteln. Im ersten und letzten Kapitel findet sich eine Einführung bzw.\ Zusammenfassung. Der restliche Teil der Arbeit kann in die folgenden drei Themenbereiche unterteilt werden: Effekte der endlichen Kernausdehnung in wasserstoffähnlichen Atomen, reelle und virtuelle Comptonstreuung, und Zwei-Photonen-Austausch-Effekte.

Ersterer Themenbereich ist von direkter Relevanz für die Bestimmung des Ladungsradius des Protons anhand von Wasserstoff und myonischem Wasserstoff. Wir leiten die Effekte der endlichen Kernausdehnung mithilfe einer dispersiven Darstellung der elektromagnetischen Formfaktoren des Protons her. Im Ergebnis offenbart sich eine Limitierung in der üblichen Beschreibung der Effekte der endlichen Kernausdehnung in Form von Ladungs- und Magnetisierungsradien. Ein Modell für die Formfaktoren des Nukleons, welches diese Beschränkungen ausnutzt um das Protonenradiusproblem zu lösen, lässt sich leicht konstruieren.

Das zweite Thema --- die Comptonstreuung --- ist wichtig für das Verständnis der Zwei-Photonen-Austausch-Effekte.
Wir wiederholen das Konzept der Dispersionsrelationen und die Summenregeln für Comptonstreuung, welche aus den grundlegenden Prinzipien der Unitarität, Kausalität und Analytizität hergeleitet werden. 
Ein neuer Satz von Summenregeln für den Beitrag des elastischen Kanals zu den quasi-elastischen Polarisierbarkeiten wird hergeleitet und innerhalb der Quantenelektrodynamik überprüft. Wir betrachten Beiträge zur Comptonstreuung bis einschlie\ss lich der übernächstführenden Ordnung  in der  SU(2) baryonischen chiralen Störungstheorie mit dem $\Delta(1232)$-Isobar als zusätzlichem Freiheitsgrad.

Im letzten Themengebiet verwenden wir die doppelt virtuelle Comptonstreuung an Kernen um die Zwei-Photonen-Austausch-Effekte in gebundenen Zuständen aus einem Lepton und einem Nukleus zu berechnen. Wir konzentrieren uns auf die führenden und nächstführenden Polarisierbarkeitsbeiträge der Ordnungen $(Z\al)^5$ und $(Z\al)^6$ zu den Spektren des myonischen Wasserstoffs, Deuteriums und Heliums. Wir präsentieren die Vorhersage der chiralen Störungstheorie in übernächster Ordnung für den Effekt der Polarisierbarkeit des Protons auf die Lamb-Verschiebung und die Hyperfeinstruktur-Aufspaltung in myonischem Wasserstoff, sowie eine erste modellunabhängige Vorhersage für den Effekt der Polarisierbarkeit des Neutrons in leichten myonischen Atomen. Inspiriert durch den $N_c$-Limes der Quantenchromodynamik betrachten wir den Effekt der $\Delta(1232)$-Resonanz auf die Hyperfeinstruktur in myonischem Wasserstoff. Wir studieren den Austausch des neutralen Pions und ein Äquivalent zur Coulomb Deformation, welche beide zur Klasse der Zwei-Photonen-Austausch-Prozesse abseits der Vorwärtsrichtung gehören. Wir entwickeln den Beitrag der Nicht-Born-Diagramme des Zwei-Photonen-Austausch zur Hyperfeinstruktur in Spinpolarisierbarkeiten, um einen detaillierten Vergleich mit empirischen Informationen zu ermöglichen.
Zum Abschluss berechnen wir den Einfluss unserer modellunabhängigen Vorhersagen für die Polarisierbarkeitsbeiträge auf die Bestimmung des Ladungs- und des Zemachradius des Protons anhand der Spektroskopie von myonischem Wasserstoff. 
}

\newpage
\thispagestyle{empty}
\cfoot[]{} 
\ofoot[]{} 
\ifoot[]{} 
\tableofcontents
\newpage
\pagenumbering{arabic}
\cfoot[]{} 
\ifoot[]{} 
\ofoot[ \thepage]{\thepage} 
\ohead[]{}
\chead[]{}
\automark[section]{chapter}
\ohead[]{\headmark}
\newpage 
\thispagestyle{empty} 
\hspace{1cm} 
\newpage
\pagenumbering{arabic}

\chapter{Introduction and Motivation} \chaplab{chap1}


Everyone knows it's simple things that matter. One may even say that some of them --- viz.\ elementary particles --- make up the matter. It is interesting to realize that $98\,\%$ of the nucleon mass (and hence, of the visible matter around us) comes from the strong interaction and not from the mass of its constituents (quarks), or equivalently, from the Higgs mechanism of mass generation. Therefore, the origin of the nucleon mass and, more generally, of low-energy nucleon structure is an important physics problem, which thus far has not been solved exactly based on the fundamental theory of the strong interaction --- quantum chromodynamics (QCD).

This thesis, roughly speaking, is an attempt to comprehend the nucleon structure effects in hydrogen-like atoms, of which muonic hydrogen ($\mu$H) is a beautiful example. The main motivation for this is the so-called
 {\it proton charge radius puzzle}. Before describing the puzzle and how the presented work contributes to it (\secref{chap1}{PRP}), we will give a brief introduction into the nucleon structure (\secref{chap1}{1.2}) and how it can be studied using atomic spectroscopy (\secref{chap1}{1.3}). 

\section{Nucleon Structure}  \seclab{1.2}

A traditional probe of nuclear and nucleon structure is the electron  scattering, shown schematically in Figure \ref{fig:ep}. Pioneered by  Robert Hofstadter in the 1950's \cite{Hofstadter:1955ae,Hofstadter:1956qs, Hofstadter:1957wk}, 
this method won him the 1961 Nobel Prize in Physics (``for his consequent discoveries concerning the structure of nucleons''). 
With electron scattering one is able to see objects whose spatial extent is comparable to the reduced de Broglie wavelength, $\lambdabar=\hbar/p$, of the incident electron. Increasing the electron beam energy thus allows to look deeper and deeper into the matter. An electron beam with momentum of $100\,\text{MeV}/c$ probes the matter at the extent of about $2 \,\text{fm}$. This is the typical momentum at which one begins to resolve the individual nucleons in a nucleus.

\begin{figure}[th]
\centering
       \includegraphics[width=8cm]{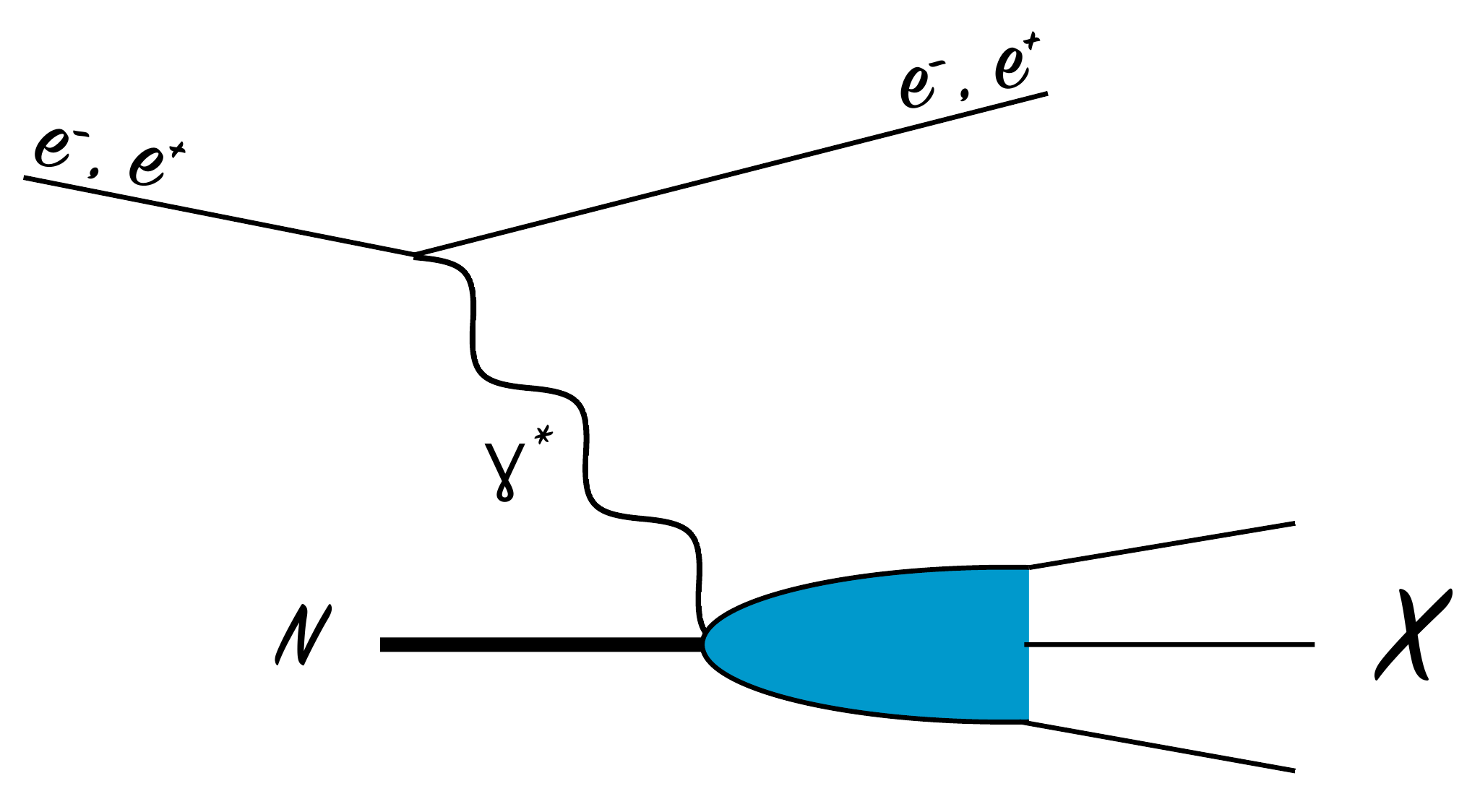}
\caption{Electron scattering.\label{fig:ep}}
\end{figure}


To resolve the constituents of the nucleons --- quarks and gluons --- one needs a beam of at least several $\,\text{GeV}/c$. The Mainz Microtron (MAMI) and the CEBAF at Jefferson Lab are prominent examples of present-day facilities
operating electron beams suited for studies of the nucleon structure. In this thesis, we will often deal with results obtained in these labs. 

Of course, the first indications that the nucleon is not elementary but has a substructure came before
the electron scattering experiments. In 1933, Otto Stern discovered an anomalously large magnetic moment of the proton \cite{Frisch1933,Estermann1933}, winning the
Nobel Prize of 1943. The proton magnetic dipole moment is about 2.79 $e/2M$ rather than simply $e/2M$ as predicted by the Dirac theory of a spin-1/2 particle, where $e$ is the charge and $M$ is the mass of the particle. The electron fit very well in the
Dirac theory prediction, the proton did not.  Today we know that the large anomalous magnetic moment of the proton, $\kappa \approx 1.79$, is qualitatively explained as the sum of magnetic moments of the constituent quarks in the naive quark model.


A more obvious observation of nucleon structure came in the early 1950's. Fermi and collaborators discovered the first nucleon excitation --- the $\Delta(1232)$-resonance ---
using pion beams \cite{Anderson:1952nw}. Shortly after came the era of electron scattering. 
The very first experiments of Hofstadter showed that the nucleon has a finite size, and hence, provided a direct proof of its
compositeness.  

The nature of the nucleon constituents was fully disclosed in the deep inelastic scattering (DIS) experiments. The first of their kind were performed at SLAC \cite{bloom1969,Breidenbach1969}. Later, experiments at higher energies were built at CERN, Fermilab (FNAL) and HERA (DESY).
The electron scattering maps out the nucleon structure functions as functions of the photon virtuality $Q^2$ and the photon energy
$\nu$. More often than not, the variable $\nu$ is traded for the dimensionless Bjorken variable, $x=Q^2/2M\nu$, which in the naive parton picture is the momentum fraction carried by a parton~\cite{Feynman:1969wa,Bjorken:1969ja}, cf.\ \secref{chap4}{SR}. The weak dependence of the nucleon structure function $f_2(x,Q^2)$ on $Q^2$ for fixed $x$ indicates the scattering off point-like constituents. The experimental verification of the Callan-Gross relation \cite{Callan:1969uq}, $2xf_1(x)=f_2(x)$, showed that the constituents are spin-$1/2$ particles. Since protons and neutrons have spin-$1/2$, one understood that the nucleons are formed by three valence quarks. The observed charge range of nucleon resonances ($\Delta^{++},\Delta^+,\Delta^0,\Delta^-$) implied that quarks have fractional charges ($\pm e/3$ or $\pm 2e/3$). Besides the valence quarks, DIS uncovered the sea quarks which, however, carry only a small momentum fraction. Altogether, the quarks carry only about a half of the nucleon momentum. The missing momentum was shown to be carried by the gluons \cite{Han:1965pf}, which are the gauge bosons of QCD.
More about the history of scattering experiments can be found in Ref.~\cite{Povh:1993jx}, whereas we now turn to the
present situation.

The main purpose of observing the elastic electron scattering is to measure the e.m.\ nucleon form factors (FFs), which can 
vaguely be 
interpreted as the Fourier transforms of the nucleon charge and magnetisation distributions in the Breit frame. The most precise data set for the electric and magnetic Sachs FFs \cite{Ernst:1960zza}, $G_E(Q^2)$ and $G_M(Q^2)$, was obtained in an outstanding measurement at the MAMI facility \cite{Bernauer:2010} in 2010.
Figure \ref{fig:ep} depicts an electron scattering process in the leading one-photon approximation, where obviously the target is probed by the exchanged virtual photon. The Rosenbluth formula \cite{Rosenbluth:1950}, cf.\ \Eqref{Rosenbluth}, allows to extract the Sachs FFs at fixed values of $Q^2$ by measuring the cross section for different scattering angles and accordingly different incident beam energies. One of the first nucleon FF measurements based on the (one-photon exchange) Rosenbluth separation was performed at Stanford \cite{Hughes:1965}.  Later, the polarization-transfer technique was proposed \cite{Dombey:1969wk,Dombey:1969wi,Akhiezer:1968ek,Akhiezer:1974em,Arnold:1980zj} and experimentally realized at CEBAF \cite{Jones:1999rz,Gayou:2001qd,Punjabi:2005wq,Puckett:2010ac}.

In the polarization-transfer experiment, the longitudinally polarized electrons are scattering off an unpolarized target and  polarize it in the process. One then measures the ratio of transverse and longitudinal polarizations of, e.g., the recoil protons in the elastic electron-proton ($ep$) scattering, which is directly proportional to the ratio of $G_{E}(Q^2)/G_{M}(Q^2)$. However, as it turns out, there is a discrepancy between the $Q^2$ behavior of the FF ratios extracted from unpolarized Rosenbluth and polarization-transfer experiments. The ratio from polarization transfer is decreasing linearly with $Q^2$, whereas the Rosenbluth ratio is roughly constant. The general belief is that the discrepancy can be explained by two-photon-exchange (TPE) effects, which are not considered in the classic Rosenbluth separation.\footnote{For reviews on the subject we refer to \citet{HydeWright:2004gh} (e.m.\ nucleon FFs), \citet{Carlson:2007aa} (TPE physics in hadronic processes), and \citet{Arrington2011782} (TPE in $ep$ scattering).} In recent years, several collaborations measured the ratio of positron-nucleon to electron-nucleon cross sections, $R_{2\gamma}$, as an indicator for TPE or multi-photon-exchange processes. Experiments at CLAS (JLab) \cite{Moteabbed:2013isu,Adikaram:2014ykv} and the \mbox{VEPP-3} storage ring (Novosibirsk) \cite{Rachek:2014fam} find evidence for a significant TPE contribution, explaining the discrepancy up to $2-3\, \text{GeV}^2/c^2$. Similarly, the OLYMPUS (DESY) experiment \cite{Henderson:2016dea} suggests that TPE is causing most of the discrepancy at low $Q^2$. However, the experimental $R_{2\gamma}$ is generally smaller than expected from theoretical calculations of hard TPE effects, thus, more measurements are needed at large $Q^2$.

We will see that the TPE is not only highly relevant  for the e.m.\ FF measurements. It also plays an important role in the physics of atomic bound states. The dominant uncertainty in the theoretical description of the $\mu$H spectrum, in both the Lamb shift (LS) and the hyperfine splitting (HFS), is given by elastic and inelastic TPE corrections, which we aim to improve upon in Chapters \ref{chap:5LS} and \ref{chap:5HFS}. The long-range force between electrically neutral atoms and molecules, referred to as dispersion force\footnote{See Ref.~\cite{Feinberg:1989ps} for a review on dispersion forces arising from two-photon, two-neutrino, and two-meson exchanges between charged and neutral systems.} \cite{London1930,Casmir:1947hx}, is induced by TPE. It outweighs the one-photon-exchange (OPE) interaction, which cancels almost completely among the positive and negative sub-charges of the systems.

\begin{figure}[t]
\centering
       \includegraphics[width=7.5cm]{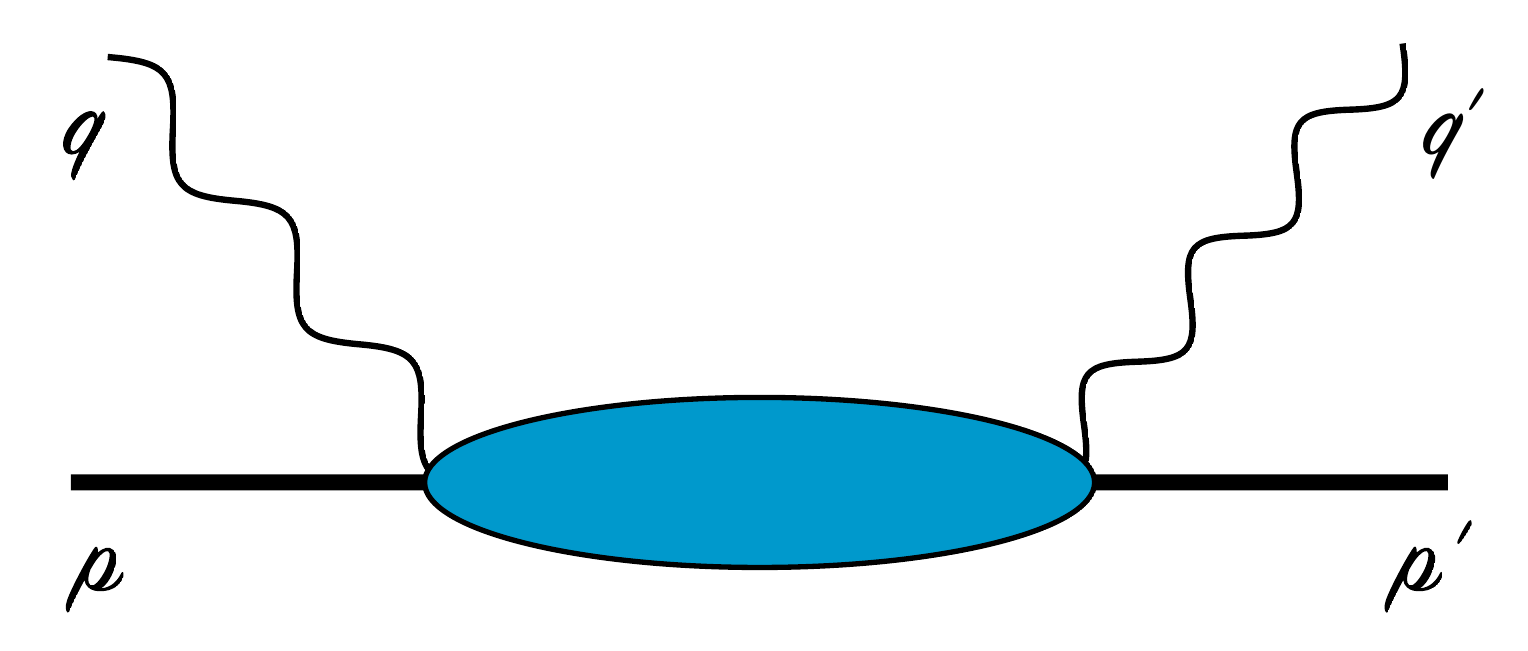}
\caption{Compton scattering.\label{fig:CS}}
\end{figure}

The electron-nucleon ($eN$) scattering is deeply connected to the process of Compton scattering (CS) off the nucleon, cf.\ \secref{chap3}{sec3.1}. CS is an elastic scattering of a photon by a target, see \Figref{CS}. Various facilities have experimental programs dedicated to real and virtual CS: the Lebedev
Institute (Moscow) \cite{Baranov:1974ec}, MUSL (Illinois) \cite{Fed91}, SAL~\cite{Hallin:1993ft}, LEGS (BNL) \cite{Leg01},  MIT-Bates \cite{Bourgeois2011}, MAX-Lab (Lund) \cite{Myers14}, MAMI \cite{Roche:2000,Olm01,Galler01,Wolf01,Cam02,Janssens:2008qe} and 
JLab \cite{Laveissiere:2004}. Besides the previously described FF measurements, which also carry information on the magnetic and electric radii as explained in \secref{chap2}{Radii}, CS gives access to the nucleon polarizabilities, which describe the nucleon response to e.m.\ fields, cf.\ \secref{chap3}{polarizabilities}.
The static scalar and spin polarizabilities are extracted from real Compton scattering (RCS). Virtual Compton scattering (VCS) and forward doubly-virtual Compton scattering (VVCS) are described by different sets of generalized polarizabilities (GPs).\footnote{For reviews on polarizabilities see, e.g.,
\citet{Guichon:1998xv} (VCS and generalized polarizabilities), \citet{Phillips:2009af} (neutron polarizabilities), \citet{Holstein:2013kia} (pion, kaon, nucleon polarizabilities) and \citet{Hagelstein:2015egb} (nucleon polarizabilities).} In this thesis, we will study RCS in \chapref{chap3} and VVCS in \chapref{chap4}.

Chiral perturbation theory (ChPT) is a low-energy effective field theory of QCD, based on hadronic fields instead of quark and gluon fields \cite{Weinberg:1978kz,Gasser:1983yg,Gasser:1987rb}. As there are no free low-energy constants, the contribution to the nucleon polarizabilities at leading order (LO) in the pion momentum, $\mathcal{O}(p^3)$, comes out as a pure prediction of ChPT; and ChPT has proven to be quite successful in reproducing the nucleon polarizabilities at low $Q$ \cite{Lensky:2008re,Len10,Lensky:2014dda,Lensky:2015awa}.\footnote{See Ref.~\cite{Geng:2013xn} for a review of recent developments in ChPT.} However, there is a problem with the longitudinal-transverse polarizabilities --- the \textit{$\delta_{LT}$ puzzle}. 
There are two different power-counting schemes, the $\delta$-  \cite{Pascalutsa:2003aa} and the $\epsilon$-expansion \cite{Hemmert:1996xg}, commonly used upon inclusion of $\Delta(1232)$ degrees of freedom (DOFs) into the ChPT framework, see \secref{chap4}{DeltaChPT}. The $\epsilon$-counting gives a prediction for the longitudinal-transverse polarizability of the proton \cite{Bernard:2012hb}, $\delta_{LT}^{(p)}$, which is in significant contradiction to the empirical information collected in the MAID isobar model, see \Figref{deltaLT}. In \chapref{chap4}, we will calculate the nucleon polarizabilities at next-to-leading order (NLO) in baryon chiral perturbation theory (BChPT) and try to clear up the $\delta_{LT}$ puzzle.

\begin{figure}[t] 
  \centering 
\begin{minipage}[t]{0.48\textwidth}
    \centering        \includegraphics[width=\textwidth]{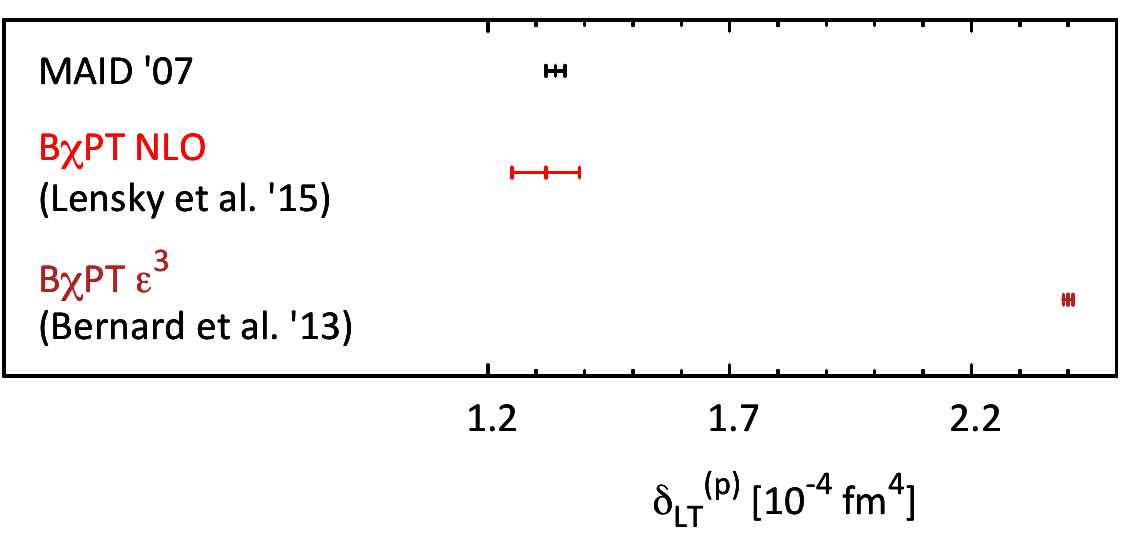}
\end{minipage}
\begin{minipage}[t]{0.50\textwidth}
    \centering 
       \raisebox{0.0cm}{ \includegraphics[width=\textwidth]{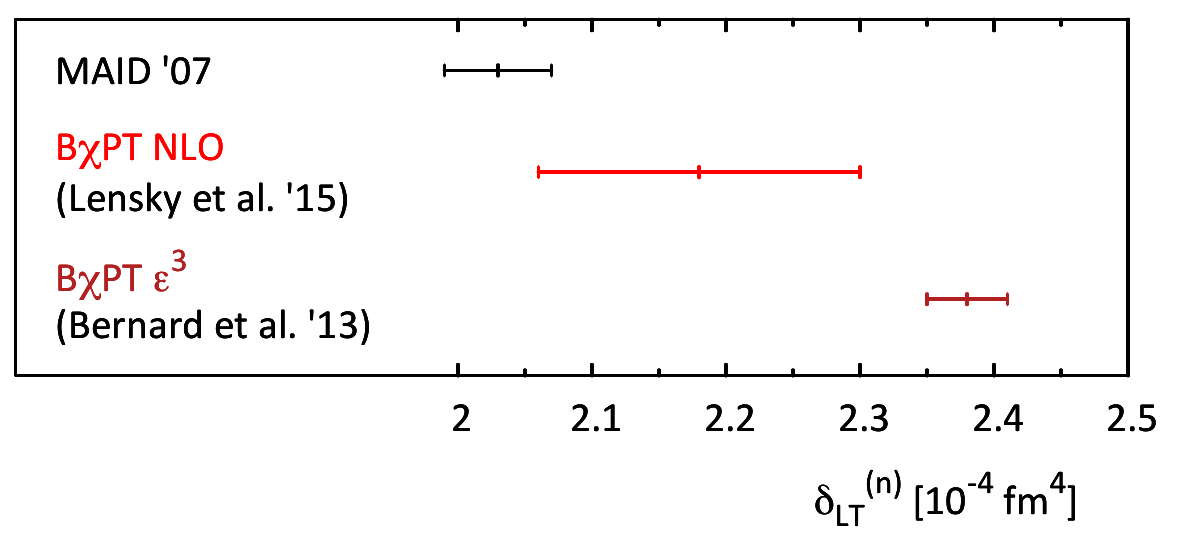}}
       \end{minipage}
 \caption{Longitudinal-transverse polarizability, $\delta_{LT}$, for the proton (left panel) and neutron (right panel), respectively. We show the baryon chiral perturbation theory predictions of Lensky et al.~\cite{Lensky:2014dda,Alarcon2017} and \citet{Bernard:2012hb}, and a result from MAID \cite{MAID}. \label{fig:deltaLT}}
\end{figure}

\section{Atomic Spectroscopy} \seclab{1.3}

Hydrogen spectroscopy has an eventful history and played a crucial role in the development of quantum physics. The experimental gain in the resolution of the hydrogen spectrum triggered further progress in the theoretical understanding and allowed for the establishment of quantum electrodynamics (QED). The non-relativistic Schr\"odinger equation allows for a description of transitions between energy levels in the hydrogen atom with different principal quantum numbers, e.g., the Balmer series. However, only the relativistic Dirac theory can describe the fine structure (FS), which is partially generated by the spin-orbit coupling. Explaining the HFS, on the other hand, requires the spin-spin coupling. The LS is induced mainly by QED loop corrections. A description of the (muonic) hydrogen spectrum and a more complete review of the interplay between theory and experiment in the history of hydrogen spectroscopy, including references, can be found in \secref{chap2}{HydrogenTheory}.

\begin{table}[tbh]
\caption{Exotic atoms and their experimental realizations. \label{exoticatoms}}
\begin{small}
\begin{tabular}{|p{2.75cm}|cc|p{5.5cm}|}
\hline
\rowcolor[gray]{.7}
\bf{simple atom}&\bf{``light compound''}&\bf{``heavy compound''}&\bf{(first) exp.\ formation/observation}\\
muonic atoms&$\mu^-$&H, D, $^3\text{He}^+$, $^4\text{He}^+$&CREMA collaboration \cite{Pohl:2010zza,Pohl1:2016xoo}\\
\rowcolor[gray]{.95}
muonium&\multicolumn{2}{c|}{$e^-$ and $\mu^+$}& \citet{Hughes:1960zz} (1960)\\
true muonium/&\multicolumn{2}{c|}{\multirow{2}{*}{$\mu^-$ and $\mu^+$}}&theoretically possible to generate\\
muononium&&& at modern $e^+e^-$ colliders \cite{Brodsky:2009gx}\\
\rowcolor[gray]{.95}
positronium&\multicolumn{2}{c|}{$e^-$ and $e^+$}&\citet{Deutsch:1951aa} (1951)\\
protonium/&\multicolumn{2}{c|}{\multirow{3}{*}{$\bar{p}$ and $p$}}&Daresbury-Mainz-TRIUMF\\
anti-protonic &&&collaboration \cite{Auld:1978hd} (1978),\\
hydrogen&&&ATHENA collaboration \cite{Zurlo:2006rk} (2006)\\
\rowcolor[gray]{.95}
anti-hydrogen&$e^+$&$\bar{p}$&PS210 Coll.\ (LEAR) \cite{Baur:1995ck} (1995)\\
pionium&\multicolumn{2}{c|}{$e^-$ and $\pi^+$}&\citet{Mundiger1989} (1989)\\
\rowcolor[gray]{.95}
true pionium&\multicolumn{2}{c|}{$\pi^-$ and $\pi^+$}&\citet{Flik:1986zz} (1986)\\
pionic hydrogen&$\pi^-$&H&\citet{Bailey:1970wr} (1970)\\
\rowcolor[gray]{.95}
&&&\citet{Davies:1979aj} (1979) and \\
\rowcolor[gray]{.95}
kaonic hydrogen&$K^-$&H& \citet{Bird:1983yb} (1983),\\
\rowcolor[gray]{.95}
&&&
KEK proton synchrotron \cite{Iwasaki:1997aa} (1997)\\
\hline
\end{tabular}
\end{small}
\end{table}

Exotic atoms are unique laboratories to study nuclear properties, perform stringent QED tests \cite{Karshenboim:1900zz,Karshenboim:2006ht} or determine fundamental constants.
The history of (simple) exotic bound states, such as muonic or pionic atoms (or ions), dates back to the first half of the last century. The first experimentally observed exotic atom in a long list of exotic atoms, see Table \ref{exoticatoms}, was positronium in 1951 \cite{Deutsch:1951aa}. The first mention of a bound positron-electron system already dates back to 1934  \cite{Mohorovicic1934}, and the name ``positronium'' was mentioned in writing for the first time in 1945 \cite{Ruark:1945aa}.

In muonic atoms or ions, one or more valence electrons are kicked out and replaced by one muon \cite{Bertin:1975qh,Ruckstuhl:1985xg,Pohl:2010zza,Pohl1:2016xoo}. The CREMA collaboration, for example, uses the $\pi$E5 beam-line of the proton accelerator at PSI to generate low-energy muons which are stopped in low-density gases to form $\mu$H, $\mu$D, $\mu^3$He$^+$ and $\mu^4$He$^+$. As the muon is about $200$ times heavier than the electron, its Bohr orbits are about $200$ times closer to the nucleus. Therefore, the hydrogen-like muonic atoms are used for having a closer look at the nucleus. The proton charge radius extraction from the $\mu$H LS demonstrates this point by an order of magnitude improvement in precision, see \secref{chap1}{PRP}.

QED tests with ordinary atoms are hindered by the nuclear structure effects. Purely leptonic atoms, on the other hand, are build from point-like components, are not subject to strong interactions and, hence, have no such limitation. Positronium and muonium are the perfect tools to verify the theory of bound-state QED \cite{Karshenboim:2003vs, Karshenboim:2006ht} and probe New Physics beyond the Standard Model \cite{Rubbia:2004ix}. The fine-structure constant $\alpha$ can, f.i., be extracted from the positronium decay rate. The upcoming re-measurement of the ground-state HFS in muonium by the MuSEUM collaboration (J-PARC) \cite{Strasser:2016smg} will improve the determinations of the muon anomaly $(g-2)_\mu$ and its mass.\footnote{See Ref.~\cite{Jungmann:2016gak} for a review on muonium spectroscopy.} Formation of true muonium or tauonium at $e^+e^-$ colliders would provide the heaviest and most compact pure QED systems \cite{Brodsky:2009gx}. 

In hadronic atoms --- pionic or kaonic atoms --- the nucleus binds a pion  \cite{Camac:1952zz,Stearns:1954aa} or kaon \cite{Burleson:1965aa}, respectively. Hadronic atoms are used for precision studies of the pion-nucleon and kaon-nucleon interactions, viz.\  scattering lengths, see Ref.\ \cite{Gasser:2007zt} for a review. Another example of exotic atoms are anti-protonic atoms \cite{Bamberger:1970nc}, such as protonium \cite{Auld:1978hd,Zurlo:2006rk} and anti-hydrogen \cite{Baur:1995ck}, see Ref.\ \cite{Backenstoss:1989id} for a review. The ongoing upgrade of experimental facilities at CERN  will allow for a first determination of the anti-proton charge radius from the LS in anti-hydrogen \cite{Crivelli:2016gog}. The $1S-2S$ transition in anti-hydrogen has been recently measured for the first time  by the ALPHA collaboration and was found to be in agreement with ordinary hydrogen \cite{Ahmadi:2016fir}.

\section{The Charge Radius Puzzle}\seclab{PRP}
The present $5.6$ standard deviations ($5.6\,\sigma$) discrepancy between the proton root-mean-square (rms) charge radius found by electron probes \cite{Mohr:2015ccw}, viz.\ $ep$ scattering and H spectroscopy,
and the $\mu$H experiment \cite{Antognini:1900ns} has attracted a lot of attention in the physics community \cite{Pohl:2013yb,Carlson:2015jba,Karshenboim:2015,Hagelstein:2015egb} and outside \cite{Bernauer:2014cwa}. The results of experiments with electrons are collected in the CODATA~'14 review \cite[Table XXIX and Eq.\ (74)]{Mohr:2015ccw}:
\begin{subequations}
\eqlab{CODATAradii}
\bea
R_{Ep}(\text{H})&=&0.8764(89)\,\mathrm{fm},\\
R_{Ep}(ep)&=&0.879(11)\,\mathrm{fm},
\eea
where a recommended charge radius is given:
\beq
R_{Ep}(\text{CODATA '14})=0.8751(61)\,\mathrm{fm}.\eqlab{REpCodata}
\eeq
\end{subequations}
As one can see from \Figref{RE}, the proton radius recommended by the CODATA task group has not changed much since the 2002 adjustment \cite{CODATA2005,Mohr:2008fa,Mohr:2012aa,Mohr:2015ccw}.\footnote{During the last analyses it was mainly the error estimate of the CODATA recommended charge radius that varied. Therefore, based on the previous CODATA '10 proton radius \cite{Mohr:2012aa}, the comparison with the $\mu$H LS measurement \cite{Antognini:1900ns} temporarily resulted in a $7.2\, \sigma$ discrepancy.} In 2010, the proton radius was extracted from the $\mu$H spectrum for the first time by the CREMA collaboration \cite{Pohl:2010zza}.\footnote{The first result being: $R_{Ep} (\mu\mathrm{H})  =  0.84184(66)$ fm \cite{Pohl:2010zza}.} In the first run, the $2P_{3/2}(f=2)-2S_{1/2}(f=1)$ transition from the $2S$-triplet state was measured, see \Figref{MuHSpec}, and had to be supplemented by theory input for the $2S$ HFS\footnote{$E_{\mathrm{HFS}}(2S)=22.8148(78) \,\mathrm{meV}$ \cite{Martynenko:2004bt} (with $R_\mathrm{Z}=1.022 \,\mathrm{fm}$ \cite{Faustov:2001pn}) was used as theory input for the $2S$ HFS.} to extract the proton radius. In 2013, the transition from the $2S$-singlet state, $2P_{3/2}(f=1)-2S_{1/2}(f=0)$, was measured in addition \cite{Antognini:1900ns}. The two transition frequencies allowed 
for an independent measurement of the classic LS and the $2S$ HFS, which subsequently led to the extraction of the proton rms charge radius and the Zemach radius \cite{Antognini:1900ns,Antognini:2012ofa}:
\begin{subequations}
\label{rmuH}
\bea
R_{Ep} (\mu\mathrm{H})  &=&  0.84087(39)\, \mathrm{fm}, \eqlab{remuH}\\
R_{\mathrm{Z}p} (\mu\mathrm{H})  &=&  1.082(37) \, \mathrm{fm}.
\eea
\end{subequations}
While the Zemach radius agrees with the value calculated from different analytic FF parametrizations \cite{Arrington:2006hm,Arrington:2007ux,Kelly:2004hm,Friedrich:2003iz}, $R_{Zp}=1.049\div1.091$ fm \cite{Carlson:2008ke},  the charge radius is significantly smaller than the CODATA value. The latter discrepancy is referred to as the \textit{proton charge radius puzzle}.

\begin{figure}
\centering
\includegraphics[scale=0.7]{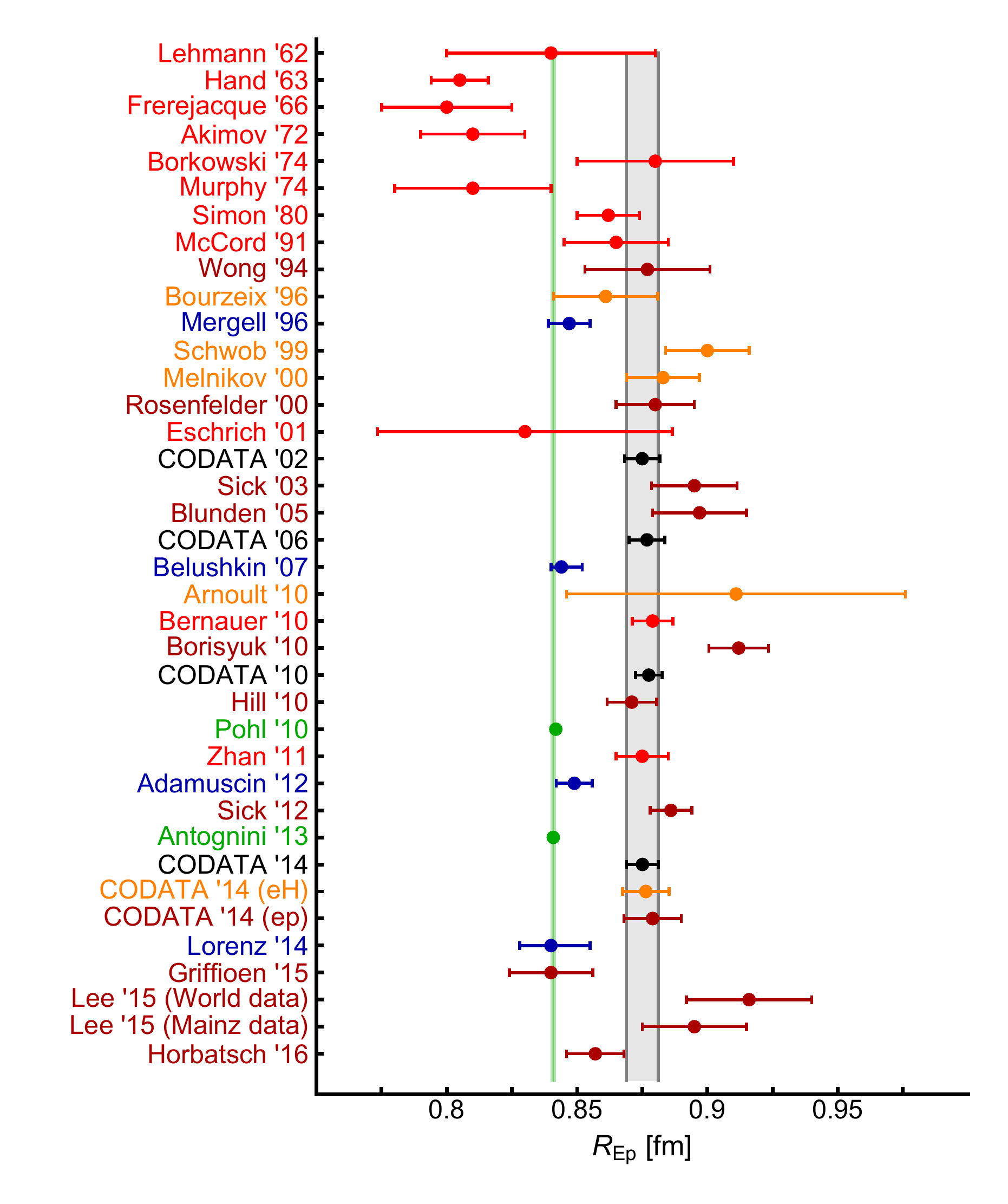}
\caption{Collection of various proton charge radius determinations. A) electron-proton scattering experiments in red: Lehmann '62 \cite{Lehmann:1962dr}, Hand '63 \cite{Hand:1963zz}, Frerejacque '66 \cite{Frerejacque:1965ic}, Akimov '72 \cite{Akimov:1972nu}, Borkowski '74 \cite{Borkowski:1974tm}, Murphy '74 \cite{Murphy:1974zz}, Simon '80 \cite{Simon1980381}, McCord '91 \cite{McCord:1991sd}, Eschrich '01 \cite{GoughEschrich:2001ji}, Bernauer '10 \cite{Bernauer:2010wm}, Zhan '11 \cite{Zhan:2011ji} (recoil polarimetry); B) re-analyses of electron-proton scattering data in dark red: Wong '94 \cite{Wong:1994sy}, Rosenfelder '00 \cite{Rosenfelder:1999cd} (Coulomb corrections), Sick '03 \cite{Sick:2003gm}, Blunden '05 \cite{Blunden:2005jv} (two-photon-exchange corrections), Borisyuk '10 \cite{Borisyuk:2009mg}, Hill '10 \cite{Hill:2010yb} ($z$ expansion), Sick '12 \cite{Sick:2012zz}, Griffioen '15 \cite{Griffioen:2015hta}, Lee '15 \cite{Lee:2015jqa}, Horbatsch '16 \cite{Horbatsch:2016ilr} (fit with chiral perturbation theory input for higher moments); C) electron-proton scattering fits within a dispersive framework in blue: Mergell '96 \cite{Mergell:1995bf}, Belushkin '07 \cite{Belushkin:2006qa}, Adamuscin '12 \cite{Adamuscin:2012zz}, Lorenz '14 \cite{Lorenz:2014yda}; D) hydrogen and deuterium spectroscopy in orange: Bourzeix '96 \cite{Bourzeix:1996zz}, Schwob '99 \cite{Schwob:1999zz}, Melnikov '00 \cite{Melnikov:1999xp}, Arnoult '10 \cite{Arnoult2010}; E) muonic-hydrogen spectroscopy in green: Pohl '10 \cite{Pohl:2010zza}, Antognini '13 \cite{Antognini:1900ns}; F) CODATA recommended charge radii in black: '02 \cite{CODATA2005}, '06 \cite{Mohr:2008fa}, '10 \cite{Mohr:2012aa}, '14 \cite{Mohr:2015ccw}. The green line is the prediction from the latest muonic-hydrogen Lamb shift measurement \cite{Antognini:1900ns} and the gray line is the CODATA '14 recommended charge radius \cite{Mohr:2015ccw}. \label{fig:RE}}
\end{figure}

Figure \ref{fig:RE} shows a collection of proton charge radius determinations in chronological order. The green and gray bands highlight the latest result from $\mu$H \cite{Antognini:1900ns} and the CODATA~'14 recommendation \cite{Mohr:2015ccw}. Historically, the proton radius puzzle seems to be a reoccurring event. \citet{Pachucki:1994ega} already referred to the $4.8 \, \sigma$ discrepancy between the Stanford [$R_{Ep}=0.805(11)$ fm] \cite{Hand:1963zz} and Mainz radii [$R_{Ep}=0.862(12)$ fm] \cite{Simon1980381} as the ``proton radius puzzle''. However, the present situation is more severe. Before the $\mu$H result became available, the later $ep$ experiments (red) and re-analyses (dark red) using standard FF fits all pointed towards a bigger proton radius in accordance with the electronic-hydrogen spectroscopy results (orange), cf.\ \Figref{RE}. Not much attention was payed to the dispersive fits \cite{Mergell:1995bf,Belushkin:2006qa,Adamuscin:2012zz,Lorenz:2014yda} (blue) indicating a smaller proton radius, see Ref.~\cite{Hoferichter:2016duk} for a recent update. After the publication of the first proton radius prediction from $\mu$H, the theory of hydrogen spectra again turned into an active field of studies and the $ep$ scattering data fits were intensely debated \cite{Bernauer:2016ziz,Sick:2017aor}. 

Meanwhile, publications proposing exotic explanations of the puzzle have been piling up \cite{TuckerSmith:2010ra,Batell:2011qq,Barger:2011mt,Carlson:2012pc,Walcher:2012qp,Mart:2013gfa,Wang:2013fma,Onofrio:2013fea,Giannini:2013bra,Karshenboim:2014tka,Carlson:2015poa,Liu:2016qwd}. Of course, a simultaneous solution of the proton radius puzzle and the muon anomalous magnetic moment, $(g-2)_\mu$, discrepancy would be most desirable. However, most of the beyond Standard Model scenarios, including extra dimensions, lepton flavor non-universality, new (dark) forces and new particles, are highly limited by other experiments.


\begin{figure}
\centering
\includegraphics[scale=1.05]{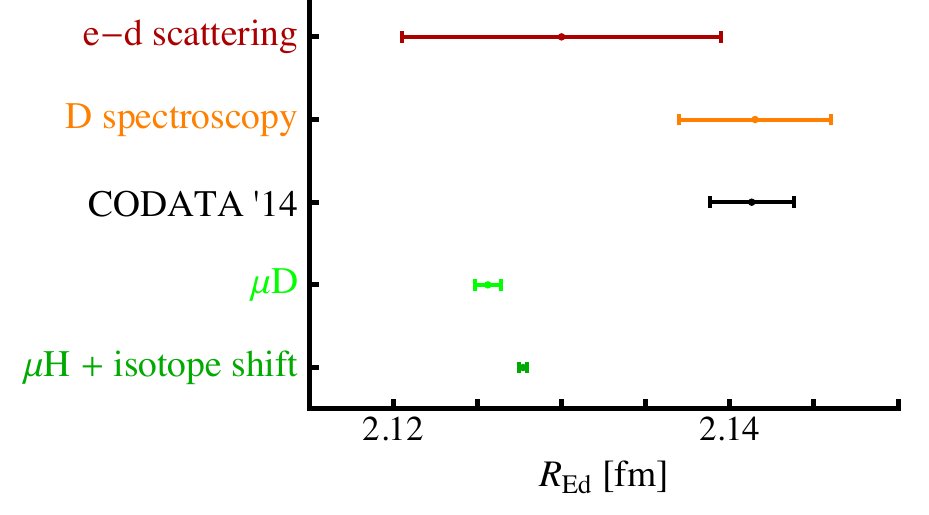}
\caption{Collection of various deuteron charge radius determinations. A) electron-deuteron scattering \cite{Sick:1998cvq}; B) deuterium spectroscopy \cite{Pohl:2016glp}; C) CODATA '14 \cite{Mohr:2015ccw}; D) muonic-deuterium Lamb shift \cite{Pohl1:2016xoo}; E) muonic hydrogen + isotope shift \cite{Antognini:1900ns}.\figlab{REd}}
\end{figure}

With the recently published deuteron charge radius extraction from the $\mu$D LS \cite{Pohl1:2016xoo,Krauth:2015nja}, 
the proton radius puzzle in fact turned into a \textit{$Z=1$ (hydrogen isotope) charge radius puzzle} \cite{Pohl:2016}, see \Figref{REd}:
\begin{subequations}
\bea
R_{Ed} (\mu\mathrm{D})  &=&  2.12562(78) \, \mbox{fm \cite{Pohl1:2016xoo}},\eqlab{rmuD}\\
R_{Ed}(\text{CODATA '14}) &=&  2.1413(25)\, \mbox{fm \cite{Mohr:2015ccw}}.
\eea
The discrepancy between the $\mu\mathrm{D}$ result and the CODATA recommended deuteron rms charge radius amounts to $6.3\,\sigma$. Ref.~\cite{Pohl:2016glp} deduced a deuteron charge radius based on D spectroscopy alone:
\beq
R_{Ed} (\text{D spectroscopy})  =  2.1415(45) \, \mbox{fm \cite{Pohl:2016glp}},\eqlab{Dspec}
\eeq
\end{subequations}
which is in agreement with the CODATA averages \cite{Mohr:2012aa,Mohr:2015ccw} but $3.5\,\sigma$ discrepant with $\mu\mathrm{D}$ \cite{Pohl1:2016xoo}. This disagreement is then uncorrelated with the proton charge radius.

Further LS measurements have been performed in muonic-helium ions, $\mu^3$He$^+$ and $\mu^4$He$^+$ \cite{Nebel:2012qla}, and the results of their analyses are soon to be published \cite{Pohl:2016,Diepold:2016cxv}. All spectroscopy experiments in muonic atoms have one thing in common, they rely on precise theory input to extract nuclear charge or Zemach radii. Comprehensive theory reviews have been put together for $\mu$H \cite{Antognini:2012ofa}, $\mu$D \cite{Krauth:2015nja} and $\mu^4$He$^+$ \cite{Diepold:2016cxv}.\footnote{A summary of the $\mu^3$He$^+$ theory is in preparation \cite{Pohl:2016}.} Comparing the measured HFS or LS transition to the theoretical expectation of the atomic spectrum as a function of nuclear radii, allows to extract these same, see Eqs.\ \eref{muHtheory} and \eref{theorypredictions} for the theoretical descriptions of the $2S$ HFS in $\mu$H and the LSs in $\mu$H, $\mu$D and $\mu^4$He$^+$.

Assuming the electron-probe experiments are correct, and so is the measured $\mu$H LS, we can translate the discrepancy between the theoretical expectation of the classical LS in $\mu$H \cite{Antognini:2012ofa}, based on the CODATA recommended charge radius \cite{Mohr:2015ccw}, and the $\mu$H experiment \cite{Antognini:1900ns} into a missing piece in the $\mu$H-theory budget of about $310\,\upmu$eV. Doing the same for the case of $\mu$D, the discrepancy translates into a missing piece in the theory budget of $409$ meV.

It was suspected that proton structure at order $(Z\al)^5$, i.e., forward TPE, could produce such an effect \cite{DeRujula:2010dp,Miller:2012}. These TPE corrections split into an elastic and a polarizability part, cf.\ \secref{chap5}{TPE}. Today, the dispersive calculations of the forward TPE corrections to the LS  \cite{Pachucki:1999zza,Martynenko:2005rc,Nevado:2007dd,Carlson:2011zd,Hill:2011wy,Birse:2012eb,Gorchtein:2013yga,Peset:2014jxa}, backed up by the BChPT prediction of Ref.~\cite{Alarcon:2013cba}, give an order of magnitude smaller result. Nevertheless, the limiting factor in the accuracy of the theoretical description of the spectra of hydrogen-like muonic atoms remains to be set by TPE effects \cite{Antognini:2012ofa,Krauth:2015nja,Diepold:2016cxv}. A major task of this thesis is to improve the predictions of the forward TPE polarizability corrections to LS and HFS in $\mu$H and calculate the subleading off-forward TPE polarizability effects in light muonic atoms (Chapters \ref{chap:5LS} and \ref{chap:5HFS}). Furthermore, we will put the de R\'ujula \cite{DeRujula:2010dp} scenario for solving the proton radius puzzle on a rigorous basis in \secref{chap2}{Exact}.

The advent of the proton radius puzzle has also triggered many new experimental programs. Equation \eref{CODATAradii} shows that the H spectroscopy data play a dominant role in the CODATA adjustment. While the CODATA average is, due to the reduced (statistical) error, in $5.6\,\sigma$ discrepancy to the $\mu$H result, half of the H spectroscopy measurements agree with the muonic result at the level of $1\,\sigma$. Besides the three LS measurements in H \cite{Newton373,Lundeen86,Hagley:1994zz}, one is often using a combination of two H transitions to extract the proton radius and the Rydberg constant $R_\infty$ simultaneously, cf.\ \Eqref{RydbergE}. The first choice is the $1S-2S$ transition \cite{Parthey:2011lfa}, which is known most precisely, and the second input could be for instance the $2S-4P$ transition.  Looking at the individual measurements, cf.\ Figure 2 of Ref.~\cite{Pohl:2013yb}, only the $2S-8D_{5/2}$ transition \cite{deBeauvoir:1997zz} disagrees with the muonic proton radius by $3\,\sigma$. Furthermore, it is striking that the biggest discrepancies occur in comparing to measurements from the LPTF and LKB Paris groups ($2S-12D$ \cite{Schwob:1999zz} and $2S-8D$ \cite{deBeauvoir:1997zz} transitions). Therefore, on second glance, the discrepancy between H and $\mu$H spectroscopy results is less pronounced, and an overlooked systematic effect in the Paris experiments could relativize the discrepancy. Only more data will be able to clarify the situation and a first promising result for the $2S-4P$ transition was communicated by the MPQ (Garching) \cite{Pohl:2016}. Also, the $n=2$ LS in H will be re-measured \cite{Hessels:2014,Horbatsch:2016aa}. Rydberg states in one-electron ions come with the promising potential for a determination of the Rydberg constant which is independent of the proton radius \cite{Jentschura:2008aa,Jentschura2010,Tan:2014vha}. Experimental efforts using neon ions in a penning trap are underway at NIST \cite{Tan2011,Brewer:2013aa}.\footnote{See Ref.~\cite{Brown:1986aa} for a review on geonium theory, that is the theory of single ions or electrons in a penning trap.} In addition, theory advances \cite{Pachucki16,Puchalski:2016ibj} have made molecular hydrogen ($H_2$) an interesting candidate for QED and proton-size tests \cite{Eikema:2016}. 

\begin{figure}[t]
\centering
\includegraphics[scale=0.85]{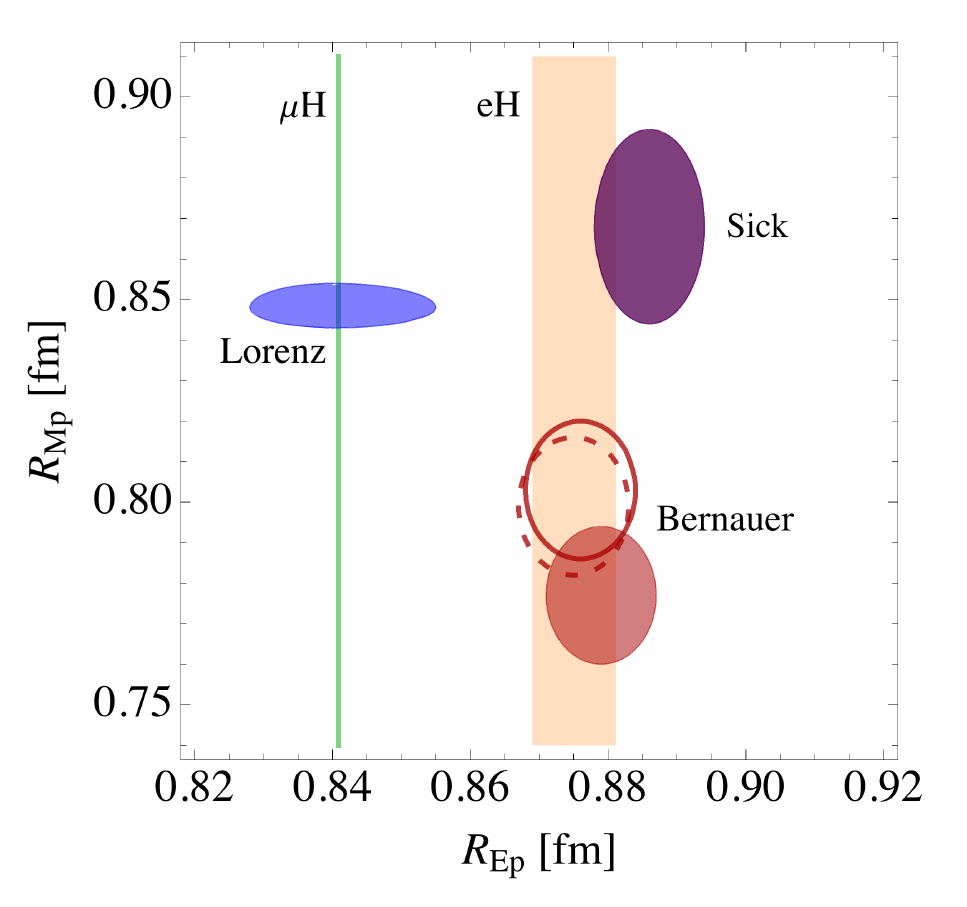}
\caption{Determinations of the proton's electric and magnetic radii. The bands correspond to the CODATA '14 recommendation \cite{Mohr:2015ccw} (orange) and the muonic-hydrogen Lamb shift result \cite{Antognini:1900ns} (green). The error ellipses show the analyses of electron-proton scattering data by \citet{Lorenz:2014yda} (blue), \citet{Sick:2012zz} (purple) and \citet{Bernauer:2014} (red). The red lines display the Bernauer fit with two-photon-exchange corrections: \cite{PhysRevC.75.038202} (solid), \cite{Arrington2011782,PhysRevC.72.034612} (dashed).\label{fig:RERM}}
\end{figure}

The hydrogen-deuterium isotope shift of the $1S-2S$ transition \cite{Parthey:2010aya} determines the (squared) deuteron-proton charge radius difference \cite{Jentschura:2011aa}:
\beq
R_{Ed}^2-R_{Ep}^2=3.82007(65)\,\mathrm{fm}^2. \eqlab{raddif}
\eeq
The proton and deuteron charge radii extracted from the isotope shift and the LSs in $\mu$D and $\mu$H, respectively, support the observation of smaller radii:
\begin{subequations}
\eqlab{isotopicR}
\bea
R_{Ep}(\mu\text{D + iso.})&=&0.8356(20)\,\mbox{fm \cite{Pohl1:2016xoo}},\eqlab{isotopicRp}\\
R_{Ed}(\mu\text{H + iso.})&=&2.12771(22)\,\mbox{fm \cite{Antognini:1900ns}}.\eqlab{isotopicRd}
\eea
\end{subequations}
The concept of isotope shifts will be also used to re-extract the $^6$He and $^8$He charge radii \cite{Wang:2004ze,Mueller:2008bj} once the $\alpha$-particle radius is extracted from the $\mu^4$He LS. And the $\mu^3$He$^+$ and $\mu^4$He$^+$ charge radii will be used to disentangle the $4\,\sigma$ discrepancy between $^3$He and $^4$He isotopic shift measurements \cite{vanRooij196,Pastor:2012aa}.

The Bernauer MAMI measurement \cite{Bernauer:2010} extended the world data set of $ep$ scattering substantially. Presently, the lower bound in the $Q^2$ range is situated at $Q^2=0.004\, \mathrm{GeV}^2/c^2$. Nevertheless, it could be that the range and accuracy of the present $ep$ scattering data is simply insufficient to make a quantitative statement on the proton charge radius. For example, two recent re-analyses \cite{Horbatsch:2015qda,Higinbotham:2015rja} concluded that $ep$ scattering is consistent with both H and $\mu$H spectroscopy, i.e., with radii in the range of $0.84$ to $0.89$ fm. The predictions obtained from $ep$ scattering highly depend on the fitting model, inclusion or neglect of high-$Q^2$ data, etc. An obvious quality criterion should be that the employed FF parametrization displays a physical behavior. In this respect, the dispersive framework is advantageous, as it incorporates the analyticity and unitarity constraints on the proton structure \cite{Mergell:1995bf,Belushkin:2006qa,Lorenz:2014yda}. To mention other ideas, it was suggested that a FF basis with analytic Fourier transform is desirable, since the  charge distribution of the proton, $\varrho_{Ep}(r)$, could be studied at the same time \cite{Sick:2011zz,Sick:2012zz,Sick:2014sra,Sick:2014kna}. A new fitting ansatz with higher moments fixed to the values predicted by ChPT  \cite{Horbatsch:2016ilr} achieved a proton radius right between the $\mu$H and CODATA values, cf.\ \Figref{RE}.
Recalling that in the analysis of $ep$ scattering the e.m.\ FFs, $G_{Ep}$ and $G_{Mp}$, are extracted simultaneously, we plot $R_{Mp}$ vs.\ $R_{Ep}$ in \Figref{RERM}, illustrating that there are also discrepancies in the determination of $R_{Mp}$. 

 For the future, new data closing the gap at low $Q^2$ are anticipated, which will make the extrapolation of the FFs to $Q^2=0$, cf.\ \Eqref{REpDeriv}, easier. The initial state radiation experiment at MAMI \cite{Mihovilovic:2014aya,Mihovilovic:2016rkr} plans to reach down to $Q^2 \sim 10^{-4}\, \mathrm{GeV}^2/c^2$ and determine the FFs with a sub-percent accuracy. The pRad (JLab) experiment \cite{Peng:2016szv} is likewise planing to measure the FFs at $Q^2 \sim 2\times10^{-4}\, \mathrm{GeV}^2/c^2$ with a sub-percent accuracy. Their magnetic-spectrometer-free setup will allow them to reach extremely low scattering angles and improve the systematical uncertainties. The MUSE muon-proton ($\mu p$) scattering experiment \cite{Kohl:2014lza} wants to directly compare $ep$ and $\mu p$ cross sections measured under the same systematic conditions and search for a possible violation of lepton flavor universality. In addition, TREK (J-PARC) will search for violation of lepton universality in stopped kaon decays \cite{Kohl:2016afc}.

Several groups are now planning to seize the idea \cite{BAKALOV1993277, Dupays:2003zz} of measuring the ground-state HFS in $\mu$H: the CREMA \cite{Pohl:2016xsr}, FAMU \cite{Bakalov:2014hda, Bakalov:2015xya, Adamczak:2016pdb} and J-PARC / Riken-RAL collaborations \cite{Sato:2014uza} (see Ref.~\cite{Pohl:2016xsr} for a comparison of experimental methods). The corresponding $1S$ transition is much narrower than the so far observed $2S$ to $2P$ transitions, therefore providing the basis to improve on the precision of the Zemach radius. In addition, the CREMA collaboration wants to measure the ground-state HFS in $\mu^3$He$^+$. We hope that our results for the TPE polarizability contribution to the HFS in $\mu$H, see \chapref{5HFS}, will be useful in this respect. 

In the future, lattice QCD (LQCD) predictions should be able to shed light upon the proton radius puzzle \cite{Alexandrou:2016hiy}. Several LQCD collaborations are studying the e.m.\ nucleon FFs: ETMC \cite{Alexandrou:2013joa,Abdel-Rehim:2015jna}, LHPC \cite{Bratt:2010jn}, PACS \cite{Yamazaki:2015vjn}, PNDME \cite{Bhattacharya:2013ehc}, Mainz \cite{Djukanovic:2016ocj} and others \cite{Green:2014xba,Green:2015wqa,Capitani:2015sba}. At present, the extraction of the proton radius from LQCD involves a fit of the lattice data points, analogous to the fitting of $ep$ scattering data. However, there are proposals to directly access the slope of the FF at zero momentum transfer. Such an approach has already been implemented to predict the Dirac radius \cite{Hasan:2016pec}. In addition, most LQCD collaborations only calculate isovector combinations of the FFs, e.g., $\half (G_{Ep}-G_{En})$, in order to avoid disconnected diagrams. The accurately known neutron charge radius is then used as input to extract the proton charge radius.

For more information on the proton charge radius puzzle and its present status, we refer to the following reviews: \citet{Pohl:2013yb}, \citet{Carlson:2015jba} and \citet{Hagelstein:2015egb}. For a complete coverage of the $\mu$H LS and HFS theory, we refer the interested reader to \citet{Antognini:1900ns}, \citet{Karshenboim:2015}, as well as Jentschura \cite{Jentschura:2010ej,Jentschura:2010ha}. The LS in light muonic atoms is reviewed in Ref.~\cite{Eides:2000xc} and Refs.~\cite{Borie:2012zz, Borie:1982ax}.

\section{Thesis Outline}

Throughout the thesis, we will introduce the theoretical foundations of CS and hydrogen-like atoms, illustrating the complex interplay between CS and hydrogen theory. Nucleon properties --- polarizabilities and  FFs --- as entering into the CS process and the TPE effects in atomic bound states are of special interest. The extraordinarily beneficial concept of dispersion relations (DRs) and CS sum rules is recapitulated and widely applied. Furthermore, the level-scheme of atomic spectra is surveyed and the finite-size effects (FSE), including the prominent charge radius term and subleading TPE corrections, are introduced.\footnote{A comprehensive review, which covers most (but not all) of the presented aspects, can be found in Ref.~\cite{Hagelstein:2015egb}.}

In \chapref{chap2}, we review the theory of hydrogen-like atoms in historical order (\secref{chap2}{QuantumNumbersSec}), specify the finite-size and polarizability effects in atomic bound states (\secref{chap2}{Radii}), and describe quantitative differences between the spectra of electronic and muonic hydrogen (\secref{chap2}{MuEComp}). The Breit potential is derived from the OPE diagram with FF dependent e.m.\ coupling to the nucleus (\secref{chap2}{BreitDerivation}). The contributions of finite-size and electronic vacuum polarization (eVP) corrections to the $\mu$H energy levels are calculated in the framework of perturbation theory and compared against the literature. From the Breit Hamiltonian we also deduce finite-size recoil effects at order $(Z\al)^5$, which are usually embedded in the nuclear-pole part of the TPE corrections and not written in terms of the first moments of the charge distributions. Our dispersive ansatz then allows us to present an alternative formulation of the FSEs (\secref{chap2}{Exact}), omitting the usual expansion in moments of charge and magnetization distributions \cite{Hagelstein:2015aa}. Working with the exact (un-expanded) formulas, we present a model of the electric Sachs FF which is able to resolve the proton radius puzzle \cite{Hagelstein:2016jgk} (\secref{chap2}{ToyModel}).

\chapref{chap3} is devoted to RCS and model-independent sum rules for the extraction of polarizabilities from photoabsorption cross sections; with the basic concepts of CS, DRs and sum rules recapitulated (\secref{chap3}{sec3.1}). We give the physical interpretation of polarizabilities (\secref{chap3}{polarizabilities}) and report on the present status of nucleon polarizabilities (\secref{chap3}{status}). The Compton contribution to photoabsorption and the associated contributions to the lowest-order scalar and spin polarizabilities are calculated at one-loop level in QED and a necessary modification of the sum rules is presented \cite{Gryniuk:2015aa,Gryniuk:2016gnm} (\secref{chap3}{elastic}). 

The transition to forward VVCS is made in \chapref{chap4}. We start with a rather detailed summary of the relevant theory (\secref{chap4}{VVCStheory}), as it will be of interest also in the subsequent Chapter. The extension of the ChPT framework to the region with $\Delta(1232)$ DOFs is motivated (\secref{chap4}{DeltaChPT}), focusing on the two common counting schemes, the $\delta$- and $\epsilon$-expansions, which give contradictory results for the longitudinal-transverse polarizability of the proton. We then calculate the tree-level $\Delta$-exchange contribution to the process of CS off the nucleon (\secref{chap4}{DeltaExchangeSec}) and the ($N\gamma^* \rightarrow \pi \Delta$) photoabsorption cross sections for pion-delta production (\secref{chap4}{DeltaCrossSectionSec}). The results are used to update the
predictions for the moments of the nucleon structure functions presented in Ref.~\cite{Lensky:2014dda} by including the $g_C$ coupling in the $N\gamma^*\rightarrow \Delta$ vertex (\secref{chap4}{DeltaPol}). Furthermore, the contribution of diagrams with photons coupling to the $\Delta$-isobar is studied as the possible origin of the $\delta_{LT}$ puzzle (\secref{chap4}{DeltaCrossSectionSec}). 

\chapref{chap5} provides the theoretical basics of forward and off-forward TPE in hydrogen-like atoms. In \secref{chap5}{TPE}, we give an extensive derivation of the forward TPE effects in atomic bound states and clarify the definition of the polarizability contribution. In addition, we compare the nucleon-pole contributions of LS and HFS to the results from OPE (\secref{chap5}{matchingOTPE}), and present an expansion of the TPE polarizability contribution in terms of polarizabilities (\secref{chap5}{PolExpTPE}). In \secref{chap5}{6.3}, we discuss the off-forward TPE in hydrogen-like bound states.

We then calculate polarizability contributions to the spectra of $\mu$H and other hydrogen-like muonic atoms. Hereby, \chapref{5LS} is dedicated to the LS and \chapref{5HFS} focuses on the HFS. Both Chapters proceed analogously. We start with the NLO BChPT prediction for the order-$\al^5$ proton-polarizability contribution, which is generated by $\pi N$-loop and $\Delta$-exchange diagrams. Our results are compared to heavy baryon chiral perturbation theory (HBChPT) and data-based dispersive approaches. Furthermore, we consider different contributions to the spectra of light muonic atoms appearing at order $(Z\alpha)^6$. For the LS, we derive the nuclear-polarizability effect at order $(Z\al)^6 \ln Z \al$ (\secref{5LS}{offTPE}). It is equivalent to the Coulomb-distortion long range polarization potential and shown to contribute non-negligibly to $\mu$D, $\mu^3$H, $\mu^3$He$^+$ and $\mu^4$He$^+$. For the HFS, the effect of neutral-pion exchange is calculated \cite{Hagelstein:2015b,Hagelstein:2015egb} (\secref{5HFS}{neutralPion}). As a highlight, we present the first theory prediction for the neutron-polarizability contribution to the HFS of light muonic atoms (\secref{5HFS}{polHFSneutron}) and an expansion of the spin-dependent non-Born TPE in terms of polarizabilities (\secref{5HFS}{polexpHFS}). In a final step, we evaluate the rms charge radii of proton and deuteron (\secref{5LS}{chap6.3}) and the proton Zemach radius  (\secref{5HFS}{newRZ})  based on the polarizability contributions found in here. 

We conclude with a summary of conclusions and an outlook (\chapref{chap7}).
\begin{subappendices}
\section{Notations and Conventions}
Unless specified, all calculations are presented in natural units, $\hbar=c=1$, and we use the following notations for the well-established parameters, along with their PDG values \cite{Agashe:2014kda,Olive:2016xmw}:
\begin{itemize}
\item[$\al$] the fine-structure constant, $\al = e^2/4\pi=1/137.035\,999\,139(31)$.
\item[$\hbar c$] conversion constant, $\hbar c=197.326 \,9788(12) \, \mathrm{MeV}\,\mathrm{fm}$.
\item[$m$]   lepton mass, $\{m_e, m_\mu \} = \{0.510\,998\,9461(31), 105.658\,3745(24)\}\,\mathrm{MeV}$.
\item[$m_\pi$] pion mass, $\{m_{\pi^0}, m_{\pi^\pm}\}= \{134.9766(6), 139.570\,18(35)\}\,\mathrm{MeV}$.
\item[$M$]   nucleon mass, $\{M_p, M_n\} = \{938.272\,0813(58), 939.565\,4133(58)\}\,\mathrm{MeV}$.
\item[$\kappa $]   nucleon anomalous magnetic moment, 
$\{\kappa_p, \kappa_n\} \simeq \{1.7929,\, -1.9130 \}$.
\end{itemize}
We introduce the Minkowski metric:
\begin{equation}g=
\left(
\begin{array}{cccc}
 1 & 0 & 0 & 0 \\
 0 & -1 & 0 & 0 \\
 0 & 0 & -1 & 0 \\
 0 & 0 & 0 & -1
\end{array}
\right),
\end{equation}
to define the four-vector scalar product:
\begin{equation}
a \cdot b= g_{\rho \sigma} \,a^\rho b^\sigma = a_\sigma b^\sigma =  a_0b_0 - \boldsymbol {a}\cdot \boldsymbol{b},
\end{equation} 
and utilize the Einstein summation convention to sum over repeated indices.
The Levi-Civita symbol is chosen as: $\epsilon_{0123}=+1=-\epsilon^{0123}$. Furthermore, we define:
\begin{itemize}
\item Pauli matrices
\begin{subequations}
\beq
\sigma_1=\left(\begin{array}{cc}0&1\\
1&0
\end{array}\right),\quad \sigma_2=\left(\begin{array}{cc}0&-i\\
i&0\end{array}\right), \quad\sigma_3=\left(\begin{array}{cc}1&0\\
0&-1
\end{array}\right),\qquad\qquad\\
\eeq
\bea
\si_i \si_j&=&\delta_{ij} \,\mathds{1}+i\sum_{k=1}^3\epsilon_{ijk}\si_k,\\
\left\{\sigma_i,\sigma_j\right\}&=&\sigma_i\sigma_j+\sigma_j\sigma_i=2\delta_{ij}\sigma_0;\qquad
\eea
\end{subequations}
\item Dirac matrices and algebra
\begin{subequations}
\beq
\gamma^0=\left(\begin{array}{cc}\mathds{1}&0\\
0&-\mathds{1}
\end{array}\right),\quad \gamma^k=\left(\begin{array}{cc}0&\sigma^k\\
-\sigma^k&0\end{array}\right), \quad\gamma^5=i\ga^0\ga^1\ga^2\ga^3=\left(\begin{array}{cc}0&\mathds{1}\\
\mathds{1}&0
\end{array}\right),
\eeq
\bea
\left\{\gamma^\mu,\gamma^5\right\}&=&0,\\
\left\{\gamma^\mu,\gamma^\nu\right\}&=&\gamma^\mu \gamma^\nu+\gamma^\nu \gamma^\mu=2g^{\mu \nu},\\
\gamma^{\mu \nu}&=&\frac{1}{2}\left[\gamma^\mu,\gamma^\nu\right]=\frac{1}{2}\left(\gamma^\mu \gamma^\nu-\gamma^\nu \gamma^\mu\right)=\gamma^\mu \ga^\nu-g^{\mu \nu}=-\frac{i}{2}\epsilon^{\mu \nu \al \be}\ga_\al \ga_\be \ga^5,\qquad\qquad\\
\gamma^{\mu \nu \al}&=&\frac{1}{2} \left(\gamma^\mu\gamma^\nu \ga^\al-\ga^\al \ga^\nu \ga^\mu \right)=-i\epsilon^{\mu \nu \al \be} \ga_\be \ga^5,\qquad\\
\gamma^{\mu \nu \al \be}&=&\frac{1}{2} \left[\gamma^{\mu \nu \al},\gamma^\be\right]=i\epsilon^{\mu \nu \al \be} \ga^5.
\eea
\end{subequations}
\end{itemize}
\end{subappendices}

\chapter{Finite-Size Effects by Dispersive Technique} \chaplab{chap2}
In the present Chapter, we derive the known expressions for the nuclear-size effects in hydrogen-like atoms by means of a dispersive technique (\secref{chap2}{BreitDerivation}). The semi-relativistic Breit potential formalism presented below is advantageous because it needs only little modification to calculate QED and electroweak corrections to the Coulomb potential (\appref{chap2}{2appVP}). We will point out a limitation of the usual accounting of FSEs in terms of the expansion in moments of charge and magnetization distributions \cite{Hagelstein:2015aa,Hagelstein:2015egb} (\secref{chap2}{Exact}), and present a toy-model which is able to resolve the proton charge radius puzzle \cite{Hagelstein:2016jgk} (\secref{chap2}{ToyModel}). Furthermore, we report finite-size recoil effects at order $(Z\al)^5$, which need to be compared to the nuclear-pole part of the TPE corrections, see \secref{chap5}{matchingOTPE}.

\section{Proton Structure in Hydrogen-Like Atoms} 
\seclab{HydrogenTheory}
We shall begin with elaborating on the spectra of hydrogen-like atoms. The synergy between theoretical and experimental efforts, mentioned in \secref{chap1}{1.3}, which lead to milestone achievements in the history of physics --- such as the establishment of quantum mechanics and quantum electrodynamics ---, shall be reviewed in more details. For simplicity, we will mainly talk about hydrogen. However, the formalism can readily be extended to hydrogen-like atoms with an arbitrary value of the nuclear charge $Z$. Effects of proton structure on the hydrogen spectrum, classified into finite-size and polarizability effects, will be our main focus. Some attention will be paid to the differences between H and $\mu$H.

\subsection{Hydrogen Spectrum} \seclab{QuantumNumbersSec}
 \footnote{In preparation of this Section, the following textbooks were used: Refs.~\cite{Foot2005,Johnson2007,Hertel2015}.}In 1911, Ernest Rutherford performed his
famous experiment of scattering $\al$-particles off a thin gold foil \cite{Rutherford:1911zz}. The ``planetary'' Rutherford model then pictured atoms as a sort of solar system, with a massive positively charged nucleus playing the role of the sun, and the light negatively charged electrons playing the role of the planets. In 1917, Rutherford used $\al$-particles to induce the nuclear fusion reaction: $\alpha^{2+} +\; ^{14}_7\text{N} \rightarrow p+\;^{17}_8\text{O}^+$. The observation of the hydrogen nucleus in the final state supported Prout's hypothesis \cite{Prout1815} that heavy atoms are just clusters of hydrogen atoms (as suggested by their masses, being multiples of the atomic weight of hydrogen). In honour of Prout's prediction, Rutherford named the hydrogen nucleus as ``proton''. 

\begin{figure}[t]
\centering
       \includegraphics[width=15.5cm]{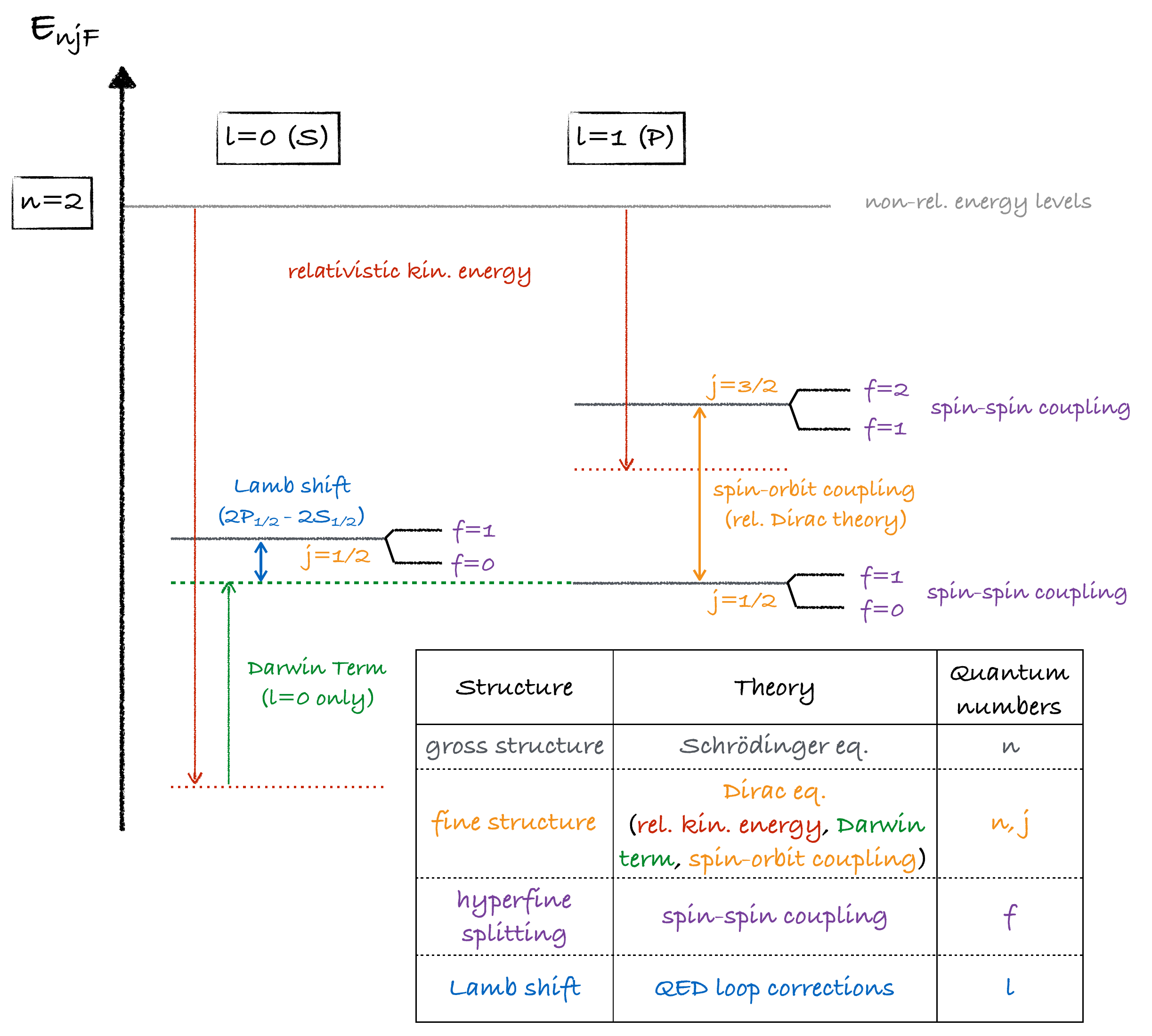}
                     \caption{Energy levels in hydrogen-like atoms (not drawn to scale). The schema ignores finite-size effects, $P$-level mixing, etc.\figlab{DiracSchrödinger}}
\end{figure}

In 1932, Sir James Chadwick \cite{Chadwick:1932ma, Chadwick:1932wcf} confirmed the existence of a neutron by similarly scattering $\al$-particles off a beryllium target (Nobel Prize 1935). The discovery of the electron dates back to early cathode ray experiments of, e.g., Sir J.\ J.\ Thomson \cite{Thomson1897}.

In 1913, Niels Bohr postulated his quantum model, which won him the Nobel Prize of 1922 \cite{Bohr:1913zba,Bohr:1913bca}. In the Bohr model, the electrons are only allowed to be on certain circular orbits with discrete values of angular momentum: $l=n \hbar$, with $\boldsymbol{l}=\boldsymbol{r}\times\boldsymbol{p}$. On these orbits, in contradiction to classical mechanics, the electrons would move without radiating. Hence, their energy is constant, and given by the Bohr formula:
\beq
E_n=-\frac{Z\al}{2an^2}, \eqlab{Bohr}
\eeq
where $n$ is the \textit{principal quantum number} describing the particular orbit, $a=1/(Z\al m_r)$ is the Bohr radius and $m_r=mM/(m+M)$ is the reduced mass of the electron-nucleus system with $m$ the electron mass and $M$ the nucleus mass. The energy levels described by \Eqref{Bohr} agree with the prediction from quantum mechanics (QM) of non-relativistic electrons without spin. They are referred to as the gross structure of the atomic spectra.

The $1/n^2$ dependence in \Eqref{Bohr} is motivated by the Rydberg formula \cite{Foot2005}, 
\beq
\frac{1}{\lambda} = Z^2 R_\infty \frac{m_r}{m}\left(\frac{1}{n^2}-\frac{1}{n^{\prime\,2}}\right), \eqlab{Rydberg}
\eeq
which was invented in 1888 to describe the ($n=2$) Balmer \cite{Balmer1885} series of spectral lines in the H emission spectrum.\footnote{The Rydberg formula also describes other spectral lines such as the later observed Lyman \cite{Lyman1906,Lyman1914}, Paschen \cite{Paschen1908}, Brackett \cite{Brackett1922} and Pfund series \cite{Pfund:24}.}
Note that the Rydberg constant, $R_\infty$, makes the (infinitely heavy) static nucleus assumption. In Eqs.~\eref{Bohr} and \eref{Rydberg}, we replaced the electron mass by the reduced mass of the electron-nucleus bound state, in order to  account for the movement of nucleus and electron around their center of mass.

A modification of the Bohr model was proposed by Arnold Sommerfeld \cite{Sommerfeld1916} in 1916. In the Sommerfeld model, the electrons are moving on elliptical orbits (now restricted by a space quantization law that imposes three quantum numbers), obeying the equations of motion of special relativity. This model allowed to generate the FS splittings and explain the splitting of spectral lines in static magnetic and electric fields, i.e., the normal Zeeman- and Stark-effects.

Werner Heisenberg won the Nobel Prize in Physics in 1932 for the ``creation of quantum mechanics'' \cite{Heisenberg:1925zz}. This marked the beginning of a new era. Heisenberg's uncertainty principle says that observables of non-commuting operators can not be simultaneously measured with arbitrary precision, e.g., the more precisely the position of a particle is determined, the less precisely its momentum can be known. Instead, particles are now described by wavefunctions, which are complex-valued probability amplitudes describing, f.i., the probability of a particle to be at a given point in space.

Erwin Schr\"odinger and Paul Dirac were awarded the Nobel Prize in Physics in 1933 for their development of QM. The non-relativistic, time-independent Schr\"odinger equation \cite{Schroedinger1926},
\beq
\hat{H}\, \Psi=E\,\Psi,
\eeq
with the quantum Hamiltonian
\beq
\hat{H}=\frac{\boldsymbol{\hat{p}}^2}{2m}+V, \eqlab{Hamiltonian}
\eeq
is a differential eigenvalue equation describing the eigenstates $\Psi$ and energy eigenvalues $E$ of a particle which moves in a potential $V$. Here, $m$ is the mass of the particle and $\boldsymbol{\hat{p}}=-i \,\nabla$ is the momentum operator. The Schr\"odinger energy eigenvalues for the Coulomb potential, $V(r)=-Z\al/r$, match \Eqref{Bohr}.\footnote{The Schr\"odinger and Dirac wave functions of the Coulomb problem are listed in the Appendices \ref{chap:chap2}.\ref{sec:SchrödingerWF} and \ref{chap:chap2}.\ref{sec:DiracWF}.}

The relativistic Dirac equation, 
\begin{subequations}
\beq
\left[\boldsymbol{\alpha} \,\boldsymbol{\hat{p}}+\beta \,m+V(r)\right]\psi = E\, \psi,
\eeq
with
\beq
\boldsymbol{\alpha}=\left(\begin{array}{cc}0&\boldsymbol{\sigma}\\
\boldsymbol{\sigma}&0\end{array}\right), \quad \beta=\left(\begin{array}{cc}\mathds{1}&0\\
0&-\mathds{1}\end{array}\right),
\eeq
\end{subequations}
the Pauli matrices $\boldsymbol{\sigma}$ and the Dirac spinors $\psi$, describes spin-1/2 particles \cite{Dirac:1928hu}. Applying the Dirac equation to the Coulomb potential, one obtains the following energy eigenvalues:
 \begin{subequations}
 \bea
 E_{nj}&=&m+m_r  \left\{\left[1+\left(\frac{Z \alpha}{n-(j+1/2)+\sqrt{(j+1/2)^2-(Z\alpha)^2}}\right)^2\right]^{-1/2}-1\right\},\qquad\eqlab{EDirac}\\
  &\approx&m+E_n \left\{1+\frac{E_n}{2 m_r}\left[3-\frac{4n}{j+1/2}\right]\right\}+\mathcal{O}(Z\alpha)^6,\eqlab{DiracEnergies6}
 \eea
 \end{subequations}
 which depend on the \textit{principal quantum number} $n$ and the \textit{electron total angular momentum quantum number} $j$.
In the last step, we expanded in the parameter $Z \alpha$, which is small for light nuclei. The expansion illustrates that the Dirac equation produces the rest energy, the energy levels from the non-relativistic limit, cf.\ $E_n$ in \Eqref{Bohr}, and higher-order relativistic corrections. The additional dependence of the energy levels on $j$ generates the FS on top of the gross structure, cf.\ \Figref{DiracSchrödinger}. The FS is induced by the spin-orbit coupling (cf.\ \Eqref{LSrelC}, $\boldsymbol{j}=\boldsymbol{l}+\boldsymbol{s}$), i.e., by the interaction of the electron magnetic dipole moment (or the electron spin\footnote{Experimental evidence for the electron spin was given by the Stern-Gerlach experiment \cite{Gerlach:1922zz}, the observation of the FS and the so-called anomalous Zeeman effect.} $\boldsymbol{s}$) and the magnetic dipole moment associated with the electron's orbital angular momentum ($\boldsymbol{l}$,  \textit{electron orbital angular momentum quantum number} $l$). A first observation of FS splittings was reported by \citet{Houston:1934aa} in a re-measurement of the Balmer series in H.

Figure \ref{fig:DiracSchrödinger} shows a further splitting of the $S_{1/2}$, $P_{1/2}$ and $P_{3/2}$ lines --- the so-called HFS.
The leading HFS of $S$-levels is given by the Fermi energy:
\beq
E_\mathrm{F}(nS)=\frac{8Z\al}{3a^3}\frac{1+\kappa}{mM}\frac{1}{n^3}.\eqlab{FermiE}
\eeq
Here and hereafter, $M$ is the mass of the nucleus, $m$ is the mass of the lepton and $\kappa$ is the anomalous magnetic moment of the nucleus. Explaining the HFS requires the spin-spin coupling (cf.\ \Eqref{sSoperator}, $\boldsymbol{f}=\boldsymbol{j}+\boldsymbol{S}$), that is the interaction between the nuclear magnetic dipole moment (or nuclear spin $\boldsymbol{S}$) and the magnetic field generated by the electron (electron total angular momentum $\boldsymbol{j}$). The dependence of the HFS on the atom's total angular momentum ($\boldsymbol{f}$, \textit{atom's total angular momentum quantum number} $f$) becomes obvious from the static potential in \Eqref{V2}. The $nS$ HFS is then the splitting between levels with quantum numbers $n$, $l=0$ and $f=0$ or $f=1$, respectively.
 
 The $1S$ HFS of H is widely known as the $21$-cm line observed in radio astronomy \cite{Purcell1951, MULLER:1951aa}. Since the transition probability of the $1S$ HFS is extremely small, it turned out to be a great tool to study the spiral structure of the Milky Way. One can gain information on the density, velocity and temperature distributions of H atoms in the universe by detecting lines with different redshifts and Doppler shifts.
 
Both the non-relativistic Schr\"odinger energies, \Eqref{Bohr}, and the relativistic Dirac energies, \Eqref{EDirac}, of the Coulomb problem are degenerate for levels with equal $n$ and $j$, hence, display no classic  ($2P_{1/2}-2S_{1/2}$) LS.  To explain the LS observed in 1947 by \citet{Lamb:1947zz}, cf.\ \Figref{DiracSchrödinger}, one needs QED.\footnote{In 1965, Sin-Itiro Tomonaga, Julian Schwinger and Richard P.\ Feynman were jointly awarded the Nobel Prize in Physics ``for their fundamental work in QED, with deep-ploughing consequences for the physics of elementary particles".} The first LS calculation was performed by \citet{Bethe:1947id} using non-relativistic QED. The necessary QED corrections involve vertex and selfenergy corrections for the electron and nucleus, respectively. In contrast to H, the $\mu$H LS is dominated by eVP. cf.\ \secref{chap2}{MuEComp}. In the follow-up Sections, we enlist all effects relevant to the spectra of hydrogen. In particular, we confront the theories of H and $\mu$H.

\subsection{Finite-Size Effects: Charge Radius and Beyond}\seclab{Radii}

The nuclear structure is long known to affect the atomic spectra of (normal) electronic and, more significantly, muonic atoms. Naturally, the nuclear structure effects are divided into two categories:
\begin{enumerate}
\item[(i)] finite-size effects (FSEs), i.e., the effects due to the elastic FFs, $G_E$ and $G_M$;
\item[(ii)] polarizability effects, which is everything else.
\end{enumerate}
In the present Chapter, we derive the OPE Breit potential. From this potential we can only obtain FSEs. An extensive derivation of the structure effects through TPE, clarifying the definition of the polarizability contribution, is delegated to \chapref{chap5}.

The FSEs are predominantly given by an upwards shift of the $n$-th $S$-level. To order $(Z\alpha)^5$, the classic LS and the $S$-level HFSs are found as (omitting recoil) \cite{Eides:2000xc}:
\begin{subequations}
\eqlab{FSEs}
\bea
E_\mathrm{LS}&\equiv &E(2P_{1/2})-E(2S_{1/2})=-\frac{Z\al}{12a^3} \left[ R_E^2 -\nicefrac{1}{2a}\, R^3_{\mathrm{F} }\right] + \mathcal{O}(Z\al)^6,\eqlab{LambShift}\\
E_{\mathrm{HFS}}(nS)&\equiv&E(nS^{f=1}_{1/2})-E(nS^{f=0}_{1/2}) =E_\mathrm{F}(nS)\left[1-\nicefrac{2}{a} \,R_{\mathrm{Z}}\right]+ \mathcal{O}(Z\al)^6,\qquad \eqlab{HFS}
\eea
\end{subequations}
where the radii are defined as follows:
\begin{itemize}
\item Charge radius (rms):
\begin{subequations}
\eqlab{REpDeriv}
\begin{align}
R_E &=  \sqrt{\langle r^2\rangle_E}, \\
\quad \langle r^2\rangle_E
&\equiv \int \! \dd \br\,  r^2  \varrho_{E}(\br) = -6 \frac{\dd}{\dd Q^2} G_E(Q^2)\Big|_{Q^2= 0}\;;
\end{align}
\end{subequations}
\item Friar radius (or, the 3\textsuperscript{rd} Zemach moment of the charge distribution) \cite{Zemach:1956,Friar:1978wv}:
\begin{subequations}
\eqlab{Friar}
\begin{align}
R_{\mathrm{F} } &=\sqrt[3]{\langle r^3\rangle_{E(2)} },\\
 \langle r^3\rangle_{E(2)} &\equiv  \int \dd \br\, r^3\, \varrho_{E(2)}(r), \quad\text{with}\quad\varrho_{E(2)}(r)=\int \dd \boldsymbol{r'} \,\varrho_E (\vert \boldsymbol{r'}-\br\vert)\,\varrho_E(\boldsymbol{r'}),\\
 &\equiv\frac{48}{\pi} \int_0^\infty \!\frac{\dd Q}{Q^4}\,
\Big[ G_E^2(Q^2) -1 +\third R^2_E\, Q^2\Big];
\end{align}
\end{subequations}
\item Zemach radius \cite{Zemach:1956}:\footnote{Note that \citet{Faustov:2001pn} use a definition of the Zemach radius which depends on the reduced mass of the hydrogen bound state. Therefore, they obtain different Zemach radii for electronic ($R_{\mathrm{Z}p}=1.025 \,\mathrm{fm}$) and muonic hydrogen ($R_{\mathrm{Z}p}=1.022 \,\mathrm{fm}$).}
\begin{subequations}
\eqlab{Zemach}
\begin{align}
R_{\mathrm{Z}}&\equiv \int \dd \br \, r\,\varrho_{\mathrm{Z}}(r)\quad \text\quad \varrho_{\mathrm{Z}}(r)=\int  \dd\boldsymbol{r'}\,\varrho_E (\vert \boldsymbol{r'}-\br\vert)\,\varrho_M(\boldsymbol{r'}),\\
&\equiv -\frac{4}{\pi}\int_0^\infty \frac{\dd Q}{Q^2}\left[\frac{G_E(Q^2)G_M(Q^2)}{1+\kappa}-1\right].
\end{align}
\end{subequations}
\end{itemize}
The formulas presented above represent the usual accounting of FSEs as an expansion in moments of charge and magnetization distributions. We will present the un-expanded formalism in \secref{chap2}{Exact} and point out the limitations of the expansion in \secref{chap2}{ToyModel}.

Presently, the determination of the proton charge radius from the $\mu$H LS relies on input for the Friar radius obtained from elastic FFs measured in $ep$ scattering. Karshenboim \cite{Karshenboim:2014maa,Karshenboim:2014vea} points out that this poses a consistency problem, as the Friar radius is in turn related to the charge radius, cf.\ \Eqref{Friar}, while the present results from $\mu$H spectroscopy and $ep$ scattering give contradictory proton charge radii, see also Ref.~\cite[Section 6.2.2]{Hagelstein:2015egb}.

\begin{figure}[h]
\centering
       \includegraphics[width=14cm]{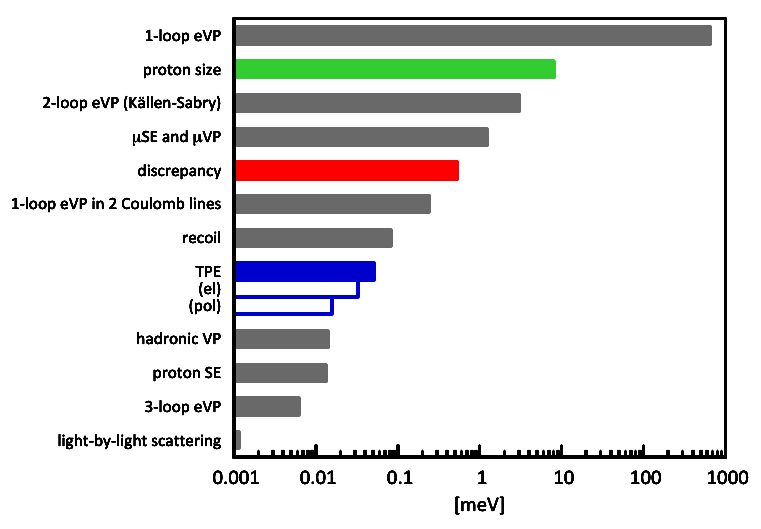}
                     \caption{Theoretical budget of the Lamb shift in muonic hydrogen \cite{Antognini:2012ofa}. The two-photon exchange is displayed in blue; we give estimates for the elastic and polarizability contributions (unfilled bars), as well as for the total two-photon-exchange contribution (solid bar). The discrepancy between theory and experiment adds up to $0.31\, \mathrm{meV}$. The theoretical uncertainty is estimated as $0.0025 \,\mathrm{meV}$, cf.\ \Eqref{LS theory}. \label{fig:Pohl}}
\end{figure}

\subsection{Electronic vs.\ Muonic Hydrogen}\seclab{MuEComp}

We now cosider the spectra of muonic and electronic atoms, with a particular focus on hydrogen. Muonic atoms are especially suited to study the nuclear size, because its $207$ times heavier mass makes the muon orbit closer to the nucleus than an electron. Hence, the Bohr radius of $\mu$H is $186$ times smaller than the Bohr radius of ordinary H. Equation \eref{FSEs} shows that the effect from the charge radius term is proportional to the reduced mass (roughly the lepton mass) of the lepton-nucleus bound state and the charge of the nucleus. Therefore, the FSEs in muonic hydrogen are bigger than in electronic hydrogen, and the FSEs in, f.i., helium are bigger than in hydrogen. In general, the NLO FSEs are suppressed by the inverse Bohr radius, i.e., not only by the fine-structure constant $\alpha$. Therefore, they can be omitted in the H theory and become appreciable in $\mu$H only.

\noindent The theoretical description of the classic LS and  $2S$ HFS in $\mu$H \cite{Antognini:2012ofa} adds-up to (in  $\mathrm{meV}$):
\begin{subequations}
\eqlab{muHtheory}
\bea
&&\hspace{-1cm}E_{\mathrm{LS}}^{\mu\text{H th.}}=206.0336(15)-5.2275(10)\,\left(\nicefrac{R_{Ep}}{\mathrm{fm}}\right)^2+E^{\mathrm{TPE}}_{\mathrm{LS}}\, , \; \text{with} \,\,  E^{\mathrm{TPE}}_{\mathrm{LS}}=0.0332(20),\quad\eqlab{LS theory}\\
&&\hspace{-1cm} E_{\mathrm{HFS}}^{\mu\text{H th.}}(2S)=22.9763(15)-0.1621(10)\,(\nicefrac{R_{\mathrm{Z}p}}{\mathrm{fm}}) +E^{\mathrm{pol.}}_{\mathrm{HFS}}\, , \; \text{with} \,\, E^{\mathrm{pol.}}_{\mathrm{HFS}}=0.0080(26),\qquad\;\eqlab{muHtheoryHFS}
\eea
\end{subequations}
where $E^{\mathrm{TPE}}_{\mathrm{LS}}$
contains the Friar radius, recoil FSEs, and 
the proton-polarizability effects; $E^{\mathrm{pol.}}_{\mathrm{HFS}}(2S)$ is the proton-polarizability effect only. From the H bound-state tabulation of Ref.~\cite{Horbatsch:2016aa}, we deduced the following theoretical $2P_{1/2}^{f=1}-2S_{1/2}^{f=0}$ LS (in units of $\upmu$eV):
\beq
E_{\mathrm{LS}}^{\text{H th.}}\approx -3.7624-0.0008\,(\nicefrac{R_{Ep}}{\mathrm{fm}})^2,
\eeq
with the Rydberg constant, $R_\infty=3\, 289\,841\,960\,355(19)$ kHz, taken from the CODATA '14 adjustment \cite{Mohr:2015ccw}.  This formula should be only considered as an approximation, as we neglected terms with logarithmic dependence on the charge radius by substituting the CODATA recommendation for $R_{Ep}$, \Eqref{REpCodata}.

Figure~\ref{fig:Pohl} illustrates the size of different contributions to the theory prediction of the $\mu$H LS. The dominating contribution comes from eVP. Already the second largest term is given by the proton size. Furthermore, there are corrections due to muonic self-energy ($\mu$SE) and muonic vacuum polarization ($\mu$VP) which are absent in H.  For the H atom, the QED corrections of the e.m.\ electron vertex generate the main part of the LS. The TPE effects can be split into an elastic finite-size part and a polarizability part. These are nuclear structure effects at order $(Z\al)^5$ and as such they are seizable in $\mu$H but can be neglected in H. On the other hand, they impose the largest uncertainty on the $\mu$H theory, cf.\ Chapters \ref{chap:5LS} and \ref{chap:5HFS}. The relative strength of all these contributions alters the $\mu$H spectrum considerably in comparison to the H spectrum.\footnote{See Ref.\ \cite{Karshenboim:2015} for a more detailed description of important differences between H and $\mu$H.} 

The $n=2$ energy levels of $\mu$H and H are shown in Figures \ref{fig:MuHSpec} and \ref{fig:EHSpec}. The configuration of the energy levels in $\mu$H differs due to the strong eVP shift of the $2S$ level, cf.\ \Eqref{UehlingNum}. Also, the proton-size has a large effect on the $2S$ level. In general, the splittings between energy levels in $\mu$H are much wider. The LS transition frequencies in $\mu$H are $4$ orders of magnitude bigger than in H. Therefore, H studies require  precision spectroscopy experiments, whereas the difficulties of the $\mu$H experiment lie in other aspects, e.g., the lifetime of the $\mu$H.

In H spectroscopy, there are ``small splitting measurements'' between levels with equal principal quantum numbers, e.g., the LS measurements, and ``big splitting measurements'' between levels with different principal quantum numbers, e.g., the $1S-2S$ transition. The $S$-levels can be roughly described as \cite{Pohl:2013yb}:
\beq
\eqlab{RydbergE}
E(nS)\simeq -\frac{R_\infty}{n^2}+\frac{L_{1S}}{n^3},
\eeq
where $L_{1S}$ is the LS of the $1S$ ground state which contains the proton charge radius effect. For the small splitting measurements, the Rydberg constant is known precisely enough from other sources to extract the charge radius. However, for the big splitting measurements it is necessary to have two H transitions and extract $R_\infty$ and $R_{Ep}$ simultaneously. Therefore, extractions of the Rydberg constant and the proton radius from H are correlated and an independent measure of the Rydberg constant would be very much appreciated \cite{Jentschura:2008aa,Jentschura2010,Tan:2014vha,Tan2011,Brewer:2013aa}.

\begin{figure}[h!] 
    \centering        \includegraphics[width=0.86\textwidth]{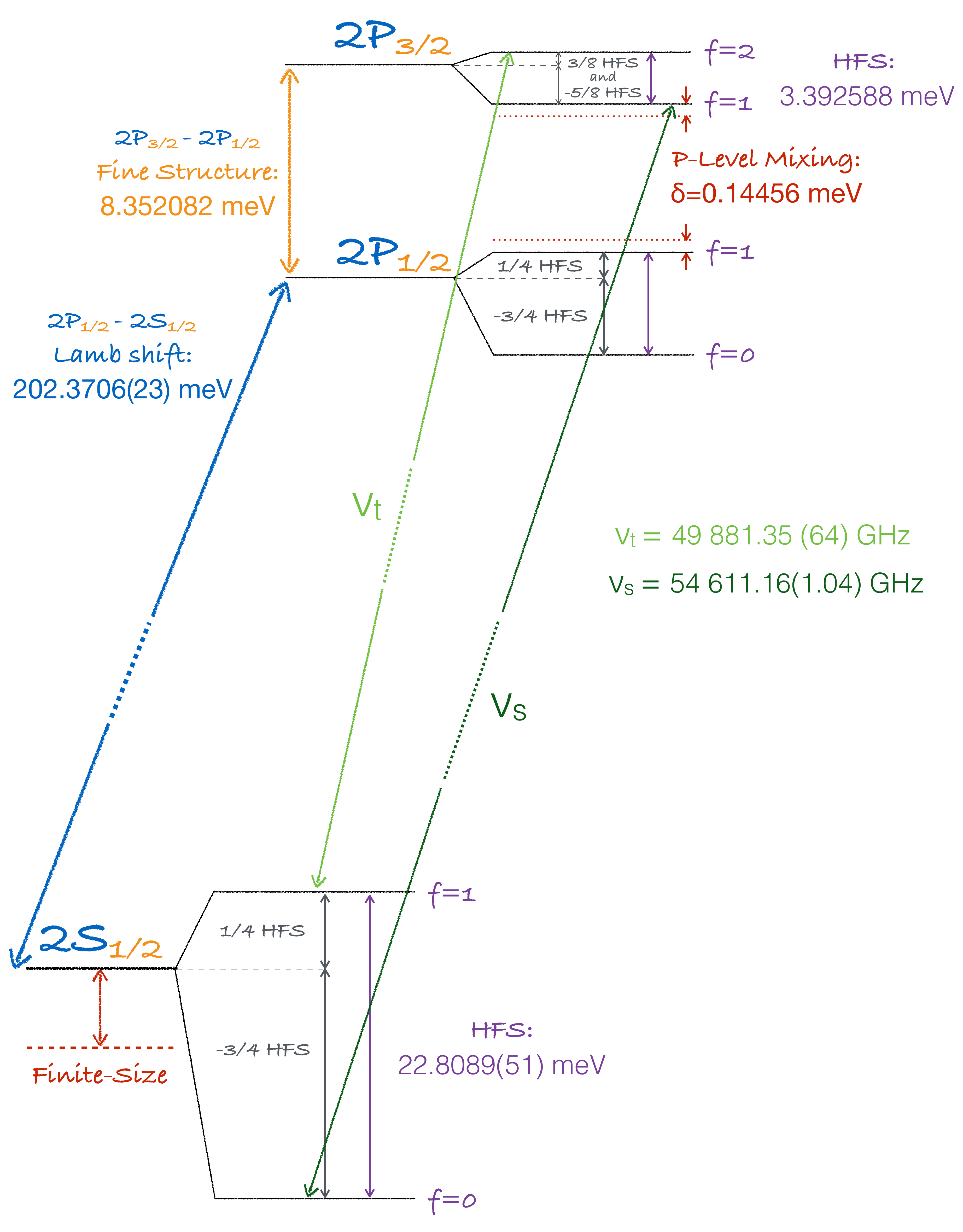}
 \caption{Spectrum of muonic hydrogen. The $2P$ fine structure, the $P_{3/2}$ hyperfine splitting and the $P$-level mixing are taken from the theory summary of Ref.\ \cite{Antognini:2012ofa}. The two transition frequencies, $\nu_t$ and $\nu_s$, are experimental results from Refs.\ \cite{Pohl:2010zza,Antognini:1900ns}. The $2S$ hyperfine splitting and the classic $2P_{1/2}-2S_{1/2}$ Lamb shift are reconstructed from the measurements and the theoretical shifts \cite{Pohl:2010zza,Antognini:1900ns,Antognini:2012ofa}.\label{fig:MuHSpec}}
\end{figure}

\begin{figure}[h!] 
    \centering        \includegraphics[width=0.83\textwidth]{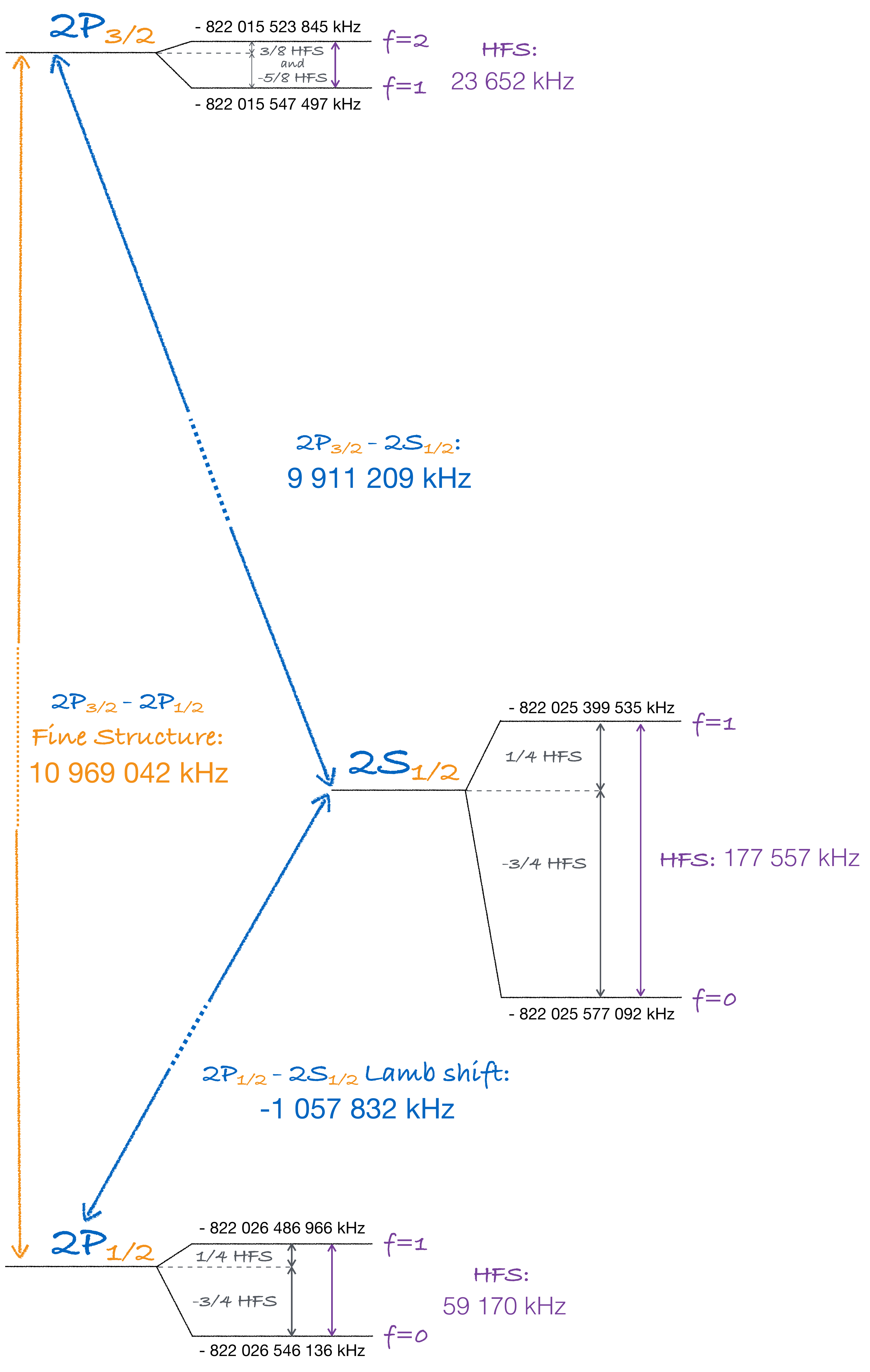}
 \caption{Spectrum of electronic hydrogen. The energy levels are extracted from Ref.\ \cite{Horbatsch:2016aa}.\label{fig:EHSpec}}
\end{figure}

\section{Breit Potential with Finite Nuclear Size} \seclab{BreitDerivation}
Effective Hamiltonians are the standard tools in state-of-the-art calculations of atomic spectra. A semi-relativistic expansion of the one-photon interaction in two-particle bound-states was first considered by \citet{Breit:1929zz,Breit:1930zza,Breit1932}. Breit Hamiltonians are for instance used to describe muonic atoms \cite{Pachucki:1996zza,Veitia:2004zz,Jentschura:2011nx,Karshenboim:2012wv}, mesonic atoms \cite{Kelkar:2007nq}, and neutron structure effects in the deuteron and one-neutron halo nuclei \cite{Nowakowski:2005dd}. 

In the present Section, we will briefly sketch the derivation of a semi-relativistic OPE Breit potential describing hydrogen-like atoms, i.e., bound states of a nucleus and a single lepton. The Breit potential will account for the finite nuclear size by incorporating the nuclear FFs. It will then be treated in the non-relativistic Schr\"odinger theory as perturbation to the Coulomb potential of a point-like charged nucleus. In comparison to Ref.\ \cite{Daza:2010rh}, which also studies the Breit potential for the hydrogen atom with proton FFs, our dispersive ansatz is more general, as it is not limited by a particular model of the FFs. Therefore, it can also be applied for QED and electroweak corrections to the Coulomb potential.


\subsection{Nuclear Form Factors}
\begin{wrapfigure}[9]{r}{0.4\textwidth} 
\vspace{-0.2cm}
\centering
       \includegraphics[width=4.5cm]{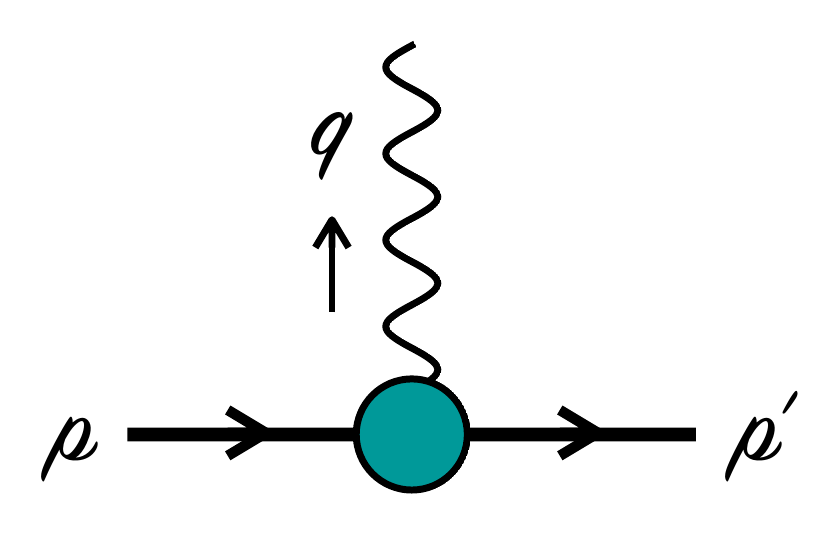}
\caption{Electromagnetic vertex.\label{fig:gpp}}
\end{wrapfigure}
The nuclear size is embedded in the FFs of the nucleus. The e.m.\ interaction vertex of a spin-1/2 particle, drawn in \Figref{gpp}, is related to the elastic FFs as follows \cite{Foldy:1952,Foldy:1952zz,Foldy:1958zz}:
\beq
\Ga^\mu =Ze \left[\ga^\mu F_1(Q^2) +\frac{1}{2Mc} \ga^{\mu\nu} q_\nu
F_2(Q^2)\right] \eqlab{protonphotonvertex},
\eeq
where $Ze$ and $M$ are the particle's charge and mass, respectively. Since the derivation involves a semi-relativistic expansion, we will keep $c$ for now. The photon momentum is defined as $q=p-p'$, i.e., as outgoing, cf.~\Figref{gpp}; $Q^2=-q^2>0$ is the space-like momentum transfer. $F_1(Q^2)$ and $F_2(Q^2)$ are the Dirac and Pauli FFs \cite{Hofstadter:1956qs,Mcallister:1956ng,Hofstadter58}, normalized to the unit charge [$F_1(0)=1$] and the anomalous magnetic moment [$F_2(0)=\kappa$]. As suggested by Yennie, Sachs and others \cite{Yennie1957,Ernst:1960zza,Sachs:1962zzc}, it is 
often convenient to use the so-called Sachs electric and magnetic FFs instead:\footnote{The e.m.\ Sachs FFs are determined by means of electron scattering and, f.i., extracted from the measured cross sections via Rosenbluth separation. The Rosenbluth formula \cite{Rosenbluth:1950} reads:
\begin{subequations}
\eqlab{Rosenbluth}
\bea
\left(\frac{\dd \sigma}{\dd \Omega}\right)&=&\left(\frac{\dd \sigma}{\dd \Omega}\right)_\text{Mott} \;\left[\frac{G_E^2(Q^2)+\tau\, G_M^2(Q^2)}{1+\tau}+2\tau\, G_M^2(Q^2) \tan ^2 \frac{\theta}{2}\right],\\
&=&\left(\frac{\dd \sigma}{\dd \Omega}\right)_\text{Mott} \;\frac{\epsilon \,G_E^2(Q^2)+\tau\, G_M^2(Q^2)}{\epsilon(1+\tau)},
\eea
\end{subequations}
with the scattering angle $\theta$, the photon polarization $\epsilon=[1+2(1+\tau)\tan ^2 \nicefrac{\theta}{2}]^{-1}$ ($0 \leq \epsilon \leq1$) and $\tau=Q^2/4(Mc)^2$. The larger the photon momentum is, the smaller is its reduced wavelength and, hence, its resolution power. For higher momentum transfers, the cross section reduces because the electron or photon can no longer see the total charge of the target but only a fraction of it. This is described by the FFs which multiply the Mott cross section. The Mott cross section is a modification of the Rutherford cross section in consideration of the electron spin:
\beq
\left(\frac{\dd \sigma}{\dd \Omega}\right)_\text{Mott}=\frac{4(Z\al)^2 E^{\prime\,2}}{Q^4}\frac{E'}{E}\left[1-\beta^2\sin ^2 \frac{\theta}{2}\right],
\eeq
where $E$ ($E'$) is the initial (final) electron energy, $\beta=v/c$ is the usual velocity ratio and the factor $\nicefrac{E'}{E}$ accounts for the target recoil. The notation of the Rosenbluth formula in terms of e.m.\ Sachs FFs is especially useful because there is no interference term between $G_E$ and $G_M$ . }
\begin{subequations}
\eqlab{SachsFFDP}
\bea
G_E(Q^2)&=&F_1(Q^2)-\tau F_2(Q^2),\\
G_M(Q^2)&=&F_1(Q^2)+F_2(Q^2),
\eea
\end{subequations}
with the dimensionless momentum transfer $\tau=Q^2/4(Mc)^2$.\footnote{See Ref.\ \cite{Hand:1963zz} for a systematic comparison of Sachs FFs versus Dirac and Pauli FFs.} They have an intuitive physical interpretation as the Fourier transforms of the charge and magnetization distributions, $\varrho_E(r)$ and $\varrho_M(r)$, in the Breit frame \cite{Ernst:1960zza,Sachs:1962zzc}:\footnote{Note that for all nuclei and also the proton, the magnetic moment is defined as $\mu=Z(\kappa+2S)$ in units of nuclear magneton $\mu_N=e/2M_p$, with $S$ being the spin of the nucleus. The $g$-factor is defined as $\kappa=(g/Z-2)S$. However, for the neutron it is $\mu=\kappa$.} 
\begin{subequations}
\bea
G_E(Q^2)
  &=&  \frac{4\pi}{Q}\int_0^\infty \dd r \,r\sin (Qr)\, \varrho_E(r),\,
\\
\frac{G_M(Q^2)}{1+\kappa} &=&\frac{4\pi}{Q}\int_0^\infty \dd r\, r\sin (Qr) \,\varrho_M(r).
\eea
\end{subequations}
Hence, their slopes at the real photon point carry information on the charge and magnetic radii, cf.\ \Eqref{REpDeriv}. Here, the densities are assumed to be spherically symmetric and, thus, are Lorentz invariant. The more extended the spatial distribution is, the stronger the FF falls off with increasing momentum transfer. In other words, the steeper the electric FF is, the bigger the proton radius. A dipole FF corresponds to an exponential distribution in coordinate space. For recent reviews on nucleon e.m.\ FFs, see Refs.\ \cite{HydeWright:2004gh,Perdrisat:2006hj}.

\subsection{One-Photon Exchange}  \seclab{OPE}


\begin{wrapfigure}[14]{r}{0.5\textwidth} 
\vspace{-0.5cm}
\centering
\includegraphics[width=4cm]{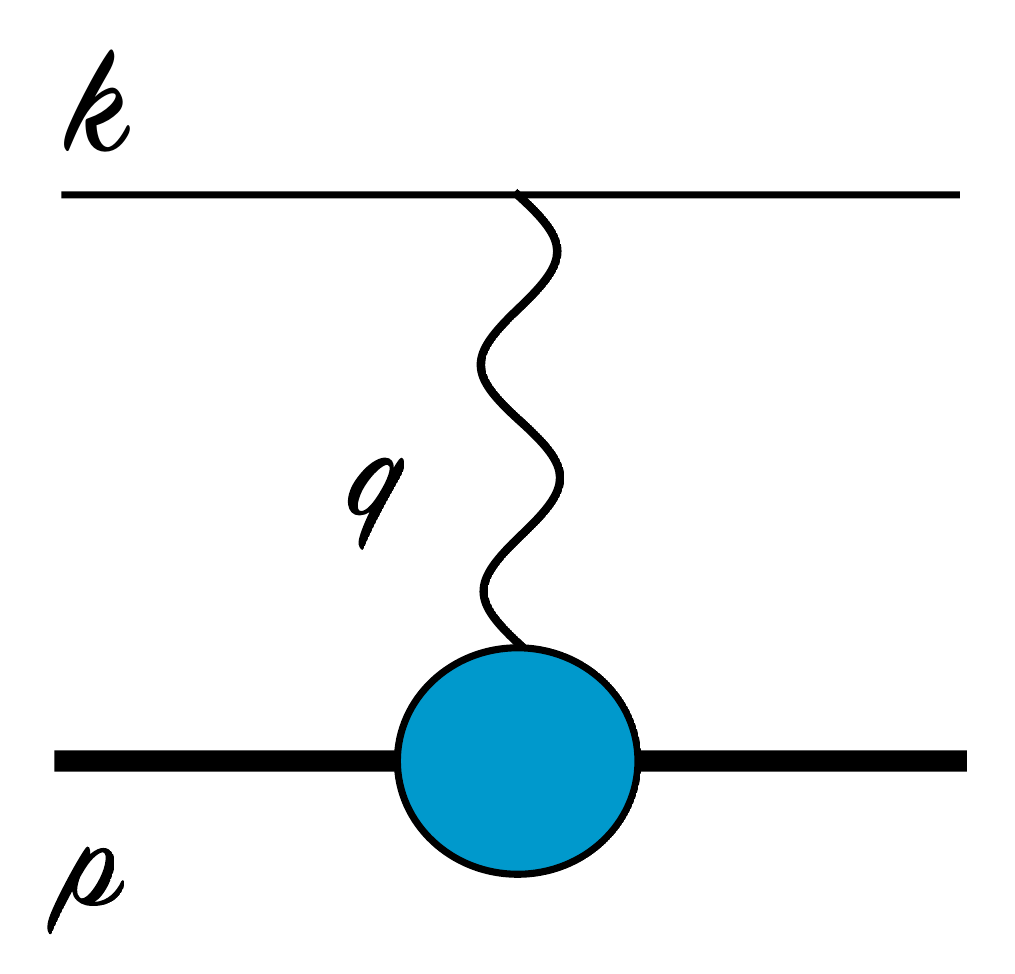}
    \caption{One-photon-exchange diagram with nuclear form factors, giving rise to finite-size effects. The horizontal lines correspond to the lepton and the nucleus (bold). \label{fig:ope}}  
\end{wrapfigure}  

Figure \ref{fig:ope} represents the OPE between a lepton and a nucleus (bold line). The four-momenta of photon, lepton and nucleus are denoted by $q$, $k$ and $p$, respectively. The outgoing particles will be indicated with primed variables ($k'$ and $p'$). The photon momentum is defined as $q=p-p'=k'-k$. Since it also contains the Coulomb interaction, the OPE is the leading contribution to the description of a hydrogen-like bound state. The associated potential is proportional to the OPE scattering amplitude. An energy prefactor stems from the non-relativistic reduction to the Breit equation. For a spin-1/2 nucleus, like the proton, we have:
\beq
 V_\mathrm{OPE} =  \big(2E_k\, 2 E_{k'} \, 2E_p\,  2E_{p'}\big)^{-1/2}\;
 \bar u(k') \! \left[-e \gamma^\mu\right] \!u(k) \,
\De_{\mu\nu}(q) \;  \overline{N}(p') \Ga^\nu(q) N(p),\eqlab{BreitPotentialScatAmp}
\eeq
with the photon propagator $\Delta_{\mu\nu}(q)$ and 
the e.m.\ vertex $\Ga^\nu(q)$, \Eqref{protonphotonvertex},
which depends on the Dirac and Pauli FFs. Replacing $F_1\rightarrow 1$ and $F_2\rightarrow 0$ gives the structureless limit of a point-like, charged nucleus, viz.\ the Coulomb potential. The lepton and nucleus spinors, $u$ and $N$, we chose to be
normalized according to:
\bea
\bar u (k)\,u(k)=2mc^2, \qquad \overline{N}(p)\, N(p)=2Mc^2.
\eea

The Dirac and Pauli FFs are assumed to fulfil the once-subtracted DRs \cite{Frazer:1960zza}:\footnote{The subtraction is made for several reasons. Firstly, it allows us to separate the Coulomb interaction. Secondly, we want to assure that the DRs are convergent. For our final result we want to substitute the Sachs FFs. Ref.\ \cite{Hand:1963zz} compares DRs for Dirac, Pauli and Sachs FFs. They find that $G_E$ needs one more subtraction than $F_2$.}
\bea
\barr F_1 (Q^2)\\ 
F_2 (Q^2) \earr  = \barr 1\\
\kappa \earr - \frac{Q^2}{\pi} \int_{t_0}^\infty \!\frac{\dd t}{ t (t+Q^2) }
\barr  \im F_1(t) \\ 
\im F_2 (t) \earr , \eqlab{subtractedDR}
\eea
with $t_0\geq0$ being the lowest particle-production threshold. $\im F_1$ and $\im F_2$ are the FF discontinuities across the branch cuts in the time-like region. The subtracted DRs are derived from the unsubtracted ones,
\bea
\eqlab{unsubtractedDR}
\barr F_1 (Q^2)\\ 
F_2 (Q^2) \earr  =  \frac{1}{\pi} \int_{t_0}^\infty \!\frac{\dd t}{t+Q^2}
\barr  \im F_1(t) \\ 
\im F_2 (t) \earr, 
\eea
by removing the FF values at the real photon point, i.e., $F_1(0)=1$ and $F_2(0)=\kappa$. Our ansatz to use the DRs is very much analogous to Schwinger's method of calculating
the Uehling potential \cite{Uehling:1935uj} of the vacuum polarization (VP) effect \cite{Schwinger1949}.
 
The photon propagator is defined as:
 \begin{subequations}
 \eqlab{PhotonProp}
 \beq
 \Delta_{\mu \nu}={\mathcal P}_{\mu\nu}/Q^2,
 \eeq
 with 
\beq
{\mathcal P}_{\mu\nu}(q,t)=g_{\mu\nu}-\frac{1}{t+\bq^2}\left(q_\mu q_\nu-\chi_\mu q_\nu-\chi_\nu q_\mu\right).
\eeq
\end{subequations}
Note that the above propagator is not only depending on the photon four-momentum but also on the photon mass $t$. The mass dependence will enter in conjunction with the dispersion integral, whereas the subtraction term goes with a massless propagator. Contracting the photon propagator \eref{PhotonProp} with the e.m.\ vertex \eref{protonphotonvertex}, and plugging in the DRs of the FFs \eref{subtractedDR}, we obtain:
\bea
\De_{\mu\nu}\Ga^\nu &=&Ze\bigg\{\frac{1}{Q^2} \,{\mathcal P}_{\mu\nu}(q,0)
\bigg[\ga^\nu+\frac{\kappa}{2Mc} \ga^{\nu\al} q_\al\bigg]\eqlab{propvert}\\
&&\qquad- \frac{1}{\pi} \int_{t_0}^\infty\! \frac{\dd t}{ t (t+Q^2) }\,
{\mathcal P}_{\mu\nu}(q,t) \bigg[\ga^\nu \im F_1(t) +\frac{1}{2Mc} \ga^{\nu\al} q_\al
\im F_2(t) \bigg]\bigg\}.\nn\qquad
\eea
In the following, it will be beneficial to choose the Coulomb gauge: $\chi=(0,\bq)$. In this gauge, the tensor ${\mathcal P}_{\mu\nu}$ reduces to:
\begin{subequations}
\eqlab{PhotonProp2}
\bea
{\mathcal P}_{00}(q,t)&=&\frac{t+Q^2}{t+\bq^2}\stackrel{t=0}{=}\frac{Q^2}{\bq^2},\\
{\mathcal P}_{0i}(q,t)={\mathcal P}_{i0}(q,t)&=&0,\\
{\mathcal P}_{ij}(q,t)&=&-\delta_{ij}+\frac{q_i \,q_j}{t+\bq^2}\stackrel{t=0}{=}-\delta_{ij}+\frac{q_i \,q_j}{\bq^2}.
\eea
\end{subequations}
The temporal and spatial components of \Eqref{propvert} then read:
\begin{subequations}
\eqlab{scattAmpPart}
\bea
\De_{00}\Ga^0 &= &Ze\bigg\{\frac{1}{\bq^2}\left[\ga^0-\frac{\kappa}{2Mc} \ga^0 \bgamma\cdot\bq\right]\eqlab{temporal}\\
&&\qquad- \frac{1}{\pi} \int_{t_0}^\infty \!\frac{\dd t}{ t (t+\bq^2) }
 \left[\ga^0 \im F_1(t) -\frac{1}{2Mc} \ga^0 \bgamma\cdot\bq
\im F_2(t) \right] \bigg\}\nn,\\
\De_{ij}\Ga^j &=&Ze\bigg\{\frac{1}{Q^2} \left(\left[\ga_i-\frac{q_i\, \bgamma\cdot\bq}{\bq^2} \right]+\frac{\kappa}{2Mc} \left[
q_0 \gamma_i\ga_0 + \gamma_{ij} q^j - \frac{q_0 q_i \bgamma\cdot\bq \,\ga^0}{\bq^2}\right]\right)
\\&&\qquad -\frac{1}{\pi} \int_{t_0}^\infty \!\frac{\dd t}{ t (t+Q^2) }\left( \left[\ga_i - \frac{q_i\, \bgamma\cdot\bq}{t+\bq^2} \right] 
\im F_1(t) \right.\nn\\
&&\qquad \qquad\left.+ \frac{1}{2Mc} \left[
q_0 \gamma_i\ga^0 + \gamma_{ij} q^j - \frac{q_0 q_i \bgamma\cdot\bq \,\ga^0}{t+\bq^2}\right]
\im F_2(t) \right)\!\bigg\}, \nn
\eea
\end{subequations}
with the photon energy $q_0=\omega/c$.\footnote{Here we made use of the following relations:
\begin{subequations}
\bea
\gamma^{0\al}q_\al&=&-\gamma^0 \bgamma\cdot\bq,\\
\gamma^{i\al}q_\al&=&\ga^i\slashed{q}-q^i=\ga^i \ga^0q_0+\gamma^{ij}q_j,\\
\gamma^{i\al}q_iq_\al&=&-\bgamma\cdot\bq \,\ga^0 q_0.
\eea
\end{subequations}}

The next step is the essential one in the Breit potential derivation:
we perform the semi-relativistic expansion. In doing the expansion of \Eqref{BreitPotentialScatAmp}, we shall neglect the dependence of the denominator on the photon energy, i.e., neglect retardation effects. Our choice for the photon gauge already achieved that the temporal component, \Eqref{temporal}, has no poles in $\w$. Expanding for infinitely $c$, we will derive a potential that is valid up to and including $\mathcal{O}(1/c^2)$. In this way, we will be free of retardation, which start at $\mathcal{O}(1/c^3)$ only. The semi-relativistic expansion of the Dirac spinors is presented in \appref{chap5}{SemRelExpDSpin}. It yields the lepton and nuclear spin vectors, $\bs$ and $\bS$, in the Breit potential. From \Eqref{SachsFFDP} it follows that:
\begin{subequations}
\bea
\im G_E(t) &=& \im F_1(t) + \nicefrac{t}{4(Mc)^2} \im F_2(t), \\
\im G_M(t) &=& \im F_1(t) + \im  F_2(t).
\eea
\end{subequations}
Inverting these equations and expanding for large $c$, one obtains:
\begin{subequations}
\bea
\im F_1(t)&=&\frac{1}{1-\nicefrac{t}{(2Mc)^2}}\big[\im G_E(t)-\nicefrac{t}{(2Mc)^2}\im G_M(t)\big],\\
&\simeq&\im G_E(t)+\nicefrac{t}{(2Mc)^2}\big[\im G_E(t)-\im G_M(t)\big],\\
\im F_2(t)&=&\frac{1}{1-\nicefrac{t}{(2Mc)^2}}\big[\im G_M(t)-\im G_E(t)\big],\\
&\simeq&\left[1+\nicefrac{t}{(2Mc)^2}\right]\big[\im G_M(t)-\im G_E(t)\big].
\eea
\end{subequations}
After the expansion, we can return to our usual convention: $c=1$.

In a next step, we will Fourier transform the momentum-space potential to obtain the coordinate-space potential. We can then identify, f.i., the angular momentum operator, which is defined as: $\boldsymbol{l}=\br \times \boldsymbol{\hat{p}}$. Further details of the Breit potential derivation are moved to \appref{chap2}{DetailedDerivation}. At this point, we just give our final result for the coordinate-space Breit potential with nuclear FFs.
\paragraph*{Coordinate-space Breit Potential with Nuclear Form Factors}
\begin{subequations}
\beq
V_\mathrm{OPE}(\br,\boldsymbol{\hat{p}},t)=\left[V_\mathrm{C}+\Delta V_{\text{rel.C}}+\Delta V_{\mathrm{Y}}+\Delta V_{\mathrm{1}}+\Delta V_{\mathrm{2}}+\Delta V_{\mathrm{3}}+\Delta V_{\mathrm{4}}+\Delta V_{\mathrm{5}}\right](\br,\boldsymbol{\hat{p}},t),
\eeq
with 
\eqlab{BreitPotentialFinal}
\begin{alignat}{3}
&V_\mathrm{C}(\br)&&=-\frac{Z\al}{r},\eqlab{VC}\\
&\Delta V_\mathrm{rel. C}(\br)&&=\frac{Z \al}{2m_r^2}\left[ \pi\,\delta(\br)+\frac{\bL \cdot \bs}{r^3}\right],\eqlab{LSrelC}\\
&\Delta V_{\mathrm{Y}}(\br,t) &&=\frac{Z\al}{\pi r}\int_{t_0}^\infty \frac{\dd t}{t} \im G_E(t) \,e^{-r\sqrt{t}},\eqlab{YukawaCoordinate}\\
&\Delta V_{\mathrm{1}}(\br,\boldsymbol{\hat{p}}) &&=-Z\al\left\{\frac{\pi\,\delta(\br)}{mM}+\frac{1}{2mM}\left(\frac{2\,\boldsymbol{\hat{p}}^2}{r}+\frac{2}{r^2}\frac{\partial}{\partial r} -\frac{l(l+1)}{r^3}\right)+\frac{1}{2M^2}\frac{\bL \cdot \bs}{r^3}\right\},\eqlab{V1}\\
&\Delta V_{\mathrm{2}}(\br) &&=Z\al\,\left\{\frac{\bL\cdot\bS}{r^3}\left[\left(\frac{1}{mM}+\frac{1}{2M^2}\right)+\left(\frac{1}{mM}+\frac{1}{M^2}\right)\kappa\right]\right.\eqlab{V2}\\
&&&\quad\qquad\left.+\frac{1+\kappa}{mM}\left[\frac{2}{3}\,\bs\cdot\bS\,4\pi \,\delta(\br)-\frac{1}{r^3}\left(\bs\cdot\bS-\frac{3 (\bs\cdot \br) (\bS \cdot \br)}{r^2}\right)\right]\right\},\nn\\
&\Delta V_{\mathrm{3}}(\br,\boldsymbol{\hat{p}},t) &&=\frac{Z\al}{\pi} \int_{t_0}^\infty \frac{\dd t}{t} \im G_E(t) \left\{-\frac{1}{8}\left(\frac{1}{m^2}+\frac{1}{M^2}\right)4\pi\,\delta(\br\,)+\frac{1}{8m_r^2}\frac{t \,e^{-r\sqrt{t}}}{r}\eqlab{V3}\right.\\
&&&\qquad+\frac{1}{2mM}e^{-r\sqrt{t}}\left[(2+r\sqrt{t})\,\frac{\boldsymbol{\hat{p}}^2}{r}+\frac{2+2r\sqrt{t}+r^2t}{r^2}\frac{\partial}{\partial r}-\frac{t^{3/2}}{4}\right]\nn\\
&&&\qquad\left.-\frac{e^{-r\sqrt{t}}}{r^3}\left(1+r\sqrt{t}\right)\left[\frac{l(l+1)}{2mM}+\left(\frac{1}{2m^2}+\frac{1}{mM}\right)\bL\cdot \bs\right]\right\},\nn\\
&\Delta V_{\mathrm{4}}(\br,t) &&=\frac{Z\al}{2\pi M^2}\int_{t_0}^\infty \frac{\dd t}{t} \im G_E(t)\,\frac{e^{-r\sqrt{t}}}{r^3}\,(1+r \sqrt{t})\,\bL\cdot\bS,\eqlab{V4}\\
&\Delta V_{\mathrm{5}}(\br,t) &&=\frac{Z\al}{\pi} \int_{t_0}^\infty \frac{\dd t}{t} \im G_M(t) \left\{-\left(\frac{1}{mM}+\frac{1}{M^2}\right)(1+r\sqrt{t})\frac{e^{-r\sqrt{t}}}{r^3} \,\bL\cdot \bS\right.\eqlab{V5}\\
&&&\quad+\frac{1}{mM}\left[\frac{3e^{-r\sqrt{t}}}{r^3}\left\{\frac{1+r\sqrt{t}+r^2t}{3}\,\bs\cdot\bS-\left(1+r\sqrt{t}+\frac{r^2 t}{3}\right)\frac{(\bs\cdot\br)(\bS\cdot\br)}{r^2}\right\}\right.\nn\\
&&&\qquad\qquad\left.\left.-\frac{2}{3}\,\bs\cdot\bS\,4\pi \,\delta(\br)\right]\right\},\nn
\end{alignat}
\end{subequations}
The result agrees with Ref.~\cite[\S83. Breit's equation, Eq.~(83.15)]{LandauLifshitz4} in the structureless limit, i.e., for $F_1=1$ and $F_2=0$, or $G_E=1$ and $G_M=1$.

As explained in \appref{chap2}{DetailedDerivation}, angular averaging with spherical harmonics, cf.\ \Eqref{angularAverage}, will allow us to rewrite terms of the type $(\bs \cdot \br)(\bS \cdot \br)$. Furthermore, we can express products of spin and/or angular momentum operators, e.g., the spin-orbit and the spin-spin coupling, in terms of quantum numbers, see Eqs.~\eref{QN1}, \eref{QN2} and \eref{QN3}. 
In the absence of external fields, the previously introduced $l,j$ and $f$ quantum numbers, cf.\ \secref{chap2}{QuantumNumbersSec}, as well as the \textit{electron} and \textit{nuclear spin quantum numbers} $s$ and $i$ are suited `good' quantum numbers to describe a given atomic state. In Eqs.~\eref{SwavePot} and \eref{PwavePot}, we then give particular potentials for $S$- and $P$-states, i.e., for $l=0$ and $l=1$, respectively. 

The Breit potential derived in here describes the static contributions and nuclear-size effects to FS, HFS and LS, see \appref{chap2}{2appRecoil}. We can split it into the unperturbed Coulomb potential and a semi-relativistic perturbation. To calculate the FSEs, we can proceed in two different ways. We either work in time-independent Schr\"odinger perturbation theory (PT), or we use the relativistic Dirac wave functions as basis for the PT framework. Our choice is to use the non-relativistic Schr\"odinger wave functions. Nevertheless, we want to briefly compare the two approaches and motivate our decision. In \appref{chap2}{WFPT}, we give the Schr\"odinger and Dirac wave functions for the Coulomb problem and review the formalism of first- and second-order (Schr\"odinger) PT for the discrete and continuous spectra.

First, let us give a proper formulation of the bound-state problem we are dealing with. The Coulomb force is the dominant interaction between a lepton and a nucleus in a hydrogen-like atom. In the center-of-mass (CM) frame, the Schrödinger Hamiltonian of the two-body problem reads:
\beq
\hat{H}= 
\frac{\boldsymbol{\hat{p}}^2}{2m}+\frac{\boldsymbol{\hat{p}}^2}{2M}+V
\eqlab{TwoBodyProblem}.
\eeq
Since the mass of the nucleus is finite, both lepton and nucleus are moving inside the atom. Therefore, it is customary to simplify the two-body problem of Newtonian mechanics into a one-body problem by replacing \cite{Grotch1967}:
\beq
m \rightarrow m_r \equiv \frac{mM}{m+M} \qquad  \text{and} \qquad M \rightarrow  \mathbb{M}\equiv m+M,
\eeq
everywhere but in the (gravitational) rest energy. We then have:
\beq
\hat{H}= 
\frac{\boldsymbol{\hat{p}}^2}{2m_r}+V.
\eeq

It is not enough to modify the potential with a semi-relativistic perturbation. As the Schr\"odinger equation is not valid for relativistic particles, the kinetic energy term of the Schr\"odinger Hamiltonian, cf.\ \Eqref{TwoBodyProblem}, has to be expanded relativistically too. Expanding the relativistic equation for the kinetic energy of a particle to lowest order in $1/c$, we find:
\beq
T=c\sqrt{\boldsymbol{\hat{p}}^2+(mc)^2}-mc^2=\frac{\boldsymbol{\hat{p}}^2}{2m}- \frac{\boldsymbol{\hat{p}}^4}{8m^3c^2}+\mathcal{O}(1/c^4).
\eeq
Hence, we need to include the following relativistic corrections to the kinetic energy operator \cite{Barker:1955zz}:
\begin{subequations}
\eqlab{RelEkinfull}
\bea
\Delta V_{\mathrm{rel. E}_\mathrm{kin}}(\boldsymbol{\hat{p}})&=&-\frac{\boldsymbol{\hat{p}}^4}{8m_r^3c^2},\eqlab{RelEkin}\\
\Delta V_{\mathrm{red. Mass}}(\boldsymbol{\hat{p}})&=&\frac{3\boldsymbol{\hat{p}}^4}{8m^2Mc^2}+\frac{3\boldsymbol{\hat{p}}^4}{8M^2mc^2},
\eea
\end{subequations}
which we wrote in an expedient way. 

For the presented OPE potential, the Schr\"odinger Hamiltonian in second non-relativistic approximation is then given by (again using $c=1$):
\begin{subequations}
\eqlab{HamiltonianPert}
\bea
\hat{H}&=&\frac{\boldsymbol{\hat{p}}^2}{2m_r}- \frac{\boldsymbol{\hat{p}}^4}{8m_r^3}+\frac{3\boldsymbol{\hat{p}}^4}{8m^2Mc^2}+\frac{3\boldsymbol{\hat{p}}^4}{8M^2mc^2}+\underbrace{V_C+V_\delta}_{V_\mathrm{OPE}}\,,\\
&=&\underbrace{\frac{\boldsymbol{\hat{p}}^2}{2m_r}+V_C}_{\hat{H}_0}+\underbrace{- \frac{\boldsymbol{\hat{p}}^4}{8m_r^3}+\frac{3\boldsymbol{\hat{p}}^4}{8m^2Mc^2}+\frac{3\boldsymbol{\hat{p}}^4}{8M^2mc^2}+V_\delta }_{\hat{H}_{\delta}}. \eqlab{HamiltonianPert2}
\eea
\end{subequations}
 The perturbative Hamiltonian, $\hat{H}_{\delta}$, for the Coulomb problem at hand consists of \Eqref{RelEkinfull} and the relativistic correction to the Coulomb potential given in \Eqref{LSrelC}. The relativistic Coulomb potential splits into the spin-orbit coupling and the so-called Darwin term. The spin-orbit coupling only affects levels with $l>0$, cf.\ \Eqref{QN1}. The Darwin term, on the other hand, is a $\delta(\br)$ potential and as such is only relevant for the $S$-states, see \Figref{DiracSchrödinger}. Treating this combined semi-relativistic perturbation in first-order Schr\"odinger PT with the Coulomb wave functions, we reproduce the Coulomb energy levels predicted by the Dirac equation, \Eqref{DiracEnergies6}, up to order $(Z\al)^6$. The fascinating interplay of spin-orbit coupling, Darwin term and relativistic kinetic energy operator achieves the FS splitting and the non-trivial degeneracy of levels with equal $n$ and $j$ quantum numbers.

In the following, we will briefly address the Dirac PT and motivate that especially for heavier atoms it makes sense to work with Dirac wave functions \cite{Borie:1982ax}. Even though some works also use relativistic Dirac wave functions to calculate corrections to the hydrogen spectrum \cite{Borie:2004fv,Carroll:2011rv,Indelicato:2012pfa}, for our aim --- the re-derivation of the FSEs by a dispersive technique --- we feel safe to use the simpler non-relativistic Schr\"odinger wave functions.

In contrast to the Schr\"odinger equation, the Dirac theory describes relativistic particles. Therefore, the relativistic corrections to the kinetic energy operator and the unperturbed potential, i.e., Eqs.~\eref{RelEkinfull} and \eref{LSrelC} in case of the Coulomb problem, do not need to be included in the perturbative potential. At first order in relativistic PT, the correction to the energy eigenvalue of the radial Dirac wave functions, $f_{nl}(r)$ and $g_{nl}(r)$, is given by (for a spherically symmetric perturbation):\footnote{For an alternative formulation of a relativistic PT see Refs.~\cite{Kutzelnigg1989_T1,Kutzelnigg1989_T2}.}
\beq
\eqlab{DiracPT}
\Delta E_{nl}=\int dr\,r^2 \left[\vert g(r)\vert^2+\vert f(r)\vert^2\right]_{nl} V_{\delta}(r).
\eeq
Accordingly, we would use $V_\delta(r)=\left[V_\mathrm{OPE}-V_\mathrm{C}-V_\mathrm{rel.C}\right](r)$ and the Coulomb wave functions, see \appref{chap2}{DiracWF}, to calculate the effect of the OPE Breit potential presented in here. The absolute square of radial Dirac wave functions, as it appears in the first-order PT energy shift of \Eqref{DiracPT}, can be expressed as a sum of the squared radial Schr\"odinger wave function and a correction:
\bea
\left[\vert g(r)\vert^2+\vert f(r)\vert^2\right]_{nl}=R^2_{nl}(r)+r^2_{nl}(r),
\eea
where for $n=2$ we find the following corrections for $S$- and $P$-levels, respectively:
\begin{subequations}
\begin{align}
&r_{20}^2(r)=-e^{-r/a}\frac{(Z\al)^2}{64a^3}\left\{ 8  \left(\frac{r}{a}-2\right)^2\left[\gamma_E+\ln \frac{r}{a}\right]+\left(\frac{r}{a}\right)^3-24\left(\frac{r}{a}\right)^2+70\,\frac{r}{a}-54\right\}+\mathcal{O}(Z\al)^4,\nn\\
&r_{21}^2(r)=e^{-r/a}\frac{(Z\al)^2}{576a^3}\left\{-24 \left(\frac{r}{a}\right)^2 \left[\gamma_E+\ln \frac{r}{a}\right]-3\left(\frac{r}{a}\right)^3+56\left(\frac{r}{a}\right)^2+18\,\frac{r}{a}+54\right\}+\mathcal{O}(Z\al)^4.\nn
\end{align}
\end{subequations}
Here, we kept $r/a$ fixed, while expanding in $Z\al$. Since the above corrections are suppressed with respect to the radial Schr\"odinger wave functions by additional powers of $Z\al$, we conclude that they can be ignored for light hydrogen-like atoms. For our purpose, it is therefore sufficient to use the non-relativistic wave functions and time-independent Schr\"odinger PT. On the contrary, for heavier atoms, i.e., large values of $Z$, the relativistic Dirac wave functions become relevant.

In \appref{chap2}{2appRecoil}, we derive finite-size and recoil effects from the potentials in Eqs.~\eref{SwavePot} and \eref{PwavePot}, and present a detailed comparison to the literature. The results are arranged into $2P$ FS, $2P_{1/2}-2S_{1/2}$ LS, $2S$ and $2P$ HFSs, and $P$-level mixing. We reproduce the well-known Fermi energy, charge radius, Friar radius and Zemach radius terms in the LS and the $2S$ HFS. All static contributions are obtained as reviewed in Ref.~\cite{Eides:2000xc}. Our result for the $2P$ FS, the $2P$ HFS and the $P$-level mixing agree with Refs.~\cite{Pachucki:1996zza,Martynenko:2006gz}. In addition, we find finite-size recoil effects at order $(Z\al)^5$. These terms are usually not derived from the Breit potential and, hence, cannot be found in this form in the literature. However, they should be included in the nuclear-pole part of the TPE corrections. Therefore, we will perform a matching of the effects from one- and two-photon exchange in \secref{chap5}{matchingOTPE}. 

The dispersive approach to the OPE Breit potential with finite nuclear size, where we use once-subtracted DRs to express the nuclear FFs through their discontinuities, allows for an easy transition to the Breit potential of OPE with one-loop leptonic VP. In \appref{chap5}{2appVP}, we outline the necessary modification of the Breit potential and calculate the one-loop eVP contributions to the $\mu$H spectrum. The results are summarized in Table \ref{eVPTab} and match the literature.

\section{Exact Finite-Size Effects} \seclab{Exact}

\subsection*{or: Breakdown of the Lamb Shift Expansion in Moments of Charge Distribution}

In the previous Section, we derived the nuclear FSEs and presented them, as usual, in terms of moments of charge, magnetization and convoluted distributions, cf.\ \Eqref{FSEs} and \secref{chap2}{2appRecoil}. In this Section, we will present an alternative formulation which is exact in the sense that it refrains from expanding in the moments. Limitations of the finite-size corrections to the LS in terms of charge radii will become apparent. The main ingredient is the Yukawa-type electric FF correction to the Coulomb potential given by the coordinate-space potential in \Eqref{YukawaCoordinate} or the momentum-space potential  in \Eqref{yukawaQ}. Particular scenarios for the breakdown will be presented in the subsequent Section.

\begin{figure}[tbh]
\centering
\includegraphics[width=8cm]{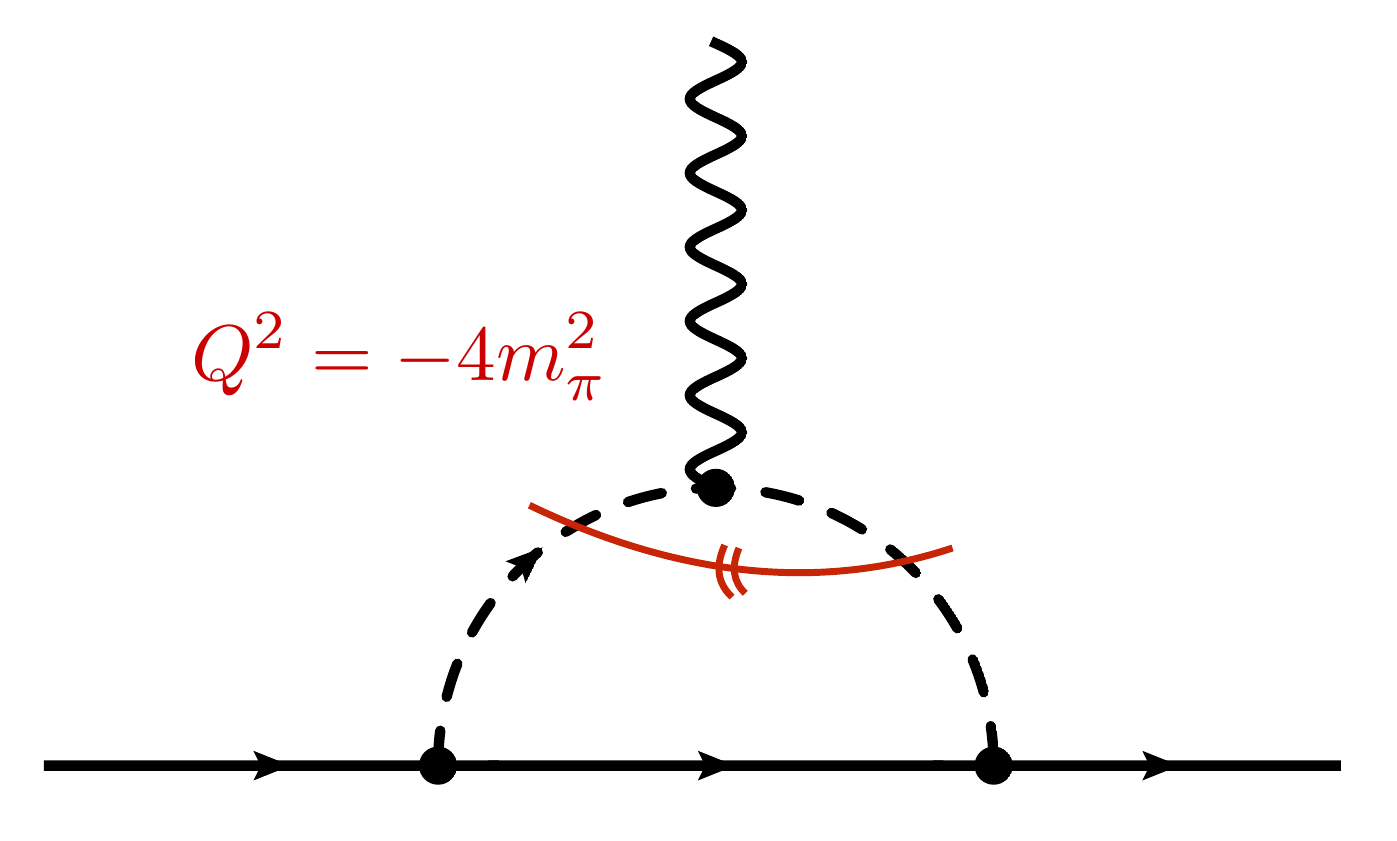}
    \caption{Correction to the electromagnetic vertex. The cut indicates that the intermediate pion-pion pair can go on-shell if $Q^2=-4m_\pi^2$.  \label{fig:FFVertexCorrPion}}  
\end{figure}  

Let us start with the classic LS as it is deduced from the coordinate-space potential of  \Eqref{YukawaCoordinate} with the non-relativistic Schr\"odinger wave functions. To first order in PT, the energy shift can be written as an integral over the imaginary part of the electric Sachs FF:
 \beq
 E^{\langle\mathrm{eFF}\rangle(1)}_\text{LS}= -\frac{(Z\al)^4 m_r^3}{2\pi} \int_{t_0}^\infty \!\!\dd t \, \frac{\im G_E (t)}{(\sqrt{t}+Z\al m_r)^4} .
\eqlab{rmsLSa}
\eeq
The imaginary part corresponds to the FF discontinuity across the branch cut in the time-like region, which starts from the lowest particle-production threshold $t_0$. Figure \ref{fig:FFVertexCorrPion} shows the lightest hadronic contribution to the nucleon FF. The produced pion-pion intermediate state is on-shell for photon virtualities of $Q^2=-4m_\pi^2$. This charged-pion production threshold represents a serious restriction for fitting $ep$ scattering data beyond $Q^2\approx-0.078$ GeV$^2$.

\begin{figure}[tbh]
\centering
\includegraphics[scale=0.65]{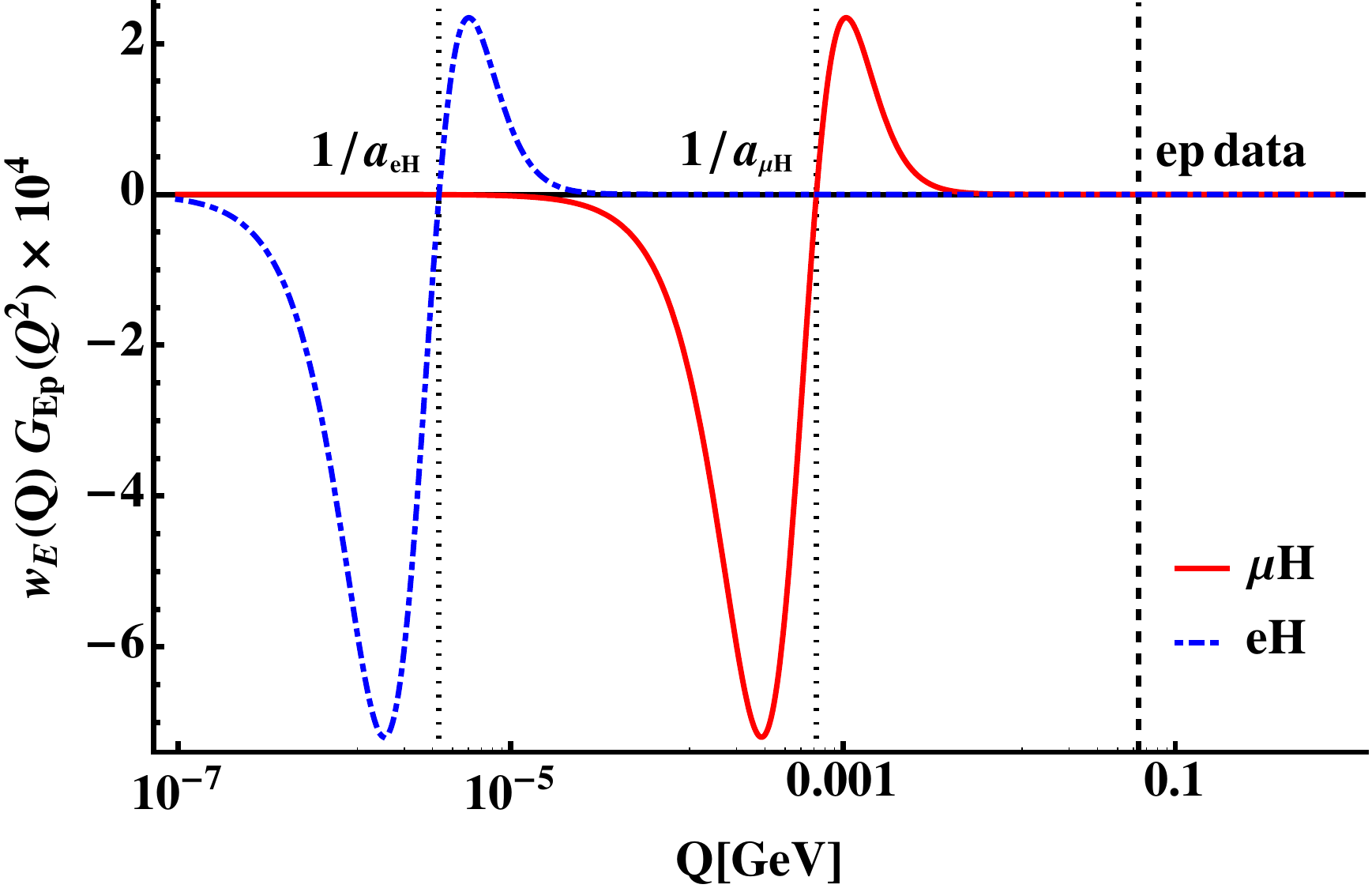}
\caption{Integrand of the first-order perturbation theory contribution to the Lamb shift [cf.\ Eqs.~\eref{wG} and \eref{iST}]
       in electronic hydrogen (blue dash-dotted line) and muonic hydrogen (red solid line) for the dipole form factor, $G_{Ep}= (1+Q^2/0.71\,\mathrm{GeV}^2)^{-2}$. The dotted vertical lines
       indicate the inverse Bohr radii of the two hydrogens, while the dashed line indicates the onset of data from
       electron-proton scattering.\label{fig:IntegrandLS}}
\end{figure}

For small inverse Bohr radii, $a^{-1}=Z\al m_r$, we expand \Eqref{rmsLSa} in the moments of the charge distribution, using the following (Lorentz-invariant) definition:
\beq
\eqlab{rmsdef}
\langle r^N\rangle_E =
\frac{(N+1)!}{\pi}\int_{t_0}^\infty\!\! \dd t \, \frac{\im  G_E(t)}{t^{N/2+1 }  },
\eeq
and arrive at:
\begin{subequations}
 \bea
 E^{\mathrm{\langle eFF\rangle }(1)}_\text{LS} &=& -\frac{(Z\al)^4 m_r^3}{12} \sum_{k=0}^\infty 
 \frac{(-Z\al m_r)^{k}}{k!} \langle r^{k+2}\rangle_E,\quad\eqlab{LSInfSeries}\\
 &\approx&-\frac{(Z\al)^4 m_r^3}{12} \left[ \langle r^2\rangle_E - Z\al m_r 
 \langle r^3\rangle_E \right].\eqlab{LSexp}
\eea
\end{subequations}
From the denominator of \Eqref{rmsLSa} one can see that the convergence radius of the power-series expansion in moments is limited
by $t_0$, i.e., the proximity of the nearest particle-production threshold. Since QED corrections to the e.m.\ interaction vertex are already separated from the nuclear FFs, we expect $t_0$ to be a hadronic scale, of which the pion mass is the lowest, see \Figref{FFVertexCorrPion}.
If this is true, the
series should converge quickly for hydrogen, and in fact, for most of the hydrogen-like systems. In \appref{chap2}{FFparam}, we check that for all popular FF parametrizations the expansion in moments is appropriate. In \secref{chap2}{ToyModel}, however, we will present a FF model which breaks down the expansion in moments of charge distribution and at the same time resolves the proton charge radius puzzle.

One of the first proposals for an explanation of the proton radius puzzle was suggested by de R\'ujula \cite{DeRujula:2010dp}, who used the expanded formulas, see \Eqref{LambShift}. As explained earlier, the subleading FSEs of order $(Z\al)^5$ come with an additional factor of the lepton-nucleus reduced mass and thus become relevant in muonic atoms only. Therefore, a large Friar radius, \Eqref{Friar}, would change the $\mu$H LS and leave the H LS almost unaffected.\footnote{At this point, one has to remember that in the analysis of the $\mu$H experiment the Friar radius is substituted from $ep$ scattering, as discussed in the last paragraph of \secref{chap2}{Radii}.} In order to generate such radius, de R\'ujula constructed an electric charge distribution from an interpolation between the charge densities of a single-pole and a dipole. The resulting charge distribution has an extended tail and, indeed, the corresponding Friar radius is large. Nevertheless, his model cannot explain the proton radius puzzle. First of all, it was shown to be incompatible with the empirical electric Sachs FF extracted
from $ep$ scattering \cite{Miller2011,Distler:2010zq}.
Furthermore, we verified that the $\mu$H LS in de R\'ujula's model
is not described correctly by the standard formulas, Eqs.~\eref{LambShift} and \eref{LSexp}, and that the infinite series of moments in \Eqref{LSInfSeries} does not provide any significant reduction of the proton radius discrepancy in this model. 

For convenience, \Eqref{rmsLSa} can be rewritten and expressed through the electric Sachs FF or the spherically-symmetric
 charge distribution. The alternative formulas read:\footnote{Starting from  
\beq
\frac{1}{\pi } \int_{t_0}^\infty\!\dd t\, W(t) \im G_E(t) = \int_0^\infty \!\dd Q\,w_E(Q) \, G_E(Q^2),
\eeq
we plug in the DR for the electric Sachs FF,
\beq
\eqlab{DRel}
G_E(Q^2) = \frac{1}{\pi} \int^\infty_{t_0} \! \! \dd t\, \frac{\im G_E(t)}{t+Q^2}\,,
\eeq
and realize that $W(t)$ is the Stieltjes integral transform of $w_E(Q)$, i.e.:
\beq
W(t) = \int_0^\infty \!\! \dd Q \, \frac{w_E(Q)}{t+Q^2}.
\eeq
We recast \Eqref{rmsLSa} in terms of $G_E(Q^2)$ by computing the inverse Stieltjes transform \cite{Schwarz:2004mv} of
\beq
W(t) =  -\frac{(Z\al)^4 m_r^3}{2(\sqrt{t}+\al m_r)^4}, 
\eeq
which is given by:
\begin{subequations}
\bea
w_E(Q)&=&  \frac{Q}{i\pi} \lim_{\varepsilon\to 0} \left\{W(-Q^2-i\varepsilon)-W(-Q^2+i\varepsilon)\right\},\\
&=&  - \frac{4 (Z\al)^5 m_r^4}{\pi}  \,
\frac{Q^2 [ (Z\al m_r)^2-Q^2]}{\left[(Z\al m_r)^2+Q^2\right]^4}.
\eea
\end{subequations}}
 \begin{subequations}
 \eqlab{wGall}
   \bea
   \eqlab{wG}
 E^{\langle \mathrm{eFF}\rangle(1)}_\text{LS} & =& \int_0^\infty \!\dd Q\,w_E(Q) \, G_E(Q^2),\\
&=& -(Z\al)^4 m_r^3 \;\frac{\pi}{3}  \int_0^\infty\! \dd r \, r^4 e^{-r/a } \varrho_E (r),\quad\eqlab{rhoLS}
 \eea
with 
\bea
\eqlab{iST}
w_E(Q)&=&  - (Z\al)^5 m_r^4\;\frac{4}{\pi}  \,
\frac{Q^2 [ (Z\al m_r)^2-Q^2]}{\left[(Z\al m_r)^2+Q^2\right]^4},
\eea
or the Laplace transform of the FF discontinuity:
 \beq
 \varrho_E (r) = \frac{1}{(2\pi)^2\, r} \int^\infty_{t_0} \! \! \dd t\, \im G_E(t)\,  e^{-r\sqrt{t}}. \eqlab{rhoE}
 \eeq
  \end{subequations}
  The expression in terms of the charge density, \Eqref{rhoLS}, is the simplest and has the most intuitive
 interpretation: the first-order LS is given by the mean-square radius cut off at the Bohr radius by the Coulomb wave function. Indeed, it is simply the LO charge radius term $\langle r^2 \rangle_E$, cf.\ \Eqref{LambShift}, replaced by 
 $\langle r^2 e^{-r/a }  \rangle_E$. Equation~\eref{wG} is slightly misleading, despite the $(Z\al)^5$ prefactor the FSE is of order $(Z\al)^4$. This is because the weighting function $w_E(Q)$, \Eqref{iST}, cannot be expanded in $\al$, as the successive integral would be infrared divergent.

\begin{figure}[tbh]
\centering
       \includegraphics[scale=0.6]{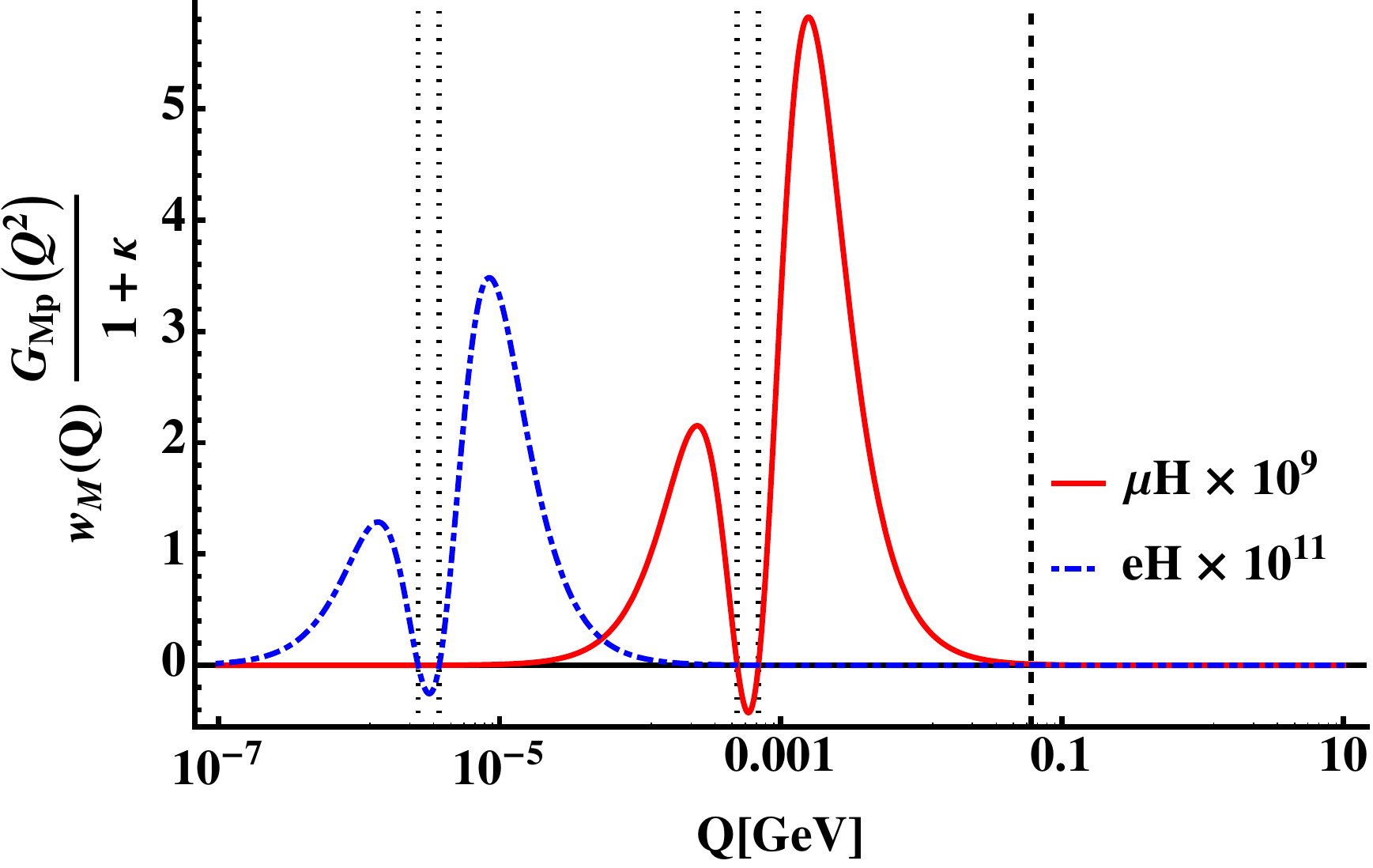}
                     \caption{Integrand of the first-order perturbation theory contribution to the $2S$ hyperfine splitting, cf.\ \Eqref{HFSexact}, in electronic hydrogen (blue dash-dotted line) and muonic hydrogen (red solid line) for the dipole form factor, $G_{Mp}= (1+\kappa)(1+Q^2/0.71\,\mathrm{GeV}^2)^{-2}$. The dotted vertical lines
       indicate scales related to the inverse Bohr radii of the two hydrogens [$1/a$ and  $1/(\sqrt{2}a)$], while the vertical dashed line indicates the onset of data from
       electron-proton scattering.\label{fig:IntegrandHFS}}
\end{figure}

The expression in terms of the FF, Eq.~(\ref{eq:wG}), can also be derived from the momentum-space potential in \Eqref{yukawaQ}. The weighting function then arises as a convolution 
of the hydrogen momentum-space wave functions, cf.\ \Eqref{ConvolutionsWFMom}. Such approach is followed for the derivation of the Wichmann-Kroll contribution 
in Refs.~\cite{Karshenboim:2010cq,Karshenboim:2010cp}.

In \Figref{IntegrandLS}, we are plotting the integrand of the first-order perturbation theory contribution to the LS, given by Eqs.~\eref{wG} and \eref{iST}. The two curves correspond to H and $\mu$H, respectively. They are plotted assuming a dipole FF for the proton. For both hydrogens one observes regions of large cancelation around their inverse Bohr radius scale. Taking $G_E$ to be constant, the cancelation would be exact and the LS would vanish. This is easiest seen from the momentum-space potential, \Eqref{yukawaQ}, which is proportional to the once-subtracted dispersion integral, cf.\ \Eqref{subtractedDR}. The cancelation regions are well separated and also well below the onset of existing $ep$ scattering data, $ Q^2 > 0.004\, \mbox{GeV}^2$. Any relatively small
 variation in the FF at these low-$Q$ scales may lead to significant effects in the LS. To define a proper charge radius, which could be determined from both the atomic and
scattering experiments, it is therefore mandatory to decompose the FFs into ``smooth'' and
 ``non-smooth'' parts. The contribution of the former can then be expanded in moments,
while the  latter must be treated exactly. In the follow-up Section, we will demonstrate how to use this fact and find a non-smooth FF modification that solves the proton charge radius puzzle by breaking the expansion of the FSEs.

Analogous to the LS case, we establish formulas for the exact FSEs in the HFS. We start from the part of the Breit potential, \Eqref{HFSpotQ}, which depends on the nuclear spin and the magnetic Sachs FF. Treating this potential to first-order in PT, we obtain the $1S$ HFS:
\begin{subequations}
\eqlab{HFS1PT1S}
 \bea 
 E^{\langle\mathrm{mFF}\rangle(1)}_{\mathrm{ HFS}} (1S)& =& -\frac{E_\mathrm{F}(1S)}{\pi}  \int_{t_0}^\infty \!\!\dd t \, \left[\frac{1}{t}-\frac{1}{[2Z\al m_r+\sqrt{t}]^2}\right]  \frac{\im G_M (t)}{1+\varkappa},\qquad\quad\\
 &=& -E_\mathrm{F}(1S)\left\{1- \frac{a^3}{2\pi} \int_0^\infty \dd Q \,Q^2\,w_{1S}(Q)\,\frac{G_M(Q^2)}{1+\varkappa}\right\},\\
&=& -E_\mathrm{F}(1S)\left\{1- \pi a^3 \int_0^\infty \dd r\, r^2 \big[R_{10}(r)\big]^2 \,\varrho_M(r)\right\},
\eea
\end{subequations}
and the $2S$ HFS:
\begin{subequations}
\eqlab{HFS1PT}
 \bea
 E^{\langle\mathrm{mFF}\rangle(1)}_{\mathrm{ HFS}} (2S)& =& -\frac{E_\mathrm{F}(2S)}{\pi}  \int_{t_0}^\infty \!\!\dd t \, \left[\frac{1}{t}-\frac{2t+(Z\al m_r)^2}{2[\sqrt{t}+Z\al m_r]^4}\right]  \frac{\im G_M (t)}{1+\varkappa},
\eqlab{HFS11}\\
&=& -E_\mathrm{F}(2S)+ \int_0^\infty \dd Q \,w_M(Q)\,\frac{G_M(Q^2)}{1+\varkappa},\eqlab{HFSexact}\\
&=& -E_\mathrm{F}(2S)\left\{1- 8\pi a^3 \int_0^\infty \dd r\, r^2 \big[R_{20}(r)\big]^2 \,\varrho_M(r)\right\},\eqlab{HFS1PTc}
\eea
\end{subequations}
 with the convoluted momentum-space wave function $w_{1S}(Q)$ given in \Eqref{ConvolutionsWFMom1S}, the weighting function
\beq
w_M(Q)=\frac{4E_\mathrm{F}}{\pi a} \frac{Q^2\left[ (Z\al m_r)^2-Q^2\right]\left[ (Z\al m_r)^2-2Q^2\right]}{\left[(Z\al m_r)^2+Q^2\right]^4},
\eeq 
the radial Coulomb wave functions as defined in \secref{chap2}{WFPT}, and the magnetization distribution
\beq
 \varrho_M (r) =\frac{1}{(2\pi)^2\, r} \int^\infty_{t_0} \! \! \dd t\, \frac{\im G_M(t)}{1+\varkappa}\,  e^{-r\sqrt{t}} .\eqlab{rhoM}
 \eeq
 
The integrand of \Eqref{HFSexact} is plotted in Fig.~\ref{fig:IntegrandHFS} with a dipole FF. Again, for small values of $Q$ one finds regions of enhancement, and again, they are below the onset of $ep$ scattering data and separated for the two hydrogen. Therefore, it is necessary to evaluate the effect of soft non-smooth FF contributions to the HFS with the exact formalism presented above. Other exact formulas for the FSEs on the $1S_{1/2}$, $2S_{1/2}$, $2P_{1/2}$ and $2P_{3/2}$ levels can be deduced from \appref{chap2}{2appVP}.

\section{A Form Factor Model Resolving the Proton Radius Puzzle} \seclab{ToyModel}

In the previous Section, we uncovered a possible limitation in the usual accounting of FSEs and introduced the exact (un-expanded) formulas for the LS and the HFS at first-order in PT. Here, we will modify customary parametrizations of the electric Sachs FF by adding tiny, non-smooth contributions in the region between $1$ and $50$ MeV. These modifications require the exact formalism, Eqs.~\eref{wG} and \eref{iST}, and are chosen in a way that the proton charge radius discrepancy vanishes. Our improved toy model, based on the fit of $ep$ scattering data by \citet{Arrington:2006hm}, was published in Ref.~\cite{Hagelstein:2016jgk}. In addition, we will present a modification of the dipole FF which is likewise breaking its convergence, see \appref{chap2}{DipoleFF}.


We assume the electric FF to separate into a smooth ($\ol G_E$) and a non-smooth part ($\widetilde G_E$), such that:
 \beq
  G_E(Q^2) = \ol G_E(Q^2) + \widetilde G_E(Q^2).\eqlab{newFF}
  \eeq 
We take a well-known FF parametrization \cite{Arrington:2006hm} for the smooth part,
\beq
\ol G_E(Q^2)=\frac{1}{1+\frac{3.478 \,Q^2}{1-\frac{0.140 \,Q^2}{1-\frac{1.311 \,Q^2}{1+\frac{1.128 \,Q^2}{1-0.233 \,Q^2}}}}},
\eeq
and describe the non-smooth part of the FF as:
\beq
\widetilde G_E(Q^2)=\frac{A\,Q_0^2\,Q^2\left[Q^2+\veps^2\right]}{\left[Q_0^2+Q^2\right]^4},\eqlab{fluc}
\eeq
where $A$, $\veps$ and $Q_0$ are real parameters. First and foremost, the fluctuation does not affect the charge: $\widetilde G_E(0)=0$. Furthermore, the functional form in \Eqref{fluc} has all poles at negative $Q^2$ (time-like region) and, hence, complies with the analyticity constraint on the FF.\footnote{This requirement can be easily seen from the DRs for the e.m.\ FFs, cf.\ Eqs.~\eref{DRel} and \eref{unsubtractedDR}.}

\begin{figure}[t] 
    \centering 
       \includegraphics[width=8.5cm]{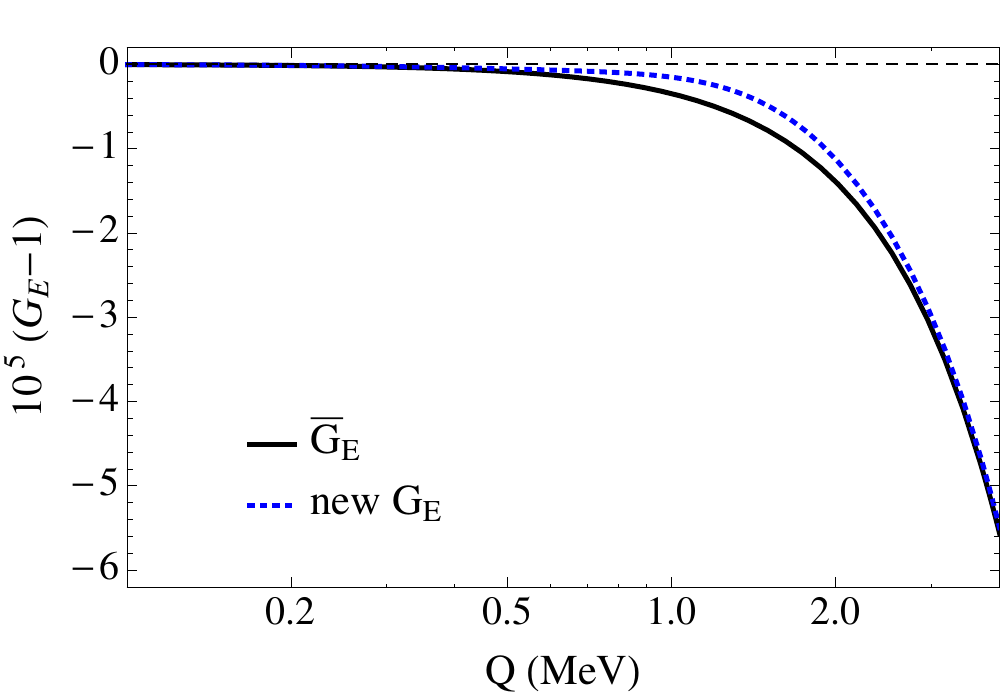}
       \caption{Modification of the Arrington and Sick form factor: The solid black curve shows the empirical form factor, $\ol G_E(Q^2)-1$, from Ref.~\cite{Arrington:2006hm}. The dotted blue curve is the modified form factor, $G_E(Q^2)-1$, discussed in the text \cite{Hagelstein:2016jgk}.}
              \label{fig:GEminus1}
\end{figure}

\noindent As one can see from \Figref{IntegrandLS}, if there is a small missing effect in the FF  responsible for the puzzle, 
it must be localised near one of the two inverse  Bohr radii, where its impact is maximized. Since the results from $ep$ scattering and H spectroscopy are in rather good agreement, it is most promising to search for the missing effect on the $\mu$H side. We fix the position of the FF fluctuation at $Q_0=1.6\, \mbox{MeV}$, where the weighting function $w_E(Q)$, Eq.~\eref{iST}, is especially sensible for $\mu$H, cf.\ Figs.~\ref{fig:IntegrandLS} and \ref{fig:correction}

We then fix the other parameters, $A$ and $\veps$, such that \Eqref{wG}, evaluated with the modified FF of \Eqref{newFF}, reproduces the empirical values for the hydrogen LSs:
 \begin{subequations}
 \eqlab{LSFS}
 \bea
&& E^{\mathrm{eFF}\,(\mathrm{exp.})}_\mathrm{LS}(e\mathrm{H}) = -0.620(11) \, \mbox{neV}, \label{LSeH}\\
&& E^{\mathrm{eFF}(\mathrm{exp.})}_\mathrm{LS} (\mu\mathrm{H})  = -3650(2) \,  \mbox{$\upmu$eV} \label{LSmuH}.
\eea 
 \end{subequations}
Note that these are not the LSs observed in experiment but only the finite-size contributions given in \Eqref{LambShift}, where for the radii we used:
\begin{subequations}
\eqlab{RadiiToyModel}
\bea
R_{Ep}(e\mathrm{H}) &=& 0.8758(77) \, \mbox{fm \cite{Mohr:2012aa}}
\eqlab{reH}
, \\
R_{Ep} (\mu\mathrm{H}) & = & 0.84087(39) \, \mbox{fm \cite{Antognini:1900ns,Antognini:2012ofa}},
\eea 
\end{subequations}
and 
\beq
\langle r^3\rangle_{Ep(2)} = 2.78(14) \,\mathrm{fm}^3 \mbox{ \cite{Borie:2012zz}}.
\eeq

    \begin{figure}[t] 
    \centering 
       \includegraphics[width=6.8cm]{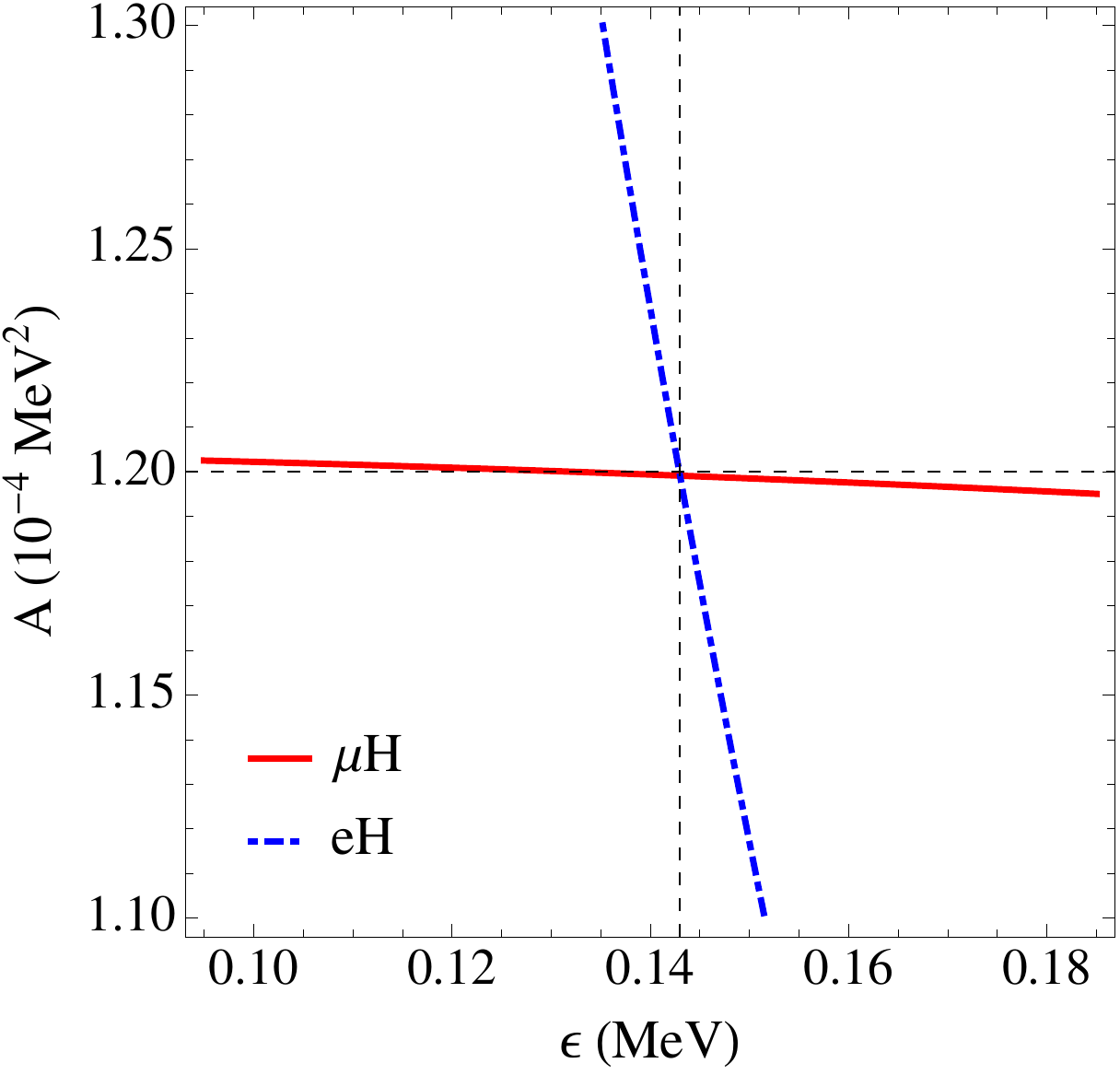}
       \caption{Modification of the Arrington and Sick form factor: Parameters of $\widetilde G_E$ for which the electronic-hydrogen (blue dot-dashed) and muonic-hydrogen (red solid) Lamb shifts of \Eqref{LSFS} 
       are reproduced. For fixed $Q_0=1.6$ MeV, we find $A=1.2\times10^{-4}$ MeV$^2$ and $\veps=0.143$ MeV as indicated by the dashed lines.}
              \label{fig:Parameter}
\end{figure}

Our parameter choice, $A=1.2\times10^{-4}$ MeV$^2$ and $\veps=0.143$ MeV, for which the modified FF complies with the H and the $\mu$H LSs, is depicted in \Figref{Parameter}. Figure \ref{fig:correction} shows the position of the fluctuation $\widetilde G_E(Q^2)$ right on top of the first extremum of the $\mu$H weighting function. Obviously, the constructed fluctuation of the FF lies almost exclusively in the region below the $ep$ data ($Q<63$ MeV), thus, is not affecting the quality of their fit.

To quantify the solution of the proton radius puzzle offered by our toy model further, we calculate the second and third moments of the modified FF, their ``would be" effect on the LS and the un-expanded
LS. The second and third moments of the FF fluctuation presented in \Eqref{fluc} are given by:\footnote{In general, the $N$-th moment of the charge (magnetization) distribution, $\varrho_E$ ($\varrho_M$), is defined as:
\beq
\langle r^N\rangle  \equiv4\pi \int_0^\infty \dd r\, r^{N+2}
\varrho(r)=
\frac{\Gamma(N+2)}{\pi}\int_{t_0}^\infty\! \dd t \, \frac{\im  G (t)}{t^{N/2+1 }  },
\eeq
where in the last step we made use of the DR for the electric  (magnetic) Sachs FF \cite{Hagelstein:2015egb}, $G_E$ ($\nicefrac{G_M}{1+\kappa}$), and Eqs.~\eref{rhoE} and \eref{rhoM}, respectively.
The density distributions are chosen to be normalized as $\langle r^0\rangle=1$. The even moments can be written as derivatives of the FFs:
\beq
\langle r^{2N}\rangle =(-1)^{N} \,\frac{(2N+1)!}{N!}\, G^{(N)}(0),
\eeq
whereas the odd moments have an integral representation:
\eqlab{oddMoments}
\bea
\langle r^{2N-1}\rangle &=&(-1)^{N}\,(2N)!\,\frac{2}{\pi}\int _0^\infty \frac{\dd Q}{Q^{2N}} \Big[ G(Q^2) -\sum_{k=0}^{N-1} \frac{Q^{2k}}{k!} G^{(k)} (0) \Big], \nn\\
&=& (-1)^{N}\,(2N)!\,\frac{2}{\pi}\int _0^\infty \frac{\dd Q}{Q^{2N}} \Big[ G(Q^2) -\sum_{k=0}^{N-1} \frac{(-Q^2)^k}{(2k+1)!} \langle r^{2k}\rangle\Big].
\eea
The latter expressions, we derived by means of the inverse Stieltjes integral transform \cite{Schwarz:2004mv}. Similar expressions hold for the Zemach moments of the convoluted charge and magnetization distributions. }
\begin{subequations}
\eqlab{oddmoments}
\bea
\widetilde{\langle r^2\rangle}_E&\equiv &-6 \frac{\dd}{\dd Q^2} \widetilde G_E(Q^2)\Big|_{Q^2= 0}
=-\frac{6 A\veps^2}{Q_0^6},\quad\\
\widetilde{\langle r^3\rangle}_E&\equiv &\frac{48}{\pi} \int_0^\infty \!\frac{\dd Q}{Q^4}\,
\left\{ \widetilde G_E(Q^2) +\sixth \widetilde{\langle r^2\rangle}_E Q^2\right\},\nn\\
&=&15 A(Q_0^2-7\veps^2)/2Q_0^7.
\eea
\end{subequations}
The results are then summarized in Table \ref{TableArrington}. One can read off that the expansion in moments breaks down for the modified FF contribution to $\mu$H, as it was anticipated. Due to the FF modification, the charge radius is slightly shrunken and the third moment increased.

 \begin{figure}[t] 
    \centering 
       \includegraphics[width=11cm]{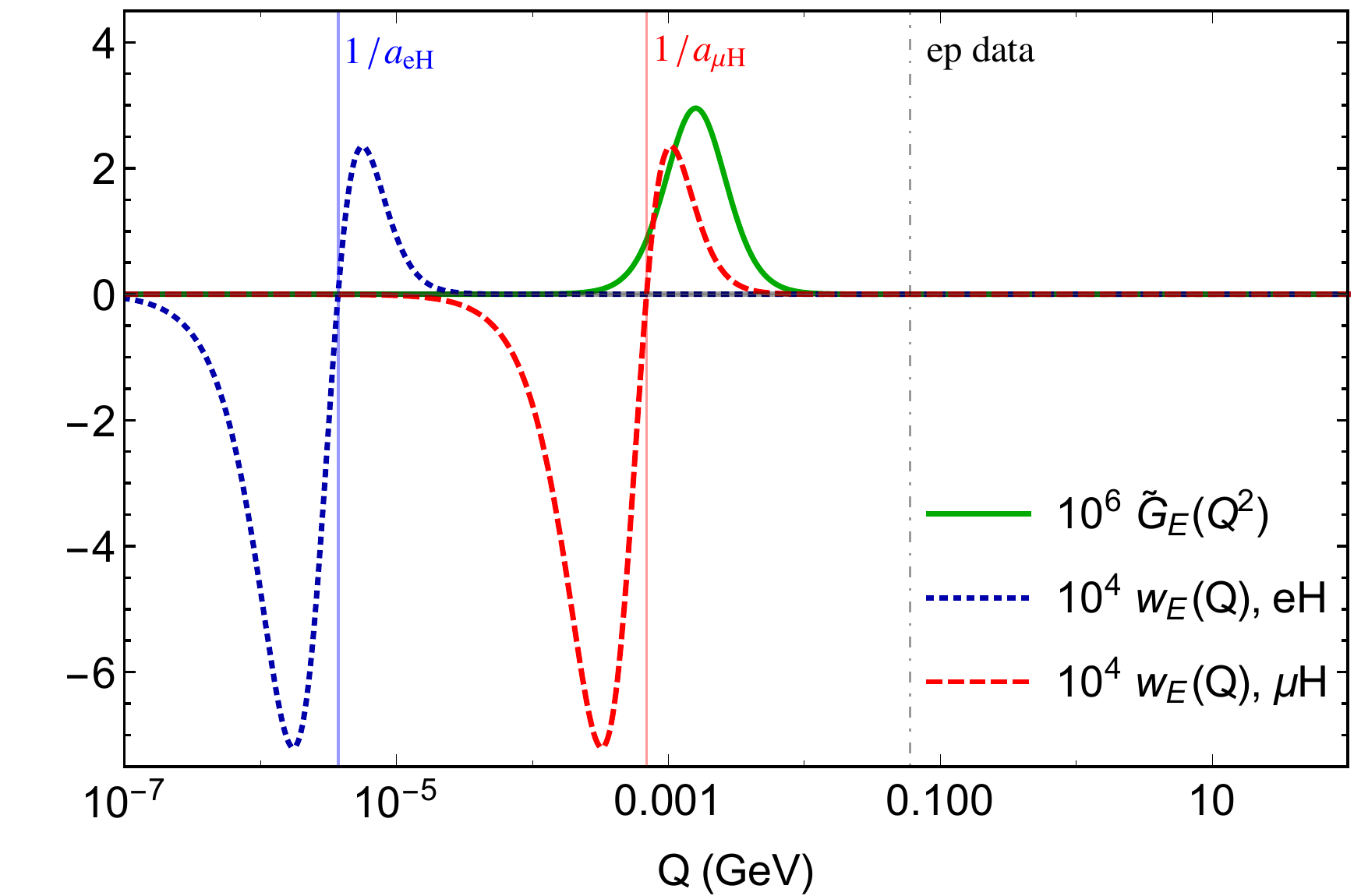}
       \caption{Modification of the Arrington and Sick form factor: The correction, $\widetilde G_E(Q^2)$, for $Q_0=1.6$ MeV, $A=1.2\times10^{-4}$ MeV$^2$ and $\veps=0.143$ MeV (solid green), and the weighting function, $w_E(Q)$, for electronic hydrogen (blue dotted) and muonic hydrogen (red dashed) as functions of $Q$.  The dot-dashed line indicates the onset of electron-proton scattering data.}
              \label{fig:correction}
\end{figure}

\noindent Let us now study the physical plausibility of the suggested FF modification.
The absolute correction to the FF is extremely tiny:
 \beq
\big\vert\widetilde G_E/\, \ol G_E\big\vert <3\times10^{-6}.
\eeq
However, since the FF at low $Q$ is approximately $1$, it was suggested by \citet{Arrington:2016rka} and \citet{KottmannAntognini} that a comparison of our correction to the deviation of the original FF from $1$ would be more meaningful. Indeed, the FF modification presented in here compares good to a point-like proton:
\beq
\big\vert\widetilde G_E/\, (\ol G_E-1)\big\vert < 0.57\,.
\eeq
In addition, we have $G_E(Q^2)\leq 1$ for all $Q$, as called for in Ref.~\cite{Arrington:2016rka}. This is shown in \Figref{GEminus1}, where the FF and the toy model FF are compared. 

 \begin{table}[htb]
 \centering
 \caption{Lamb shift and moments corresponding to the modified Arrington and Sick form factor, with $Q_0=1.6$ MeV, $A=1.2\times10^{-4}$ MeV$^2$ and $\veps=0.143$ MeV.}
 \label{TableArrington}
 \begin{small}
\begin{tabular}{|c|c|c|c|c|}
\hline
 \rowcolor[gray]{.7}
&{\bf Eq.}&$\boldsymbol{\ol G_E}$&$\boldsymbol{\widetilde G_E}$&$\boldsymbol{ G_E}$\\
\hline
$\langle r^2\rangle_E \, [\mbox{fm}^2]$&\eref{REpDeriv}&$(0.9014)^2$&$-(0.1849)^2$&$(0.8823)^2$\\
$\langle r^3\rangle_E \,[\mbox{fm}^3]$&\eref{oddmoments}&$(1.052)^3$&$(8.539)^3$&$(8.544)^3$\\
 \rowcolor[gray]{.95}
Lamb-shift, exact & \eref{wG} & && \\
 \rowcolor[gray]{.95}
$E_\mathrm{LS}^{\mathrm{eFF}(1)}(\mathrm{H}) [\text{neV}]$&&$-0.6569$&$0.0370 $&$-0.6200$\\
 \rowcolor[gray]{.95}
$E_\mathrm{LS}^{\mathrm{eFF}(1)}(\mu\mathrm{H})[\upmu\text{eV}]$&&$-4202$&$552$&$-3650$\\
Lamb-shift, expanded & \eref{LSexp} & && \\
$E_\mathrm{LS}^{\mathrm{eFF}(1)}(\mathrm{H})[\text{neV}]$ &&$-0.6569$&$0.0371$&$-0.6198$\\
$E_\mathrm{LS}^{\mathrm{eFF}(1)}(\mu\mathrm{H})[\upmu\text{eV}]$&&$-4202$&$11542$&$7340$\\
\hline
\end{tabular}
\end{small}
\end{table}

Of course, we do not insist that the presented toy models, see also \appref{chap2}{DipoleFF}, have anything to do with reality. We merely want to demonstrate how a tiny non-smooth contribution to the proton electric Sachs FF, localised at low $Q$, may shrink the convergence radius of the FF and invalidate the expansion of the LS  in moments of charge distribution. The variety of scenarios presented in here and in Refs.~\cite{Gryniuk:2015aa,Gryniuk:2016gnm}, either based on the FF parametrization of \citet{Arrington:2006hm} or the dipole FF, should emphasize how easy it is to find a fluctuation suited to resolve the proton radius puzzle. New physics and the inclusion of new light particles might be able to provide a physical justification for such a non-smooth contribution to the FF. Until then, we can only warn against a too optimistic view of uncertainties in the charge radius extractions. Similar non-smooth corrections might also affect the magnetic FF and the e.m.\ lepton vertex. Likewise, the expansion of the HFS in moments should be viewed with caution. 

In the next Section, we will summarize our results obtained from the Breit potential with nuclear FFs. A short outlook will be given and possible candidates for non-smooth contributions to the FFs are nominated.

\section{Summary and Conclusion}

  \begin{figure}[b] 
    \centering 
       \includegraphics[scale=0.55]{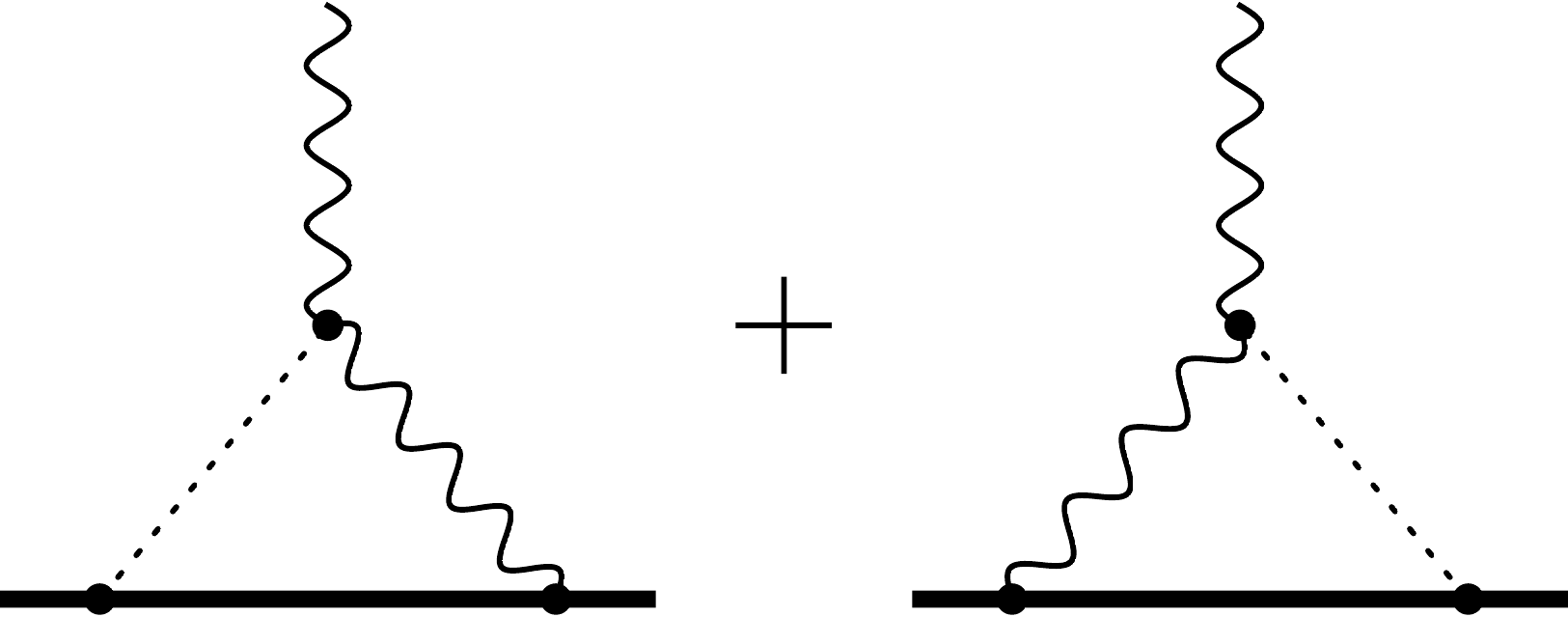}
       \caption{Proton vertex with axion exchange.}
              \label{fig:Axion}
\end{figure}

In the present Chapter, we derived the Breit potential from OPE with nuclear FF dependent e.m.\ vertex. The coordinate-space potential is given in \Eqref{BreitPotentialFinal} and the momentum-space potential is given in \Eqref{BreitMomentum}. Due to our dispersive ansatz, only little modifications of the Breit potential are needed to calculate QED or electroweak corrections instead of finite-size corrections. We replaced the FF discontinuities by the imaginary part of one-loop eVP and presented a re-evaluation of the Uehling potential and further eVP corrections (\appref{chap2}{2appVP}). In \appref{chap2}{2appRecoil}, we calculated finite-size and recoil corrections to the spectrum of Coulomb energies. For the first time, we wrote down finite-size recoil effects at order $(Z\al)^5$, which are proportional to the first moments of the charge and convoluted charge distributions. Such effects are preferably covered by the nuclear-pole part of the TPE, as we will explain in \secref{chap5}{matchingOTPE}.


The usual accounting of FSEs involves an expansion in moments of charge and magnetization distributions, cf.\ \Eqref{FSEs}. Meaning, the FSEs are expressed in terms of the charge radius, the Friar radius, the Zemach radius, etc. In \secref{chap2}{Exact}, we presented the alternative (un-expanded) formulas for the first-order PT contributions of the nuclear finite size to the LS, Eqs.~\eref{rmsLSa} and \eref{wGall}, and the HFS, Eqs.~\eref{HFS1PT1S} and \eref{HFS1PT}. Furthermore, we have shown a limitation of the usual accounting of FSEs and illustrated it with the help of two toy models in \secref{chap2}{ToyModel}. 

To conclude, the standard expansion of the hydrogenic LS in the moments of  charge distribution is only valid provided
the convergence radius of the Taylor expansion of $G_E(Q^2)$ is much larger than
the inverse Bohr radius of the given hydrogen-like system. A very small fluctuation in the FF around the inverse Bohr radius scale
may invalidate the expansion. In order to define a proper hadronic charge radius, one has to decompose the FF into ``smooth'' and ``non-smooth'' parts, where the non-smooth parts must be treated exactly.

So far, we only developed toy models with non-smooth contribution to the FF but did not provide a physical explanation for them. However, there are several conceivable origins. The required combination of mass and coupling constant basically rules out the option of an axion exchange at the proton vertex, see \Figref{Axion}. However, a weak contribution to the lepton vertex, similar to a muon decay, is not yet excluded, see \Figref{MuonDecay}. Searches for a physical implementation remain an essential task for the future. 

  \begin{figure}[h!] 
    \centering 
       \includegraphics[scale=0.4]{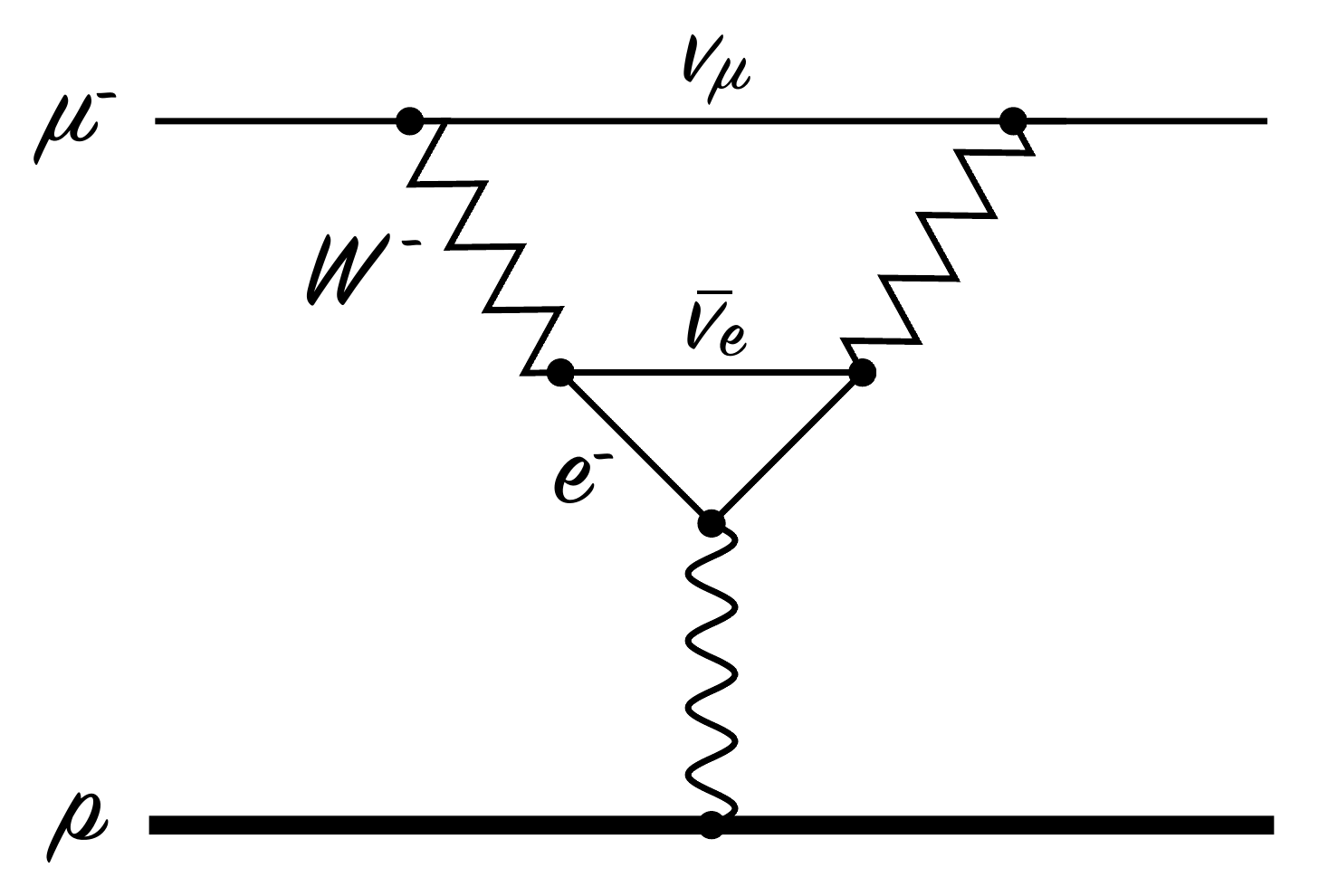}
       \caption{Muon vertex with weak decay.}
              \label{fig:MuonDecay}
\end{figure} 

\begin{subappendices}
\section{Details of the Breit Potential Derivation}  \seclab{DetailedDerivation}

Here, we will give more details on how the derivation of the Breit potential, \Eqref{BreitPotentialScatAmp}, proceeds after the semi-relativistic expansion, cf. \secref{chap2}{BreitDerivation}. We continue starting from the following momentum-space Breit potential.
\paragraph*{Momentum-space Breit Potential with Nuclear Form Factors}
\begin{subequations}
\eqlab{BreitMomentum}
\beq
V_\mathrm{OPE}(\bq,\bp,t)=\left[V_\mathrm{C}+\Delta V_{\text{rel.C}}+\Delta V_{\mathrm{Y}}+\Delta V_{\mathrm{1}}+\Delta V_{\mathrm{2}}+\Delta V_{\mathrm{3}}+\Delta V_{\mathrm{4}}+\Delta V_{\mathrm{5}}\right](\bq,\bp,t),
\eeq
with
\begin{small}
\begin{alignat}{3}
&V_{\mathrm{C}}(\bq) &&= -\frac{4\pi Z\al}{\bq^2},\eqlab{Coulomb}\\
&\Delta V_{\text{rel.C}}(\bq,\bp) &&= 4\pi Z \al\left\{\frac{1}{8m_r^2}-\frac{1}{2m_r^2}\frac{i \bs \cdot \bq\times \bp}{\bq^2}\right\},\eqlab{relC}\\
&\Delta V_{\mathrm{Y}}(\bq,t) &&=4Z\al \int_{t_0}^\infty\frac{\dd t}{t(t+\bq^2)}\,\im G_E(t),\eqlab{yukawaQ}\\
&\Delta V_{\mathrm{1}}(\bq,\bp) &&=4\pi Z \al\left\{\frac{1}{2M^2}\frac{i \bs \cdot \bq\times \bp}{\bq^2}-\frac{1}{4Mm}-\frac{1}{mM}\left[\frac{\bp^2}{\bq^2}-\left(\frac{\bq\cdot \bp}{\bq^2}\right)^2\right]\right\},\\
&\Delta V_{\mathrm{2}}(\bq,\bp) &&=4\pi Z \al\left\{\frac{1+\kappa}{mM}\left[\bs \cdot \bS-\frac{(\bs \cdot \bq)(\bS \cdot \bq)}{\bq^2}\right]\right.\\
&&&\left.\qquad\qquad-\frac{i \bS \cdot \bq\times \bp}{\bq^2}\left[\left(\frac{1}{m M}+\frac{1}{2M^2}\right)+\kappa\left(\frac{1}{m M}+\frac{1}{M^2}\right)\right]\right\},\nn\\
&\Delta V_{\mathrm{3}}(\bq,\bp,t) &&=4Z \al \int_{t_0}^\infty\frac{\dd t}{t(t+\bq^2)}\im G_E(t)\left\{\frac{1}{4mM}\left[(2\bp+\bq)^2-\frac{(2\bq\cdot\bp+\bq^2)^2}{t+\bq^2}\right]\right.\\
&&&\left.\hspace{3cm}+i \bs \cdot \bq\times \bp\left(\frac{1}{2m^2}+\frac{1}{m M}\right)-\frac{\bq^2}{8}\left(\frac{1}{m^2}+\frac{1}{M^2}\right)\right\},\nn\\
&\Delta V_{\mathrm{4}}(\bq,\bp,t) &&=-\frac{2Z \al}{M^2} \int_{t_0}^\infty\frac{\dd t}{t(t+\bq^2)}\im G_E(t)\,i\bS \cdot \bq\times \bp,\\
&\Delta V_{\mathrm{5}}(\bq,\bp,t) &&=4Z \al \int_{t_0}^\infty\frac{\dd t}{t(t+\bq^2)}
\im G_M(t)\left\{i \bS \cdot \bq\times \bp\left(\frac{1}{mM}+\frac{1}{M^2}\right)\right.\\
&&&\left.\hspace{5cm}-\frac{\bq^2}{mM}\left[\bs \cdot \bS-\frac{(\bs \cdot \bq)(\bS \cdot \bq)}{\bq^2}\right]\right\}.\nn
\end{alignat}
\end{small}
\end{subequations}
Here, we made use of the Lagrange identity, 
\beq
(\boldsymbol{a}\times\boldsymbol{b})\cdot(\boldsymbol{c}\times\boldsymbol{d})=(\boldsymbol{a}\cdot\boldsymbol{c})(\boldsymbol{b}\cdot\boldsymbol{d})-(\boldsymbol{b}\cdot\boldsymbol{c})(\boldsymbol{a}\cdot\boldsymbol{d}),
\eeq
to separate the spin-spin coupling $\bs \cdot \bS$.\footnote{In particular, we used: \beq
(\boldsymbol{s}\times\boldsymbol{q})\cdot(\boldsymbol{S}\times\boldsymbol{q})=(\bs\cdot \bS)\,\bq^2-(\bs \cdot \bq)(\bS \cdot \bq).
\eeq} Our result agrees with Ref.~\cite[\S83. Breit's equation, Eq.~(83.9)]{LandauLifshitz4} in the structureless limit, i.e., $F_1=1$, $F_2=0$. For an easier handle of the individual terms, we split the Breit potential into several sub-potentials. They, respectively, do or do not depend on the nuclear spin and the electric or magnetic FF discontinuities.
 Equation~\eref{Coulomb} gives the well-known Coulomb potential. All other potentials will be treated as perturbation to the Coulomb potential, hence, they are denoted as $\Delta V$. Furthermore, we identified the relativistic corrections to the Coulomb potential, cf.\ \Eqref{relC}, and a Yukawa-type correction, cf.\ \Eqref{yukawaQ}.

We then perform a Fourier transformation to obtain the coordinate-space Breit potential from the momentum-space Breit potential:
\beq
V(\bp,\bq,t)\xrightarrow{\mathcal{F}.\mathcal{T}\!.\,}V(\boldsymbol{\hat{p}},\br,t): \qquad V(\boldsymbol{\hat{p}},\br,t)=\frac{1}{(2\pi)^3}\int\! \dd \bq \, e^{i \bq \cdot \br}\, V(\bp,\bq,t).
\eeq
Because of the massive Coulomb gauge, the potential has an additional dependence on $t$. A list of useful Fourier transformations is provided with Table \ref{FTtab}, cf.\ Refs.~\cite[p.\ 180]{BetheSalpeter} and \cite[p.\ 339-340]{LandauLifshitz4} for the case of $t=0$. For the coordinate-space potential it is crucial to realize the singularities at the origin, i.e., the $\delta(\br)$ terms.

\renewcommand{\arraystretch}{1.75}
\begin{table} [tbh]
\centering
\caption{List of useful Fourier transformations: $4\pi f(\bq,t) \xrightarrow{\mathcal{F}.\mathcal{T}\!.\,} F(\br,t)$. \label{FTtab}}
\begin{small}
\begin{tabular}{|c|cx{2pt}c|c|}
  \hline
  \rowcolor[gray]{.7}
$\boldsymbol{f(\bq,t)}$& $\boldsymbol{F(\br,t)}$ &$\boldsymbol{f(\bq)}$& $\boldsymbol{F(\br)}$ \\
\hline
&&  $1$&$4\pi\,\delta(\br)$\\
    \rowcolor[gray]{.95}
    $ \frac{1}{t+\bq^2}$& $\frac{e^{-r\sqrt{t}}}{r}$&  $ \frac{1}{\bq^2}$& $ \frac{1}{r}$ \\
 $\frac{\bq}{t+\bq^2}$&$\frac{i(1+r\sqrt{t})e^{-r\sqrt{t}}\br}{r^3}$&                 $\frac{\bq}{\bq^2}$&$\frac{i\br}{r^3}$\\
     \rowcolor[gray]{.95}
      $\frac{1}{(t+\bq^2)^2}$&$-\frac{e^{-r\sqrt{t}}}{2\sqrt{t}}$  & &    \\
               $\frac{\bq}{(t+\bq^2)^2}$&$\frac{ie^{-r\sqrt{t}}\br}{2r}$&$\frac{\bq}{(\bq^2)^2}$&$\frac{i\br}{2r}$\\
                    \rowcolor[gray]{.95}
                   $\frac{(\bq \cdot \boldsymbol{a})(\bq \cdot \boldsymbol{b})}{(t+\bq^2)^2}$&$\frac{e^{-r\sqrt{t}}}{2r}\left[(\boldsymbol{a}\cdot \boldsymbol{b})-(1+r\sqrt{t})\frac{\br\cdot(\br \cdot \boldsymbol{a})\boldsymbol{b}}{r^2}\right]$&    $\frac{(\bq \cdot \boldsymbol{a})(\bq \cdot \boldsymbol{b})}{(\bq^2)^2}$&$\frac{1}{2r}\left[(\boldsymbol{a}\cdot \boldsymbol{b})-\frac{\br\cdot(\br \cdot \boldsymbol{a})\boldsymbol{b}}{r^2}\right]$\\
  \multirow{2}{*}{ $\frac{(\bq \cdot \boldsymbol{a})(\bq \cdot \boldsymbol{b})}{t+\bq^2}$}& $\frac{e^{-r\sqrt{t}}}{r^3}\left[(1+r\sqrt{t})(\boldsymbol{a}\cdot \boldsymbol{b})-3(1+r\sqrt{t}+\frac{r^2t}{3})\frac{\br\cdot(\br \cdot \boldsymbol{a})\boldsymbol{b}}{r^2}\right]$&    \multirow{2}{*}{$\frac{(\bq \cdot \boldsymbol{a})(\bq \cdot \boldsymbol{b})}{\bq^2}$}&$\frac{1}{r^3}\left[(\boldsymbol{a}\cdot \boldsymbol{b})-3\frac{\br\cdot(\br \cdot \boldsymbol{a})\boldsymbol{b}}{r^2}\right]$\\
 &$+\frac{4\pi}{3}(\boldsymbol{a}\cdot \boldsymbol{b})\,\delta(\br)$&&$+\frac{4\pi}{3}(\boldsymbol{a}\cdot \boldsymbol{b})\,\delta(\br)$\\
      \hline
\end{tabular}
     \end{small}
\end{table}
\renewcommand{\arraystretch}{1.3}

The momentum operator, $\boldsymbol{\hat{p}}=-i\nabla_{\br}$, has to be written to the right of all other factors, since it is an operator acting on the wave function only. We use the following replacement:
\beq
\frac{\br (\br \cdot \boldsymbol{\hat{p}}) \cdot \boldsymbol{\hat{p}}}{r^2}\, \psi_{nlm}(\br)
= \bigg[ \boldsymbol{\hat{p}}^2  + \frac{2}{r^2} i \br \cdot \boldsymbol{\hat{p}}
-\frac{l(l+1)}{r^2}  \bigg] \psi_{nlm}(\br), 
\eeq
which follows from the radial Schr\"odinger equation,\footnote{In spherical coordinates, one defines the Nabla operator as:
\beq
\boldsymbol{\nabla}= \boldsymbol{\hat e_r} \frac{\partial}{\partial r}+\frac{\boldsymbol{\hat e_\theta}}{r} \frac{\partial}{\partial \theta}+\frac{\boldsymbol{\hat e_\phi}}{r \sin \theta} \frac{\partial}{\partial \phi},
\eeq
and the Laplace operator as:
\beq
\Delta f=\frac{1}{r^2}\frac{\partial}{\partial r}\left(r^2 \frac{\partial f}{\partial r}\right)+\frac{1}{r^2 \sin \theta}\frac{\partial}{\partial \theta}\left(\sin \theta\frac{\partial f}{\partial \theta}\right)+\frac{1}{r^2 \sin^2 \theta}\frac{\partial^2 f}{\partial \phi}.
\eeq
It is worth to note that:
\beq
\Delta\, 1/r =-4\pi \, \delta (\br).
\eeq
}
\beq
\bigg[ -\frac{\pa^2}{\pa r^2} - \frac{2}{r} \frac{\pa}{\pa r} +
\frac{l(l+1)}{r^2} - 2m_r \left(\frac{\al}{r} + E_n \right)\bigg] R_{nl}(r) = 0.
\eeq
We now recall the spin and angular momentum operators. The nuclear and lepton spin operators are denoted by $\bs$ and $\bS$, respectively; the corresponding quantum numbers will be $s$ for the lepton and $i$ for the nucleus.  The lepton orbital angular momentum is defined as $\bL=\br \times \boldsymbol{\hat{p}}$. The lepton total angular momentum is $\bJ=\bL+\bs$, and in turn, the atom's total angular momentum is $\boldsymbol{f}=\boldsymbol{j}+\boldsymbol{S}$. Substituting the operators with their eigenvalues,
 \bea
&&\bs^2 \to s(s+1)\overset{s=\nicefrac{1}{2}}{=}\mbox{\small{$\frac{3}{4}$}},\quad \bS^2\to i(i+1)\overset{i=\nicefrac{1}{2}}{=}\mbox{\small{$\frac{3}{4}$}}, \quad \bL^2\to  l(l+1) , \nn\\
 &&\qquad\quad  (\bL + \bs)^2 \to j(j+1) , \quad ( \bJ+\bS)^2 \to f(f+1),\nn
 \eea
 we find:
 \bea
 \eqlab{QN1}
\bL \cdot \bs&=& \half \big[ (\bL + \bs)^2 - \bL^2 -\bs^2\big],\\
&\rightarrow& \half \big[ j(j+1) - l(l+1) -\mbox{\small{$\frac{3}{4}$}}\big],\nn\\
&\rightarrow&\begin{cases}
   0 & \text{for}\;l=0,\, j=1/2, \\
   -1      & \text{for}\;l=1,\, j=1/2,\\
   1/2      & \text{for}\;l=1,\, j=3/2.\\
  \end{cases}\nn
 \eea
 We next use the replacements \cite[Eq.~(22.8)]{BetheSalpeter}:
\beq
\bs \rightarrow \bJ\frac{\overline{(\bs \cdot \bJ)}}{\overline{\bJ^2}} \quad \text{and}\quad \bL \rightarrow \bJ\frac{\overline{(\bL \cdot \bJ)}}{\overline{\bJ^2}},
\eeq
where the bar denotes eigenvalues. These relations are applicable if one calculates the expectation values between states with equal quantum numbers $s$, $l$ and $j$. In this way, we find:
\begin{subequations}
\eqlab{QN2}
\bea
\bs \cdot\bS&\rightarrow& \bS\cdot\bJ\,\frac{\overline{(\bs \cdot \bJ)}}{\overline{\bJ^2}},\eqlab{sSoperator}\\
&\rightarrow&  \frac{\left[f(f+1)-i(i+1)-j(j+1)\right]\left[j(j+1)+s(s+1)-l(l+1)\right]}{4j(j+1)},\nn\\
&\rightarrow&\begin{cases}
\frac{1}{2}\left[f(f+1)-i(i+1)-\frac{3}{4}\right]\overset{i=\nicefrac{1}{2}}{=}\frac{1}{2}\left[f(f+1)-\frac{3}{2}\right]& \text{for}\;l=0,\,j=1/2,\\
   -\frac{1}{6}\left[f(f+1)-i(i+1)-\frac{3}{4}\right] \overset{i=\nicefrac{1}{2}}{=}-\frac{1}{6}\left[f(f+1)-\frac{3}{2}\right] & \text{for}\;l=1,\,j=1/2, \\
  \frac{1}{6}\left[f(f+1)-i(i+1)-\frac{15}{4}\right] \overset{i=\nicefrac{1}{2}}{=}\frac{1}{6}\left[f(f+1)-\frac{9}{2}\right]      & \text{for}\;l=1,\,j=3/2,
  \end{cases}\nn\\
\bL \cdot\bS&\rightarrow& \bS\cdot\bJ\,\frac{\overline{(\bL \cdot \bJ)}}{\overline{\bJ^2}},\\
&\rightarrow&  \frac{\left[f(f+1)-i(i+1)-j(j+1)\right]\left[j(j+1)+l(l+1)-s(s+1)\right]}{4j(j+1)},\qquad\nn\\
&\rightarrow&\begin{cases}
0& \text{for}\;l=0,\,j=1/2,\\
   \frac{2}{3}\left[f(f+1)-i(i+1)-\frac{3}{4}\right] \overset{i=\nicefrac{1}{2}}{=}\frac{2}{3}\left[f(f+1)-\frac{3}{2}\right] & \text{for}\;l=1,\,j=1/2, \\
  \frac{1}{3}\left[f(f+1)-i(i+1)-\frac{15}{4}\right] \overset{i=\nicefrac{1}{2}}{=}\frac{1}{3}\left[f(f+1)-\frac{9}{2}\right]     & \text{for}\;l=1,\,j=3/2.\nn
  \end{cases}
\eea
\end{subequations}
In addition, we make use of the formulas \cite[Eq.~(A.32)]{BetheSalpeter}:
\begin{subequations}
\eqlab{angularAverage}
\bea
\label{BSA32}
r^2\delta_{ij}-3x_i x_j &\rightarrow& -\frac{\overline{r^2}}{4l(l+1)-3}\left[2 l(l+1)\delta_{ij}-3\{\bL_i,\bL_j\}\right],\\&=&\begin{cases}
   -\overline{r^2}\,\{\bL_i,\bL_j\} & \text{for}\;l=0, \\
   -\frac{\overline{r^2}}{5}\left[4\delta_{ij}-3\{\bL_i,\bL_j\}\right]      & \text{for}\;l=1,\nn
  \end{cases}
\eea
and \cite{Martynenko:2006gz}:
\bea
r^2\delta_{ij}-x_i x_j &\rightarrow&\frac{\overline{r^2}}{4l(l+1)-3}\left[2 \left\{l(l+1)-1\right\}\delta_{ij}+\{\bL_i,\bL_j\}\right], \\&=&\begin{cases}
    \frac{\overline{r^2}}{3}\left[2 \delta_{ij}-\{\bL_i,\bL_j\}\right] 
    & \text{for}\,l=0, \\
   \frac{\overline{r^2}}{5}\left[2 \delta_{ij}+\{\bL_i,\bL_j\}\right]      & \text{for}\,l=1,\nn
  \end{cases}
\eea
\end{subequations}
which follow from angular averaging with the spherical harmonics, \Eqref{sphericalHarmonics}, representing the angular part of the Coulomb wave functions. We obtain:
\begin{subequations}
\eqlab{QN3}
\bea
&&\bs\cdot\bS\,-\frac{3(\bs\cdot \br)(\bS\cdot \br)}{r^2} \\
&&\rightarrow \begin{cases}
0& \text{for}\;l=0,\,j=1/2,\\
   -\frac{2}{3}\left[f(f+1)-i(i+1)-\frac{3}{4}\right]\overset{i=\nicefrac{1}{2}}{=}-\frac{2}{3}\left[f(f+1)-\frac{3}{2}\right] & \text{for}\;l=1,\,j=1/2, \\
 \frac{1}{15}\left[f(f+1)-i(i+1)-\frac{15}{4}\right]\overset{i=\nicefrac{1}{2}}{=} \frac{1}{15}\left[f(f+1)-\frac{9}{2}\right]      & \text{for}\;l=1,\,j=3/2,
  \end{cases}\nn\\
&&\bs\cdot\bS\,-\frac{(\bs\cdot \br)(\bS\cdot \br)}{r^2}\\
&&\rightarrow\begin{cases}
\frac{1}{3}\left[f(f+1)-i(i+1)-\frac{3}{4}\right]\overset{i=\nicefrac{1}{2}}{=}\frac{1}{3}\left[f(f+1)-\frac{3}{2}\right]& \text{for}\;l=0,\,j=1/2,\\
   \frac{1}{5}\left[f(f+1)-i(i+1)-\frac{3}{4}\right]\overset{i=\nicefrac{1}{2}}{=}-\frac{1}{3}\left[f(f+1)-\frac{3}{2}\right] & \text{for}\;l=1,\,j=1/2, \\
 \frac{2}{15}\left[f(f+1)-i(i+1)-\frac{15}{4}\right]\overset{i=\nicefrac{1}{2}}{=}\frac{2}{15}\left[f(f+1)-\frac{9}{2}\right]      & \text{for}\;l=1,\,j=3/2.
  \end{cases}\qquad\qquad\quad\nn
\eea
\end{subequations}

To calculate the $P$-level mixing, see \appref{chap2}{PMixing}, one needs to evaluate matrix elements of products of spin and angular momentum operators between states with different lepton total angular momenta, $j=1/2$ and $j=3/2$. Utilizing the common Clebsch-Gordan coefficients, we expand the $P$-states in a product basis:
\beq
\vert (ls)jS;f M_f\rangle=\!\!\!\!\sum_{m_l,m_s,M_j,m_S}\!\!\!\!C^{\,l\,s\,j}_{\,m_l\,m_s\,M_j}\,C^{\,j\,S\,f}_{\,M_j\,m_S\,M_f}\vert l, m_l\rangle\,\vert s, m_s\rangle\,\vert S, m_S\rangle.\qquad
\eeq Furthermore, we use the following general relations for angular momentum operators:
\begin{subequations}
\bea
J_+\vert J, m_J\rangle&=&\sqrt{(J+m_J+1)(J-m_J)}\vert J, m_J+1\rangle,\\
J_-\vert J, m_J\rangle&=&\sqrt{(J-m_J+1)(J+m_J)}\vert J, m_J-1\rangle.
\eea
where
\bea
J_+&=&J_x+i J_y,\\
J_-&=&J_x-i J_y.
\eea
\end{subequations}
The relevant matrix elements are then listed in Table~\ref{PLevelMixingTab}.

\renewcommand{\arraystretch}{1.75}
\begin{table} [t]
\centering
\caption{$P$-level mixing matrix elements: $\langle 2P_j \left\vert\hat{O}\right\vert 2P_{j'} \rangle$.\label{PLevelMixingTab}}
\begin{small}
\begin{tabular}{|c|c|c|c|c|}
\hline
 \rowcolor[gray]{.7}
$\boldsymbol{\hat O}$&$\boldsymbol{j=\nicefrac{1}{2},\,j'=\nicefrac{1}{2}}$&$\boldsymbol{j=\nicefrac{1}{2},\,j'=\nicefrac{3}{2}}$&$\boldsymbol{j=\nicefrac{3}{2},\,j'=\nicefrac{1}{2}}$&$\boldsymbol{j=\nicefrac{3}{2},\,j'=\nicefrac{3}{2}}$\\
\hline
$\bL \cdot \bs$&$-1$&$0$&$0$&$\frac{1}{2}$\\
 \rowcolor[gray]{.95}
$\bL \cdot \bS$&$\frac{1}{3}$&$-\frac{\sqrt{2}}{3}$&$-\frac{\sqrt{2}}{3}$&$-\frac{5}{6}$\\
$\bS \cdot \bs$&$-\frac{1}{12}$&$\frac{\sqrt{2}}{3}$&$\frac{\sqrt{2}}{3}$&$-\frac{5}{12}$\\
 \rowcolor[gray]{.95}
$(\bL \cdot \bs)\, (\bL \cdot \bS)$&$-\frac{1}{3}$&$\frac{\sqrt{2}}{3}$&$-\frac{1}{3 \sqrt{2}}$&$-\frac{5}{12}$\\
$(\bL \cdot \bS)\, (\bL \cdot \bs)$&$-\frac{1}{3}$&$-\frac{1}{3 \sqrt{2}}$&$\frac{\sqrt{2}}{3}$&$-\frac{5}{12}$\\
\hline
\end{tabular}
\end{small}
\end{table}
\renewcommand{\arraystretch}{1.3}

A general expression for the final coordinate-space Breit potential is presented with \Eqref{BreitPotentialFinal}. Here, we will given the $S$- and $P$-wave potentials ($s=1/2$ and $i=1/2$).

\paragraph*{$\boldsymbol{S}$- and $\boldsymbol{P}$-Waves Coordinate-space Breit Potential with Nuclear Form Factors}
\begin{itemize}
\item $S$-wave potential ($l=0$)
\begin{subequations}
\eqlab{SwavePot}
\begin{small}
\begin{alignat}{3}
&\Delta V_\mathrm{rel. C.}(\br)&&=\frac{Z \al}{2m_r^2} \,\pi\delta(\br),\\
&\Delta V_{\mathrm{Y}}(\br,t) &&=\frac{Z\al}{\pi r}\int_{t_0}^\infty \frac{\dd t}{t} \im G_E(t) \,e^{-r\sqrt{t}},\\
&\Delta V_{\mathrm{1}}(\br,\boldsymbol{\hat{p}}) &&=-\frac{Z\al}{mM}\left\{\pi\,\delta(\br)+\frac{\boldsymbol{\hat{p}}^2}{r}+\frac{1}{r^2}\frac{\partial}{\partial r}\right\}\stackrel{(*)}{=}\frac{Z\al}{mM}\left\{\pi\,\delta(\br)-\frac{\boldsymbol{\hat{p}}^2}{r}\right\},\\
&\Delta V_{\mathrm{2}}(\br) &&=\frac{Z\al}{3}\frac{1+\kappa}{mM}\left[f(f+1)-\frac{3}{2}\right]4\pi\,\delta(\br)\eqlab{FermiPot},\\
&\Delta V_{\mathrm{3}}(\br,\boldsymbol{\hat{p}},t) &&=\frac{Z\al}{\pi} \int_{t_0}^\infty \frac{\dd t}{t} \im G_E(t) \left\{-\frac{1}{8}\left(\frac{1}{m^2}+\frac{1}{M^2}\right)4\pi\,\delta(\br\,)+\frac{t}{8m_r^2}\frac{e^{-r\sqrt{t}}}{r}\right.\\
&&&\qquad\qquad\left.+\frac{e^{-r\sqrt{t}}}{2mM}\left[(2+r\sqrt{t})\,\frac{\boldsymbol{\hat{p}}^2}{r}+\frac{2+2r\sqrt{t}+r^2t}{r^2}\frac{\partial}{\partial r}-\frac{t^{3/2}}{4}\right]\right\},\nn\\
&&&\stackrel{(*)}{=}\frac{Z\al}{\pi} \int_{t_0}^\infty \frac{\dd t}{t} \im G_E(t) \left\{-\frac{1}{8}\left(\frac{1}{m_r^2}+\frac{2}{mM}\right)4\pi\,\delta(\br\,)+\frac{1}{8m_r^2}\frac{t \,e^{-r\sqrt{t}}}{r}\right.\\
&&&\qquad\qquad\left.+\frac{e^{-r\sqrt{t}}}{2mM}\left[(2+r\sqrt{t})\,\frac{\boldsymbol{\hat{p}}^2}{r}+\frac{t^{3/2}}{4}\right]\right\},\nn\\
&\Delta V_{\mathrm{4}}(\br) &&=0,\\
&\Delta V_{\mathrm{5}}(\br,t) &&=\frac{Z\al}{3\pi mM} \int_{t_0}^\infty \frac{\dd t}{t} \im G_M(t) \left[f(f+1)-\frac{3}{2}\right]\left\{\frac{t e^{-r\sqrt{t}}}{r}-4\pi \delta(\br)\right\},\eqlab{HFSpotQ}
\end{alignat}
\end{small}
\end{subequations}
\item $P$-wave potential ($l=1$) for states with equal $j$ (upper case $j=1/2$, lower case $j=3/2$)
\begin{subequations}
\eqlab{PwavePot}
\begin{small}
\begin{alignat}{3}
&\hspace{-0.2cm}\Delta V_\mathrm{rel. C.}(\br)&&\hspace{-0.05cm}=\frac{Z \al}{4m_r^2}\frac{1}{r^3}\left[j(j+1)-\frac{11}{4}\right],\\
&\hspace{-0.2cm}\Delta V_{\mathrm{Y}}(\br,t) &&\hspace{-0.05cm}=\frac{Z\al}{\pi r}\int_{t_0}^\infty \frac{\dd t}{t} \im G_E(t) \,e^{-r\sqrt{t}},\\
&\hspace{-0.2cm}\Delta V_{\mathrm{1}}(\br,\boldsymbol{\hat{p}}) &&\hspace{-0.05cm}=Z\al\left[\frac{1}{r^3}\left\{\frac{1}{mM}-\frac{1}{4M^2}\left[j(j+1)-\frac{11}{4}\right]\right\}-\frac{1}{mM}\left(\frac{\boldsymbol{\hat{p}}^2}{r}+\frac{1}{r^2}\frac{\partial}{\partial r}\right)\right],\\
&&&\hspace{-0.05cm}\stackrel{(*)}{=}Z\al\left[\frac{1}{r^3}\left\{\frac{1}{mM}-\frac{1}{4M^2}\left[j(j+1)-\frac{11}{4}\right]\right\}-\frac{1}{mM}\frac{\boldsymbol{\hat{p}}^2}{r}\right],\\
&\hspace{-0.2cm}\Delta V_{\mathrm{2}}(\br) &&\hspace{-0.05cm}=Z\al\,\left\{ \frac{1}{r^3}\left[\left(\frac{1}{mM}+\frac{1}{2M^2}\right)+\left(\frac{1}{mM}+\frac{1}{M^2}\right)\kappa\right]\right. \bL \cdot \bS\eqlab{V284e}\\
&&&\hspace{-0.05cm}\quad\qquad\left.+\frac{1+\kappa}{5mM}\frac{1}{r^3}\big[4 \,\bs\cdot \bS-3\left\{\bL \cdot \bs,\bL \cdot \bS\right\}\!\big]\right\},\nn\\
&&&\hspace{-0.05cm}=Z\al\,\left\{ \frac{1}{r^3}\left[\left(\frac{1}{mM}+\frac{1}{2M^2}\right)+\left(\frac{1}{mM}+\frac{1}{M^2}\right)\kappa\right]\right. \times{\frac{2}{3}\left[f(f+1)-\frac{3}{2}\right] \atop \frac{1}{3}\left[f(f+1)-\frac{9}{2}\right]  }\\
&&&\hspace{-0.05cm}\quad\qquad\left.+\frac{1+\kappa}{mM}\frac{1}{r^3}\times{\frac{2}{3}\left[f(f+1)-\frac{3}{2}\right] \atop \frac{1}{15}\left[\frac{9}{2}-f(f+1)\right] }\right\},\nn\\
&\hspace{-0.2cm}\Delta V_{\mathrm{3}}(\br,\boldsymbol{\hat{p}},t) &&\hspace{-0.05cm}=\frac{Z\al}{\pi} \int_{t_0}^\infty \frac{\dd t}{t} \im G_E(t) \Bigg[\frac{1}{8m_r^2}\frac{t \,e^{-r\sqrt{t}}}{r}\\
&&&\hspace{-0.05cm}\qquad+\frac{1}{2mM}e^{-r\sqrt{t}}\left\{(2+r\sqrt{t})\,\frac{\boldsymbol{\hat{p}}^2}{r}+\frac{2+2r\sqrt{t}+r^2t}{r^2}\frac{\partial}{\partial r}-\frac{t^{3/2}}{4}\right\}\nn\\
&&&\hspace{-0.05cm}\qquad-\frac{e^{-r\sqrt{t}}}{r^3}\left(1+r\sqrt{t}\right)\left\{\frac{1}{mM}+\frac{1}{2}\left(\frac{1}{2m^2}+\frac{1}{mM}\right)\left[j(j+1)-\frac{11}{4}\right]\right\}\!\Bigg],\nn\\
&&&\hspace{-0.05cm}\stackrel{(*)}{=}\frac{Z\al}{\pi} \int_{t_0}^\infty \frac{\dd t}{t} \im G_E(t) \Bigg[\frac{1}{8m_r^2}\frac{t \,e^{-r\sqrt{t}}}{r}+\frac{e^{-r\sqrt{t}}}{2mM}\left\{(2+r\sqrt{t})\,\frac{\boldsymbol{\hat{p}}^2}{r}+\frac{t^{3/2}}{4}\right\}\\
&&&\hspace{-0.05cm}\qquad-\frac{e^{-r\sqrt{t}}}{r^3}\left(1+r\sqrt{t}\right)\left\{\frac{1}{mM}+\frac{1}{2}\left(\frac{1}{2m^2}+\frac{1}{mM}\right)\left[j(j+1)-\frac{11}{4}\right]\right\}\!\Bigg],\nn\\
&\hspace{-0.2cm}\Delta V_{\mathrm{4}}(\br,t) &&\hspace{-0.05cm}=\frac{Z\al}{2\pi M^2}\int_{t_0}^\infty \frac{\dd t}{t} \im G_E(t)\,\frac{e^{-r\sqrt{t}}}{r^3}\,(1+r \sqrt{t})\times{\frac{2}{3}\left[f(f+1)-\frac{3}{2}\right] \atop \frac{1}{3}\left[f(f+1)-\frac{9}{2}\right]  } ,\\
&\hspace{-0.2cm}\Delta V_{\mathrm{5}}(\br,t) &&\hspace{-0.05cm}=\frac{Z\al}{\pi} \int_{t_0}^\infty \frac{\dd t}{t} \im G_M(t) \,\frac{e^{-r\sqrt{t}}}{r^3}\left\{-\left(\frac{1}{mM}+\frac{1}{M^2}\right)(1+r\sqrt{t}) \,\bL\cdot \bS\right.\\
&&&\hspace{-0.05cm}\quad\left.-\frac{1}{5mM}\left((1+r\sqrt{t})\big[4 \,\bs\cdot \bS-3\left\{\bL \cdot \bs,\bL \cdot \bS\right\}\!\big]-r^2t\big[2\,\bs\cdot \bS+\left\{\bL \cdot \bs,\bL \cdot \bS\right\}\!\big]\right)\right\},\nn\\
&&&\hspace{-0.05cm}=\frac{Z\al}{\pi} \int_{t_0}^\infty \frac{\dd t}{t} \im G_M(t) \,\frac{e^{-r\sqrt{t}}}{r^3}\left\{-\left(\frac{1}{mM}+\frac{1}{M^2}\right)(1+r\sqrt{t}) \,\times{\frac{2}{3}\left[f(f+1)-\frac{3}{2}\right] \atop \frac{1}{3}\left[f(f+1)-\frac{9}{2}\right]  }\right.\nn\\
&&&\hspace{-0.05cm}\quad\left.+\frac{1}{mM}\left((1+r\sqrt{t})
\times{\frac{2}{3}\left[\frac{3}{2}-f(f+1)\right] \atop \frac{1}{15}\left[f(f+1)-\frac{9}{2}\right] }
+r^2t \times{\frac{1}{3}\left[\frac{3}{2}-f(f+1)\right] \atop \frac{2}{15}\left[f(f+1)-\frac{9}{2}\right] }\right)\right\}.
\end{alignat}
\end{small}
\end{subequations}
\end{itemize}
The asterix refers to symmetrized potentials which can only be applied to first order in PT.\footnote{The symmetrized potentials are derived by partial integration:
\beq
\int_0^\infty \!\!\dd r\, f(r) \,\frac{1}{2}\left[ R_{n'l} \frac{\pa}{\pa r} R_{nl}
+ R_{nl} \frac{\pa}{\pa r} R_{n'l}\right] = - \frac{1}{2} f(r) R_{n'l} R_{nl}\Big|_{r=0}
- \frac{1}{2} \int_0^\infty \!\!\dd r\,  f^\prime (r)\, R_{n'l}  R_{nl}.
\eeq 
}

\section{Coulomb Wave Functions and Perturbation Theory} \seclab{WFPT}
In the present Appendix, we state the non-relativistic Schrödinger and the relativistic Dirac wave functions of the Coulomb potential, $V_C=-Z\al/r$, and introduce the framework of Schrödinger PT.\footnote{Note that we introduced the reduced mass of the lepton-nucleus system in order to correct for the nuclear motion or, in other words, the finite mass of the nucleus, as explained below \Eqref{TwoBodyProblem}. Also, our definitions of the spherical harmonics, Legendre polynomials and Laguerre polynomials comply with \textsc{Wolfram Mathematica}. } 

\addtocontents{toc}{\protect\setcounter{tocdepth}{0}}
\subsection{Non-Relativistic Schrödinger Wave Functions} \seclab{SchrödingerWF}
The Schrödinger equation for the spherically symmetric Coulomb problem can be solved by separation of variables. Accordingly, the non-relativistic Schrödinger Coulomb wave functions are written as a product of radial wave functions and spherical harmonics:
\beq
\Psi_{nlm}(\br)=R_{nl}(r)\, Y_{lm}(\theta, \phi),
\eeq
where $n$, $l$, $m$ are the \textit{principal}, \textit{orbital angular momentum} and \textit{magnetic quantum numbers}, respectively.  The wave functions are all normalized to unity:
 \beq
 \int_0^\infty \dd r \,r^2 R_{nl}^2(r) =1, \quad \int_0^{2\pi} \dd \phi \int_0^{\pi} \dd \theta \,\sin \theta\, Y_{lm}^2=1, \quad \int \dd \br \,\Psi_{nlm}^2=1.
 \eeq 
 The spherical harmonics are defined as:
  \beq
  \eqlab{sphericalHarmonics}
Y_{lm}(\theta, \phi)=\sqrt{\frac{(2l+1)}{4\pi}\frac{(l-m)!}{(l+m)!}}\,P_{lm}(\cos \theta)\,e^{im\phi},
\eeq
where $P_{lm}(x)$ are the associated Legendre polynomials given by the Rodrigues formula:\footnote{The Legendre polynomials are orthogonal functions which fulfil the second-order differential equation:
\beq
(1-x^2)\,\frac{\dd^2 P_{lm}(x)}{\dd x^2}-2x\, \frac{\dd P_{lm}(x)}{\dd x}+\left[l(l+1)-\frac{m^2}{1-x^2}\right]P_{lm}(x)=0.
\eeq }
\beq
P_{lm}(x)=\frac{(-1)^m}{2^l \, l!}\,(1-x^2)^{m/2}\,\frac{\dd^{l+m}}{\dd x^{l+m}}\,(x^2-1)^l.
\eeq
For $m=0$, the spherical harmonics of $S$- and $P$-waves read:
\begin{subequations}
\bea
Y_{00}(\theta, \phi)&=&\sqrt{\frac{1}{4\pi}},\\
Y_{10}(\theta, \phi)&=&\sqrt{\frac{3}{4\pi}} \cos \theta.
\eea
\end{subequations}
A three-dimensional polar plot with radius $\vert Y_{lm}(\theta, \phi)\vert^2$ displays a spherical surface for $l=0$, a dump-bell shaped surface for $l=1$ and $m=0$, and a donut shaped surface for $l=1$ and $m=\pm 1$.

For the radial part, we distinguish wave functions for the discrete and continuous spectra.
\subsubsection*{Discrete Wave Functions}
The discrete radial Coulomb wave functions can be given in a general form:
 \begin{subequations}
 \eqlab{WFdisc}
 \begin{align}
 R_{nl} (r) & =   \frac{2}{n^2 a^{3/2}} 
 \left[\frac{2 r}{na}\right]^l e^{\nicefrac{-r}{na}}\,\frac{1}{(2l+1)!}\,\sqrt{\frac{(n+l)!}{(n-l-1)!}}\,{}_1F_1\left(-n+l+1, 2l+2,  \nicefrac{2r}{na}\right),\\
 &=\frac{2}{n^2 a^{3/2}} 
 \left[\frac{2 r}{na}\right]^l e^{\nicefrac{-r}{na}}\,\sqrt{\frac{(n-l-1)!}{(n+l)!}}{}\,L_{n-l-1}^{2l+1}\left(\nicefrac{2r}{na}\right),
 \end{align}
 \end{subequations}
in terms of the associated Laguerre polynomials $L^k_j(x)$ or the confluent hypergeometric functions of the first kind (${}_1F_1$, Kummer's function of the first kind), which are related by:
\beq
L^k_j(x)=\frac{(k+j)!}{k!\,j!}\,{}_1F_1\left(-j, k+1, x\right).
\eeq
The associated Laguerre polynomials are orthogonal functions fulfilling a second-order differential equation of the type:
\beq
x \,\frac{\dd^2 L^k_j(x)}{\dd x^2}+(k+1-x)\,\frac{\dd L^k_j(x)}{\dd x}+j \,L^k_j(x)=0,
\eeq
and their Rodrigues formula reads:
\beq
L^k_j(x)=\frac{e^x x^{-k}}{j!}\frac{\dd^j}{\dd x^j}(e^{-x} x^{k+j}).
\eeq
The confluent hypergeometric function of the first kind can be written as:\footnote{Ref.~\cite[Eqs.~(3.5) and (3.16)]{BetheSalpeter} uses a different definition for the associated Laguerre polynomials, they therefore have: 
\beq
 R_{nl} (r)  = -\frac{2}{n^2 a^{3/2}} 
 \left[\frac{2 r}{na}\right]^l   e^{\nicefrac{-r}{na}}\sqrt{\frac{(n-l-1)!}{\left[(n+l)!\right]^3}}\, \mathcal{L}^{2l+1}_{n+l} \left(\nicefrac{2r}{na}\right). \\
\eeq}
\beq
{}_1F_1(a,b,x)=\frac{\Gamma(b)}{\Gamma(b-a)\Gamma(a)}\int_0^1 \dd t\, e^{xt} \,t^{a-1}(1-t)^{b-a-1}.
\eeq
Again, the lowest $S$- and $P$-level wave functions are of special interest:
 \begin{subequations}
 \bea
 R_{10} (r) & = & \frac{2}{a^{3/2}} \, e^{-r/a}, \\
 R_{20} (r) & = & \frac{1}{\sqrt{2} \, a^{3/2}} \left(1- \frac{r}{2a} \right) 
 e^{-r/2a},\\
 R_{21} (r) & = & \frac{1}{2\sqrt{6} \, a^{3/2}}\, \frac{r}{a}
\, e^{-r/2a}.
 \eea
 \end{subequations}

\subsubsection*{Continuous Wave Functions}
To transit to the continuous spectrum, we replace the factor of $1/n^{3/2}$ contained in each wave function, cf.\ \Eqref{WFdisc}, by:
\beq
\frac{1}{n^{3/2}}\rightarrow\sqrt{\frac{2\pi k}{1-\exp(-\frac{2\pi}{k})}},
\eeq
and otherwise substitute $n\rightarrow 1/ik$. The continuous radial wave function of a lepton with energy:
\beq
E_k=\frac{Z \al\,k^2}{2a},
\eeq
 moving in the Coulomb field of a nucleus with charge $Ze$ is then given by:
 \bea
 R_{kl} (r) & = & \frac{1}{a^{3/2}}\left[\frac{2 k r}{a}\right]^le^{\nicefrac{-ikr}{a}}\frac{1}{(2l+1)!}\,\sqrt{\frac{4\pi k}{\sinh \frac{\pi}{k}}} \,e^{\nicefrac{\pi}{2k}}\left[\prod_{s=1}^l
 \sqrt{s^2+\frac{1}{k^2}}\right]  \\
 && \,{}_1F_1\Big(i/k+l+1,\, 2l+2, \, \nicefrac{2ikr}{a}\Big). \nn
 \eea
The wave functions are normalized in the $k$-scale:\footnote{See Ref.~\cite[Sect.~4]{BetheSalpeter} for other choices of normalization conditions.} 
\beq
\int_0^\infty \dd r \, r^2 R_{kl}(r)\int_{k-\Delta k}^{k+\Delta k}\dd k' \,R_{k'l(r)}=1,
\eeq
The pertinent wave functions are:
 \begin{subequations}
 \bea
  R_{k0} (r) & = &\frac{1}{a^{3/2}}\sqrt{\frac{4\pi k}{\sinh (\pi/k)}} \, e^{\pi/2k} \, e^{-\nicefrac{ikr}{a}} \,
  {}_1F_1(i/k+1,\, 2, \, \nicefrac{2ikr}{a}),\\
 R_{k1} (r) & = &\frac{1}{3a^{3/2}}\sqrt{\frac{4\pi k}{\sinh (\pi/k)}}  \, e^{\pi/2k} \, r \, e^{-\nicefrac{ikr}{a}} \,
 \sqrt{1+k^2}\, {}_1F_1(i/k+2,\, 4, \, \nicefrac{2ikr}{a}).\qquad
 \eea
  \end{subequations}
  
\subsection{Schrödinger Perturbation Theory}
\renewcommand{\arraystretch}{1.75}
\begin{table}[t!]
\caption{Matrix elements of auxiliary potentials with discrete non-relativistic Schr\"odinger wave functions.\label{1PTallpotentials}}
\centering
\begin{small}
\begin{tabular}{|c|c|c|c|}
\hline
 \rowcolor[gray]{.7}
$\boldsymbol{V}$&$\boldsymbol{\langle1S|V|1S\rangle}$&$\boldsymbol{\langle2S|V|2S\rangle}$&$\boldsymbol{\langle2P|V|2P\rangle}$\\
\hline
$1/r$&$\frac{1}{a}$&$\frac{1}{4a}$&$\frac{1}{4a}$\\
 \rowcolor[gray]{.95}
$e^{-r\sqrt{t}}/r$&$\frac{4}{a \left(2+a \sqrt{t}\right)^2}$&$\frac{1+2 a^2 t}{4 a \left(1+a \sqrt{t}\right)^4}$&$\frac{1}{4 a \left(1+a \sqrt{t}\right)^4}$\\
$4\pi \delta(\br)$&$\frac{4}{a^3}$&$\frac{1}{2a^3}$&$0$\\
 \rowcolor[gray]{.95}
$\exp[-r\sqrt{t}]$&$\frac{8}{\left(2+a \sqrt{t}\right)^3}$&$\frac{1-a \sqrt{t}+a^2 t}{\left(1+a \sqrt{t}\right)^5}$&$\frac{1}{\left(1+a \sqrt{t}\right)^5}$\\
$1/r^3$&---&---&$\frac{1}{24 a^3}$\\
 \rowcolor[gray]{.95}
$(1+r\sqrt{t})\,e^{-r\sqrt{t}}/r^3$&---&---&$\frac{1+3 a \sqrt{t}}{24 a^3 \left(1+a \sqrt{t}\right)^3}$\\
$\boldsymbol{\hat{p}}^2/r$&$\frac{3}{a^3}$&$\frac{7}{16a^3}$&$\frac{5}{48 a^3}$\\
 \rowcolor[gray]{.95}
$(2+r\sqrt{t})\,e^{-r\sqrt{t}}\,\boldsymbol{\hat{p}}^2/r$&$\frac{8 \left(6+8 a \sqrt{t}+3 a^2 t\right)}{a^3 \left(2+a \sqrt{t}\right)^3}$&$\frac{7+33 a \sqrt{t}+60 a^2 t+52 a^3 t^{3/2}+24 a^4 t^2}{8 a^3 \left(1+a \sqrt{t}\right)^5}$&$\frac{5+19 a \sqrt{t}+20 a^2 t}{24 a^3 \left(1+a \sqrt{t}\right)^5}$\\
$\boldsymbol{\hat{p}}^4$&$\frac{5}{a^4}$&$\frac{13}{16a^4}$&$\frac{7}{48a^4}$\\
\hline
\end{tabular}
\end{small}
\end{table}
\renewcommand{\arraystretch}{1.3}

  \renewcommand{\arraystretch}{1.75}
\begin{table}[t]
\centering
\caption{(Symmetric) mixed matrix elements of auxiliary potentials for the discrete spectrum: $\langle nlm \vert V \vert n'lm\rangle$ with $n>n'$.}
\label{2PTdiscreteSYM}
\begin{small}
\centering
\begin{tabular}{|c|c|c|}
\hline
 \rowcolor[gray]{.7}
$\boldsymbol{V}$&$\boldsymbol{n'l}$&$\boldsymbol{\langle nlm \vert V \vert n'lm\rangle \equiv \langle n'lm \vert V \vert nlm\rangle}$\\
\hline
\multirow{3}{*}{$e^{-r\sqrt{t}}/r$}&1S&$\frac{4}{a}\frac{\sqrt{n}}{n^2\left(1+a\sqrt{t}\right)^2-1}\exp\left[-2n \arctanh \frac{1}{n\left(1+a\sqrt{t}\right)}\right]$\\
&2S&$-\frac{4\sqrt{2n}}{a}\frac{4-n^2\left(3+4a^2t\right)}{\left[n^2\left(1+2a\sqrt{t}\right)^2-4\right]^2}\exp\left[-2n \arctanh \frac{2}{n\left(1+2a\sqrt{t}\right)}\right]$\\
&2P&$\frac{16\sqrt{2}}{\sqrt{3}\,a}\frac{\sqrt{n^3\left(n^2-1\right)}}{\left[n^2\left(1+2a\sqrt{t}\right)^2-4\right]^2}\exp\left[-2n \arctanh \frac{2}{n\left(1+2a\sqrt{t}\right)}\right]$\\
  \rowcolor[gray]{.95}& 1S& $\frac{4}{\sqrt{n^3}\,a^3}$\\
 \rowcolor[gray]{.95}\multirow{-2}{*}{$4\pi \delta\left(\br\right)$}& 2S& $\frac{\sqrt{2}}{\sqrt{n^3}\,a^3}$\\
\multirow{3}{*}{$e^{-r\sqrt{t}}$}&1S&$\frac{8\sqrt{n^5t}\,a}{\left[n^2\left(1+a\sqrt{t}\right)^2-1\right]^2}\exp\left[-2n \arctanh \frac{1}{n\left(1+a\sqrt{t}\right)}\right]$\\
&2S&$-\frac{32\sqrt{2n^5t}\,a\left[4-n^2\left(5-4a\sqrt{t}+4a^2t\right)\right]}{\left[n^2\left(1+2a\sqrt{t}\right)^2-4\right]^3}\exp\left[-2n \arctanh \frac{2}{n\left(1+2a\sqrt{t}\right)}\right]$\\
&2P&$\frac{256\sqrt{2}\,a}{\sqrt{3}}\frac{\sqrt{t\,n^7\left(n^2-1\right)}}{\left[n^2\left(1+2a\sqrt{t}\right)^2-4\right]^3}\exp\left[-2n \arctanh \frac{2}{n\left(1+2a\sqrt{t}\right)}\right]$\\
 \rowcolor[gray]{.95}
$1/r^3$&2P&$\frac{1}{4\sqrt{6n\left(n^2-1\right)}\,a^3}\left\{1+3\exp\left[-2n \arctanh \frac{2}{n}\right]\right\}$\\
$\frac{1+r\sqrt{t}}{r^3}e^{-r\sqrt{t}}$&2P&$\frac{1}{4\sqrt{6n\left(n^2-1\right)}\,a^3}\left\{1-\frac{12-n^2\left(3-12a\sqrt{t}-4a^2t\right)}{\left[n^2\left(1+2a\sqrt{t}\right)^2-4\right]}\exp\left[-2n \arctanh \frac{2}{n\left(1+2a\sqrt{t}\right)}\right]\right\}$\\
 \rowcolor[gray]{.95}
&1S&$\frac{8}{\sqrt{n^3}\,a^4}\left\{1-\frac{2n^2}{n^2-1}\exp \left[-2n\arctanh \frac{1}{n}\right]\right\}$\\
 \rowcolor[gray]{.95}&2S&$\frac{2\sqrt{2}}{\sqrt{n^3}\,a^4}\left\{1-\frac{8n^2\left(n^2-2\right)}{\left[n^2-4\right]^2}\exp \left[-2n\arctanh \frac{2}{n}\right]\right\}$\\
 \rowcolor[gray]{.95}\multirow{-3}{*}{$\boldsymbol{\hat{p}}^4$}&2P&$\frac{\sqrt{2}}{\sqrt{3n\left(n^2-1\right)}\,a^4}\left\{1+\frac{16+32n^2-21n^4}{\left[n^2-4\right]^2}\exp \left[-2n\arctanh \frac{2}{n}\right]\right\}$\\
\hline
\end{tabular}
\end{small}
\end{table}
\renewcommand{\arraystretch}{1.3}

The concept of perturbation theory (PT) is very useful and widely applied to provide estimates for complex physical problems which cannot be calculated exactly or whose solution is too time-consuming. QED corrections to, f.i., the anomalous magnetic moment of the electron or muon can be arranged in a perturbative expansion in the fine-structure constant $\al\approx\nicefrac{1}{137}$. In ChPT, all Feynman diagrams contributing to a particular process can be classified into leading and subleading orders with the aid of power-counting, cf.\ \secref{chap4}{powercounting}. In other words, the fundamental idea of PT is to work out a perturbative series in a small parameter and quantify the absolute importance of certain contributions in the full result.

In the following, we will briefly review the Schrödinger PT and work out approximate solutions for complex quantum mechanical systems based on known solutions for simpler eigenvalue problems. As shown in \Eqref{HamiltonianPert2}, we expect the Schrödinger Hamiltonian to split into an unperturbed Hamiltonian, $\hat{H}_0$, and a weak perturbation, $\hat{H}_{\delta}$. Here, the unperturbed Hamiltonian corresponds to the Coulomb problem, which is solved by the wave functions in \Eqref{WFdisc} and the energies in \Eqref{Bohr}. 

To first order in time-independent PT, the energy shift of the $nl$-level due to a perturbation of the Coulomb potential is  given by:
 \beq
 \Delta E^{\langle \de \rangle (1)}_{nl} \equiv \left<nlm|\,  V_\de \, | nlm\right> 
 = \int_0^{\infty}\! \dd \boldsymbol{r} \,\Psi_{nlm}^*(\br)\, \hat{H}_{\delta} \,\Psi_{nlm}(\br).
 \eeq
For a spherically  symmetric correction $V_\de (r)$, this simplifies to:
 \beq
\Delta E^{\langle \de \rangle (1)}_{nl}  = \frac{1}{2\pi^2}
 \int_0^\infty\! \dd Q\, Q^2\, w_{nl}(Q)\, 
 V_\de (Q)
 = \int_0^{\infty}\! \dd r\, r^2 R_{nl}^2(r) \,  V_\de (r), \eqlab{1PTsphsym}
 \eeq
 where the momentum-space expression contains
 the convolution of the momentum-space wave functions:
 \beq
  w_{nl}(Q) = 
  \int \dd \bp \,
\vfi_{nlm}^\ast (\bp+\bQ) \, \vfi_{nlm} (\bp).
\eqlab{wfconvolution}
 \eeq
The explicit forms
of the coordinate-space Schrödinger wave functions are given in \appref{chap2}{SchrödingerWF} and the convoluted momentum-space wave functions are given in \Eqref{ConvolutionsWFMom}. Relevant matrix elements are listed in Table \ref{1PTallpotentials}.

At second-order in time-independent PT, the energy shift follows as the sum of perturbations of the discrete and continuous spectra:
   \beq
 \Delta E^{\langle \de \rangle(2)}_{nl} = \sum_{n'\neq n} \frac{\left|\left<nlm|\, \hat{H}_{\delta}\, | n'lm\right>\right|^2}{E_n^{(0)}-E_{n'}^{(0)} }+\frac{1}{2\pi}\int_0^\infty\dd k\,\frac{\left\vert \langle nl\left\vert \hat{H}_{\delta}\right\vert k l\rangle\right\vert^2}{E_n-E_k}.
 \eeq
Some relevant matrix elements are listed in Tables \ref{2PTdiscreteSYM} and \ref{2PTcontinuousSYM}, where in the latter we give the limit of large $k$.\footnote{For the evaluation of matrix elements we rely on the detailed mathematical Appendices of Ref.~\cite[Appendices \S d-\S f]{LandauLifshitz3}.
Among other things, we use \cite[Eqs.~(f.9)-(f.10)]{LandauLifshitz3}:
\bea
&& \int_0^{\infty}\! \dd r\, r^{\ga -1} \, e^{-\la r}
\,  {}_1F_1(\al,\, \ga, \, k r)\,  {}_1F_1(\al',\, \ga, \, k' r ) \qquad 
\\&&=\, \Gamma(\ga)\, \la^{\al+\al'-\ga} \,(\la-k)^{-\al} 
(\la-k')^{-\al'} \, {}_2 F_1\Big(\al,\al',\ga,\frac{kk'}{(\la-k)(\la-k')}\Big),
\nn
\eea
and \cite[Eqs.~(f.1)-(f.2)]{LandauLifshitz3}:
\beq
\int_0^{\infty}\! \dd r\, r^\nu e^{-\la r} \, {}_1F_1(\al,\, \ga, \, k r)
= \Gamma(\nu+1)\, \la^{-\nu-1} {}_2F_1(\al,\, \nu+1,\, \ga, \, k/\la),
\eeq
for $\re \nu >-1$ and $\re \lambda > \vert \re k \vert$ (or $\re \lambda > 0$, if $\al$ is a negative integer).}

\renewcommand{\arraystretch}{1.75}
\begin{table}[h]
\centering
\caption{(Symmetric) mixed matrix elements of auxiliary operators for the continuous spectrum: $c_k\langle klm \vert V \vert nlm\rangle$ with $c_k=\sqrt{\frac{1-\exp[-2\pi /k]}{2\pi k}}$.}
\label{2PTcontinuousSYM}
\begin{small}
\begin{tabular}{|c|c|c|}
\hline
 \rowcolor[gray]{.7}
$\boldsymbol{V}$&$\boldsymbol{nl}$&$\boldsymbol{c_k\langle klm \vert V \vert nlm\rangle \equiv c_k\langle nlm \vert V \vert klm\rangle}$\\
\hline
\multirow{3}{*}{$e^{-r\sqrt{t}}/r$}&1S&$\frac{4}{a}\frac{1}{k^2+\left(1+a\sqrt{t}\right)^2}\exp\left[-\frac{2}{k} \arctan \frac{k}{1+a\sqrt{t}}\right]$\\
&2S&$\frac{4\sqrt{2}}{a}\frac{3+4\left(a^2t+k^2\right)}{\left[4k^2+\left(1+2a\sqrt{t}\right)^2\right]^2}\exp\left[-\frac{2}{k} \arctan \frac{2k}{1+2a\sqrt{t}}\right]$\\
&2P&$-\frac{16\sqrt{2}}{\sqrt{3}\,a}\frac{\sqrt{1+k^2}}{\left[4k^2+\left(1+2a\sqrt{t}\right)^2\right]^2}\exp\left[-\frac{2}{k} \arctan \frac{2k}{1+2a\sqrt{t}}\right]$\\
 \rowcolor[gray]{.95}
&1S&$\frac{4}{a^3}$\\
 \rowcolor[gray]{.95}\multirow{-2}{*}{$4\pi \delta\left(\br\right)$}&2S&$\frac{\sqrt{2}}{a^3}$\\
\multirow{3}{*}{$e^{-r\sqrt{t}}$}&1S&$\frac{8a\sqrt{t}}{\left[k^2+\left(1+a\sqrt{t}\right)^2\right]^2}\exp\left[-\frac{2}{k} \arctan \frac{k}{1+a\sqrt{t}}\right]$\\
&2S&$\frac{32\sqrt{2 t}\,a\left[5-4\left(a\sqrt{t}-a^2t-k^2\right)\right]}{\left[4k^2+\left(1+2a\sqrt{t}\right)^2\right]^3}\exp\left[-\frac{2}{k} \arctan \frac{2k}{1+2a\sqrt{t}}\right]$\\
&2P&$-\frac{256\sqrt{2}\,a}{\sqrt{3}}\frac{\sqrt{t\left(1+k^2\right)}}{\left[4k^2+\left(1+2a\sqrt{t}\right)^2\right]^3}\exp\left[-\frac{2}{k} \arctan \frac{2k}{1+2a\sqrt{t}}\right]$\\
 \rowcolor[gray]{.95}
$1/r^3$&2P&$-\frac{1}{4\sqrt{6\left(1+k^2\right)}\,a^3}\left\{1+3 \exp \left[-\frac{2}{k}\arctan 2k\right]\right\}$\\
$\left(1+r\sqrt{t}\right)e^{-r\sqrt{t}}/r^3$&2P&$-\frac{1}{4\sqrt{6\left(1+k^2\right)}\,a^3}\left\{1+\left[3-\frac{8\left(3a\sqrt{t}+2a^2t\right)}{4k^2+\left(1+2a\sqrt{t}\right)^2}\right]\exp\left[-\frac{2}{k} \arctan \frac{2k}{1+2a\sqrt{t}}\right]\right\}$\\
 \rowcolor[gray]{.95}
&1S&$\frac{8}{a^4}\left\{1-\frac{2}{1+k^2}\exp\left[-\frac{2}{k} \arctan k\right]\right\}$\\
 \rowcolor[gray]{.95}&2S&$\frac{2\sqrt{2}}{a^4}\left\{1-\frac{8\left(1+2k^2\right)}{\left[1+4k^2\right]^2}\exp\left[-\frac{2}{k} \arctan 2k\right]\right\}$\\
 \rowcolor[gray]{.95}\multirow{-3}{*}{$\boldsymbol{\hat{p}}^4$}&2P&$-\frac{\sqrt{2}}{\sqrt{3\left(1+k^2\right)}\,a^4}\left\{1+\left[1-\frac{22+40k^2}{\left[1+4k^2\right]^2}\right]\exp\left[-\frac{2}{k} \arctan 2k\right]\right\}$\\
\hline
\end{tabular}
\end{small}
\end{table}
\renewcommand{\arraystretch}{1.3}
  
  \subsection{Relativistic Dirac Wave Functions} \seclab{DiracWF}
  
  The Dirac wave functions for a spherically symmetric potential can be written as bispinors,
\beq
\psi_{jlm}=\left(
\begin{matrix}
\varphi_{jlm}\\
\chi_{jlm}
\end{matrix}
\right),
\eeq
where the small and large components are given by:
\begin{subequations}
\bea
\varphi_{jlm}&=&i\, g(r) \,\Omega_{jl'm}(\boldsymbol{\hat{r}}),\\
\chi_{jlm}&=&-f(r) \,\Omega_{jl'm}(\boldsymbol{\hat{r}}),
\eea
\end{subequations}
with $l'=2j-l$.
The radial functions are normalized to unity,
\beq
\int dr\, r^2 \left[\vert g(r)\vert^2+\vert f(r)\vert^2\right]=1,\\
\eeq
and for the Coulomb potential read as follows \cite{Akhiezer}:
\begin{subequations}
\bea
g(r)&=&-\frac{\sqrt{\Gamma(2\gamma+n_r+1)}}{\Gamma(2\gamma+1)\sqrt{n_r!}} \sqrt{\frac{1+\frac{\mathcal{E}}{m_r}}{4N(N-\chi)}}\left(\frac{2}{Na}\right)^{3/2}e^{-\nicefrac{r}{Na}}\left(\frac{2r}{Na}\right)^{\gamma-1}\\
&&\left\{n_r\, {}_1F_1\left[-n_r+1,2\gamma+1,\nicefrac{2r}{Na}\right]-(N-\chi)\,{}_1F_1\left[-n_r,2\gamma+1,\nicefrac{2r}{Na}\right]\right\}\nn,\qquad\quad\\
f(r)&=&-\frac{\sqrt{\Gamma(2\gamma+n_r+1)}}{\Gamma(2\gamma+1)\sqrt{n_r!}} \sqrt{\frac{1-\frac{\mathcal{E}}{m_r}}{4N(N-\chi)}}\left(\frac{2}{Na}\right)^{3/2}e^{-\nicefrac{r}{Na}}\left(\frac{2r}{Na}\right)^{\gamma-1}\\
&&\left\{n_r\, {}_1F_1\left[-n_r+1,2\gamma+1,\nicefrac{2r}{Na}\right]+(N-\chi)\,{}_1F_1\left[-n_r,2\gamma+1,\nicefrac{2r}{Na}\right]\right\}\nn.
\eea
\end{subequations}
The spinor spherical harmonics are defined as:
\beq
\left[\Omega_{jlM}(\boldsymbol{\hat{r}})\right]^\mu=C^{J\,M}_{l \,(M-\mu)\,1/2\,\mu}\,Y_{l\,(M-\mu)}(\boldsymbol{\hat{r}}),
\eeq
with $\mu=\pm \nicefrac{1}{2}$ and the Clebsch-Gordan coefficients $C_{\cdots}^{\,\cdots}$ as defined in Ref.~\cite{Akhiezer}.
Furthermore, we introduced the new quantum number $\chi$:
\begin{subequations}
\beq
\chi=\mp(j+1/2)=\left\{\begin{matrix}-(l+1)&j=l+1/2\\l&j=l-1/2\end{matrix}\right. ,
\eeq
the kinetic energy, cf.\ \Eqref{EDirac},
\beq
\mathcal{E}=\frac{m_r}{\sqrt{1+(Z\al)^2/(\gamma+n_r)^2}}
\eeq
and,
\bea
n_r&=&n-|\chi|,\\
\gamma&=&\sqrt{\chi^2-Z^2\al^2},\\
N&=&\sqrt{n^2-2n_r(|\chi|-\gamma)}.
\eea
\end{subequations}

\addtocontents{toc}{\protect\setcounter{tocdepth}{4}}

\section{Finite-Size and Recoil Effects}\seclab{2appRecoil}

\addtocontents{toc}{\protect\setcounter{tocdepth}{0}}

In the following, we discuss the finite-size and recoil effects as deduced from the nuclear FF dependent Breit potential presented in \Eqref{BreitPotentialFinal}. We start by decoding the characteristic structures of the hydrogen spectrum displayed in Figures \ref{fig:DiracSchrödinger}, \ref{fig:MuHSpec} and \ref{fig:EHSpec}, and postpone quantitative expressions for the $2P$ FS, the $P$-level mixing, the $2P_{1/2}-2S_{1/2}$ LS and the HFS to Appendices \ref{chap:chap2}.\ref{sec:2PFineStructure}-\ref{chap:chap2}.\ref{sec:HFSSec}.

Neglecting the $Q^2$ dependence of the nuclear FFs, i.e., replacing $F_1\rightarrow1$ and $F_2\rightarrow \kappa$, or equivalently, $G_E\rightarrow1$ and $G_M\rightarrow1+\kappa$ in \Eqref{BreitPotentialFinal}, we obtain the static limit of the Breit potential. In this limit, the non-vanishing potentials are $\Delta V_{\mathrm{rel. C.}}$, $\Delta V_{\mathrm{rel. E}_\mathrm{kin}}$, $\Delta V_1$ and $\Delta V_2$.\footnote{In the structureless limit, we in addition have $\kappa \rightarrow 0$, and hence, a change in $\Delta V_2$.} The remaining potentials, $\Delta V_Y$, $\Delta V_3$, $\Delta V_4$ and $\Delta V_5$, describe the proton structure corrections. We will first go through the static corrections and subsequently discuss the FSEs.

The unperturbed Coulomb potential, \Eqref{VC}, generates the well-known gross structure as given in \Eqref{Bohr}, cf.\ also the second term in \Eqref{DiracEnergies6}. Together with the relativistic corrections to the Coulomb Schrödinger equation, i.e., the sum of the potentials in Eqs.~\eref{LSrelC} and \eref{RelEkin}, the Coulomb Dirac energy is reproduced up to order $(Z\al)^6$:
\beq
\langle 2l_{j} \left\vert \Delta V_{\mathrm{rel. C.}}+\Delta V_{\mathrm{rel. E}_\mathrm{kin}} \right\vert 2l_{j}  \rangle=\frac{Z \al}{128 a^3}\frac{1}{m_r^2}\left[3-\frac{8}{j+1/2}\right], \eqlab{newDirac}
\eeq
cf.\ third term in \Eqref{DiracEnergies6}. Obviously, \Eqref{newDirac} introduces a FS splitting of, f.i., the $2P$ levels, while the $2P_{1/2}$ and $2S_{1/2}$ levels of the classic LS are degenerate.

Treating Eqs.~\eref{FermiPot} and \eref{V284e} at first-order in PT, we reproduce the LO HFSs. For $S$-levels that is the well-known Fermi energy, cf.\ \Eqref{FermiE}. The HFSs of $2P_{1/2}$ and $2P_{3/2}$ levels, cf.\ Eqs.~\eref{P12HFS} and \eref{P32HFS}, are obtained as in Ref.~\cite[Eqs.~(86) and (87)]{Pachucki:1996zza}. Furthermore, we checked that \Eqref{V284e} agrees with Ref.~\cite[Eq.~(24)]{Martynenko:2006gz}.

The total static contribution to the $2P$ FS is generated by the relativistic corrections to the Coulomb potential and the potential $\Delta V_1$. Neglecting the lepton anomalous magnetic moment, our result, \Eqref{2PFScom}, agrees with Ref.~\cite[Eq.(80)]{Pachucki:1996zza} and Ref.~\cite[Eq.~(3)]{Martynenko:2006gz} up to order $(Z\al)^6$.

The total static contribution to the classic LS is generated by the kinetic energy operators of $\Delta V_\mathrm{red.Mass}$ and the potential $\Delta V_1$. The LS effect is of nuclear recoil type, cf.\ \Eqref{staticLS}, and was predicted long time ago by Ref.~\cite{Barker:1955zz}, see also Ref.~\cite[Eq.~(47)]{Pachucki:1996zza}.

In conclusion, we reproduce all the static contributions to the hydrogen spectrum as reviewed in Ref.~\cite[Eq.~(3.4)]{Eides:2000xc} and the leading contributions to the HFS. We now turn our attention to the FSEs. The Yukawa-type electric Sachs FF perturbation, \Eqref{YukawaCoordinate}, and the magnetic Sachs FF perturbation, \Eqref{V5}, were discussed in details  in \secref{chap2}{Exact}, where we pointed out a limitation in the usual accounting of FSEs in terms of electric and magnetic radii. Nevertheless, below we will apply the standard procedure of expanding in moments of charge and magnetization distributions, as we want to compare to the literature. The results are derived in first-order perturbation theory (1PT) and second-order perturbation theory (2PT) of the continuous spectrum. The exact results are expanded in the small parameter $Z\al$ and the electromagnetic radii are identified as in \Eqref{rmsdef}. For brevity, we will truncate the presented expressions at order $(Z \al)^5$. In doing so, the higher order terms from $P$-waves can be neglected. 

Combining Eqs.~\eref{VY1PT} and \eref{VY1PT}, we reproduce the well-known NLO FSEs in the LS, \Eqref{LambShift}, given by the rms charge radius and the Friar radius. Combining Eqs.~\eref{V5RZ}, \eref{V2RZ} and \eref{V5RZ2PT}, we reproduce the well-known Zemach radius contribution to the HFS, cf.\ Eqs.~\eref{HFS} and \eref{ZemachTerm}. In addition, we find recoil FSEs of order $(Z\al)^5$, see Eqs.~\eref{V3int} and \eref{recoilFSE1}. In \secref{chap5}{matchingOTPE}, we will try to match the latter to the nucleon-pole contribution of TPE.

Next, we study the $P$-level mixing indicated in \Figref{MuHSpec} and described by the parameter $\delta$. We proceed analogously to Ref.~\cite{Pachucki:1996zza} and keep the notations used therein. We form a matrix for an effective Hamiltonian in the basis of the states $2^1P_{1/2}$, $2^3P_{1/2}$, $2^3P_{3/2}$, $2^5P_{3/2}$:
\beq
H=\left[\begin{matrix}
-\frac{3}{4}\beta_1&&&\\
&\frac{1}{4}\beta_1&\beta_2&\\
&\beta_2&-\frac{5}{8}\beta_3+\gamma&\\
&&&\frac{3}{8}\beta_3+\gamma\\
\end{matrix}\right].
\eeq
Here, $\beta_1$ and $\beta_3$ are the static $2P$ HFSs in Eqs.~\eref{beta1} and \eref{beta3}. The non-vanishing mixed matrix elements of $2^3P_{1/2}$- and $2^3P_{3/2}$-states, originating from the static $\Delta V_2$ potential and the $\Delta V_5$ potential with eVP, cf.\ Eqs.~\eref{beta2part1} and \eref{VP5Pmix} (last row in Table \ref{eVPTab}), add up to:
\beq
\beta_2=-0.797 \mbox{ meV}.
\eeq 
The dominant static and eVP contributions to the $2P$ FS, cf.\ Eqs.~\eref{2PFScom}, \eref{3eVP2P} and \eref{3eVP2P32}, amount to:
\beq
\gamma=8.332\mbox{ meV}.
\eeq
To obtain the eigenvalues in the chosen basis, we are left to diagonalise the matrix $H$.\footnote{A matrix of the form:
\beq
M=\left[\begin{matrix}
b_1&m\\
m&b_2\\
\end{matrix}\right],
\eeq
has the eigenvalues $\nicefrac{1}{2}\left[b_1+b_2\pm\sqrt{(b_2-b_1)^2+4m^2}\right]$.} In accordance with Ref.~\cite[Eq.~(93)]{Pachucki:1996zza} and Ref.~\cite{Martynenko:2006gz}, we find a shift of $2^3P$ levels with\footnote{Note that, in contrast to Refs.~\cite{Pachucki:1996zza, Martynenko:2006gz}, we neglected the anomalous magnetic moment of the muon.}
\beq
\delta=0.14530\mbox{ meV}.
\eeq
The inclusion of the eVP correction presented in \eref{VP5Pmix}, changes $\delta$ at the level of $10^{-4}$ meV.

\subsection[$2P$ Fine Structure]{$\boldsymbol{2P}$ Fine Structure}  \seclab{2PFineStructure}
The static contributions to the $2P$ FS are:
\begin{itemize}
\item $\Delta V_{\mathrm{rel. C.}}$ (1PT):
\begin{subequations}
\begin{align}
\langle 2P_{1/2} \left\vert \Delta V_{\mathrm{rel. C.}} \right\vert 2P_{1/2} \rangle&=-\frac{Z \al}{48a^3}\frac{1}{m_r^2},\\
\langle 2P_{3/2} \left\vert \Delta V_{\mathrm{rel. C.}} \right\vert 2P_{3/2} \rangle&=\frac{Z \al}{96a^3}\frac{1}{m_r^2};
\end{align}
\end{subequations}
\item $\Delta V_1$ (1PT):
\begin{subequations}
\begin{align}
\langle 2P_{1/2} \left\vert \Delta V_{1} \right\vert 2P_{1/2} \rangle&=-\frac{Z \al}{16a^3}\left[\frac{1}{mM}-\frac{1}{3}\frac{1}{M^2}\right],\\
\langle 2P_{3/2} \left\vert \Delta V_{1} \right\vert 2P_{3/2} \rangle&=-\frac{Z \al}{16a^3}\left[\frac{1}{mM}+\frac{1}{6}\frac{1}{M^2}\right].
\end{align}
\end{subequations}
\end{itemize}
Summing up, the static $2P$ FS evaluates to:
\begin{align}
&\langle 2P_{3/2} \left\vert \Delta V_{\mathrm{rel. C.}}+\Delta V_1 \right\vert 2P_{3/2}\rangle-\langle 2P_{1/2} \left\vert \Delta V_{\mathrm{rel. C.}}+\Delta V_1 \right\vert 2P_{1/2}\rangle\nn\\
&=\frac{Z\al}{32a^3}\left[\frac{1}{m_r^2}-\frac{1}{M^2}\right]=8.329\mbox{ meV}.\eqlab{2PFScom}
\end{align}
The nuclear finite-size contribution to the $2P$ FS from $\Delta V_3$ starts at order $(Z\al)^6$ only and can be neglected.

\subsection[$P$-Level Mixing]{$\boldsymbol{P}$-Level Mixing} \seclab{PMixing}
The mixed matrix element from the static potential reads:
\begin{itemize}
\item $\Delta V_{2}$ (1PT):
\beq
\langle 2^3P_{1/2} \left\vert \Delta V_{2} \right\vert 2^3P_{3/2} \rangle=-\frac{\sqrt{2}Z \al}{144 \,a^3}\left[\frac{1+\kappa}{mM}+\frac{1+2\kappa}{M^2}\right]=-0.796 \mbox{ meV}\equiv \beta_2. \eqlab{beta2part1}
\eeq
\end{itemize}
The nuclear finite-size contributions to the $P$-level mixing from $\Delta V_4$ and $\Delta V_5$ start at order $(Z\al)^6$ only and can be neglected.

\subsection{Lamb Shift}
The static contributions to the $2P_{1/2}-2S_{1/2}$ LS are:
\begin{itemize}
\item $\Delta V_{\mathrm{rel. C.}}+\Delta V_{\mathrm{rel. E}_\mathrm{kin}}$ (1PT):
\begin{subequations}
\begin{align}
\langle 2S_{1/2} \left\vert \Delta V_{\mathrm{rel. C.}}+\Delta V_{\mathrm{rel. E}_\mathrm{kin}} \right\vert 2S_{1/2} \rangle&=\frac{Z \al}{a^3}\frac{1}{m_r^2}\left[\frac{1}{16}-\frac{13}{128}\right]=-\frac{5Z \al}{128a^3}\frac{1}{m_r^2},\\
\langle 2P_{1/2} \left\vert \Delta V_{\mathrm{rel. C.}}+\Delta V_{\mathrm{rel. E}_\mathrm{kin}} \right\vert 2P_{1/2} \rangle&=-\frac{Z \al}{a^3}\frac{1}{m_r^2}\left[\frac{1}{48}+\frac{7}{384}\right]=-\frac{5Z \al}{128a^3}\frac{1}{m_r^2};
\end{align}
\end{subequations}
\item $\Delta V_\mathrm{red.Mass}$ (1PT):
\beq
\langle 2P_{1/2} \left\vert \Delta V_\mathrm{red.Mass} \right\vert 2P_{1/2} \rangle-\langle 2S_{1/2}\left\vert \Delta V_\mathrm{red.Mass} \right\vert 2S_{1/2} \rangle=-\frac{Z \al}{4a^3}\frac{1}{mM};
\eeq
\item $\Delta V_{1}$ (1PT):
\beq
\langle 2P_{1/2} \left\vert \Delta V_{1} \right\vert 2P_{1/2} \rangle-\langle 2S_{1/2}\left\vert \Delta V_{1} \right\vert 2S_{1/2} \rangle =\frac{Z \al}{4a^3}\left[\frac{1}{mM}+\frac{1}{12M^2}\right].
\eeq
\end{itemize}
Summing up, the static $2P_{1/2}-2S_{1/2}$ LS evaluates to:
\beq
\langle 2P_{1/2} \left\vert \Delta V_\mathrm{red.Mass}+\Delta V_{1} \right\vert 2P_{1/2} \rangle-\langle 2S_{1/2}\left\vert V_\mathrm{red.Mass}+\Delta V_{1} \right\vert 2S_{1/2} \rangle =\frac{Z \al}{48a^3}\frac{1}{M^2}.\eqlab{staticLS}
\eeq
The finite-size contributions to the $2P_{1/2}-2S_{1/2}$ LS are:\footnote{We make use of the following relations:
\begin{subequations}
\bea
\langle r\rangle_E-\frac{1}{2}\langle r\rangle_{E(2)}&=&\frac{1}{\pi^2} \int_{t_0}^\infty \frac{\dd t}{t}\int_{t_0}^\infty \frac{\dd t'}{t'}\frac{\im G_E(t)\im G_E(t')}{\sqrt{t}+\sqrt{t'}},\\
\langle r^3\rangle_E-\frac{1}{2}\langle r^3\rangle_{E(2)}&=&-\frac{12}{\pi^2} \int_{t_0}^\infty \frac{\dd t}{t^{3/2}}\int_{t_0}^\infty \frac{\dd t'}{t'^{3/2}}\frac{\im G_E(t)\im G_E(t')}{\sqrt{t}+\sqrt{t'}}.\qquad
\eea
\end{subequations}}
\begin{itemize}
\item $\Delta V_Y$ (1PT):
\beq
\langle 2S_{1/2} \left\vert \Delta V_Y \right\vert 2S_{1/2}\rangle=\frac{Z \al}{a^3}\left[\frac{\langle r^2\rangle_E}{12}-\frac{\langle r^3\rangle_E}{12a}\right]+\mathcal{O}(Z\al)^6\eqlab{VY1PT};
\eeq
\item $\Delta V_Y$ (interference at 2PT):
\beq
 E^{\langle \Delta V_Y \rangle\langle \Delta V_Y \rangle(2)}_{2S_{1/2}}=\frac{Z \al}{12 a^4}\left[\langle r^3\rangle_E-\frac{1}{2}\langle r^3\rangle_{E(2)}\right]+\mathcal{O}(Z\al)^6;\eqlab{VY2PT}
 \eeq
\item $\Delta V_3$ (1PT):
\beq
\langle 2S_{1/2} \left\vert \Delta V_{3} \right\vert 2S_{1/2} \rangle=\frac{Z\al}{a^4}\left[\frac{3}{8mM}-\frac{1}{8m_r^2}\right]\langle r\rangle_E+\mathcal{O}(Z\al)^6;\eqlab{V31PT}
\eeq
\item $\Delta V_Y$ and $\Delta V_\mathrm{rel.C}$ (interference at 2PT):
\beq
E^{\langle \Delta V_Y \rangle\langle \Delta V_\mathrm{rel.C} \rangle(2)}_{2S_{1/2}}=-\frac{Z\al}{8a^4}\frac{1}{m_r^2}\langle r\rangle_E+\mathcal{O}(Z\al)^6;
\eeq
\item $\Delta V_Y$ and $\Delta V_{\mathrm{rel. E}_\mathrm{kin}}$ (interference at 2PT):
\beq
E^{\langle \Delta V_Y \rangle\langle \Delta V_{\mathrm{rel. E}_\mathrm{kin}} \rangle(2)}_{2S_{1/2}}=\frac{Z\al}{4a^4}\frac{1}{m_r^2}\langle r\rangle_E+\mathcal{O}(Z\al)^6;
\eeq
\item $\Delta V_Y$ and $\Delta V_{\mathrm{red.Mass}}$ (interference at 2PT):
\beq
E^{\langle \Delta V_Y \rangle\langle \Delta V_{\mathrm{red.Mass}} \rangle(2)}_{2S_{1/2}}=-\frac{3Z\al}{4a^4}\frac{1}{mM}\langle r\rangle_E+\mathcal{O}(Z\al)^6;
\eeq
\item $\Delta V_Y$ and $\Delta V_1$ (interference at 2PT):
\beq
 E^{\langle \Delta V_Y \rangle\langle \Delta V_1 \rangle(2)}_{2S_{1/2}}=\frac{Z\al}{4a^4}\frac{1}{mM}\langle r\rangle_E+\mathcal{O}(Z\al)^6;\eqlab{V1int}
 \eeq
 \item $\Delta V_Y$ and $\Delta V_3$ (interference at 2PT):
 \beq
 E^{\langle \Delta V_Y \rangle\langle \Delta V_3 \rangle(2)}_{2S_{1/2}} =\frac{Z\al}{8a^4}\left[\langle r\rangle_E-\frac{1}{2}\langle r\rangle_{E(2)}\right]\left[\frac{2}{m_r^2}-\frac{5}{mM}\right]+\mathcal{O}(Z\al)^6.\eqlab{V3int}
 \eeq
\end{itemize}
Equations~\eref{VY1PT} and \eref{VY2PT} add up to the well-known non-recoil FSEs, cf.\ \Eqref{LambShift}. Summing up Eqs.~\eref{V31PT}-\eref{V1int}, we obtain a recoil contribution proportional to the first moment of the charge distribution:
\beq
E^{\langle  \Delta V_{3} \rangle(1)+\langle \Delta V_Y \rangle\langle \Delta V_\mathrm{rel.C}+\Delta V_{\mathrm{rel. E}_\mathrm{kin}}+\Delta V_{\mathrm{red.Mass}}+\Delta V_1 \rangle(2)}_{2S_{1/2}}=-\frac{Z\al}{8a^4}\frac{1}{mM}\langle r\rangle_E+\mathcal{O}(Z\al)^6.\eqlab{recoilFSE1}
\eeq
Furthermore, \Eqref{V3int} gives finite-size recoil effects proportional to the characteristic difference of first moments of charge and convoluted charge distributions:
\beq
\langle r\rangle_E-\frac{1}{2}\langle r\rangle_{E(2)}=\frac{2}{\pi}\int_0^\infty \frac{\dd Q}{Q^2} \left[G_E(Q^2)-1\right]^2.
\eeq
\subsection{Hyperfine Splitting} \seclab{HFSSec}
The static contributions to the HFSs are:
\begin{itemize}
\item $\Delta V_2$ (1PT):
\begin{subequations}
\begin{align}
\langle nS_{1/2} \left\vert \Delta V_{2} \right\vert nS_{1/2} \rangle_{(f=1)-(f=0)}&=\frac{4Z\al}{3a^3}\frac{1+\kappa}{mM}\frac{1}{n^3}\left[\frac{1}{2}+\frac{3}{2}\right]\equiv E_\mathrm{F}(nS),\\
\langle 2P_{1/2}  \left\vert \Delta V_{2} \right\vert 2P_{1/2}  \rangle_{(f=1)-(f=0)}&=\frac{Z \al}{18a^3}\left[\frac{1}{2}+\frac{3}{2}\right]\left\{\frac{1+\kappa}{mM}+\frac{1+2\kappa}{4M^2}\right\}\eqlab{P12HFS},\\&=7.953\,\mathrm{meV}\equiv\beta_1\eqlab{beta1},\\
\langle 2P_{3/2}  \left\vert \Delta V_{2} \right\vert 2P_{3/2}  \rangle_{(f=2)-(f=1)}&=\frac{Z \al}{18a^3}\left[\frac{3}{2}+\frac{5}{2}\right]\left\{\frac{1}{5}\frac{(1+\kappa)}{mM}+\frac{1+2\kappa}{8M^2}\right\}\eqlab{P32HFS},\\&=3.392\,\mathrm{meV}\equiv\beta_3\eqlab{beta3};
\end{align}
\end{subequations}
\end{itemize}
The finite-size contributions to the $2S$ HFS are:
\begin{itemize}
\item $\Delta V_5$ (1PT):
\begin{subequations}
\eqlab{V5RZ}
\begin{align}
\langle 2S_{1/2} \left\vert \Delta V_{5} \right\vert 2S_{1/2} \rangle_{(f=1)-(f=0)}&\approx\frac{8E_\mathrm{F}(2S)}{a}\frac{1}{\pi}\int_0^\infty \frac{\dd k}{k^2}\left[\frac{G_M(-k^2)}{1+\kappa}-1\right],\\
&=-2E_\mathrm{F}(2S) \,\frac{\langle r\rangle_M}{a}+\mathcal{O}(Z\al)^6;
\end{align}
\end{subequations}
\item $\Delta V_Y$ and $\Delta V_2$ (interference at 2PT):
\begin{subequations}
\eqlab{V2RZ}
\begin{align}
 E^{\langle \Delta V_Y \rangle\langle \Delta V_2 \rangle(2)}_\mathrm{HFS}(2S)&\approx\frac{8E_\mathrm{F}(2S)}{a}\frac{1}{\pi}\int_0^\infty \frac{\dd k}{k^2}\left[G_E(-k^2)-1\right],\\
 &=-2E_\mathrm{F}(2S)\frac{\langle r\rangle_E}{a}+\mathcal{O}(Z\al)^6;
\end{align}
\end{subequations}
\item $\Delta V_Y$ and $\Delta V_5$ (interference at 2PT):
\begin{subequations}
\eqlab{V5RZ2PT}
\begin{align}
 E^{\langle \Delta V_Y \rangle\langle \Delta V_5 \rangle(2)}_\mathrm{HFS}(2S)&\approx\frac{8E_\mathrm{F}(2S)}{a}\frac{1}{\pi}\int_0^\infty \frac{\dd k}{k^2}\left[G_E(-k^2)-1\right]\left[\frac{G_M(-k^2)}{1+\kappa}-1\right],\\
 &=-2E_\mathrm{F}(2S)\,\frac{\left[R_\text{Z}-\langle r\rangle_E-\langle r\rangle_M\right]}{a}+\mathcal{O}(Z\al)^6.
\end{align}
\end{subequations}
\end{itemize}
Summing up, the finite-size contribution to the $2S$ HFS evaluates to:
\beq
 E^{\langle \Delta V_5 \rangle+\langle \Delta V_Y \rangle\langle \Delta V_2+\Delta V_5 \rangle}_\mathrm{HFS}(2S)=-2E_\mathrm{F}(2S)\,\frac{R_\text{Z}}{a}+\mathcal{O}(Z\al)^6.
\eeq
The nuclear finite-size contributions to the $P$-level HFSs from $\Delta V_4$ and $\Delta V_5$ start at order $(Z\al)^6$ only and can be neglected.

\addtocontents{toc}{\protect\setcounter{tocdepth}{4}}

\section{Vacuum Polarization Contributions} \seclab{2appVP}

In the following, we will motivate how one derives the Breit potential for OPE with VP, cf.\ \Figref{VP}. Electromagnetic gauge invariance constrains the VP tensor
to the well-known form:
\beq
\Pi^{\mu\nu}(q)= (g^{\mu\nu}q^2 -q^\mu q^\nu )\, \Pi(q^2),
\eeq
where the scalar VP function satisfies a once-subtracted DR:
\beq
\Pi(Q^2) = -\frac{Q^2}{\pi} \int^\infty_{t_0} \!  \frac{\dd t}{t}\, \frac{\im \, \Pi(t)}{t+Q^2}, \eqlab{VPDR}
\eeq
with $q^2=-Q^2$ and $t_0=4m^2$ being the (lowest) pair-production threshold for the particles in the VP loop. 

\begin{figure}[tbh]
\centering
       \includegraphics[scale=0.65]{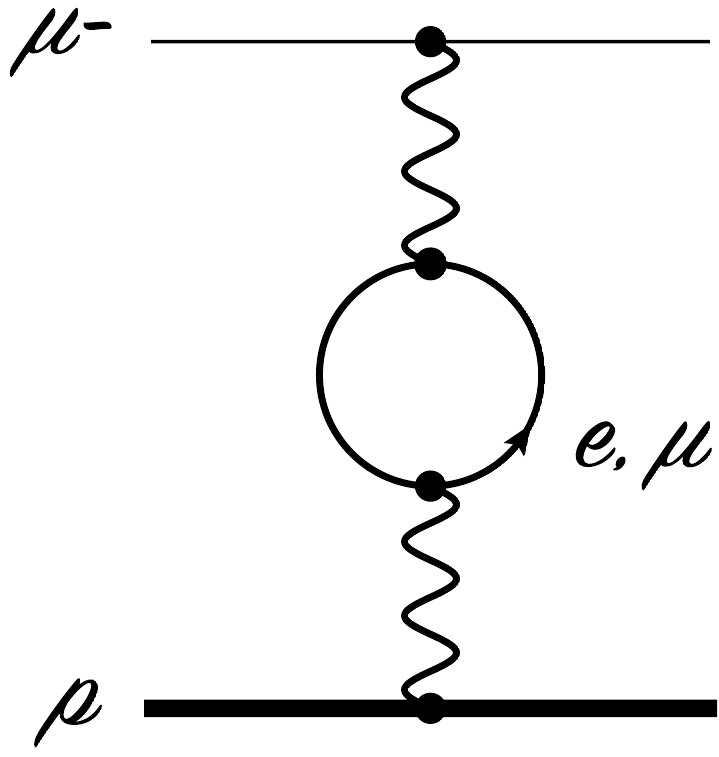}
                     \caption{One-loop vacuum-polarization correction.\label{fig:VP}}
\end{figure}

As shown in \Figref{VP}, the $\mu$H bound state receives contributions from eVP and $\mu$VP. 
We will limit ourselves to the eVP, as it dominates the $\mu$H LS, cf.\ \secref{chap2}{MuEComp}. The calculation of the $\mu$VP effects proceeds analogously. The hadronic VP contribution to the $\mu$H LS is discussed in the literature, see Refs.~\cite{Friar:1998wu,Martynenko:2001qf}.

Modifying the photon propagator, given in Eqs.~\eref{PhotonProp} and \eref{PhotonProp2}, with a VP insertion, we obtain:
\bea
\Delta_{\mu \nu} &\rightarrow& \Delta_{\mu \al}(q)\,\Pi^{\al \be}(q)\,\Delta_{\be \nu}(q)=\Delta_{\mu \nu} \frac{\Pi(q^2)}{q^2}.
\eea
Since the nuclear FFs satisfy a once-subtracted DR similar to \Eqref{VPDR}, we can deduce the VP Breit potential from the nuclear FF dependent Breit potentials in \Eqref{YukawaCoordinate} and Eqs.~\eref{V3}-\eref{V5}, cf.\ \secref{chap2}{OPE}, by replacing:
\beq
G_E(Q^2) \rightarrow \Pi (Q^2), \qquad G_M(Q^2) \rightarrow (1+\kappa)\,\Pi (Q^2).
\eeq
At one-loop level in spinor QED, we can then rely on the well-known expression for leptonic VP:
\beq
\eqlab{1loopVP}
\im \, \Pi^{(1)}(4m^2 t) = -\frac{\al}{3}\left(1+\frac{1}{2t}\right)\sqrt{1-\frac{1}{t} }. 
\eeq

In what follows, we give exact expressions for the contributions of eVP to the $1S$, $2S$ and $2P$ levels in hydrogen-like atoms. In Table \ref{eVPTab}, we quantify the effects in $\mu$H and compare to the literature values. All results are obtained in first-order PT. The lowest-order eVP correction is given by the Uehling potential \cite{Uehling:1935uj}:
\begin{subequations}
\eqlab{UehlingNum}
\bea
E^{\,\mathrm{eVP} \langle \Delta V_Y\rangle}_{1S_{1/2}}&=&\frac{4Z\al}{a}\frac{1}{\pi}\int_{1}^\infty\frac{\dd t}{t} \frac{1}{\left[2+\sqrt{t}/\lambda\right]^2} \,\im \Pi(4m_e^2t),\\
E^{\,\mathrm{eVP}\langle \Delta V_Y\rangle}_{2S_{1/2}}&=&\frac{Z\al}{4a}\frac{1}{\pi}\int_{1}^\infty\frac{\dd t}{t} \frac{1+2\, t/\lambda^2}{\left[1+\sqrt{t}/\lambda\right]^4}\, \im \Pi(4m_e^2t),\qquad \eqlab{UehlingNum2S}\\
E^{\,\mathrm{eVP}\langle \Delta V_Y\rangle}_{2P}&=&\frac{Z\al}{4a}\frac{1}{\pi}\int_{1}^\infty\frac{\dd t}{t} \frac{1}{\left[1+ \sqrt{t}/\lambda\right]^4} \,\im \Pi(4m_e^2t), \eqlab{UehlingNum2P}
\eea
\end{subequations}
where we introduced $\lambda =1/2am_e $. As one can see from Table \ref{eVPTab} and \Figref{Pohl}, \Eqref{UehlingNum2S} represents the dominant contribution to the LS in $\mu$H and is responsible for the rearrangement of $2P_{1/2}$- and $2S_{1/2}$-levels as compared to the H spectrum, see Figures \ref{fig:MuHSpec} and \ref{fig:EHSpec}. An additional but weaker eVP contribution to the LS and the $2P$ FS is described by:
\begin{subequations}
\begin{align}
E^{\,\mathrm{eVP} \langle \Delta V_3\rangle}_{1S_{1/2}}& = \frac{2Z\al}{ a^3}\frac{1}{\pi}\int_{1}^\infty\frac{\dd t}{t}\left\{\frac{4+3\sqrt{t}/\lambda}{mM}-\frac{1+\sqrt{t}/\lambda}{m_r^2}\right\}\frac{\im \Pi(4m_e^2t)}{\left[2+\sqrt{t}/\lambda\right]^2},\\
E^{\,\mathrm{eVP} \langle \Delta V_3\rangle}_{2S_{1/2}}& = \frac{Z\al}{16 a^3}\frac{1}{\pi}\int_{1}^\infty\frac{\dd t}{t}\left\{\frac{1}{mM}\left(5+18\sqrt{t}/\lambda+22\, t/\lambda^2+12\,t^{3/2}/\lambda^3\right)\right.\nn\\
&\left.\quad-\frac{1}{2m_r^2}\left(2+8\sqrt{t}/\lambda+11\,t/\lambda^2+8\,t^{3/2}/\lambda^3\right)\right\}\frac{\im \Pi(4m_e^2t)}{\left[1+\sqrt{t}/\lambda\right]^4}, \eqlab{3eVP2S}\\
E^{\,\mathrm{eVP} \langle \Delta V_3\rangle}_{2P_{1/2}}& = \frac{Z\al}{96 a^3}\frac{1}{\pi}\int_{1}^\infty\frac{\dd t}{t}\left\{\frac{3\, t/\lambda^2}{m_r^2}+\frac{2}{m^2}\left(1+4\sqrt{t}/\lambda+3\,t/\lambda^2\right)\right.\nn\\
&\left.\quad+\frac{2}{mM}\left(5+14\sqrt{t}/\lambda+6\, t/\lambda^2\right)\right\}\frac{\im \Pi(4m_e^2t)}{\left[1+\sqrt{t}/\lambda\right]^4},\eqlab{3eVP2P}\\
E^{\,\mathrm{eVP} \langle \Delta V_3\rangle}_{2P_{3/2}}& = \frac{Z\al}{96 a^3}\frac{1}{\pi}\int_{1}^\infty\frac{\dd t}{t}\left\{\frac{3\, t/\lambda^2}{m_r^2}-\frac{1}{m^2}\left(1+4\sqrt{t}/\lambda+3\,t/\lambda^2\right)\right.\nn\\
&\left.\quad+\frac{1}{mM}\left(4+4\sqrt{t}/\lambda-6\, t/\lambda^2\right)\right\}\frac{\im \Pi(4m_e^2t)}{\left[1+\sqrt{t}/\lambda\right]^4}.\eqlab{3eVP2P32}
\end{align}
\end{subequations}
The modification of the magnetic Sachs FF potential, $\Delta V_5$, will give relevant contributions to the HFSs:
\begin{subequations}
\eqlab{VP5all}
\bea
&&E^{\,\mathrm{eVP} \langle \Delta V_5\rangle}_\mathrm{HFS}(1S) = -4E_\mathrm{F}(1S)\,\frac{1}{\pi}\int_{1}^\infty\frac{\dd t}{t}\frac{1+\sqrt{t}/\lambda}{\left[2+\sqrt{t}/\lambda\right]^2}\im \Pi(4m_e^2t),\\
&&E^{\,\mathrm{eVP} \langle \Delta V_5\rangle}_\mathrm{HFS}(2S) = -\frac{E_\mathrm{F}(2S)}{2\pi}\int_{1}^\infty\frac{\dd t}{t}\frac{2+8\sqrt{t}/\lambda+11\,t/\lambda^2+8\,t^{3/2}/\lambda^3}{\left[1+\sqrt{t}/\lambda\right]^4}\im \Pi(4m_e^2t),\eqlab{VP52S}\qquad\qquad\quad\\
&&E^{\,\mathrm{eVP} \langle \Delta V_5\rangle}_\mathrm{HFS}(2P_{1/2}) =- \frac{Z\al}{18 a^3}\frac{1+\kappa}{\pi}\int_{1}^\infty\frac{\dd t}{t}\left\{\frac{1}{mM}\left(2+8\sqrt{t}/\lambda+9\,t/\lambda^2\right)\nn\right.\\
&&\left.\hspace{3.4cm}+\frac{1}{M^2}\left(
1+4\sqrt{t}/\lambda+3\,t/\lambda^2\right)\right\} \frac{\im \Pi(4m_e^2t)}{\left[1+\sqrt{t}/\lambda\right]^4},\eqlab{VP52P}\\
&&E^{\,\mathrm{eVP} \langle \Delta V_5\rangle}_\mathrm{HFS}(2P_{3/2}) =- \frac{Z\al}{90 a^3}\frac{1+\kappa}{\pi}\int_{1}^\infty\frac{\dd t}{t}\left\{\frac{4}{mM}\left(1+4\sqrt{t}/\lambda\right)\nn\right.\\
&&\left.\hspace{3.4cm}+\frac{5}{M^2}\left(
1+4\sqrt{t}/\lambda+3\,t/\lambda^2\right)\right\} \frac{\im \Pi(4m_e^2t)}{\left[1+\sqrt{t}/\lambda\right]^4},\eqlab{VP52P3}\\
&&\langle 2^3P_{1/2} \left\vert \Delta V^{\,\mathrm{eVP}}_{5} \right\vert 2^3P_{3/2} \rangle =\frac{Z\al}{72\sqrt{2} a^3}\frac{1+\kappa}{\pi}\int_{1}^\infty\frac{\dd t}{t}\left\{\frac{1}{mM}\left(7+28\sqrt{t}/\lambda+51\,t/\lambda^2\right)\nn\right.\\
&&\left.\hspace{4.6cm}+\frac{2}{M^2}\left(
1+4\sqrt{t}/\lambda+3\,t/\lambda^2\right)\right\} \frac{\im \Pi(4m_e^2t)}{\left[1+\sqrt{t}/\lambda\right]^4}.\eqlab{VP5Pmix}
\eea
\end{subequations}
In the last line, we also give the effect on the $P$-level mixing. The effect from the eVP analogue of $\Delta V_4$ on the $2P$ HFS can be neglected, as it is strongly suppressed by the nuclear mass:
\bea
E^{\,\mathrm{eVP} \langle \Delta V_4\rangle}_{2P}& = &\frac{Z\al}{a^3}\frac{1}{36M^2}\frac{1}{\pi}\int_{1}^\infty\frac{\dd t}{t}\frac{1+3\sqrt{t}/\lambda}{\left[1+\sqrt{t}/\lambda\right]^3}\im \Pi(4m_e^2t),
\eea
and the associated $P$-level mixing would be further suppressed by an additional factor of $-(2\sqrt{2})^{-1}$. However, it is worth noting that it produces the same HFS for $2P_{1/2}$ and $2P_{3/2}$.

Table \ref{eVPTab} gives an overview of the above eVP effects evaluated for $\mu$H. For the Uehling potential, we explicitly checked that the difference of Eqs.~\eref{UehlingNum2P} and \eref{UehlingNum2S} matches the formula presented in Ref.~\cite[Eq.\ (16)]{Pachucki:1996zza}. Also, we agree with the Breit potential of Ref.~\cite{Jentschura:2011nx}. The comparison of our numerical results for the FS and HFS with Refs.~\cite{Martynenko:2006gz,Pachucki:1996zza}, see Table \ref{eVPTab}, is not exact because we omit the anomalous magnetic moment of the muon. To summarize, the dispersive approach to the OPE Breit potential with VP insertion reproduces all the known one-loop leptonic VP contributions to the spectra of hydrogen-like atoms.

\renewcommand{\arraystretch}{1.5}
\begin{table}[h]
\centering
\caption{Numerical results for the electronic vacuum polarization contributions to the spectrum of muonic hydrogen.
}
\label{eVPTab}
\begin{small}
\begin{tabular}{|c|c|c|c|}
\hline
 \rowcolor[gray]{.7}
 \hline
&&{\bf our result}&\\
 \rowcolor[gray]{.7}
\multirow{-2}{*}{{\bf eVP contribution}}&\multirow{-2}{*}{{\bf Eqs.}}& {\bf [meV]}&\multirow{-2}{*}{{\bf literature value [meV]}}\\
\hline
&&& $205.0074$ \cite[Table I]{Pachucki:1999zza} \\
\multirow{-2}{*}{$E^{\,\mathrm{eVP}\langle \Delta V_Y\rangle}_{\mathrm{LS}}$}&\multirow{-2}{*}{\eref{UehlingNum2S}, \eref{UehlingNum2P}}&\multirow{-2}{*}{$205.0074$}&and \cite[Eq.\ (16)]{Pachucki:1996zza}\\
 \rowcolor[gray]{.95}
$E^{\,\mathrm{eVP}\langle \Delta V_3\rangle}_{\mathrm{LS}}$&\eref{3eVP2S}, \eref{3eVP2P}&$-0.0277$&$-0.0277$ \cite[$\delta E^{(1)}$ from Table I]{Jentschura:2011nx}\\
$E^{\,\mathrm{eVP}\langle \Delta V_3\rangle}_{\mathrm{FS}}$&\eref{3eVP2P}, \eref{3eVP2P32}&$0.0030$&$0.005$ \cite[Eq.\ (82)]{Pachucki:1996zza} \\
 \rowcolor[gray]{.95}
$E^{\,\mathrm{eVP} \langle \Delta V_5\rangle}_\mathrm{HFS}(2S_{1/2})$&\eref{VP52S}&$0.0482$&$0.0481$ \cite[Eq.~(18)]{Martynenko:2004bt}\\
$E^{\,\mathrm{eVP} \langle \Delta V_5\rangle}_\mathrm{HFS}(2P_{1/2})$&\eref{VP52P}&$0.0039$&$0.0038$ \cite[Eq.~(34)]{Martynenko:2006gz}\\
 \rowcolor[gray]{.95}
$E^{\,\mathrm{eVP} \langle \Delta V_5\rangle}_\mathrm{HFS}(2P_{3/2})$&\eref{VP52P3}&$0.0005$& $0.0005$ \cite[Eq.~(35)]{Martynenko:2006gz}\\
$\langle 2^3P_{1/2} \left\vert \Delta V^\mathrm{eVP}_{5} \right\vert 2^3P_{3/2} \rangle$&\eref{VP5Pmix}&$-0.0006$&\\
\hline
\end{tabular}
\end{small}
\end{table}
\renewcommand{\arraystretch}{1.3}

\section{Analysing Common Nucleon Form Factor Parametrizations}\seclab{FFparam}

We apply six different parametrizations for the electric and magnetic Sachs FFs of the proton to calculate various e.m.\ radii and contributions to the $\mu$H spectrum.\footnote{Similarly, Refs.~\cite{Carroll:2011de, Indelicato:2012pfa} compare exponential, uniform, Yukawa-, Fermi- and Gaussian-type charge distributions and evaluate, f.i., the Zemach radius and the expanded FSEs.} The electric FFs of the proton are shown in \Figref{DipoleFFComp}, together a parametrization from a simultaneous fit of all e.m.\ nucleon FFs \cite{Bradford:2006yz}, cf.\ $E_7$, and a Pad\'e approximation in $Q$ \cite{Bosted:1995}, cf.\ $E_8$, and compared to the simple dipole FF:
\beq
F_0(Q^2)=\left(\frac{\Lambda ^2}{\Lambda ^2+Q^2}\right)^2,
\eeq
with $\Lambda=0.71\text{GeV}$. There are two chain-fraction fits and  four Pad\'e approximations with the dimensionless momentum transfer $\tau=\frac{Q^2}{4M^2}$:
\begin{enumerate}
\item Chain-fraction fit by \citet{Arrington:2006hm}:
\begin{small}
\bea
E_1(Q^2)&=&\frac{1}{1+\frac{3.44 \,Q^2}{1-\frac{0.178 \,Q^2}{1-\frac{1.212 \,Q^2}{1+\frac{1.176 \,Q^2}{1-0.284 \,Q^2}}}}},\nn\\
M_1(Q^2)/\mu&=&\frac{1}{1+\frac{3.173 \,Q^2}{1-\frac{0.314 \,Q^2}{1-\frac{1.165 \,Q^2}{1+\frac{5.619 \,Q^2}{1-1.087 \,Q^2}}}}};\nn
\eea
\end{small}
\item Chain-fraction fit by \citet{Arrington:2006hm} with TPE corrections \cite{PhysRevC.72.034612}:
\begin{small}
\bea
E_2(Q^2)&=&\frac{1}{1+\frac{3.478 \,Q^2}{1-\frac{0.140 \,Q^2}{1-\frac{1.311 \,Q^2}{1+\frac{1.128 \,Q^2}{1-0.233 \,Q^2}}}}},\nn\\
M_2(Q^2)/\mu&=&\frac{1}{1+\frac{3.224 \,Q^2}{1-\frac{0.313 \,Q^2}{1-\frac{0.868 \,Q^2}{1+\frac{4.278 \,Q^2}{1-1.102 \,Q^2}}}}};\nn
\eea
\end{small}
\item Pad\'e approximation in $Q^2$ by \citet{Kelly:2004hm}:
\begin{small}
\bea
E_3(Q^2)&=&\frac{1-0.24 \,\tau }{1+10.98 \,\tau+12.82 \,\tau ^2+21.97 \,\tau ^3 },\nn\\
M_3(Q^2)/\mu&=&\frac{1+0.12 \,\tau }{1+10.97 \,\tau+18.86 \,\tau ^2+6.55 \,\tau ^3 }; \nn
\eea
\end{small}
\item Pad\'e approximation in $Q^2$ by \citet{Arrington:2007ux}:
\begin{small}
\begin{subequations}
\eqlab{AMT}
\bea
E_4(Q^2)&=&\frac{1+3.439 \,\tau-1.602 \,\tau ^2+0.068 \,\tau ^3}{1+15.055 \,\tau+48.061 \,\tau ^2+99.304 \,\tau ^3+0.012 \,\tau ^4+8.650 \,\tau ^5},\qquad\\
M_4(Q^2)/\mu&=&\frac{1-1.465 \,\tau+1.260 \,\tau ^2+0.262 \,\tau ^3}{1+9.627 \,\tau+11.179 \,\tau ^4+13.245 \,\tau ^5};
\eea
\end{subequations}
\end{small}
\item Pad\'e approximation in $Q^2$ by \citet{Alberico:2008sz}:
\begin{small}
\bea
E_5(Q^2)&=&\frac{1-0.19 \,\tau }{1+11.12 \,\tau+15.16 \,\tau ^2+21.25 \,\tau ^3},\nn\\
M_5(Q^2)/\mu&=&\frac{1+1.09 \,\tau }{1+12.31 \,\tau+25.57 \,\tau ^2+30.61 \,\tau ^3};\nn
\eea
\end{small}
\item Pad\'e approximation in $Q^2$ by \citet{Venkat:2010by}:
\begin{small}
\bea
E_6(Q^2)&=&\frac{1+2.909\,66 \,\tau-1.115\,422\,29 \,\tau ^2 +3.866\,171\times 10^{-2} \,\tau ^3}{1+14.518\,7212 \,\tau+40.883\,33 \,\tau ^2 +99.999\,998 \,\tau ^3+4.579\times 10^{-5}  \,\tau ^4+10.358\,0447 \,\tau ^5},\nn\\
M_6(Q^2)/\mu&=&\frac{1-1.435\,73 \,\tau+1.190\,520\,66 \,\tau ^2+0.254\,558\,41 \,\tau ^3 }{1+9.707\,036\,81 \,\tau+3.7357\times10^{-4} \,\tau ^2 +6\times10^{-8} \,\tau ^3+9.952\,7277  \,\tau ^4+12.797\,7739 \,\tau ^5};\nn
\eea
\end{small}
\item Pad\'e approximation in $Q^2$ by \citet{Bradford:2006yz}:\\
The e.m.\ nucleon FFs are parametrized as:
\beq
G(Q^2)=\frac{\sum_{k=0}a_k \tau^k}{1+\sum_{k=0}b_k \tau^k},\eqlab{functionalForm}
\eeq
with the fit parameters given in Table \ref{param}.
\renewcommand{\arraystretch}{1.75}
\begin{table}[h]
\centering
 \begin{scriptsize}
\caption{Fit parameters of Ref.~\cite{Bradford:2006yz} corresponding to the functional form given in \Eqref{functionalForm}.\label{param}}
\begin{tabular}{| c| c| c| c| c| c| c| c|}
\hline
 \rowcolor[gray]{.7}
{\bf Sachs FFs} & $a_0$& $a_1$& $a_2$& $b_1$& $b_2$& $b_3$& $b_4$\\
\hline
$G_{Ep}$&1&$-0.0578\pm0.166$&&$11.1\pm0.217$&$13.6\pm1.39$&$33.0\pm 8.95$&\\
 \rowcolor[gray]{.95}
$G_{Mp}/\mu_p$&1&$0.150\pm0.0312$&&$11.1\pm0.103$&$19.6\pm0.281$&$7.54\pm0.967$&\\
$G_{En}$&0&$1.25\pm0.368$&$1.30\pm1.99$&$-9.86\pm6.46$&$305\pm28.6$&$-758\pm77.5$&$802\pm156$\\
 \rowcolor[gray]{.95}
$G_{Mn}/\mu_n$&1&$1.81\pm0.402$&&$14.1\pm0.597$&$20.7\pm2.55$&$68.7\pm14.1$&\\
\hline
\end{tabular}
 \end{scriptsize}
\end{table}
\renewcommand{\arraystretch}{1.3}

\end{enumerate}

\renewcommand{\arraystretch}{1.75}
\begin{table}[h!]
 \centering
 \caption{Difference between various proton form factor parametrizations.}
 \label{TableDipole}
 \begin{scriptsize}
\begin{tabular}{|c|c|c|c|c|c|c|c|c|}
\hline
 \rowcolor[gray]{.7}
&{\bf Eq.}& $\boldsymbol{F_\mathrm{Dipole}}$&$\boldsymbol{E_1, M_1}$&$\boldsymbol{E_2, M_2}$&$\boldsymbol{E_3, M_3}$&$\boldsymbol{E_4, M_4}$&$\boldsymbol{E_5, M_5}$&$\boldsymbol{E_6, M_6}$\\
\hline
$R_E$ [fm]&\eref{REpDeriv}&$0.8112$&$0.8965$&$0.9014$&$0.8628$&$0.8779$&$0.8662$&$0.8776$\\
 \rowcolor[gray]{.95}
$\sqrt[3]{\langle r^3_E \rangle}$ 
[fm]&\eref{oddmoments}&$0.917$&$1.045$&$1.053$&$0.994$&$1.019$&$0.996$&$1.020$\\
$R_\mathrm{F}$ [fm]&\eref{Friar}&$1.265$&$1.425$&$1.434$&$1.362$&$1.391$&$1.365$&$1.392$\\
 \rowcolor[gray]{.95}
$R_\mathrm{Z}$ [fm]&\eref{Zemach}&$1.025$&$1.091$&$1.097$&$1.069$&$1.081$&$1.078$&$1.081$\\
$\Delta E^\mathrm{eFF(1)}_\mathrm{LS}$ exp.\ [$\upmu$eV]&\eref{LSexp}&$-3406$&$-4156$&$-4202$&$-3851$&$3986$&$-3882$&$-3984$\\
 \rowcolor[gray]{.95}
$\Delta E^\mathrm{eFF(1)}_\mathrm{LS}$ exact [$\upmu$eV]&\eref{wG}&$-3406$&$-4156$&$-4202$&$-3851$&$3986$&$-3882$&$-3984$\\
$\Delta E_\mathrm{HFS}^\mathrm{Z}$ [$\upmu$eV]&\eref{ZemachTerm}&$-164.12$&$-174.81$&$-175.77$&$-171.20$&$173.08$&$-172.72$&$-173.14$\\
 \rowcolor[gray]{.95}
$\Delta E_\mathrm{HFS}^{\,\mathrm{recoil}}$ [$\upmu$eV]&\eref{recoilHFS}&$19.13$&$19.04$&$18.98$&$19.05$&$19.03$&$19.00$&$19.01$\\
\hline
\end{tabular}
\end{scriptsize}
\end{table}
\renewcommand{\arraystretch}{1.3}

    \begin{figure}[thb] 
    \centering 
       \includegraphics[width=11cm]{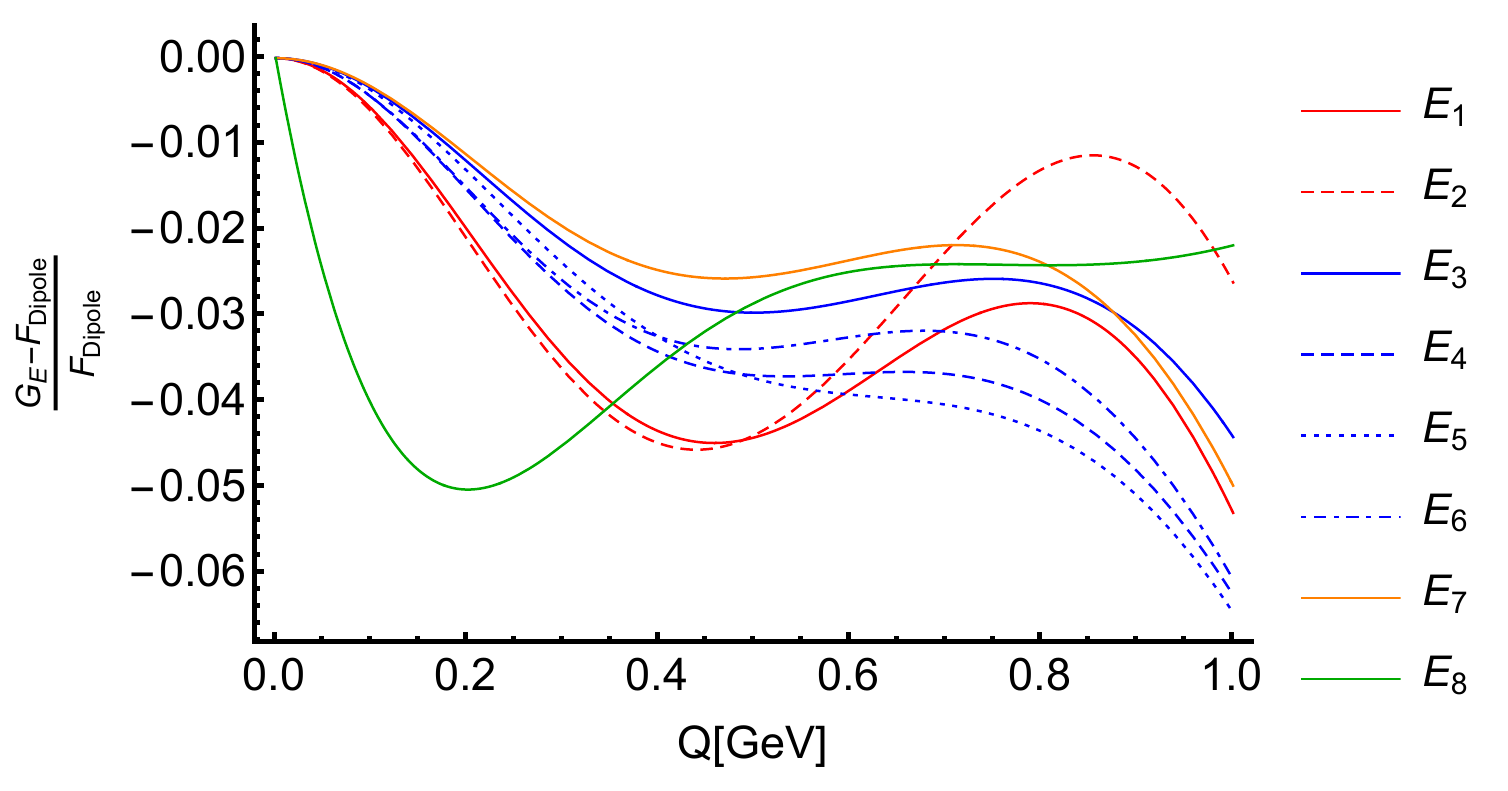}
       \caption{Various fits of the electric Sachs form factor in comparison to the dipole form factor.}
              \label{fig:DipoleFFComp}
\end{figure}

\section{Toy Model Based on the Dipole Form Factor}\seclab{DipoleFF}

Analogously to \secref{chap2}{ToyModel}, we now want to derive another toy model starting from the simple dipole FF:
\beq
\ol G_E(Q^2)=\left(\frac{\Lambda^2}{Q^2+\Lambda^2}\right)^2, \quad \text{with}\;\Lambda^2=0.71\,\text{GeV}^2.
\eeq
Even though, the dipole FF can only describe the $ep$ scattering data very roughly, it has two features interesting to us. While most modern FF fits display unphysical poles, the dipole FF has second-order poles at $Q=\pm i \Lambda$, i.e., poles on the imaginary axis of the $Q$ plane, as expected from analyticity constraints. Furthermore, the dipole FF gives a small proton charge radius: $R_{Ep}=0.8112$ fm. Therefore, in this case, we are required to construct a fluctuation that enhances the radius.

    \begin{figure}[t] 
    \centering 
       \includegraphics[width=7cm]{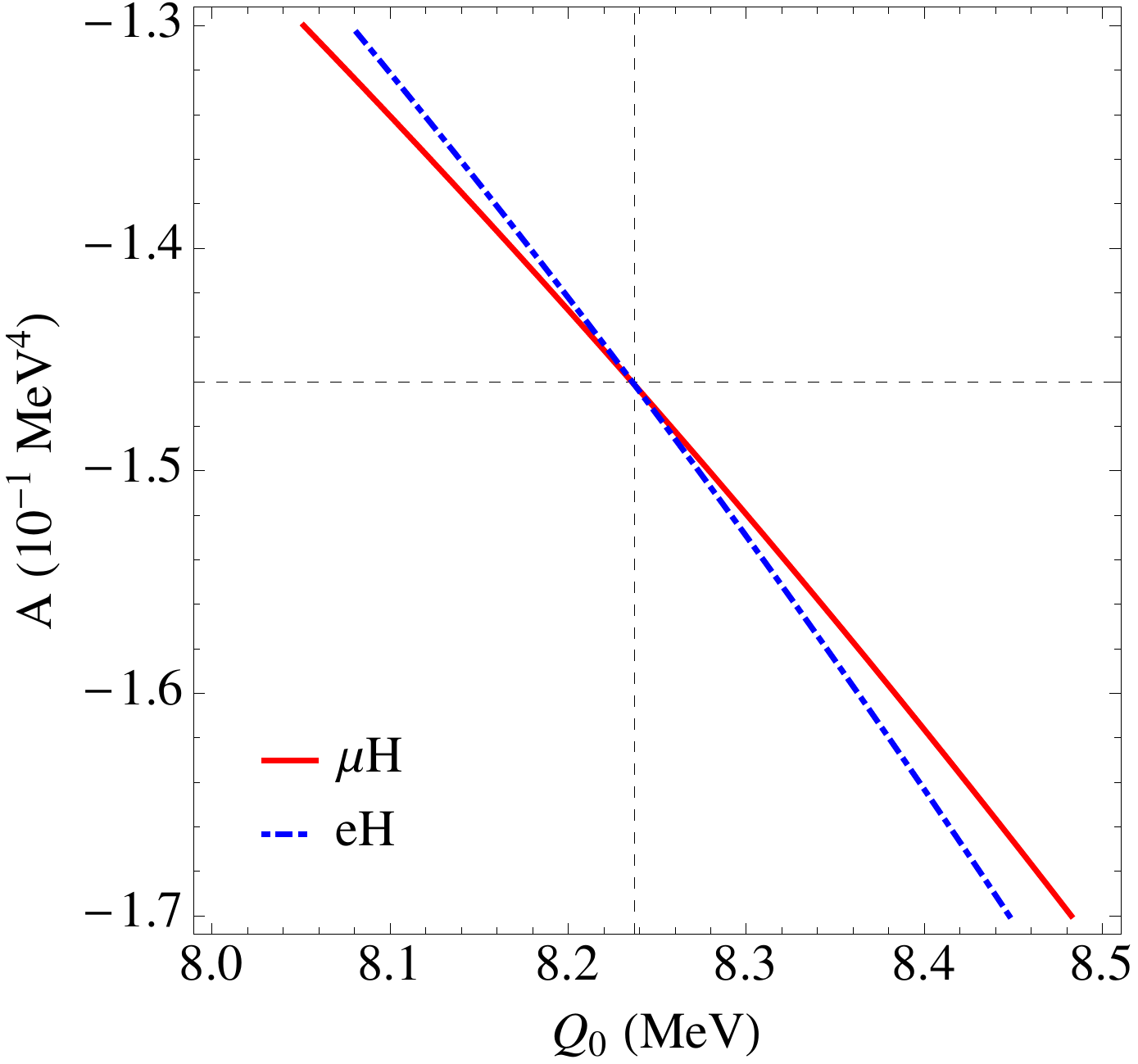}
       \caption{Modification of the dipole form factor: Parameters of $\widetilde G_E$ for which the electronic-hydrogen (blue dot-dashed) and muonic-hydrogen (red solid) Lamb shifts of \Eqref{LSFS} 
       are reproduced. We chose $A=-0.146$ MeV$^4$ and $Q_0=8.237$ MeV, as indicated by the dashed lines.}
              \label{fig:DipoleParameter}
\end{figure}

\begin{table}[h!]
 \centering
 \caption{Lamb shift and moments corresponding to the modified dipole form factor, with $A=-0.146$ MeV$^4$ and $Q_0=8.237$ MeV.}
 \label{TableDipole}
 \begin{small}
\begin{tabular}{|c|c|c|c|c|}
\hline
 \rowcolor[gray]{.7}
&{\bf Eq.}&$\boldsymbol{\ol G_E}$&$\boldsymbol{\widetilde G_E}$&$\boldsymbol{ G_E}$\\
\hline
$\langle r^2\rangle_E \, [\mbox{fm}^2]$&\eref{REpDeriv}&$(0.8112)^2$&$(0.3305)^2$&$(0.8760)^2$\\
$\langle r^3\rangle_E \,[\mbox{fm}^3]$&\eref{oddmoments}&$(0.917)^3$&$(2.969)^3$&$(2.998)^3$\\
 \rowcolor[gray]{.95}
Lamb-shift, exact & \eref{wG} & && \\
 \rowcolor[gray]{.95}
$E_\mathrm{LS}^{\mathrm{eFF}(1)}(e\mathrm{H}) [\text{neV}]$&&$-0.532$&$-0.088$&$-0.620$\\
 \rowcolor[gray]{.95}
$E_\mathrm{LS}^{\mathrm{eFF}(1)}(\mu\mathrm{H})[\upmu\text{eV}]$&&$-3406$&$-243$&$-3650$\\
Lamb-shift, expanded & (\ref{eq:LSexp}) & && \\
$E_\mathrm{LS}^{\mathrm{eFF}(1)}(e\mathrm{H})[\text{neV}]$ &&$-0.532$&$-0.088$&$-0.620$\\
$E_\mathrm{LS}^{\mathrm{eFF}(1)}(\mu\mathrm{H})[\upmu\text{eV}]$&&$-3406$&$-90$&$-3496$\\
\hline
\end{tabular}
\end{small}
\end{table}

The non-smooth part of the FF is now inspired by the weighting function,
\beq
\widetilde G_E(Q^2)=\frac{A\,Q^2\left[ Q_0^2-Q^2\right]}{\left[Q_0^2+Q^2\right]^4},
\eeq
and has fourth-order poles at $Q=\pm i Q_0$. Again, the charge
remains unchanged as $\widetilde G_E(0)=0$. This correction to the FF is described by two free parameters for its strength, $A$, and location, $Q_0$. A stable solution is found for the parameter pair: $A=-0.146$ MeV$^4$ and $Q_0=8.237$ MeV, cf.\ Fig.~\ref{fig:DipoleParameter}. The contribution of such a correction to the second and third moments is given by:
 \begin{subequations}
\bea
\widetilde{\langle r^2\rangle}_E&\equiv &-6 \frac{\dd}{\dd Q^2} \widetilde G_E(Q^2)\Big|_{Q^2= 0}
=-\frac{6 A}{Q_0^6},\\
\widetilde{\langle r^3\rangle}_E&\equiv &\frac{48}{\pi} \int_0^\infty \!\frac{\dd Q}{Q^4}\,
\left\{ \widetilde G_E(Q^2) +\sixth \widetilde{\langle r^2\rangle}_E Q^2\right\},\nn\\
&=&-\frac{60 A}{Q_0^7}.
\eea
\end{subequations}

Table \ref{TableDipole} summarizes the numerical values of these moments as well as the expanded and un-expanded LSs, respectively. Again, the LS expansion in moments is broken for the case of $\mu$H. Since the dipole FF itself gives a smaller charge radius than either values in \Eqref{RadiiToyModel} and the associated LSs are larger than the experimental values quoted in \Eqref{LSFS}, the correction has to decrease the integrand of \Eqref{wGall} and strengthen the negative slope of the FF at $Q=0$. Figure \ref{fig:Dipolecorrection} shows that our fitted correction exactly meets these requirements. Figure \ref{fig:DipoleGEminus1} compares the dipole FF and the modified dipole FF. Again, the fluctuation is very tiny:
  \beq
\big\vert\widetilde G_E/\, \ol G_E\big\vert <2.5 \times 10^{-6},
\eeq
also in comparison to the FF deviation from $1$:
 \beq
\big\vert\widetilde G_E/\, (\ol G_E-1)\big\vert <0.17.
\eeq

  \begin{figure}[h!] 
    \centering 
       \includegraphics[width=10cm]{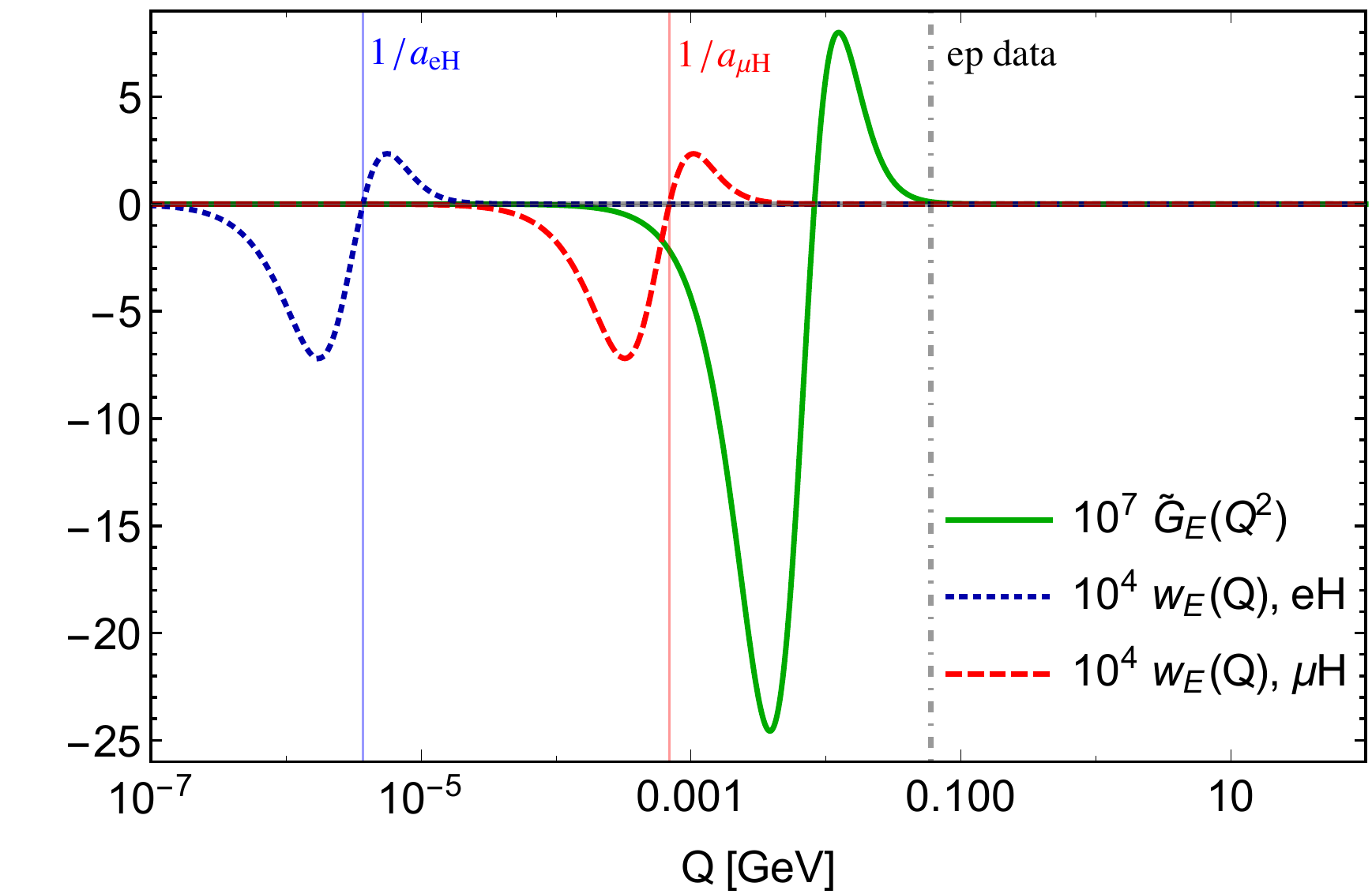}
       \caption{Modification of the dipole form factor: The correction, $\widetilde G_E(Q^2)$, with $A=-0.146$ MeV$^4$ and $Q_0=8.237$ MeV (solid green), and the weighting function, $w_E(Q)$, for electronic hydrogen (blue dotted) and muonic hydrogen (red dashed) as functions of $Q$. The dot-dashed line indicates the onset of electron-proton scattering data.}
              \label{fig:Dipolecorrection}
\end{figure}
 
   \begin{figure}[h!] 
    \centering 
       \includegraphics[width=11cm]{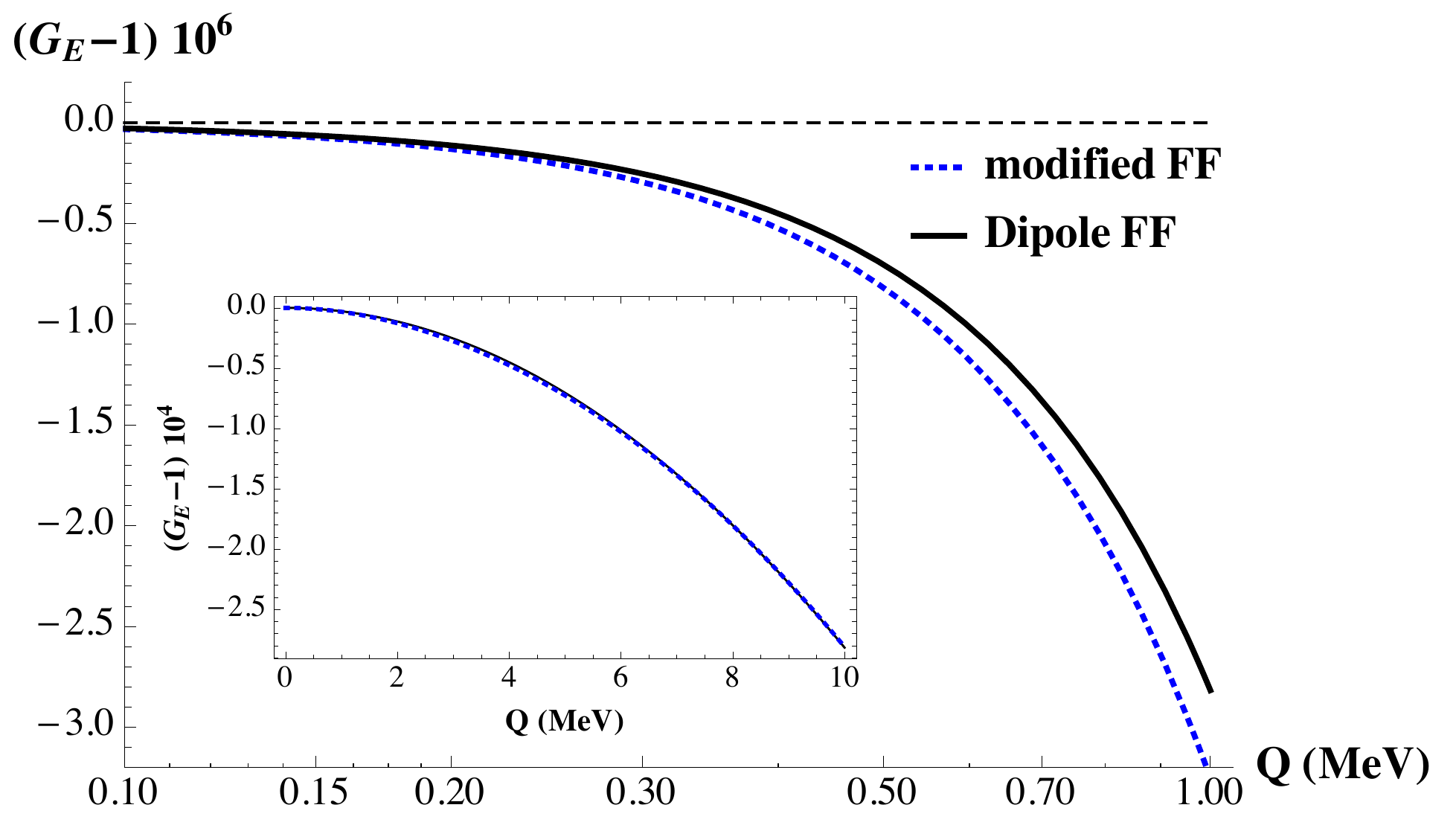}
       \caption{Modification of the dipole form factor: The solid black curve shows the dipole form factor, $\ol G_E(Q^2)-1$. The dotted blue curve shows the modified dipole form factor, $G_E(Q^2)-1$, discussed in the text.}
              \label{fig:DipoleGEminus1}
\end{figure}

\end{subappendices}

\chapter{Compton Scattering and Polarizabilities}  \chaplab{chap3}

In this Chapter, we classify the CS processes (\secref{chap3}{sec3.1}) and give a general introduction into the concepts of polarizabilities (\secref{chap3}{polarizabilities}) and model-independent sum rules (\secref{chap3}{RCSSR}). After that, we will focus on the RCS while delegating
 the case of VVCS to \chapref{chap4}. The status of our knowledge of the lowest-order nucleon polarizabilities is reviewed in \secref{chap3}{status}. In \secref{chap3}{elastic}, we will study the Compton contribution to photoabsorption and the associated CS sum rules in scalar and spinor one-loop QED. A modification of the sum rules which deals with the infrared divergences has been published in Refs.~\cite{Gryniuk:2015aa,Gryniuk:2016gnm}.

\section{Basic Principles} \seclab{sec3.1}

Figure \ref{fig:CS} shows a CS process --- an  absorption and subsequent emission of a photon  by a target. The particles in the initial and final states are the same, and their initial (final) momenta are denoted by $q(q')$ for the photon and $p(p')$ for the target.  The photons can be real, i.e., $q^2=0=q^{\prime\,2}$, or virtual. In VCS, the initial photon is virtual and the final photon is real, $\gamma^*\,p\rightarrow \gamma\,p$. In VVCS, both photons are virtual.

\begin{figure}[h]
\centering
       \includegraphics[scale=0.55]{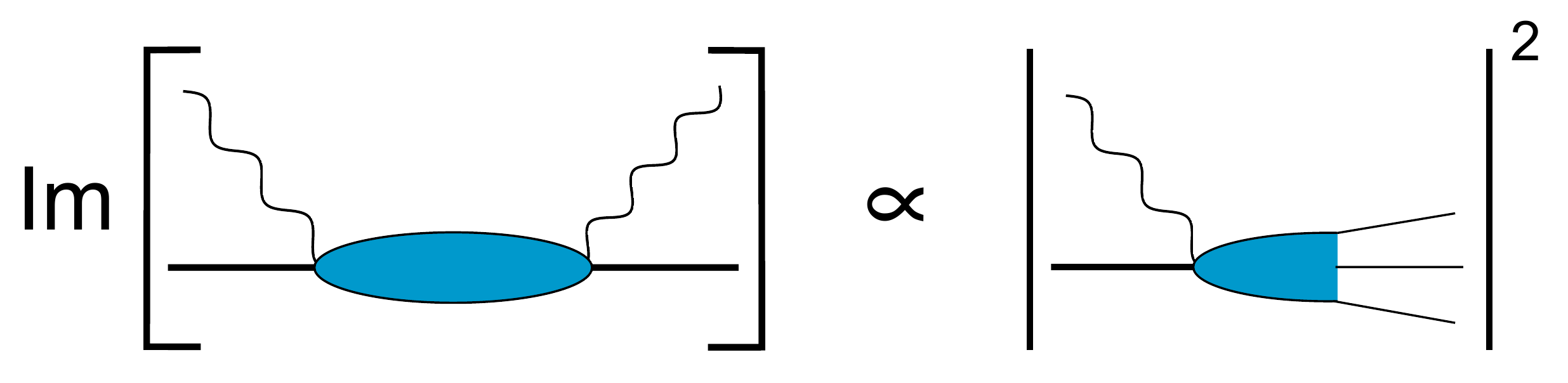}
\caption{Illustration of the optical theorem, relating the imaginary part of the forward Compton scattering amplitude to the total photoabsorption cross section.
\label{fig:OT}}
\end{figure}

Of special interest is the forward limit, where $p=p'$ and $q=q'$.  Accordingly, the Mandelstam invariant $t=(q-q')^2=(p-p')^2$ vanishes.
In this case, unitarity leads to the  optical theorem (see Ref.~\cite{Newton1976} for a review of the optical theorem and its modern application in scattering theory).
It expresses the imaginary part of the forward CS amplitude through the total photoabsorption cross section, as is graphically depicted in \Figref{OT}: on the left-hand side (lhs) we have the CS amplitude and on the right-hand side (rhs) we have the squared photoabsorption cross section. The exact formula representation depends on the choice of a photon flux factor, see Ref.~\cite{Drechsel:2002ar}, and is postponed to Eqs.~\eref{OptT} and \eref{VVCSunitarity}.

Photoabsorption of a (virtual) photon (on, e.g., a nucleon:  $\gamma^* N\rightarrow X$)
 can be divided into two categories: \textit{elastic} and \textit{inelastic}. Elastic is the process which leaves the target intact. 
The elastic photoabsorption on the nucleon is shown in \Figref{scat} (a).\footnote{For real photons, the elastic photoabsorption cross section is given by: $\gamma N \rightarrow \gamma N$.} 
In the case of inelastic photoabsorption, cf.\ \Figref{scat} (b),  other particles appear. For photoabsorption on the nucleon these can be pions, nucleon excitations such as the $\Delta(1232)$, etc. One the CS side, the elastic corresponds with the (nucleon-)\textit{pole} and inelastic with the \textit{non-pole} contributions. Examples of those
are illustrated by, respectively, Figures~\ref{fig:Born} and \ref{fig:DeltaExchange}. 

\begin{figure}[t]
\centering
\begin{minipage}{0.49\textwidth}
\centering
       \includegraphics[width=4cm]{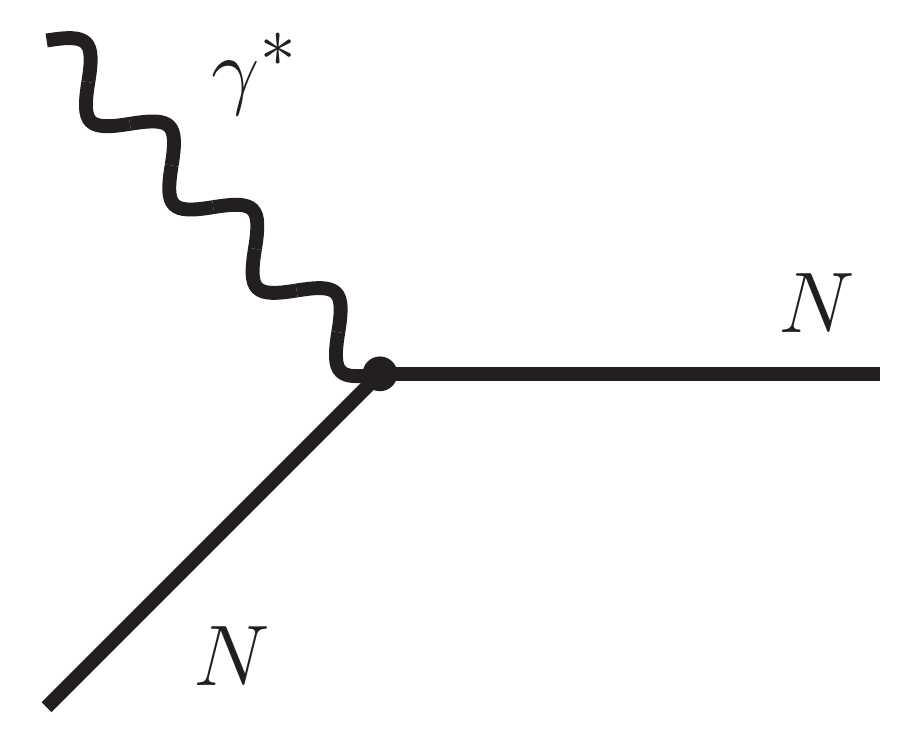}

      (a)
    \end{minipage}\hfill
\begin{minipage}{0.49\textwidth}
\centering
  \includegraphics[width=4cm]{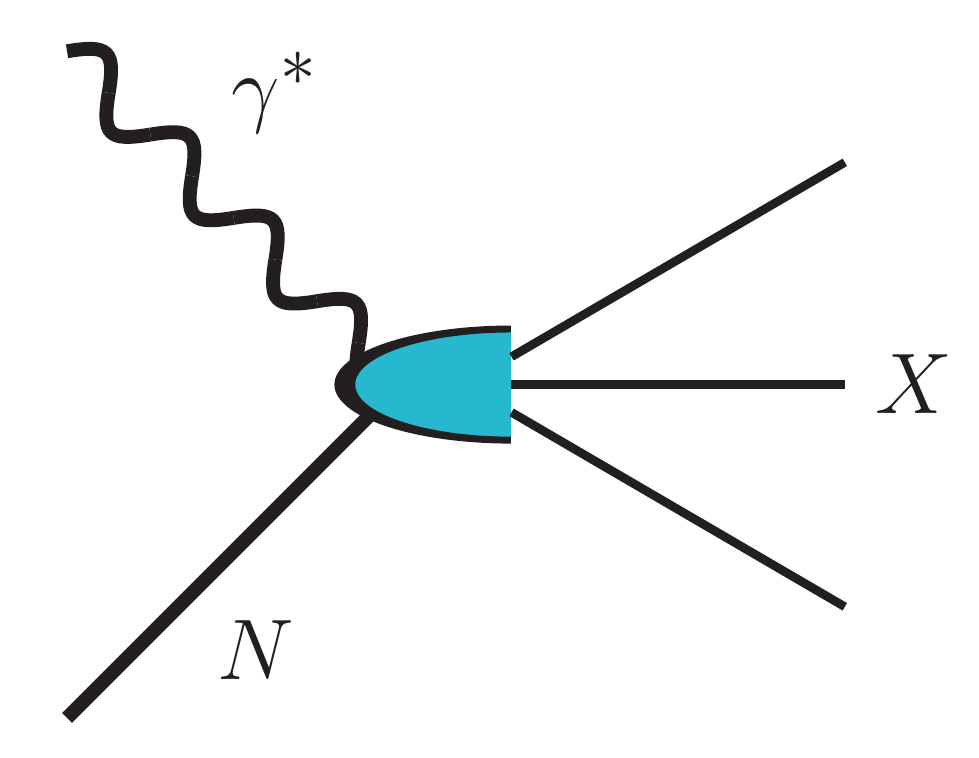}
  
  (b)
\end{minipage}
\caption{(a) ``Elastic'' and (b) ``inelastic'' photoabsorption cross sections. (a) is related to the ``nucleon-pole'' part of the Compton scattering amplitude, whereas (b) is related to the ``non-pole'' part.
\label{fig:scat}}
\end{figure}

\begin{figure}[b]
  \centering
  \includegraphics[scale=0.3]{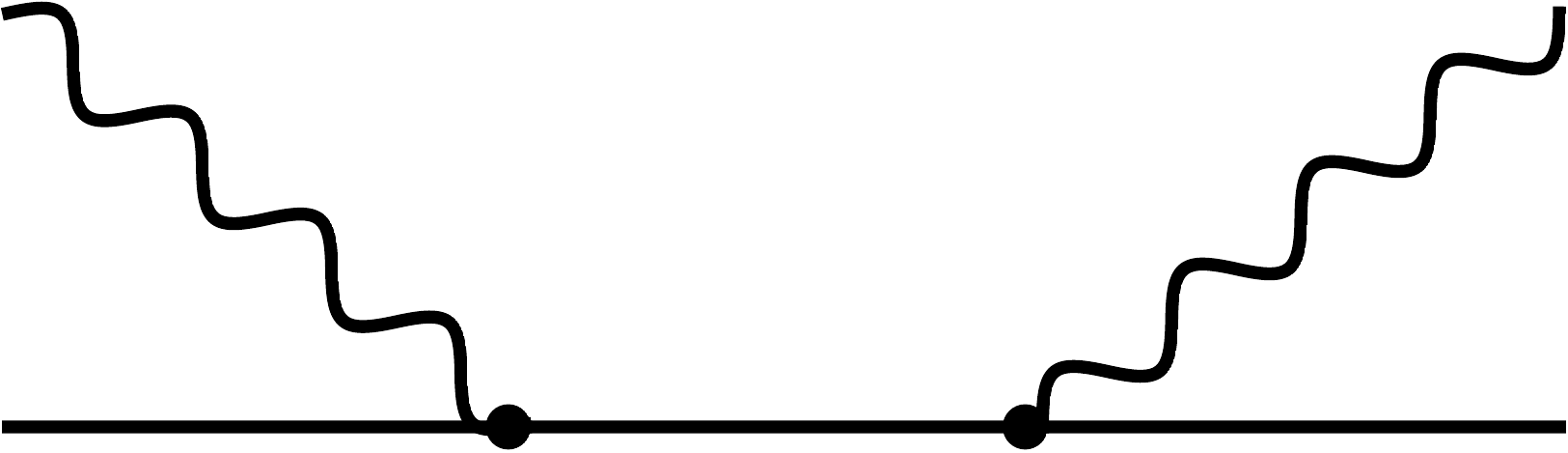}\hspace{0.5cm}
 \includegraphics[scale=0.3]{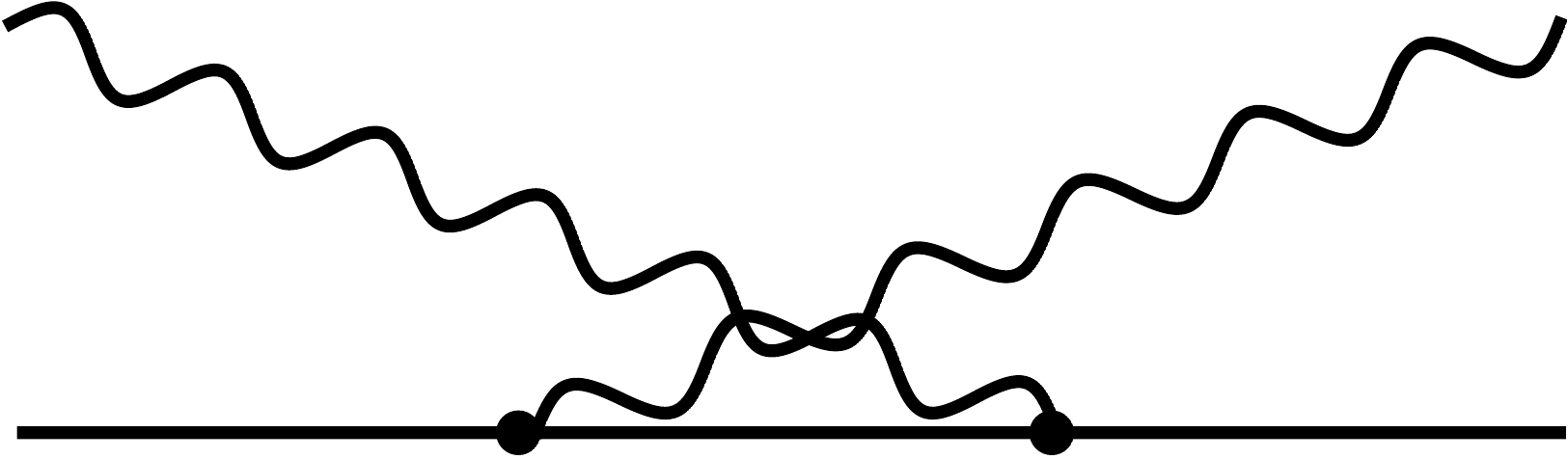}
  	\caption{Tree-level Compton scattering.}  
	\figlab{Born}
\end{figure} 

\subsection{Polarizabilities} \seclab{polarizabilities} 
Polarizabilities depend on the inner structure of the particle. A classic example are the scalar dipole polarizabilities. Imagine a composite particle immersed in an homogenous electric $\boldsymbol{E}$ or magnetic $\boldsymbol{H}$ field: its charged constituents will be displaced, forming electric and magnetic dipoles proportional to the strength of the field:
\begin{subequations}
\begin{eqnarray}
\boldsymbol{d}_{\mathrm{ind.}}&=&4\pi \,\alpha_{E1}\,\boldsymbol{E},\\
\boldsymbol{\mu}_{\mathrm{ind.}}&=&4\pi \,\beta_{M1}\,\boldsymbol{H}.
\end{eqnarray}
\end{subequations}
The proportionality coefficients are the so-called electric and magnetic dipole polarizabilities $\alpha_{E1}$ and $\beta_{M1}$. They reflect the mobility of the constituents.

However, not all polarizabilities can be interpreted as easily as the dipole polarizabilities, and therefore, a rigorous definition for the term ``polarizability'' is needed.
Generally speaking, the polarizabilities of, f.i., the proton provide information on the masses, charges and interactions of the proton's constituents. They are probed in CS because photonic probes from, e.g., $ep$ scattering are stronger than any static e.m.\ field available in a laboratory. It is therefore advantageous to find a definition of polarizabilities in the language of CS. As explained in \secref{chap3}{sec3.1}, the CS amplitude separates into elastic ``pole'' and inelastic ``non-pole'' contributions. Equivalently, it separates into the contribution of (tree-level) Born diagrams, see \Figref{Born}, and non-Born diagrams. Writing down a schematic  equation, we have (for CS off the nucleon):
\beq
\boxed{
\text{nucleon-pole}}+\boxed{\text{inelastic}}=\boxed{\text{Born}}+\boxed{\text{non-Born}},\eqlab{TextEq}
\eeq
where it is important to understand that the nucleon-pole part and the Born part are not necessarily the same, cf.\ \Eqref{BornElasticDiff}. We then define a \textit{polarizability} as anything that stems from the non-Born part of the CS process. On the other hand, the Born diagrams describe the charge and anomalous magnetic dipole moment of the nucleon, as well as its Dirac and Pauli radii.

The electric and magnetic fields are embedded in the e.m.\ field strength tensor, 
$F_{\mu\nu} =\pa_\mu A_\nu - \pa_\nu A_\mu$, as $E_i = F_{0i}$ and $H_i = \half \eps_{ijk} F_{jk}$. The energy response of a composite system to an external e.m.\ field can be described by an effective Hamiltonian~\cite{Bab98,Holstein:1999uu}:
\begin{subequations}
\bea 
\mathcal{H}_{\mathrm{eff}}^{(2)} &=& -4\pi\left[\half\,\alpha_{E1} \,\boldsymbol{E}^2 + 
\half \,\beta_{M1}\, \boldsymbol{H}^2 \right], 
\eqlab{H2eff}\\
\mathcal{H}_{\mathrm{eff}}^{(3)} & = & -4\pi \left[
\half\, \gamma_{E1E1}\,\boldsymbol{\sigma}\cdot(\boldsymbol{E}\times\dot{\boldsymbol{E}})+\half\, \gamma_{M1M1}\,\boldsymbol{\sigma}\cdot(\boldsymbol{H}\times\dot{\boldsymbol{H}}) 
\right.\\
&&\qquad\left.-\gamma_{M1E2}\,E_{ij}\sigma_{i}H_{j}+\gamma_{E1M2}\,H_{ij}\sigma_{i}E_{j}\right],
\eqlab{H3eff}\quad\nn\\
\mathcal{H}_{\mathrm{eff}}^{(4)} &=& - 4\pi
    \left[\half \,\alpha_{E1\nu}\, \dot{\boldsymbol{E}}^2 + \half\, \beta_{M1\nu}\, \dot{\boldsymbol{H}}^2
  +\mbox{$\frac{1}{12}$} \,\alpha_{E2}\, E_{ij}^2 + \mbox{$\frac{1}{12}$}\,  \beta_{M2}\, H_{ij}^2\right],\eqlab{H4eff}
\eea 
\eqlab{Heff}
\end{subequations}
with the Pauli matrices $\boldsymbol{\sigma}$ representing the spin of the composite particle. Here, the superscript denotes the number of spacetime derivatives of the photon 
field, see Ref.~\cite{Drechsel:2002ar} for $\mathcal{H}_{\mathrm{eff}}^{(5)}$. The higher-order scalar polarizabilities, i.e., the quadrupole polarizabilities $\al_{E2}$ and $\beta_{M2}$, and the leading dispersive contributions to the dipole polarizabilities denoted as $\al_{E1\nu}$ and $\be_{M1\nu}$, will not be of further interest in this Chapter. We will meet them briefly in \chapref{chap4}. For now, we will focus on the lowest-order dipole and spin polarizabilities.

The spin polarizabilities \cite{Ragusa:1993rm}, entering in $\mathcal{H}_{\mathrm{eff}}^{(3)}$, describe the coupling of the particles spin to the e.m.\ moments induced by an external field. Their label ($\gamma_{Xl\,Yl'}$) indicates the multipolarity of the initial ($Xl$) and final ($Yl'$) photon, respectively. Therefore, $\gamma_{E1E1}$ and $\gamma_{M1M1}$ describe dipole excitations, whereas $\gamma_{E1M2}$ and $\gamma_{M1E2}$ describe photon scattering off the target with a change of the photons angular momentum by one unit. The prominent forward spin polarizability (FSP) is defined as a linear combination of the lowest spin polarizabilities:
\begin{equation}
\gamma_0=-(\gamma_{E1,E1}+\gamma_{M1,M1}+\gamma_{E1,M2}+\gamma_{M1,E2}).\eqlab{FSPdef}
\end{equation}

The nucleon polarizabilities are measured in
units of $\mathrm{fm}^{n+1}$, where
$n$ is the order at which they appear in the effective Hamiltonian. Nuclei are usually easier to polarize, therefore having much bigger polarizabilities than nucleons.\footnote{For a recent reviews on CS off protons and light nuclei, see Ref.\ \cite{Griesshammer:2012we}.}
 A feature that will be of importance in \secref{5LS}{offTPE}. Below, we will explain how to extract the most prominent polarizabilities from either the low-energy
expansion (LEX) of the RCS amplitudes or the RCS sum rules. The generalized polarizabilities of VVCS will be discussed in  \secref{chap4}{SR}. 
 
\subsection{Sum Rules and other Model-Independent Relations}\seclab{SRderivation}
For a spin-1/2 target, the CS helicity amplitude can be written as: 
\begin{equation}
\eqlab{helamp}
T_{\la_\ga'\la_N'\la_\ga \la_N} 
= \ol N_{\la_N'} (\bp') \, \boldsymbol{\eps}^\ast_{\la_\ga'}(q')\cdot T(q',q,P)\cdot \boldsymbol{\eps}_{\la_\ga}(q)
\, N_{\la_N}  (\bp)\,  ,
\end{equation}
with the Dirac spinors $N$, the photon polarization vectors $\boldsymbol{\eps}$ and the Compton tensor $T^{\mu \nu}$. The helicities of the incoming (outgoing) photon and nucleon are denoted by $\la_\ga(\la_\ga')$ and $\la_N (\la_N')$, respectively. $P=\half(p+p')$ is the sum of incoming and outgoing nucleon four-momenta, and the spinors are normalized according to:
\bea
&& \ol N_{\la_N'}( \boldsymbol{p} )\, N_{\la_N} (\boldsymbol{p})
= 2M \delta_{\la_N' \la_N}. \eqlab{spinorNorm}
\eea

In the forward limit, the RCS tensor is given by two independent scalar amplitudes:
\beq
\label{covT}
T^{\mu\nu}(p,q) = - \left[ g^{\mu\nu} f(\nu)  + \ga^{\mu\nu} g(\nu)
\right].
\eeq
Here, $f$ is a spin-independent and $g$ is a spin-dependent amplitude. They are functions of the photon lab-frame energy $\nu$. Due to causality and analyticity, they fulfil the following DRs:\footnote{For the derivation of a DR, see Ref.~\cite[Appendix B]{Hagelstein:2015egb}.}
\begin{subequations}
\eqlab{DRRCS}
\begin{eqnarray}
\re f(\nu) &=&-\,\frac{\alpha}{M} \,+\, \frac{2\nu^2}{\pi}\fint_0^\infty\!\frac{\dd \nu'}{\nu'} \frac{\im f(\nu')}{\nu^{\prime \, 2}-\nu^2},\label{KKf}\\
\re g(\nu) &=& \frac{2\nu}{\pi}\fint_0^\infty\! \dd \nu'\,\frac{\im g(\nu')}{\nu^{\prime \, 2}-\nu^2}. \label{KK}
\end{eqnarray}
\end{subequations}
Obviously, the DR for the spin-independent amplitude is once subtracted, where the subtraction equals the so-called Thomson term:
\beq
f(0)=-\,\frac{\alpha}{M}. \eqlab{ThomsonTerm}
\eeq
Here and in the rest of this Chapter, we for simplicity choose the charge of the target as $e$, i.e., $Z=1$.

The optical theorem states the following relations between the imaginary parts of the scalar amplitudes and the photoabsorption cross sections:
\begin{subequations}
\eqlab{OptT}
\begin{eqnarray}
\im f(\nu) &=& \frac{\nu}{8\pi} \left[\sigma_{1/2}(\nu)+\sigma_{3/2}(\nu)\right]
\equiv \frac{\nu}{4\pi}\,\si_T(\nu) , \\
\im g(\nu) &=& \frac{\nu}{8\pi}
\left[\sigma_{1/2}(\nu)-\sigma_{3/2}(\nu)\right]
\equiv \frac{\nu}{4\pi}\,\si_{TT}(\nu), \quad
\eqlab{unitarity}
\end{eqnarray}
\end{subequations}
where the subscript on the cross sections denotes the total helicity of the $\gamma N$ state. Plugging \Eqref{OptT} into the rhs of \Eqref{DRRCS}, we arrive at:
\begin{subequations}
\eqlab{DRwCS}
\begin{eqnarray}
\re f(\nu) &=&-\,\frac{\alpha}{M} \,+\, \frac{\nu^2}{2\pi^2}\fint_0^\infty\!\dd \nu' \frac{\si_T(\nu')}{\nu^{\prime \, 2}-\nu^2},\eqlab{KKf}\\
\re g(\nu) &=& \frac{\nu}{2\pi^2}\fint_0^\infty\! \dd \nu'\,\frac{\nu' \si_{TT}(\nu')}{\nu^{\prime \, 2}-\nu^2}.
\end{eqnarray}
\end{subequations}
Now, the rhs is expressed through the unpolarized and the helicity-difference photoabsorption cross section, which can be measured in experiment.

Another important piece in the sum rule derivation is the LEX of the scalar amplitudes \cite{Bab98}:
\begin{subequations}
\eqlab{LEX}
\bea 
f(\nu)  &=&  -\frac{\al}{M}
+  \left[\al_{E1} +\be_{M1} \right]\, \nu^2 +\left[\alpha_{E\nu} + \beta_{M\nu} + \nicefrac{1}{12}\,(\alpha_{E2} + \beta_{M2})\right]\,\nu^4
+\mathcal{O}(\nu^6), \qquad\qquad \eqlab{LET1}\\
 g(\nu) &=&  -\frac{\al \,\kappa^2 }{2M^2}\,\nu + \gamma_0 \,\nu^3+ \bar \gamma_0 \,\nu^5
+\mathcal{O}(\nu^7),\eea 
\end{subequations}
where $\bar \gamma_0$ is a higher-order FSP. The $\mathcal{O}(\nu^0)$ term in \Eqref{LET1} --- the Thomson term --- represents the low-energy theorem (LET) of RCS \cite{Thirring:1950fj,Low:1954kd,Gell_Mann:1954kc}. Replacing the lhs of \Eqref{DRwCS} by \Eqref{LEX}, one can read off a CS sum rule for each order in the photon energy. 

At lowest order in the spin-dependent case, the famous Gerasimov-Drell-Hearn (GDH) sum rule occurs \cite{Drell:1966jv,Gerasimov:1965et}:
\beq
\label{IGDH}
I_{\mathrm{GDH}} \equiv -\int_{\nu_0}^\infty\!\dd\nu\,\frac{\si_{TT}(\nu)}{\nu}
= \frac{\pi^2\alpha}{M^2}\,\kappa^2.
\eeq
It dates back to 1966 and was experimentally verified for the nucleons by the GDH collaboration of MAMI and ELSA \cite{Helbing:2003pv}. At the next two orders, one finds the FSP sum rules \cite{Drechsel:2002ar}:
\begin{subequations}
\eqlab{FSP}
\bea
\gamma_0 &=& \frac{1}{2\pi^2} \int_{\nu_0}^\infty\!\dd\nu\,\frac{\si_{TT}(\nu)}{\nu^3}, \\
\bar{\gamma_0} &=& \frac{1}{2\pi^2} \int_{\nu_0}^\infty\!\dd\nu\,\frac{\si_{TT}(\nu)}{\nu^5}\,.
\eea
\end{subequations}
From the spin-independent amplitude, one derives the Baldin sum rule \cite{Baldin:1960}:
\beq
\eqlab{BaldinSR}
\alpha_{E1} + \beta_{M1} =\frac{1}{2\pi^2} \int_{\nu_0}^\infty \!\dd\nu \,\frac{\si_T(\nu)}{\nu^2},
\eeq
and a fourth-order sum rule: 
\beq
\eqlab{4thSR}
\alpha_{E\nu} + \beta_{M\nu} + \nicefrac{1}{12}\,(\alpha_{E2} + \beta_{M2}) =\frac{1}{2\pi^2} \int_{\nu_0}^\infty \!\dd\nu \,\frac{\si_T(\nu)}{\nu^4}.
\eeq

\seclab{RCSSR}
\section{Present Status of Nucleon Polarizabilities} \seclab{status}

In the present Section, we want to review the state-of-the-art knowledge on the nucleon polarizabilities. We will mainly focus on the lowest-order polarizabilities, $\al_{E1}$, $\be_{M1}$ and $\ga_0$, of the proton and neutron, respectively.

Figures \ref{fig:alphabeta_p}, \ref{fig:alphabeta_n} and \ref{fig:alphaVSbeta} summarize the situation for the electric and magnetic dipole polarizabilities. Figure \ref{fig:gamma0} shows predictions for the FSP. On the theory side, we list predictions from BChPT, HBChPT and LQCD. Furthermore, there are a number of experimental results. If no other empirical information is available, we compare to the predictions of the MAID isobar model. Last but not least, we present evaluations of the CS sum rules introduced above.

\begin{figure}[h!] 
  \centering 
\begin{minipage}[t]{0.49\textwidth}
    \centering 
       \includegraphics[width=\textwidth]{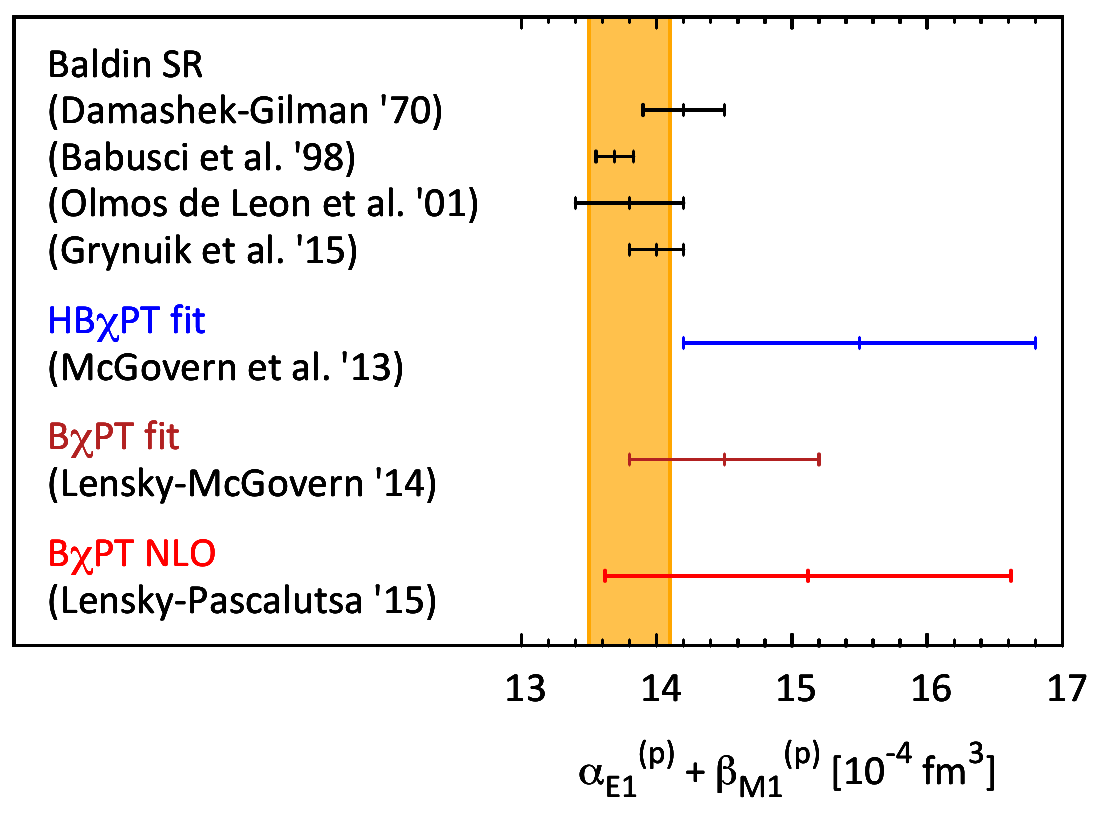}
\end{minipage}
\begin{minipage}[t]{0.49\textwidth}
    \centering 
      \raisebox{0.04cm}{ \includegraphics[width=\textwidth]{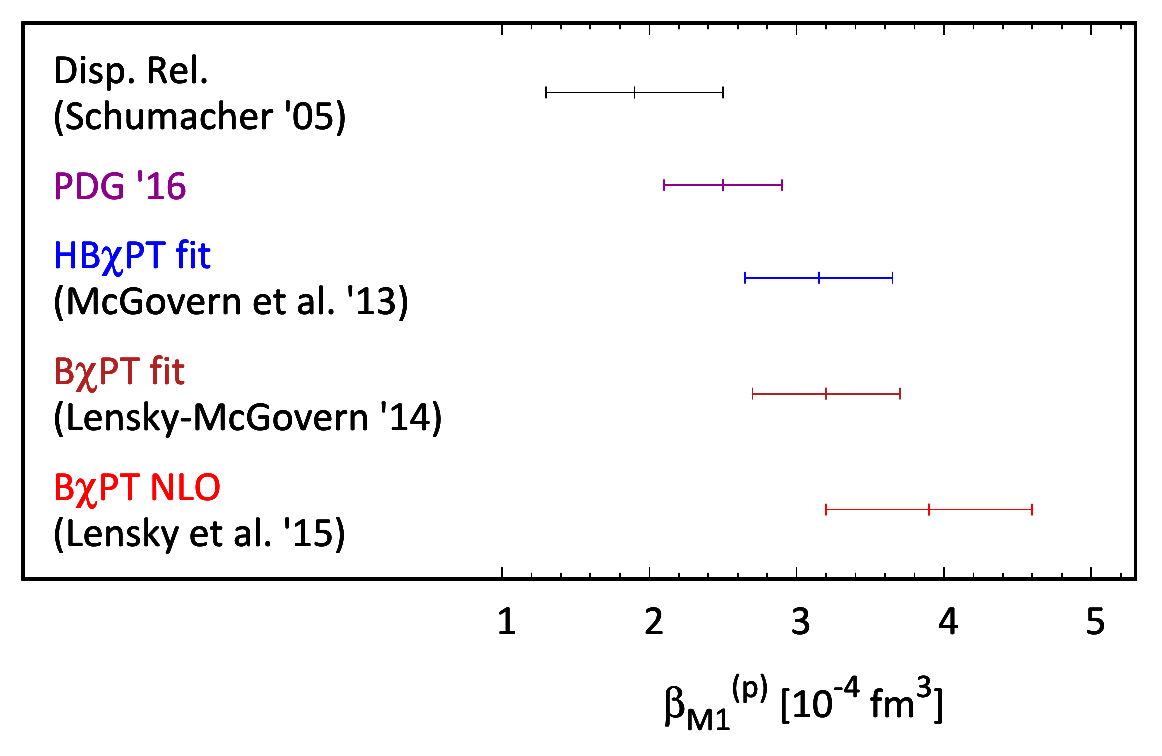}}
       \end{minipage}
 \caption{Left panel: sum of the electric and magnetic dipole polarizabilities of the proton. Right panel: the magnetic dipole polarizability of the proton. The orange band is a weighted average over Baldin sum rule evaluations \cite{Damashek:1969xj,Babusci,Olm01,Gryniuk:2015aa}. The dispersion relation prediction for $\beta_{M1}^{(p)}$ can be found in the review of  \citet{Schumacher:2005an}. The heavy baryon chiral perturbation theory fit is from Ref.~\cite{McG13} and the baryon chiral perturbation theory fit is from Ref.~\cite{Lensky:2014efa}. ``Lensky-Pascalutsa '15'' refers to Ref.~\cite{Lensky:2014dda}, whereas ``Lensky et al.\ '15'' refers to Ref.~\cite{Lensky:2015awa}. \label{fig:alphabeta_p}}
\end{figure}

 \begin{figure}[h!] 
  \centering 
\begin{minipage}[t]{0.52\textwidth}
    \centering 
 \includegraphics[width=\textwidth]{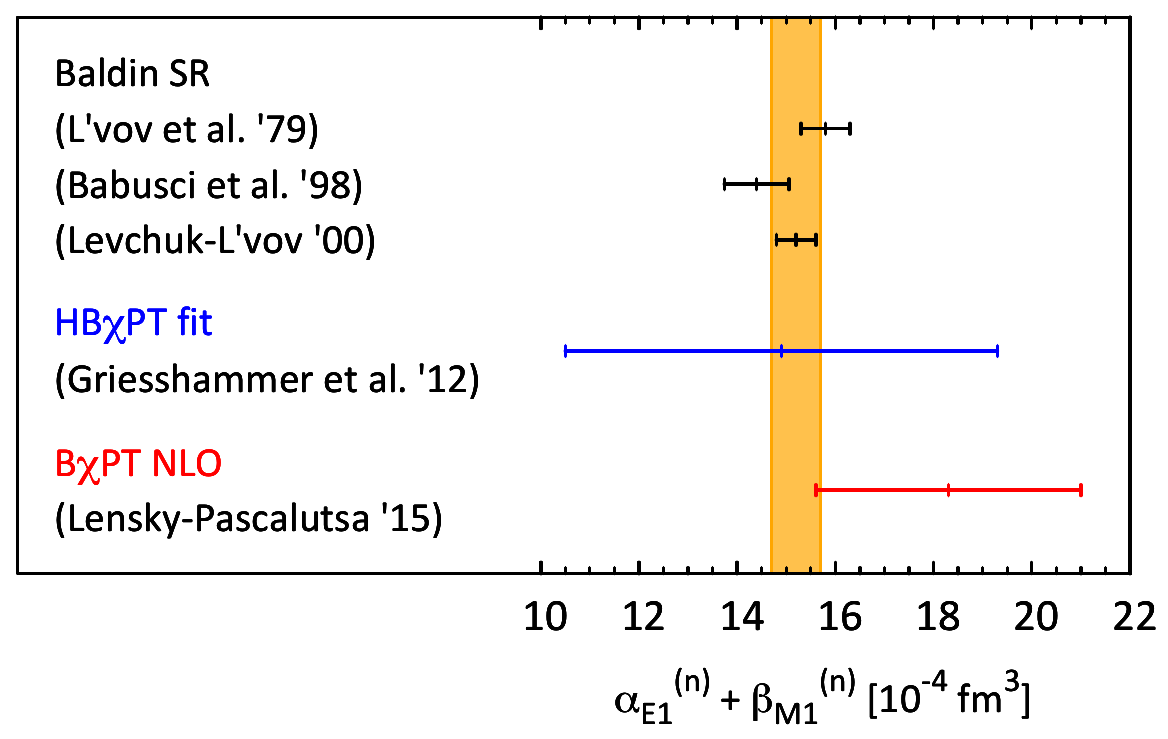}
\end{minipage}
\begin{minipage}[t]{0.46\textwidth}
    \centering 
     \raisebox{0.02cm}{  \includegraphics[width=\textwidth]{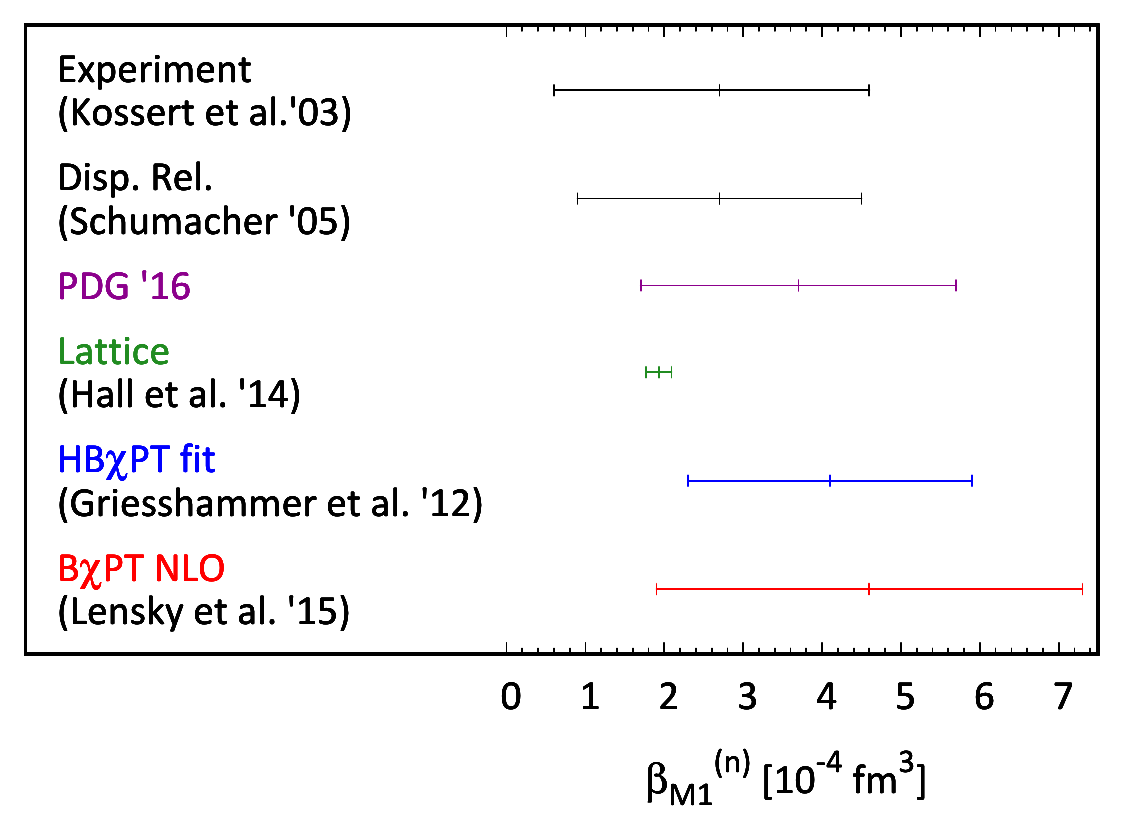}}
       \end{minipage}
 \caption{Left panel: sum of the electric and magnetic dipole polarizabilities of the neutron. Right panel: the magnetic dipole polarizability of the neutron. The orange band is a weighted average over Baldin sum rule evaluations \cite{Lvov:1979, Babusci,Levchuk:1999zy}.
The experimental results for $\beta_{M1}^{(n)}$ are from Refs.~\cite{Kossert:2002jc,Kossert:2002ws} and \cite{Schumacher:2005an}. The lattice prediction is from Ref.~\cite{Hall:2013dva}. The heavy baryon chiral perturbation theory fit is from Ref.~\cite{Griesshammer:2012we}. ``Lensky-Pascalutsa '15'' refers to Ref.~\cite{Lensky:2014dda}, whereas ``Lensky et al.\ '15'' refers to Ref.~\cite{Lensky:2015awa}.\label{fig:alphabeta_n}} 
\end{figure} 

 \begin{figure}[t] 
  \centering 
\begin{minipage}[t]{0.49\textwidth}
    \centering 
       \raisebox{0.00cm}{\includegraphics[width=0.935\textwidth]{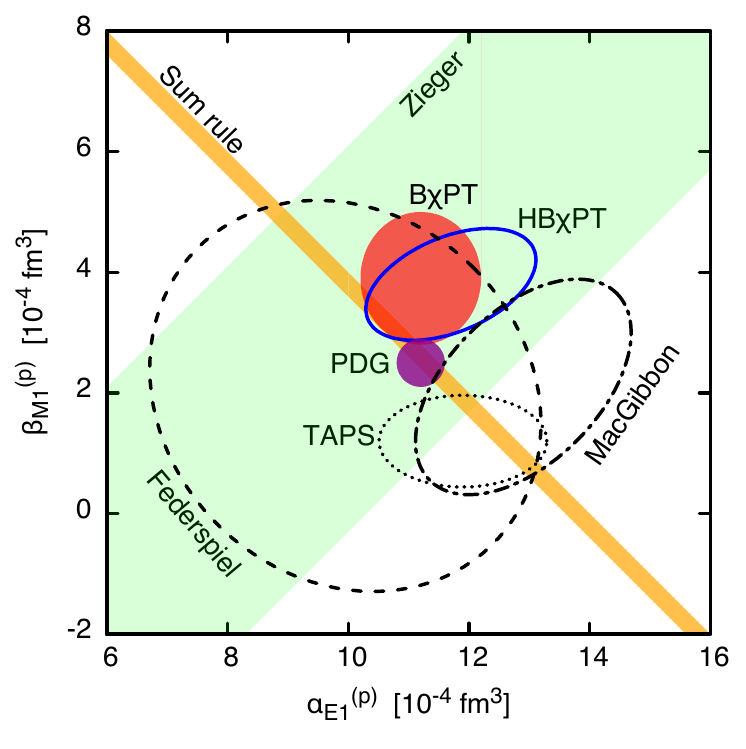}}
\end{minipage}
\begin{minipage}[t]{0.49\textwidth}
    \centering 
       \raisebox{0.1cm}{\includegraphics[width=0.935\textwidth]{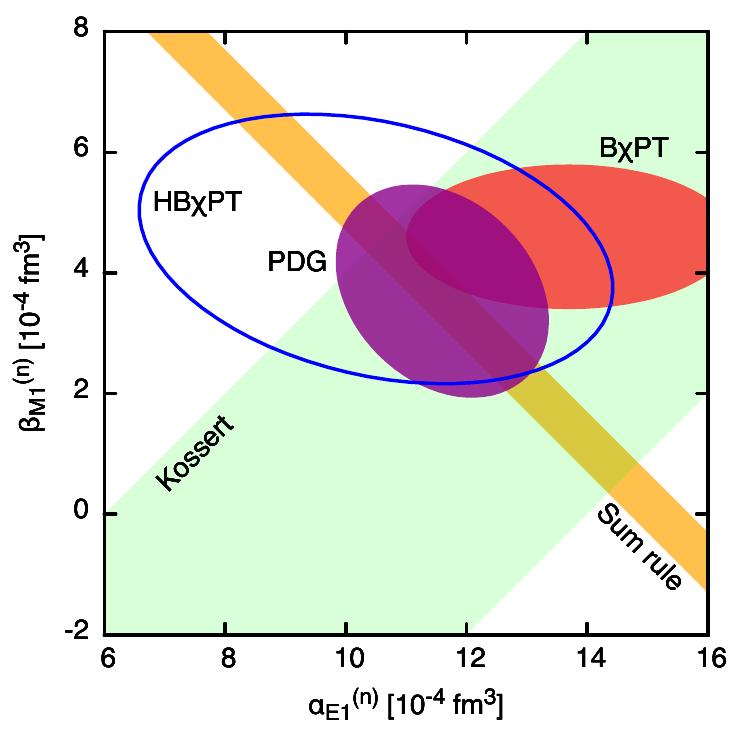}}
       \end{minipage}
 \caption{Plot of $\alpha_{E1}$ versus $\beta_{M1}$ for the proton (left panel) and neutron (right panel), respectively. The references agree mostly with Figures \ref{fig:alphabeta_p} and  \ref{fig:alphabeta_n}. The baryon chiral perturbation theory result is from Ref.~\cite{Lensky:2015awa}. The light green bands show experimental constraints on the difference of dipole polarizabilities, cf.\ Refs.~\cite{Kossert:2002jc,Kossert:2002ws,Zieger:1992jq}. In addition, we show other experimental results for the proton polarizabilities from Refs.~\cite{MacGibbon95, Fed91,Olm01}.\label{fig:alphaVSbeta}}
\end{figure}

The latest (2016) PDG average yields the following values for the dipole polarizabilities of the proton \cite{Olive:2016xmw}:
\begin{subequations}
\bea
\eqlab{PDG2016}
\al_{E1}^{(p)}&=&\left[11.2\pm0.4\right]\times10^{-4}\,\mbox{fm}^3,\\
\be_{M1}^{(p)}&=&\left[2.5\pm0.4\right]\times10^{-4}\,\mbox{fm}^3,
\eea
and the neutron:
\bea
\al_{E1}^{(n)}&=&\left[11.8\pm1.1\right]\times10^{-4}\,\mbox{fm}^3,\\
\be_{M1}^{(n)}&=&\left[3.7\pm1.2\right]\times10^{-4}\,\mbox{fm}^3.
\eea
\end{subequations}
The error on the neutron polarizabilities is larger, due to the lack of a ``free'' neutron target. In a recent re-analysis of photoabsorption data and CS sum rules, we found \cite{Gryniuk:2015aa,Gryniuk:2016gnm}:
\begin{subequations}
\bea
\left[\al_{E1}+\be_{M1}\right]^{(p)}&=&\left[14.0\pm0.2\right]\times10^{-4}\,\mbox{fm}^3,\\
\gamma_0^{(p)}&=&-\left[92.9\pm10.5\right]\times10^{-6}\,\mbox{fm}^4,\\
\left[\alpha_{E\nu} + \beta_{M\nu} + \nicefrac{1}{12}\,(\alpha_{E2} + \beta_{M2})\right]^{(p)}&=&\left[6.04\pm0.03\right]\times10^{-4}\,\mbox{fm}^5,\\
\bar \gamma_0^{(p)}&=&\left[48.8\pm8.2\right]\times10^{-6}\,\mbox{fm}^6.
\eea
\end{subequations}

 \begin{figure}[h!] 
  \centering 
       \includegraphics[width=0.75\textwidth]{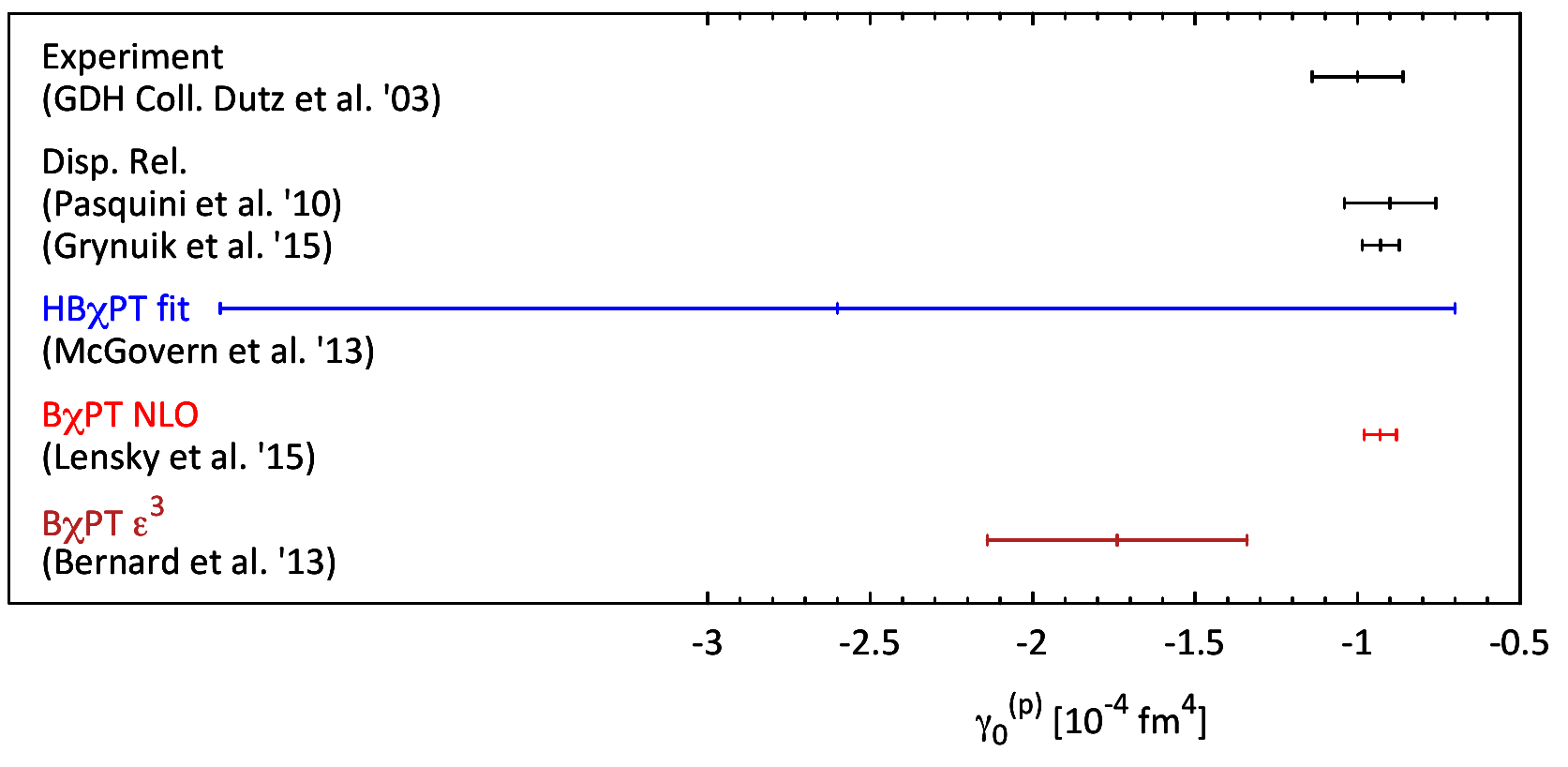}
       \includegraphics[width=0.75\textwidth]{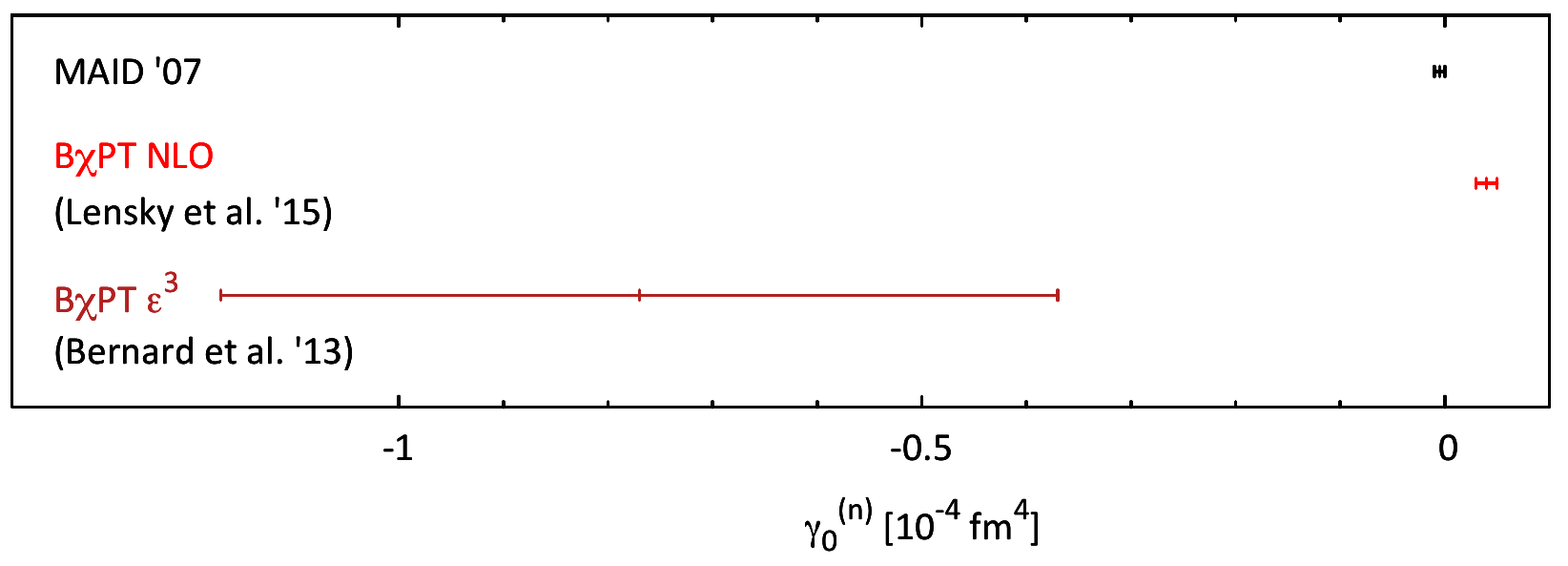}
 \caption{Forward spin polarizability, $\ga_0$, of the proton (top panel) and neutron (bottom panel), respectively. Shown are the  experimental value from the GDH collaboration \cite{Dutz:2003mm}, the sum rule 
 results of Refs.~\cite{Pasquini:2010zr, Gryniuk:2015aa}, the prediction from the MAID isobar model \cite{MAID}, the heavy baryon chiral perturbation theory fit of Refs.~\cite{McG13}, and the baryon chiral perturbation theory predictions of Lensky et al.~\cite{Lensky:2014dda,Alarcon2017} and \citet{Bernard:2012hb}. \label{fig:gamma0}}
\end{figure} 

In general, the situation of the polarizabilities presented in the summary figures looks quite promising. In the future, LQCD will increase its predictive significance. The tension in the value of the proton's magnetic dipole polarizability might be reduced by new measurement techniques, which are independent of the Baldin sum rule. Also, a new result has been obtained from the $\Sigma_3$ beam asymmetry \cite{Sokhoyan:2016yrc}. For the FSP, the results from \citet{McG13} and \citet{Bernard:2012hb} attract attention with their huge error bars. However, the agreement between ($\delta$-counting) NLO BChPT \cite{Lensky:2014dda,Alarcon2017} and the empirical results is quite satisfactory. In contrast, the situation for the longitudinal-transverse polarizability, $\delta_{LT}$, is much more ambiguous, as we will discuss in \chapref{chap4}, where we meet the $\delta_{LT}$ puzzle, see also \Figref{deltaLT}. More details on CS theory and the present status of nucleon polarizabilities can be found in Ref.~\cite{Hagelstein:2015egb}.

\section{Compton Contribution to Photoabsorption} \seclab{elastic}

In the following Section, we calculate the Compton contribution to photoabsorption and the associated contributions to the Baldin sum rule, \Eqref{BaldinSR}, and the FSP sum rules, \Eqref{FSP}, at one-loop level in spinor QED. We will regularize the occurring divergences by presenting an appropriate reformulation of the CS sum rules.\footnote{This work was published in Refs.~\cite{Gryniuk:2015aa,Gryniuk:2016gnm}. Ref.~\cite{Gryniuk:2016gnm} covers both the spin-independent and the spin-dependent case in spinor QED. In Ref.  \cite{Gryniuk:2015aa}, we consider the spin-independent case in scalar QED.} 

First, we want to clarify our terminology. As outlined, we want to study the Compton contribution to photoabsorption off a spin-1/2 particle, f.i., a nucleon. Hereby, we mean the contribution of the ``elastic'' photoabsorption cross sections with $\gamma N\rightarrow \gamma N$. At first glance, this definition of ``elastic'' does not agree with the one presented in \secref{chap3}{sec3.1} and \Figref{scat}. Nevertheless, they are consistent. Previously, we considered the case of virtual photons. In that case, we have $\gamma^*N \rightarrow N$ as the simplest cross section channel. For real photons, however, this process is forbidden by the kinematics and the elastic channel only allows for pairs of photons and target particles. In general, the elastic photoabsorption cross section includes diagrams of the type $\gamma^*N \rightarrow N$ and  $\gamma N \rightarrow \gamma N$. 

\noindent On the rhs of the sum rules, the photoabsorption cross sections enter. The unpolarized and the helicity-difference cross section can be deduced from the helicity amplitudes. These amplitudes were introduced in \Eqref{helamp}, and in general, there are six independent ones for the CS process. At $\mathcal{O}(\al^2)$, the cross sections are given by the Born diagrams shown in \Figref{Born}. The tree-level helicity amplitudes can be found in Refs.~\cite{Tsai:1972sg,Milton:1972sh}. The cross sections read \cite{Holstein:2005db}: 
\begin{subequations}
\eqlab{crossSectionTree}
\begin{align}
&\sigma_T^{(2)}(y)=\frac{2\pi\al^2}{M^2}\left\{\frac{1+y}{y^3}\left[\frac{2y(1+y)}{1+2y}-\ln(1+2y)\right]+\frac{1}{2y}\ln(1+2y)-\frac{1+3y}{(1+2y)^2}\right\},\\
&\sigma_{TT}^{(2)}(y)=\frac{\pi\al^2}{M^2y}\left\{\left[1+\frac{1}{y}\right]\ln(1+2y)-2\left[1+\frac{y^2}{(1+2y)^2}\right]\right\},
\end{align}
\end{subequations}
with $y=\nu/M$. In the low-energy limit, the helicity-difference cross section is vanishing, whereas the total unpolarized cross section reproduces the Thomson cross section:
\beq
  \si_T^{(2)}(0)  = 8\pi \al^2/3M^2,
 \eeq
a result that is unaltered by loop corrections.

\begin{figure}[tbh]
  \centering
    \includegraphics[width=0.3\columnwidth]{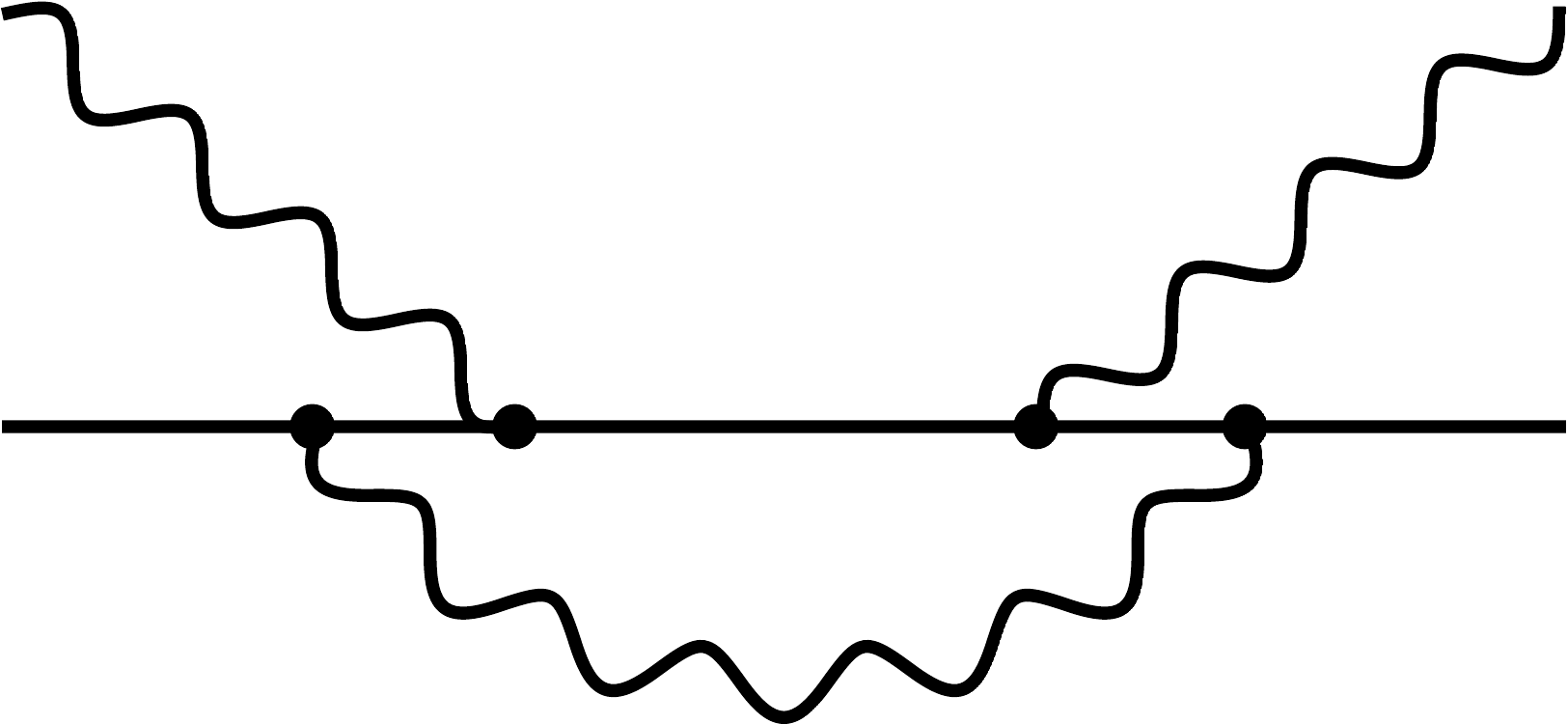}\hspace{5mm}
      \includegraphics[width=0.3\columnwidth]{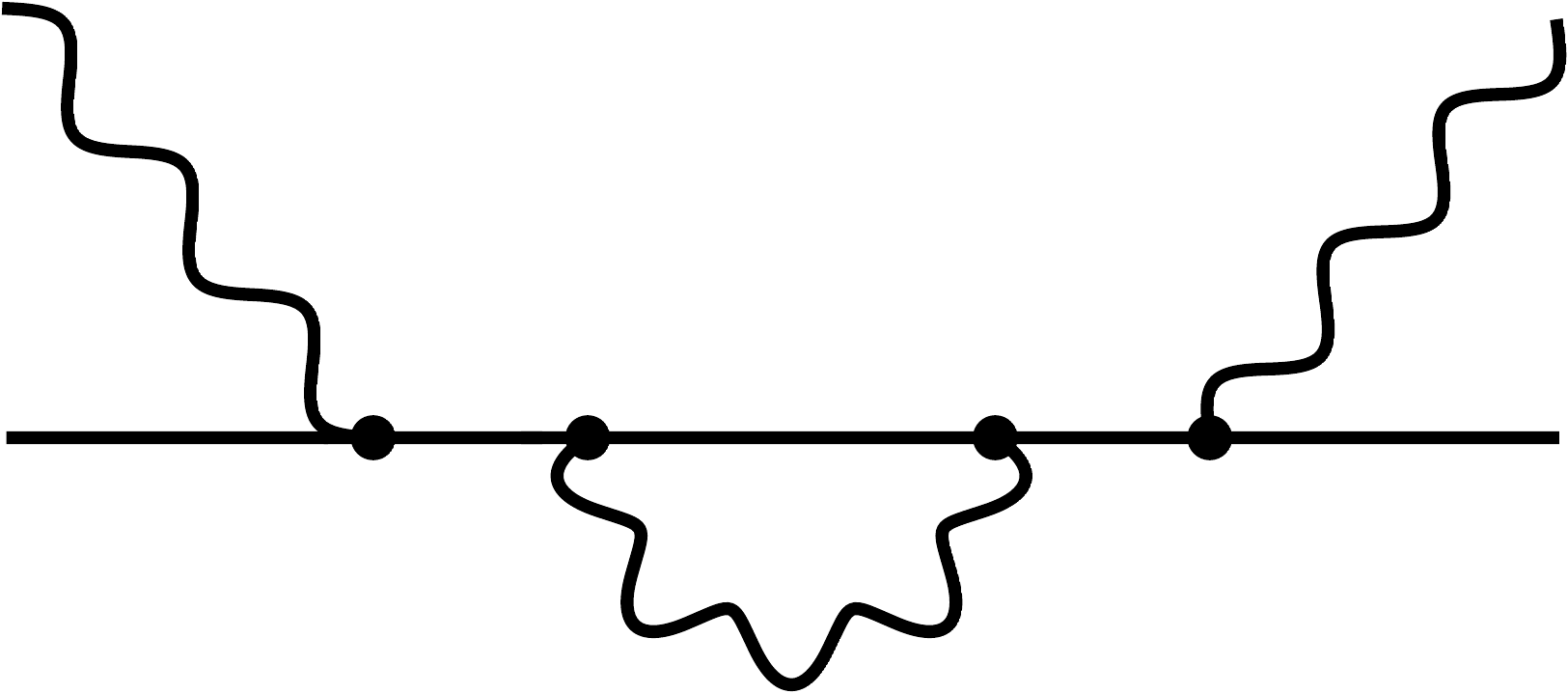}\\[3mm]
          \includegraphics[width=0.3\columnwidth]{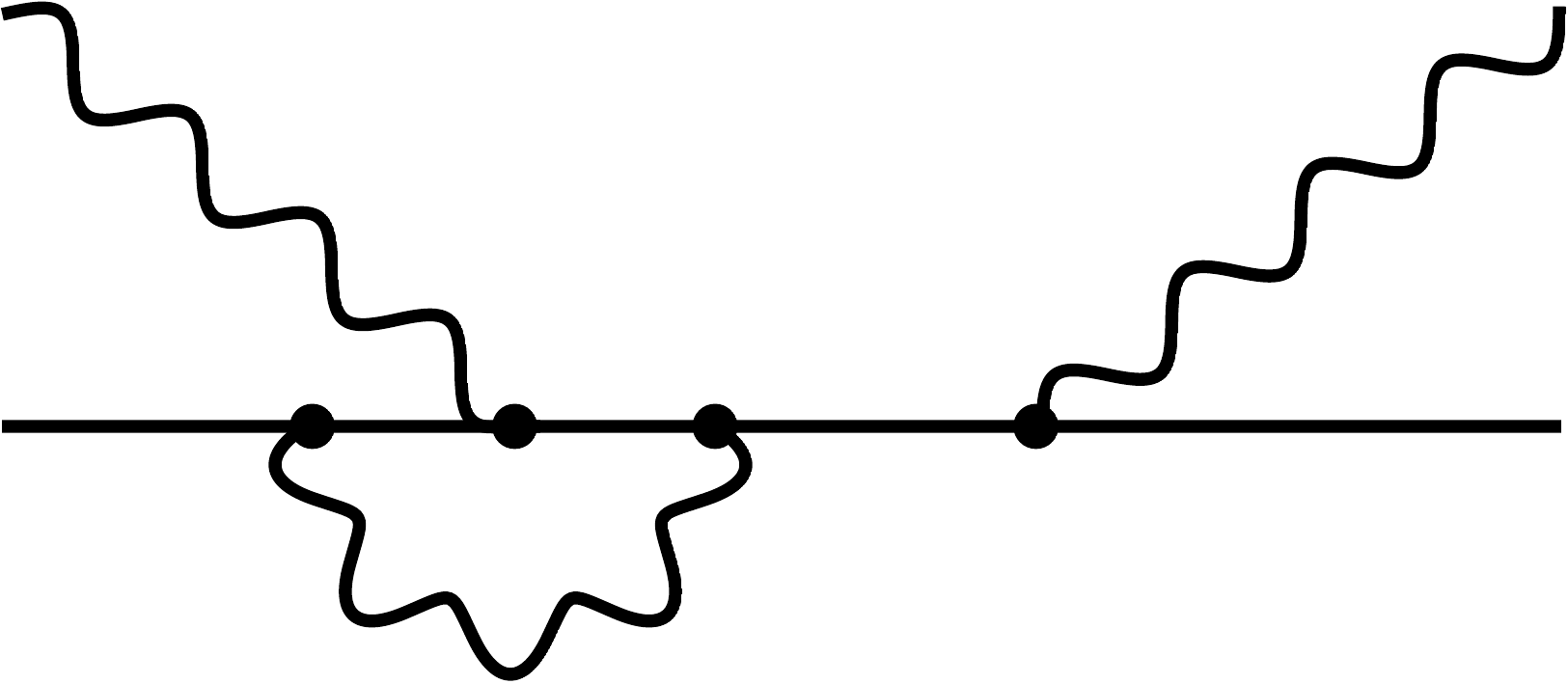}\hspace{5mm}
      \includegraphics[width=0.3\columnwidth]{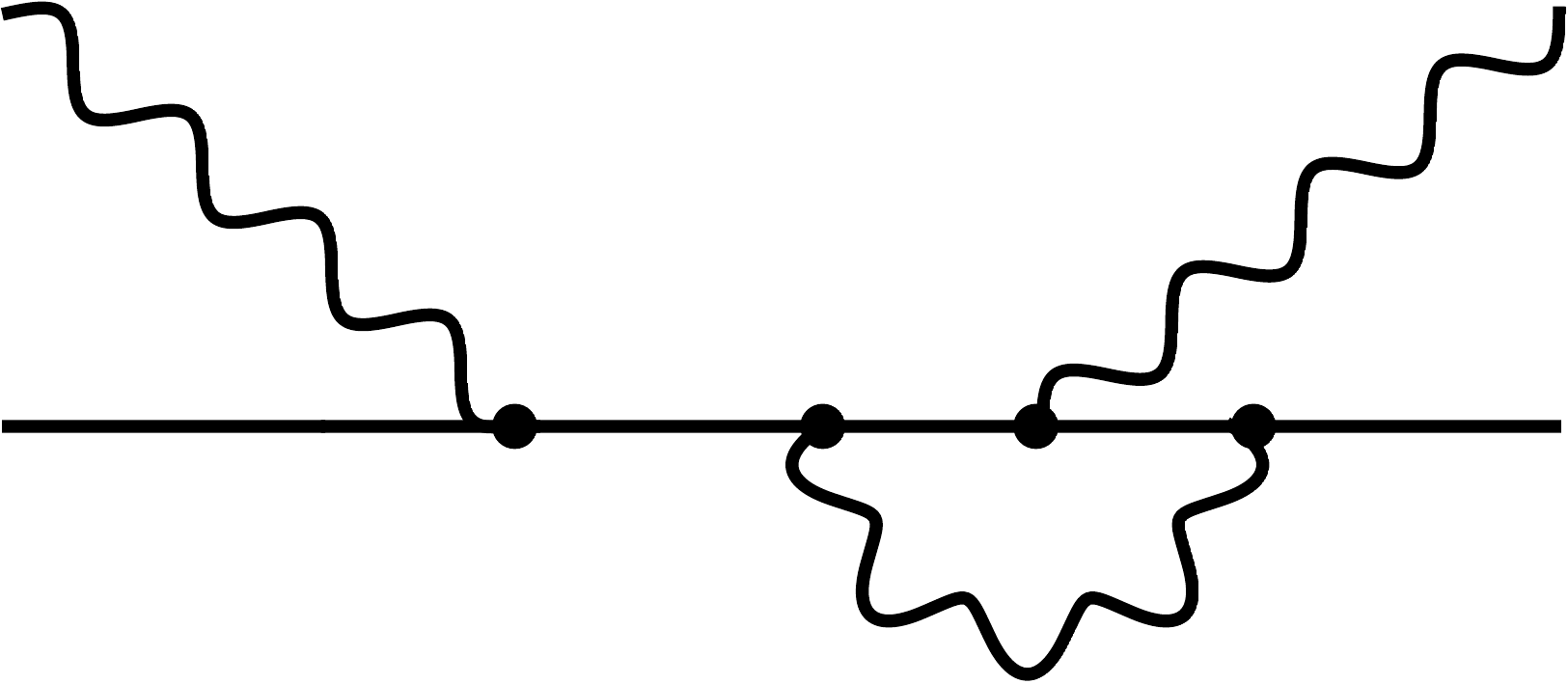}\\[3mm]
 	\caption{One-loop diagrams contributing to the forward Compton scattering. Diagrams obtained from these by crossing of the photon lines are included too.}  
	\label{fig:IMdiagrams}
\end{figure} 

On the lhs of the sum rules, the forward CS amplitudes enter. The one-loop diagrams contributing to the CS at $\mathcal{O}(\al^2)$ in spinor QED are shown in \Figref{IMdiagrams}. In the forward limit, we have $q=q'$. Therefore, only the helicity amplitudes without spin-flip are non-vanishing: $T_{+1\,+\nicefrac{1}{2}\,+1\,+\nicefrac{1}{2}}$ and $T_{-1\,+\nicefrac{1}{2}\,-1\,+\nicefrac{1}{2}}$. They can be used to reconstruct the spin-dependent and spin-independent scalar amplitudes of forward CS: 
\begin{subequations}
\bea
f&=&\frac{1}{4M}\left[T_{+1\,+\nicefrac{1}{2}\,+1\,+\nicefrac{1}{2}} +T_{-1\,+\nicefrac{1}{2}\,-1\,+\nicefrac{1}{2}}\right],\\
g&=&\frac{1}{4M}\left[T_{+1\,+\nicefrac{1}{2}\,+1\,+\nicefrac{1}{2}}-T_{-1\,+\nicefrac{1}{2}\,-1\,+\nicefrac{1}{2}}\right].
\eea
\end{subequations}
For the tree-level amplitudes, cf.\ \Figref{Born}, the spinor QED calculation yields: 
\beq
f^{(1)}(\nu) = -\al/M \quad \text{and} \quad g^{(1)}(\nu) =0, \eqlab{TT2}
\eeq
where the superscript indicates the order of $\al$. At the next order, cf.\ \Figref{IMdiagrams}, we obtain:
\begin{subequations}
\eqlab{LoopAmps}
\bea
f^{(2)}(y)&=&\frac{\al^2}{4\pi M}\left\{\frac{24 y ^2\left(1-3 y ^2\right)+\pi ^2 \left(4 y ^4+8 y ^3-9 y ^2-2 y +2\right)}{6 y ^2 \left(1-4 y ^2\right)}-\frac{4 y ^2 \left(4 y ^2-3\right)}{\left(4 y ^2-1\right)^2}\,\ln 2y\right.\qquad\quad\nn\\
&&\left.-\frac{y ^2-2 y -2}{y ^2}\left[ \ln 2 y \ln(1+2y)+\mathrm{Li}_2\left(-2y\right)\right]+\frac{y ^2+2 y -2}{y ^2}\,\mathrm{Li}_2\left(1-2y\right)\right\}\nn\\
&&+\frac{i M y}{4\pi}\, \sigma_T^{(2)}(y),\\
g^{(2)}(y)&=&\frac{\al^2}{4\pi M}\left\{\frac{12 y ^2+\pi ^2 \left(4 y ^3-4 y ^2-y +1\right)}{6 y  \left(4 y ^2-1\right)}-\frac{16 y ^3}{\left(4 y ^2-1\right)^2}\,\ln 2y\right.\nn\\
&&\left.-\frac{y +1}{y }\left[ \ln 2 y \ln(1+2y)+\mathrm{Li}_2\left(-2y\right)\right]-\frac{y -1}{y }\,\mathrm{Li}_2\left(1-2y\right)\right\}\nn\\
&&+\frac{i M y}{4\pi} \,\sigma_{TT}^{(2)}(y).
\eea 
\end{subequations}
This result was deduced from the one-loop helicity amplitudes of Refs.~\cite{Tsai:1972sg,Milton:1972sh}.

The optical theorem implies that:
\beq
\im f^{(2)} (\nu)=\nu\,\sigma_T^{(2)}(\nu)/4\pi\quad \text{and} \quad \im g^{(2)} (\nu)=\nu\,\Delta\sigma_{TT}^{(2)}(\nu)/4\pi.
\eeq
 This can be easily verified by comparing Eqs.~\eref{crossSectionTree} and \eref{LoopAmps}. Also, we have checked that the one-loop amplitudes indeed
satisfy the DRs:
\begin{subequations}
\eqlab{DRspinor}
\bea
\eqlab{fDR}
f^{(2)}(\nu)&=&\frac{2\nu^2}{\pi}\int_0^\infty \! \dd\nu' \,\frac{\im f^{(2)}(\nu')}{\nu'\left(\nu^{\prime \, 2}-\nu^2 - i0^+\right)},\\
g^{(2)}(\nu)&=&\frac{2\nu}{\pi}\int_0^\infty \! \dd\nu' \,\frac{\im g^{(2)}(\nu')}{\nu^{\prime \, 2}-\nu^2 - i0^+}.
\eea
\end{subequations}
Remember that the DR for the spin-independent amplitude, \Eqref{KKf}, needs one subtraction. This subtraction is usually taken at $\nu=0$ and corresponds to the Thomson term, cf.\ Eqs.~\eref{ThomsonTerm} and \eref{TT2}. Therefore, the subtraction in \Eqref{fDR} corresponds to $f^{(2)}(0)=0$.

So far, we have verified the optical theorem and the DRs for the scalar amplitudes at $\mathcal{O}(\al^2)$ in spinor QED.
We are now left with the LEX. Expanding the real part of Eqs.~\eref{DRspinor} and \eref{LoopAmps} for small photon energies,  we find:
\begin{subequations}
\eqlab{dontmatch}
\begin{align}
 &\frac{\alpha^2}{\pi M} \bigg(\frac{11+48 \ln \frac{2\nu}{M}}{18M^2 } \nu^2+\frac{7 (257+1140\ln \frac{2\nu}{M})}{450M^4}\nu ^4+\frac{68(107+672 \ln \frac{2\nu}{M})}{441M^6 }\nu ^6  + \ldots \bigg)\qquad\quad\nn
\\
&\quad= \frac{1}{2\pi^2} \sum_{n=1}^\infty \nu^{2n} \int_0^\infty \!\dd\nu' \,\frac{\sigma_T^{(2)}(\nu')}{\nu^{\prime\, 2n}},\qquad\\
&\frac{\alpha^2}{\pi M} \bigg(\frac{ 37+60  \ln \frac{2\nu}{M}}{18 M^3}\nu ^3+\frac{64 (29+105  \ln \frac{2\nu}{M})}{225 M^5} \nu ^5
+\frac{18 (89+504 \ln \frac{2\nu}{M})}{49 M^7 } \nu ^7 +\ldots \bigg)\nn
\\
&\quad= -\frac{1}{2\pi^2} \sum_{n=1}^\infty \nu^{2n-1} \int_0^\infty \!\dd\nu' \,\frac{ \sigma_{TT}^{(2)}(\nu')}{\nu^{\prime\, 2n-1}}.\qquad
\end{align}
\end{subequations}
We can now see that on both sides the coefficients diverge in the infrared. However, there is an apparent mismatch:  
they are logarithmically divergent on the lhs
 and power-divergent on the rhs. The reason for the appearance of divergences is that the cross section of real-photon absorption has no threshold in the elastic channel, thus, starts from $\nu=0$. To match the sides of \Eqref{dontmatch} exactly at
 each order of $\nu$, we subtract all the power divergences on the
 rhs and 
regularize the dispersion integral with an infrared cutoff equal to $\nu$:
\begin{subequations}
\begin{align}
\eqlab{goodscalarLEX}
& \frac{\alpha^2}{\pi M} \bigg(\frac{11+48 \ln \frac{2\nu}{M}}{18M^2 } \nu^2+\frac{7 (257+1140\ln \frac{2\nu}{M})}{450M^4}\nu ^4 + \ldots \bigg)\nn\\
&\quad= \frac{1}{2\pi^2} \sum_{n=1}^\infty \nu^{2n} 
\int_\nu^\infty \!\dd\nu' \,\frac{\sigma_T^{(2)}(\nu')-\sum\limits_{k=0}^{2(n-1)}\frac{1}{k!}\frac{\dd^k\sigma_T^{(2)}(\nu)}{\dd\nu^k}\Big|_{\nu=0}\,\nu^{\prime\, k}}{\nu^{\prime\, 2n}},\\
&\frac{\alpha^2}{\pi M} \bigg(\frac{ 37+60  \ln \frac{2\nu}{M}}{18 M^3}\nu ^3+\frac{64 (29+105  \ln \frac{2\nu}{M})}{225 M^5} \nu ^5 +\ldots \bigg)\nn\\
&\quad= -\frac{1}{2\pi^2} \sum_{n=2}^\infty \nu^{2n-1} 
\int_\nu^\infty \!\dd\nu' \,\frac{\sigma_{TT}^{(2)}(\nu')-\sum\limits_{k=0}^{2n-3}\frac{1}{k!}\frac{\dd^k\sigma_{TT}^{(2)}(\nu)}{\dd\nu^k}\Big|_{\nu=0}\,\nu^{\prime\, k}}{\nu^{\prime\, 2n-1}}.\eqlab{noGDH}
\end{align}
\end{subequations}
The $\mathcal{O}(\nu)$ term was omitted in \Eqref{noGDH} , since the GDH sum rule only differs from zero starting from $\mathcal{O}(\al^3)$.  As desired, both sides are  now identical at each order of $\nu$. This is nontrivial, at least for the analytic terms; the logs are fairly easily obtained from the non-regularized rhs of the low-energy expanded DR, cf.\ Ref.~\cite{Holstein:2005db}.

Applying these modifications to all orders in $\al$, we find that the proper
LEX of the ``elastic'' part of the amplitudes
reads as:
\begin{subequations}
\eqlab{LEXdiv}
\bea
\eqlab{elLEX}
 f_{\mathrm{el}}(\nu)&=& -\,\frac{\al}{M} + \frac{1}{2\pi^2}\sum_{n=1}^\infty \nu^{2n} 
\!\int_\nu^\infty\! \dd\nu' \,\frac{\sigma_T(\nu')-\bar\sigma_T^{[n]}(\nu')}{\nu^{\prime\, 2n}}\,,\\
 g_{\mathrm{el}}(\nu)&=& - \frac{1}{2\pi^2}\sum_{n=1}^\infty \nu^{2n-1} 
\!\int_\nu^\infty\! \dd\nu' \,\frac{\sigma_{TT}(\nu')-\ol{\sigma}_{TT}^{[n]}(\nu')}{\nu^{\prime\, 2n-1}}\,,
\eea
\end{subequations}
where the bar denotes the infrared subtractions:
\begin{subequations}
\bea
\bar\sigma_T^{[n]}(\nu') &\equiv& \sum\limits_{k=0}^{2(n-1)}\frac{1}{k!}\frac{\dd^k\sigma_T(\nu)}{\dd\nu^k}\Big|_{\nu=0}\,\nu^{\prime\, k},\\
\ol{\sigma}_{TT}^{[n]}(\nu')&\equiv&\begin{cases}
0&n=1,\\
 \sum\limits_{k=0}^{2n-3}\frac{1}{k!}\frac{\dd^k\sigma_{TT}(\nu)}{\dd\nu^k}\Big|_{\nu=0}\,\nu^{\prime\, k}&n>1.
\end{cases}
\eea
\end{subequations}
In this way, we define sum rules for the Compton contribution to the ``quasi-static'' 
 polarizabilities:
\begin{subequations}
\eqlab{newSRs}
\begin{align}
&(\alpha_{E1}+\beta_{M1})_{\mathrm{el}} =\frac{1}{2\pi^2}\int_\nu^\infty \!\dd\nu' \,\frac{\sigma_T(\nu')-\sigma_T(0)}{\nu^{\prime\,2}},\\
&(\gamma_0)_{\mathrm{el}} =\frac{1}{2\pi^2} \int_\nu^\infty\! \dd\nu' \,\frac{\sigma_{TT}(\nu')-\sigma_{TT}'(0)\,\nu'}{\nu^{\prime\,3}},\\
&(\bar{\gamma_0})_{\mathrm{el}} =\frac{1}{2\pi^2} \int_\nu^\infty \!\dd\nu' \,\frac{\sigma_{TT}(\nu')-\sigma_{TT}'(0)\,\nu'-\sigma_{TT}''(0)\,\nicefrac{\nu^{\prime\,2}}{2}- \sigma_{TT}'''(0)\,\nicefrac{\nu^{\prime\,3}}{6}}{\nu^{\prime\,5}}.\qquad\quad
\end{align}
\end{subequations}
Plugging in the tree-level cross sections from \Eqref{crossSectionTree}, we obtain: 
\begin{subequations}
\bea
(\alpha_{E1}+\beta_{M1})_{\mathrm{el}} &=& \frac{\alpha^2}{18\pi M^3}\left(11+48 \ln \frac{2\nu}{M}\right),\\
(\gamma_0)_{\mathrm{el}} &=&-\frac{\alpha^2}{18\pi M^4}\left(37+60  \ln \frac{2\nu}{M}\right),\\
(\bar{\gamma_0})_{\mathrm{el}} &=&-\frac{64\,\alpha^2}{225\pi M^6}\left(29+105  \ln \frac{2\nu}{M}\right),
\eea
\end{subequations}
what obviously matches the corresponding terms in the LEX of the one-loop amplitudes. Thereby, we proved that the newly presented sum rules, \Eqref{newSRs}, for the Compton contribution are working correctly in the case of one-loop spinor QED. Similarly, we verified them in scalar QED \cite{Gryniuk:2015aa}.
\section{Summary and Conclusion}
In the present Chapter, we introduced the nucleon polarizabilities as observed in CS and presented a derivation of the well-known RCS sum rules. The unique feature of CS sum rules is that they relate the CS amplitudes and polarizabilities to weighted integrals of the photoabsorption cross sections with respect to the photon energy. Their model independence gives them a high predictive power in accessing the nucleon polarizabilities. They allow to obtain empirical results based on measured cross sections. Furthermore, they connect tree-level cross sections to one-loop diagrams, thus, are providing a computational simplification.

In \secref{chap3}{elastic}, we studied the Compton contribution to photoabsorption and the associated CS sum rules \cite{Gryniuk:2015aa,Gryniuk:2016gnm} as  an academic exercise. For the elastic channel, the LEX of the amplitudes and the dispersion integrals yield logarithmic and power divergences. A proper definition for the divergent pieces was achieved by introducing an infrared cutoff on the dispersion integral and by making infrared subtractions on the photoabsorption cross sections, cf.\ \Eqref{LEXdiv}. Sum rules for the Compton contribution to the quasi-static
 polarizabilities, cf.\ \Eqref{newSRs}, were presented and verified at one-loop level in scalar and spinor QED.

\chapter[Forward Doubly-Virtual Compton Scattering]{Forward Doubly-Virtual Compton Scattering} \chaplab{chap4}

In the previous Chapter we introduced the CS process and polarizabilities and focused on the RCS and static polarizabilities. In this Chapter we consider the case of forward VVCS, which is
relevant the subsequent calculations of the TPE effects in hydrogen-like atoms. 

The theory of forward VVCS is summarized in \secref{chap4}{VVCStheory}. The LEX of the Compton amplitudes and the sum rules for GPs will be of particular interest (\secref{chap4}{SR}). We will then introduce the framework of ChPT as the low-energy effective field theory of our choice (\secref{chap4}{DeltaChPT}). In view of the $\delta_{LT}$ puzzle, we will put special attention on the inclusion of the spin-3/2 $\Delta(1232)$-isobar and the two prominent power-counting schemes: the $\delta$- and $\epsilon$-expansion. In the following, we will calculate the tree-level $\Delta$-exchange contribution to VVCS (\secref{chap4}{DeltaExchangeSec}) and the ($N\gamma^*\rightarrow \pi \Delta$) photoabsorption cross sections for pion-delta production (\secref{chap4}{DeltaCrossSectionSec}) in BChPT with $\delta$-expansion. We will determine the contribution of the $\Delta$-resonance to the nucleon polarizabilities and review the status of the $\delta_{LT}$ puzzle.

\section{Generalities} \seclab{VVCStheory}

\subsection{Lorentz Structure}
Figure \ref{fig:frwCS} shows the process of CS in forward kinematics, i.e., with equal initial and final photon (target) momenta. In the lab frame, 
\beq
p=(M,\boldsymbol{0}), \qquad q=(\nu,\boldsymbol{q}),
\eeq
forward CS depends on two variables: the photon lab-frame energy $\nu$ and the photon virtuality $Q^2=-q^2>0$.

 The forward VVCS amplitude allows for the tensor decomposition into four independent scalar amplitudes:
\bea
\hspace{-0.5cm}T^{\mu \nu}(q,p) & = &  
\left( -g^{\mu\nu}+\frac{q^{\mu}q^{\nu}}{q^2}\right)
T_1(\nu, Q^2) +\frac{1}{M^2} \left(p^{\mu}-\frac{p\cdot
q}{q^2}\,q^{\mu}\right) \left(p^{\nu}-\frac{p\cdot
q}{q^2}\, q^{\nu} \right) T_2(\nu, Q^2)\nn\quad\\
&&-   \frac{1}{M}\gamma^{\mu \nu \al} q_\al \,S_1(\nu, Q^2)  -  
\frac{1}{M^2} \Big( \gamma^{\mu\nu} q^2 + q^\mu \gamma^{\nu\al} q_\al  -  q^\nu \gamma^{\mu\al}
q_\al \Big) S_2(\nu, Q^2).
\eqlab{fVVCS}
\eea
This form explicitly obeys the e.m.\ current conservation: $q_\mu T^{\mu\nu} = 0=q_\nu T^{\mu\nu}$.

\begin{figure}[t]
\centering
       \includegraphics[width=8.5cm]{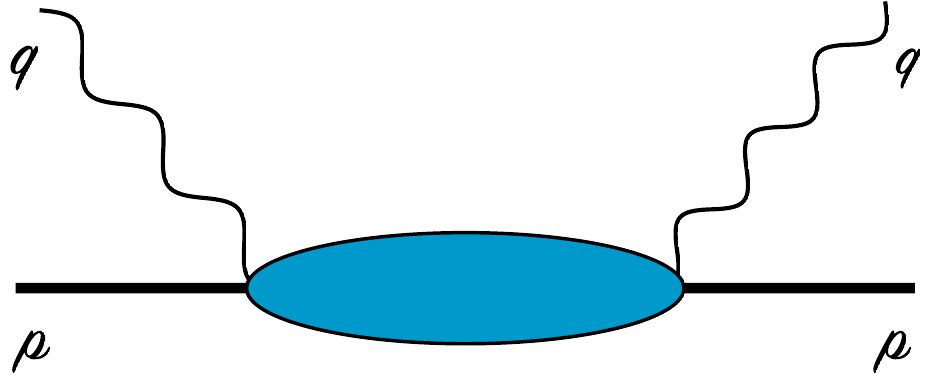}
\caption{Compton scattering in forward kinematics.\label{fig:frwCS}}
\end{figure}

\noindent The amplitudes $T_1$ and $T_2$ are spin-independent, whereas the amplitudes $S_1$ and $S_2$ are spin-dependent. 
Recall that the forward RCS, obtained from VVCS in the limit $Q^2\to 0$, is described by two scalar amplitudes, see Eq.~\ref{covT}. The relation of the RCS and VVCS amplitudes at $Q^2=0$ is as follows:
\beq
f(\nu)=\frac{1}{4\pi}T_1(\nu,0), \qquad g(\nu)=\frac{\nu }{4\pi M} S_1(\nu,0).
\eeq
Omitting terms which vanish upon contraction with the photon polarization vectors, i.e.\ the ones containing $q^\mu$ or $q^\nu$, the symmetric and antisymmetric parts of the second-rank Compton tensor,
\beq
T^{\mu \nu} (q,p) = \left[T^{\mu \nu}_S+T^{\mu \nu}_A\right] (q,p) ,
\eeq
read:
\begin{subequations}
\bea
T^{\mu \nu}_S(q,p) & = & -g^{\mu\nu}\,
T_1(\nu, Q^2)  +\frac{p^{\mu} p^{\nu} }{M^2} \, T_2(\nu, Q^2), \eqlab{VVCS_TS}\\
T^{\mu \nu}_A (q,p) & = &-\frac{1}{M}\gamma^{\mu \nu \al} q_\al \,S_1(\nu, Q^2) +
\frac{Q^2}{M^2}  \gamma^{\mu\nu} S_2(\nu, Q^2).
\eea
\end{subequations}
As we will see in \chapref{chap5}, the symmetric, nucleon-spin independent part of the Compton amplitude contributes to the LS, and the antisymmetric, nucleon-spin dependent part of the amplitude contributes to the HFS.

Explicit expressions for the leading tree-level VVCS amplitudes, cf.\ \Figref{Born}, and the corresponding contributions to the structure functions are presented in \chapref{chap5}, where we connect the forward TPE effect in hydrogen-like atoms to the forward VVCS. 
The elastic nucleon-pole part is discussed in \secref{chap5}{NucleonPole} and the Born part is given \secref{chap5}{1334}. In the following, we will introduce the GPs probed with virtual photons. 

\subsection{Low-Energy Expansion}

The LEX of the relativistic amplitudes goes as, up to $\mathcal{O}(\nu^4,\nu^2Q^2,Q^4)$ \cite{Drechsel:1998zm,Drechsel:2002ar,Pascalutsa:2014zna}:
\begin{subequations}
\eqlab{LEXVVCS}
\begin{align}
&\frac{1}{4\pi}\Big[T_1-T_1^\mathrm{pole}\Big](\nu,Q^2)=-
\, \frac{ Z^2 \al}{M}+\left[\frac{Z\al}{3M}\langle r^2\rangle_1 +\beta_{M1}\right]Q^2+ \eqlab{T1LEX}\\
&\hspace{4.2cm}+ \left[\alpha_{E1}(Q^2)+\beta_{M1}(Q^2)\right]\,\nu^2+\ldots ,\nn\\
&\frac{1}{4\pi}\Big[T_2-T_2^\mathrm{pole}\Big](\nu,Q^2)= \left(\alpha_{E1}+\beta_{M1}\right)Q^2+ \ldots .\eqlab{T2LEX}\\
&\frac{1}{4\pi}\Big[S_1-S_1^\mathrm{pole}\Big](\nu,Q^2)=
 \frac{Z^2 \al \,\kappa^2}{2M }  \left[-1+\frac{1}{3}Q^2\langle r^2 \rangle_2 \right]
+   M  \gamma_0(Q^2) \,  \nu^2 \eqlab{S1lex}\\ 
&\hspace{4.1cm}+\,  M\, Q^2 \Big\{\gamma_{E1 M2} - 3 M \alpha \big[ P^{\prime(M1, M1)1} (0)+  P^{\prime (L1, L1)1}(0) \big]\Big\}+\ldots,\nn\\
&\hspace{3.7cm}=\frac{8\pi \al}{M}I_1(Q^2)\\
&\hspace{4.1cm}+\left\{\frac{8\pi \al}{M}\frac{1}{Q^2}\left[I_A(Q^2)-I_1(Q^2)\right]+4\pi M \delta_{LT}(Q^2)\right\}\nu^2+\cdots,\nn\\
&\frac{\nu}{4\pi} \Big[S_2-S_2^\mathrm{pole}\Big](\nu, 0) =- M^2 \nu^2  \Big[ \ga_0 +\ga_{E1E1}\eqlab{S2lex}
\\
&\hspace{3.9cm}- 3 M \alpha \big\{ P^{\prime(M1, M1)1} (0)-  P^{\prime (L1, L1)1}(0) \big\}\Big](Q^2)+\ldots. \nn
\end{align}
\end{subequations}
In \chapref{chap5}, we want to apply the VVCS formalism to the TPE corrections in atomic bound states. Therefore, we already here introduced the nuclear charge $Z$.
Besides polarizabilities, the anomalous magnetic moment of the nucleus, the Thomson term, the Dirac mean-squared radius, $\langle r^2\rangle_1=-6\,\nicefrac{\dd}{\dd Q^2}\,F_1(Q^2)\vert_{Q^2=0}$, and the Pauli mean-squared radius, $\langle r^2\rangle_2=-6\,\nicefrac{\dd}{\dd Q^2}\,F_2(Q^2)\vert_{Q^2=0}$, enter the non-pole parts as part of VVCS LET, cf.\ Refs.~\cite{Thirring:1950fj,Low:1954kd,Gell_Mann:1954kc}. 

The sum of dipole polarizabilities in \Eqref{T1LEX} is $Q^2$ dependent, while the sum of dipole polarizabilities in \Eqref{T2LEX} is not. The latter are the static polarizabilities  introduced previously, cf.\ Sections~\ref{chap:chap3}.\ref{sec:polarizabilities} and \ref{chap:chap3}.\ref{sec:SRderivation}. The former are the so-called GPs of VVCS, which are valid for finite momentum transfer. These GPs are all the polarizabilities entering the pure $\nu$-expansion of the amplitudes. Furthermore, we find the momentum derivatives of the generalized VCS polarizabilities, which will not be of interest to our studies.\footnote{For CS off the nucleon, the momentum derivatives of the VCS GPs are given by:
\bea
\eqlab{GPslope}
&&  P^{\prime\,  (M1, M1)1}(0) \pm  P^{\prime\,  (L1, L1)1}(0) 
 \equiv \frac{\dd}{\dd \bq^2 } \Big[
 P^{ (M1, M1)1}(\bq^2) \pm  P^{ (L1, L1)1}(\bq^2 )  \Big]_{\bq^2 =0},
 \eea
with $\bq^2$ being the absolute value of the initial photon's CM three-momentum. The superscript indicates the multipolarities, $L1(M1)$ denoting electric (magnetic) dipole transitions of the initial and final photons, and ``$1$'' implying that these transitions involve a spin-flip of the nucleon, cf.\  Refs.~\cite{Guichon:1995, Guichon:1998xv}.} $I_1$ and $I_A$ are the generalized GDH integrals and will be defined in Eqs.~\eref{I1def} and \eref{IA}.

The relativistic amplitudes are related to a set of non-relativistic scalar amplitudes in the following way:
\begin{subequations}
\bea
f_T(\nu,Q^2)&=&T_1(\nu,Q^2),\\
f_L(\nu,Q^2)&=&-T_1(\nu,Q^2)+\frac{\nu^2+Q^2}{Q^2}T_2(\nu,Q^2),\\
g_{TT}(\nu,Q^2)&=&\frac{\nu}{M}\left[S_1(\nu,Q^2)-\frac{Q^2}{M \nu}S_2(\nu,Q^2)\right],\quad\\
g_{LT}(\nu,Q^2)&=&\frac{Q}{M}\left[S_1(\nu,Q^2)+\frac{\nu}{M}S_2(\nu,Q^2)\right].
\eea
\end{subequations}
The CS amplitude can be written in terms of these amplitudes as:
\bea
\label{Eq:T-Compt-definition}
\hspace{-0.5cm}\boldsymbol{\eps}^\ast \!\cdot T(\nu,Q^2)\cdot \boldsymbol{\eps}&=&f_{L}(\nu,Q^2) +(\boldsymbol{\eps}^{\, *} \cdot \boldsymbol{\epsilon}\,) \,f_{T}(\nu,Q^2)  \\
&&+  i \boldsymbol{\sigma}\cdot (\boldsymbol{\epsilon}^{\, *} \times \boldsymbol{\epsilon}\,)\, g_{TT}(\nu,Q^2) -  i \boldsymbol{\sigma}\cdot [(\boldsymbol{\epsilon}^{\, *} - \boldsymbol{\epsilon}\,)\times \boldsymbol{q}]\; g_{LT}(\nu,Q^2).\nn
\eea
with the photon polarization vectors $\boldsymbol{\eps}$ and the Pauli matrices $\boldsymbol{\sigma}$. The LEX  of the non-relativistic amplitudes is given by:
\begin{subequations}
\bea
\bar{f}_T(\nu,Q^2)&=&4\pi \left\{Q^2 \beta_{M1}+ \left[\alpha_{E1}(Q^2)+\beta_{M1}(Q^2)\right]\nu^2+\cdots\right\},\qquad\,\eqlab{fTLEX}\\
\bar{f}_L(\nu,Q^2)&=&4\pi \,\left\{\alpha_{E1}+\alpha_{L}(Q^2)\,\nu^2\right\}Q^2+\cdots,\eqlab{fLLEX}\\
\bar{g}_{TT}(\nu,Q^2)&=&4\pi\, \gamma_0(Q^2)\, \nu^3+\cdots,\eqlab{gTTLEX}\\
\bar{g}_{LT}(\nu,Q^2)&=&4\pi\, \delta_{LT}(Q^2) \,\nu^2\,Q+\cdots,\eqlab{gLTLEX}
\eea
\end{subequations}
where the bar denotes the non-Born part. As one can see, the non-relativistic amplitudes are very convenient for reading off the polarizabilities.

\subsection{Dispersion Relations, Unitarity and Sum Rules} \seclab{SR}
All invariant CS amplitudes fulfil DRs:\footnote{Since the Born part of $S_2$, \Eqref{S2Born}, has a pole for the subsequent limits of $Q^2\rightarrow 0$ and $\nu\rightarrow 0$, it is advisable to use a DR for $\nu S_2$ instead.}
\begin{subequations}
\eqlab{DRVVCS}
\bea
\eqlab{T1DR}
T_1 ( \nu, Q^2) &=&\frac{2}{\pi} \int_{\nu_{\mathrm{el}}}^\infty \dd \nu' \, 
\frac{\nu' \im T_1 ( \nu', Q^2)}{\nu^{\prime\,2} -\nu^2 - i 0^+}\nn\\
&=& \frac{8\pi Z^2\al}{M} \int_{0}^1 
\frac{\dd x}{x} \, 
\frac{f_1 (x, Q^2)}{1 - x^2 (\nu/\nu_{\mathrm{el}})^2 - i 0^+},\\
\eqlab{T2DR}
T_2 ( \nu, Q^2) &=& \frac{2}{\pi} \int_{\nu_{\mathrm{el}}}^\infty \dd \nu' \, 
\frac{\nu' \im T_2 ( \nu', Q^2)}{\nu^{\prime\,2} -\nu^2 - i 0^+} \nn\\
&=&\frac{16\pi Z^2\al M}{Q^2} \int_{0}^1 
\!\dd x\, 
\frac{f_2 (x, Q^2)}{1 - x^2 (\nu/\nu_{\mathrm{el}})^2  - i 0^+},\\
\eqlab{S1DR}
S_1 ( \nu, Q^2) &=&  \frac{2}{\pi} \int_{\nu_{\mathrm{el}}}^\infty \dd \nu' 
\, \frac{\nu' \im S_1 ( \nu', Q^2)}{\nu^{\prime\,2} -\nu^2 - i0^+ }\nn\\
&=&\frac{16\pi Z^2\al M}{Q^2} \int_{0}^1 
\!\dd x\, 
\frac{g_1 (x, Q^2)}{1 - x^2 (\nu/\nu_{\mathrm{el}})^2  - i 0^+},\\
\eqlab{S2DR}
\nu S_2 ( \nu, Q^2) &=&  \frac{2}{\pi} \int_{\nu_{\mathrm{el}}}^\infty \dd \nu' 
\, \frac{\nu^{\prime\,2} \im S_2 ( \nu', Q^2)}{\nu^{\prime\,2} -\nu^2 - i0^+ }\nn\\
&=&\frac{16\pi Z^2\al M^2}{Q^2} \int_{0}^1 
\!\dd x\, 
\frac{g_2 (x, Q^2)}{1 - x^2 (\nu/\nu_{\mathrm{el}})^2  - i 0^+},
\eea
\end{subequations}
where in the last step we plugged in the optical theorem:
\begin{subequations}
\eqlab{VVCSunitarity}
\bea
\im T_1(\nu,Q^2)&=&\frac{4\pi^2 Z^2\al }{M}f_1(x,Q^2)=\sqrt{\nu^2+Q^2}\,\sigma_T(\nu,Q^2), \eqlab{ImT1} \\
\im T_2(\nu,Q^2)&=&\frac{4\pi^2 Z^2\al}{\nu}f_2(x,Q^2)=\frac{Q^2 }{\sqrt{\nu^2+Q^2}}\left[\sigma_T+\sigma_L\right](\nu,Q^2), \eqlab{ImT2}\\
\im S_1(\nu,Q^2) &=& \frac{4\pi^2 Z^2\alpha}{\nu} \, g_1(x,Q^2) = 
\frac{M  \nu }{\sqrt{\nu^2+Q^2}}\left[\frac{Q}{\nu}\sigma_{LT}  + \sigma_{TT}\right](\nu,Q^2), \eqlab{ImS1}\\
\im S_2(\nu,Q^2) & =&  \frac{4\pi^2 Z^2\alpha M}{\nu^2} \, g_2(x, Q^2)  
= \frac{M^2}{\sqrt{\nu^2+Q^2}}\left[\frac{\nu}{Q}\sigma_{LT}  - \sigma_{TT}\right](\nu,Q^2), \qquad \quad\eqlab{ImS2}
\eea
\end{subequations}
which in the physical region ($x\in [0,1]$, with $x=\nu_\mathrm{el}/\nu$ being the Bjorken variable and $\nu_\mathrm{el}=Q^2/2M$) relates the absorptive parts of the forward VVCS amplitudes to 
the nucleon structure functions $f_1$, $f_2$, $g_1$ and $g_2$ (functions of $x$ and $Q^2$) or the photoabsorption cross sections $\si_T$, $\si_L$, $\si_{TT}$ and $\si_{LT}$ (functions of $\nu$ and $Q^2$). The optical theorem for RCS was given in \Eqref{OptT} with a photon flux factor that corresponds to $K(\nu)=\nu$ \cite{Drechsel:2002ar}. In the case of virtual CS, we modify the photon flux factor and use Gilman's definition $K(\nu,Q^2)=\sqrt{\nu^2+Q^2}$ \cite{Gilman:1967sn}.\footnote{Note that this choice is different to the one in Ref.~\cite{Hagelstein:2015egb}.} The cross section $\sigma_{LT}$ 
describes a simultaneous helicity
change of the photon (from longitudinal to transverse) and the nucleon (spin-flip) such that the
total helicity is conserved. 
The other cross sections are the usual combinations of helicity cross sections: $\sigma_T=\nicefrac12\, (\sigma_{1/2}+\sigma_{3/2})$ and $\sigma_{TT}=\nicefrac12\, (\sigma_{1/2}-\sigma_{3/2})$ for transversely polarized photons, and $\sigma_L=\nicefrac12\, (\sigma_{1/2}+\sigma_{-1/2})$ for longitudinal photons, where the subscript on the rhs denotes the total helicity of the $\gamma^\ast N$ state. Note that $\sigma_L$ and $\sigma_{LT}$ are vanishing in the real photon limit.

Just as in \Eqref{KKf}, the high-energy asymptotic of the spin-independent structure function $f_1(x,Q^2)$ prevent the convergence of an unsubtracted DR. Therefore,  $T_1$ requires a once-subtracted DR:
\begin{subequations}
\eqlab{T1Subtrgen}
\bea
T_1 ( \nu, Q^2) &=& T_1(0,Q^2) +\frac{2\nu^2}{\pi} \int_{\nu_{\mathrm{el}}}^\infty \dd \nu'
\frac{\im T_1 ( \nu', Q^2)}{\nu'( \nu^{\prime\,2} -\nu^2 - i 0^+)}\\
&=&T_1(0,Q^2) +\frac{32\pi Z^2\al M\nu^2}{Q^4}\int_{0}^1 
\,\dd x \, 
\frac{x f_1 (x, Q^2)}{1 - x^2 (\nu/\nu_{\mathrm{el}})^2 - i 0^+}\eqlab{T1Subtr},\quad
\eea
\end{subequations}
where we need to make a subtraction at all values of $Q^2$, see discussion of the $T_1(0,Q^2)$ subtraction function in  Ref.~\cite[Section 5.5.1]{Hagelstein:2015egb}. Similar expressions hold for the non-relativistic set of amplitudes.

Restricting the integration in \Eqref{DRVVCS} to $x\in [0,x_0]$, with $x_0$ being the inelastic threshold, we isolate the inelastic scattering region. Replacing the lhs of the inelastic dispersion integrals with the LEX of the VVCS amplitudes with the nucleon-pole part subtracted, cf.\ \Eqref{LEXVVCS}, one can derive a number of VVCS sum rules. 
\begin{itemize}
\item generalized Baldin sum rule \cite{Drechsel:2002ar}:
\begin{subequations}
\bea
\al_{E1}(Q^2)+\beta_{M1}(Q^2)&=&\frac{8 Z^2 \al M}{Q^4}\int_0^{x_0}\dd x\, x\, f_1(x,Q^2),\eqlab{alphabetaf1}\\
&=&\frac{1}{2\pi^2}\int_{\nu_0}^\infty \frac{\dd \nu}{\nu^3}\,\sqrt{\nu^2+Q^2}\,\sigma_{T}(\nu,Q^2);
\eea
\end{subequations}
\item generalized FSP sum rule \cite{Drechsel:2002ar}: 
\begin{subequations}
\eqlab{g0gen}
\bea
\gamma_0(Q^2)&=&\frac{16 Z^2\al M^2}{Q^6}\int_0^{x_0}\dd x\, x^2 \Big[g_1-\left(\nicefrac{2Mx}{Q}\right)^2g_2\Big](x,Q^2),\\
&=&\frac{1}{2\pi^2}\int_{\nu_0}^\infty\frac{\dd \nu}{\nu^4}\,\sqrt{\nu^2+Q^2}\,\sigma_{TT}(\nu,Q^2);
\eea
\end{subequations}
\item longitudinal polarizability \cite{Lensky:2014dda}:
\begin{subequations}
\bea
\al_{L}(Q^2)&=&\frac{4Z^2 \al M}{Q^6}\int_0^{x_0}\dd x\, \Big[\left\{1+\left(\nicefrac{2Mx}{Q}\right)^2\right\}f_2-2x f_1\Big](x,Q^2),\\
&=&\frac{1}{2\pi^2}\int_{\nu_0}^\infty \frac{\dd \nu}{Q^2\nu^3}\,\sqrt{\nu^2+Q^2}\,\sigma_{L}(\nu,Q^2);
\eea
\end{subequations}
\item longitudinal-transverse polarizability \cite{Drechsel:2002ar}:
\begin{subequations}
\eqlab{dLTgen}
\bea
\delta_{LT}(Q^2)&=&\frac{16Z^2 \al M^2}{Q^6}\int_0^{x_0}\dd x\, x^2 \big[g_1+g_2\big](x,Q^2),\\
&=&\frac{1}{2\pi^2}\int_{\nu_0}^\infty  \frac{\dd \nu}{Q\nu^3}\,\sqrt{\nu^2+Q^2}\,\sigma_{LT}(\nu,Q^2);
\eea
\end{subequations}
\item generalized GDH integral (zeroth moment of $g_1$):
\beq
I_1(Q^2)=\frac{2Z^2M^2}{Q^2}\int_0^{x_0}\dd x\, g_1(x,Q^2);\eqlab{I1def}\\
\eeq
\item generalized GDH integral:
\begin{subequations}
\eqlab{IAdef}
\bea
I_A(Q^2)&=&\frac{2Z^2M^2}{Q^2}\int_0^{x_0}\dd x\, \Big[g_1-\left(\frac{2Mx}{Q}\right)^2g_2\Big](x,Q^2),\eqlab{IA}\\
&=&\frac{M^2}{4\pi^2\al}\int_{\nu_0}^\infty \frac{\dd \nu}{\nu^2}\,\sqrt{\nu^2+Q^2}\,\sigma_{TT}(\nu,Q^2);
\eea
\end{subequations}
\item Burkhardt-Cottingham (BC) sum rule  \cite{Burkhardt:1970ti}:
\begin{subequations}
\bea
0&=&\int_0^{1}\dd x\, g_2(x,Q^2),\eqlab{BCdef}\\
I_2(Q^2)&=&\frac{2Z^2M^2}{Q^2}\int_0^{x_0}\dd x\, g_2(x,Q^2)=\frac{Z^2}{4}F_2(Q^2)G_M(Q^2);\eqlab{I2def}
\eea
\end{subequations}
\item second moment of the higher twist part of $g_2$:
\begin{subequations}
\bea
\ol d_2(Q^2)&=&Z^2\int_0^{x_0}\dd x\,x^2\Big[ g_2(x,Q^2)+g_2^\mathrm{WW}(x,Q^2)\Big],\\
&=&Z^2\int_0^{x_0}\dd x\,x^2\big[3 g_2(x,Q^2)+2g_1(x,Q^2)\big];
\eea
\end{subequations}
\end{itemize}
The GPs of VVCS can then be extracted from either the LEX of the CS amplitudes or the CS sum rules listed above. Note that the generalized GDH integrals $I_1$ and $I_A$ are no pure polarizabilities, as will be explained in \secref{chap5}{1334}.

The VVCS GP $\ol d_2$ is defined as the second moment of the higher twist part of the structure function $g_2$. The twist-2 part of $g_2$ can be expressed through $g_1$ by the Wandzura-Wilczek relation \cite{Wandzura:1977qf}:
\beq
\eqlab{WWrelation}
g_2^\mathrm{WW}(x,Q^2)=-g_1(x,Q^2)+\int_x^1 \frac{\dd x'}{x'} \,g_1(x',Q^2).
\eeq
This polarizability goes as $Q^6$ for low $Q$ and can be related to the longitudinal transverse polarizability and the generalized GDH integrals:
\bea
\bar d_2(Q^2)=\frac{Q^4}{8M^4}\left\{\frac{M^2 Q^2}{\al}\delta_{LT}(Q^2)+\left[I_1(Q^2)-I_A(Q^2)\right]\right\}.
\eea
It is an interesting quantity in connection to the concept of
color polarizability \cite{Filippone:2001ux}. 

\section[Chiral Perturbation Theory Calculation] {Chiral Perturbation Theory Calculation} \seclab{DeltaChPT}
ChPT is an effective field theory of QCD at energies well below 1 GeV. The seminal ChPT papers of Weinberg \cite{Weinberg:1978kz}, Gasser and Leutwyler~\cite{Gasser:1983yg,Gasser:1987rb}
deal with pions as the Goldstone bosons of spontaneous chiral symmetry breaking in QCD. The key observation is that the coupling of Goldstone bosons is proportional to their momentum, hence, at
low momenta the coupling is weak and a perturbative expansion is possible. The breakdown scale is set by the scale of spontaneous chiral symmetry breaking $\Lambda_{\chi\mathrm{SB}}\sim 4f_\pi \approx 1$ GeV, where
$f_\pi$ is the pion decay constant. In reality ChPT breaks down somewhat earlier, as seen, e.g., in $\pi \pi$ scattering where the $\si(600)$- and $\rho(775)$- meson excitations set the limiting scale of a perturbative expansion \cite{Colangelo:2001df,Caprini:2005an}.
Nonetheless, ChPT has proven to be very useful in studying the low-energy strong interaction. The literature on the subject is immense and we have to quickly narrow down the discussion to the case at hand, i.e., 
the nucleon CS. 

For our calculation of the nucleon VVCS, we will be using the SU(2) BChPT, which is the manifestly Lorentz-invariant variant of ChPT in the single-baryon sector (see, e.g., 
Ref.~\cite[Sec.\ 4]{Pascalutsa:2006up}). Let us note right away that, at least in some cases, the predictions of BChPT and HBChPT differ substantially. This is, for instance, the case for the longitudinal-transverse polarizability of the proton and the proton-polarizability contribution to the LS, see Sections \ref{chap:chap4}.\ref{sec:DeltaCrossSectionSec}, \ref{chap:5LS}.\ref{sec:HBcompLS} and \ref{chap:5HFS}.\ref{sec:HBcompHFS}.

The $\Delta(1232)$ is the lowest nucleon resonance with the excitation energy $\varDelta=M_\Delta-M_N\approx 294$ MeV, which is not much higher than the pion mass. Therefore, it is customary to include 
the delta as an explicit DOF in the chiral effective Lagrangian. Thus, the relevant fields are: 
the pion scalar iso-vector $\pi^a(x)$, the nucleon spinor iso-doublet $\mathcal{N}(x)$, the $\Delta(1232)$ vector-spinor iso-quartet $\Delta_\mu(x)$, and the photon vector field $A_\mu(x)$.  
The terms of the chiral effective Lagrangian relevant for us  are the following.
\beq
\mathcal{L}=\mathcal{L}^{(1)}_{\mathcal{N}}+\mathcal{L}^{(1)}_{\Delta }+\mathcal{L}^{(1)}_{\pi\Delta \mathcal{N}}+\mathcal{L}^{(2)}_\pi+\mathcal{L}^{(2)\,\mathrm{nm}}_{\gamma N \Delta}+\mathcal{L}^{(1)\,\mathrm{nm}}_{\gamma\Delta\Delta}, \qquad\quad \eqlab{Lsum}
\eeq
where the superscript denotes the order of the Lagrangian reflected by the number of comprised small quantities (pion mass, momentum and factors of $e$). They read ~\cite{Ledwig:2011cx}:
\begin{subequations}
\bea
\mathcal{L}^{(1)}_\mathcal{N}&=&\ol{\mathcal{N}} \big(\slashed{D}-M_N\big)\mathcal{N}-\frac{g_A}{2f_\pi}\ol{\mathcal{N}} \tau^a \left(\slashed{D}^{ab} \pi^b\right)\gamma_5\, \mathcal{N},\eqlab{piNN}\\
\mathcal{L}^{(1)}_\Delta&=&\ol{\Delta}_\mu \left(i \gamma^{\mu \nu \lambda}D_\lambda -M_\Delta \gamma^{\mu \nu}\right) \Delta_\nu+\frac{H_A}{2f_\pi M_\Delta}\,\varepsilon^{\mu \nu \al \lambda}\,\ol \Delta_\mu \mathscr{T}^{\,a} \left(D_\al \Delta_\nu\right)D_\lambda^{ab}\pi^b,\qquad\quad\\
\mathcal{L}^{(1)}_{\pi\Delta \mathcal{N}}&=&\frac{i h_A}{2f_\pi M_\Delta} \ol{\mathcal{N}} \,T^a\gamma^{\mu \nu \lambda}\left(D_\mu \Delta_\nu\right)\left(D_\lambda^{ab}\pi^b\right)+\text{h.c},\\
\mathcal{L}^{(2)}_\pi&=&\frac{1}{2} \Big(D_\mu^{ab} \pi^b\Big)\Big(D^\mu_{ac} \pi^c\Big)-\frac{1}{2}m_\pi^2 \pi_a\pi^a,
\eea
\end{subequations}
where the covariant derivatives,\footnote{Note that $D_\mu \mathcal{N}$ is of $\mathcal{O}(p^0)$, while $\big(\slashed{D}-M_N\big)\mathcal{N}$ is of $\mathcal{O}(p)$, cf.\ Ref.~\cite[Eq.~(5.23)]{Scherer2012}.}
\begin{subequations}
\bea
D_\mu^{ab} \pi^b&=&\delta^{ab} \partial_\mu \pi^b+ieQ_\pi^{ab}A_\mu\pi^b,\\
D_\mu \mathcal{N}&=&\partial_\mu \mathcal{N}+ieQ_N A_\mu\mathcal{N}+\frac{i}{4f_\pi^2}\,\varepsilon^{abc}\tau^a \pi^b (\partial_\mu \pi^c),\\
D_\mu \Delta_\nu&=& \partial_\mu \Delta_\nu+ie Q_\Delta A_\mu \Delta_\nu+\frac{i}{2f_\pi^2} \,\varepsilon^{abc}\, \mathscr{T}^a\pi^b(\partial_\mu \pi^c),\qquad\quad
\eea
\end{subequations}
are used and one defines the particle charges through:
\begin{subequations}
\bea
Q_\pi^{ab}&=&-i \varepsilon^{ab3},\\
Q_N&=&\half \!(1+\tau^3),\\
Q_\Delta&=&\half \!(1+3 \mathscr{T}^3).
\eea
\end{subequations}
The isospin $1/2$ to $3/2$ and the isospin $3/2$ to $3/2$ transition matrices can be found in Ref.~\cite[Appendix A]{Ledwig:2011cx} and Ref.~\cite{Pascalutsa:2005vq}. They commute with the Dirac matrices. The Lagrangian for the $\gamma^*N \rightarrow\Delta$ transition will be of special interest in \secref{chap4}{DeltaExchangeSec}, it can be found in Ref.~\cite{Pascalutsa:2005ts}:\footnote{Note that here we corrected a typo appearing in Refs.~\cite{Pascalutsa:2005ts,Pascalutsa:2005vq,Pascalutsa:2006up}.}
\bea
\eqlab{nmGammaNDeltaLag}
 \mathcal{L}^{(2) \,\mathrm{nm}}_{\gamma N \Delta}&=&\frac{3e}{2M_N(M_N+M_\Delta)}\Bigg[\bar{N} T_3\Big\{i g_M (\partial_\mu \Delta_\nu) \tilde{F}^{\mu \nu}-g_E  \gamma_5 (\partial_\mu \Delta_\nu) F^{\mu \nu}\eqlab{LagrangianGND}\\
 &&+i \frac{g_\mathrm{C}}{M_\Delta}\gamma_5 \gamma^\al (\partial_\al \Delta_\nu-\partial_\nu \Delta_\al)\partial_\mu F^{\mu \nu}\Big\}+\Big\{g_E (\partial_\mu \bar\Delta_\nu)\gamma_5 F^{\mu \nu}\nn\\
 &&-i g_M (\partial_\mu \bar\Delta_\nu) \tilde{F}^{\mu \nu}+i \frac{g_\mathrm{C}}{M_\Delta}(\partial_\al \bar\Delta_\nu-\partial_\nu \bar\Delta_\al)\gamma^\al\gamma_5  \partial_\mu F^{\mu \nu}\Big\}T_3^\dagger N\Bigg].\nn
 \eea
Furthermore, we partially include the non-minimal coupling of the photon to the $\Delta(1232)$ resonance \cite{Ledwig:2011cx}:
\beq
\mathcal{L}^{(1)\,\mathrm{nm}}_{\gamma\Delta\Delta}=\frac{e}{M_\Delta}\ol \Delta_\mu \left(i\kappa_1F^{\mu \nu}-\kappa_2 \gamma_5 \tilde F^{\mu \nu}\right) \Delta_\nu,
\eeq
with $\kappa_1=1=\kappa_2$ as in $N=2$ supergravity. From the above Lagrangians we derive the Feynman rules listed in \appref{chap4}{FeynmanRules}.

\begin{table}[t]
 \centering
 \caption{The parameters used in the chiral perturbation theory calculations.}
 \label{LEC}
 \begin{small}
\begin{tabular}{|c|c|c|}
\hline
 \rowcolor[gray]{.7}
{\bf  ChPT parameter}& {\bf  value} &  {\bf source}\\
 \hline
   $m_\pi$ & 139.57 MeV &\\
    \rowcolor[gray]{.95}
$f_\pi$&$92.21$ MeV&pion decay $\pi^+\rightarrow \mu^+ \nu_\mu$ \cite{Olive:2016xmw}\\
  $M_N$ & 938.27 MeV &\\
     \rowcolor[gray]{.95}
$g_A$&$1.27$&neutron decay $n\rightarrow p\,e^-\, \bar\nu_e$ \cite{Olive:2016xmw}\\
 $M_\Delta$ & 1232 MeV &\\
    \rowcolor[gray]{.95}
&&$P_{33}$ partial wave in $\pi N$ scattering\\
   \rowcolor[gray]{.95}
\multirow{-2}{*}{$h_A$}&\multirow{-2}{*}{$2.85$}&$\Delta(1232)$ decay width \cite{Pascalutsa:2006up,Pascalutsa:2004je,Pascalutsa:2005nd} \\
$g_M$&$2.97$ &\\
$g_E$&$-1.0$&\\
$g_\mathrm{C}$&$-2.6$&\multirow{-3}{*}{pion electroproduction $e^- N \rightarrow e^- N \pi$ \cite{Pascalutsa:2005vq}}\\
\hline
\end{tabular}
\end{small}
\end{table}

In ChPT, the coupling strengths of the different interactions are embedded in the so-called low-energy constants (LECs). These coupling constants can be fitted to various experiments. Table \ref{LEC} lists the LECs appearing in the ChPT Lagrangian at the presented order. The pion and neutron decay constant, $f_\pi$ and  $g_A$, are determined from the respective decays. $g_A$ is also known as the axial coupling of the nucleon. In the limit of a large number of colors ($N_c$), it is related to $h_A=\nicefrac{3}{\sqrt{2}}\,g_A$ and $H_A=\nicefrac{9}{5}\,g_A$.  Because the $\Delta(1232)$-resonance is dominating the $P_{33}$-state in the partial wave analysis of $\pi N$ scattering, one can deduce the $\Delta(1232)$ decay width ($\Gamma_\Delta=0.115$ GeV) and in turn extract the coupling constant $h_A$ from $\pi N$ scattering. The magnetic, electric and Coulomb couplings of the $\gamma^*N\Delta$ interaction are extracted from pion electroproduction.\footnote{Fits of only $g_M$ and $g_E$, neglecting $g_C$, can be found in Refs.~\cite{Pascalutsa:2003aa,Pascalutsa:2003zk}.} Since all LECs are known from other processes, the description of forward VVCS at $\mathcal{O}(p^{7/2})$ in ChPT comes out as 
a prediction.

\begin{table}[tbh]
 \centering
 \caption{Power-counting for the diagrams contributing to doubly-virtual Compton scattering.}
 \label{PC}
 \begin{small}
\begin{tabular}{|c|c|cc|cc|}
\hline
 \rowcolor[gray]{.7}
 & & \multicolumn{2}{c|}{\bf $\boldsymbol{\delta}$-expansion}&\multicolumn{2}{c|}{\bf $\boldsymbol{\epsilon}$-expansion}\\
  \rowcolor[gray]{.7}
\multirow{-2}{*}{\bf contribution }& \multirow{-2}{*}{\bf Figure}& $p\sim m_\pi$ &  $p\sim \Delta$& $p\sim m_\pi$ &  $p\sim \Delta$\\
 \hline
Born&\ref{fig:Born}&$p$&$p$&\multicolumn{2}{c|}{$p$}\\
  \rowcolor[gray]{.95}
$\Delta$-exchange&\ref{fig:DeltaExchange}&$p^{7/2}$&$p$&\multicolumn{2}{c|}{$p^3$}\\
$\pi N$-loops&\ref{fig:PiNLoopDiags}&$p^3$&$p^3$&\multicolumn{2}{c|}{$p^3$}\\
  \rowcolor[gray]{.95}
&\ref{fig:DeltaOldPlot}&&&\multicolumn{2}{c|}{}\\
  \rowcolor[gray]{.95}
\multirow{-2}{*}{$\pi \Delta$-loops}&\ref{fig:DeltaNewPlot} (first row)&\multirow{-2}{*}{$p^{7/2}$}&\multirow{-2}{*}{$p^3$}&\multicolumn{2}{c|}{\multirow{-2}{*}{ $p^3$}}\\
$\pi \Delta$-loops&
\ref{fig:DeltaNewPlot} (second row)&$p^4$&$p^3$&\multicolumn{2}{c|}{$p^3$}\\
  \rowcolor[gray]{.95}
$\pi \Delta$-loops&\ref{fig:DeltaNewPlot} (third row)&$p^{9/2}$&$p^3$&\multicolumn{2}{c|}{$p^3$}\\
\hline
\end{tabular}
\end{small}
\end{table}

\subsection[Power-Counting Schemes: $\delta$- and $\epsilon$-Expansion]{Power-Counting Schemes: $\boldsymbol{\delta}$- and $\boldsymbol{\epsilon}$-Expansion}\seclab{powercounting}

ChPT is a perturbative description for QCD at low energies, which is based on a small parameter, e.g., $p/\Lambda_{\chi\mathrm{SB}}$. The power-counting assigns an order to every Feynman diagram and thereby defines which Feynman diagrams need to be included at a given order in the perturbative chiral expansion of a given process. There are two prominent power-counting schemes for ChPT with $\Delta$ DOFs: the $\delta$- and the $\epsilon$-expansion. Ref.~\cite{Pascalutsa:2006up} provides an extensive review on the e.m.\ excitation of the $\Delta$(1232)-resonance with special focus on the proper formulation of ChPT with inclusion of spin-3/2 fields and the chiral expansion in the resonance region. In what follows, we will first describe the $\delta$-expansion applied by us. Afterwards, we will highlight the differences to the $\epsilon$-expansion.

In ChPT with pion and nucleon fields, the order, $\mathcal{O}(p^{n})$, of a Feynman diagram with $L$ loops, $N_\pi$ ($N_N$) pion (nucleon) propagators, and $V_k$ vertices from $k$-th order Lagrangians is given by \cite{Gasser:1987rb}:
\beq
n=4L-2N_\pi-N_N+\sum_k k\, V_k.
\eeq
The vertices given in \appref{chap4}{FeynmanRules} follow from the first- and second-order Lagrangians as:
\begin{itemize}
\item $k=1$: $\Ga_{\pi N N}$, $\Ga^\mu_{\gamma NN}$, $\Ga_{N\Delta \pi}^\al$, $\Ga_{ N\gamma\pi\Delta }^{\al \mu}$, $\Ga^{\al \be}_{\pi \Delta \Delta}$, $\Ga^{\al \be \mu}_{\gamma \Delta \Delta}$;
\item $k=2$: $\Ga^\mu_{\gamma\pi\pi}$, $\Ga_{\ga\gamma\pi\pi}^{\mu\nu}$, $\Ga_{\gamma N\Delta}^{\al\mu}$.
\end{itemize}
In ChPT with additional $\Delta$-fields, another scale appears. Accordingly, there are two different small parameters: $\epsilon=m_\pi/\Lambda_{\chi\mathrm{SB}}$ and 
$\delta=\varDelta/\Lambda_{\chi\mathrm{SB}}$, with $\varDelta=M_\Delta-M_N$. Despite that, for the power-counting it is easier to expand in one rather than two small parameters. Therefore, the $\delta$- and $\epsilon$-expansion power-countings relate the two parameters:
\begin{itemize}
\item $\delta$-expansion \cite{Pascalutsa:2003aa}: $\epsilon\sim\delta^2$;
\item $\eps$-expansion \cite{Hemmert:1996xg}: $\epsilon\sim\delta$.
\end{itemize}

    \begin{figure}[t] 
    \centering 
       \includegraphics[width=13cm]{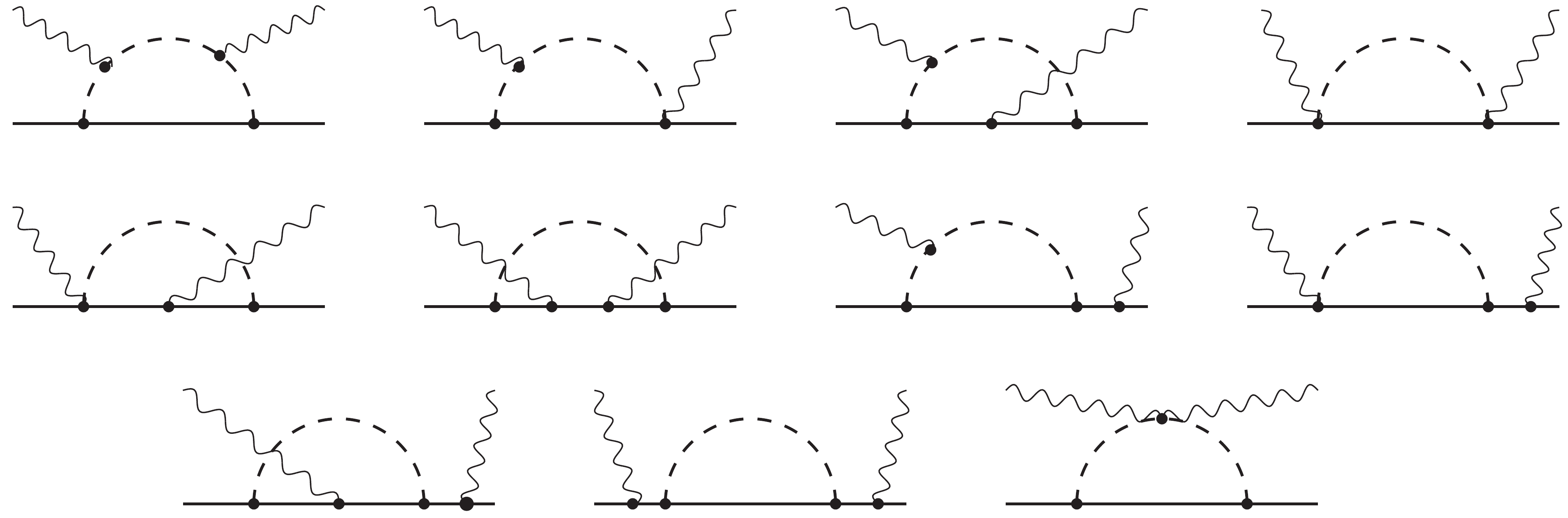}
       \caption{$\pi N$-loop contribution to Compton scattering amplitudes at $\mathcal{O}(p^3)$ in the low-energy domain of the $\delta$-expansion. Pion and nucleon propagators are denoted by 
       dashed and solid lines, respectively.
       Diagrams obtained from these by crossing and time-reversal are included too. Figure taken from Ref.~\cite{Lensky:2014dda}.}
              \label{fig:PiNLoopDiags}
\end{figure}
 
   \begin{figure}[t] 
    \centering 
       \includegraphics[width=12cm]{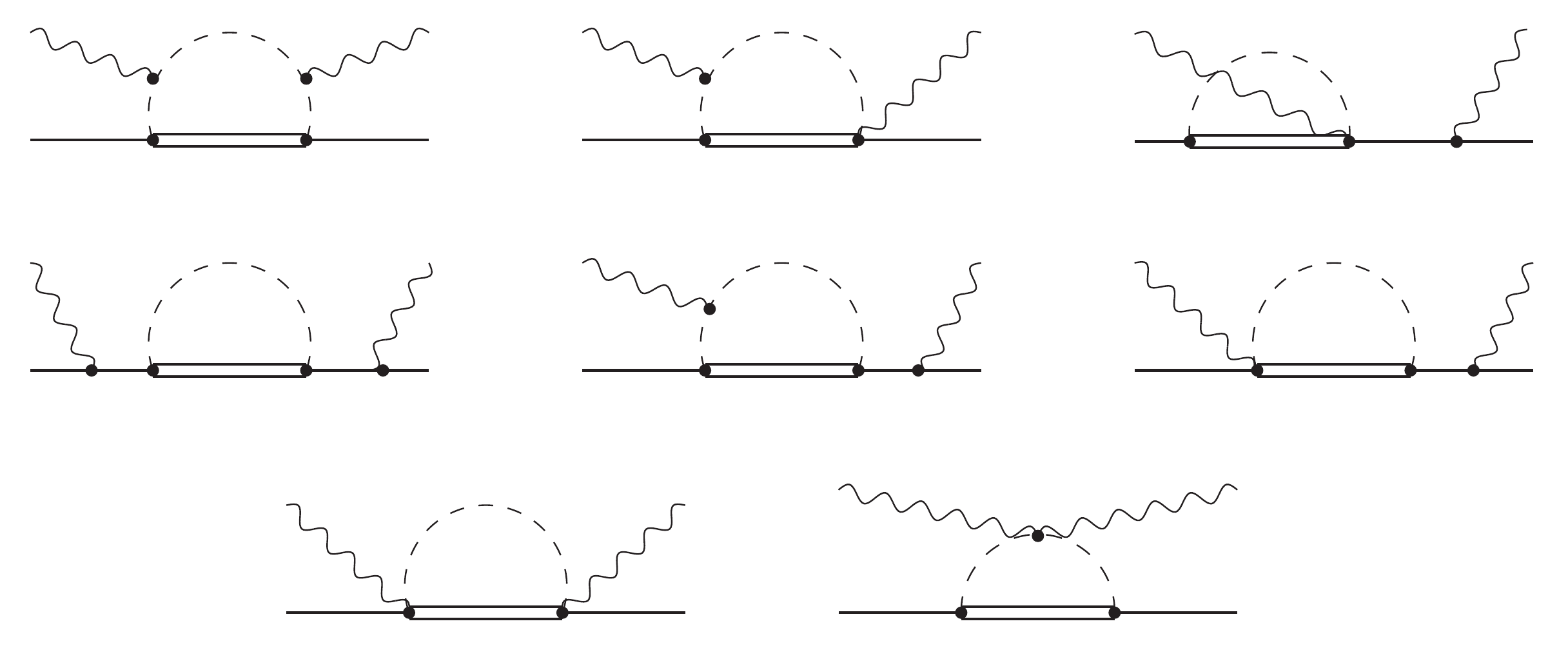}
       \caption{$\pi\Delta$-loop  amplitudes at $\mathcal{O}(p^{7/2})$ in the low-energy domain of the $\delta$-expansion. Delta propagators are denoted by double lines.
       Diagrams obtained from these by crossing and time-reversal are included too. Figure taken from Ref.~\cite{Lensky:2014dda}.}
              \label{fig:DeltaOldPlot}
\end{figure}

\noindent The $\delta$-expansion furthermore distinguishes one-delta-reducible (1$\Delta$R) propagators and one-delta-irreducible (1$\Delta$I) propagators to incorporate a higher weighting for 1$\Delta$R graphs in the resonance region. The $\delta$-expansion then defines \cite{Pascalutsa:2003aa}:
\begin{subequations}
\beq
n_\delta=\begin{cases}n-\nicefrac{1}{2}N_\Delta&p\sim m_\pi,\\
n-3N_\text{1$\Delta$R}-N_\text{1$\Delta$I}&p\sim\Delta,
\end{cases}
\eeq
with
\beq
N_\Delta=N_\text{1$\Delta$R}+N_\text{1$\Delta$I},
\eeq
\end{subequations}
and $N_\text{1$\Delta$R}$ ($N_\text{1$\Delta$I}$) being the number of 1$\Delta$R (1$\Delta$I) propagators, and assigns a size $\mathcal{O}(p^{n_\delta})$ to each graph in the chiral expansion.

Obviously, the $\Delta(1232)$ is supposed to play a dominant role in the resonance region, while it is suppressed at low energies.
Nevertheless, the $\eps$-expansion treats nucleon- and delta-propagators everywhere in the same way.  Therefore, the $\eps$-expansion overestimates the contribution of the $\Delta(1232)$ at low energies and underestimates its contribution in the resonance region. 

In the $\eps$-expansion, all diagrams shown in Figures~\ref{fig:DeltaExchange}, \ref{fig:DeltaOldPlot}, \ref{fig:DeltaNewPlot} and \ref{fig:PiNLoopDiags} are of $\mathcal{O}(p^3)$. In the low-energy region, the $\delta$-expansion assigns the $\mathcal{O}(p^3)$ to the $\pi N$-loop diagrams in \Figref{PiNLoopDiags} and the $\mathcal{O}(p^{7/2})$ to the tree-level $\Delta$-exchange and the $\pi \Delta$-loop diagrams in Figures~\ref{fig:DeltaExchange} and \ref{fig:DeltaOldPlot}. In Table \ref{PC}, we list all the diagrams considered in this thesis and give their respective orders in the $\delta$-expansion for both the low-energy and the resonance region. We will come back to the differences between the two power-counting schemes in \secref{chap4}{DeltaCrossSectionSec}, where we study the $\pi \Delta$-loop graphs in view of the $\delta_{LT}$ puzzle, but first we discuss the $\Delta$-exchange in CS off the nucleon.

\section[Compton Scattering off the Nucleon with $\Delta$-Exchange]{Compton Scattering off the Nucleon with $\boldsymbol{\Delta}$-Exchange} \seclab{DeltaExchangeSec}

The  $\Delta(1232)$-resonance is an almost perfect elastic $\pi N$-resonance, i.e., in $99.4\, \%$ of the cases  it decays into $\pi N$ \cite{Olive:2016xmw}. On the contrary, the chances that the $\Delta(1232)$ decays into $\gamma N$ are only $0.55-0.65\, \%$ \cite{Olive:2016xmw}. Nonetheless, the tree-level diagrams with $\Delta$-exchange are of importance to the CS off the nucleon due to their 1$\Delta$R propagator.
\vspace{0.5cm}
\begin{figure}[bh]
\centering
       \includegraphics[scale=0.525]{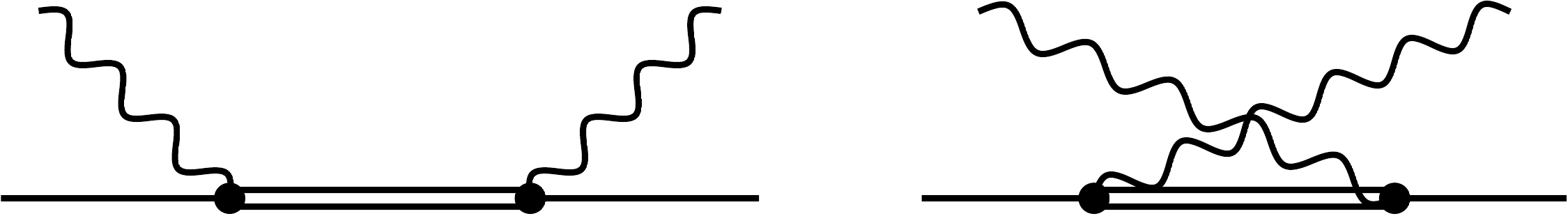}
\caption{$\Delta$-exchange contribution to Compton scattering. These graphs are of $\mathcal{O}(p^{7/2})$ in the low-energy domain of the $\delta$-expansion.
\label{fig:DeltaExchange}}
\end{figure}

\subsection{Compton Amplitudes and Structure Functions} \seclab{3.1CSDelta}
The $\Delta$-exchange diagrams, shown in \Figref{DeltaExchange}, contribute to the CS process at LO in the resonance region and at NNLO in the low-energy region. They give identical contributions to the CS off the proton and the neutron. We calculate them in ChPT (or more precisely BChPT) based on the Lagrangian in \Eqref{LagrangianGND}. The delta propagator and the $\Ga_{\gamma N \Delta}$ vertex are given in \appref{chap4}{FeynmanRules}. The imaginary part of the tree-level diagrams is easily calculated, as it stems from the propagator of the $s$-channel diagram only. We rewrite the scalar part of the propagator as: 
\begin{subequations}
\bea
\lim_{\lambda \rightarrow 0^+}\,
\frac{1}{s-M_\Delta^2+i \lambda}&=&\lim_{\lambda \rightarrow 0^+}\,\frac{1}{2M}\frac{1}{\nu-\nu_\Delta+i\lambda},\\
&=&\lim_{\lambda \rightarrow 0^+}\,\frac{1}{2M}\frac{\nu-\nu_\Delta-i\lambda}{(\nu-\nu_\Delta)^2+\lambda^2},\\
&=&\frac{1}{2M}\frac{1}{\nu-\nu_\Delta}-\frac{i\pi}{2M}\delta(\nu-\nu_\Delta),
\eea
\end{subequations}
where in the last step we identified the nascent $\delta$-function:
\beq
n_\lambda(x)=\frac{\lambda}{\pi}\frac{1}{x^2+\lambda^2}, \quad \text{with} \quad \delta(x)=\lim_{\lambda \rightarrow 0^+} n_\lambda(x). \eqlab{Lorentzian}
\eeq
The threshold for delta production is at lab-frame photon energies of
\beq
\nu_\Delta=\frac{M_\Delta^2-M^2+Q^2}{2M},
\eeq
where we denote the nucleon and delta masses by $M$ and $M_\Delta$, respectively.

\noindent We decompose our results for the CS amplitudes in the following way:
\begin{subequations}
\eqlab{DeltaAmps}
\bea
T_1(\nu,Q^2)&=&T_1(0,Q^2)+T_1^{\Delta\mathrm{-pole}}(\nu,Q^2)+\widetilde T_1(\nu,Q^2)+\frac{i\,4\pi^2 \al}{M}\, f_1(\nu,Q^2),\qquad\\
T_2(\nu,Q^2)&=&T_2^{\Delta\mathrm{-pole}}(\nu,Q^2)+\widetilde T_2(\nu,Q^2)+\frac{i\,4\pi^2 \al}{\nu} \,f_2(\nu,Q^2),\\
S_1(\nu,Q^2)&=&S_1^{\Delta\mathrm{-pole}}(\nu,Q^2)+\widetilde S_1(\nu,Q^2)+\frac{i\,4\pi^2 \al}{\nu} \,g_1(\nu,Q^2),\\
S_2(\nu,Q^2)&=&S_2^{\Delta\mathrm{-pole}}(\nu,Q^2)+\widetilde S_2(\nu,Q^2)+ \frac{i\,4\pi^2 \al M}{\nu^2} \,g_2(\nu,Q^2).
\eea
\end{subequations}
The real parts of the amplitudes are given in \appref{chap4}{VVCSDeltaAmp}. Terms which are proportional to:
\beq
\frac{1}{[s-M_\Delta][u-M_\Delta]}=\frac{1}{4M^2}\frac{1}{\nu_\Delta^2-\nu^2},
\eeq
are denoted by $T_i^{\Delta\mathrm{-pole}}$ and $S_i^{\Delta\mathrm{-pole}}$, see \Eqref{Dpole}. In addition, we find the $\widetilde T_i$ and $\widetilde S_i$ terms which are free of poles in $\nu$, see \Eqref{noDpole}. They emerge as:
\beq
\frac{\nu^{n+2}}{[s-M_\Delta][u-M_\Delta]}=\frac{1}{4M^2} \left(\frac{\nu_\Delta^2 \nu^n}{\nu_\Delta^2-\nu^2}-\nu^n\right),
\eeq
where in the second term the $\Delta$-pole canceled out.

The imaginary parts of the amplitudes are related to structure functions through the optical theorem, cf.\   \Eqref{VVCSunitarity}. As we are studying CS off the nucleon, we have $Z=1$. The contribution to the nucleon structure functions from delta production then reads:
\begin{subequations}
\eqlab{structurefunc}
\begin{alignat}{3}
&f_1(\nu,Q^2)&&=\frac{1}{2M_+^2}\left[g_M^2 \vert \bq \vert ^2 (\nu +M_+)+\frac{g_E^2\, (\nu -\varDelta ) \left(M\nu -Q^2\right)^2}{M^2}+\frac{g_\mathrm{C}^2 \,Q^4 s (\nu -\varDelta )}{M^2 M_\Delta^2}\right.\\
&&&\qquad\qquad-\frac{g_M g_E\, \vert \bq \vert ^2 \left(M\nu -Q^2\right)}{M}+\frac{g_M g_\mathrm{C}\, \vert \bq \vert ^2 Q^2}{M}\nn\\
&&&\qquad\qquad+\left.\frac{2 g_E g_\mathrm{C} \,Q^2 \left(M\nu -Q^2\right) (-M_\Delta(M+\nu)+s)}{M^2 M_\Delta}\right]\delta\!\left(\nu-\nu_\Delta\right),\nn\\
&f_2(\nu,Q^2)&&=\frac{\nu Q^2}{2 M M_+^2}\left[g_M^2 (\nu +M_+)+g_E^2 (\nu -\varDelta )-\frac{g_\mathrm{C}^2 \,Q^2 (\varDelta -\nu )}{M_\Delta^2}\right.\\
&&&\qquad\qquad\quad\left.-\frac{g_M g_E \left(M\nu -Q^2\right)}{M}+\frac{g_M g_\mathrm{C} \,Q^2}{M}\right]\delta\!\left(\nu-\nu_\Delta\right),\qquad\nn\\
&g_1(\nu,Q^2)&&=-\frac{\nu}{4M_+^2}\left[g_M^2 \nu  (\nu +M_+)+\frac{g_E^2 \left(\nu  M-Q^2\right) \left(M(\nu-\varDelta)-Q^2\right)}{M^2}\right.\eqlab{g1Delta}\\
&&&\qquad \qquad\quad-\frac{g_\mathrm{C}^2 \,Q^4 (M_\Delta M-s)}{M^2 M_\Delta^2} -\frac{g_M g_E \left(M_\Delta Q^2+4 \nu \left(M\nu-Q^2\right)\right)}{M}\nn\\
&&&\qquad \qquad\quad -\frac{g_M g_\mathrm{C}\, Q^2 \left(-4 \nu  M_\Delta+M \nu -Q^2\right)}{M M_\Delta}\nn\\
&&&\qquad \qquad\quad \left.+\frac{g_E g_\mathrm{C}\, Q^2 \left(\nu  M^2+\varDelta  \left(Q^2-  s\right)\right)}{M^2 M_\Delta}\right]\delta\!\left(\nu-\nu_\Delta\right),\nn\\
&g_2(\nu,Q^2)&&=\frac{\nu^2}{4M_+^2}\left[g_M^2 (\nu +M_+)+\frac{ g_E^2 \,\varDelta\left(\nu  M-Q^2\right)}{M^2}+\frac{g_\mathrm{C}^2 \,Q^2 (\nu  M_\Delta M-\varDelta  s)}{M^2 M_\Delta^2}\right.\eqlab{g2Delta}\\
&&&\qquad \qquad\quad-\frac{g_M g_E \left(-\nu  M_\Delta+4\left(M \nu  - Q^2\right)\right)}{M}\nn\\
&&&\qquad \qquad\quad-\frac{g_E g_\mathrm{C} \left(M \vert \bq \vert ^2\left(M-2 M_\Delta \right)+\nu  M_+ Q^2-Q^4+Q^2 s+\varDelta  \nu  s\right)}{M^2 M_\Delta}\nn\\
&&&\qquad \qquad\quad \left.+\frac{g_M g_\mathrm{C} \left(4 M_\Delta Q^2+\nu \left(M \nu-  Q^2\right)\right)}{M M_\Delta}\right]\delta\!\left(\nu-\nu_\Delta\right).\eqlab{g2Delta}\nn
\end{alignat}
\end{subequations}
Alternatively, we can convert to the Bjorken variable and replace the $\delta$-function in the following way:
\beq
\delta\! \left(\nu-\nu_\Delta\right)=\frac{Q^2}{2M\nu_\Delta^2}\,\delta\!\left(x-\nicefrac{Q^2}{2M\nu_\Delta}\right). \eqlab{dfunc}
\eeq
For brevity, we introduced the following shorthands:
\begin{subequations}
\eqlab{shorthands}
\bea
\varDelta&=&M_\Delta - M,\\
M_+&=&M_\Delta + M,\\
\vert \bq \vert&=&\sqrt{\nu^2+Q^2},\\
Q_\pm&=&\sqrt{(M_\Delta\pm M)^2+Q^2}, \\
\omega_\pm&=&(M_\Delta^2-M^2\pm Q^2)/2M_\Delta.
\eea
\end{subequations}
The $\Delta$-production threshold in the lab frame is related to the $\Delta$-production threshold in the CM frame (or delta rest-frame) as: $\nu_\Delta=(\nicefrac{M_\Delta}{M})\, \omega_+$. In \appref{chap4}{HelicityCS}, we derive the $\Delta$-production helicity cross sections and confirm our results for the nucleon structure functions.

We verified that exploiting the DRs in Eqs.~\eref{T2DR}-\eref{S2DR} and the once-subtracted DR for the amplitude $T_1$, cf.\ \Eqref{T1Subtr}, the structure functions \eref{structurefunc} reproduce the $\Delta$-pole part \eref{Dpole} of the VVCS amplitudes. It is important to emphasize that the nucleon structure functions in \Eqref{structurefunc} can not reproduce the non-pole contributions to the VVCS amplitudes, cf.\  \Eqref{noDpole} \cite{BlundenYang}. To describe the non-pole contributions in a dispersive framework, we define the following structure functions:
\begin{subequations}
\eqlab{npstrucfunc}
\bea
\widetilde f_1(x,Q^2)&=&\frac{M x}{8 \pi \al} \,\widetilde T_1(Q^2)\, \delta (x),\\
\widetilde f_2(x,Q^2)&=&\frac{M x}{8 \pi \al} \,\widetilde T_2(Q^2)\, \delta (x),\\
\widetilde g_1(x,Q^2)&=&\frac{Q^2}{16 \pi \al M} \,\widetilde S_1(Q^2)\, \delta (x),
\eea
\end{subequations}
which reproduce Eqs.~\eref{T1nonpole}-\eref{S1nonpole} as plugged into the DRs, cf.\ Eqs.~\eref{T2DR}, \eref{S1DR} and \eref{T1Subtr}, respectively.

In \Eqref{S2DR}  we wrote a DR for $\nu S_2$ rather than $S_2$. This will be convenient later, because the Born contribution to the CS off the nucleon has a pole in the $S_2$ amplitude, \Eqref{S2Born}, for the subsequent limits of $Q^2\rightarrow 0$ and $\nu\rightarrow 0$. It is important to understand that the non-pole contribution of $\nu S_2$ can not be deduced from \Eqref{nonpoleS2}. We instead derive the following expression:
\bea
\widetilde{\nu S_2}(\nu,Q^2)&=&\frac{2\pi \al }{M M_+^2}\bigg[g_E^2 \,M_\Delta \varDelta\, \omega_-+\frac{g_M^2\, M Q_+^2}{2}+\frac{g_\mathrm{C}^2\,Q^2(Q^2-\varDelta^2)}{2M_\Delta}\eqlab{nuS2np}\\
&&\qquad\qquad+g_E g_M \,M_\Delta (M_\Delta \omega_+-4 M \omega_-)-g_E g_\mathrm{C}\, \varDelta (2Q^2+M \omega_+)\nn\qquad\\
&&\qquad\qquad+g_M g_\mathrm{C} \,Q^2(4M-\omega_+)\bigg]+\frac{\widetilde S_2(\nu,Q^2)}{\nu}\left[\frac{M_\Delta^2 \,\omega_+^2}{M^2}+\nu^2\right],\nn
\eea
which has not only terms proportional to $\nu^2$ but also has terms constant in $\nu$. The $\nu$-independent part of \Eqref{nuS2np} can be described by: 
\beq
\widetilde g_{2,a}(x,Q^2)=\frac{Q^2 }{16 \pi  \al M^2}\Bigg[ \widetilde{\nu S_2}\Big\vert_{\nu\rightarrow0}\Bigg] \delta (x), \eqlab{S2npg}
\eeq
as plugged into \Eqref{S2DR}. The part of \Eqref{nuS2np} proportional to $\nu^2$ can be described based on:
\beq
\widetilde g_{2,b}(x,Q^2)=\frac{Q^6 }{64 \pi \al M^4} \frac{1}{x^2}\left[\frac{\widetilde S_2(Q^2)}{\nu}\right] \delta (x),\eqlab{S2npgp2}
\eeq
and the dispersion integral on the rhs of:
\beq
\widetilde{\nu S_2}(\nu,Q^2)-\widetilde{\nu S_2}(0,Q^2)=\frac{64 \pi \al M^4\nu^2}{Q^6}  \int_{0}^{x_0} 
\!\dd x\,
\frac{x^2 \,\tilde g_{2,b} (x, Q^2)}{1 - x^2 (\nu/\nu_{\mathrm{el}})^2}.
\eeq

\subsection[Jones-Scadron Form Factors and the Large-$N_c$ Limit]{Jones-Scadron Form Factors and the Large-$\boldsymbol{N_c}$ Limit}\seclab{Jones}

In \secref{chap4}{3.1CSDelta}, we calculated the $\Delta$-exchange contribution to forward VVCS within the framework of BChPT. We now want to make a connection to the pion electroproduction\footnote{In the OPE approximation, pion electroproduction is equivalent to pion photoproduction with virtual photons.} experiments, $\gamma^*N \rightarrow \pi N$, at the resonance position, $s=M_\Delta^2$, see \Figref{PionEP}. The $\gamma^* N \leftrightarrow \Delta$ transition, depicted by the blob in \Figref{PionEP}, corresponds to a e.m.\ decay of the lowest nucleon resonance, $\Delta(1232)$ with spin and parity $3/2^+$, into the nucleon ground state, $N$ with spin and parity $1/2^+$. The selection rules for this transition require, firstly, a conservation of parity, and secondly, compliance with the triangle inequality:
\beq
\vert J_i-J_f\vert \leq \ell \leq J_i+J_f, 
\eeq
imposed on the orbital momentum $\ell$ of the photon relative to the target nucleon. Since the $E$-even and $M$-odd transitions are parity conserving:
\begin{subequations}
\begin{alignat}{3}
\Delta P&=(-1)^\ell& \quad E\ell\text{ transition},\\
\Delta P&=(-1)^{\ell+1}& \quad M\ell\text{ transition},
\end{alignat}
\end{subequations}
the e.m.\ $\Delta(1232)$ decay can be described by magnetic dipole ($M1$), electric quadrupole ($E2$) and Coulomb quadrupole ($C2$) transitions, where the Coulomb transition is a electric transition with longitudinal photons. Naturally, one expects higher multipolarities to be suppressed. A rule of thumb also says that the probabilities of $M \ell$ and $E(\ell+1)$ transitions are roughly equal. However, as we will see later, the nucleon-to-delta transition is dominantly of magnetic dipole type.

\begin{figure}[t]
\centering
       \includegraphics[width=8.7cm]{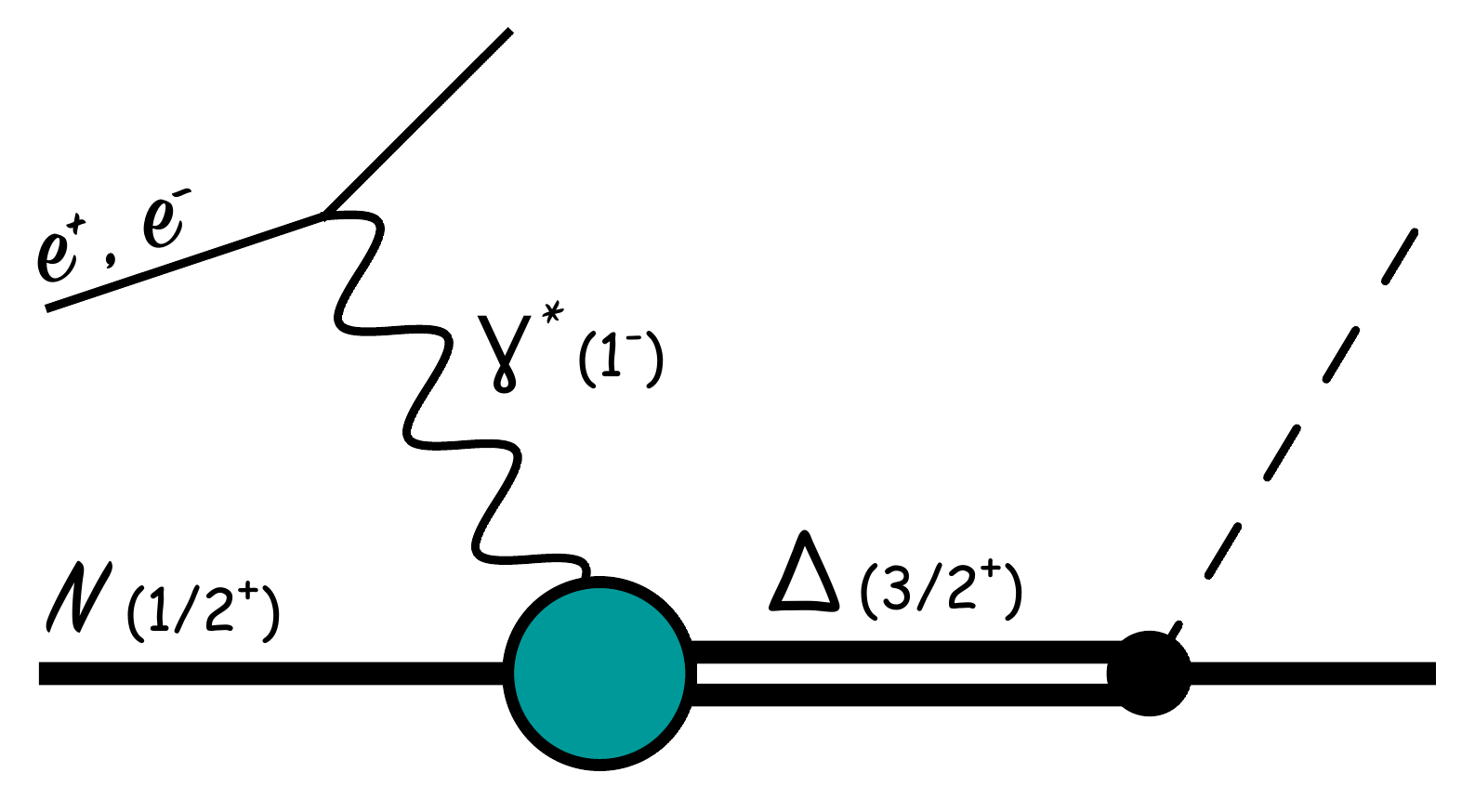}
\caption{Pion electroproduction with intermediate $\Delta$-exchange.\label{fig:PionEP}}
\end{figure}

\noindent Let us now switch to the notation of pion production. The final state has a pion with spin and parity $0^-$. It can be described by an orbital angular momentum $l$ of the pion relative to the recoiling nucleon. Including the intrinsic parity of the pion, the parity of the final $\pi N$-state is given by $(-1)^{l+1}$. The pion electroproduction can then be expanded in the $M_{1+}^{(3/2)}$,  $E_{1+}^{(3/2)}$ and  $S_{1+}^{(3/2)}$ multipoles, cf.\ Ref.~\cite[Section 2.3 and Table 3]{Drechsel:1992pn}. Here, the superscript denotes the isospin. The subscript denotes the $l=1$ partial wave and the ``$+$'' indicates that spin and orbital angular momentum of the nucleon are parallel.

The magnetic ($g_M$), electric ($g_E$) and Coulomb ($g_\mathrm{C}$) couplings are per definition related to the magnetic $(G_M^*)$, electric $(G_E^*)$ and Coulomb $(G_\mathrm{C}^*)$ nucleon-to-delta transition FFs of Jones and Scadron \cite{Jones:1972ky}:
\begin{subequations}
\eqlab{48}
\bea
g_M&=&G_M^*(Q^2)-G_E^*(Q^2),\\
g_E&=&-\frac{Q_+^2}{\omega_-^2+Q^2}\left[\frac{\omega_-}{M_\Delta}G_E^*(Q^2)+\frac{Q^2}{2M_\Delta^2}G_\mathrm{C}^*(Q^2)\right],\\
g_\mathrm{C}&=&\frac{Q_+^2}{\omega_-^2+Q^2}\left[G_E^*(Q^2)-\frac{\omega_-}{2M_\Delta}G_\mathrm{C}^*(Q^2)\right].
\eea
\end{subequations}
Another systematic approach to approximate the strong interaction is the $1/N_c$ expansion \cite{tHooft:1973alw,Witten:1979kh}. This perturbative expansion of QCD has the advantage that it is based on a parameter which is small at all energy scales. On the contrary, ChPT is restricted to low energies. As a direct consequence of QCD, the baryons are static in the large-$N_c$ limit,
\beq
M=\mathcal{O}(N_c)\quad \text{and} \quad M_{\Delta}=\mathcal{O}(N_c),
\eeq
and the baryon sector has an exact contracted $\text{SU}(2N_f)$ spin-flavor symmetry, with $N_f$ being the number of light quark flavors.\footnote{As pointed out in Ref.~\cite{Cohen:2002sd}, the excitation energy of the delta is vanishing in the large-$N_c$ limit:
\beq
\varDelta=\mathcal{O}(N_c^{-1}),
\eeq
hence, the $1/N_c$ expansion triggers an unphysical region where $\varDelta \ll m_\pi$. Therefore, the chiral limit ($m_\pi\rightarrow 0$) and the large-$N_c$ limit do not commute and one expects the former to dominate.} In the following, we will present an alternative approach to the $\Delta$-exchange, where we relate our prediction to empirical observables by means of large-$N_c$ relations.  

   \begin{figure}[htb]
\centering
\includegraphics[scale=0.81]{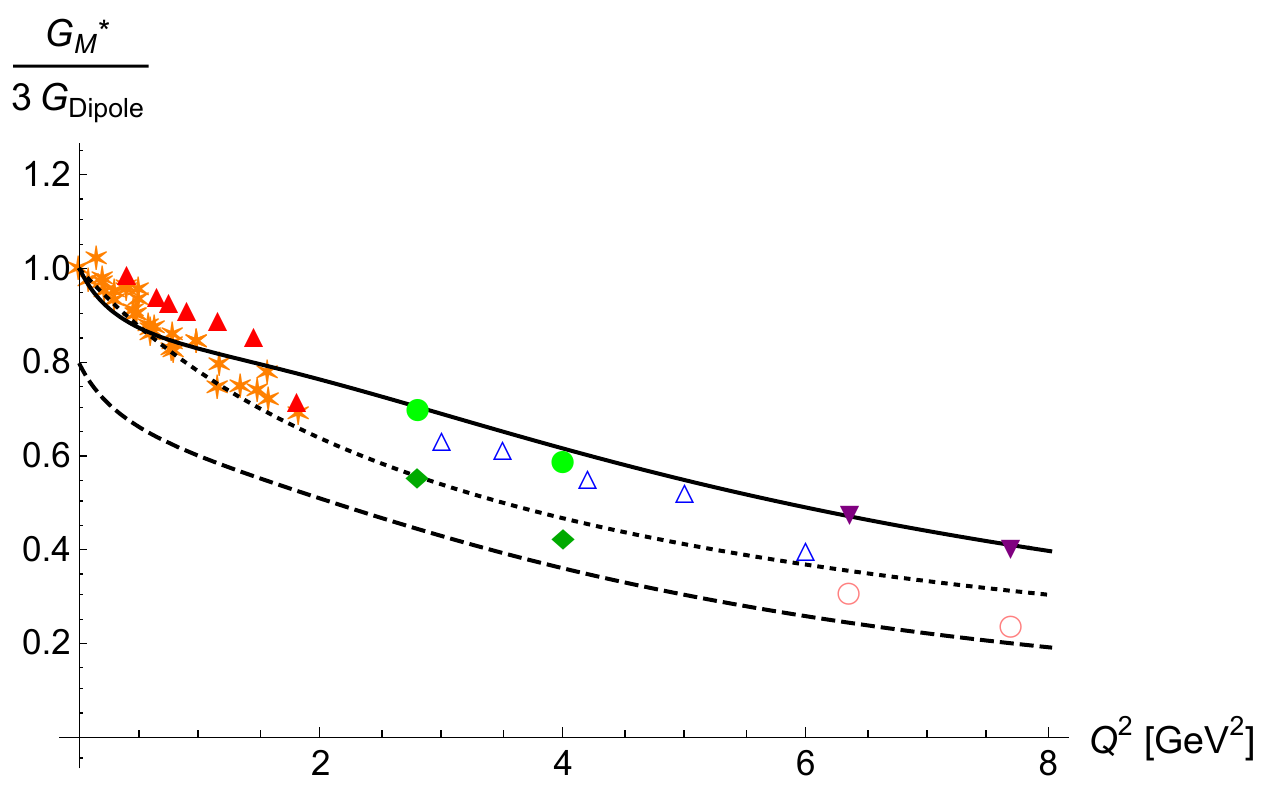}\hspace{1cm}
\caption{$G_M^*(Q^2)$ as a function of $Q^2$, normalized to the dipole form factor $G_D(Q^2)$ (multiplied by a factor $3$). Equation \eref{GM*} (normalized to $G_M^*(0)=3$) is evaluated for the dipole form factor (dotted curve) and the form factor parametrizations of \citet{Bradford:2006yz} (solid curve). The dashed curve shows \Eqref{GM*new} with the form factor parametrizations of Ref.~\cite{Bradford:2006yz}. The data are from JLab: CLAS with red pyramids \cite{Joo:2001tw}, blue open pyramids \cite{Ungaro:2006df} and purple triangles \cite{Aznauryan:2009mx}; Hall C with pink light green circles \cite{Frolov:1998pw} and pink open circles \cite{Villano:2009sn}.}
\figlab{gMdata}
\end{figure}

\noindent The nucleon-to-delta transition FFs can be connected to the e.m.\ nucleon properties via large-$N_c$ relations:\footnote{Note that \Eqref{GE*0} follows from a relation between the $N \rightarrow \Delta$ quadrupole moment and the neutron charge radius: $Q_{p\rightarrow \Delta^+}=\nicefrac{\langle r^2\rangle_{En} }{\sqrt{2}}$.}
\begin{subequations}
\bea
G_M^*(0)&=&\frac{\kappa_V}{\sqrt{2}} \quad \mbox{\cite{Jenkins:1994md}}, \\
G_E^*(0)&=&\frac{M^2-M_\Delta^2}{12\sqrt{2}}\left(\frac{M}{M_\Delta}\right)^{3/2}\langle r^2\rangle_{En} \quad\mbox{\cite{Buchmann:2002mm}},\eqlab{GE*0}\\
G_\mathrm{C}^*(0)&=&\frac{4M_\Delta^2}{M_\Delta^2-M^2}\,G_E^*(0)\quad \mbox{\cite{Pascalutsa:2007wz}},
\eea
\end{subequations}
where we introduced the isovector anomalous magnetic moment of the nucleon: $\kappa_V=\kappa_p-\kappa_n\simeq 3.7$ \cite{Olive:2016xmw}. An extension of these relations to finite momentum transfer is modeled in Ref.~\cite{Pascalutsa:2007wz}:
\begin{subequations}
\eqlab{PascalutsaLNC}
\bea
\eqlab{GM*old}
G_M^*(Q^2)&=&\frac{1}{\sqrt{2}}\left[F_{2p}(Q^2)-F_{2n}(Q^2)\right],\\
G_E^*(Q^2)&=&\left(\frac{M}{M_\Delta}\right)^{3/2}\frac{\varDelta M_+}{2\sqrt{2}\,Q^2}G_{En}(Q^2),\eqlab{GM*old2}\\
G_\mathrm{C}^*(Q^2)&=&\frac{4M_\Delta^2}{\varDelta M_+}G_E^*(Q^2),\eqlab{GM*old3}
\eea
\end{subequations}
where $F_{2p}$ and $F_{2n}$ are the Pauli FFs of the proton and neutron, respectively, and $G_{En}$ is the electric Sachs FF of the neutron. 
We furthermore use the fact that $F_{2p}(Q^2)=-F_{2n}(Q^2)$ in the  large-$N_c$ limit\footnote{Ref.~\cite{Luty:1994ub} performs a simultaneous expansion in $1/N_c$ and $m_s$, the mass of the strange quark, and finds: $3\mu_n+2\mu_p=0$.} and modify the relation in \Eqref{GM*old}:
\beq
\eqlab{GM*}
G_M^*(Q^2)=\sqrt{2} \,C_M^* F_{2p}(Q^2),
\eeq
with $C_M^*=\frac{3.02}{\sqrt{2}\,\kappa_p}$, to reproduce the empirical value of $G_M^*(0)\simeq3.02$ \cite{Tiator:2003xr}.

Starting from the ChPT structure functions, cf.\ \Eqref{structurefunc}:
\beq
F_{\mathrm{ChPT}}(\nu,Q^2)=c_\mathrm{MM}\, g_M^2+c_\mathrm{EE} \,g_E^2+c_\mathrm{CC}\, g_\mathrm{C}^2+c_\mathrm{ME}\, g_M g_E+c_\mathrm{MC} \,g_M g_\mathrm{C}
+c_\mathrm{EC} \,g_E g_\mathrm{C},\eqlab{ChPTSF}
\eeq
with $F$ being either $f_1$, $f_2$, $g_1$ or $g_2$, and $c$'s functions of $\nu$ and $Q^2$, we set up  a model for the nucleon structure functions based on the Jones-Scadron transition FFs:
\bea
F_{\mathrm{JS}}(\nu,Q^2)&=&G_M^{*2}\left\{C_\mathrm{MM}+C_\mathrm{ME}\, R_\mathrm{EM}+C_\mathrm{MC} \,R_\mathrm{SM}+C_\mathrm{EE} \,R_\mathrm{EM}^2\right.\\
&&\qquad\quad\left.+C_\mathrm{EC} \,R_\mathrm{EM}R_\mathrm{SM}+C_\mathrm{CC}\, R_\mathrm{SM}^2
\right\}.\nn
\eea
Here, the $Q^2$ dependence is in the magnetic Jones-Scadron FF and the multipole ratios $R_\mathrm{EM}=\mathrm{E2}/\mathrm{M1}$ and $R_\mathrm{SM}=\mathrm{C2}/\mathrm{M1}$ measured in pion electroproduction. The $C$'s are functions of $\nu$ and $Q^2$, they are related to the coefficients in \Eqref{ChPTSF} through \Eqref{48} in the following way:
\begin{subequations}
\eqlab{capC}
\bea
C_\mathrm{MM}&=&c_\mathrm{MM},\\
C_\mathrm{ME}&=&2\left[c_\mathrm{MM}+\frac{2M_\Delta}{Q_-^2}\left(\omega_-\,c_\mathrm{ME}-M_\Delta c_\mathrm{MC}\right)\right],\\
C_\mathrm{MC}&=&\frac{8M_\Delta^2}{Q_+Q_-^3}\left[Q^2c_\mathrm{ME}+M_\Delta \omega_-\,c_\mathrm{MC}\right],\\
C_\mathrm{EE}&=&c_\mathrm{MM}+\frac{4M_\Delta}{Q_-^2}\Big[\omega_-\,c_\mathrm{ME}-M_\Delta c_\mathrm{MC}\\
&&+\frac{4M_\Delta}{Q_-^2}\left(\omega_-^2c_\mathrm{EE}-M_\Delta \omega_-\, c_\mathrm{EC}+M_\Delta^2 c_\mathrm{CC}\right)\Big]\nn,\\
C_\mathrm{EC}&=&\frac{8M_\Delta^2}{Q_+Q_-^3}\Big\{Q^2c_\mathrm{ME}+M_\Delta \omega_-\,c_\mathrm{MC}\\
&&+\frac{4M_\Delta}{Q_-^2}\left[2Q^2\omega_-\, c_\mathrm{EE}+M_\Delta (\omega_-^2-Q^2)c_\mathrm{EC}-2M_\Delta^2 \omega_-\, c_\mathrm{CC}\right]\!\Big\}\nn,\nn\\
C_\mathrm{CC}&=&\frac{64M_\Delta^4}{Q_+^2Q_-^6}\left[Q^4 c_\mathrm{EE}+M_\Delta \omega_-\left(Q^2 c_\mathrm{EC}+M_\Delta \omega_-\,c_\mathrm{CC}\right)\right].
\eea
\end{subequations}

The multipole ratios are naturally small and go as $\mathcal{O}(1/N_c^2)$ \cite{Jenkins:2002rj}. They can be written in terms of the Jones-Scadron FFs as \cite{Pascalutsa:2007wz}:
\begin{subequations}
\eqlab{multipoles}
\bea
R_\mathrm{EM}(Q^2)&=&-\frac{G_E^*(Q^2)}{G_M^*(Q^2)},\eqlab{REM}\\
R_\mathrm{SM}(Q^2)&=&-\frac{Q_+Q_-}{4M_\Delta^2}\frac{G_\mathrm{C}^*(Q^2)}{G_M^*(Q^2)}.\eqlab{RSM}
\eea
\end{subequations}
For $Q^2=0$, these large-$N_C$ expressions coincide: $R_\mathrm{EM}=R_\mathrm{SM}$. In \Figref{ratios}, we show the multipole ratios, cf.\ \Eqref{multipoles}, as described by the large-$N_c$ relations in Eqs.~\eref{GM*old2}, \eref{GM*old3} and \eref{GM*}, and compare to experimental data. For the e.m.\ FFs we make the same choice as Ref.~\cite{Pascalutsa:2007wz} and apply the parametrizations of \citet{Bradford:2006yz}, cf.\ \Eqref{functionalForm} and Table \ref{param}. We present two curves for different values of $C_M^*$. The dashed curve is chosen to reproduce the empirical value of $G_M^*(0)$ and the solid curve corresponds to Ref.~\cite[Fig.~1]{Pascalutsa:2007wz}. Both curves describe $R_\mathrm{EM}$ equally good. For $R_\mathrm{SM}$ we observe a slight deviation of the dashed curve as compared to the data. However, since the nucleon-to-delta transition is anyway dominated by the magnetic dipole, cf.\ \Figref{I12}, we tend to neglect the $R_\mathrm{SM}$ multipole ratio and replace $R_\mathrm{EM}(Q^2)$ by its static value $R_\mathrm{EM}(0)$, cf.\ discussion below \Eqref{approxg1} and \secref{5HFS}{Depole}. Therefore, the description of $R_\mathrm{SM}$ is still acceptable.
 
      \begin{figure}[h!] 
  \centering 
       \includegraphics[width=14cm]{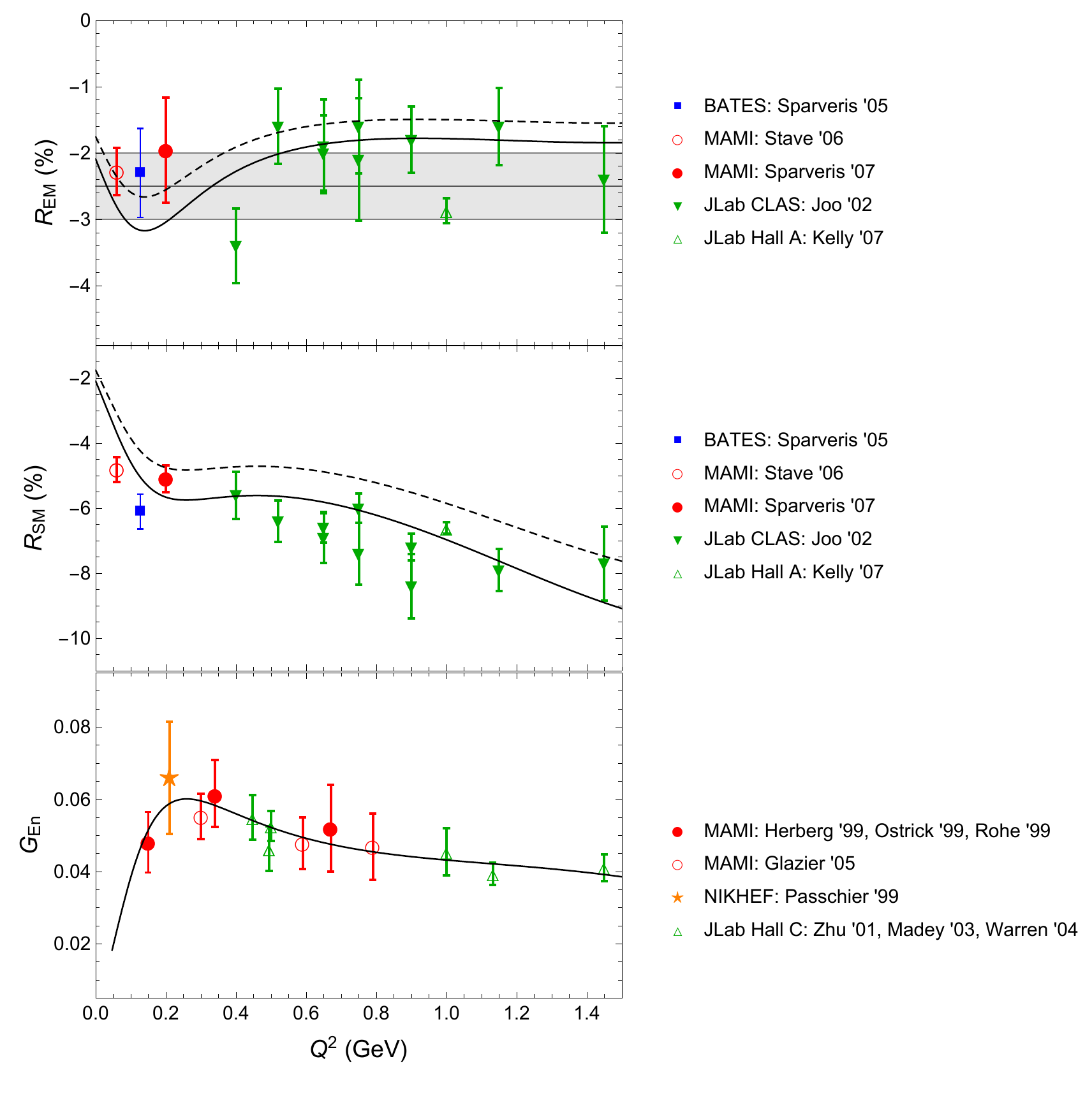}
       \caption{Multipole ratios of the nucleon-to-delta transition: $R_\mathrm{EM}$ (upper panel) and $R_\mathrm{SM}$ (middle panel), cf.\ \Eqref{multipoles}. The data are taken from Refs.~\cite{Sparveris:2004jn,Stave:2006ea,Sparveris:2006uk,Joo:2001tw,Kelly:2005jj,Kelly:2005jy,Herberg:1999ud,Ostrick:1999xa,Rohe:1999sh,Glazier:2004ny,Passchier:1999cj,Zhu:2001md,Madey:2003av,Warren:2003ma}. The gray band corresponds to the experimental value for the $R_\mathrm{EM}$ multipole ratio at the real-photon point: $R_\mathrm{EM}(0)=-2.5\pm0.5 \%$ \cite{Yao:2006px}. The curves are based on Eqs.~\eref{GM*old2}, \eref{GM*old3} and \eref{GM*} with $C_M^*=1$ (solid) and $C_M^*=\frac{3.02}{\sqrt{2}\,\kappa_p}$ (dashed).  The neutron electric Sachs FF $G_{En}$ (lower panel) is taken from Ref.~\cite{Bradford:2006yz}.}
              \label{fig:ratios}
\end{figure}

Along the same line, see \secref{5HFS}{Depole}, it will later be useful to establish:
\begin{subequations}
\eqlab{GM*RatioF2}
\bea
G_M^*(Q^2)&=&\frac{\sqrt{2}\,M_+}{Q_+}\frac{F_{2p}(Q^2)}{\sqrt{1-6\,R_\mathrm{EM}(0)}},\eqlab{GM*new}\\
G_E^*(Q^2)&=&\left(\frac{M}{M_\Delta}\right)^{3/2}\frac{\varDelta M_+^2}{2\sqrt{2}\,Q^2Q_+}\frac{G_{En}(Q^2)}{\sqrt{1-6\,R_\mathrm{EM}(0)}},\\
G_\mathrm{C}^*(Q^2)&=&\frac{4M_\Delta^2}{M_\Delta^2-M^2}G_E^*(Q^2),
\eea
\end{subequations}
without changing the multipole ratios.\footnote{Strictly speaking, the multipole ratios depend on our choice of $C_M*$, cf.\ \Eqref{GM*}. For $C_M^*=\frac{3.02}{\sqrt{2}\,\kappa_p}$ we have $R_\mathrm{EM}(0)=R_\mathrm{SM}(0)=-0.0176$,
while $C_M^*=1$ gives $R_\mathrm{EM}(0)=R_\mathrm{SM}(0)=-0.0209$.} In \Figref{gMdata}, we show the magnetic Jones-Scadron FF.
We refrain from showing the experimental error bars, as they are very small for most of the References. We compare \Eqref{GM*} for the dipole FF and the FF fit from  Ref.~\cite{Bradford:2006yz}. Both FFs give a good description of the spread of data. Furthermore, we utilize the set of nucleon FFs presented in Ref.~\cite{Bradford:2006yz} to plot \Eqref{GM*new}. The resulting curve has an offset towards lower values of $G_M^*$ over the whole $Q^2$ range. This is due to the fact that we no longer fix the static value to the experimental $G_M^*(0)\simeq3.02$ \cite{Tiator:2003xr}. A motivation for \Eqref{GM*new} shall be postponed to Sections \ref{chap:chap4}.\ref{sec:DeltaPol} and \ref{chap:5HFS}.\ref{sec:Depole}.
 
\subsection[Contribution of the $\Delta$-Exchange to Nucleon Polarizabilities]{Contribution of the $\boldsymbol{\Delta}$-Exchange to Nucleon Polarizabilities} \seclab{DeltaPol}

We will now evaluate the contribution of the $\Delta$-exchange to the GPs introduced in \secref{chap4}{SR}. We first present predictions derived in the pure ChPT framework. We then give expressions for the generalized GDH integrals in terms of the magnetic Jones-Scadron FF and the multipole ratios of pion electroproduction, as derived from the large-$N_c$ relations presented in the above \secref{chap4}{Jones}. In \Figref{PolarizabilitiesGrid}, we show the pure BChPT prediction for the nucleon GPs of VVCS along with the data driven description of the $\Delta$-exchange contribution to GPs through nucleon FFs. 

Let us start by presenting our BChPT results for the $Q^2$ dependence of the nucleon polarizabilities up to $\mathcal{O}(Q^2)$, see Eqs.~\eref{alphabetaQ2}-\eref{d2momQ2}. Following Ref.~\cite{Pascalutsa:2005vq}, we introduce a dipole on the magnetic coupling to take into account the vector-meson diagram shown in Ref.~\cite[Fig. 2(d)]{Pascalutsa:2005vq}:
 \beq
g_M\rightarrow \frac{g_M}{(1+Q^2/\Lambda^2)^2}, \eqlab{gMDipole}
\eeq
Inclusion of this FF is important to reproduce pion electroproduction data, where the usual choice would be $\Lambda=\sqrt{0.71 \,\mathrm{GeV}^2}$. In \appref{chap4}{Q2regpol}, we will compare to  results with a general momentum-cutoff in the $\Ga_{\gamma N \Delta}$ vertex function.

\paragraph*{Electric and Magnetic Dipole Polarizabilities}\mbox{}\\[0.2cm]
\begin{subequations}
\eqlab{alphabetaQ2}
\bea
\al_{E1}\,(Q^2=0)&=&-\frac{e^2 g_E^2}{2 \pi  M_+^3},\\
\be_{M1}\,(Q^2=0)&=&\frac{e^2 g_M^2}{2 \pi  \varDelta  M_+^2},\\
\frac{\dd \left[\alpha_{E1}(Q^2)+\beta_{M1}(Q^2)\right]}{\dd Q^2}\Bigg\vert_{\substack{Q^2=0\\ \Lambda \rightarrow \infty}}&=&-\frac{e^2}{\pi M_+^2}\left(\frac{g_M^2}{\varDelta^2}\left[\frac{1}{M_+}-\frac{1}{2\varDelta}\right]\right.\\
&&+\frac{g_M g_E}{M }\left[\frac{1}{4\varDelta^2}-\frac{1}{\varDelta M_+}+\frac{1}{4M_+^2}\right]\nn\\
&&\left.-\frac{g_E^2}{4M M_+}\left[\frac{1}{\varDelta}-\frac{5}{M_+}\right]-\frac{ g_M g_C}{2  \varDelta  M M_+}+\frac{g_E g_C}{M M_+^2}\right).\nn
\eea
\end{subequations}

\noindent With running $g_M$ coupling:
\bea
\frac{\dd \left[\alpha_{E1}(Q^2)+\beta_{M1}(Q^2)\right]}{\dd Q^2}\Bigg\vert_{Q^2=0}&=&\frac{\dd \left(\alpha_{E1}(Q^2)+\beta_{M1}(Q^2)\right)}{\dd Q^2}\Bigg\vert_{\substack{Q^2=0\\ \Lambda \rightarrow \infty}}-\frac{2e^2}{\pi \varDelta M_+^2}\frac{g_M^2}{\Lambda^2}.\qquad
\eea

\paragraph*{Longitudinal Polarizability}\mbox{}\\[0.2cm]
\begin{subequations}
\bea
\al_L\,(Q^2=0)&=&\frac{e^2 M_\Delta^2}{\pi M_+^3}\left(\frac{ g_E^2}{\varDelta  M M_+^2}-\frac{g_C^2}{2 M_\Delta^4}+\frac{ g_E g_C}{M M_\Delta^2 M_+}\right),\\
\frac{\dd\, \al_L(Q^2)}{\dd Q^2}\Bigg\vert_{\substack{Q^2=0\\ \Lambda \rightarrow \infty}}&=&\frac{e^2 M_\Delta^3}{\pi \varDelta M_+^4}\left(\frac{2g_E^2}{\varDelta^2 M_+^2}\left[\frac{2}{M_\Delta}-\frac{1}{M}\right]-\frac{g_C^2}{ M_\Delta^4}\left[\frac{1}{M}-\frac{3}{2M_\Delta}\right]\right.\\
&&\left.+\frac{g_E g_C}{\varDelta M_\Delta^2 M_+}\left[\frac{5}{M_\Delta}-\frac{3}{M}\right]\right).\nn
\eea
\end{subequations}
\paragraph*{Forward Spin Polarizability}\mbox{}\\[0.2cm]
\begin{subequations}
\bea
\gamma_0(Q^2=0)&=&-\frac{e^2}{4\pi M_+^2}\left(\frac{g_M^2}{\varDelta^2}+\frac{g_E^2}{M_+^2}-\frac{4g_M g_E}{M_+\varDelta}\right),\\
\frac{\dd \gamma_0(Q^2)}{\dd Q^2}\Bigg\vert_{\substack{Q^2=0\\ \Lambda \rightarrow \infty}}&=&-\frac{e^2}{\pi M_+^2 \varDelta}\left(\frac{g_M^2}{\varDelta}\left[\frac{1}{4\varDelta^2}-\frac{1}{\varDelta M_+}+\frac{1}{2 M_+^2}\right]\right.\\
&& \left.+\frac{g_E^2}{2 M_+^2}\left[\frac{1}{2\varDelta}-\frac{3}{M_+}\right]-\frac{g_M g_E}{M_+}\left[\frac{1}{\varDelta^2}-\frac{5}{\varDelta M_+}+\frac{1}{M_+^2}\right]+\frac{2 g_M g_C}{\varDelta M_+^2}-\frac{g_E g_C}{M_+^3}\right).\nn
\eea
\end{subequations}
\noindent With running $g_M$ coupling:
\beq
\frac{\dd \gamma_0(Q^2)}{\dd Q^2}\Bigg\vert_{Q^2=0}=\frac{\dd \gamma_0(Q^2)}{\dd Q^2} \Bigg\vert_{\substack{Q^2=0\\ \Lambda \rightarrow \infty}}+\frac{e^2}{\pi M_+^2 \varDelta}\frac{1}{\Lambda^2}\left(\frac{g_M^2}{ \varDelta}-\frac{2g_M g_E}{ M_+}\right).
\eeq

\paragraph*{Longitudinal-Transverse Polarizability}\mbox{}\\[0.2cm]
\begin{subequations}
\bea
\delta_{LT}(Q^2=0)&=&\frac{e^2 M_\Delta}{4\pi M_+^3}\left(\frac{g_E^2 }{M M_+}+\frac{g_M g_E }{\varDelta M}-\frac{g_E g_C}{M_\Delta^2}\right),\\
\frac{\dd\, \delta_{LT}(Q^2)}{\dd Q^2}\Bigg\vert_{\substack{Q^2=0\\ \Lambda \rightarrow \infty}}&=&\frac{e^2  M_\Delta \varDelta}{4
\pi M M_+^2}\left(\frac{g_E^2 }{\varDelta^2 M_+^2}\left[\frac{1}{\varDelta}-\frac{4}{M_+}\right]-\frac{g_C^2}{\varDelta M_\Delta^2 M_+^2}\right.\\
&&+\frac{g_M g_E }{\varDelta^2 M_+}\left[\frac{1}{\varDelta^2}-\frac{3}{\varDelta M_+}+\frac{1}{M_+^2}\right]\nn\\
&&\left.+\frac{g_M g_C}{\varDelta M_\Delta^2 }\left[\frac{1}{2\varDelta^2}-\frac{2}{\varDelta M_+}+\frac{1}{2 M_+^2}\right]-\frac{g_E g_C}{2M_\Delta^2 M_+^2}\left[\frac{7}{\varDelta}+\frac{1}{M_+}\right]\right).\nn
\eea
\end{subequations}
\noindent With running $g_M$ coupling:
\beq
\frac{\dd\, \delta_{LT}(Q^2)}{\dd Q^2}\Bigg\vert_{Q^2=0}=\frac{\dd\, \delta_{LT}(Q^2)}{\dd Q^2} \Bigg\vert_{\substack{Q^2=0\\ \Lambda \rightarrow \infty}}-\frac{e^2 M_\Delta }{2 
\pi\varDelta M M_+^3}\frac{g_M g_E }{ \Lambda^2}.
\eeq
\paragraph*{Generalized GDH Integral $\boldsymbol{I_A(Q^2)}$} \mbox{}\\[0.2cm]
\begin{subequations}
\bea
I_A(Q^2=0)&=&0\eqlab{IAstatic},\\
\frac{\dd\, I_A(Q^2)}{\dd Q^2}\Bigg\vert_{\substack{Q^2=0\\ \Lambda \rightarrow \infty}}&=&-\frac{M^2}{M_+^2}\left(\frac{g_M^2}{2 \varDelta ^2}+\frac{g_E^2}{M M_+}-\frac{2 g_M g_E}{\varDelta  M_+}-\frac{g_E g_C}{M_\Delta M_+}\right).
\eea
\end{subequations}

\paragraph*{Zeroth Moment of $\boldsymbol{g_1}$} \mbox{}\\[0.2cm]
\begin{subequations}
\bea
I_1(Q^2=0)&=&0\eqlab{I1static},\\
\frac{\dd\, I_1(Q^2)}{\dd Q^2}\Bigg\vert_{\substack{Q^2=0\\ \Lambda \rightarrow \infty}}&=&-\frac{M_\Delta M^2}{2M_+^3}\left(\frac{g_E^2}{M M_\Delta}-\frac{g_M g_E}{ \varDelta  M}
-\frac{g_E g_C}{M_\Delta^2 }\right).
\eea
\end{subequations}
\paragraph*{Second Moment of the Higher Twist Part of $\boldsymbol{g_2}$}\mbox{}\\[0.2cm]
\begin{subequations}
\eqlab{d2momQ2}
\bea
\bar d_2(Q^2=0)&=&\frac{\dd\, \bar d_2(Q^2)}{\dd Q^2}\Bigg\vert_{\substack{Q^2=0\\ \Lambda \rightarrow \infty}}=\frac{\dd^2\, \bar d_2(Q^2)}{\dd (Q^2)^2}\Bigg\vert_{\substack{Q^2=0\\ \Lambda \rightarrow \infty}}=0,\\
\frac{1}{6}\frac{\dd^3\, \bar d_2(Q^2)}{\dd( Q^2)^3}\Bigg\vert_{\substack{Q^2=0\\ \Lambda \rightarrow \infty}}&=&\frac{M_\Delta}{16M^2M_+^2}\left(\frac{g_M^2}{ \varDelta ^2 M_\Delta}+\frac{g_E^2}{M_+^2}\left[\frac{3 }{ M}+\frac{1}{M_\Delta}\right]\right.\\
&&\left.+\frac{g_M g_E}{\Delta M_+ }\left[\frac{3}{ M}-\frac{4}{M_\Delta}\right]-\frac{3 g_E g_C}{ M_\Delta^2 M_+}\right).\nn
\eea
\end{subequations}

\begin{table}[h!]
 \centering
 \caption{$\Delta$-exchange contribution to the nucleon polarizabilities from real Compton scattering.}
 \label{RCSPolTab}
 \begin{small}
\begin{tabular}{|c|c|cc|c|}
\hline
    \rowcolor[gray]{.7}
 &&\multicolumn{2}{c|}{\bf empirical values}&\\
    \rowcolor[gray]{.7}
\multirow{-2}{*}{\bf     polarizability}&\multirow{-2}{*}{\bf   $\boldsymbol{\Delta}$-exchange contr.}&{\bf proton}&{\bf neutron}&\multirow{-2}{*}{\bf \cellcolor[gray]{.7}  dimension}\\
 \hline
 $\al_{E1}+\be_{M1}$&$7.040$&$14.0(2)\mbox{\,\cite{Gryniuk:2015aa}}$&$14.40(66)\mbox{\,\cite{Babusci}}$&$10^{-4}\,\text{fm}^3$\\
   \rowcolor[gray]{.95}
  $\al_{L}$&$0.002$&$2.32\mbox{\,\cite{MAID}}$&$3.32\mbox{\,\cite{MAID}}$&$10^{-4}\,\text{fm}^5$\\
   $\ga_0$&$-2.844$&$-0.929(105)\mbox{\,\cite{Gryniuk:2016gnm}}$&$-0.005\mbox{\,\cite{MAID}}$&$10^{-4}\,\text{fm}^4$\\
      \rowcolor[gray]{.95}
    $\delta_{LT}$&$-0.160$&$1.34\mbox{\,\cite{MAID}}$&$2.03\mbox{\,\cite{MAID}}$&$10^{-4}\,\text{fm}^4$\\
\hline
\end{tabular}
\end{small}
\end{table}

\noindent In Table \ref{RCSPolTab}, we summarize our BChPT predictions for the static nucleon polarizabilities, i.e., the polarizabilities at the real photon point, obtained with a dipole FF on the $g_M$ coupling (black curves in \Figref{PolarizabilitiesGrid}). We compare our result for the $\Delta$-exchange contribution to empirical values for the absolute polarizabilities known from RCS. One observes that the $\Delta$-exchange contribution is especially important for the sum of dipole polarizabilities and the FSP. The generalized GDH integrals are not included in the Table, because their values in the real photon limit are proportional to the anomalous magnetic moments of the nucleon, cf.\ Eq.~\ref{IGDH}. Nevertheless, the $\Delta$-exchange is important for the characteristics of the $Q^2$ behavior of the GDH integrals $I_A^{(p)}$, $I_A^{(n)}$ and $I_1^{(n)}$, as one can see from \Figref{PolarizabilitiesGrid} and Ref.~\cite[Figs.~5 and 6]{Lensky:2014dda} (as well as Ref.~\cite{Alarcon2017}).

   \begin{figure}[t] 
    \centering\hspace{2cm}
       \includegraphics[width=12cm]{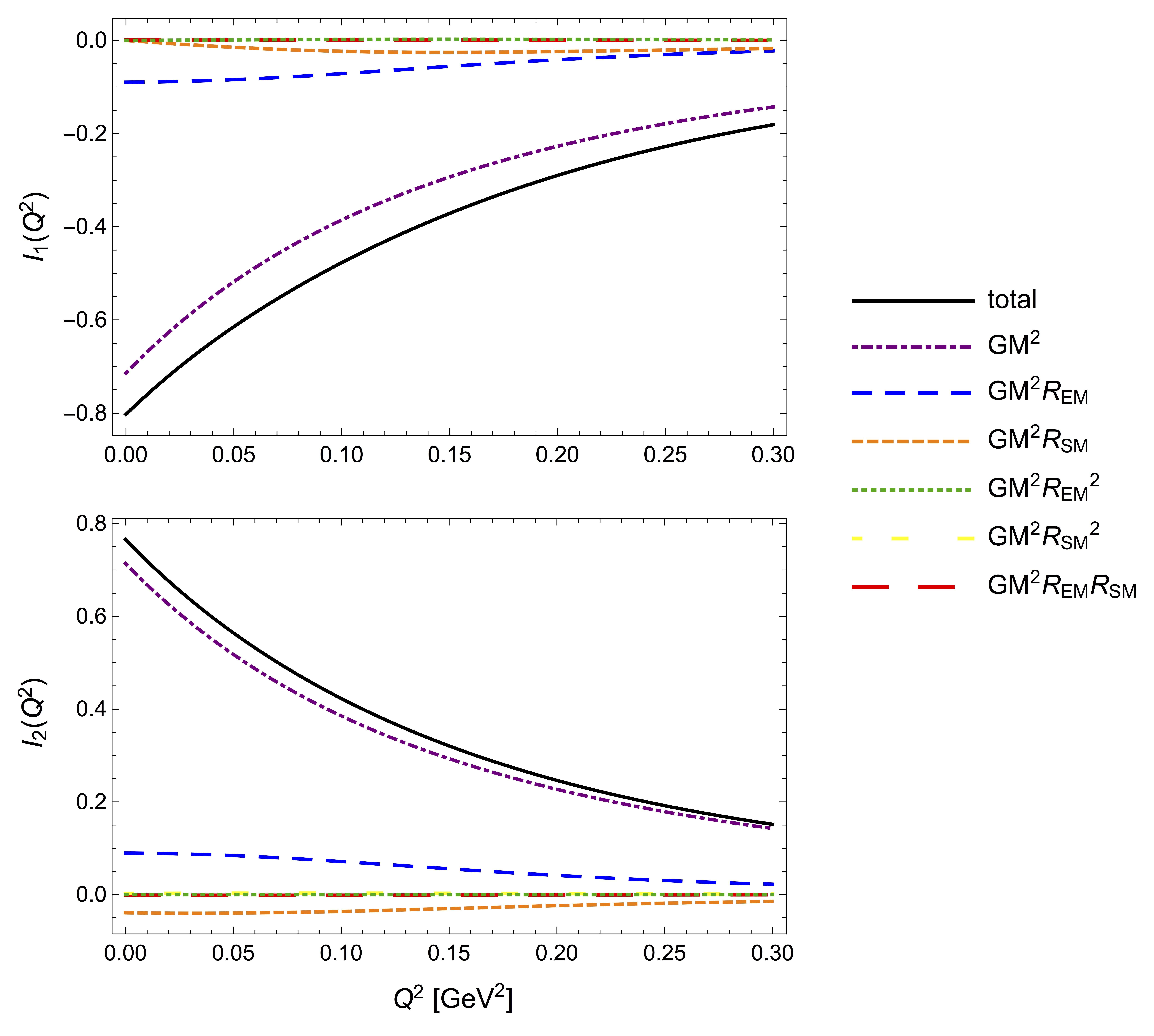}
           \caption{$\Delta$-pole contribution to the zeroth moments of the spin-dependent structure functions.}
              \label{fig:I12}
\end{figure}

Following the procedure outlined in \secref{chap4}{Jones}, cf.\ Eqs.~\eref{ChPTSF}-\eref{capC}, the polarizabilities given in Eqs.~\eref{alphabetaQ2}-\eref{d2momQ2} can be easily expressed in terms of Jones-Scadron FFs. The results are indicated in \Figref{PolarizabilitiesGrid} by the red long dash-dotted curves. Here, we also give expressions for the $\Delta$-pole contributions to the generalized GDH integrals and the BC sum rule. \Figref{I12} shows the $\Delta$-pole contributions to the zeroth moments of the spin-dependent structure functions. The plot visualizes the impact strength of the various combinations of Jones-Scadron FFs or multipole ratios, respectively. Clearly, the magnetic dipole contribution is dominating. The interference of the electric and the magnetic Jones-Scadron FFs gives the second larges contribution, while all other contributions are basically negligible. The spin-dependent structure functions, Eqs.~\eref{g1Delta} and \eref{g2Delta}, at the $\Delta$-resonance position can be approximated by:
\beq
g_1(x_\Delta,Q^2)\approx-\frac{Q^2Q_+^2}{16M^2M_+^2}G_M^{*2}(Q^2)\left[1-6R_\mathrm{EM}(0) \right]\approx-g_2(x_\Delta,Q^2), \eqlab{approxg1}
\eeq
where we neglected $R_\mathrm{SM}$ and $R_\mathrm{EM}^2$ and approximated $R_\mathrm{EM}(Q^2)\approx R_\mathrm{EM}(0)$, as motivated in \Figref{I12}. Analytic expressions for the $\Delta$-pole contributions to the generalized GDH integrals and the zeroth moment of the $g_2$ structure function are given below:
\begin{subequations}
\bea
I^{\Delta\mathrm{-pole}}_1(Q^2)&=&-\frac{G_M^{*2}(Q^2)Q_+^2}{8M_+^2}\Big\{1-6\,R_\mathrm{EM}(Q^2)-3\,R_\mathrm{EM}^2(Q^2)\eqlab{I1pole}\\
&&\hspace{2.5cm}-\frac{4MQ^2}{Q_+Q_-\omega_+}\,R_\mathrm{SM}(Q^2)\left[1+3\,R_\mathrm{EM}(Q^2)\right]\Big\},\nn\qquad\\
I^{\Delta\mathrm{-pole}}_A(Q^2)&=&-\frac{G_M^{*2}(Q^2)Q_-^2Q_+^4}{32M_\Delta^2\omega_+^2M_+^2}\Big\{1-6\,R_\mathrm{EM}(Q^2)-3\,R_\mathrm{EM}^2(Q^2)\Big\},\\
I^{\Delta\mathrm{-pole}}_2(Q^2)&=&\frac{G_M^{*2}(Q^2)Q_+^2}{8M_+^2}\Big\{1-6\,R_\mathrm{EM}(Q^2)-3\,R_\mathrm{EM}^2(Q^2)\eqlab{I2pole}\\
&&\hspace{2.25cm}+\frac{4M_\Delta^2 \omega_+}{MQ_+Q_-}\,R_\mathrm{SM}(Q^2)\left[1+3\,R_\mathrm{EM}(Q^2)\right]\Big\},\nn
\eea
\end{subequations}
where the corresponding static values ($Q^2=0$) are:
\begin{subequations}
\begin{align}
I^{\Delta\mathrm{-pole}}_1(0)&=I^{\Delta\mathrm{-pole}}_A(0)=-\frac{G_M^{*2}(0)}{8}\Big\{1-6\,R_\mathrm{EM}(0)-3\,R_\mathrm{EM}^2(0)\Big\},\\
I^{\Delta\mathrm{-pole}}_2(0)&=\frac{G_M^{*2}(0)}{8}\left\{1-6\,R_\mathrm{EM}(0)-3\,R_\mathrm{EM}^2(0)+\frac{2M_\Delta}{M}\,R_\mathrm{SM}(0)\left[1+3\,R_\mathrm{EM}(0)\right]\right\}.  
\end{align}
\end{subequations}
Here, it is worth to point out that the $\Delta$-pole and the non-pole contributions cancel exactly in the static values of all the integrals, i.e., in Eqs.~\eref{IAstatic}, \eref{I1static}, and furthermore: $I_2(Q^2=0)=0$. 

The vanishing contribution of the $\Delta$-exchange to $I_2$ is expected, as it is a pure polarizability contribution, see \secref{chap5}{BCHFS} for a proof. The same requirement is fulfilled by the polarizability contributions coming from the $\pi N$- and $\pi \Delta$-loop diagrams, cf.\ Figures \ref{fig:DeltaOldPlot} and \ref{fig:PiNLoopDiags}. In \Figref{BCPlot}, we show that the $\Delta$-pole and the elastic FF contributions to the $I_2$ integral, \Eqref{BCdef}, cancel partially. Here, the solid red line corresponds to the contribution of the approximate $g_2$ structure function, cf.\ \Eqref{approxg1}, and the light red line corresponds to \Eqref{I2pole}. In both cases, we are using the large-$N_c$ relations from \Eqref{GM*RatioF2}.

   \begin{figure}[t] 
    \centering       \includegraphics[width=10.5cm]{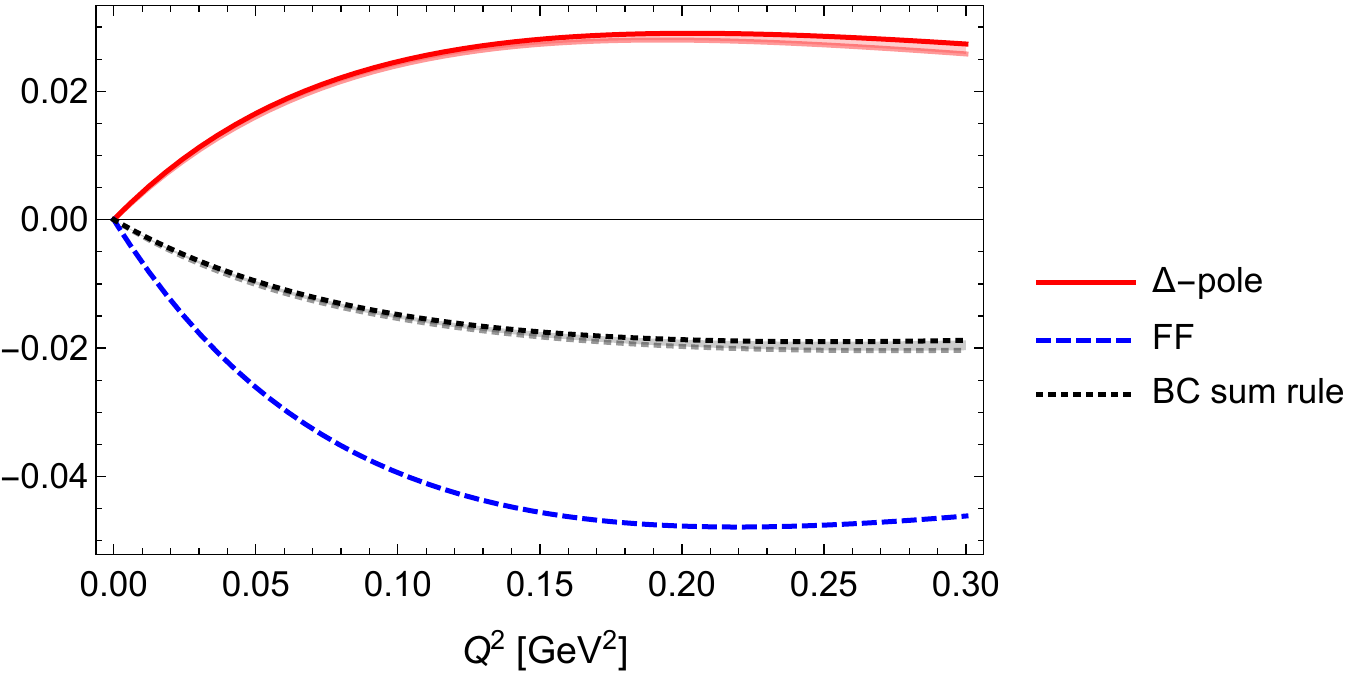}
           \caption{Different contributions to the Burkhardt-Cottingham sum rule: $\int_0^1 \dd x\, g_2(x,Q^2)$. For the elastic form factors, we use the parametrizations of Ref.~\cite{Bradford:2006yz}. The $\Delta$-pole contribution is plotted based on Eqs.~\eref{approxg1}, \eref{GM*RatioF2} and the same form factors.}
              \label{fig:BCPlot}
\end{figure}

Figure \ref{fig:I1Plot} shows the generalized GDH integral, $I_1(Q^2)$, and the Pauli FF, $-\nicefrac{1}{4}\,F_2^2(Q^2)$. At the real photon point, the generalized GDH integral passes into the GDH integral, $I_1(Q^2=0)=-\kappa^2/4$, in line with the squared Pauli FF. As the nucleon-to-delta transition is dominantly of magnetic dipole type, cf.\ \Figref{I12}, and the magnetic Jones-Scadron FF can be expressed through the Pauli FF by large-$N_c$ relations, cf.\ \Eqref{GM*}, it is not far to seek that the $\Delta$-pole contribution to $I_1$ and the Pauli FF will look alike. Using Eqs.~\eref{GM*RatioF2} and \eref{approxg1}, we construct a perfect agreement: $I^{\Delta\mathrm{-pole}}_1(Q^2)=-\nicefrac{1}{4}\,F_2^2(Q^2)$, where we are neglecting $R_\mathrm{SM}$, etc. In \Figref{I1Plot}, this is illustrated with the orange long dashed line, indicating $I^{\Delta\mathrm{-pole}}_1(Q^2)$, and the gray dotted line, indicating $-\nicefrac{1}{4}\,F_2^2(Q^2)$. The light orange line shows the full $\Delta$-pole contribution as given in \Eqref{I1pole}, i.e., fully including the electric and Coulomb transitions. In each case, we plotted $I_1$ with the Jones-Scadron FFs of \Eqref{GM*RatioF2}. Again, it becomes obvious that the magnetic Jones-Scadron FF makes up the major part of $I^{\Delta\mathrm{-pole}}_1$. The relevance of the $\Delta$-pole contribution will be explained in \secref{chap5}{Depole}.

   \begin{figure}[t] 
    \centering       \includegraphics[width=15cm]{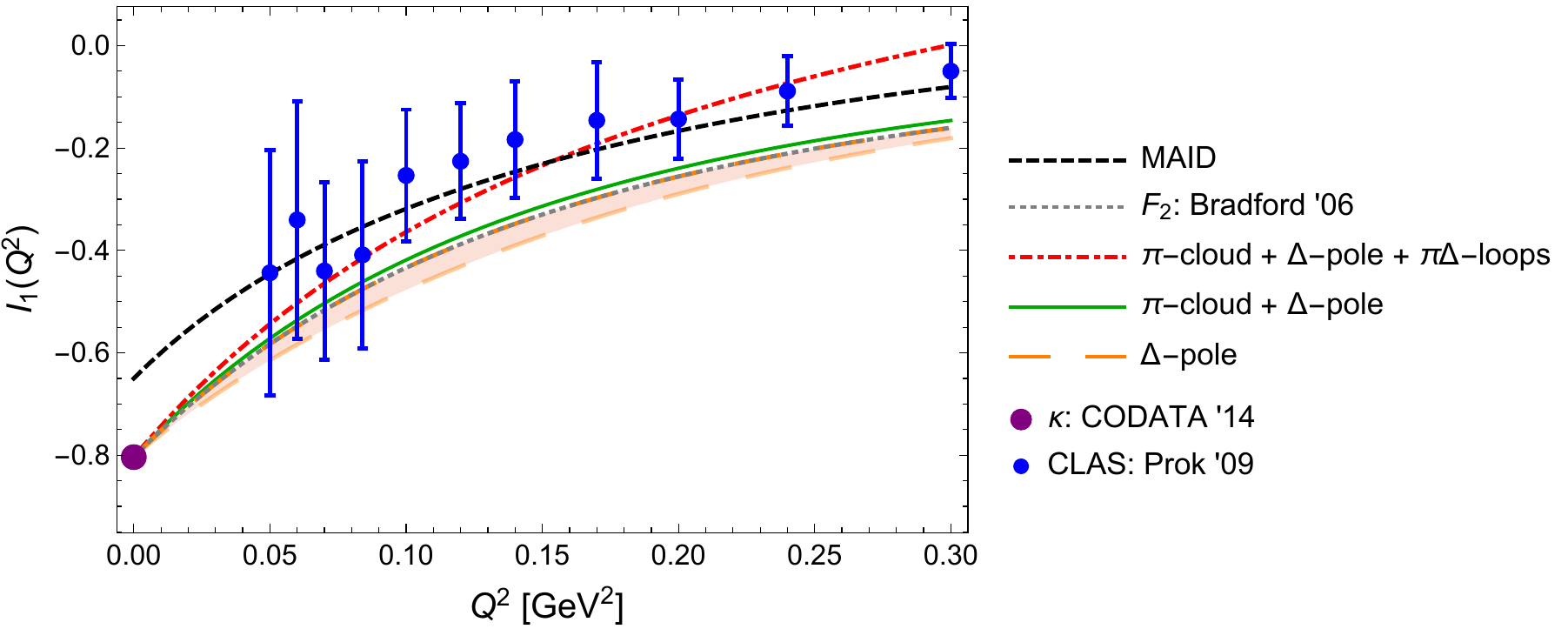}
           \caption{Comparison of $I_1(Q^2)$ and $-\nicefrac{1}{4}\,F_2^2(Q^2)$ as functions of $Q^2$.
    The black short dashed line is the MAID prediction for $I_1$ \cite{MAID}. The blue dots are deduced from CLAS data \cite{Prok:2008ev} and the purple dot at the real-photon point complies with $-\kappa^2/4$ \cite{Mohr:2012aa}. The gray dotted line is the Pauli FF parametrization of Ref.~\cite{Bradford:2006yz}. The orange long dashed line is the $\Delta$-pole contribution to $I_1$ as given in \Eqref{I1pole}; the green solid line is the sum of $\pi$-cloud and $\Delta$-pole contributions \cite{Lensky:2014dda}; the red dash-dotted line in addition includes the $\pi \Delta$-loop contribution \cite{Lensky:2014dda}.
    }
              \label{fig:I1Plot}
\end{figure}

In \Figref{PolarizabilitiesGrid}, we compare our results for the nucleon GPs, which we derived from either ChPT or large-$N_c$ relations. The large-$N_c$ relations are useful as they express the Jones-Scadron FFs, describing the e.m.\ nucleon-to-delta transition, through nucleon FFs, which are measured extensively. In \secref{chap4}{Jones}, we presented two sets of large-$N_c$ relations. In the following, we applied Eqs.~\eref{GM*old2}, \eref{GM*old3} and \eref{GM*} with $C_M^*=\frac{3.02}{\sqrt{2}\,\kappa_p}$ to derive the complete $\Delta$-exchange contribution to the nucleon polarizabilities (red long dash-dotted curves in \Figref{PolarizabilitiesGrid}). On the other hand, we used \Eqref{GM*RatioF2} to derive the contribution to the polarizabilities which originates from the $\Delta$-pole part\footnote{For $I_1$ we do not plot the $\Delta$-pole part. Instead, we refer to Figures~\ref{fig:I12} and \ref{fig:I1Plot}} (orange short dash-dotted curves in \Figref{PolarizabilitiesGrid}), e.g., from Eqs.~\eref{structurefunc} or \eref{Dpole}. In both cases, we substituted the FF parametrizations of Ref.~\cite{Bradford:2006yz}. 

Focusing for now on the full BChPT result (black curve) and the large-$N_c$ description of the $\Delta$-exchange (red curve), we find a very similar description of the sum of dipole polarizabilities and the FSP. Also for the generalized GDH integrals, one observes a similar evolution in both approaches. The only difference is in a stronger curvature of the BChPT result with increasing $Q^2$. For the longitudinal and the longitudinal-transverse polarizabilities, one observes an upwards shift of the large-$N_c$ curves. This follows from the fact that the ChPT coupling constants,  \newpage

   \begin{figure}[h!] 
    \centering 
       \includegraphics[width=14.5cm]{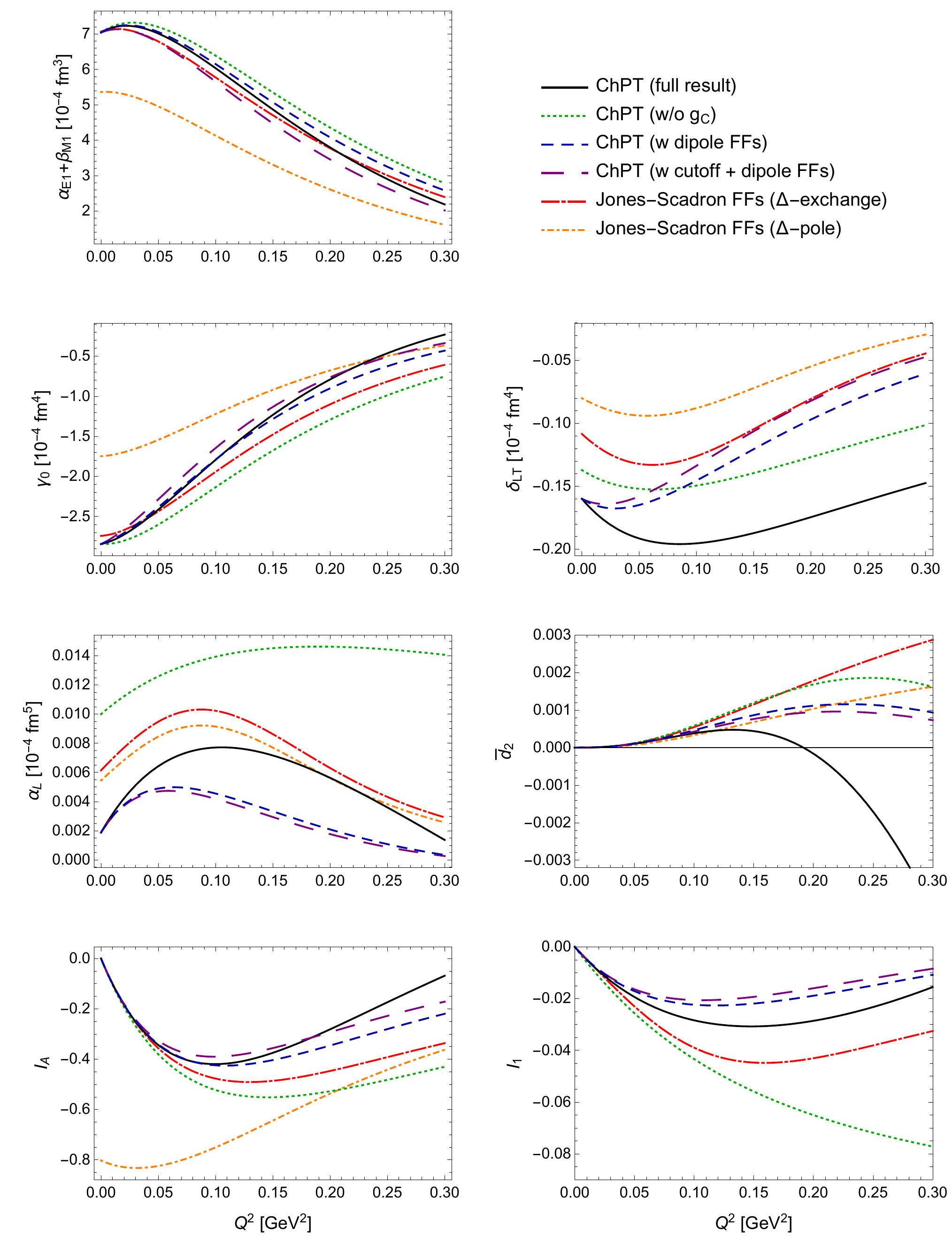}
           \caption{The sum of electric and magnetic dipole polarizabilities $\left[\alpha_{E1}+\beta_{M1}\right] (Q^2)$, the forward spin polarizability $\gamma_0(Q^2)$, the longitudinal-transverse polarizability $\delta_{LT} (Q^2)$, the longitudinal polarizability $\alpha_L (Q^2)$ and the generalized integrals $I_1 (Q^2)$, $I_A (Q^2)$ and $\ol d_2(Q^2)$ as functions of $Q^2$. The legend is given in the upper right corner, further information is given in the text. \vspace{1cm}}
              \label{fig:PolarizabilitiesGrid}
\end{figure}

\newpage

   \begin{figure}[t] 
    \centering 
       \includegraphics[width=12.5cm]{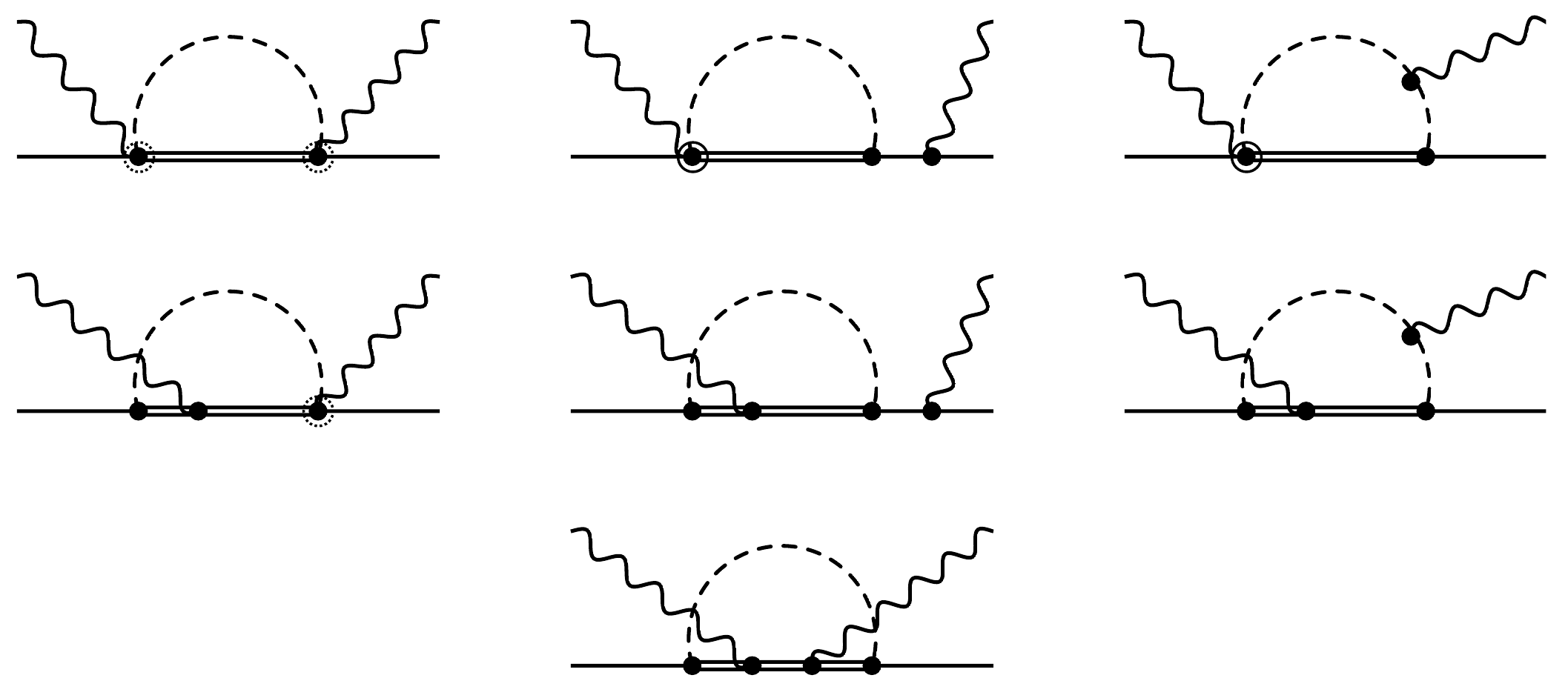}
       \caption{$\pi\Delta$-loop amplitudes with photon coupling minimally to the delta at $\mathcal{O}(p^{7/2})$ (first row), $\mathcal{O}(p^4)$ (second row) and  $\mathcal{O}(p^{9/2})$ (third row) in the low-energy domain of the $\delta$-expansion. Diagrams obtained from these by crossing and time-reversal are included too.}
              \label{fig:DeltaNewPlot}
\end{figure}

   \begin{figure}[t] 
    \centering 
       \includegraphics[width=15.5cm]{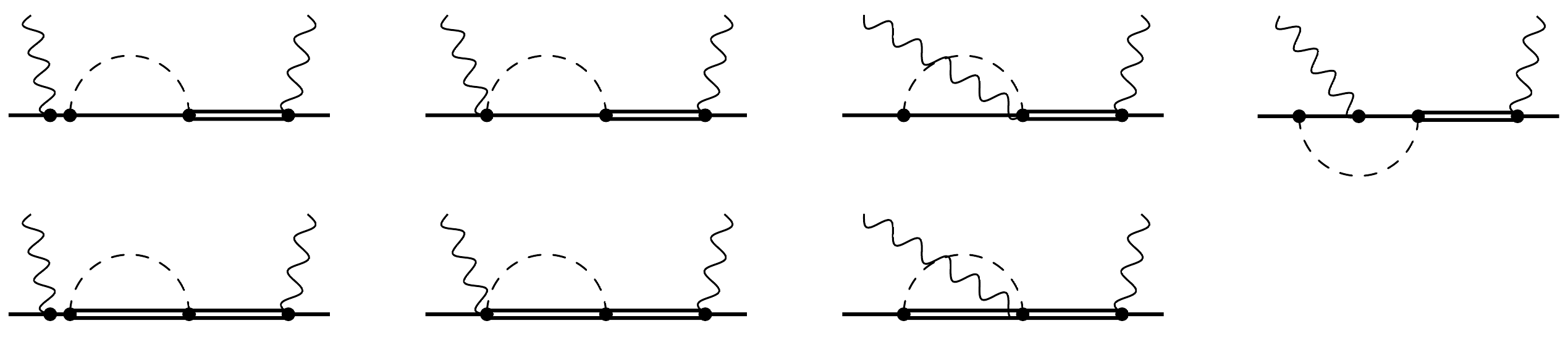}
       \caption{One-delta-irreducible $\pi N$- and $\pi\Delta$-loop amplitudes at
       $\mathcal{O}(p^{9/2})$ (first row) and $\mathcal{O}(p^5)$ (second row) in the low-energy domain of the $\delta$-expansion. Diagrams obtained from these by crossing and time-reversal are of the same orders.}
              \label{fig:MissingDiagrams}
\end{figure}

\noindent given in Table \ref{LEC}, and the so to say large-$N_c$ couplings do not agree at the real photon point. Evaluating \Eqref{48}, one finds: $g_M(Q^2=0)\approx 2.97$,  $g_E(Q^2=0)\approx -0.78$ and $g_C(Q^2=0)= 0$. The limit of the running magnetic coupling was fixed in accordance with experiment by our choice of $C_M^*$. The electric couplings resemble on another. Nevertheless, one can observe a slight offset in the FSP. The biggest deviation is in the Coulomb couplings. As only the static values of $\al_L$ and $\delta_{LT}$ depend on $g_C$, this is where the biggest differences are seen.

Ref.~\cite{Lensky:2014dda} studied moments of nucleon structure function at NLO in BChPT. For this part of the thesis, our motivation was to extend the former calculation by including the Coulomb coupling. Strictly speaking, the Lagrangian with $g_E$ and $g_C$ couplings, cf.\ \Eqref{nmGammaNDeltaLag}, is attributed to order $k=3$ and not $k=2$ \cite{Pascalutsa:2006up}. This is due to the fact that the contained $\gamma_5$ matrix mixes small and large components of the Dirac spinors, cf.\ Ref.~\cite[Eq.~(3.33)]{Krause:1990xc}. Therefore, the $\Delta$-exchange diagram belongs to several orders in the chiral expansion. It has a dominant contribution proportional to $g_M^2$, interference terms proportional to $g_M g_E$ or $g_M g_C$, and terms originating purely from $k=3$ Lagrangian, e.g., proportional to $g_E^2$, $g_E g_C$ or $g_C^2$. In the $\delta$-expansion, these terms are of $\mathcal{O}(p^{7/2})$, $\mathcal{O}(p^{9/2})$ and $\mathcal{O}(p^{11/2})$, respectively. Likewise, they are of $\mathcal{O}(p)$, $\mathcal{O}(p^2)$ and $\mathcal{O}(p^3)$ in the first resonance region. In \Figref{PolarizabilitiesGrid}, we compare the previous results including the magnetic and electric couplings (green dotted curve) \cite{Lensky:2014dda} and our improved result including the Coulomb coupling (black solid curve). Despite its higher order in the power counting, the contribution of the Coulomb coupling gives significant effects, especially for $\al_L$ and $\delta_{LT}$.

\noindent The remaining curves show different regularizations for the higher $Q^2$ behavior. For one thing, we introduce a vector-meson type of dependence on $g_E$ and $g_C$ analogue to \Eqref{gMDipole} (blue short dashed curve) \cite{Pascalutsa:2005vq}. For another thing, in addition to the dipole FFs on the coupling constants, we introduce an overall prefactor which cuts off higher momenta, cf.\ \Eqref{Q2cutoff} (purple short dashed curve) \cite{Pascalutsa:2006up}.

\section[Compton Scattering off the Nucleon with $\pi \Delta$-Loops]{Compton Scattering off the Nucleon with $\boldsymbol{\pi \Delta}$-Loops} \seclab{DeltaCrossSectionSec}

Within the $\delta$-counting, the diagrams shown in \Figref{DeltaOldPlot} enter the chiral expansion of the VVCS amplitude at NNLO; they are of $\mathcal{O}(p^{7/2})$. The diagrams in the first row of \Figref{DeltaNewPlot} contribute to the VVCS amplitude at the same order.\footnote{The dashed circles on the vertices of the far left diagram in the first row should indicate that either both vertices have the photon minimally coupling to the delta, or, one photon is coupling to the pion and the other is coupling to the delta. The diagram with both photons coupling to the pion is already shown in \Figref{DeltaOldPlot}. Similarly, the dashed circle on the vertex of the far left diagram in the second row indicates that the photon can couple to the pion or the delta. This is for instance allowed in the $\gamma\,n \rightarrow \Delta^\pm \pi^\mp$ channels.} The remaining diagrams, shown in the second and third row of \Figref{DeltaNewPlot}, are of higher order. They contribute at $\mathcal{O}(p^{4})$ and $\mathcal{O}(p^{9/2})$, respectively. However, they are required from e.m.\ gauge invariance and for the renormalization program.
\citet{Lensky:2014dda} bypassed  the calculation of the latter diagrams, cf.\ \Figref{DeltaNewPlot} second and third row, by includes only the subset of pion-delta loops which would give a non-vanishing contribution to the case of a neutral delta, i.e., the diagrams shown in \Figref{DeltaOldPlot}.\footnote{The structures shown in the first row of \Figref{DeltaNewPlot} can be found also in \Figref{DeltaOldPlot}.} They followed a procedure outlined previously for the case of RCS \cite[Section 3]{Len10} to make this subclass of diagrams gauge invariant and effectively include the lower-order contributions of the one-loop graphs with minimal coupling of photons to the delta. In this way, they achieved a BChPT prediction for the amplitude of CS off the nucleon at NNLO and the nucleon polarizabilities at NLO in the low-energy domain of the $\delta$-expansion. 

The $\eps$-expansion, however, counts all diagrams in Figures~\ref{fig:DeltaOldPlot} and \ref{fig:DeltaNewPlot} as $\mathcal{O}(p^3)$, cf.\ Table \ref{PC}. Working in the $\eps$-expansion, \citet{Bernard:2012hb} obtain a larger value for the longitudinal-transverse polarizability of the proton (and neutron), cf.\ \Figref{deltaLT}, which is in significant contradiction to the empirical information and the result from Ref.~\cite{Lensky:2014dda}. The claim is that the difference is due to the diagram with two photons coupling minimally to the delta inside the chiral loop, see third row of \Figref{DeltaNewPlot}. To solve the $\delta_{LT}$ puzzle, it will be enlightening to repeat the calculation of Ref.~\cite{Bernard:2012hb} and consider the diagrams shown in Figures~\ref{fig:DeltaOldPlot} and \ref{fig:DeltaNewPlot} entirely, i.e., not only the lower-order contributions. Such project has been initiated by V.~Lensky et al.\ \cite{Alarcon2017} and results are underway. As part of this thesis, the $\pi\Delta$-production cross sections were calculated, which serve as a cross check for the $\pi\Delta$-loop amplitudes \cite{Alarcon2017}. In \secref{chap4}{deltaCS}, we will outline the calculation of the cross sections and present our results graphically. In \secref{chap4}{DeltaCrossSectionSec}, we will provide some newly developed insights into the $\delta_{LT}$ puzzle.

The last thing we want to point out is that, while the diagrams in Figures~\ref{fig:Born}, \ref{fig:PiNLoopDiags}, \ref{fig:DeltaOldPlot}, \ref{fig:DeltaExchange} and \ref{fig:DeltaNewPlot} give the complete $\mathcal{O}(p^3)$ VVCS amplitude in the $\epsilon$-expansion, they do not give a full amplitude in the $\delta$-expansion. In the low-energy domain of the $\delta$-expansion, this set of diagrams is incomplete and lacks the diagrams shown in the first row of \Figref{MissingDiagrams} to give the VVCS amplitude at $\mathcal{O}(p^{9/2})$. Also, in the first resonance region the diagrams will not be able to fully describe the CS amplitude at $\mathcal{O}(p^3)$, because the $\mathcal{O}(p^2)$ diagrams shown in \Figref{MissingDiagrams} are missing. On the other hand, all diagrams in  \Figref{MissingDiagrams} are of higher order in the $\epsilon$-expansion, in particular, they are of $\mathcal{O}(p^4)$. Therefore,  extending the calculation of Ref.~\cite{Lensky:2014dda} as described above gives us no more explanatory power in the $\delta$-expansion power-counting. It will only help to understand the origin of the $\delta_{LT}$ puzzle.

\subsection[$\pi \Delta$-Production Cross Sections]{$\boldsymbol{\pi \Delta}$-Production Cross Sections}\seclab{deltaCS}
  \begin{figure}[tbh] 
    \centering 
       \includegraphics[width=15cm]{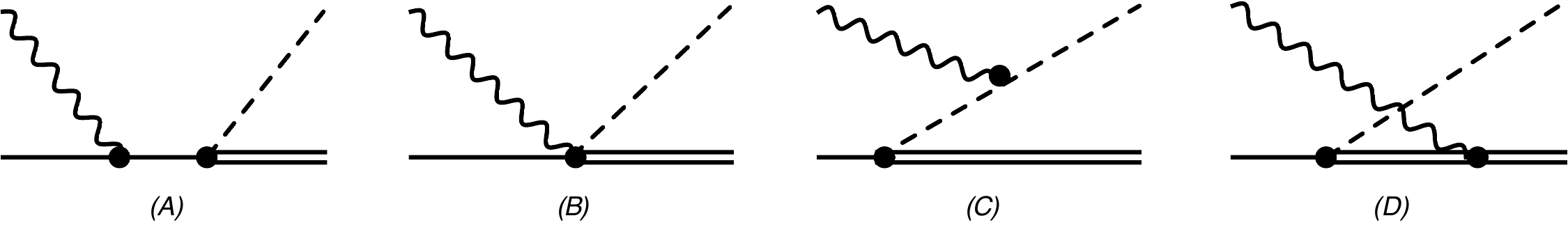}
       \caption{$\pi \Delta$-production photoabsorption cross sections.}
              \label{fig:PiDeLoopDiags}
\end{figure}
\noindent In the following, we study the cross section of $\gamma^* N \rightarrow \pi \Delta$. In general, the cross section for two-body scattering, i.e., $1+2\rightarrow 3+\dots +n$, is defined as:
\beq
\dd \sigma=(2\pi)^4 \delta^{(4)}\left(p_f-p_i\right)\vert \mathcal{M}_{fi}\vert^2 \frac{1}{4I} \prod_a \frac{\dd^3 p_a'}{(2\pi)^3 2 E_a'}, \eqlab{generalCS}
\eeq
where the subscript $i$ ($f$) stands for initial (final) particles,  $I^2=(p_1 \cdot p_2)^2-m_1^2 m_2^2$, and the integration is over the final phase space. For the two-to-two scattering at hand, we can write
\beq
\dd \sigma=\frac{1}{32\pi}\frac{\vert \bq \vert \vert \boldsymbol{p}' \vert}{M^2\left(\nu^2+Q^2\right)}\vert \mathcal{M}_{fi}\vert^2 \; \dd\! \cos \theta,\eqlab{2CS2}
\eeq
where $\theta$ is the scattering angle and $I^2=M^2(\nu^2+Q^2)$. Here, we eliminated four integrations by help of the energy and momentum conservation delta functions, and evaluated the azimuthal integration with a factor of $2\pi$.

For the calculation, we chose the CM kinematics:
\begin{subequations}
\bea
p&=&\left(E,0,0,-\vert \bq \vert\right),\eqlab{pmom}\\
q&=&\left(\omega,0,0,\vert \bq \vert\right),\\
p_\Delta&=&\left(E_\Delta,\vert \boldsymbol{p}' \vert \sin \theta,0,\vert \boldsymbol{p}' \vert \cos \theta\right),\\
p_\pi&=&\left(E_\pi,-\vert \boldsymbol{p}' \vert \sin \theta,0,-\vert \boldsymbol{p}' \vert \cos \theta\right).
\eea
\end{subequations}
All particles need to be on the mass shell and the photon virtuality is given by $q^2=-Q^2$.
Later, we can transit to the lab-frame variables with a set of replacement rules:
\begin{subequations}
\bea
E&=&\frac{s+Q^2-M \nu}{\sqrt{s}},\\
\omega&=&\frac{M\nu-Q^2}{\sqrt{s}},\\
E_\Delta&=&\frac{s+M_\Delta^2-m_\pi^2}{2\sqrt{s}},\\
E_\pi&=&\frac{s+m_\pi^2-M_\Delta^2}{2\sqrt{s}},\\
\vert \bq \vert&=&\sqrt{Q^2+\frac{\left(Q^2-M \nu\right)^2}{s}},\\
\vert \boldsymbol{p}' \vert&=&\sqrt{\frac{\left(s+m_\pi^2-M_\Delta^2\right)^2}{4s}-m_\pi^2}\,,
\eea
\end{subequations}
where $s=M^2+2M\nu-Q^2$ is the usual Mandelstam variable.

For VVCS, the photon can be either transverse or longitudinal polarized. The polarization vectors for a space-like photon read \cite{Hagelstein:2015egb}:
\begin{subequations}
\eqlab{polvecs}
\bea
\eps_+&=&-\nicefrac{1}{\sqrt{2}}\,\left(0,1,i,0\right),\eqlab{epsTp}\\
\eps_-&=&\nicefrac{1}{\sqrt{2}}\,\left(0,1,-i,0\right),\eqlab{epsTm}\\
\eps_0&=&\nicefrac{1}{Q} \,\left(\vert \bq \vert,0,0,\omega\right),
\eea
\end{subequations}
where the subscript indicates the helicity of the photon. For the nucleon momentum given in \Eqref{pmom}, the spin-1/2 Dirac spinors $N_{\sigma_N}$ follow from \Eqref{generalDS} as:
\begin{subequations}
\eqlab{Nspins}
\bea
N_{1/2}(p)&=& \sqrt{E+M}\,\left(1,0,-\frac{\vert \bq \vert}{E+M},0\right),\\
N_{-1/2}(p)&=& \sqrt{E+M}\,\left(0,1,0,\frac{\vert \bq \vert}{E+M}\right),
\eea
\end{subequations}
where the subscript again indicates the helicity. The Rarita-Schwinger vector-spinors describing the spin-3/2 delta are introduced in \appref{chap4}{RSVS}.\footnote{Note that we are using different normalizations in \appref{chap4}{RSVS}.} Here, the nucleon spinors are normalized according to \Eqref{spinorNorm} and the delta vector-spinors are normalized as:
\beq
\ol U_\mu (p_\Delta, \lambda'_\Delta) \,U^\mu(p_\Delta, \lambda_\Delta)=-2M_\Delta \,\delta_{\lambda'_\Delta\lambda_\Delta}.\eqlab{RSVSnorm}
\eeq 
The photon polarization vectors obey the transversality condition:
\beq
q\cdot \eps_{\lambda_\gamma}(q)=0,
\eeq
and the orthogonality condition:
\beq
\eps^*_{\lambda'_\gamma}(q) \cdot \eps_{\lambda_\gamma}(q)=(-1)^{\lambda_\gamma} \,\delta_{\lambda'_\gamma \, \lambda_\gamma}.
\eeq

The diagrams contributing to the $\pi \Delta$-production process are shown in \Figref{PiDeLoopDiags}. They can be calculated with the Feynman rules collected in \appref{chap4}{FRAppendix}. Since \Eqref{2CS2} requires the squared matrix element, the spin-energy projection operator in \Eqref{SEprojec} is useful. It combines the Rarita-Schwinger vector spinors in the right way and performs the necessary sum over all possible helicities in the final state.

For the cross sections, we distinguish 5 individual channels:
\begin{enumerate}
\item channel: $\gamma p \rightarrow \Delta^{++} \pi^-$,
\item channel: $\gamma p \rightarrow \Delta^{+} \pi^0$,
\item channel: $\gamma p \rightarrow \Delta^{0} \pi^+$,
\item channel: $\gamma n \rightarrow \Delta^{+} \pi^-$,
\item channel: $\gamma n \rightarrow \Delta^{-} \pi^+$.
\end{enumerate}
Furthermore, we have three different combinations of helicity cross sections for each channel: $\sigma_T$, $\sigma_{TT}$ and $\sigma_L$, as explained in \secref{chap4}{SR}. In addition, the possibility of a spin flip of the nucleon gives rise to the longitudinal transverse cross section $\sigma_{LT}$. In that case, we in the initial state have a longitudinal photon with helicity $\lambda_\gamma=0$ ($\epsilon_0$) and a nucleon with helicity $\lambda_N=\mp 1/2$ ($N_{\mp1/2}$). In the final state, this changes into a transverse photon with helicity $\lambda_\gamma'=\pm 1$ ($\epsilon_\pm$) and a nucleon with inverted helicity $\lambda_N'=\pm 1/2$ ($N_{\pm1/2}$). The VVCS amplitude is then given by:
\beq
T(\nu,Q^2)=\sqrt{2}\, g_{LT}(\nu,Q^2),
\eeq
as one can read of from Eq.~\ref{Eq:T-Compt-definition}.
Therefore, in order to derive the longitudinal transverse cross section, we need to divide the squared matrix element by $\sqrt{2}$.

Our results for the cross sections are shown in Figures \ref{fig:protonTplot}-\ref{fig:neutronLTplot}, see \appref{chap4}{CSplots}. We show $\sigma_T$, $\sigma_{TT}$, $\sigma_L$ and $\sigma_{LT}$ as functions of the lab-frame photon energy $\nu$ for different values of the photon virtuality $Q^2$. For each helicity cross section, we show a grid plot of the different channels and the combined result for the proton and neutron, respectively. In \Figref{protonTplot}, we include data bins from RCS \cite{Wu:2005wf}. Especially for the neutral-delta channel, our BChPT prediction gives cross sections which are much higher than the data points.

To check our results, we used the cross sections ($\sigma_T$, $\sigma_{TT}$ and $\sigma_{LT}$) to obtain the higher-order polarizabilities, e.g., $\ol \gamma_0$, and compared to the polarizabilities extracted from the real part of the $\pi \Delta$-loop amplitudes \cite{Lensky:2014dda} shown in Figures~\ref{fig:DeltaOldPlot} and \ref{fig:DeltaNewPlot}. As part of this thesis, we also calculated the imaginary part of some of the $\mathcal{O}(p^{7/2})$ $\pi \Delta$-loop spin-independent amplitudes and verified the optical theorem for $\sigma_T$ and $\sigma_L$. For the SE diagram (far left diagram in the second row of \Figref{DeltaOldPlot}), we in addition compared to Ref.~\cite{Pascalutsa:2005nd} and checked the optical theorem for all cross sections.


In Figures \ref{fig:CSpComp} and \ref{fig:CSnComp}, we compare $\sigma_T$, $\sigma_{TT}$, $\sigma_L$ and $\sigma_{LT}$ from different orders in the chiral expansion for the proton and neutron, respectively. On one hand, we show the $\pi \Delta$-production cross sections which generate the $\pi \Delta$-loop diagrams contributing to the CS at $\mathcal{O}(p^{7/2})$ in the low-energy domain of the $\delta$-expansion (black solid curves). In that case, the matrix element for the cross sections only contains the diagrams (A)-(C) in \Figref{PiDeLoopDiags}. On another hand, we show the cross sections related to the CS amplitude at $\mathcal{O}(p^{9/2})$ in the $\delta$-expansion and $\mathcal{O}(p^{3})$ in the $\epsilon$-expansion (blue dashed curves). Here we include all diagrams in \Figref{PiDeLoopDiags} into the cross section calculation. Finally, we illustrate the influence of the CS diagram with two photons minimally coupling to the delta inside the pion loop by showing the cross section associated to the CS amplitude at $\mathcal{O}(p^4)$ (red dotted curves). It thus excludes the contribution of the diagram  (D) squared.

For $\sigma_T$ and $\sigma_{TT}$ we observe that the $\mathcal{O}(p^4)$ and $\mathcal{O}(p^{9/2})$ curves are very close, hence the $\mathcal{O}(p^{9/2})$
contribution is very small.  
For $\sigma_L$ and $\sigma_{LT}$, however, we observe significant effects 
of both $\mathcal{O}(p^4)$ and $\mathcal{O}(p^{9/2})$ contributions. 
At $\mathcal{O}(p^{9/2})$, the absolute value of the cross sections is considerably reduced, if not almost vanishing.  We may conclude that the contribution of the CS diagram with three delta propagators (i.e., two photons coupling to the delta) is larger if longitudinal photons are involved.

Unfortunately, we have not yet been able to evaluate these $\pi \Delta$-effects on the controversial $\delta_{LT}$ polarizability, because of the ultraviolet divergence.  
For example, the value at $Q^2=0$, given as 
\beq
\delta_{LT}(0)=\lim_{Q \rightarrow0}\,\frac{1}{2\pi^2}\int_0^\infty\frac{\dd \nu}{\nu^2 Q}\,\sigma_{LT}(\nu,Q^2),
\eeq
will diverge upon substituting the $\pi \Delta$-production cross section we obtained, because of its bad (unphysical) high-energy behaviour. In ChPT this problem
is taken care of by renormalization, whereas here we can use subtractions of the dispersion integrals. The correspondence of the renormalization of
the loop calculation and of the dispersive calculation is an open issue for this particular case. We hope to sort out this issue in the near future.


\section{Comparison of ChPT and Empirical Nucleon Structure Functions}\seclab{comp5}

In order to visualize our BChPT prediction for the nucleon structure functions and compare to empirical information, we need to replace the $\delta$-functions contained in the $\Delta$-production cross sections, cf.\ Eqs.~\eref{dfunc} and \eref{structurefunc}, by nascent $\delta$-functions. We will use a Lorentz function, \Eqref{Lorentzian}, which is similar to the Breit-Wigner resonance parametrization \cite[Eq.~(46.55)]{Agashe:2014kda}. The Lorentz function is normalized to ensure: $\int_{-\infty}^\infty \dd x \,n_\lambda (x)=1$.
Starting from \Eqref{structurefunc}, we first replace:
\beq
\delta\!\left(x-\nicefrac{Q^2}{2M\nu_\Delta}\right)=\frac{\left[2M \nu_\Delta\right]^2}{Q^2}\,\delta\!\left(s-M_\Delta^2\right).
\eeq
Afterwards, we substitute the $\delta$-function as:
\beq
\delta\!\left(s-M_\Delta^2\right)\rightarrow \frac{1}{\pi}\frac{ M_\Delta\Gamma_\Delta}{\left(s-M_\Delta^2\right)^2+(M_\Delta\Gamma_\Delta)^2}
\eeq
To describe the $\Delta(1232)$-resonance, we then use the pole position $M_\Delta=1.210$ GeV and the width $\Gamma_\Delta=0.1$ MeV \cite{Agashe:2014kda}.

  \begin{figure}[t] 
    \centering 
       \includegraphics[width=12cm]{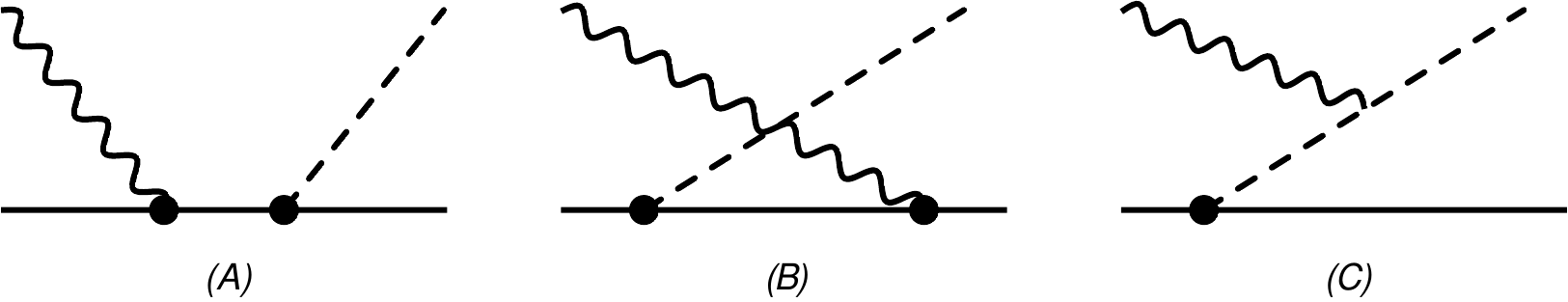}
       \caption{$\pi N$-production photoabsorption cross sections.}
              \label{fig:PiNLoopDiagsCS}
\end{figure}

We limit the comparison to the spin-dependent proton structure function $g_1(x,Q^2)$. The plots of ChPT and empirical nucleon structure functions are moved to \appref{chap4}{nucleonSFplots}. Our BChPT prediction consists of the $\Delta$-production cross sections presented in \secref{chap4}{3.1CSDelta} and above and the $\pi N$-production cross sections \cite{Lensky:2014dda}. For the $\Delta$-production cross sections, we use the ChPT couplings in Table \ref{LEC} and a dipole on the magnetic coupling, cf.\ \Eqref{gMDipole}, to include the effect of vector-meson diagrams. The $\pi N$-production cross sections, see \Figref{PiNLoopDiagsCS}, were calculated in Ref.~\cite{Lensky:2014dda} and reproduced in the course of this thesis.

 If available, we show low-$Q^2$ data from different experiments \cite{Adeva:1999pa,Fatemi:2003yh}. Furthermore, we compare to two different parametrizations of the spin-dependent structure functions of the proton. 
We show the JLab parametrization provided to us by K.~Griffioen \cite{Prok:2008ev,Dharmawardane:2006zd} and a parametrization provided to us by S.~Simula \cite{Simula:2001iy,Simula:2002tv}. While the Simula parametrization is lacking some recent data for the low-$Q^2$ region, the JLab parametrization includes data as low as  $Q^2 = 0.0452 \,\text{GeV}^2$. Nevertheless, the Simula parametrization proves to be very useful for our purposes, as its composition of background and resonance contributions is easy to access. This allows us to plot different ingredients to the structure function parametrization separately, e.g., the contribution of the $\Delta(1232)$-resonance, the combined resonant contribution of $\Delta(1232)$, $P_{11}(1440)$, $D_{13}(1520)$, $S_{11}(1535)$, $S_{11}^*(1650)$, $D_{15}+F_{15}(1680)$, $D_{33}(1700)$, $F_{35}(1905)$ and $F_{37}(1950)$, and the ``background'' contribution. 

Remember that, by the standard DRs, the imaginary part of the $\Delta$-exchange CS amplitude is only related to the $\Delta$-pole part of the $\Delta$-exchange. For the non-pole part we deduced the structure functions in Eqs.\ \eref{npstrucfunc}, \eref{S2npg} and \eref{S2npgp2}. Since we are plotting the $\Delta$-production contribution to the structure functions as given in \Eqref{structurefunc}, we observe an unphysical behavior at low $x$. This behavior would be canceled by the $\delta(x)$ structure functions in Eqs.\ \eref{npstrucfunc}, \eref{S2npg} and \eref{S2npgp2}. It is another manifestation of the cancelations between $\Delta$-pole and non-pole contributions, which were also observed for the $I_1$ and $I_2$ integrals in \secref{chap4}{DeltaPol}. Unfortunately, we can not plot the missing part of the structure functions as the parameters of the Lorentz function are unknown. We therefore put a gray shade over the low-$x$ region in all plots and draw the readers attention to the $\Delta$-resonance region.

\section{Summary and Conclusion}
In the present Chapter, we have considered the forward VVCS off the nucleon and the $Q^2$ dependence of the  nucleon polarizabilities, a.k.a.\ symmetric GPs.
Our main aim was to calculate and study the VVCS amplitudes and GPs in the framework of ChPT.  

We have discussed the different power-counting schemes for ChPT with explicit $\Delta(1232)$ DOFs
(see \secref{chap4}{powercounting}) and made a comprehensive analysis of 
the $\Delta$-exchange contribution. More specifically, we have extended the calculation of Ref.~\cite{Lensky:2014dda} by including the Coulomb 
($g_C$) coupling. We have verified the optical theorem with the imaginary part of the $\Delta$-exchange amplitudes, cf.\ \Eqref{structurefunc}, and the $\Delta$-production cross sections, cf.\ \appref{chap4}{HelicityCS}. The $\Delta$-exchange amplitudes split into a $\Delta$-pole part and a non-pole part; the resulting amplitudes are given in \appref{chap4}{VVCSDeltaAmp}. The $\Delta$-pole part is obtained from the  $\Delta$-production cross sections with the help of DRs. For the non-pole part, we write down the structure functions which do not peak at the pole position, i.e., are not proportional to $\delta\!\left(x-\nicefrac{Q^2}{2M\nu_\Delta}\right)$, but behave as $\delta(x)$, see Eqs.\ \eref{npstrucfunc}, \eref{S2npg} and \eref{S2npgp2}. The GPs of VVCS with $\Delta$-exchange are listed in Eqs.\ \eref{alphabetaQ2}-\eref{d2momQ2} and \appref{chap4}{Q2regpol}, where we compare different regularizations of their high-$Q^2$ behavior. In \Figref{PolarizabilitiesGrid}, we have plotted the resulting symmetric GPs and studied the effect of  the $g_C$ coupling. The strongest effect is observed in $\al_L$ and $\delta_{LT}$, where even the value at the real photon point is affected. 

In addition to the $\Delta$-production cross sections, we reproduced the $\pi N$-production cross sections from Ref.~\cite{Lensky:2014dda} and derived the $\pi \Delta$-production cross sections. We have shown that the polarizability and Born contributions to the BC sum rule are vanishing independently, see \secref{chap5}{BCHFS}. For the $\Delta$-exchange, the $\pi N$-loop and the $\pi\Delta$-loop CS amplitudes we confirmed that the polarizability contributions to the BC sum rule are equal to zero. 

The calculated $\pi \Delta$-production cross sections are shown in \appref{chap4}{CSplots}. They have been helpful to verify the $\pi \Delta$-loop CS amplitudes calculated by Lensky et al.\ \cite{Lensky:2014dda} at $\mathcal{O}(p^{9/2})$ in the  $\delta$-expansion. However, we have gone further and included the diagrams with photons coupling minimally to the delta. Hence, we also consider the diagram with both photons coupling to delta inside the pion loop, which is thought to be responsible for the $\delta_{LT}$ puzzle. 

The $\delta_{LT}$ puzzle refers to the significant discrepancy between the BChPT calculations of 
Bernard et al.\ \cite{Bernard:2012hb} and Lensky et al.\ \cite{Lensky:2014dda}. Our present calculation seems to have all the necessary ingredients
to sort this puzzle out. However, thus far we have not obtained the  contribution of the controversial diagram to $\delta_{LT}$ due to the ultraviolet
divergence of the relevant dispersion integral.

In \secref{chap4}{comp5} and \appref{chap4}{nucleonSFplots}, we compared our BChPT prediction for the nucleon structure functions to empirical information. We have shown that the $\Delta$-resonance peak in the structure functions is underestimated by BChPT. In the future, we plan to improve the high-energy asymptotics of the cross sections by some kind of ultraviolet completion, which will, in particular, allow us to obtain the result for $\delta_{LT}$.

Besides our ChPT results, we presented the $\Delta$-exchange contribution to the nucleon polarizabilities in terms of Jones-Scadron FFs, see \secref{chap4}{Jones}. We reviewed how the Jones-Scadron FFs are related to the e.m.\ nucleon FFs by large-$N_c$ relations. We contrasted the $\Delta$-exchange effect with the $\Delta$-pole effect. In \Eqref{GM*RatioF2}, we have set up the basics for a model which we want to implement in \chapref{5HFS} to calculate the polarizability contribution to the HFS in $\mu$H. The characteristic feature of the model is that we have: $I_1^{\Delta-\mathrm{pole}}(Q^2)=\frac14 F_2^2(Q^2)$, see \Figref{I1Plot}.

In \appref{chap4}{FRAppendix}, we derive Feynman rules from the ChPT Lagrangian given in \secref{chap4}{DeltaChPT}. In \secref{chap4}{RSVS}, we found a general expression for the Rarita-Schwinger vector-spinors of spin-3/2 particles. 

The VVCS amplitudes presented in this Chapter will be used in the remaining part of this thesis to calculate the TPE effects in $\mu$H.

\begin{subappendices}

\section{Feynman Rules} \seclab{FRAppendix}
\addtocontents{toc}{\protect\setcounter{tocdepth}{0}}
\seclab{FeynmanRules}
\subsection{Vertices}
The Feynman rules for the vertices are derived from the Lagrangians in \secref{chap4}{DeltaChPT}.

\begin{itemize}
\item e.m.\ vertex for the pion: $\gamma^*\,\pi\rightarrow \pi$\\[0.5cm]
\begin{minipage}{4cm}{\centering\includegraphics[scale=0.45]{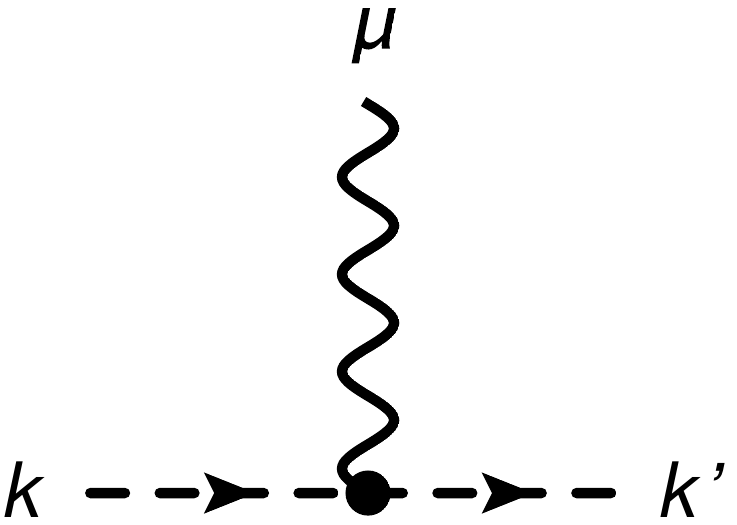}} 
\end{minipage}\hfill
\begin{minipage}{9cm}\begin{align}
\Ga^\mu_{\gamma\pi\pi}(k,k')=-e_\pi(k+k')^\mu\nn
\end{align}\end{minipage}
\\[0.5cm]
\item e.m.\ seagull vertex for the pion: $\gamma^*\,\pi\rightarrow \gamma^*\,\pi$\\[0.5cm]
\begin{minipage}{4cm}{\centering\includegraphics[scale=0.45]{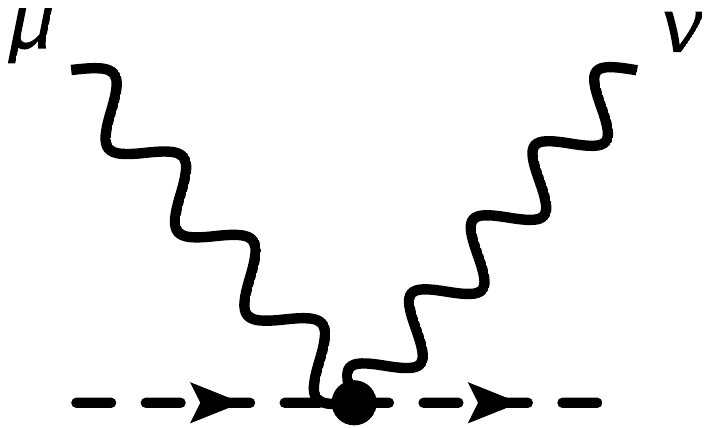}} 
\end{minipage}\hfill
\begin{minipage}{9cm}\begin{align}
\Ga^{\mu\nu}_{\gamma\gamma\pi\pi}=2e^2g^{\mu\nu}\nn
\end{align}\end{minipage}
\\[0.5cm]
\item e.m.\ vertex for the nucleon: $\gamma^*\,N\rightarrow N$\\[0.5cm]
\begin{minipage}{4cm}{\centering\includegraphics[scale=0.45]{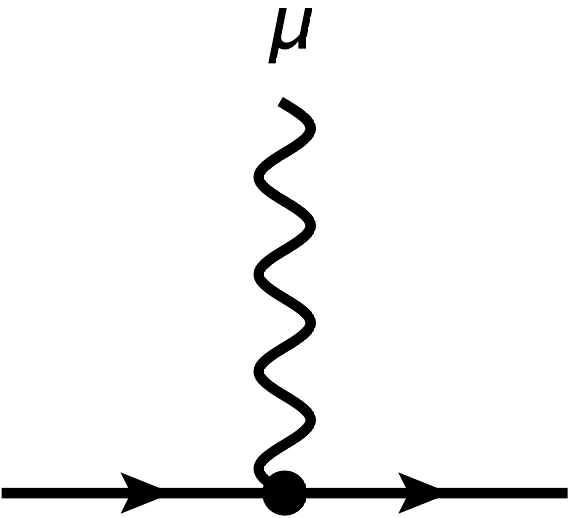}} 
\end{minipage}\hfill
\begin{minipage}{9cm}\begin{align}
\Ga^\mu_{\gamma NN}=-e_N\gamma^\mu\nn
\end{align}\end{minipage}
\\[0.5cm]
\item e.m.\ vertex for the delta: $\gamma^*\,\Delta\rightarrow\Delta$\\[0.5cm]
\begin{minipage}{4cm}{\centering\includegraphics[scale=0.45]{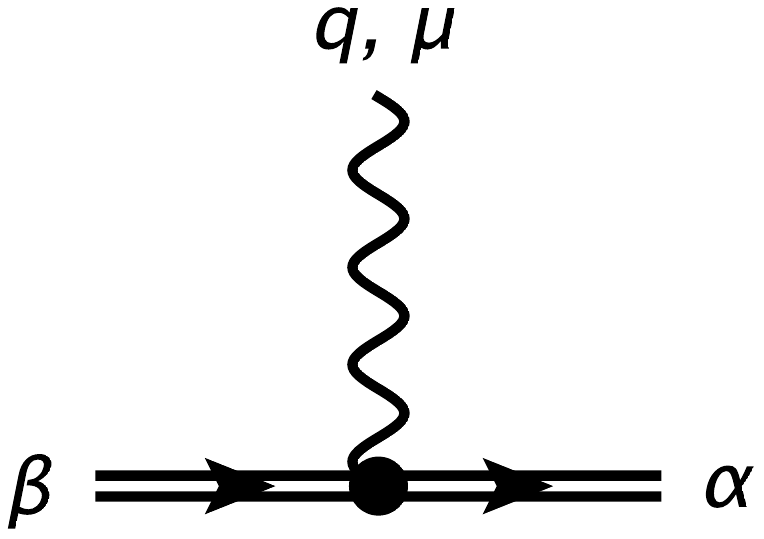}} 
\end{minipage}\hfill
\begin{minipage}{9cm}\begin{align}
\Ga^{\al \be \mu}_{\gamma \Delta \Delta}(q)&=-e_\Delta \gamma^{\al \be \mu}+\frac{e_\Delta}{M_\Delta}\left[\kappa_1 \left(q^\al g^{\be \mu}-q^\be g^{\al \mu}\right)\right.\nn\\
&\left.\quad-\kappa_2 \gamma^{\al \be \mu \rho}q_\rho\right]\nn
\end{align}\end{minipage}
\\[0.5cm]
\item pion-nucleon vertex: $N\rightarrow N \,\pi$\\[0.5cm]
\begin{minipage}{4cm}{\centering\includegraphics[scale=0.45]{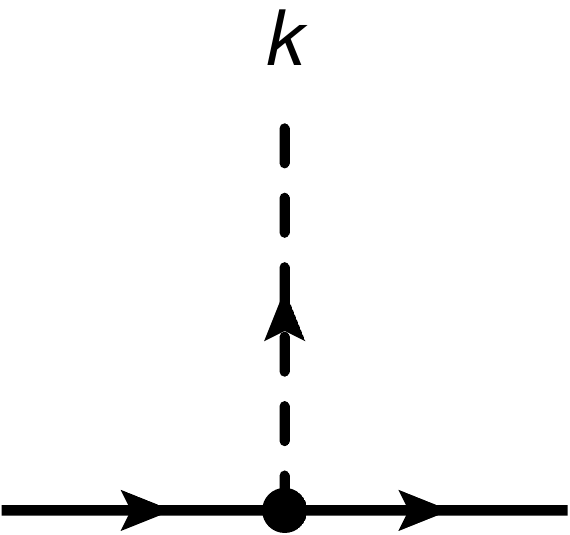}} 
\end{minipage}\hfill
\begin{minipage}{9cm}\begin{align}
\Ga_{\pi N N}(k)=-\frac{ig_A}{2f_\pi} \,\slashed{k}\,\gamma_5 \begin{cases}1 & p\rightarrow p\, \pi^0,\\
\sqrt{2}& p\rightarrow n\, \pi^+, \; n\rightarrow p\,\pi^-,\\
-1&n\rightarrow n\,\pi^0; \end{cases}\nn
\end{align}\end{minipage}
\\[0.5cm]
\item pion-delta vertex: $\Delta\rightarrow \Delta \,\pi$\\[0.5cm]
\begin{minipage}{4cm}{\centering\includegraphics[scale=0.45]{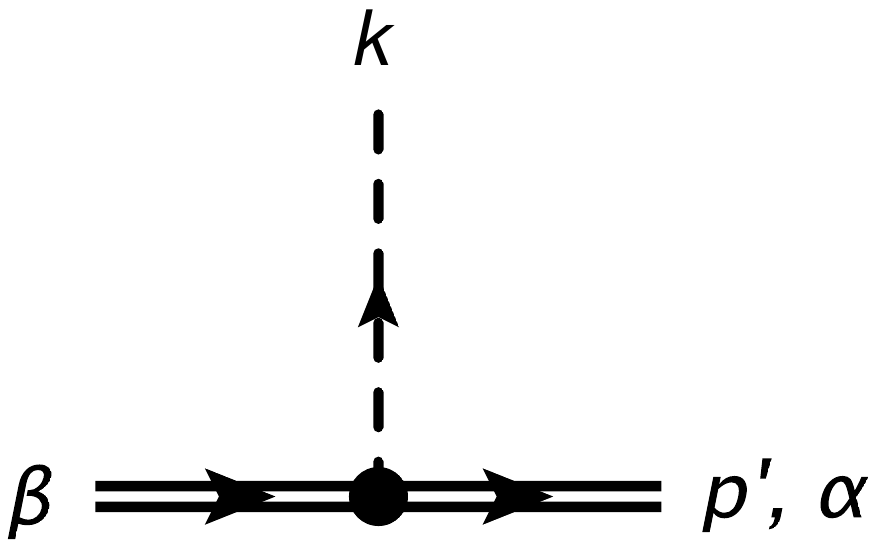}} 
\end{minipage}\hfill
\begin{minipage}{9cm}\begin{align}
\Ga^{\al \be}_{\pi \Delta \Delta}(p',k)=\frac{H_A}{2f_\pi M_\Delta}\eps^{\al \be  \rho \sigma} k_\rho \,p'_\sigma \begin{cases}
1& \Delta^{++}\rightarrow \Delta^{++}\pi^0,\\
\sqrt{2/3}& \Delta^{++}\rightarrow \Delta^{+}\pi^+,\\
\sqrt{2/3}& \Delta^{+}\rightarrow \Delta^{++}\pi^-,\\
1/3& \Delta^{+}\rightarrow \Delta^{+}\pi^0,\\
2\sqrt{2/3}& \Delta^{+}\rightarrow \Delta^{0}\pi^+,\\
2\sqrt{2/3}& \Delta^{0}\rightarrow \Delta^{+}\pi^-,\\
-1/3& \Delta^{0}\rightarrow \Delta^{0}\pi^0,\\
\sqrt{2/3}& \Delta^{0}\rightarrow \Delta^{-}\pi^+,\\
\sqrt{2/3}& \Delta^{-}\rightarrow \Delta^{0}\pi^-,\\
-1& \Delta^{-}\rightarrow \Delta^{-}\pi^0;\\
\end{cases} \nn
\end{align}\end{minipage}
\\[0.5cm]
\item pionic nucleon-to-delta transition vertex: $N \rightarrow \Delta \, \pi$\\[0.5cm]
\begin{minipage}{4cm}{\centering\includegraphics[scale=0.45]{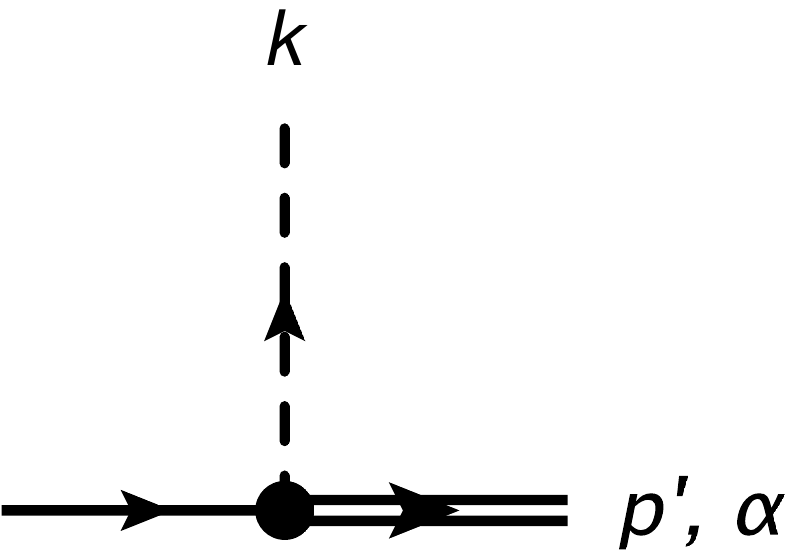}} 
\end{minipage}\hfill
\begin{minipage}{9cm}\begin{align}
\Ga_{N\Delta \pi}^\al(p',k)=\frac{ih_A}{2f_\pi M_\Delta} \gamma^{\rho \sigma \al} k_\rho p'_\sigma \begin{cases}
-1& p \leftrightarrow \Delta^{++}\,\pi^-,\\
\sqrt{2/3}& p \leftrightarrow \Delta^{+}\,\pi^0,\\
1/\sqrt{3}& p \leftrightarrow \Delta^{0}\,\pi^+,\\
-1/\sqrt{3}& n \leftrightarrow \Delta^{+}\,\pi^-,\\
\sqrt{2/3}& n \leftrightarrow \Delta^{0}\,\pi^0,\\
1& n \leftrightarrow \Delta^{-}\,\pi^+;
\end{cases}\nn
\end{align}\end{minipage}
\\[0.5cm]
\newpage
\item nucleon-to-delta transition vertex: $\gamma^* N \rightarrow \Delta$\\[0.5cm]
\begin{minipage}{4cm}{\centering\includegraphics[scale=0.45]{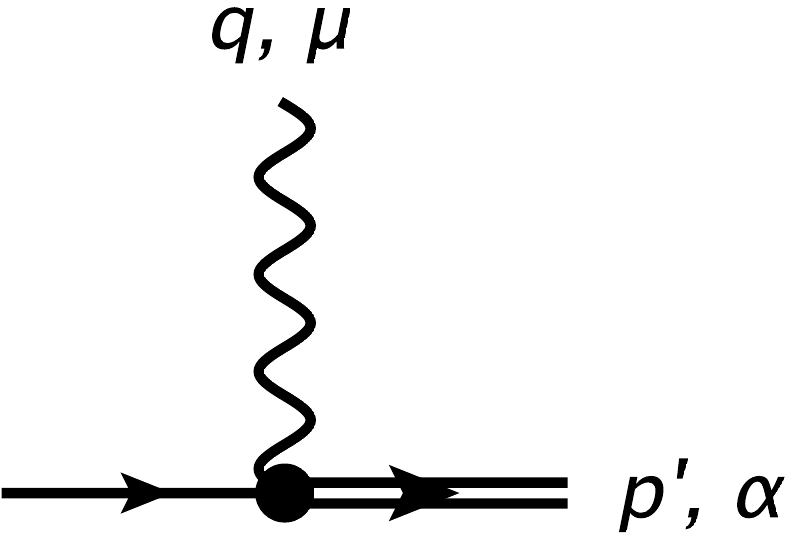}} 
\end{minipage}\hfill
\begin{minipage}{9cm}\begin{align}
\Ga_{\Delta \rightarrow \gamma N}^{\al \mu}(p',q)&=-\sqrt{\frac{3}{2}}\frac{e}{M\left(M+M_\Delta\right)}\left\{g_M \gamma^{\al \mu \kappa \lambda}p'_\kappa q_\lambda\right.\nn\\
&\quad+g_E(p'\cdot q\, g^{\al \mu}-q^\al p^{\prime\mu})-\frac{g_\mathrm{C}}{M_\Delta} \left(q^2 g^{ \al \mu}\slashed{p}'\right.\nn\\
&\quad\left.\left.-q^2 p^{\prime\mu} \gamma^\al +p'\cdot q \,q^\mu \gamma^\al-q^\al q^\mu \slashed{p}'\right)\right\}\gamma_5,\qquad\nn\\
\Ga_{\gamma N \rightarrow \Delta}^{\al \mu}(p',q)&=\sqrt{\frac{3}{2}}\frac{e}{M\left(M+M_\Delta\right)}\left\{g_M \gamma^{\al \mu \kappa \lambda}p'_\kappa q_\lambda\right.\nn\\
&\quad+g_E (p'\cdot q\, g^{\al \mu}-q^\al p^{\prime\mu})+\frac{g_\mathrm{C}}{M_\Delta} \left(q^2 g^{ \al \mu}\slashed{p}'\right.\nn\\
&\quad\left.\left.-q^2 p^{\prime\mu} \gamma^\al +p'\cdot q \,q^\mu \gamma^\al-q^\al q^\mu \slashed{p}'\right)\right\}\gamma_5;\qquad\nn
\end{align}\end{minipage}
\\[0.5cm]
\item 4-point vertex with photon coupling minimally to the pion: $\gamma^*\,N \rightarrow \Delta \pi$\\[0.5cm]
\begin{minipage}{4cm}{\centering\includegraphics[scale=0.45]{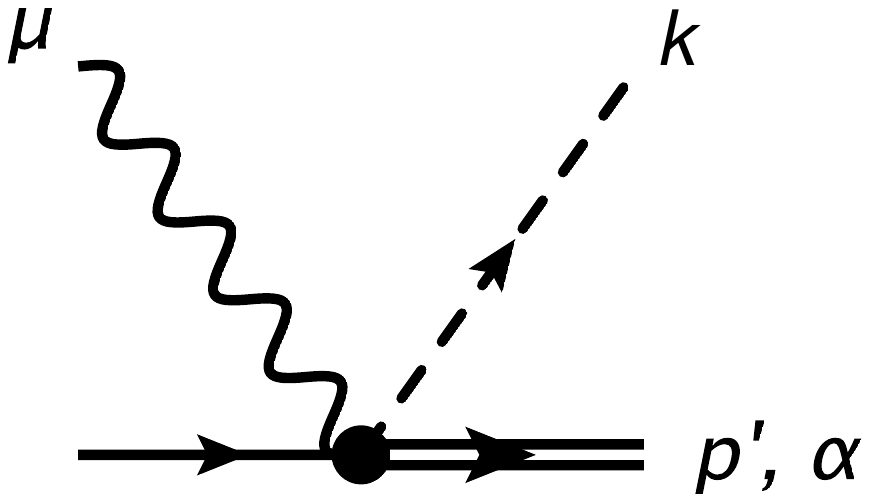}} 
\end{minipage}\hfill
\begin{minipage}{9cm}\begin{align}
\Ga_{ N\gamma  \pi \Delta}^{\al \mu}(p')&=\frac{ih_A e}{2f_\pi M_\Delta} \gamma^{\rho \al \mu}p'_\rho \times\nn\\
&\quad\times\begin{cases} 
1 & \gamma \,p \leftrightarrow \Delta^{++} \,\pi^-,\, \gamma \, n \leftrightarrow \Delta^{-} \,\pi^+,\\
1/\sqrt{3}& \gamma\, p \leftrightarrow \Delta^{0} \,\pi^+, \gamma \, n \leftrightarrow \Delta^{+} \,\pi^-;
\end{cases}\nn
\end{align}\end{minipage}
\\[0.5cm]
\item 4-point vertex with photon coupling minimally to the delta: $\gamma^*\,N \rightarrow \Delta \pi$\\[0.5cm]
\begin{minipage}{4cm}{\centering\includegraphics[scale=0.45]{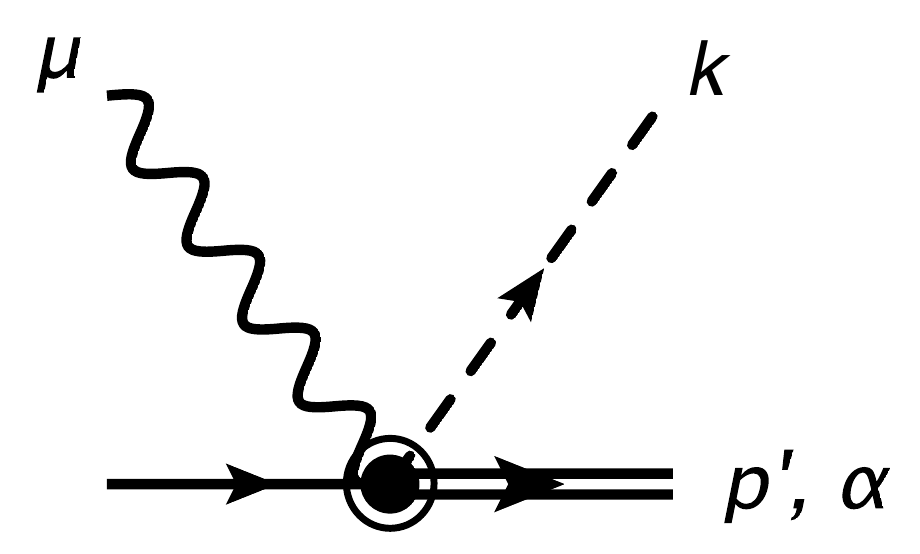}} 
\end{minipage}\hfill
\begin{minipage}{9cm}\begin{align}
\Ga_{N\gamma \Delta \pi}^{\al \mu}(k)=\frac{ih_A e_\Delta}{2f_\pi M_\Delta} \gamma^{\rho \al \mu}k_\rho\begin{cases}
1& \gamma \,p \leftrightarrow \Delta^{++} \,\pi^-,\\
-\sqrt{2/3}&\gamma\, p \leftrightarrow \Delta^{+} \,\pi^0,\\
1/\sqrt{3}&\gamma \, n \leftrightarrow \Delta^{+} \,\pi^-,\\
-1&\gamma \, n \leftrightarrow \Delta^{-} \,\pi^+.
\end{cases}\nn
\end{align}\end{minipage}
\\[0.5cm]
\end{itemize}
\subsection{Propagators}
\begin{itemize}
\item pion propagator:\\[0.5cm]
\begin{minipage}{4cm}{\centering\includegraphics[scale=0.45]{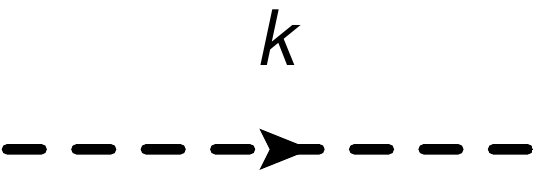}} 
\end{minipage}\hfill
\begin{minipage}{9cm}\begin{align}
S_\pi(k)=\frac{1}{k^2-M_\pi^2+i 0^+}\nn
\end{align}\end{minipage}
\\[2.5cm]
\item nucleon propagator:\\[0.5cm]
\begin{minipage}{4cm}{\centering\includegraphics[scale=0.45]{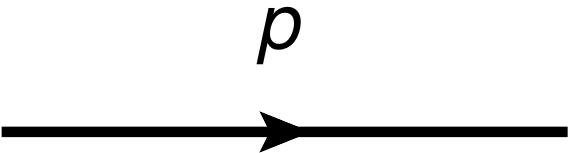}} 
\end{minipage}\hfill
\begin{minipage}{9cm}\begin{align}
S_N(p)=\frac{\slashed{p}+M_N}{p^2-M_N^2+i 0^+}\nn
\end{align}\end{minipage}
\\[0.5cm]
\item delta propagator \cite{Pascalutsa:2003aa}:\\[0.5cm]
\begin{minipage}{4cm}{\centering\includegraphics[scale=0.45]{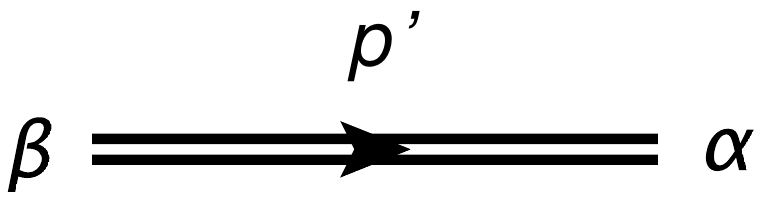}} 
\end{minipage}\hfill
\begin{minipage}{9cm}\begin{align}
S^{\al \be}_\Delta(p')&=\frac{\slashed{p}'+M_\Delta}{p^{\prime\,2}-M_\Delta^2+i 0^+}\left[-g^{\al \be}+\frac{1}{3} \gamma^\al \gamma^\be\right.\nn\\
&\left.\quad+\frac{1}{3M_\Delta}\left(\gamma^\al p^{\prime\,\beta}-\gamma^\beta p^{\prime\,\al}\right)+\frac{2}{3M_\Delta^2}p^{\prime\,\al}p^{\prime\,\beta}\right].\nn
\end{align}\end{minipage}
\\[0.5cm]
\end{itemize}
\addtocontents{toc}{\protect\setcounter{tocdepth}{4}}

\section{Rarita-Schwinger Vector-Spinors} \seclab{RSVS}
\addtocontents{toc}{\protect\setcounter{tocdepth}{0}}

The $\Delta(1232)$-resonance is a spin-3/2 particle. It can be described by a vector-spinor $U_\mu(p,\lambda)$, which satisfies the Rarita-Schwinger equation \cite{Rarita:1941mf}:
\beq
\left(\gamma^\al p_\al - M_\Delta\right)U_\mu(p,\lambda)=0.
\eeq
Here, $p$ is the four momentum of the delta and $\lambda$ is its helicity.
In addition, $U_\mu(p,\lambda)$ has to fulfil the following set of supplementary conditions:
\beq
\gamma^\mu U_\mu(p,\lambda)=0, \quad p^\mu U_\mu(p,\lambda)=0.
\eeq
The proper Rarita-Schwinger vector-spinors can be constructed as \cite{GarciaRavelo:2002nd}:
\beq
U_\mu(p,\lambda)=\sum_{\al \be}C^{\,1\;\nicefrac{1}{2}\;\nicefrac{3}{2}}_{\,\al\,\be\,\lambda}\, \epsilon_\mu (p,\al)\, u(p,\be),
\eeq
where $\epsilon_\mu$ is a polarization vector, $u$ is a Dirac spinor and the coefficient is a conventional Clebsch-Gordan coefficient. The summation runs over $\al$ and $\be$, the helicities of the polarization vector and the Dirac spinor, respectively.

In the following, we construct Rarita-Schwinger vector-spinors for different kinematics. In this Section, we chose the spin-3/2 spinors to be normalized according to $\ol U_\mu(p_\Delta) U^\mu(p_\Delta)=-1$.\footnote{In other words, we for simplicity remove a factor of $\sqrt{2M_\Delta}$ in each vector-spinor as compared to the rest of this thesis.} Accordingly, the Dirac spinors need to be normalized as $\bar u(p_\Delta) u(p_\Delta)=1$. The Dirac spinors are defined by:
\beq
u(p,\beta)=A\left(\begin{array}{ccc}
\varphi_0\\
\frac{\boldsymbol{\sigma}\cdot \bp}{ E+ M}\varphi_0
\end{array}\right), \qquad \text{with} \qquad A=\sqrt{\frac{E+M}{2M}},
\eeq
where $\varphi_0=(1,0)$ stands for the spin projection $\beta=+1/2$ along the $z$-axis and $\varphi_0=(0,1)$ for the spin projection $\beta=-1/2$.
They take the general form:
\beq
u(p)=\sqrt{\frac{E+M}{M}}\left(\frac{\chi_1}{\sqrt{2}},\frac{\chi_2}{\sqrt{2}},\frac{\chi_2 \,p_1-i \chi_2\, p_2+\chi_1\, p_3}{\sqrt{2}\,(E+M)},\frac{\chi_1\, p_1+i \chi_1 \,p_2-\chi_2 \,p_3}{\sqrt{2}\,(E+M)}\right), \eqlab{generalDS}
\eeq
whereby $p=\left(E,p_1,p_2,p_3\right)$ and $p^2=M^2$. For the transverse spin-1 polarization vectors we use Eqs.~\eref{epsTp} and \eref{epsTm}.
\subsection{Rest Frame}
\begin{subequations}
\begin{small}
\eqlab{RSVSrest}
\begin{align}
U(p,\lambda=3/2)&=\Big\{\Big(0,0,0,0\Big),\Big(-1/\sqrt{2},0,0,0\Big),\Big(-i/\sqrt{2},0,0,0\Big),\Big(0,0,0,0\Big) \Big\},\\
U(p,\lambda=1/2)&=\Big\{\Big(0,0,0,0\Big),\Big(0,-1/\sqrt{6},0,0\Big),\Big(0,-i/\sqrt{6},0,0\Big),\Big(\sqrt{2/3},0,0,0\Big) \Big\},\qquad\\
U(p,\lambda=-1/2)&=\Big\{\Big(0,0,0,0\Big),\Big(1/\sqrt{6},0,0,0\Big),\Big(-i/\sqrt{6},0,0,0\Big),\Big(0,\sqrt{2/3},0,0\Big) \Big\},\\
U(p,\lambda=-3/2)&=\Big\{\Big(0,0,0,0\Big),\Big(0,1/\sqrt{2},0,0\Big),\Big(0,-i/\sqrt{2},0,0\Big),\Big(0,0,0,0\Big) \Big\},
\end{align}
\end{small}
\end{subequations}
with $p=(M_\Delta,0,0,0)$ and $\epsilon_0=\left(0,0,0,1\right)$.
\subsection{Momentum Along $z$-Direction}
\begin{subequations}
\begin{footnotesize}
\begin{alignat}{3}
&U(p,\lambda=3/2)&&=\Big\{\Big(0,0,0,0\Big),\Big(-\frac{1}{2}\sqrt{\frac{E_\Delta+M_\Delta}{M_\Delta}},0,-\frac{\vert\bp\vert}{2M_\Delta}\sqrt{\frac{M_\Delta}{E_\Delta+M_\Delta}},0\Big),\\
&&&\qquad\Big(-\frac{i}{2}\sqrt{\frac{E_\Delta+M_\Delta}{M_\Delta}},0,-\frac{i\vert\bp\vert}{2M_\Delta}\sqrt{\frac{M_\Delta}{E_\Delta+M_\Delta}},0\Big),\Big(0,0,0,0\Big) \Big\},\quad\nn\\
&U(p,\lambda=1/2)&&=\Big\{\Big(\frac{\vert\bp\vert}{\sqrt{3}M_\Delta}\sqrt{\frac{E_\Delta+M_\Delta}{M_\Delta}},0,\frac{\vert\bp\vert^2}{\sqrt{3}M_\Delta^2}\sqrt{\frac{M_\Delta}{E_\Delta+M_\Delta}},0\Big),\\
&&&\qquad\Big(0,-\frac{1}{2\sqrt{3}}\sqrt{\frac{E_\Delta+M_\Delta}{M_\Delta}},0,\frac{\vert\bp\vert}{2\sqrt{3} M_\Delta}\sqrt{\frac{M_\Delta}{E_\Delta+M_\Delta}}\Big),\nn\\
&&&\qquad\Big(0,-\frac{i}{2\sqrt{3}}\sqrt{\frac{E_\Delta+M_\Delta}{M_\Delta}},0,\frac{i\vert\bp\vert}{2\sqrt{3}M_\Delta}\sqrt{\frac{M_\Delta}{E_\Delta+M_\Delta}}\Big),\qquad\nn\\
&&&\qquad\Big(\frac{E_\Delta}{\sqrt{3}M_\Delta}\sqrt{\frac{E_\Delta+M_\Delta}{M_\Delta}},0,\frac{E_\Delta \vert\bp\vert}{\sqrt{3}M_\Delta^2}\sqrt{\frac{M_\Delta}{E_\Delta+M_\Delta}},0\Big) \Big\},\nn\\
&U(p,\lambda=-1/2)&&=\Big\{\Big(0,\frac{\vert\bp\vert}{\sqrt{3}M_\Delta}\sqrt{\frac{E_\Delta+M_\Delta}{M_\Delta}},0,-\frac{\vert\bp\vert^2}{\sqrt{3}M_\Delta^2}\sqrt{\frac{M_\Delta}{E_\Delta+M_\Delta}}\Big),\\
&&&\qquad,\Big(\frac{1}{2\sqrt{3}}\sqrt{\frac{E_\Delta+M_\Delta}{M_\Delta}},0,\frac{\vert\bp\vert}{2\sqrt{3} M_\Delta}\sqrt{\frac{M_\Delta}{E_\Delta+M_\Delta}},0\Big),\nn\\
&&&\qquad\Big(-\frac{i}{2\sqrt{3}}\sqrt{\frac{E_\Delta+M_\Delta}{M_\Delta}},0,-\frac{i\vert\bp\vert}{2\sqrt{3}M_\Delta}\sqrt{\frac{M_\Delta}{E_\Delta+M_\Delta}},0\Big),\qquad\nn\\
&&&\qquad\Big(0,\frac{E_\Delta}{\sqrt{3}M_\Delta}\sqrt{\frac{E_\Delta+M_\Delta}{M_\Delta}},0,-\frac{E_\Delta \vert\bp\vert}{\sqrt{3}M_\Delta^2}\sqrt{\frac{M_\Delta}{E_\Delta+M_\Delta}}\Big) \Big\},\nn\\
&U(p,\lambda=-3/2)&&=\Big\{\Big(0,0,0,0\Big),\Big(0,\frac{1}{2}\sqrt{\frac{E_\Delta+M_\Delta}{M_\Delta}},0,-\frac{\vert\bp\vert}{2M_\Delta}\sqrt{\frac{M_\Delta}{E_\Delta+M_\Delta}}\Big),\\
&&&\qquad\Big(0,-\frac{i}{2}\sqrt{\frac{E_\Delta+M_\Delta}{M_\Delta}},0,\frac{i\vert\bp\vert}{2M_\Delta}\sqrt{\frac{M_\Delta}{E_\Delta+M_\Delta}}\Big),\Big(0,0,0,0\Big) \Big\},\nn
\end{alignat}
\end{footnotesize}
\end{subequations}
with $p=(E_\Delta,0,0,\vert\bp\vert)$ and $\epsilon_0=\frac{1}{M_\Delta}\left(\vert\bp\vert,0,0,E_\Delta\right)$

\subsection{Arbitrary Momentum}
\begin{subequations}
\eqlab{RaritaSchwingerArbitrary}
\begin{scriptsize}
\begin{alignat}{3}
&U(p,\lambda=3/2)&&=\Big\{\Big(-\frac{\vert\bp\vert \sin \varphi}{2M_\Delta}\sqrt{\frac{E_\Delta+M_\Delta}{M_\Delta}},0,-\frac{\vert\bp\vert^2 \sin 2 \varphi}{4M_\Delta^2}\sqrt{\frac{M_\Delta}{E_\Delta+M_\Delta}},-\frac{\vert\bp\vert^2 \sin^2 \varphi}{2M_\Delta^2}\sqrt{\frac{M_\Delta}{E_\Delta+M_\Delta}}\Big),\\
&&&\qquad\Big(-\frac{1}{2M_\Delta}\left(E_\Delta\sin^2 \varphi+M_\Delta\cos^2\varphi\right)\sqrt{\frac{E_\Delta+M_\Delta}{M_\Delta}},0,\nn\\
&&&\qquad \quad-\frac{\vert\bp\vert \cos \varphi}{2M_\Delta^2}\left(E_\Delta\sin^2 \varphi+M_\Delta\cos^2\varphi\right)\sqrt{\frac{M_\Delta}{E_\Delta+M_\Delta}},\nn\\
&&&\qquad\quad-\frac{\vert\bp\vert \sin \varphi}{2M_\Delta^2}\left(E_\Delta \sin^2 \varphi+M_\Delta\cos^2\varphi\right)\sqrt{\frac{M_\Delta}{E_\Delta+M_\Delta}}\Big),\nn\\
&&&\qquad\Big(-\frac{i}{2}\sqrt{\frac{E_\Delta+M_\Delta}{M_\Delta}},0,-\frac{i\vert\bp\vert \cos \varphi}{2M_\Delta}\sqrt{\frac{M_\Delta}{E_\Delta+M_\Delta}},-\frac{i\vert\bp\vert \sin \varphi}{2M_\Delta}\sqrt{\frac{M_\Delta}{E_\Delta+M_\Delta}}\Big),\nn\\
&&&\qquad\Big(\frac{\left(M_\Delta-E_\Delta\right)\sin 2 \varphi}{4M_\Delta}\sqrt{\frac{E_\Delta+M_\Delta}{M_\Delta}},0,\frac{\vert\bp\vert\left(M_\Delta-E_\Delta\right)\cos^2\varphi\sin\varphi}{2M_\Delta^2}\sqrt{\frac{M_\Delta}{E_\Delta+M_\Delta}},\nn\\
&&&\qquad\quad\frac{\vert\bp\vert\left(M_\Delta-E_\Delta\right)\cos\varphi\sin^2\varphi}{2M_\Delta^2}\sqrt{\frac{M_\Delta}{E_\Delta+M_\Delta}}\Big) \Big\},\nn\\
&U(p,\lambda=1/2)&&=\Big\{\Big(\frac{\vert\bp\vert \cos \varphi}{\sqrt{3}M_\Delta}\sqrt{\frac{E_\Delta+M_\Delta}{M_\Delta}},-\frac{\vert\bp\vert \sin \varphi}{2\sqrt{3}M_\Delta}\sqrt{\frac{E_\Delta+M_\Delta}{M_\Delta}},\\
&&&\qquad\quad\frac{\vert\bp\vert^2\left( 1+3\cos 2\varphi\right)}{4\sqrt{3}M_\Delta^2}\sqrt{\frac{M_\Delta}{E_\Delta+M_\Delta}},\frac{\sqrt{3}\vert\bp\vert^2 \sin 2\varphi}{4M_\Delta^2}\sqrt{\frac{M_\Delta}{E_\Delta+M_\Delta}}\Big),\nn\\
&&&\qquad\Big(\frac{(E_\Delta-M_\Delta) \cos \varphi \sin \varphi}{\sqrt{3}M_\Delta}\sqrt{\frac{E_\Delta+M_\Delta}{M_\Delta}},-\frac{\left(E_\Delta\sin^2 \varphi+M_\Delta\cos^2\varphi\right)}{2\sqrt{3}M_\Delta}\sqrt{\frac{E_\Delta+M_\Delta}{M_\Delta}},\nn\\
&&&\qquad\quad\frac{\vert\bp\vert \left(E_\Delta(1+3 \cos 2 \varphi)-6M_\Delta \cos^2 \varphi\right)\sin \varphi }{4\sqrt{3}M_\Delta^2}\sqrt{\frac{M_\Delta}{E_\Delta+M_\Delta}},\nn\\
&&&\qquad\quad\frac{\vert\bp\vert \left(3E_\Delta\sin^2 \varphi+M_\Delta (\cos^2 \varphi-2\sin^2 \varphi)\right)\cos \varphi }{2\sqrt{3}M_\Delta^2}\sqrt{\frac{M_\Delta}{E_\Delta+M_\Delta}}\Big),\nn\\
&&&\qquad\Big(0,-\frac{i}{2\sqrt{3}}\sqrt{\frac{E_\Delta+M_\Delta}{M_\Delta}},-\frac{i\vert\bp\vert \sin \varphi}{2\sqrt{3}M_\Delta}\sqrt{\frac{M_\Delta}{E_\Delta+M_\Delta}},\frac{i\vert\bp\vert \cos \varphi}{2\sqrt{3}M_\Delta}\sqrt{\frac{M_\Delta}{E_\Delta+M_\Delta}}\Big),\nn\\
&&&\qquad\Big(\frac{\left(M_\Delta\sin^2 \varphi+E_\Delta\cos^2\varphi\right)}{\sqrt{3}M_\Delta}\sqrt{\frac{E_\Delta+M_\Delta}{M_\Delta}},\frac{\left(M_\Delta-E_\Delta\right)\sin 2 \varphi}{4\sqrt{3}M_\Delta}\sqrt{\frac{E_\Delta+M_\Delta}{M_\Delta}},\nn\\
&&&\qquad\quad\frac{\vert\bp\vert \left(E_\Delta(1+3 \cos 2 \varphi)-6M_\Delta \sin^2 \varphi\right)\cos \varphi }{4\sqrt{3}M_\Delta^2}\sqrt{\frac{M_\Delta}{E_\Delta+M_\Delta}},\nn\\
&&&\qquad\quad\frac{\vert\bp\vert\left(M_\Delta(2 \sin^2 \varphi-\cos^2 \varphi)+3 E_\Delta \cos ^2\varphi\right)\sin \varphi}{2\sqrt{3}M_\Delta^2}\sqrt{\frac{M_\Delta}{E_\Delta+M_\Delta}}\Big) \Big\},\nn\\
&U(p,\lambda=-1/2)&&=\Big\{\Big(\frac{\vert\bp\vert \sin \varphi}{2\sqrt{3}M_\Delta}\sqrt{\frac{E_\Delta+M_\Delta}{M_\Delta}},\frac{\vert\bp\vert \cos \varphi}{\sqrt{3}M_\Delta}\sqrt{\frac{E_\Delta+M_\Delta}{M_\Delta}},\\
&&&\qquad\quad\frac{\sqrt{3}\vert\bp\vert^2 \sin 2\varphi}{4M_\Delta^2}\sqrt{\frac{M_\Delta}{E_\Delta+M_\Delta}},-\frac{\vert\bp\vert^2\left( 1+3\cos 2\varphi\right)}{4\sqrt{3}M_\Delta^2}\sqrt{\frac{M_\Delta}{E_\Delta+M_\Delta}}\Big),\nn\\
&&&\qquad\Big(\frac{\left(E_\Delta\sin^2 \varphi+M_\Delta\cos^2\varphi\right)}{2\sqrt{3}M_\Delta}\sqrt{\frac{E_\Delta+M_\Delta}{M_\Delta}},\frac{(E_\Delta-M_\Delta) \cos \varphi \sin \varphi}{\sqrt{3}M_\Delta}\sqrt{\frac{E_\Delta+M_\Delta}{M_\Delta}},\nn\\
&&&\qquad\quad\frac{\vert\bp\vert \left(3E_\Delta\sin^2 \varphi+M_\Delta (\cos^2 \varphi-2\sin^2 \varphi)\right)\cos \varphi }{2\sqrt{3}M_\Delta^2}\sqrt{\frac{M_\Delta}{E_\Delta+M_\Delta}},\nn\\
&&&\qquad\quad-\frac{\vert\bp\vert \left(E_\Delta(1+3 \cos 2 \varphi)-6M_\Delta \cos^2 \varphi\right)\sin \varphi }{4\sqrt{3}M_\Delta^2}\sqrt{\frac{M_\Delta}{E_\Delta+M_\Delta}}\Big),\nn\\
&&&\qquad\Big(-\frac{i}{2\sqrt{3}}\sqrt{\frac{E_\Delta+M_\Delta}{M_\Delta}},0,-\frac{i\vert\bp\vert \cos \varphi}{2\sqrt{3}M_\Delta}\sqrt{\frac{M_\Delta}{E_\Delta+M_\Delta}},-\frac{i\vert\bp\vert \sin \varphi}{2\sqrt{3}M_\Delta}\sqrt{\frac{M_\Delta}{E_\Delta+M_\Delta}}\Big),\nn\\
&&&\qquad\Big(\frac{\left(E_\Delta-M_\Delta\right)\sin 2 \varphi}{4\sqrt{3}M_\Delta}\sqrt{\frac{E_\Delta+M_\Delta}{M_\Delta}},\frac{\left(M_\Delta\sin^2 \varphi+E_\Delta\cos^2\varphi\right)}{\sqrt{3}M_\Delta}\sqrt{\frac{E_\Delta+M_\Delta}{M_\Delta}},\nn\\
&&&\qquad\quad\frac{\vert\bp\vert\left(M_\Delta(2 \sin^2 \varphi-\cos^2 \varphi)+3 E_\Delta \cos ^2\varphi\right)\sin \varphi}{2\sqrt{3}M_\Delta^2}\sqrt{\frac{M_\Delta}{E_\Delta+M_\Delta}},\nn\\
&&&\qquad\quad-\frac{\vert\bp\vert \left(E_\Delta(1+3 \cos 2 \varphi)-6M_\Delta \sin^2 \varphi\right)\cos \varphi }{4\sqrt{3}M_\Delta^2}\sqrt{\frac{M_\Delta}{E_\Delta+M_\Delta}}\Big) \Big\},\nn\\
&U(p,\lambda=-3/2)&&=\Big\{\Big(0,\frac{\vert\bp\vert \sin \varphi}{2M_\Delta}\sqrt{\frac{E_\Delta+M_\Delta}{M_\Delta}},\frac{\vert\bp\vert^2 \sin^2 \varphi}{2M_\Delta^2}\sqrt{\frac{M_\Delta}{E_\Delta+M_\Delta}},-\frac{\vert\bp\vert^2 \sin 2 \varphi}{4M_\Delta^2}\sqrt{\frac{M_\Delta}{E_\Delta+M_\Delta}}\Big),\\
&&&\qquad\Big(0,\frac{1}{2M_\Delta}\left(E_\Delta \sin^2 \varphi+M_\Delta\cos^2\varphi\right)\sqrt{\frac{E_\Delta+M_\Delta}{M_\Delta}},\nn\\
&&&\qquad\quad\frac{\vert\bp\vert \sin \varphi}{2M_\Delta^2}\left(E_\Delta \sin^2 \varphi+M_\Delta\cos^2\varphi\right)\sqrt{\frac{M_\Delta}{E_\Delta+M_\Delta}},\nn\\
&&&\qquad\quad-\frac{\vert\bp\vert \cos \varphi}{2M_\Delta^2}\left(E_\Delta \sin^2 \varphi+M_\Delta\cos^2\varphi\right) \sqrt{\frac{M_\Delta}{E_\Delta+M_\Delta}}\Big),\nn\\
&&&\qquad\Big(0,-\frac{i}{2}\sqrt{\frac{E_\Delta+M_\Delta}{M_\Delta}},-\frac{i\vert\bp\vert \sin \varphi}{2M_\Delta}\sqrt{\frac{M_\Delta}{E_\Delta+M_\Delta}},\frac{i\vert\bp\vert \cos \varphi}{2M_\Delta}\sqrt{\frac{M_\Delta}{E_\Delta+M_\Delta}}\Big),\nn\\
&&&\qquad\Big(0,\frac{\left(E_\Delta-M_\Delta\right)\cos \varphi\sin \varphi}{2M_\Delta}\sqrt{\frac{E_\Delta+M_\Delta}{M_\Delta}},\frac{\vert\bp\vert\left(E_\Delta-M_\Delta\right)\cos\varphi\sin^2\varphi}{2M_\Delta^2}\sqrt{\frac{M_\Delta}{E_\Delta+M_\Delta}},\nn\\
&&&\qquad\quad\frac{\vert\bp\vert\left(M_\Delta-E_\Delta\right)\cos^2\varphi\sin\varphi}{2M_\Delta^2}\sqrt{\frac{M_\Delta}{E_\Delta+M_\Delta}}\Big) \Big\},\nn
\end{alignat}
\end{scriptsize}
\end{subequations}
with $p=(E_\Delta,\vert\bp\vert \sin \varphi,0,\vert\bp\vert \cos \varphi)$. 

From \Eqref{RaritaSchwingerArbitrary} we derive the spin-energy projection operator:
\begin{subequations}
\eqlab{SEprojec}
\bea
\Sigma^{\mu \nu}(p)&=&\sum_\lambda U^\mu(p,\lambda)\,\ol U^\nu(p,\lambda),\\
&=&\frac{M_\Delta +\slashed{p}}{2M_\Delta}\left\{-g^{\mu \nu}+\frac{1}{3}\gamma^\mu \gamma^\nu+\frac{2}{3M_\Delta^2}\, p^\mu p^\nu+\frac{p^\nu \gamma^\mu-p^\mu \gamma^\nu}{3M_\Delta}\right\},
\eea
\end{subequations}
which is relevant for the calculation of the $\pi\Delta$-production cross sections in \secref{chap4}{deltaCS}.
\addtocontents{toc}{\protect\setcounter{tocdepth}{4}}
\section[Compton Scattering off the Nucleon with $\Delta$-Exchange]{Compton Scattering off the Nucleon with $\boldsymbol{\Delta}$-Exchange}

\addtocontents{toc}{\protect\setcounter{tocdepth}{0}}

\subsection{Compton Scattering Amplitudes}\seclab{VVCSDeltaAmp}
In \secref{chap4}{3.1CSDelta}, we discussed the process of CS off the Nucleon with $\Delta$-exchange. The corresponding nucleon structure functions were given in \Eqref{structurefunc}. Here, we present our results for the real part of the  tree-level CS amplitudes, cf.\ \Figref{DeltaExchange}, were we use the shorthands defined in \Eqref{shorthands}. In \Eqref{DeltaAmps}, we distinguished the $\Delta$-pole and the non-pole terms, and also the $T_1$ subtraction function. The respective terms are provided below.

Omitting the prefactor $\left[(s-M_\Delta^2)(u-M_\Delta^2)\right]^{-1}$, the $\Delta$-pole contributions read:
\begin{subequations}
\eqlab{Dpole}
\begin{alignat}{3}
&T_1^{\Delta\mathrm{-pole}}(\nu,Q^2)&&\propto\frac{2 \pi \al \nu^2 Q_-^2}{M M_\Delta M_+^2 \omega_+}\Big[g_M^2 Q_+^4+4 g_E^2 M_\Delta^2 \omega_-^2+4 g_\mathrm{C}^2 Q^4\\
&&&\quad-2 g_M g_E M_\Delta Q_+^2 \omega_-+2 g_M g_\mathrm{C} Q^2 Q_+^2-8 g_E g_\mathrm{C} M_\Delta Q^2 \omega_-\Big]\nn,\\
&T_2^{\Delta\mathrm{-pole}}(\nu,Q^2)&&\propto\frac{8\pi \al M_\Delta Q^2\omega_+}{M M_+^2}\Big[g_M^2 Q_+^2+g_E^2 Q_-^2+\frac{g_\mathrm{C}^2 Q^2 Q_-^2}{M_\Delta^2}\\
&&&\quad-2 g_M g_E M_\Delta \omega_-+2 g_M g_\mathrm{C} Q^2\Big]\nn,\\
&S_1^{\Delta\mathrm{-pole}}(\nu,Q^2)&&\propto-\frac{4\pi \al M_\Delta^2 \omega_+^2}{M M_+^2}\left[g_M^2 Q_+^2+\frac{ g_E^2  \omega_- \left(\varDelta ^2-Q^2\right)}{\omega_+}+\frac{2 \varDelta  g_\mathrm{C}^2 Q^4}{M_\Delta^2 \omega_+}\right.\\
&&&\quad-\frac{2 g_M g_E \left(M_\Delta M Q^2+\varDelta ^2 M_+^2-Q^4\right)}{M_\Delta\omega_+}+2g_M g_\mathrm{C} Q^2\left\{4-\frac{M \omega_-}{M_\Delta \omega_+}\right\}\quad\nn\\
&&&\quad\left.-\frac{2g_E g_\mathrm{C} Q^2 ( \omega_- (2 M_\Delta-M)- \varDelta   M)}{M_\Delta \omega_+}\right],\nn\\
&S_2^{\Delta\mathrm{-pole}}(\nu,Q^2)&&\propto\frac{2\pi \al  M \nu}{M_+^2}\left[2 g_M^2 M Q_+^2+4 g_E^2 M_\Delta \varDelta \omega_--\frac{2 g_\mathrm{C}^2 Q^2 \left(\varDelta ^2-Q^2\right)}{M_\Delta}\right.\\
&&&\quad+4 g_M g_E M_\Delta \left(M_\Delta \omega_+-4 M \omega_-\right)\nn\\
&&&\quad+\frac{g_M g_\mathrm{C} \left(16 M_\Delta M Q^2+\varDelta ^2 M_+^2-Q^4\right)}{M_\Delta}\nn\\
&&&\quad\left.+\frac{g_E g_\mathrm{C} \left(M_\Delta^4-6 M_\Delta^2 Q^2-M^4+2 M_\Delta M^3-2 M_\Delta^3 M+6 M_\Delta M Q^2+Q^4\right)}{M_\Delta}\right].\nn
\end{alignat}
\end{subequations}
The non-pole contributions are given by:
\begin{subequations}
\eqlab{noDpole}
\bea
\widetilde T_1(\nu,Q^2)&=&-\frac{4\pi \al  \nu^2}{M M_+^2}\left[g_M^2+g_E^2-g_M g_E\right],\eqlab{T1nonpole}\\
\widetilde T_2(\nu,Q^2)&=&-\frac{4\pi \al Q^2}{M M_+^2}\left[g_M^2+g_E^2-g_M g_E+\frac{g_\mathrm{C}^2 Q^2}{M_\Delta^2}\right],\\
\widetilde S_1(\nu,Q^2)&=&\frac{\pi \al }{M M_+^2}\left[g_M^2 Q_+^2+g_E^2 \left(\varDelta ^2-3 Q^2\right)+\frac{4 g_\mathrm{C}^2 Q^4}{M_\Delta^2}-8 g_M g_E M_\Delta \omega_-\right.\eqlab{S1nonpole}\\
&&\qquad\qquad\left.-\frac{2 g_M g_\mathrm{C} Q^2 (M-4 M_\Delta)}{M_\Delta}+\frac{2 g_E g_\mathrm{C} Q^2 (3 M-2 M_\Delta)}{M_\Delta}\right],\qquad\nn\\
\widetilde S_2(\nu,Q^2)&=&-\frac{2\pi \al M \nu}{M_\Delta M_+^2}\big[g_M +g_E \big]g_\mathrm{C}\,,\eqlab{nonpoleS2}
\eea
\end{subequations}
and the $T_1$ subtraction function is:
\bea
T_1(0,Q^2)&=&\frac{4\pi \al Q^4}{M_\Delta M_+ \omega_+}\left[\frac{g_M^2}{Q^2}-\frac{g_E^2 \varDelta}{M^2 M_+}-\frac{g_\mathrm{C}^2 \varDelta \left(M^2-Q^2\right)}{M^2 M_\Delta^2 M_+}+\frac{g_M g_E}{M M_+}\right.\eqlab{T1su}\\
&&\hspace{2.3cm}+\frac{g_M g_\mathrm{C}}{M M_+}+\left.\frac{2 g_E g_\mathrm{C} \left(M\varDelta +Q^2\right)}{M^2 M_\Delta M_+}\right].\nn
\eea

\subsection[$\Delta$-Production Cross Section from Helicity Amplitudes]{$\boldsymbol{\Delta}$-Production Cross Section from Helicity Amplitudes} \seclab{HelicityCS}
\begin{figure}[tbh]
\centering
       \includegraphics[scale=0.15]{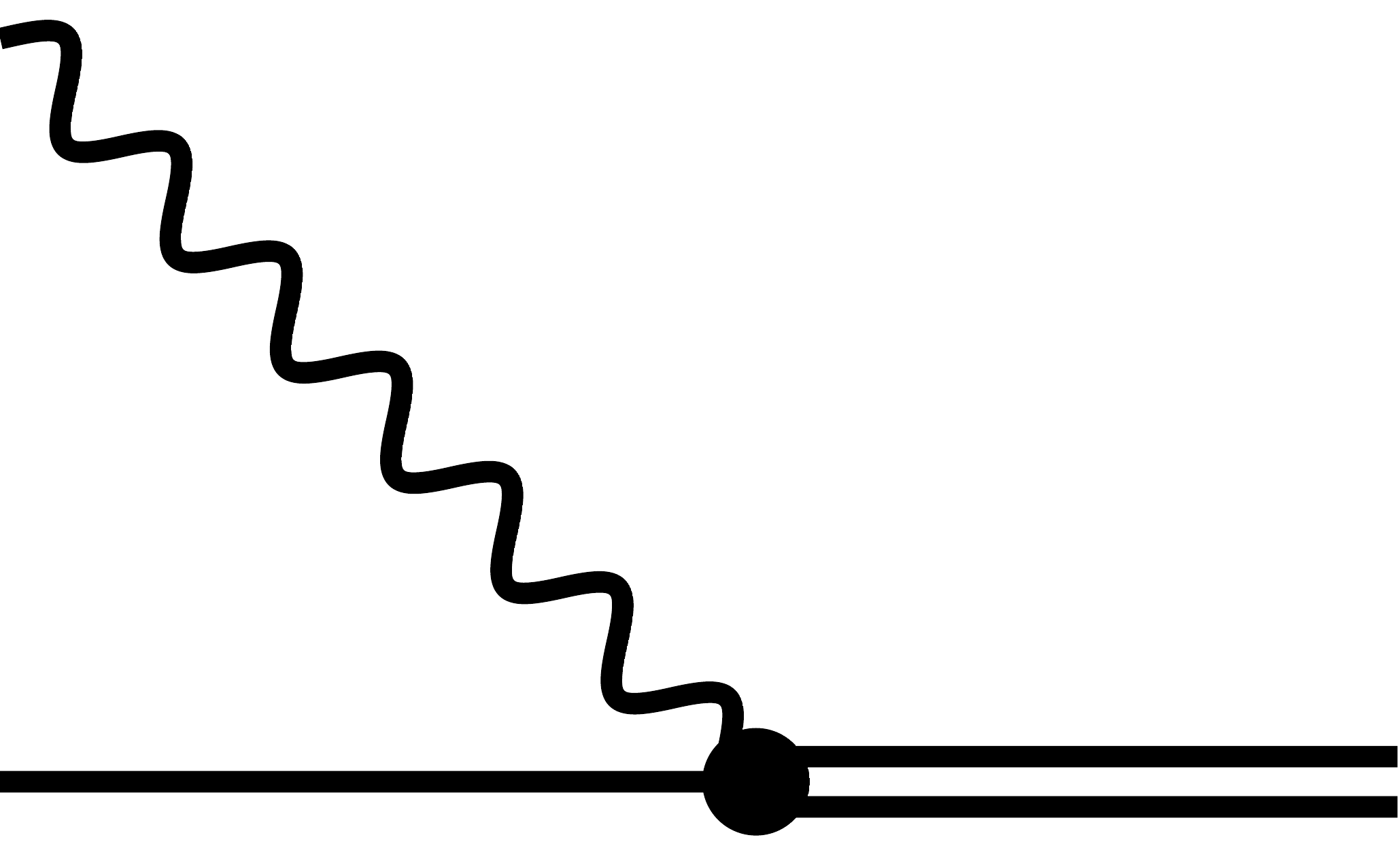}
\caption{$\Delta$-production mechanism.\label{fig:DeltaCS}}
\end{figure}
\noindent The general formula for a scattering cross section is given in \Eqref{generalCS}.
For the $\Delta$-production process, $\gamma^* N \rightarrow \Delta$, this becomes:
\beq
\dd \sigma= \frac{\pi}{4M_\Delta^2\sqrt{\nu^2+Q^2}}\,\vert \mathcal{M}_{fi}\vert^2\,\delta\left(M_\Delta-\left(E_p+\omega\right)\right) \delta\left(\bp_\Delta-\left(\bp+\bq\right)\right) \dd \bp_\Delta, \eqlab{CSDeltarest}
\eeq
where $I=\vert\bq\vert \left(E_p+\omega\right)=M_\Delta \sqrt{\nu^2+Q^2}$ and the matrix element corresponds to the diagram in \Figref{DeltaCS}.
It is convenient to work in the rest frame of the $\Delta$:
\bea
p=\left(E_p,0,0,-\vert \bq \vert\right),\quad q=\left(\omega,0,0,\vert \bq \vert\right),\quad
p'=\left(M_\Delta,0,0,0\right).
\eea
Employing the on-shell conditions on the nucleon and delta four-momenta, as well as the energy and momentum conservation stored in the $\delta$-functions, we obtain:
\bea
E_p&=&\frac{M_\Delta^2+M^2+Q^2}{2M_\Delta},\\
\omega&=&M_\Delta-\frac{M_\Delta^2+M^2+Q^2}{2M_\Delta},\\
\vert \bq \vert&=& \sqrt{Q^2+\left(M_\Delta-\frac{M_\Delta^2+M^2+Q^2}{2M_\Delta}\right)^2}\,.
\eea
The Rarita-Schwinger vector-spinors for the rest kinematics are derived in \Eqref{RSVSrest}, where we need to multiply by a factor of $\sqrt{2M_\Delta}$ to achieve the normalization in \Eqref{RSVSnorm}. The photon polarization vectors and the nucleon spinors are the same as in Eqs.~\eref{polvecs}-\eref{Nspins}. 

We then derive the following helicity amplitudes $T_{\la_\Delta'\la_\ga \la_N} $:
\begin{subequations}
\bea
T_{\nicefrac{3}{2}\;1\;-\nicefrac{1}{2}}&=&-\frac{e\,\sqrt{3} \left[G_M^*+G_E^*\right]Q_-Q_+^2}{2\sqrt{2} M(M+M_\Delta)},\eqlab{T32}\\
T_{\nicefrac{1}{2}\;1\;\nicefrac{1}{2}}&=&\frac{e\left[3\,G_E^*-G_M^*\right]Q_-Q_+^2}{2\sqrt{2} M(M+M_\Delta)},\eqlab{T12}\\
T_{\nicefrac{1}{2}\;0\;-\nicefrac{1}{2}}&=&\frac{e \,G_C^* \,Q Q_+ \vert \bq \vert}{M(M+M_\Delta)},\eqlab{T12L}\\
T_{-\nicefrac{1}{2}\;0\;\nicefrac{1}{2}}&=&-\frac{e \,G_C^* \,Q Q_+ \vert \bq \vert}{M(M+M_\Delta)}.\eqlab{Tm12L}
\eea
\end{subequations}
The results are conform with Ref.~\cite[Section 2.1.1.]{Pascalutsa:2006up}. Evaluating \Eqref{CSDeltarest} with the helicity amplitudes from Eqs.~\eref{T32} and \eref{T12}, we obtain the cross sections for total helicities of $3/2$ and $1/2$, respectively. Eqs.~\eref{T12L} and \eref{Tm12L} contain longitudinal photons, they produce cross sections with total helicities $-1/2$ and $1/2$. $\sigma_T$, $\sigma_{TT}$ and $\sigma_L$ follow as described in \secref{chap4}{SR}. For $\sigma_{LT}$ we produce the spin-flip of the nucleon by combining Eqs.~\eref{T12} and \eref{Tm12L}. 

The above derivation of the $\Delta$-production cross sections is useful to have as a cross-check for the structure functions in \Eqref{structurefunc}, which were calculated from the $\Delta$-exchange contribution to the CS off the nucleon. Indeed, the $\Delta$-production cross sections presented in here with the photon flux factor $\sqrt{\nu^2+Q^2} \,M_\Delta/M$ agree with our previous results.

\subsection[$Q^2$ Dependence of Nucleon Polarizabilities]{$\boldsymbol{Q^2}$ Dependence of Nucleon Polarizabilities}\seclab{Q2regpol}
We replace the prefactor \cite{Pascalutsa:2006up}:
\beq
\frac{e}{M\left(M+M_\Delta\right)}\rightarrow\frac{e\,(M+M_\Delta)}{M\left[(M+M_\Delta)^2+Q^2\right]},\eqlab{Q2cutoff}
\eeq
to introduce a cutoff on higher momentum transfers. The values of the GPs at the real photon point are not affected by this regularisation, cf.\ \secref{chap4}{DeltaPol}. For the slopes of the  GPs we notwithstanding obtain Eqs.~\eref{85}-\eref{88}.

\begin{itemize}
\item sum of electric and magnetic dipole polarizabilities:
\begin{align}
\eqlab{85}
\hspace{-0.7cm}\frac{\dd \left[\alpha_{E1}(Q^2)+\beta_{M1}(Q^2)\right]}{\dd Q^2}\Big\vert_{Q^2=0}=&\;\frac{e^2 \varDelta}{\pi M_+^2}\left(\frac{g_M^2}{\varDelta^2}\left[\frac{1}{2 \varDelta ^2 }-\frac{1}{\varDelta  M_+}-\frac{1}{  M_+^2}\right]\right.\\
&+\frac{g_E^2}{M M_+}\left[\frac{1}{4 \varDelta^2 }-\frac{3}{4 \varDelta M_+}-\frac{1 }{2 M_+^2}\right]+\frac{ g_M g_C}{2  \varDelta^2  M M_+}\nn\\
&\left.-\frac{g_M g_E}{\varDelta M}\left[\frac{1}{4 \varDelta ^2}-\frac{1}{\varDelta  M_+}+\frac{1}{4 M_+^2}\right]-\frac{g_E g_C}{\varDelta M M_+^2}\right)\nn
\end{align}
\item longitudinal polarizability:
\begin{align}
\hspace{-1cm}\frac{\dd\, \al_L(Q^2)}{\dd Q^2}\Big\vert_{Q^2=0}=&\;\frac{e^2 M_\Delta^3}{\pi M_+^4}\left(\frac{g_E^2}{\varDelta^2 M_+^2}\left[\frac{2}{\varDelta M_\Delta }-\frac{4}{M M_+}\right]\right.\\
&\left.+\frac{g_C^2}{M_\Delta^4}\left[\frac{1}{2 \varDelta  M_\Delta}-\frac{1}{M M_+}\right]+\frac{g_E g_C}{M M_\Delta^3}\left[\frac{1}{\varDelta ^2}-\frac{4}{\varDelta  M_+}-\frac{2}{M_+^2}\right]\right)\nn
\end{align}
\item forward spin polarizability:
\begin{align}
\hspace{-0.3cm}\frac{\dd \gamma_0(Q^2)}{\dd Q^2}\Big\vert_{Q^2=0}=&\,-\frac{e^2}{\pi M_+^2}\left(\frac{g_M^2}{\varDelta^3}\left[\frac{1}{4 \varDelta }-\frac{1}{M_+}\right]+\frac{g_E^2}{M_+^2}\left[\frac{1}{4 \varDelta ^2}-\frac{3}{2 \varDelta  M_+}-\frac{1}{2 M_+^2}\right]\right.\\
&\left.-\frac{g_M g_E}{\varDelta M_+}\left[\frac{1}{\varDelta ^2}-\frac{5}{\varDelta M_+}-\frac{1}{M_+^2}\right]+\frac{2 g_M g_C}{\varDelta ^2 M_+^2}-\frac{g_E g_C}{\varDelta M_+^3}\right)\nn
\end{align}
\item longitudinal-transverse polarizability:
\begin{align}
\eqlab{88}
\hspace{-1cm}\frac{\dd\, \delta_{LT}(Q^2)}{\dd Q^2}\Big\vert_{Q^2=0}=&\;\frac{e^2 M_\Delta \varDelta}{4
\pi M M_+^2}\left(\frac{g_E^2 }{\varDelta M_+^2 }\left[\frac{1}{\varDelta ^2}-\frac{4}{\varDelta  M_+}-\frac{2}{M_+^2}\right]-\frac{g_C^2}{\varDelta M_\Delta^2 M_+^2}\right.\\
&+\frac{g_M g_E }{\varDelta^2 M_+}\left[\frac{1}{\varDelta ^2}-\frac{3}{\varDelta  M_+}-\frac{1}{M_+^2}\right]\nn\\
&\left.+\frac{g_M g_C}{\varDelta  M_\Delta^2 }\left[\frac{1}{2 \varDelta^2 }-\frac{2}{\varDelta M_+}+\frac{1 }{2 M_+^2}\right]-\frac{g_E g_C}{M_\Delta^2 M_+^2}\left[\frac{5}{2 \varDelta}+\frac{3 }{2 M_+}\right]\right).\nn
\end{align}
\end{itemize}
\addtocontents{toc}{\protect\setcounter{tocdepth}{4}}

\section[Plots of $\pi \Delta$-Production Cross Sections]{Plots of $\boldsymbol{\pi \Delta}$-Production Cross Sections} \seclab{CSplots}

Hereby we present the results of our tree-level calculation of the $\pi\Delta$-production cross sections.

    \begin{figure}[h!] 
    \centering 
      \vskip-2mm
       \includegraphics[width=13.5cm]{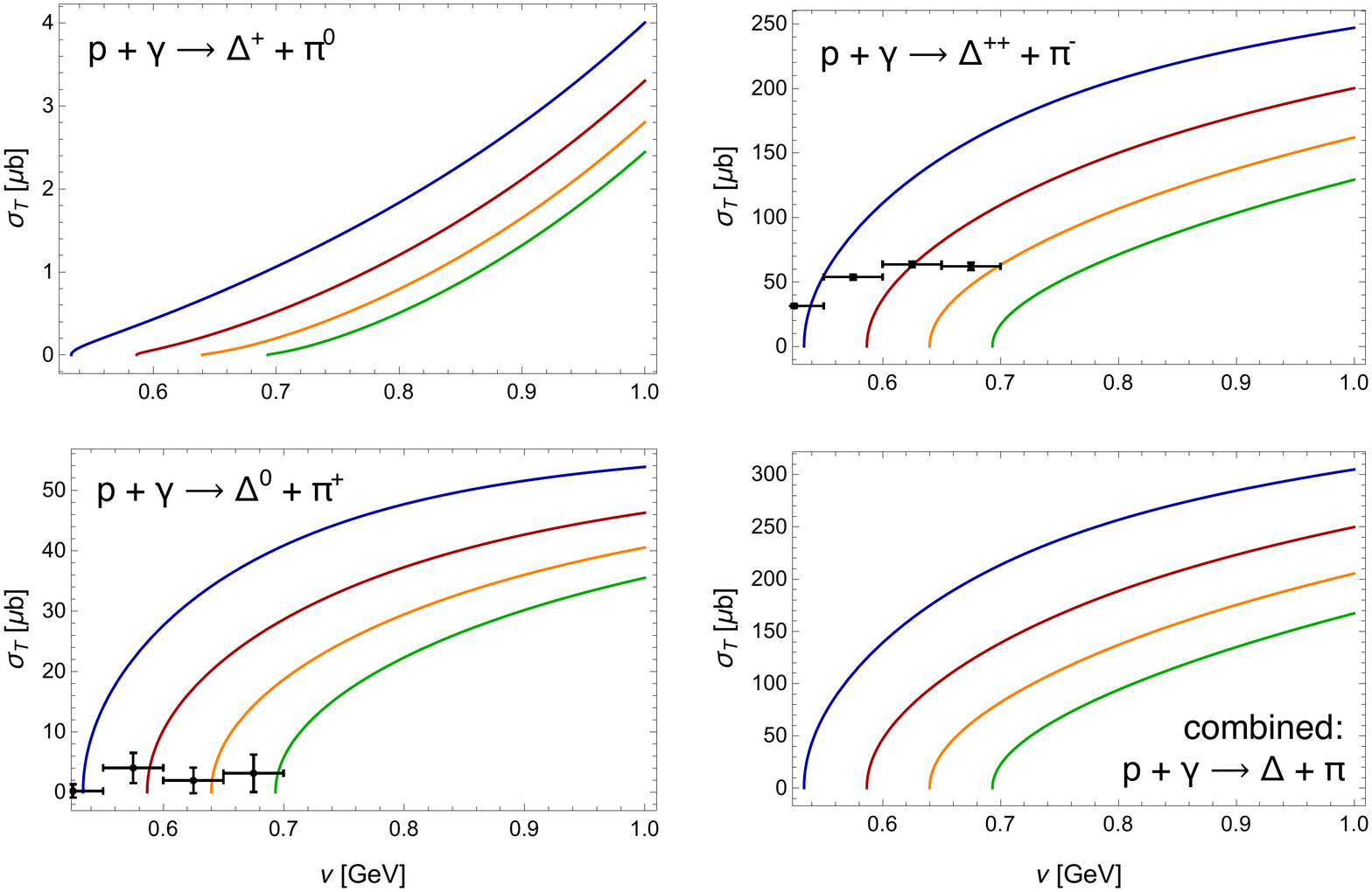}
       \vskip-5mm
       \caption{Total cross section for pion-delta electroproduction on the proton. The blue, red, orange and green curves correspond to $Q^2=0,\,0.1,\,0.2,\,0.3$ GeV$^2$. Data points are from~Ref.~\cite[Table 17-18]{Wu:2005wf}.}
              \label{fig:protonTplot}
\end{figure}

    \begin{figure}[h!] 
    \centering 
     \vskip-5mm
       \includegraphics[width=13.5cm]{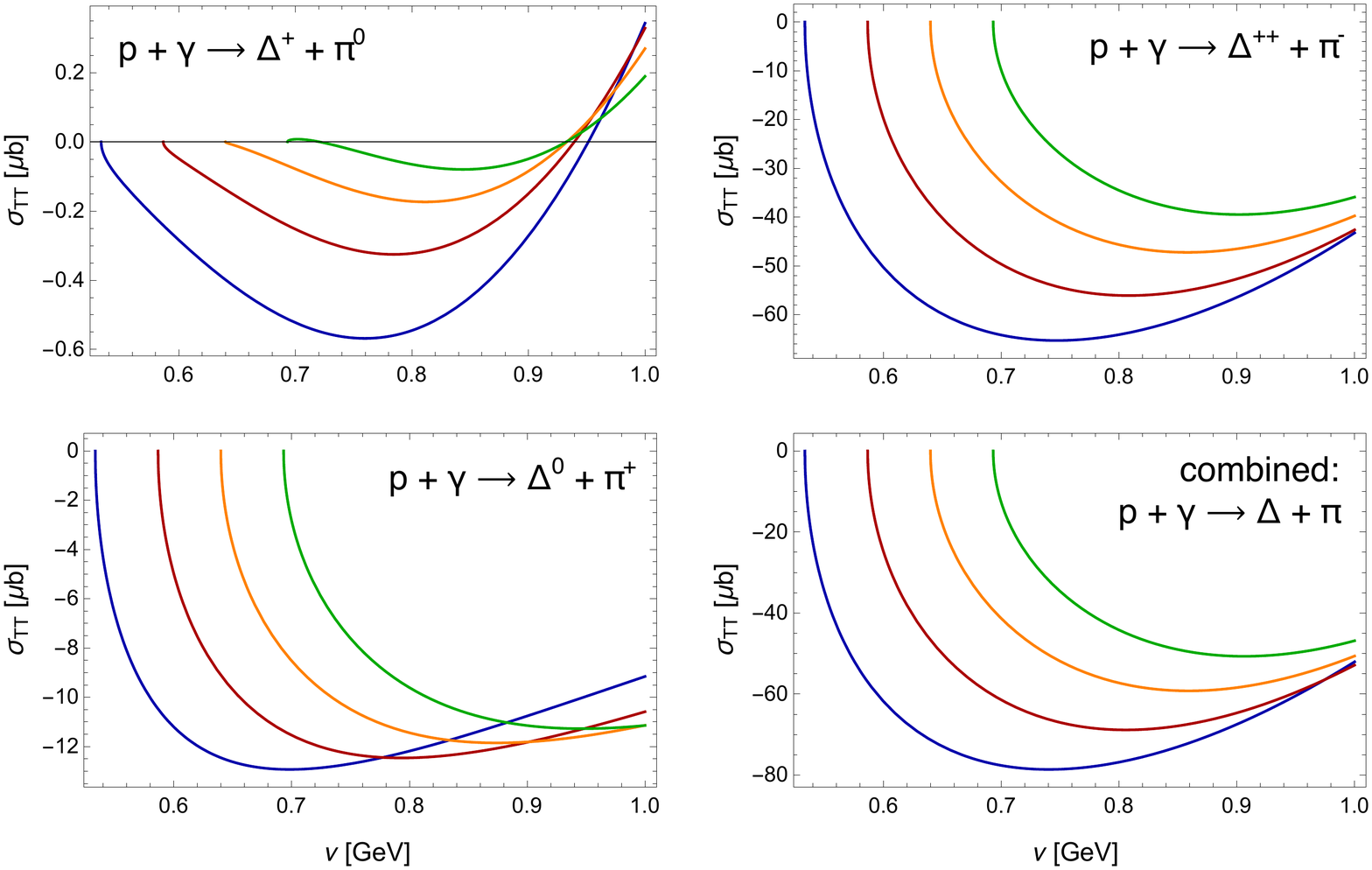}
       \vskip-5mm
       \caption{Polarized cross section $\sigma_{TT}$ for pion-delta electroproduction on the proton. Legend for the curves is the same as in \Figref{protonTplot}.}
              \label{fig:protonTTplot}
\end{figure}

    \begin{figure}[h!] 
    \centering 
    \vskip-5mm
       \includegraphics[width=13.5cm]{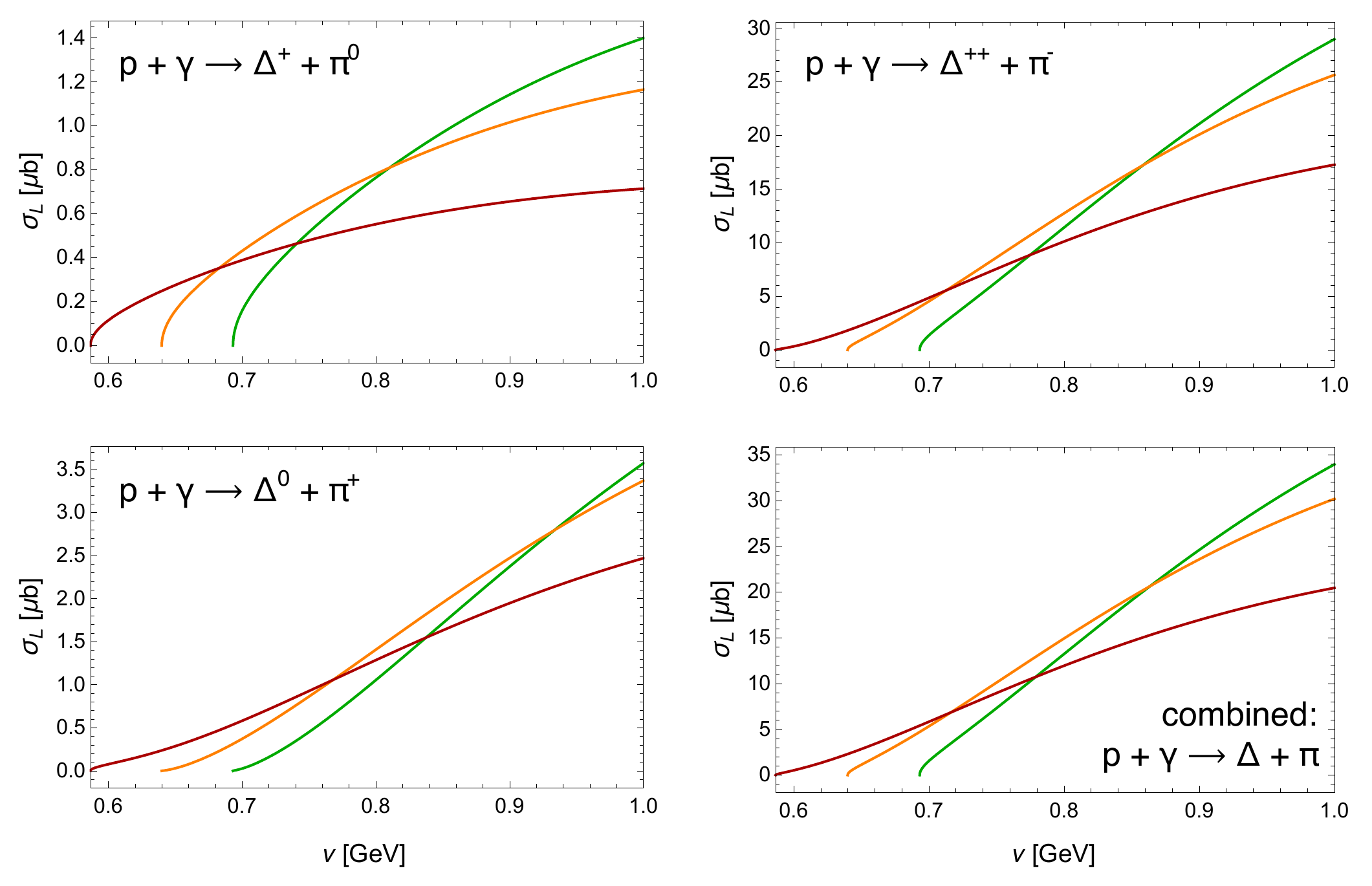}
       \caption{Longitudinal unpolarized cross section $\sigma_{L}$ for pion-delta electroproduction on the proton. The red, orange and green curves correspond to $Q^2=0.1,\,0.2,\,0.3$ GeV$^2$, respectively.}
              \label{fig:protonLplot}
\end{figure}

    \begin{figure}[h!] 
    \centering 
      \vskip-5mm
       \includegraphics[width=13.5cm]{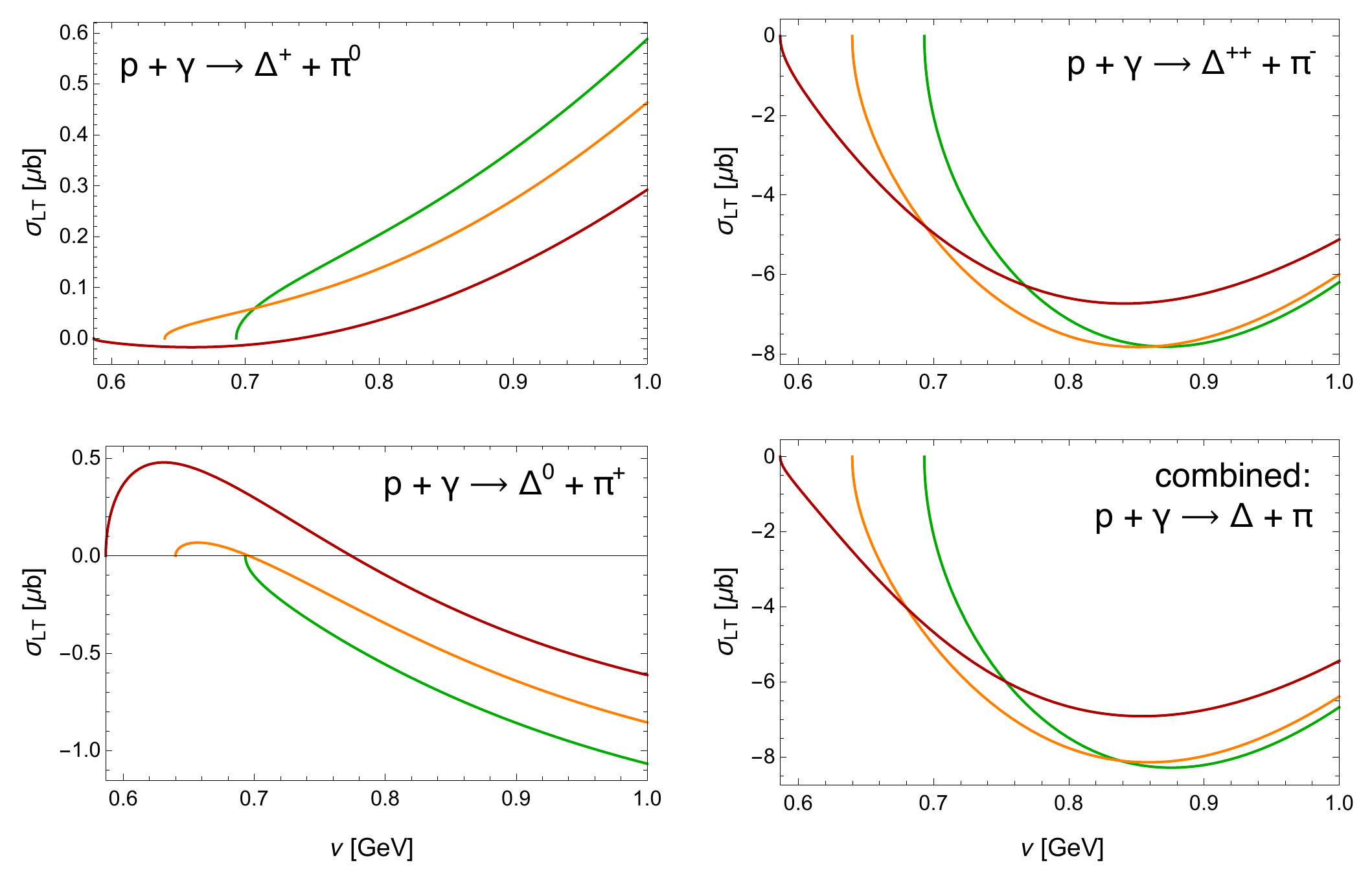}
       \caption{Longitudinal-transverse polarized cross section $\sigma_{LT}$ for pion-delta electroproduction on the proton. Legend for the curves is the same as in \Figref{protonLplot}. }
              \label{fig:protonLTplot}
\end{figure}

    \begin{figure}[h!] 
    \centering 
       \vskip-5mm
       \includegraphics[width=13.5cm]{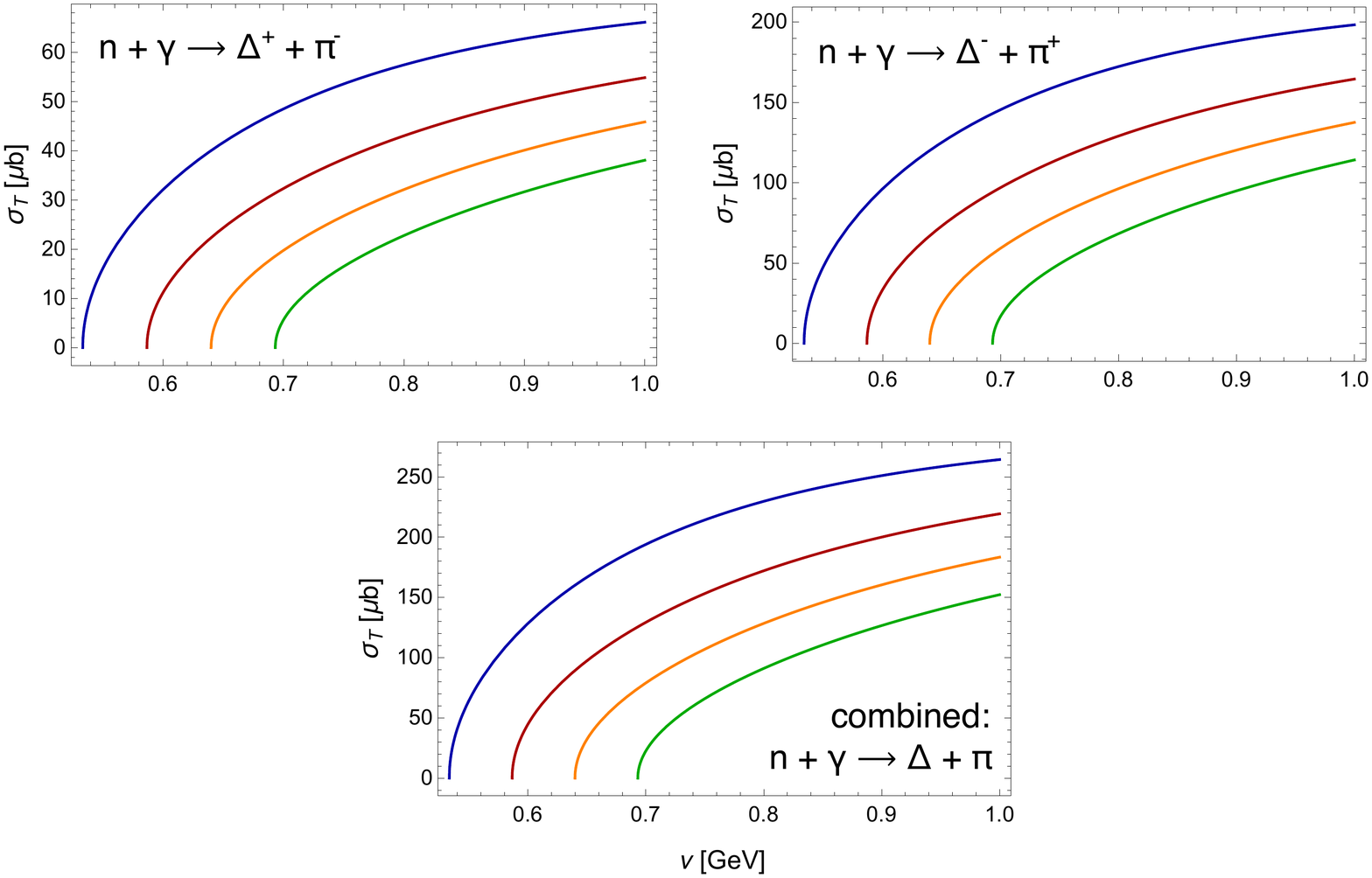}
       \vskip-5mm
       \caption{Unpolarized total cross section $\sigma_T$ for pion-delta electroproduction on the neutron. Legend for the curves is the same as in \Figref{protonTplot}.}
              \label{fig:neutronTplot}
\end{figure}

    \begin{figure}[h!] 
    \centering 
       \vskip-5mm
       \includegraphics[width=13.5cm]{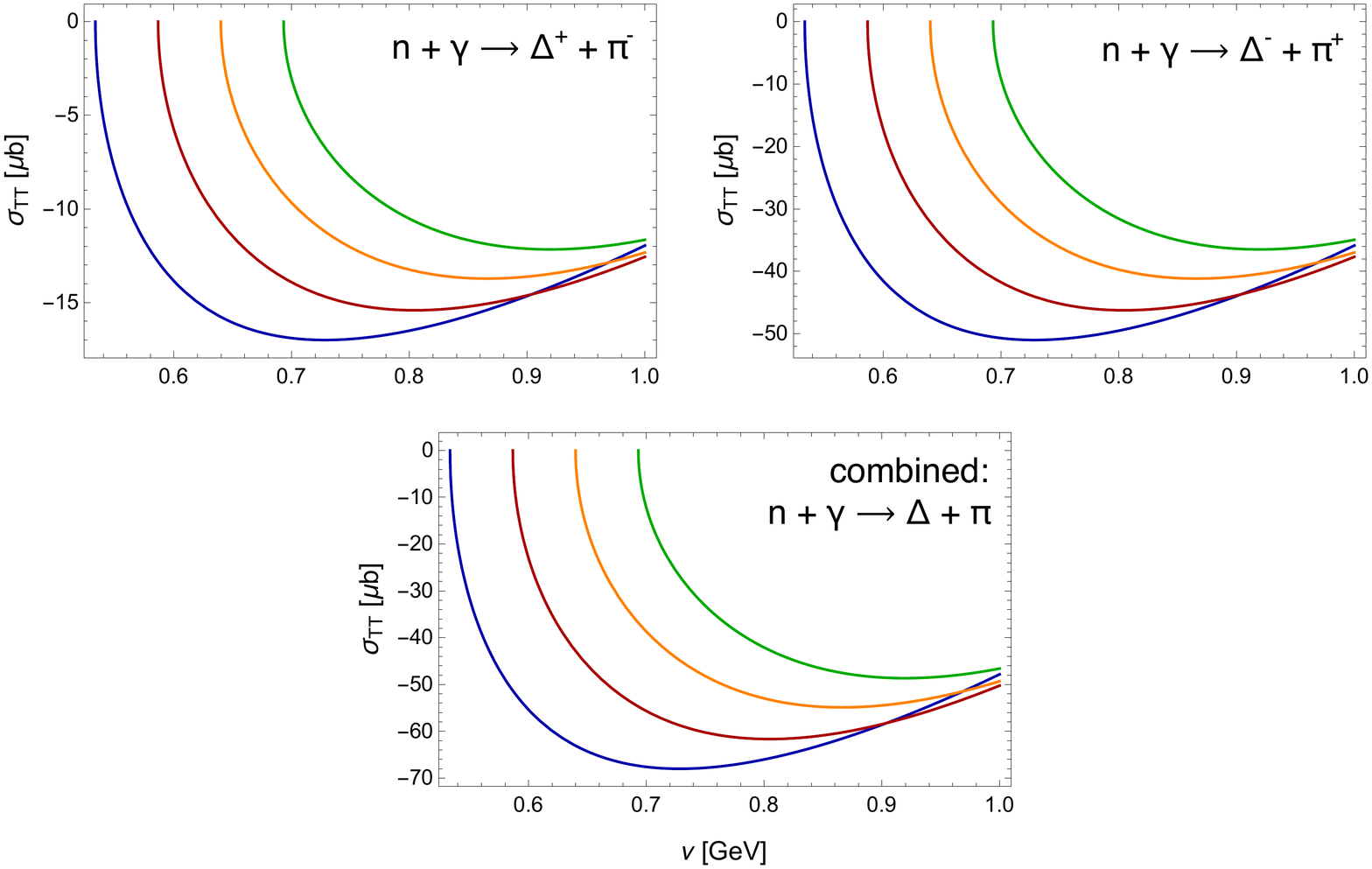}
       \vskip-5mm
       \caption{Polarized cross section $\sigma_{TT}$ for pion-delta electroproduction on the neutron. Legend for the curves is the same as in \Figref{protonTplot}.}
              \label{fig:neutronTTplot}
\end{figure}

    \begin{figure}[h!] 
    \centering 
      \vskip-5mm
       \includegraphics[width=13.5cm]{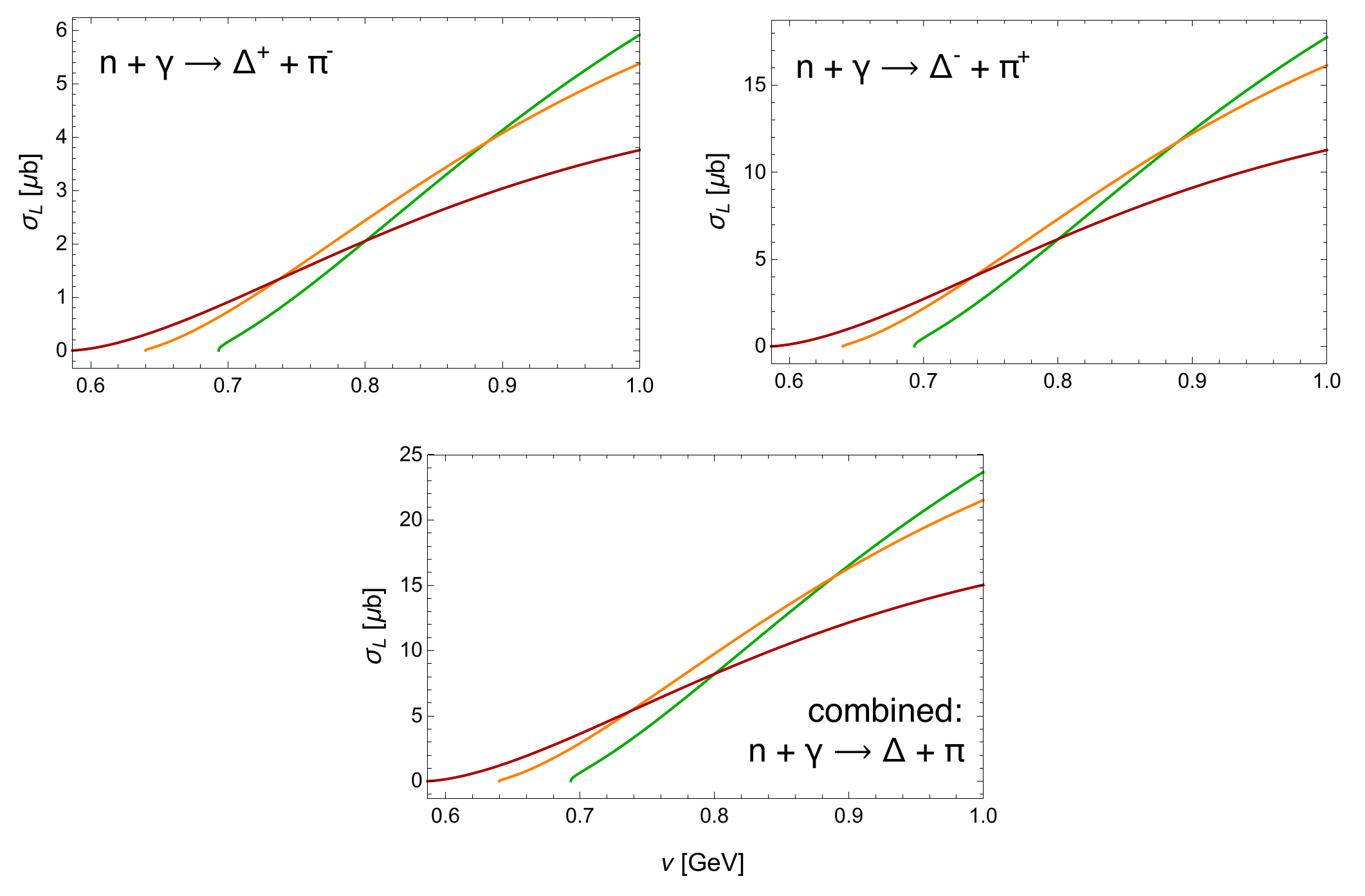}
       \caption{Longitudinal unpolarized cross section $\sigma_{L}$ for pion-delta electroproduction on the neutron. Legend for the curves is the same as in \Figref{protonLplot}.}
              \label{fig:neutronLplot}
\end{figure}

    \begin{figure}[h!] 
    \centering 
      \vskip-5mm
       \includegraphics[width=13.5cm]{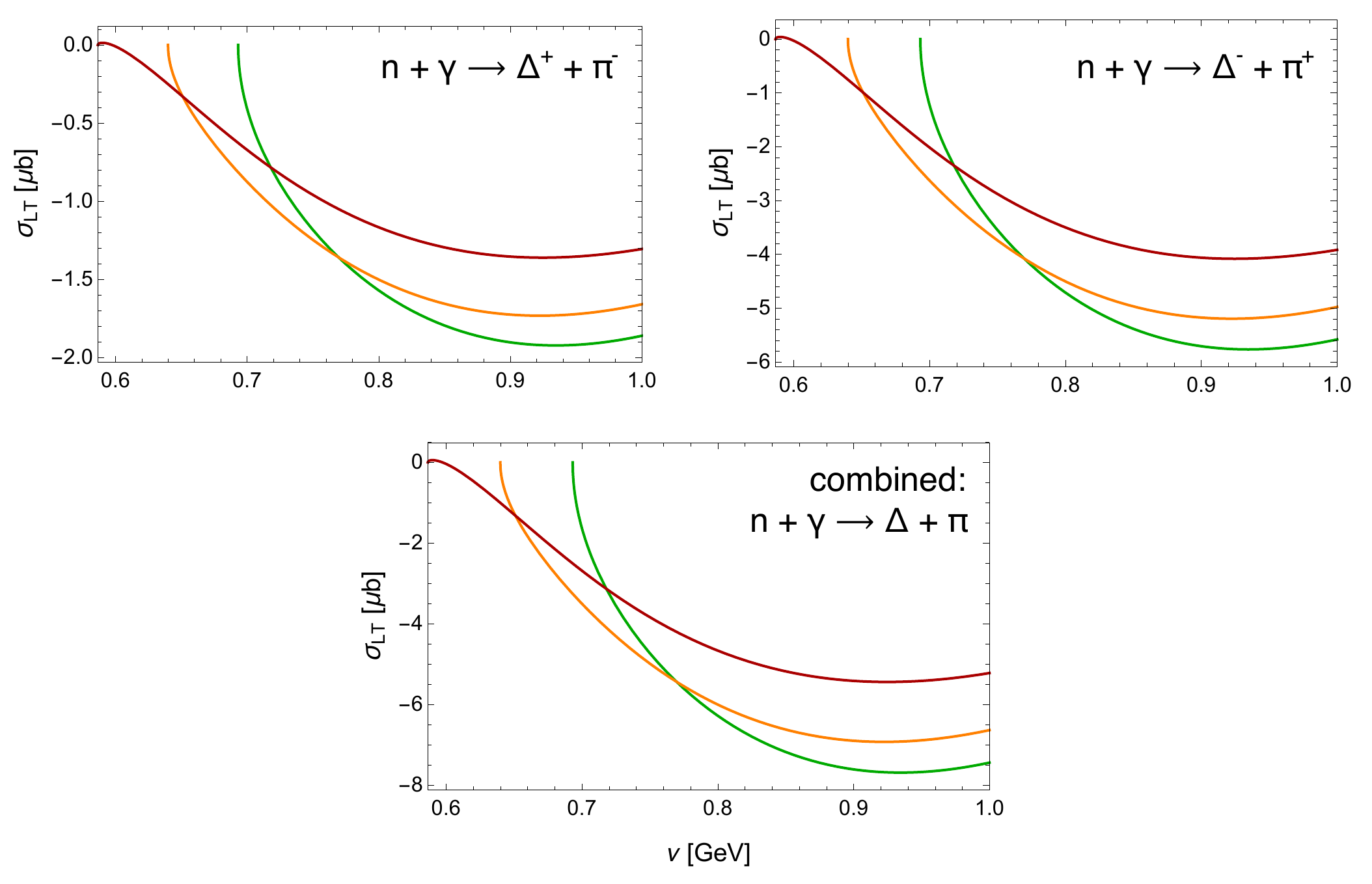}
       \caption{Longitudinal-transverse polarized cross section $\sigma_{LT}$ for pion-delta electroproduction on the neutron. Legend for the curves is the same as in \Figref{protonLplot}. }
              \label{fig:neutronLTplot}
\end{figure}

    \begin{figure}[h!] 
    \centering 
       \vskip-5mm
       \includegraphics[width=13.5cm]{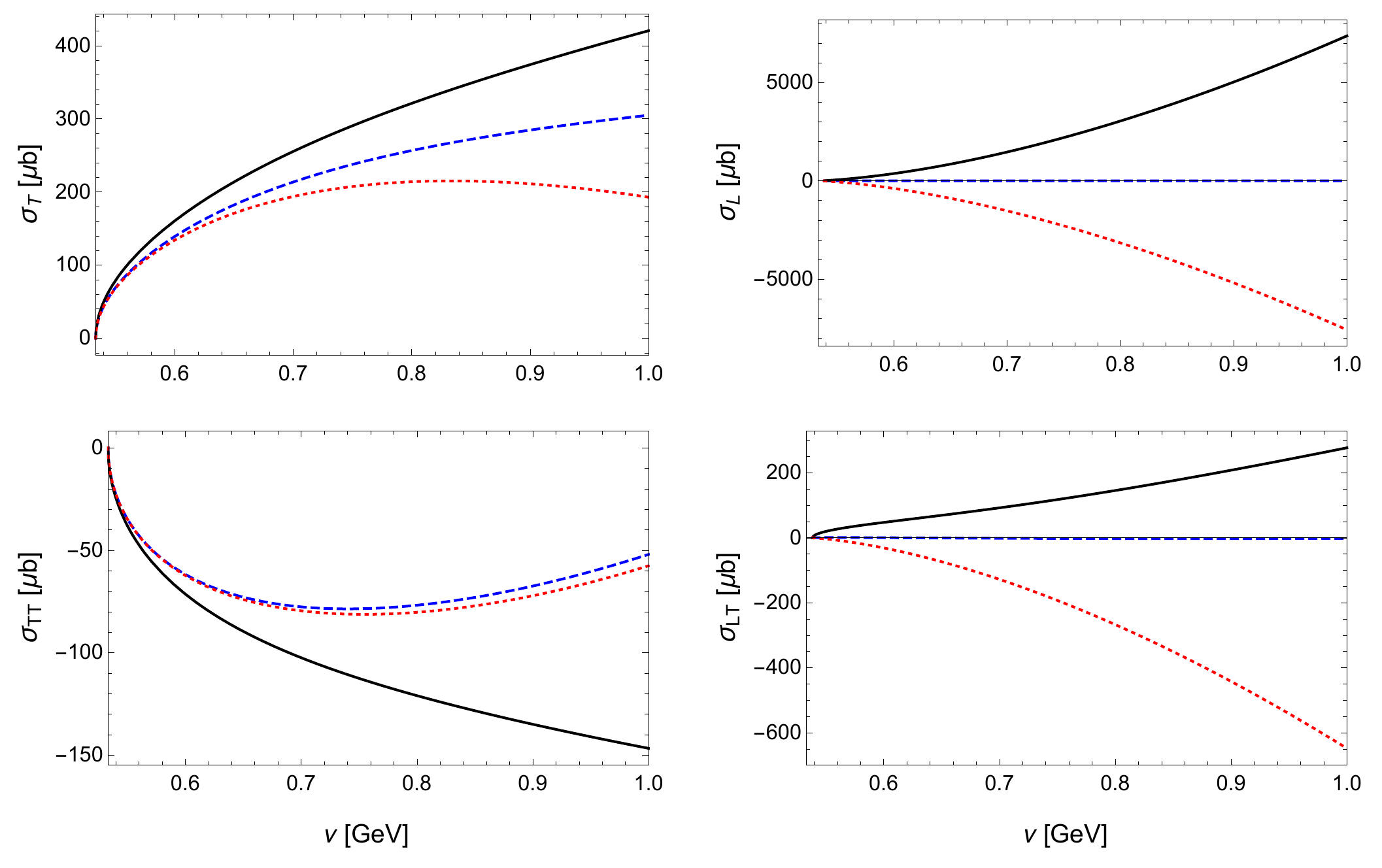}
       \caption{$\pi\Delta$-electroproduction cross sections for the proton. The cross sections are related to the CS amplitudes of $\mathcal{O}(p^{7/2})$ (black solid curves), $\mathcal{O}(p^4)$ (red dotted curves) and $\mathcal{O}(p^{9/2})$ (blue dashed curves) in the low-energy domain of the $\delta$-expansion. For $\sigma_T$ and $\sigma_{TT}$ we used $Q^2=0$ and for $\sigma_L$ and $\sigma_{LT}$ we used $Q^2=0.01\,\mathrm{GeV}^2$.}
              \label{fig:CSpComp}
\end{figure}

    \begin{figure}[t] 
    \centering 
    \vskip-5mm
       \includegraphics[width=13.5cm]{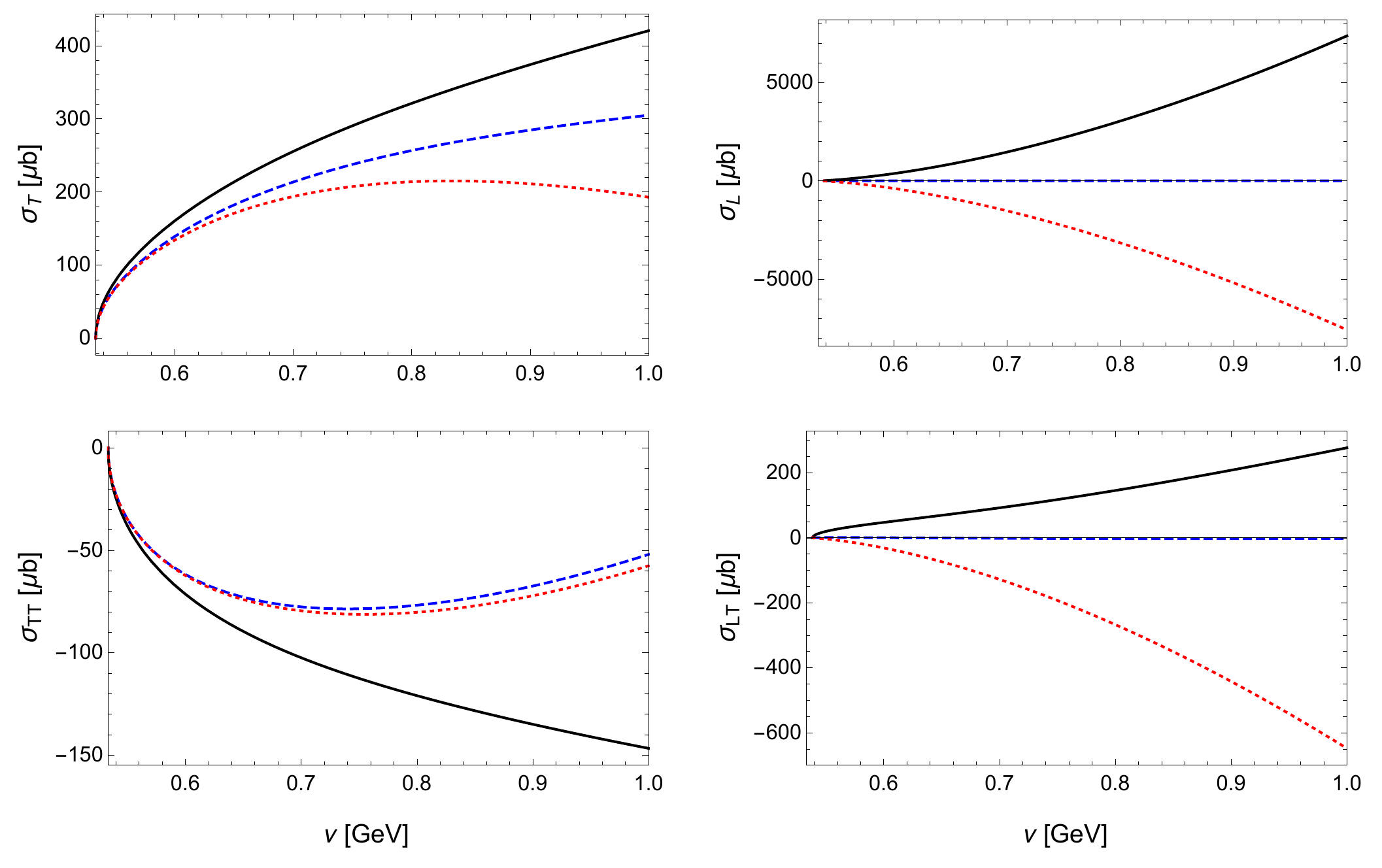}
       \caption{$\pi\Delta$-electroproduction cross sections for the neutron. Legend for the curves is the same as in \Figref{CSpComp}. }
              \label{fig:CSnComp}
\end{figure}

\newpage
\clearpage
\section{Plots of Nucleon Structure Functions}\seclab{nucleonSFplots}
Here we confront our BChPT calculation of the proton spin structure function $g_1(x,Q^2)$ with  experimental data and the most commonly used empirical 
parametrizations. 

    \begin{figure}[h!] 
    \centering 
       \includegraphics[width=16cm]{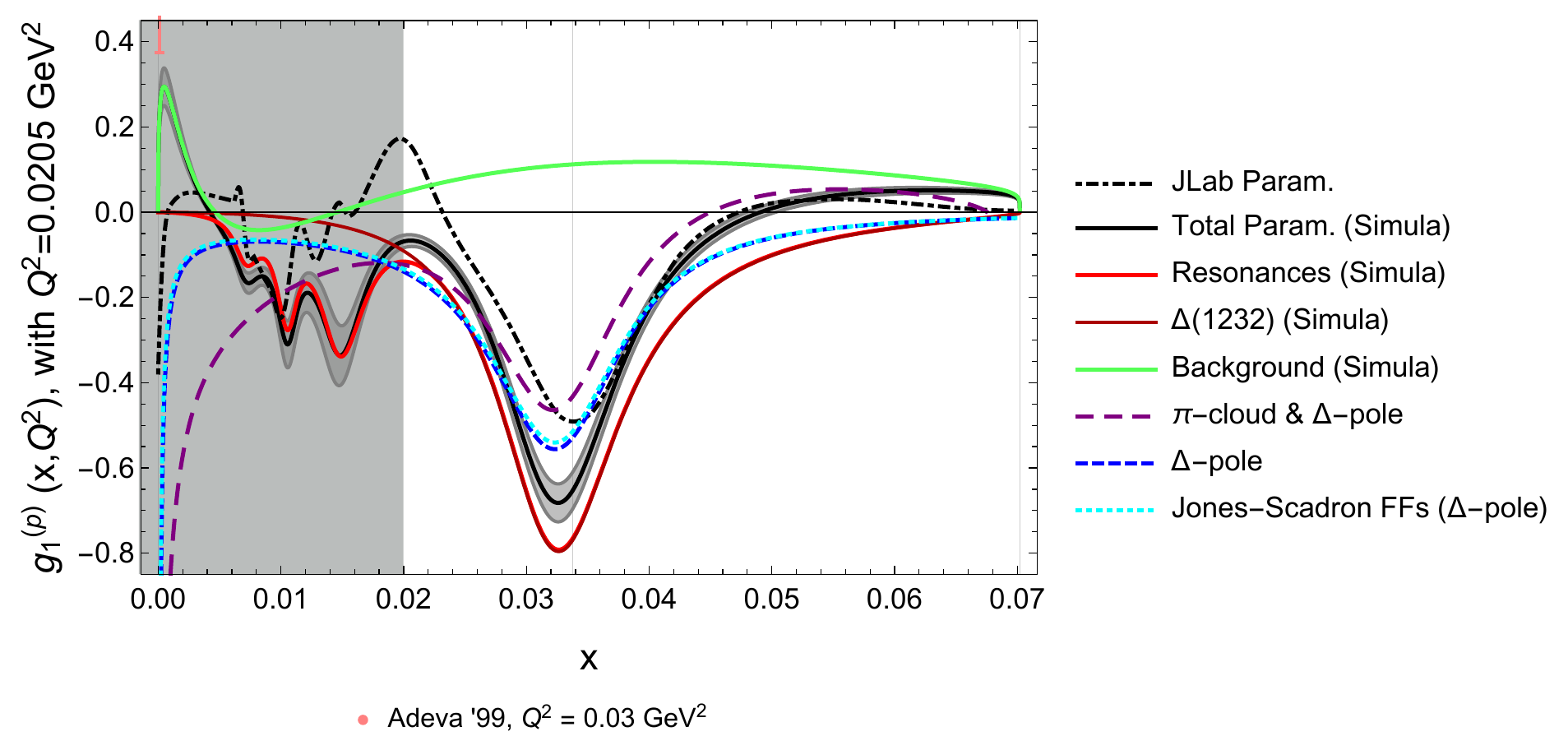}
       \caption{Spin structure function $g_1$ of the proton at $Q^2=0.0205\,\text{GeV}^2$}
              \label{fig:g1p1}
\end{figure}

    \begin{figure}[h!] 
    \centering 
       \includegraphics[width=16cm]{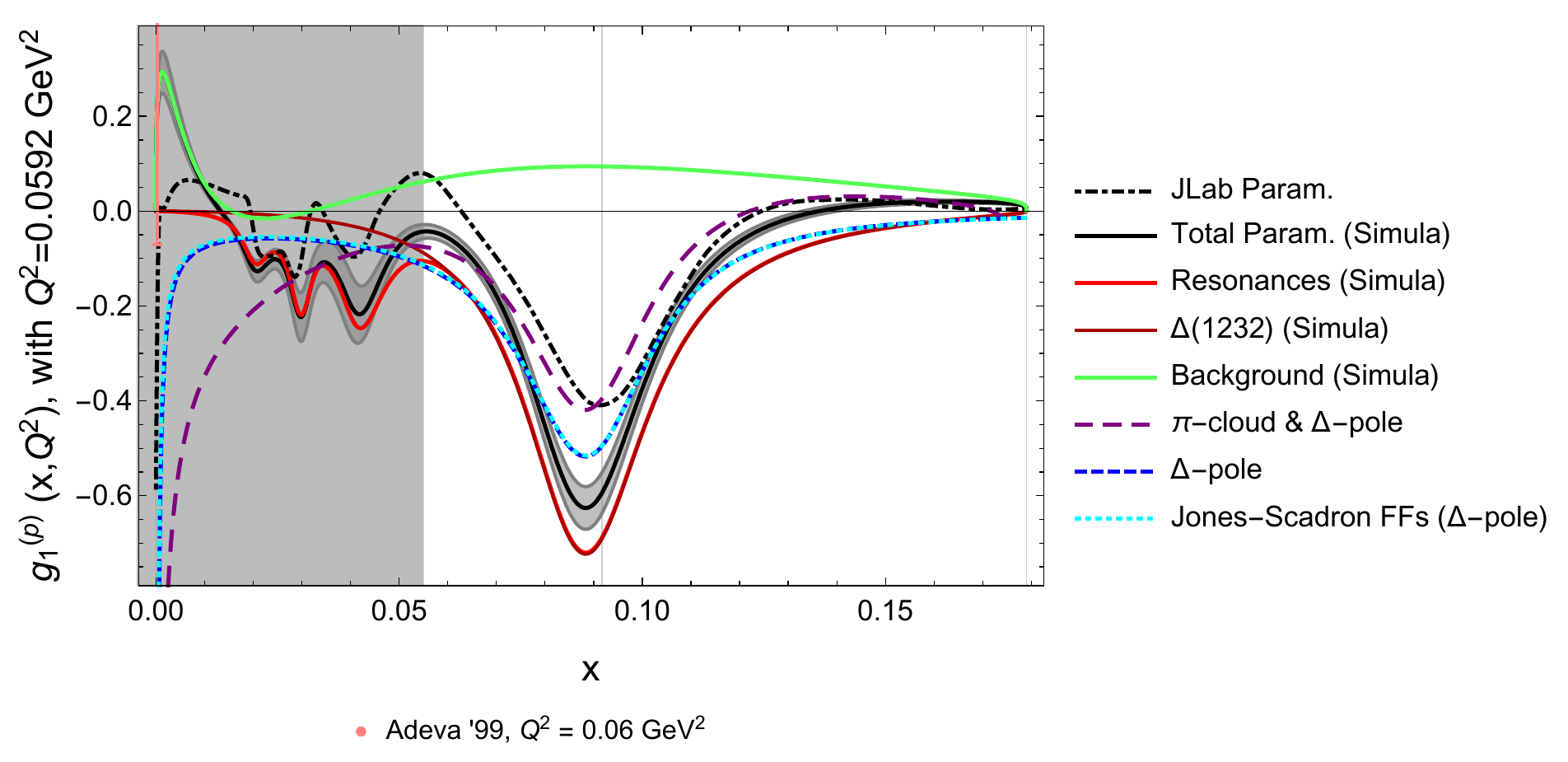}
       \caption{Spin structure function $g_1$ of the proton at $Q^2=0.0592\,\text{GeV}^2$}
              \label{fig:g1p2}
\end{figure}

    \begin{figure}[h!] 
    \centering 
       \includegraphics[width=16cm]{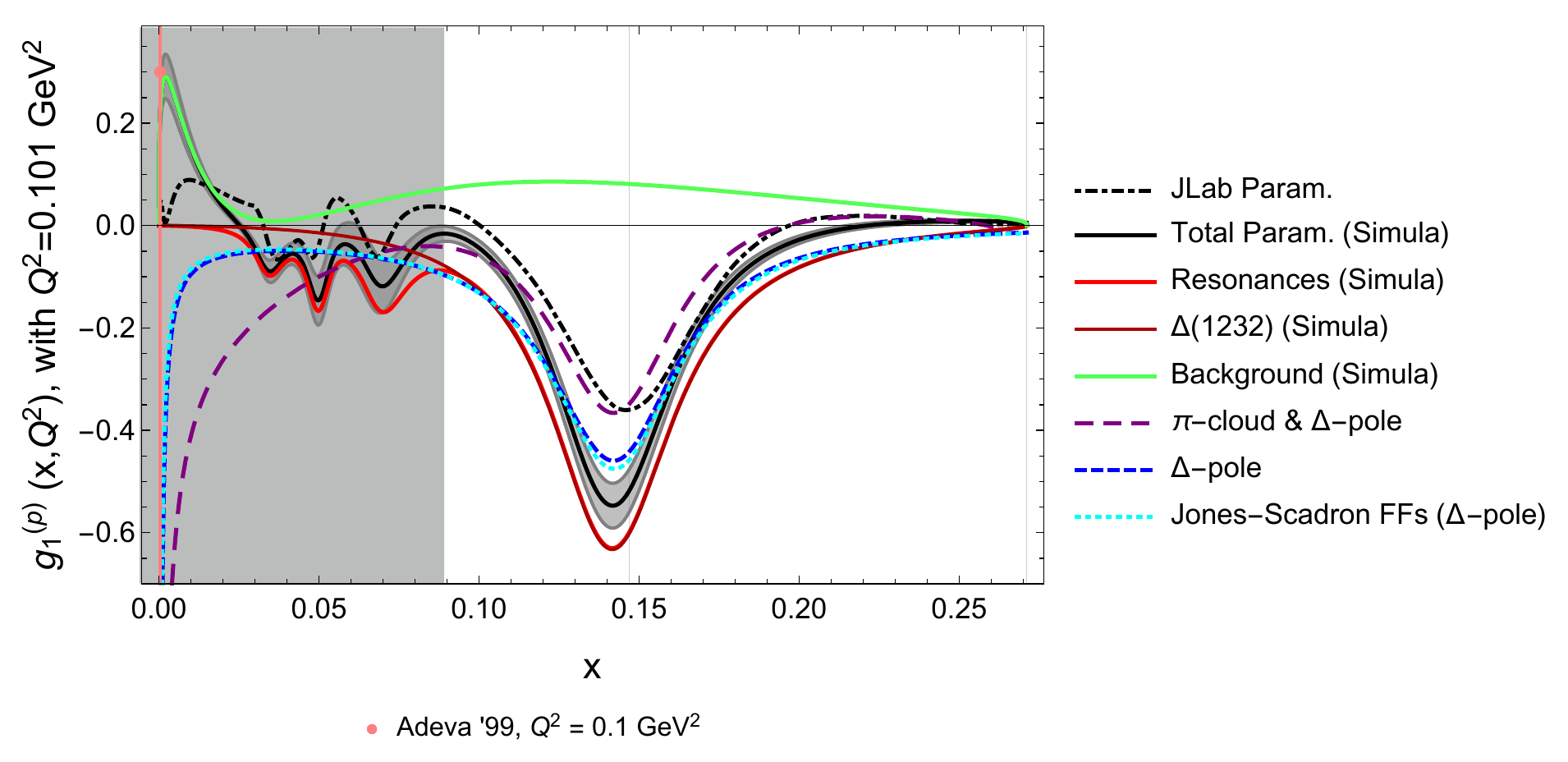}
       \caption{Spin structure function $g_1$ of the proton at $Q^2=0.101\,\text{GeV}^2$}
              \label{fig:g1p3}
\end{figure}

    \begin{figure}[h!] 
    \centering 
       \includegraphics[width=16cm]{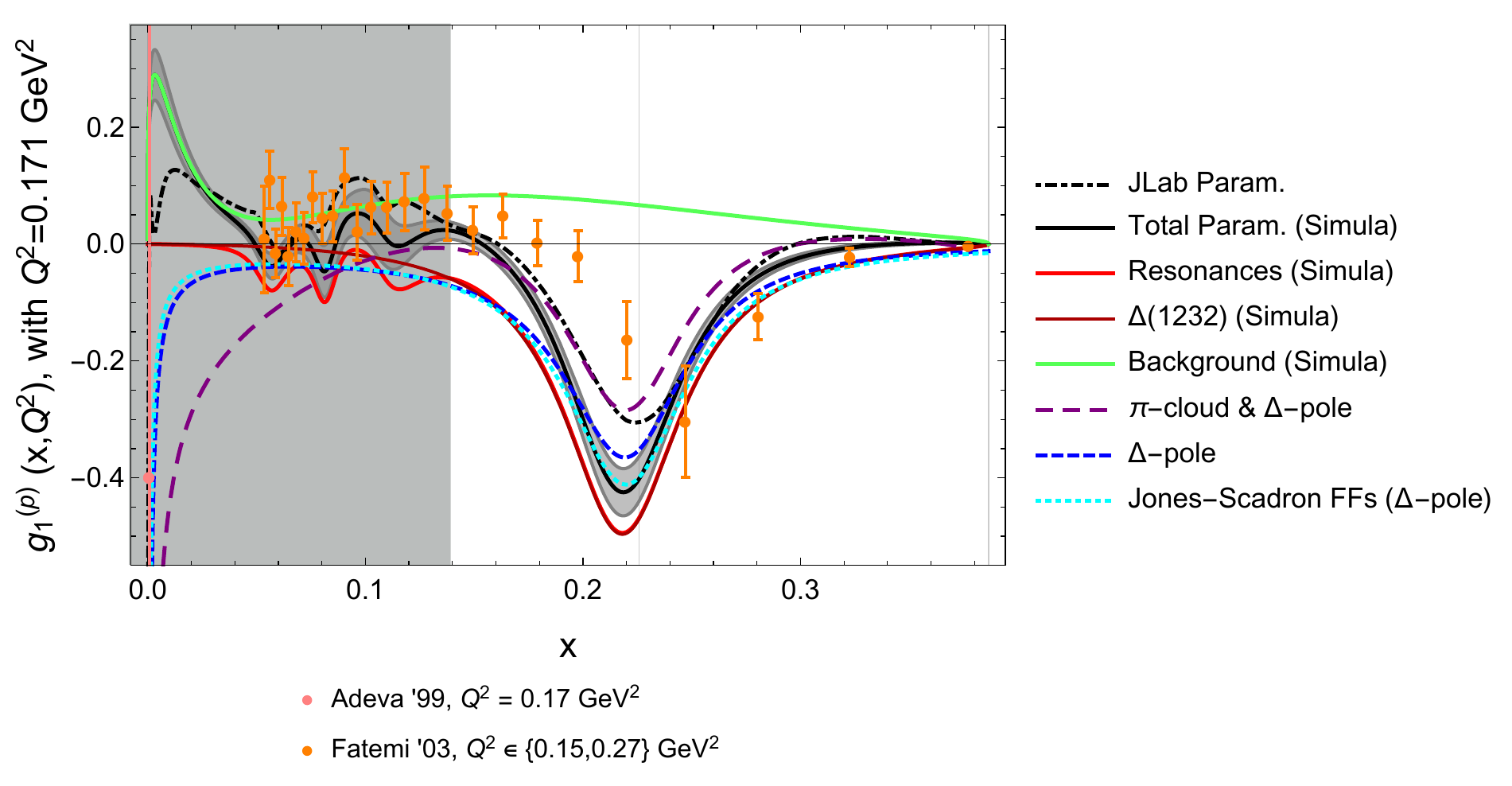}
       \caption{Spin structure function $g_1$ of the proton at $Q^2=0.171\,\text{GeV}^2$}
              \label{fig:g1p4}
\end{figure}

    \begin{figure}[h!] 
    \centering 
       \includegraphics[width=16cm]{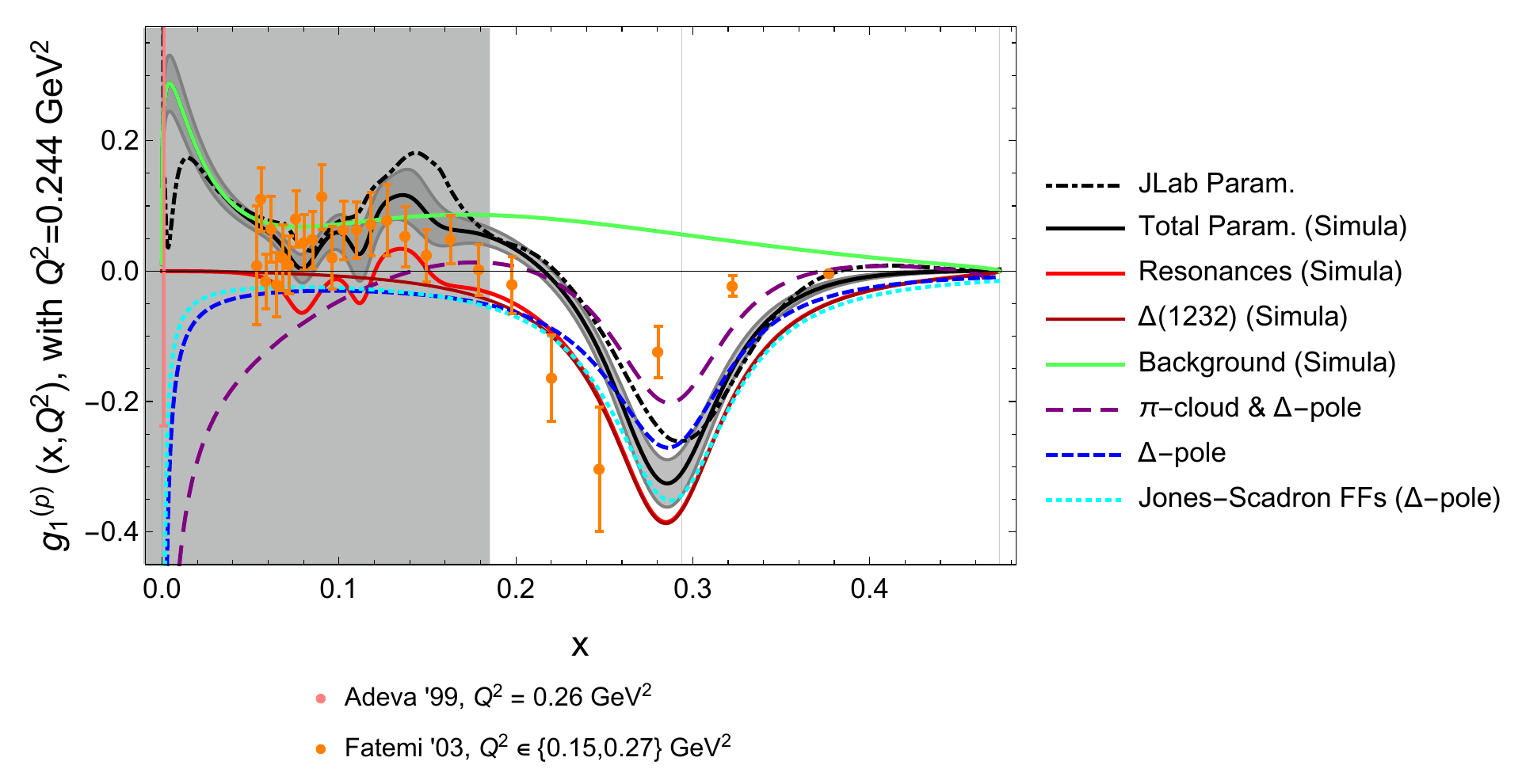}
       \caption{Spin structure function $g_1$ of the proton at $Q^2=0.244\,\text{GeV}^2$}
              \label{fig:g1p5}
\end{figure}

    \begin{figure}[h!] 
    \centering 
       \includegraphics[width=16cm]{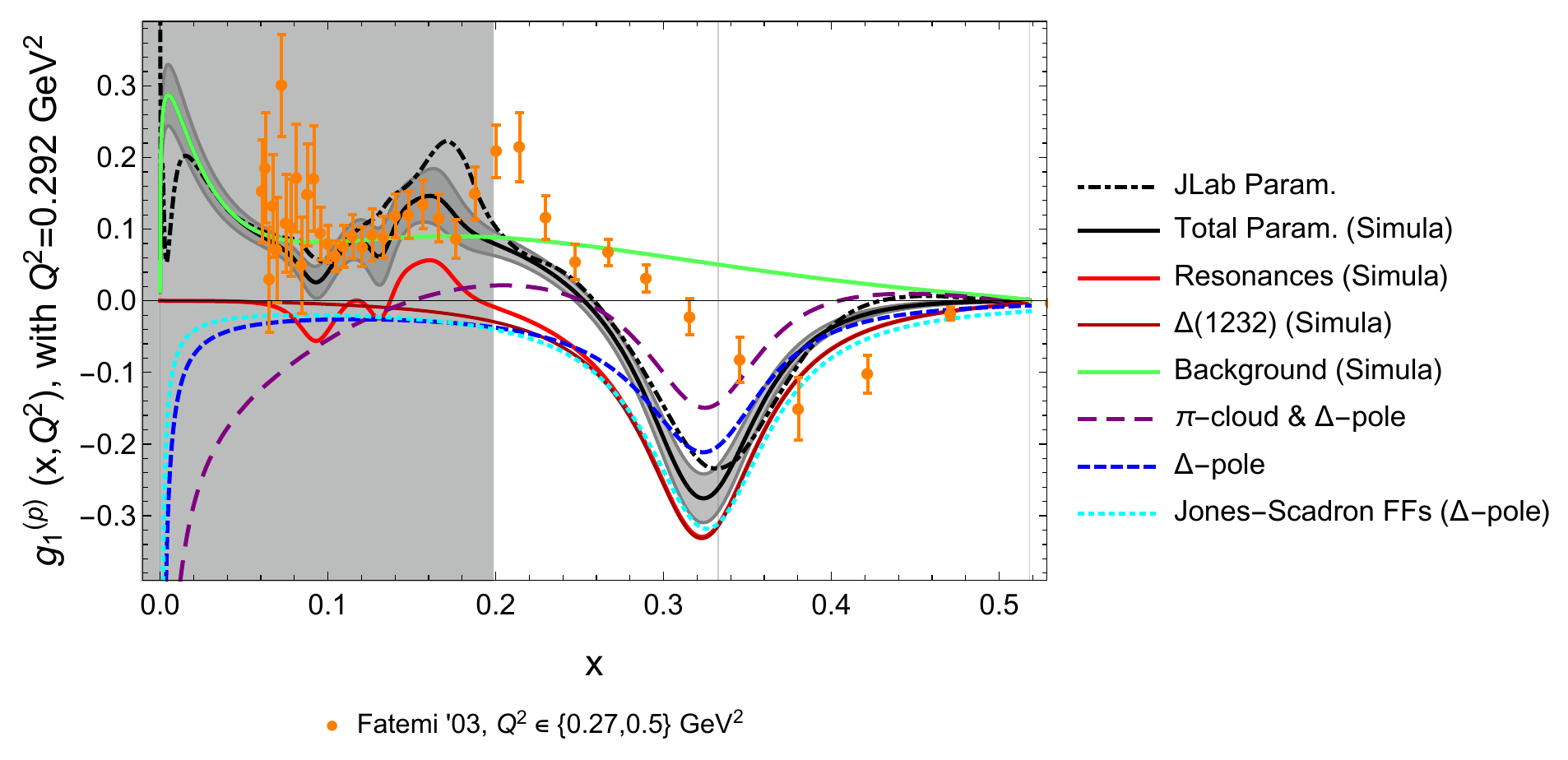}
       \caption{Spin structure function $g_1$ of the proton at $Q^2=0.292\,\text{GeV}^2$}
              \label{fig:g1p6}
\end{figure}

\end{subappendices}
\chapter{Two-Photon Exchange in Hydrogen-Like Atoms}\chaplab{chap5}

The rest of the thesis (Chapters \ref{chap:chap5}-\ref{chap:5HFS}) is devoted to the TPE corrections, with
particular focus on the proton-polarizability effects in $\mu$H. Numerical results for the polarizability contributions to the LS and HFS will be given in Chapters \ref{chap:5LS} and \ref{chap:5HFS},
respectively. 

In this Chapter we present an extensive derivation of the structure effects through the forward TPE (\secref{chap5}{TPE}). We will show how the TPE effects can be subdivided into either ``elastic'' and ``inelastic'' (\secref{chap5}{SepElinEl}), or, ``Born'' and ``polarizability'' contributions (\secref{chap5}{1334}). Final expressions for the LS and HFS are given in Sections \ref{chap:chap5}.\ref{sec:LSformulasTPE} and \ref{chap:chap5}.\ref{sec:HFSformulasTPE}, respectively. In \secref{chap5}{matchingOTPE}, we partially match the  effects from one- and two-photon exchange, e.g.: the gross structure, the Fermi energy, the charge radius, Friar radius and Zemach radius terms.
In \secref{chap5}{PolExpTPE}, we expand the TPE formulas for small values of Bjorken $x$ and express the TPE polarizability contribution to the HFS in terms of polarizabilities. We will return to this formalism in \secref{5HFS}{polexpHFS} and apply it to interpret our results for the HFS.

The forward TPE effects are of order $(Z\al)^5$.  In \secref{chap5}{6.3}, we turn our attention to the off-forward TPE effects. The off-forward TPE is suppressed by an addition factor of $Z\al$. Nevertheless, certain enhancement mechanism leave the possibility for a significant off-forward TPE effect in atomic bound states. In particular, we will study the neutral-pion exchange \cite{Hagelstein:2015b,Hagelstein:2015egb} (\secref{5HFS}{neutralPion}). Also, we consider the nuclear-polarizability corrections to the LS at order $(Z\al)^6 \ln Z \al$ (\secref{5LS}{offTPE}). In \appref{5LS}{offVVCSBorn}, we calculate the Born and elastic off-forward VVCS amplitudes.

\begin{figure}[thb]
\centering
\includegraphics[scale=0.85]{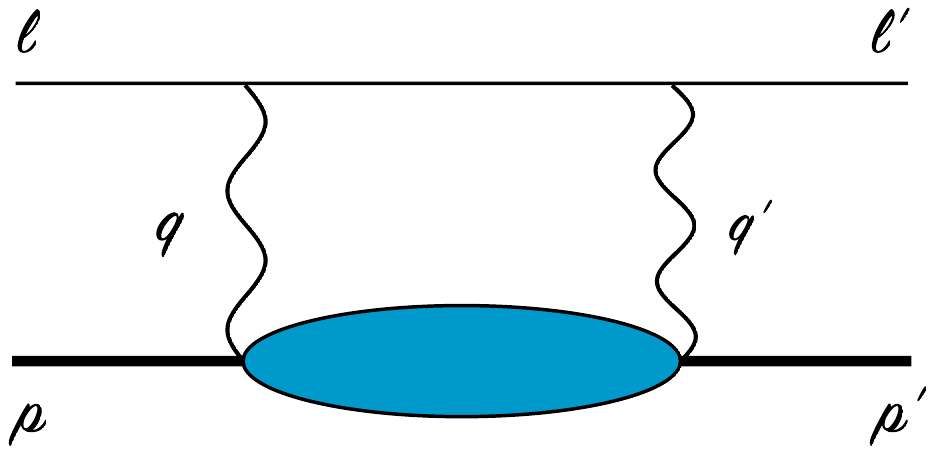}
\caption{Two-photon-exchange diagram in general kinematics: the horizontal lines correspond to the lepton and the nucleus (bold), where the ``blob'' can be understood as doubly-virtual Compton scattering.}
\label{TPEFRW}
\end{figure}

\section{Structure Effects Through Forward Two-Photon Exchange} \seclab{TPE}

The leading effect of the proton structure on hydrogen is the finite-size (charge-radius) effect arizing at order $(Z\al)^4$. 
The subleading effects [$(Z\al)^5$, etc.] are best described by considering the TPE, shown in Fig.~\ref{TPEFRW}. 
Obviously, the main uknown ingredient therein is the VVCS process, obtained from TPE by removing the lepton line.
In \chapref{chap4}, we formulated the VVCS process in terms of CS amplitudes and structure functions, and presented a BChPT prediction for VVCS off the nucleon at $\mathcal{O}(p^{7/2})$ in the low-energy domain of the $\delta$-expansion \cite{Lensky:2014dda,Alarcon2017}. In the following, we will incorporate that calculation into the TPE correction to the spectra of hydrogen-like atoms.

\subsection{Two-Photon Exchange in Terms of Compton Amplitudes}
Figure \ref{TPEFRW} shows the TPE diagram in general kinematics. To compute the TPE effect to order $(Z\al)^5$ it sufficient
to consider the forward kinematics, where $q'=q$ (and hence, $p'=p$, $l'=l$, $t=0$), see \Figref{TPE}. In this case
the TPE correction is expressed in terms of the forward VVCS amplitudes. 
The TPE-induced shift of the $nS$-level is then given by \cite{Carlson:2011zd}:
\beq
\Delta E^\mathrm{TPE}(nS)= 8\pi \al m \,\phi_n^2\,
\frac{1}{i}\int_{-\infty}^\infty \!\frac{\dd\nu}{2\pi} \int \!\!\frac{\dd \bq}{(2\pi)^3}   \frac{\left(Q^2-2\nu^2\right)T_1(\nu,Q^2)-(Q^2+\nu^2)\,T_2(\nu,Q^2)}{Q^4(Q^4-4m^2\nu^2)},\eqlab{VVCS_LS}
\eeq
where $\phi_n^2=1/(\pi n^3 a^3)$ is the wavefunction at the origin, and $\nu=q_0$, $Q^2 = \bq^2 -q_0^2$. The TPE correction to the $nS$ HFS is given by \cite{Carlson:2011af}:
\bea
\frac{E^\mathrm{TPE}_{\mathrm{HFS}}(nS)}{E_\mathrm{F}(nS)}&=&\frac{4m}{Z(1+\kappa)}\frac{1}{i}\int_{-\infty}^\infty \!\frac{\dd\nu}{2\pi} \int \!\!\frac{\dd \bq}{(2\pi)^3}\,\frac{1}{Q^4-4m^2\nu^2}\times \eqlab{VVCS_HFS} \\
&&\qquad\qquad\times\left\{\frac{\left(2Q^2-\nu^2\right)}{Q^2}S_1(\nu,Q^2)+\frac{3\nu}{M}S_2(\nu,Q^2)\right\}\nn,
\eea
where $E_\mathrm{F}$ is the Fermi energy, \Eqref{FermiE}, and $\kappa$ is the anomalous magnetic moment of the nucleus. Again, $m$ refers to the lepton mass ($m_e$ or $m_\mu$, respectively, for H and $\mu$H), $M$ is the mass of the nucleus, $m_r$ is the reduced mass and $a=(Z \al m_r)^{-1}$ is the Bohr radius of the lepton-nucleus system. Clearly, the underlying coordinate-space potentials are proportional to $\delta(\br)$. From Eqs.~\eref{VVCS_LS} and \eref{VVCS_HFS} one can see that the LS is ``softer'', i.e., has a weaker dependence on the region of high photon virtualities, and the HFS is ``harder''.

For further evaluation, it is convenient to perform a Wick rotation, so as to change the integration over $q_0$ to the imaginary axis, i.e., $q_0\rightarrow iQ_0$. This is straighforward at zero energy ($p \cdot l=mM$), whereas 
at finite energy one needs to take
care of the poles moving across the imaginary $q_0$-axis, cf.\ Ref.~\cite{Pascalutsa:1999pv}. 
After the Wick rotation, the integration four-momentum
is Euclidean and we can evaluate it in hyperspherical
coordinates, see \appref{chap5}{App:Wick}. The discussed coordinate transformations are performed in the following for the LS:
\bea
\Delta E^\mathrm{TPE}(nS)
&=&\frac{\al}{2\pi^2m}\,\phi_n^2\,\int_0^\infty\! \frac{\dd Q}{Q}  \int_0^{\pi}\! \dd \chi\;\sin^2\!\chi \,\times\eqlab{LSAngular}\\
&&\times\frac{\left(1+2\cos^2 \!\chi\right)T_1(iQ\cos \chi,Q^2)-\sin^2\!\chi\, T_2(iQ\cos \chi,Q^2)}{\tau_l+\cos^2 \!\chi},\nn
\eea
and the HFS:

\bea
\frac{E^\mathrm{TPE}_{\mathrm{HFS}}(nS)}{E_\mathrm{F}(nS)}&=&
\frac{1}{4\pi^3Z(1+\kappa)m} \int_0^\infty\! \dd Q \,Q \int_0^{\pi}\! \dd \chi\;\sin^2\!\chi \,\times\eqlab{HFSS1S2}\\
&& \times\frac{\left\{(2+\cos^2\!\chi)\,S_1(iQ\cos \chi,Q^2)+\frac{3iQ\cos \chi}{M}\,S_2(iQ\cos \chi,Q^2)\right\}}{\tau_l+ \cos^2 \!\chi},\nn
\eea
where $\tau_l=Q^2/4m^2$.
\subsection{Master Formulae in Terms of Proton Structure Functions} \seclab{MasterFormulae}
The integral over $\nu = iQ\cos\chi$ can be done after
substituting the DRs for the VVCS amplitudes, see \Eqref{DRVVCS}.  Introducing the ``lepton velocity''
$v_l = \sqrt{1+1/\tau_l}$, we obtain
the following expressions for the $S$-level shift:
\bea
\Delta E^\mathrm{TPE}(nS)&=&\frac{16(Z\al)^2m}{M}\,\phi_n^2\,\int_0^\infty \frac{\dd Q}{Q^3}\,\int_0^1\dd x\,\frac{1}{v_l+\sqrt{1+x^2\tau^{-1}}}\Bigg\{ \frac{f_1(x,Q^2)}{x} - \frac{f_2(x,Q^2)}{2\tau}\nn\\
&&+
\frac{1}{(1+v_l)(1+\sqrt{1+x^2\tau^{-1}})}\bigg[\frac{2f_1(x,Q^2)}{x}+ \frac{f_2(x,Q^2)}{2\tau}\bigg] \Bigg\},
\eqlab{LSMaster}
\eea
and the HFS:
\bea
\frac{E^\mathrm{TPE}_{\mathrm{HFS}}(nS)}{E_\mathrm{F}(nS)}&=&\frac{16Z\al m M}{\pi (1+\kappa)}\int_0^\infty\frac{\dd Q}{Q^3}\int_0^1\dd x\,\frac{1}{v_l+ \sqrt{1+x^2\tau^{-1}}}  \Bigg\{3\,g_2(x,Q^2)\nn
\\
&& + \left[1+\frac{1}{2(v_l+1)(1+ \sqrt{1+x^2\tau^{-1}})}\right]2\,g_1(x,Q^2) \Bigg\},\eqlab{fullHFS}\eea
with $\tau=Q^2/4M^2$. These are the master formulae containing all the structure
effects to order $(Z\alpha)^5$. As we will show in \secref{chap5}{matchingOTPE}, they also contain the charge radius term and the Fermi energy. Note that here we substituted the unsubtracted DR for $T_1$, \Eqref{T1DR}; only in the following we will perform the required subtraction, cf.\ \Eqref{T1Subtrgen}. Also, we substituted the unsubtracted DR for $\nu S_2$, \Eqref{S2DR}, and hence, the contribution of the BC sum rule, \Eqref{BCdef}, is still included.

In the subsequent Sections, we will split into the elastic and inelastic contributions (\secref{chap5}{ElasInelasContr}), construct the polarizability contribution (\secref{chap5}{1334}), separate $T_1(0,Q^2)$ and eliminate the contribution proportional to the BC sum rule.
\subsection{Separating Nucleon-Pole and Inelastic Contributions} \seclab{SepElinEl}
The TPE can be divided into an ``elastic'' and an ``inelastic'' part. Here, we are mainly interested in the ``inelastic'' contribution, which is formed by excited intermediate states, e.g., by the $\Delta(1232)$-resonance excitation.  \seclab{ElasInelasContr}
\begin{figure}[h]
\centering
\begin{minipage}{0.49\textwidth}
\centering
       \includegraphics[width=6.7cm]{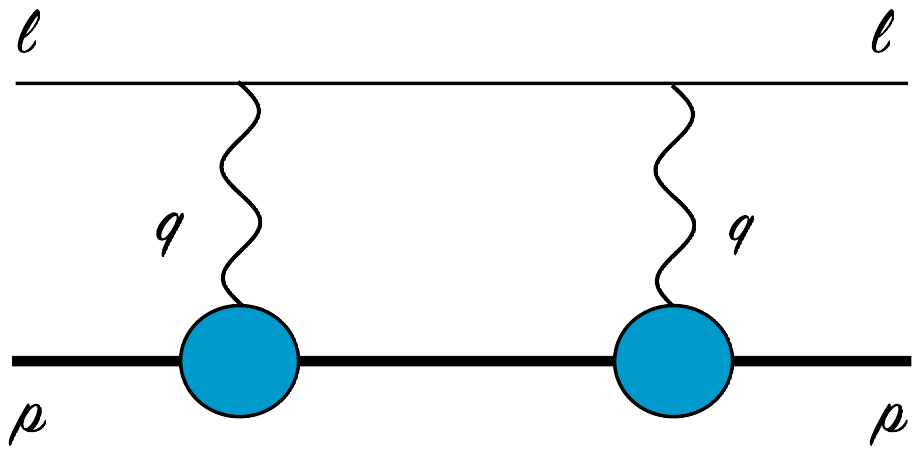}

      (a)
    \end{minipage}\hfill
\begin{minipage}{0.49\textwidth}
\centering
  \includegraphics[width=6.7cm]{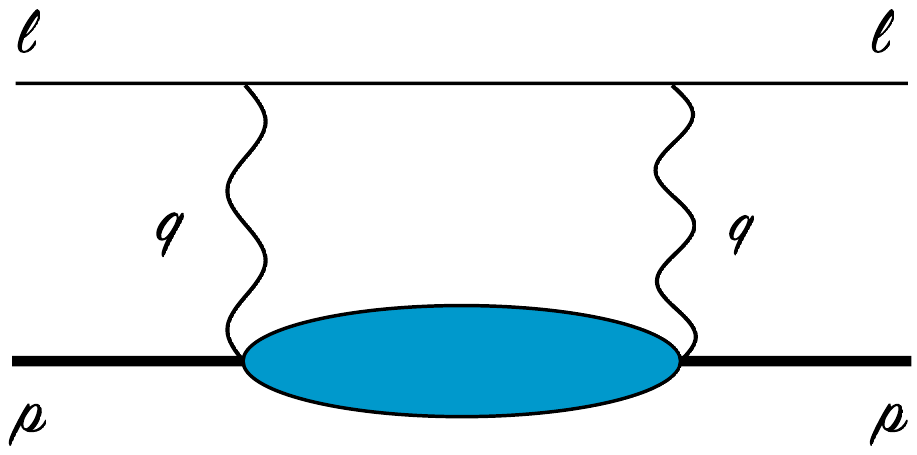}
  
  (b)
\end{minipage}
\caption{Two-photon-exchange diagrams in forward kinematics: the horizontal lines correspond to the lepton and the nucleus (bold). (a) Elastic contribution to the two-photon-exchange diagram. (b) Inelastic contribution to the two-photon-exchange diagram, where the ``blob'' represents all possible excitations. The crossed diagrams are not drawn.\label{fig:TPE}}
\end{figure}
\addtocontents{toc}{\protect\setcounter{tocdepth}{0}}
\subsubsection{Nucleon-Pole Contribution} \seclab{NucleonPole}
The nucleon-pole contribution shown in Fig.~\ref{fig:TPE} (a) can be expressed through the elastic structure functions shown in \Figref{scat} (a). The structure functions can be written in terms of the elastic FFs as:
\begin{subequations}  
\eqlab{elstructure}
\bea 
f_1^{\mathrm{el}}(x,Q^2) & =  & \frac{1}{2}\,  G^2_M(Q^2)\,  \delta( 1 - x ),\eqlab{f1elastic} \\
f_2^{\mathrm{el}}(x,Q^2) & = & \frac{1}{1 + \tau} \, \big[G^2_E(Q^2) + \tau G^2_M(Q^2) \big]\, \delta( 1 - x), \eqlab{f2elastic}\\
g_1^{\mathrm{el}}(x,Q^2) & = &  \frac{1}{2}\,  F_1(Q^2) \,G_M(Q^2) \, \delta(1 - x), \eqlab{g1elastic}\\
g_2^{\mathrm{el}}(x,Q^2) & = & -  \frac{\tau}{2}\, F_2(Q^2)\, G_M(Q^2)\,  \delta(1 - x)\eqlab{g2elastic},
\eea
\end{subequations}
where the elastic Dirac and Pauli FFs are related to the e.m.\ Sachs FFs in the following way:
\begin{subequations}  
\bea
F_1(Q^2)&=&\frac{1}{1+\tau}\left[G_E(Q^2)+\tau G_M(Q^2)\right], \\
F_2(Q^2)&=&\frac{1}{1+\tau}\left[G_M(Q^2)- G_E(Q^2)\right],
\eea
\end{subequations}
see also \Eqref{SachsFFDP}.
Substituting the elastic structure functions
into the above expressions for the S-level shift and the HFS,
the nucleon-pole contribution is found as:
\begin{subequations}
\eqlab{elTPEpole}
\bea
\Delta E^{\mathrm{pole}}(nS)&=&\frac{8(Z\al)^2 m}{M}\,\phi_n^2\,\int_0^\infty \frac{\dd Q}{Q^3}\,\Bigg\{\left[\tau+\frac{3+2\tau}{(1+v_l)(1+v)}\right]G_M^2(Q^2)\nn\\
&&-\frac{1}{\tau}\left[1-\frac{1}{(1+v_l)(1+v)}\right]G_E^2(Q^2) \Bigg\}\,\frac{1}{(1+\tau)(v_l+v)}, \\
\frac{E_{\mathrm{HFS}}^{\mathrm{pole}}(nS)}{E_\mathrm{F}(nS)}&=&\frac{16Z\al m M}{\pi (1+\kappa)}\int_0^\infty\frac{\dd Q}{Q^3}\frac{G_M(Q^2)}{v_l+ v }   \Bigg\{ \left[1+\frac{1}{2(1+v_l)(1+v)}\right]F_1(Q^2)\qquad\qquad \nn\\
&& - \frac{3\tau}{2}F_2(Q^2) \Bigg\},\quad
\eqlab{elasticHFS}
\eea
\end{subequations}
where $v=\sqrt{1+\tau^{-1}}$. Equivalently, one can plug the nucleon-pole part of the VVCS amplitudes \cite{Drechsel:2002ar},
\begin{subequations}
\eqlab{poles}
\bea
T_1^{\mathrm{pole}}(\nu, Q^2) &=&\frac{4\pi Z^2\alpha}{M}\frac{\nu_{\mathrm{el}}^2\,  G_M^2(Q^2)}{\nu_{\mathrm{el}}^2-\nu^2
-i 0^+ }, 
\eqlab{T1pole}\\
T_2^{\mathrm{pole}}(\nu, Q^2)&=&\frac{8\pi Z^2\alpha\,  \nu_{\mathrm{el}} }{\nu_{\mathrm{el}}^2-\nu^2
-i 0^+ }\frac{G^2_E(Q^2) + \tau G^2_M(Q^2)}{1 + \tau},
\eqlab{T2pole}
\\
S_1^{\mathrm{pole}}(\nu, Q^2) &=& \frac{4 \pi Z^2\alpha \, \nu_{\mathrm{el}}}{\nu_{\mathrm{el}}^2-\nu^2
-i 0^+ } F_1(Q^2)\, G_M(Q^2),\\
\left[\nu S_2\right]^{\mathrm{pole}}(\nu, Q^2)&=&-\frac{2 \pi Z^2\alpha \nu_\mathrm{el}^2}{\nu_{\mathrm{el}}^2-\nu^2
-i 0^+ } F_2(Q^2)\, G_M(Q^2),\eqlab{S2pole}
\eea
\end{subequations}
into Eqs.~\eref{LSAngular}  and \eref{HFSS1S2}.

\subsubsection{Inelastic Contribution}
We refer to the remaining TPE effect, shown in Fig.~\ref{fig:TPE} (b), as the inelastic contribution. The diagram in Fig.~\ref{fig:TPE} (b) has no nucleon-pole but excited intermediate states. Therefore, it can be described by the inelastic structure functions shown in \Figref{scat} (b). They start from the lowest particle-production threshold, $\nu_0=Q^2/ 2Mx_0$, which is effectively set by the pion production: 
\beq
\eqlab{PionProductionThreshold}
\nu_\pi=m_\pi+(m_\pi^2+Q^2)/(2M). 
\eeq
The inelastic contribution to the $S$-level shift reads as:
\bea
\Delta E^\mathrm{inel.}(nS)&=&\frac{2\al m}{\pi}\,\phi_n^2\,\int_0^\infty \frac{\dd Q}{Q^3}\frac{v_l+2}{(1+v_l)^2}\, \left[T_1(0,Q^2)-\frac{4\pi Z^2 \al}{M}G_M^2(Q^2)\right]\eqlab{LSMasterSub}\\
&&-32(Z\al)^2Mm\,\phi_n^2\,\int_0^\infty \frac{\dd Q}{Q^5}\,\int_0^{x_0}\dd x\,\frac{1}{(1+v_l)(1+\sqrt{1+x^2\tau^{-1}})}\times\nn\\
&&\times\Bigg\{\frac{2x}{(1+v_l)(1+\sqrt{1+x^2\tau^{-1}})}\left[2+\frac{3+v_l\sqrt{1+x^2\tau^{-1}}}{v_l+\sqrt{1+x^2\tau^{-1}}}\right]f_1(x,Q^2)\nn\qquad\;\\
&&+\left[1+\frac{v_l\sqrt{1+x^2\tau^{-1}}}{v_l+\sqrt{1+x^2\tau^{-1}}}\right]f_2(x,Q^2)\Bigg\},\nn
\eea
where we performed the necessary subtraction on $T_1$. The applied once-subtracted DR for the non-nucleon-pole part of $T_1$,
\bea
T_1^\mathrm{non-pole} ( \nu, Q^2) &=& T_1(0,Q^2)-\frac{4\pi Z^2 \al}{M}G_M^2(Q^2)\eqlab{npT1Subtr}\\
&&+\frac{32\pi Z^2\al M\nu^2}{Q^4}\int_{0}^1 
\,\dd x \, 
\frac{x f_1 (x, Q^2)}{1 - x^2 (\nu/\nu_{\mathrm{el}})^2 - i 0^+},\nn
\eea
follows from Eqs.~\eref{T1Subtr} and \eref{T1pole}.
The inelastic contribution to the
HFS simply reads as:
\bea
\frac{E^\mathrm{inel.}_{\mathrm{HFS}}(nS)}{E_\mathrm{F}(nS)}&=&\frac{16Z\al m M}{\pi (1+\kappa)}\int_0^\infty\frac{\dd Q}{Q^3}\int_0^{x_0}\dd x\,\frac{1}{v_l+ \sqrt{1+x^2\tau^{-1}}} \,\Bigg\{ 3g_2(x,Q^2) \\
&&+\left[1+\frac{1}{2(v_l+1)(1+ \sqrt{1+x^2\tau^{-1}})}\right]2g_1(x,Q^2)\Bigg\}.\nn
\eea

\subsubsection{Ambiguity of the Nucleon-Pole Contribution}
Note that in the above subsections, the elastic contribution to the HFS is defined through the nucleon-pole parts of $S_1$ and $\nu S_2$, respectively. Considering the nucleon-pole parts of $S_1$ and $S_2$, the decomposition into elastic and inelastic shifts slightly:
\begin{subequations}
\bea
\frac{E_{\mathrm{HFS}}^{\mathrm{pole}}(nS)}{E_\mathrm{F}(nS)}&\rightarrow&\frac{E_{\mathrm{HFS}}^{\mathrm{pole}}(nS)}{E_\mathrm{F}(nS)}+\frac{6Z\al m}{\pi(1+\kappa)M}\int_0^\infty\frac{\dd Q}{Q}\frac{F_2(Q^2)G_M(Q^2)}{v_l+1},\\
\frac{E_{\mathrm{HFS}}^{\mathrm{inel.}}(nS)}{E_\mathrm{F}(nS)}&\rightarrow&\frac{E_{\mathrm{HFS}}^{\mathrm{inel.}}(nS)}{E_\mathrm{F}(nS)}-\frac{6Z\al m}{\pi(1+\kappa)M}\int_0^\infty\frac{\dd Q}{Q}\frac{F_2(Q^2)G_M(Q^2)}{v_l+1}.
\eea
\end{subequations}
\addtocontents{toc}{\protect\setcounter{tocdepth}{4}}
\subsection{Rearrangement into Born and Polarizability Contributions}\seclab{1334} 
In practice, we are interested in the polarizability part of the TPE effect, given by the non-Born part of the VVCS amplitude, and the remaining TPE Born diagrams, cf.\ Fig.~\ref{fig:TPE} (a). In \secref{chap3}{polarizabilities}, we gave a definition of the term ``polarizability''. We need to recall that the elastic nucleon-pole part and the Born part of CS are not necessarily the same, cf.\ \Eqref{TextEq}. Since the nucleon-pole and Born parts of the Compton amplitudes are not equivalent, slight modifications are needed to rearrange the previously derived ``elastic'' and ``inelastic'' contributions into the ``Born'' and ``polarizability'' contributions of forward TPE. 

In order to get a numerical estimate for the TPE effects in the hydrogen spectrum, the final expressions can be evaluate based on empirical information, i.e., parametrizations of the elastic FFs and structure functions, or theoretical predictions for the dominant contributions to the VVCS structure functions, e.g., tree-level CS, pion-nucleon loops, $\Delta$-exchange, et cetera. However, the BC sum rule, \Eqref{BCdef}, is in general not evaluating to zero for separate diagrams or kinematic regions. Besides, the parametrized structure functions might not fulfil the BC sum rule satisfactorily. Therefore, it is convenient to remove terms proportional to the BC sum rule from both the Born and polarizability contributions. As we will see below, this is achieved in an intriguing way.

The Born part of the VVCS amplitudes, given by the tree-level diagrams shown in \Figref{Born}, is well known \cite{Drechsel:2002ar}:\footnote{In \appref{chap5}{offVVCSBorn}, we give analogue formulas for off-forward VVCS.}
\begin{subequations}
\eqlab{T12Born}
\bea
T_1^{\mathrm{Born}}(\nu, Q^2) &=&\frac{4\pi Z^2\alpha}{M}\bigg[\frac{Q^4\,G_M^2(Q^2)}{Q^4-4M^2\nu^2}-F_1^2(Q^2)\bigg],
\eqlab{T1Born}\\
T_2^{\mathrm{Born}}(\nu, Q^2)&=&\frac{16\pi Z^2\alpha M Q^2}{Q^4-4M^2\nu^2}\bigg[F_1^2 (Q^2)+\frac{Q^2}{4M^2} F_2^2(Q^2)\bigg],
\eqlab{T2Born}
\\
S_1^{\mathrm{Born}}(\nu, Q^2) &=& \frac{2\pi Z^2\alpha}{M}
\bigg[\frac{4M^2 Q^2\,G_M(Q^2)F_1(Q^2)}{Q^4-4M^2\nu^2}-F_2^2(Q^2)\bigg]\,, \eqlab{S1Born}\\
S_2^{\mathrm{Born}}(\nu, Q^2)&=&
-\, \frac{8 \pi Z^2\alpha M^2 \nu}{Q^4-4M^2\nu^2}\,G_M(Q^2) F_2(Q^2).
\eqlab{S2Born}
\eea
\end{subequations}
Note that we defined a DR for $\nu S_2$ rather than $S_2$, \Eqref{S2DR}, because $S_2$ has a pole in the subsequent limits of $Q^2\rightarrow 0$ and $\nu\rightarrow 0$. The Born part of $\nu S_2$ follows from \Eqref{S2Born} in a straight forward way.
A comparison with \Eqref{poles} shows that the nucleon-pole and Born parts of the VVCS amplitudes are related in the following way:
\begin{subequations}
\eqlab{BornElasticDiff}
\bea
T_1^{\mathrm{pole}}(\nu, Q^2) &=&T_1^{\mathrm{Born}}(\nu, Q^2) + \frac{4\pi Z^2\al}{M}F_1^2(Q^2), 
\eqlab{T1poleBorn}\\
T_2^{\mathrm{pole}}(\nu, Q^2)&=&T_2^{\mathrm{Born}}(\nu, Q^2) ,
\eqlab{T2poleBorn}
\\
S_1^{\mathrm{pole}}(\nu, Q^2) &=& S_1^{\mathrm{Born}}(\nu, Q^2) + \frac{2\pi Z^2\al}{M}F_2^2(Q^2),\eqlab{S1poleBorn}\\
\left[\nu S_2\right]^{\mathrm{pole}}(\nu, Q^2)&=&\nu S_2^{\mathrm{Born}}(\nu, Q^2)-2\pi Z^2 \al F_2(Q^2)\, G_M(Q^2).\eqlab{S2poleBorn}
\eea
\end{subequations}
From these expressions one can deduce the difference between the inelastic and polarizability parts of the amplitudes, exploiting 
\beq
T^\mathrm{pole}+T^\mathrm{inel.}=T^\mathrm{Born}+\ol T,
\eeq
where $\ol T$ denotes the non-Born part.
To derive the TPE polarizability effect, it is then useful to write down once-subtracted DRs for the non-Born part of the VVCS amplitudes:
\begin{subequations}
\bea
\ol T_1(\nu,Q^2)&=&T_1(0,Q^2)+\frac{4\pi Z^2 \al}{M}\Bigg\{\left[F_1^2(Q^2)-G_M^2(Q^2)\right]\nn\\
&&+\frac{8M^2\nu^2}{Q^4}\int_{0}^{x_0}
\,\dd x \, 
\frac{x f_1 (x, Q^2)}{1 - x^2 (\nu/\nu_{\mathrm{el}})^2 - i 0^+}\Bigg\},\eqlab{T1nBsub}\\
\ol S_1(\nu,Q^2)&=&\frac{2\pi Z^2\al}{M} \Bigg\{ \left[F_2^2(Q^2)+4I_1(Q^2)/Z^2\right]\nn\\
&&+\frac{32 M^4\nu^2}{Q^6}  \int_{0}^{x_0} 
\!\dd x\,
\frac{x^2 g_1 (x, Q^2)}{1 - x^2 (\nu/\nu_{\mathrm{el}})^2  - i 0^+}\Bigg\},\eqlab{S1subtrDR}\\
\nu \ol S_2(\nu,Q^2)&=&2\pi Z^2\al\Bigg\{-F_2(Q^2)G_M(Q^2)+4I_2(Q^2)/Z^2\nn\\
&&+ \frac{32 M^4\nu^2}{Q^6}  \int_{0}^{x_0} 
\!\dd x\,
\frac{x^2 g_2 (x, Q^2)}{1 - x^2 (\nu/\nu_{\mathrm{el}})^2  - i 0^+} \Bigg\},\quad\nn\\
&=&\frac{64\pi Z^2\al M^4\nu^2}{Q^6}  \int_{0}^{x_0} 
\!\dd x\,
\frac{x^2 g_2 (x, Q^2)}{1 - x^2 (\nu/\nu_{\mathrm{el}})^2  - i 0^+},\eqlab{S2subtrDR}
\eea
\end{subequations}
where we identified the generalized GDH integrals defined in Eqs.~\eref{I1def} and \eref{I2def}. Note that $I_1$, $I_A$ and $I_2$ are no pure polarizabilities, as will be explained in \secref{chap5}{PolExpTPE} and \Eqref{genGDHnonpole}.

In \Eqref{T1nBsub}, we introduced the conversion term of \Eqref{T1poleBorn}, $4\pi Z^2\al F_1^2(Q^2)/M$, and at the same time removed the nucleon-pole part from the subtraction function, $T_1^\mathrm{pole}(0,Q^2)=4\pi Z^2\al \,G_M^2(Q^2)/M$, cf.\ \Eqref{T1pole}. This is equivalent to removing the Born part from the subtraction function, $T_1^\mathrm{Born}(0,Q^2)$, cf.\ \Eqref{T1Born}. Equation~\eref{S2subtrDR} benefits from a cancelation between the conversion term of \Eqref{S2poleBorn}, $-2\pi Z^2 \al F_2^2(Q^2)G_M^2(Q^2)$, and the sum rule subtraction, $\propto \int_0^{x_0}\dd x\, g_2(x,Q^2)$. This becomes obvious by writing out the elastic part of the BC sum rule, cf.\ \Eqref{I2def}.
In the same way, one finds:
\beq
\left[\nu S_2\right]^\mathrm{Born}(\nu,Q^2)=\frac{64\pi Z^2\al M^4\nu^2}{Q^6}  \int_{x_0}^1 
\!\dd x\,
\frac{x^2 g_2 (x, Q^2)}{1 - x^2 (\nu/\nu_{\mathrm{el}})^2  - i 0^+}.
\eeq
Accordingly, terms proportional to the BC sum rule are removed in what follows and do not affect the TPE effect in the HFS. 
\subsection{Two-Photon Exchange in the Lamb Shift} \seclab{LSformulasTPE}
\addtocontents{toc}{\protect\setcounter{tocdepth}{0}}
\subsubsection{Born Contribution}
The Born contribution to the LS is given by:
\bea
\Delta E^{\mathrm{Born}}(nS)&=&\frac{8(Z\al)^2 m}{M}\,\phi_n^2\,\int_0^\infty \frac{\dd Q}{Q^3}\,\Bigg(-\frac{v_l+2}{(1+v_l)^2}\,F_1^2(Q^2)+\frac{1}{(1+\tau)(v_l+v)}\times\eqlab{LSBorn}\\
&&\times\Bigg\{\left[\tau+\frac{3+2\tau}{(1+v_l)(1+v)}\right]G_M^2(Q^2)-\frac{1}{\tau}\left[1-\frac{1}{(1+v_l)(1+v)}\right]G_E^2(Q^2) \Bigg\}\Bigg).\quad\nn
\eea
Nevertheless, it is common to subtract the order-$(Z\al)^4$ effect of the charge radius, cf.\ \Eqref{LambShift}, and the contribution of a static, structureless nucleus \cite{Carlson:2011zd}:
\bea
\Delta E^{\mathrm{Born}}(nS)&=&8(Z\al)^2\phi_n^2\int_0^\infty \frac{\dd Q}{Q^2} \Bigg[4m_rG_E'(0)+\frac{m}{M}\frac{1}{Q}\Bigg(-\frac{v_l+2}{(1+v_l)^2}\,\left(F_1^2(Q^2)-1\right)\nn\qquad\\
&&+\frac{1}{(1+\tau)(v_l+v)}\Bigg\{\left[\tau+\frac{3+2\tau}{(1+v_l)(1+v)}\right]\left(G_M^2(Q^2)-1\right)\nn\\
&&-\frac{1}{\tau}\left[1-\frac{1}{(1+v_l)(1+v)}\right]\left(G_E^2(Q^2)-1\right) \Bigg\}\Bigg)\Bigg],\qquad\eqlab{LSBornsub}
\eea
with $G_E'=\dd G_E(Q^2)/\dd Q^2$ and $G_E'(0)=-R_E^2/6$, cf.\ \Eqref{LSrewrite5}. In \secref{chap5}{matchingOTPE}, we will discuss how one identifies the order-$(Z\al)^4$ effects, which at first glance are of order $(Z\al)^5$. 

Furthermore, we can isolate the Friar radius contribution:
\beq
\Delta E^{\mathrm{Friar}}(nS)= - 16(Z\al)^2m_r\,\phi_n^2\int_0^\infty \frac{\dd Q}{Q^4}\left[G_E^2(Q^2)-1+\third  R_E^2 \,Q^2\right]=-\frac{Z\al}{3a^4n^3}R_\mathrm{F}^3.
\eeq
Obviously, if the Friar radius is substituted from $ep$ scattering, 
its dependence on the charge radius 
generates a consistency problem \cite{Karshenboim:2014maa,Karshenboim:2014vea,Karshenboim:2015bwa}. The remaining Born contribution is then of recoil type:
\bea
\Delta E^{\mathrm{recoil}}(nS)&=& -8(Z\al)^2\phi_n^2 \,\frac{m}{M}\int_0^\infty \frac{\dd Q}{Q^3} \Bigg[\frac{v_l+2}{(1+v_l)^2}\,\left(F_1^2(Q^2)-1\right)\\
&&-\frac{1}{(1+\tau)(v_l+v)}\left[\tau+\frac{3+2\tau}{(1+v_l)(1+v)}\right]\left(G_M^2(Q^2)-1\right)\nn\\
&&+\Bigg\{\frac{1}{\tau(1+\tau)(v_l+v)}\left[1-\frac{1}{(1+v_l)(1+v)}\right]-\frac{2M^2}{Q(M+m)} \Bigg\}\left(G_E^2(Q^2)-1\right)\Bigg].\nn
\eea
Ref.~\cite{Karshenboim:2015bwa} in addition distinguishes two classes of elastic proton structure effects: finite-size recoil effects and effects generated by the anomalous magnetic moment of the proton. This separation is useful because the effect of the anomalous magnetic moment, entering through the magnetic Sachs FF, can be calculated with higher accuracy than the FSEs.
 
\subsubsection{Polarizability Contribution}
The polarizability contribution to the LS is given by:
\bea
\Delta E^{\mathrm{pol.}}(nS)&=&\frac{2\al m}{\pi}\,\phi_n^2\,\int_0^\infty \frac{\dd Q}{Q^3}\frac{v_l+2}{(1+v_l)^2}\, \left\{T_1(0,Q^2)+\frac{4\pi Z^2 \al}{M}\left[F_1^2(Q^2)-G_M^2(Q^2)\right]\right\}\nn\qquad\\
&&-32(Z\al)^2Mm\,\phi_n^2\,\int_0^\infty \frac{\dd Q}{Q^5}\,\int_0^{x_0}\dd x\frac{1}{(1+v_l)(1+\sqrt{1+x^2\tau^{-1}})}\times\nn\\
&&\times\Bigg\{\left[1+\frac{v_l\sqrt{1+x^2\tau^{-1}}}{v_l+\sqrt{1+x^2\tau^{-1}}}\right]f_2(x,Q^2)\nn\\
&&+\frac{2x}{(1+v_l)(1+\sqrt{1+x^2\tau^{-1}})}\left[2+\frac{3+v_l\sqrt{1+x^2\tau^{-1}}}{v_l+\sqrt{1+x^2\tau^{-1}}}\right]f_1(x,Q^2)\Bigg\}.\eqlab{LSMasterSub}
\eea
This we can split into the contribution of the polarizability part of the subtraction function of the $T_1$ DR: 
\beq
\Delta E^{\mathrm{subtr.}}(nS)=8\al m\,\phi_n^2\,\int_0^\infty \frac{\dd Q}{Q^3}\frac{v_l+2}{(1+v_l)^2}\, \bigg\{\frac{T_1(0,Q^2)}{4\pi}+\frac{Z^2 \al}{M}\left[F_1^2(Q^2)-G_M^2(Q^2)\right]\bigg\},\eqlab{subpol}
\eeq
and a contribution of inelastic structure functions:
\bea
\Delta E^{\mathrm{inel.}}(nS)&=&-32(Z\al)^2Mm\,\phi_n^2\,\int_0^\infty \frac{\dd Q}{Q^5}\,\int_0^{x_0}\dd x\frac{1}{(1+v_l)(1+\sqrt{1+x^2\tau^{-1}})}\times\eqlab{inelasticpol}\\
&&\times\Bigg\{\left[1+\frac{v_l\sqrt{1+x^2\tau^{-1}}}{v_l+\sqrt{1+x^2\tau^{-1}}}\right]f_2(x,Q^2)\nn\\
&&+\frac{2x}{(1+v_l)(1+\sqrt{1+x^2\tau^{-1}})}\left[2+\frac{3+v_l\sqrt{1+x^2\tau^{-1}}}{v_l+\sqrt{1+x^2\tau^{-1}}}\right]f_1(x,Q^2)\Bigg\}.\qquad\nn
\eea
\subsubsection[$T_1$ Subtraction Function]{$\boldsymbol{\ol T_1}$ Subtraction Function} \seclab{subtrfuncsec}
The subtraction function of the $T_1$ DR, cf.\ $T_1(0,Q^2)$ in Eqs.~\eref{T1Subtrgen} and \eref{T1nBsub}, is not exactly known. Strictly speaking, the Born part of the subtraction can be deduced from \Eqref{T1Born}, while the polarizability part, cf.\ \Eqref{subpol}, is the unknown. It has to be either modelled or calculated in a theoretical framework such as ChPT.  It was suggested that one could make up a model where the effect is large enough to resolve the proton radius puzzle~\cite{Miller:2012}, but none of the realistic calculations
have corroborated this claim. In particular, the ChPT calculations demonstrate a very moderate size for these effects, cf.\ \cite{Alarcon:2013cba}.
As we shall see below, all of the dispersive models  find the polarizability part of the subtraction to give a contribution of the order of 
few $\upmu$eV, well below 300  $\upmu$eV needed to resolve the puzzle. 

In the limit of small momentum transfers, $\ol T_1(0,Q^2)$ is given by:
\beq
\lim_{Q^2\rightarrow0}\frac{\ol T_1(0,Q^2)}{Q^2}=4\pi\,\beta_{M1}.\eqlab{T1betaM1}
\eeq
Since the calculation of the LS comprises an integral over $Q^2$, the subtraction is required as a full function of $Q^2$. For the $Q^2$ dependence of the magnetic polarizability, \citet{Pachucki:1999zza} proposed a dipole parametrization:
\beq
\beta _{M1}(Q^2)=\beta_{M1}\,\frac{\Lambda^8}{(\Lambda^2+Q^2)^4},
\eeq
with $\Lambda^2=0.71 \text{ GeV}^2$. This assumption is also adopted in following-up papers \cite{Martynenko:2005rc, Birse:2012eb} by other authors.

Working in HBChPT, Ref.~\cite{Birse:2012eb} obtains a model-independent result for $\ol T_1(0,Q^2)$, valid up to and including $\mathcal{O}(Q^4)$. Their low-$Q$ prediction is matched to a $1/Q^2$ behavior at large momentum transfers, which is again establish by a dipole ansatz. \citet{Carlson:2011zd} estimate the higher-$Q$ behavior  by calculating pion-loop contributions with scalar two-pion coupling to the nucleus, as the dominating low-mass intermediate states. The logarithmic asymptotics found thereby differ from the previously shown dipole forms \cite{Pachucki:1999zza,Martynenko:2005rc,Birse:2012eb}. Nevertheless, the dominant $1/Q^2$ 
behavior can be given a reason from arguments based on quark counting rules \cite{Birse:2012eb} or the operator product expansion \cite{Hill:2011wy}. Another empirical estimate can be found in Ref.~\cite{Tomalak:2015hva}.

As outlined above, a critical point for the prediction of TPE effects in the LS is the $T_1$ subtraction function, and especially its polarizability part, which is closely connected to the magnetic dipole polarizability. Therefore, it is crucial to separate the VVCS and TPE amplitudes into Born and non-Born pieces, as favoured by Ref.~\cite{Birse:2012eb}. In contrast, the separation into elastic and inelastic pieces \cite{Carlson:2011zd} was shown to be not unique \cite{WalkerLoud:2012}, and, as pointed out by Ref.~\cite{Birse:2012eb}, inconsistent with the standard definition of the magnetic dipole polarizability through the non-Born CS amplitude, cf.\ \Eqref{T1betaM1}.

\addtocontents{toc}{\protect\setcounter{tocdepth}{4}}
\subsection{Two-Photon Exchange in the Hyperfine Splitting} \seclab{HFSformulasTPE}
The LO HFS of the $n$-th $S$-level is given by the Fermi energy, \Eqref{FermiE}. The subleading contributions to the HFS can be divided
into QED, electroweak and structure corrections:
\beq
E_{\mathrm{HFS}}(nS)=\left[1+\Delta_\mathrm{QED}+\Delta_\mathrm{weak}+\Delta_\mathrm{structure}\right]E_\mathrm{F}(nS). \eqlab{HFSwoEF}
\eeq
We are interested in the proton-structure correction, which in turn splits into three terms: Zemach radius, recoil, and polarizability contributions:
\beq
\Delta_\mathrm{structure}=\Delta_\mathrm{Z}+\Delta_\mathrm{recoil}+\Delta_\mathrm{pol.}\,. \eqlab{FS_HFS}
\eeq
Let us now specify the decomposition of the structure-dependent correction into the three terms of \Eqref{FS_HFS}. An examination of different decompositions of the TPE effect can be found in Ref.~\cite{Carlson:2011af}. The formalism presented by us is consistent with the choice of \citet{Carlson:2011af}.
\addtocontents{toc}{\protect\setcounter{tocdepth}{0}}
\subsubsection{Born Contribution}
As stated earlier, the master formulae in \secref{chap5}{MasterFormulae} contain all the structure effects to order $(Z \al)^5$, i.e., also the Fermi energy, which has to be subtracted in the following. The TPE Born contribution to the HFS splits into the Zemach radius contribution \cite{Zemach:1956}:
\beq
\Delta_\mathrm{Z}=\frac{8Z\al m_r}{\pi}\int_0^\infty \frac{\dd Q}{Q^2} \left[\frac{G_E(Q^2)G_M(Q^2)}{1+\kappa}-1\right]\equiv-2Z\al m_r R_\mathrm{Z},\eqlab{ZemachTerm} 
\eeq
and a recoil-type of correction:
\bea
\Delta_\mathrm{recoil}&=&\frac{Z \al}{\pi (1+\kappa)}\int_0^\infty \frac{\dd Q}{Q}\left\{\frac{8m M}{v_l+ v }\frac{G_M(Q^2)}{Q^2}   \left(2F_1(Q^2)+\frac{F_1(Q^2)+3F_2(Q^2)}{(v_l+1)(v+1)}  \right)\right.\qquad\quad\nn\\
&&\left.\hspace{2.75cm}-\frac{8m_r\,G_M(Q^2)G_E(Q^2)}{Q}-\frac{m}{M}\frac{5+4v_l}{(1+v_l)^2}F_2^2(Q^2)\right\}\eqlab{recoilHFS}.
\eea
In contrast to the Zemach radius term, the recoil corrections are not zero in the static limit of the elastic FFs.
\subsubsection{Polarizability Contribution} \seclab{polHFS}
In the polarizability contribution, we separate contributions due to the spin-dependent structure functions $g_1$ and $g_2$:
\begin{subequations}
\eqlab{POL}
\beq
\Delta_\mathrm{pol.}=\frac{Z\al m}{2\pi (1+\kappa) M}\left[\delta_1+\delta_2\right]=\Delta_1+\Delta_2,
\eeq
with:
\begin{align}
\delta_1&=2\int_0^\infty\frac{\dd Q}{Q}\left(\frac{5+4v_l}{(v_l+1)^2}\left[4I_1(Q^2)/Z^2+F_2^2(Q^2)\right]+\frac{8M^2}{Q^2}\int_0^{x_0}\dd x\, g_1(x,Q^2)\right.\eqlab{Delta1a}\\
& \quad \left.\left\{ \frac{4}{v_l+ \sqrt{1+x^2\tau^{-1}}}\left[1+\frac{1}{2(v_l+1)(1+ \sqrt{1+x^2\tau^{-1}})}\right]-\frac{5+4v_l}{(v_l+1)^2} \right\}\right),\nn\\
&=2\int_0^\infty\frac{\dd Q}{Q}\left(\frac{5+4v_l}{(v_l+1)^2}\left[4I_1(Q^2)/Z^2+F_2^2(Q^2)\right]-\frac{32M^4}{Q^4}\int_0^{x_0}\dd x\, x^2g_1(x,Q^2)\right.\eqlab{Delta1b}\\
& \quad  \left.\left\{ \frac{1}{(v_l+ \sqrt{1+x^2\tau^{-1}})(1+ \sqrt{1+x^2\tau^{-1}})(1+v_l)}\left[4+\frac{1}{1+ \sqrt{1+x^2\tau^{-1}}}+\frac{1}{v_l+1}\right] \right\}\right)
,\nn\qquad\\
\delta_2&=96M^2\int_0^\infty\frac{\dd Q}{Q^3}\int_0^{x_0}\dd x\,g_2(x,Q^2) \left\{\frac{1}{v_l+ \sqrt{1+x^2\tau^{-1}}}-\frac{1}{v_l+1} \right\}.\eqlab{Delta2}
\end{align}
\end{subequations}
Expanding $\delta_2$, \Eqref{Delta2}, in $x$ confirms that the BC sum rule is absent. Equation~\eref{Delta1a} is based on the master formula, cf.\ Eq.~\eref{fullHFS}, whereas \Eqref{Delta1b} is derived using a once-subtracted DR for $S_1$, cf.\ \Eqref{S1subtrDR}. The advantage of  \Eqref{Delta1b} lies in the complete separation of the zeroth moment, $I_1(Q^2)$, from higher moments of the structure function $g_1(x,Q^2)$.

\subsubsection{The Burkhardt-Cottingham Sum Rule}\seclab{BCHFS}
As described above, the contribution of the BC sum rule to the HFS is removed because it is known to equal zero. This is especially important for the dispersive TPE evaluations, since the parametrizations of the spin-dependent $g_2$ structure function do not have this feature build-in. As we will prove below, the Born and polarizability parts of the BC sum rule are supposed to vanish separately. This is even confirmed to be true for individual sets of non-Born diagrams, e.g., by our BChPT calculations of $\pi N$-loop \cite{Lensky:2014dda} and $\Delta$-exchange diagrams. 

The BC sum rule \cite{Burkhardt:1970ti} is derived from the unsubtracted DR for the amplitude $\nu S_2$, \Eqref{S2DR}, by taking the limit of $\nu \rightarrow 0$. It tells us that the integral over $x\in [0,1]$ of the spin-dependent structure function $g_2(x,Q^2)$ is equal to zero, see \Eqref{BCdef}. Plugging in the elastic part of the structure function, \Eqref{g2elastic}, the inelastic part of the BC integral, $I_2(Q^2)$, equals $\nicefrac{Z^2}{4}\,F_2(Q^2)\,G_M(Q^2)$, see \Eqref{I2def}. It is non-trivial that the Born and polarizability contributions to the BC sum rule vanish independently, as we would like to prove in the following.

Let us write the BC sum rule as:\footnote{The generalized GDH integral $I_1$ can be expressed in a similar way:
\beq 
I_1(Q^2)=\lim_{\nu \rightarrow 0}\;\frac{M S_1(\nu,Q^2)}{8\pi \al}.
\eeq}
\beq
\lim_{\nu \rightarrow 0}\;\frac{\nu S_2(\nu,Q^2)}{8\pi \al}=\frac{2Z^2M^2}{Q^2}\int_0^1 \dd x\, g_2(x,Q^2)=0.
\eeq
We can now split the $\nu S_2$ amplitude as shown in \Eqref{TextEq}. Splitting into elastic and inelastic parts gives the well-known \Eqref{I2def}. Let us instead consider the Born part of the $\nu S_2$ amplitude. It is related to the elastic part through \Eqref{S2poleBorn}. With Eqs.~\eref{S2pole} and \eref{S2poleBorn}, we find:
\begin{subequations}
\bea
\lim_{\nu \rightarrow 0}\;\frac{\nu S_2^\mathrm{Born}}{8\pi \al}&=&\lim_{\nu \rightarrow 0} \left\{\frac{\left[\nu S_2\right]^{\mathrm{pole}}}{8\pi \al}+\frac{Z^2}{4}\, F_2(Q^2)\, G_M(Q^2)\right\},\\
&=&\lim_{\nu \rightarrow 0} \left\{-\frac{Z^2}{4}\, F_2(Q^2)\, G_M(Q^2)+\frac{Z^2}{4}\, F_2(Q^2)\, G_M(Q^2)\right\}=0.\qquad\quad
\eea
\end{subequations}
In other words, the Born contribution to the BC sum rule is vanishing. Writing down the full BC sum rule:
\beq
\lim_{\nu \rightarrow 0}\;\nu S_2=\lim_{\nu \rightarrow 0}\;\left\{\nu S_2^\mathrm{Born}+\nu \ol{S_2}\right\}=0,
\eeq
the non-Born or polarizability contribution obviously also has to vanish: 
\beq
\lim_{\nu \rightarrow 0}\;\nu \ol{S_2}=0.
\eeq

\addtocontents{toc}{\protect\setcounter{tocdepth}{4}}
\section{Matching One- and Two-Photon Exchange}\seclab{matchingOTPE}

One ought to be careful
in matching the elastic TPE contribution, cf.\ \Eqref{elTPEpole}, to the standard FSEs, see \chapref{chap2} and \Eqref{FSEs}. In the heavy-nucleus limit, we obtain:
\begin{subequations}
\eqlab{NRelastic}
\bea
\Delta E^{\mathrm{pole}}(nS)&\approx& 
- 16(Z\al)^2m_r\,\phi_n^2\,\int_0^\infty \frac{\dd Q}{Q^4}\,
G_E^2 (Q^2), 
\eqlab{NRelasticLS}
\\
E_{\mathrm{HFS}}^{\mathrm{pole}}(nS)&\approx&\frac{64(Z\al)^2 m_r}{3mM}\,\phi_n^2
 \int_0^\infty\frac{\dd Q}{Q^2}\, G_M(Q^2)  \,G_E(Q^2) .
\eqlab{NRelasticHFS}
\eea
\end{subequations}
The correct matching is achieved by regularizing the
infrared divergences with the convoluted momentum-space wave functions, see \Eqref{ConvolutionsWFMom}. This procedure should not change the ultraviolet part of the $Q$-integration.
The
asymptotic behavior of the convoluted momentum-space wave functions is described by:
\beq
w_{nS}(Q) \stackrel{Q\rightarrow \infty}{=}\frac{16 \pi}{aQ^4}\phi_n^2.
\eeq
Hence, we need to replace $\phi_n^2\to a w_{nS}(Q)\,Q^4/16\pi$ in \Eqref{NRelastic}.

For the LS, we first take out the Friar radius, \Eqref{Friar}, and perform the regularization only afterwards:
\begin{subequations}
\eqlab{LSrewrite5}
\bea
\Delta E^{\mathrm{pole}}(nS)&\approx& 
- 16(Z\al)^2m_r\,\phi_n^2\bigg\{\int_0^\infty \frac{\dd Q}{Q^4}
\left[1-\third  R_E^2 \,Q^2\right]\eqlab{LSrewrite5a}\\
&&+\underbrace{\int_0^\infty \frac{\dd Q}{Q^4}\left[G_E^2(Q^2)-1+\third  R_E^2 \,Q^2\right]}_{=\frac{\pi}{48}R_\mathrm{F}^3}\bigg\},\nn\\
&=&-\frac{Z\al}{\pi} \int_0^\infty  \dd Q \,w_{nS}(Q) \left[1-\third  R_E^2 \,Q^2\right]-\frac{Z\al}{3a^4n^3}R_\mathrm{F}^3,\\
&=&-\frac{Z\al}{2an^2}+\frac{2Z\al}{3(an)^3}R_E^2-\frac{Z\al}{3a^4n^3}R_\mathrm{F}^3.
\eea
\end{subequations}
In the last step, we used:
\begin{subequations}
\bea
\frac{2\pi}{ (a n)^3}&=&\int_0^\infty  \dd Q\,Q^2\,w_{nS}(Q),\eqlab{normQ}\\
\frac{\pi}{2a n^2}&=&\int_0^\infty  \dd Q\,w_{nS}(Q),
\eea
\end{subequations}
where \Eqref{normQ} follows from \Eqref{1PTsphsym} with $V_\delta(Q)=1$, thus, it is equivalent to the normalization of the coordinate-space wave functions. In a fascinating manner, the TPE correctly reproduces the gross structure, \Eqref{Bohr}, as well as the charge and Friar radius contributions to the LS, \Eqref{LambShift}:
\beq
\Delta E^{\mathrm{pole}}(2S)=-\frac{Z\al}{8a}+\frac{Z\al}{12a^3}R_E^2-\frac{Z\al}{24a^4}R_\mathrm{F}^3.
\eeq
To avoid double counting, we have removed the order-$(Z\al)^4$ FSE and the contribution of a static, structureless nucleus from the LS formula in \Eqref{LSBornsub}. In the heavy-nucleus limit, this corresponds to the first line in \Eqref{LSrewrite5a}.

For the HFS, we take out the Zemach radius, \Eqref{Zemach}, and identify the Fermi energy, \Eqref{FermiE}:
\begin{subequations}
\bea
E_{\mathrm{HFS}}^{\mathrm{pole}}(nS)&\approx&\frac{64(Z\al)^2 m_r}{3mM}\,\phi_n^2
 \int_0^\infty\frac{\dd Q}{Q^2}\,G_M(Q^2)  \,G_E(Q^2) ,\\
 &=&\frac{8E_\mathrm{F}(nS)}{a\pi}
\bigg\{ \underbrace{\int_0^\infty\frac{\dd Q}{Q^2}\left[\frac{G_M(Q^2)  \,G_E(Q^2)}{1+\kappa} -1\right]}_{=-\frac{\pi}{4}R_\mathrm{Z}}+\int_0^\infty\frac{\dd Q}{Q^2}\bigg\},\\
&=&E_\mathrm{F}(nS)\left[\frac{1}{2\pi^2}\int_0^\infty \dd Q\, Q^2 \,\frac{w_{nS}(Q)}{\phi_n^2}-\frac{2}{a}R_\mathrm{Z}\right],\\
&=&E_\mathrm{F}(nS)\left[1-\nicefrac{2}{a}\,R_\mathrm{Z}\right].
\eea
\end{subequations}
This yields the correct Fermi energy and Zemach radius contributions, see \Eqref{HFS}. Again, we subtract the order-$(Z\al)^4$ term from the TPE in \Eqref{ZemachTerm} to avoid double counting.

The recoil
and polarizability corrections are infrared-safe and require no regularization. Unfortunately, it is not possible to match the recoil FSEs from one- and two-photon exchange. In practice, one uses the results from forward TPE discussed in this Chapter. This has several reasons. Most importantly, the Breit potential in \chapref{chap2} is based on an e.m.\ interaction vertex for on-shell nucleons and nuclei, \Eqref{protonphotonvertex}. This is not sufficient, as the intermediate state in Fig.~\ref{fig:TPE} (a) can be off-shell \cite{Carlson:2008ke, Karshenboim:2015bwa}. The dispersive approach to the TPE, see \secref{chap5}{MasterFormulae}, bypasses this issue. It uses DRs to express the VVCS amplitudes in terms of structure functions or cross sections, cf.\ \Eqref{DRVVCS}. Per definition, the cross sections have real particles in the final state. Also, the TPE formalism, in contrast to the Breit potential, does not neglect retardation.

\section{Forward Two-Photon Exchange in Terms of Polarizabilities} \seclab{PolExpTPE}

In this Section, we will present the TPE polarizability effects in Eqs.~\eref{LSMasterSub} and \eref{POL} as an expansion in moments
of structure functions. The expansion of the auxiliary functions for small $x$ goes as:
\begin{subequations}
\bea 
\frac{1}{1+ \sqrt{1+x^2\tau^{-1}}} &=& \frac{1}{2}\left[1-\frac{x^2}{4\tau}+\frac{x^4}{8\tau^2}\right]+
\mathcal{O}(x^6),\\
\frac{1}{v_l+ \sqrt{1+x^2\tau^{-1}}} &=& 
\frac{1}{1+v_l}\left[ 1 - \frac{1}{2(1+v_l)} \frac{x^2}{\tau}+ \frac{(v_l+3)}{8(1+v_l)^2}\frac{x^4}{\tau^2}  \right]+
\mathcal{O}(x^6).\qquad
\eea
\end{subequations}
Expanding the HFS formula \eref{fullHFS} up to and including $\mathcal{O}(x^4)$, we obtain a reasonable description of the main characteristics of the weighting function:
\bea
\frac{E_{\mathrm{HFS}}(nS)}{E_\mathrm{F}(nS)}&=&
\frac{8Z\al m M}{\pi (1+\kappa)}\int_0^\infty\frac{\dd Q}{Q^3}\frac{1}{(v_l+1)} \times\eqlab{ExpBC}\\
&&\times\int_0^1\dd x\left\{3\left[2-\frac{1}{(v_l+1)}\frac{x^2}{\tau}+\frac{3+vl}{4(1+v_l)^2}\frac{x^4}{\tau^2}\right] g_2(x,Q^2)\right.\nn\\
&&\left.+\frac{1}{(v_l+1)} \left[5+4v_l -\frac{11+9v_l}{4(v_l+1)}\frac{x^2}{\tau}+\frac{17+20v_l+5v_l^2}{8(v_l+1)^2}\frac{x^4}{\tau^2}\right]g_1(x,Q^2) \right\},\nn\qquad\quad
\eea
where one can read of the still present contribution of the BC sum rule, cf.\ \Eqref{BCdef}. Limiting ourselves
to the third moments of the structure functions,
we for the LS have:
\bea
\Delta E^\mathrm{pol.}(nS)&=&\frac{2\al m}{\pi}\,\phi_n^2\,\int_0^\infty \frac{\dd Q}{Q^3}\frac{v_l+2}{(1+v_l)^2}\, \bigg\{T_1(0,Q^2)+\frac{4\pi Z^2 \al}{M}\left[F_1^2(Q^2)-G_M^2(Q^2)\right]\bigg\}\nn\\
&&-16(Z\al)^2Mm\,\phi_n^2\,\int_0^\infty \frac{\dd Q}{Q^5}\,\frac{1}{ (v_l+1)^2}\times\nn\\
&&\qquad\times\int_0^{x_0}\dd x\,\Bigg\{\left[\frac{3 v_l+5}{v_l+1}\,x-\left(\frac{v_l+2}{v_l+1}\right)^2\frac{x^3}{\tau}\right] f_1(x,Q^2)\nn\\
&&\qquad+\Bigg[1+2v_l-\frac{3 v_l+1}{4(v_l+1)}\frac{x^2}{\tau}\Bigg]\,f_2(x,Q^2)\Bigg\}.\eqlab{polLSExp}
\eea
For the HFS, we have:
\begin{subequations}
\eqlab{polHFSExp}
\bea
\frac{E^\mathrm{pol.}_{\mathrm{HFS}}(nS)}{E_\mathrm{F}(nS)}&=&
\frac{Z\al m}{\pi(1+\kappa)M}\int_0^\infty\frac{\dd Q}{Q}\frac{1}{(v_l+1)^2}\Bigg\{(5+4v_l)\,F_2^2(Q^2)\eqlab{polHFSExpa}\\
&&+\frac{1}{Z^2(v_l+1)}\Bigg[(1+3v_l)\,I_A(Q^2)+(19+33v_l+16v_l^2)\,I_1(Q^2)\nn\\
&&-\frac{M^2Q^2}{2\al}(11+9v_l)\,\delta_{LT}(Q^2)\Bigg] \Bigg\},\nn
\\
&=&
\frac{\al m}{\pi Z(1+\kappa)M }\int_0^\infty\frac{\dd Q}{Q}\frac{1}{(1+v_l)^2} \Bigg\{(5+4v_l)\left[Z^2F_2^2(Q^2)+ 4I_1(Q^2)\right]\qquad\qquad\eqlab{polHFSExpb}\\
&&
-\frac{6 M^2Q^2}{\al}\,\delta_{LT}(Q^2)+\frac{1+3v_l}{1+v_l}\Bigg(\frac{ M^2Q^2}{2\al}\,\gamma_0(Q^2)\nn\\
&&+\frac{32Z^2M^6}{Q^6}\int_0^{x_0}\dd x\, x^4\, g_2(x,Q^2)\Bigg)\Bigg\},
\nn
\eea
\end{subequations}
where we identified the generalized spin polarizabilities defined in Eqs.~\eref{g0gen} and \eref{dLTgen}-\eref{IAdef}. In Eqs.~\eref{polHFSExpa} and \eref{polHFSExpb}, we use either the FSP $\ga_0$, or the generalized GDH integral $I_A$. Here, we only expand until $\mathcal{O}(x^3)$. Therefore, we need to subtract the fourth moment of $g_2$ contained in $\ga_0$, cf.\ \Eqref{polHFSExpb}.

Strictly speaking, $I_1$ and $I_A$ are no pure polarizabilities, but also contain the elastic Pauli FF:
\beq
I_1^\mathrm{non-pol.}(Q^2)=I_A^\mathrm{non-pol.}(Q^2)=-\frac{Z^2}{4}F_2^2(Q^2).\eqlab{genGDHnonpole}
\eeq
This feature follows from the difference between inelastic and polarizability contributions. With \Eqref{genGDHnonpole} in mind, it becomes obvious that the total expressions, Eqs. \eref{polLSExp} and \eref{polHFSExp}, are pure polarizability contributions.

Let us now take a closer look at \Eqref{genGDHnonpole} and derive it properly. The generalized GDH integrals are related in the following way:
\beq
I_A(Q^2)=I_1(Q^2)-\frac{8Z^2M^4}{Q^4}\int_0^{x_0}\dd x \,x^2 g_2(x,Q^2).\eqlab{IAI1diff}
\eeq
Plugging the elastic $g_2$ structure function, \Eqref{g2elastic}, into the $S_2$ DR: 
\beq
S_2(\nu,Q^2)=\frac{64\pi Z^2\al M^4\nu}{Q^6}  \int_{0}^{1} 
\!\dd x\,
\frac{x^2 g_2 (x, Q^2)}{1 - x^2 (\nu/\nu_{\mathrm{el}})^2  - i 0^+},\eqlab{S2pureDR}
\eeq
and comparing with the Born part of $S_2$, \Eqref{S2Born}, we verify that $S_2^\mathrm{pole}(\nu,Q^2)=S_2^\mathrm{Born}(\nu,Q^2)$. Accordingly, we have $S_2^\mathrm{inel.}(\nu,Q^2)=\ol S_2(\nu,Q^2)$, viz.\ the inelastic moments of $g_2$ are pure polarizabilities.
In the limit of $\nu\rightarrow 0$, the $S_2$ DR reduces to:
\beq
\frac{S_2(\nu,Q^2)}{\nu}\bigg\vert_{\nu\rightarrow 0}=\frac{64\pi Z^2 \al M^4}{Q^6}\int_0^1\dd x \,x^2 g_2(x,Q^2).
\eeq
The inelastic part of this integral is proportional to the difference between the generalized GDH integrals $I_1$ and $I_A$ given in \Eqref{IAI1diff}: 
\beq
I_A(Q^2)=I_1(Q^2)-\frac{Q^2}{8\pi \al} \frac{\ol S_2(\nu,Q^2)}{\nu}\bigg\vert_{\nu\rightarrow 0}\,.
\eeq 
Thus, the non-polarizability parts of $I_1$ and $I_A$ are equivalent.

Splitting the $S_1$ DR, \Eqref{S1DR}, at the inelastic threshold, identifying $I_1$, \Eqref{I1def}, and plugging in the elastic $g_1$ structure function, \Eqref{g1elastic}, we have:
\beq
S_1(0,Q^2)=\frac{8 \pi  \al}{M}\left[I_1(Q^2)+\frac{Z^2M^2}{Q^2}F_1(Q^2)G_M(Q^2)\right].\eqlab{S10}
\eeq
We know that the $S_1$ amplitude can be decomposed into a Born and a non-Born part. The Born contribution to $S_1(0,Q^2)$ can be read off from \Eqref{S1Born} as:
\beq
S_1^\mathrm{Born}(0,Q^2)=\frac{2\pi Z^2 \al}{M}\left[\frac{G_M(Q^2)F_1(Q^2)}{\tau}-F_2^2(Q^2)\right].
\eeq
We are then left with non-Born part, which is a polarizability and will be denote by $\ol S_1(0,Q^2)$. Replacing $S_1(0,Q^2)=S_1^\mathrm{Born}(0,Q^2)+\ol S_1(0,Q^2)$ in \Eqref{S10}, we find:
\beq
I_1(Q^2)=\frac{M}{8\pi \al}\ol S_1(0,Q^2)-\frac{Z^2}{4}F_2^2(Q^2),
\eeq
thus confirming \Eqref{genGDHnonpole}.

In a last step, we distinguish contributions to \Eqref{polHFSExp}, which originate from the spin-dependent structure functions $g_1$:
\begin{subequations}
\eqlab{polD1}
\bea
\Delta_1&=&\frac{Z\al m}{\pi(1+\kappa)M}\int_0^\infty\frac{\dd Q}{Q}\frac{1}{(v_l+1)^2}\bigg[(5+4v_l)\,F_2^2(Q^2)+\frac{1}{(v_l+1)}\times\eqlab{polD1a}\\
&&\times\left\{(31+45v_l+16v_l^2)\,I_1(Q^2)-(11+9v_l)\,\left[\frac{M^2Q^2}{2\al}\,\delta_{LT}(Q^2)+I_A(Q^2)\right]\right\} \!\bigg],\qquad\nn \\
&=&\frac{Z\al m}{\pi(1+\kappa)M}\int_0^\infty\frac{\dd Q}{Q}\frac{1}{(v_l+1)^2}\bigg\{(5+4v_l)\,\left[F_2^2(Q^2)+4I_1(Q^2)\right]\eqlab{polD1b}\\
&&-\frac{11+9v_l}{(v_l+1)}\left[\frac{ M^2Q^2}{2\al}\,\gamma_0(Q^2)+\frac{32M^6}{Q^6}\int_0^{x_0}\dd x\, x^4\, g_2(x,Q^2)\right]\!\bigg\},\nn
\eea
\end{subequations}
and $g_2$:
\begin{subequations}
\eqlab{polD2}
\bea
\Delta_2&=&-\frac{12Z\al m}{\pi(1+\kappa)M}\int_0^\infty\frac{\dd Q}{Q}\frac{1}{(v_l+1)^2}\left[I_1(Q^2)-I_A(Q^2)\right],\eqlab{polD2a}\\
&=&-\frac{12Z\al m}{\pi(1+\kappa)M}\int_0^\infty\frac{\dd Q}{Q}\frac{1}{(v_l+1)^2}\left\{\frac{ M^2Q^2}{2\al}\,\left[\delta_{LT}(Q^2)-\gamma_0(Q^2)\right]\right.\eqlab{polD2b}\\
&&\left.-\frac{32M^6}{Q^6}\int_0^{x_0}\dd x\, x^4\, g_2(x,Q^2)\right\},\nn
\eea
\end{subequations}
respectively. Again, we give two alternative sets of equations depending on our choice of polarizabilities. In \secref{5HFS}{polexpHFS}, we will return to the polarizability expansion of the HFS TPE effect and apply it to interpret our results.

\section{Off-forward Two-Photon Exchange in Hydrogen-Like Bound States} \seclab{6.3}
In the following, we turn to the more general case of off-forward TPE. 
The off-forward TPE in the lepton-nucleus bound state is shown in Fig.~\ref{TPEFRW}. The incoming (outgoing) nucleus and lepton four-momenta are denoted by $p$ ($p'$) and $l$ ($l'$), respectively. The Feynman diagram evidently comprises off-forward VVCS off a nucleus, where the four-momentum of the absorbed (emitted) photon is labeled with $q$ ($q'$). 

With respect to the forward TPE diagram, the off-forward diagram is suppressed by an additional factor of $Z \al$. Nevertheless, off-forward TPE processes might be not negligible, since the two-photon cut in the $t$-channel can generate a logarithmic enhancement of the off-forward TPE effect. Hence, our main focus will be on calculating the $(Z\al)^6 \ln Z\al$ nuclear-polarizability effect. We study the contribution of the lowest-order nuclear polarizabilities to off-forward VVCS and TPE, where the main interest is in the nuclear dipole polarizabilities. We derive the corresponding perturbative potential with a dispersive approach, which can be treated in PT. Further discussions of the (numerical) effects in LS and HFS are postponed to \secref{5LS}{offTPE} and \appref{5HFS}{6.5}. In \secref{5HFS}{neutralPion}, we evaluate a similar off-forward TPE process, viz.\ the neutral-pion exchange with two-photon coupling between lepton and pion.

The TPE matrix element satisfies a once-subtracted DR:
\beq
\mathscr M(p_t^2)=\mathscr M(0)+\frac{p_t^2}{\pi}\int_0^\infty \frac{\dd t}{t} \frac{\im \mathscr M (t)}{t-p_t^2-i0^+},\eqlab{DRM}
\eeq
where the $t$-channel momentum transfer is: $p_t=q'-q=p-p'=l'-l$.
Let us have a look at the different terms appearing in \Eqref{DRM}. The subtraction term, $\mathscr M(0)$, corresponds to the forward TPE ($q=q'$), which was  discussed in Sections \ref{chap:chap5}.\ref{sec:TPE} and \ref{chap:chap5}.\ref{sec:PolExpTPE}. It produces a $\delta(\boldsymbol{r})$-potential, as we will see in the following. We, on the other hand, are interested in the off-forward TPE, given by the remaining integral over $t$.

\noindent The coordinate-space potential is the Fourier transform of the momentum-space potential. Assuming vanishing retardation, the momentum-space potential associated to the TPE is equal to $V(\vert\boldsymbol{p_t}\vert)=\mathscr{M}(p_t^2=-\vert\boldsymbol{p_t}\vert^2)$. From \Eqref{DRM}, we derive the once-subtracted coordinate-space TPE potential:
\beq
V(r)=\mathscr M (0)\, \delta (\boldsymbol{r})-\frac{1}{\pi} \int_0^\infty \dd t \, \im \mathscr M(t)\,\left[\frac{\delta (\boldsymbol{r})}{t}-\frac{e^{-r\sqrt{t}}}{4\pi r}\right].\quad\eqlab{potentialTPEoff}
\eeq
Besides the $\delta(\boldsymbol{r})$-potential, a Yukawa-like potential $\propto \nicefrac{e^{-r\sqrt{t}}}{r}$ emerges.

The above potential will be treated in first-order of Schrödinger PT. The relevant theory is summarized in \appref{chap2}{WFPT}. There we list, f.i., the non-relativistic Coulomb wave functions, \Eqref{WFdisc}, and  give a formula for the energy correction induced by a spherically symmetric coordinate-space potential, \Eqref{1PTsphsym}. The information on the lepton-nucleus system of interest is contained in the Bohr radius and the reduced mass $m_r=M m/(M+m)$. For $M \gg m$, the reduced mass is approximately equal to the lepton mass: $m_r\approx m$.

Again, we are only interested in the off-forward process, represented by the integral term in \Eqref{potentialTPEoff}. The $\delta (\boldsymbol{r})$-potential leads to an upward shift of the $S$-levels, cf.\ Table \ref{1PTallpotentials}:
\beq
\langle nl\vert \delta (\boldsymbol{r})\vert nl\rangle=\frac{\delta_{l0}}{\pi (an)^3},
\eeq
 with $l$ the orbital angular momentum.
The Yukawa-type potential contributes to all orbitals, i.a., $S$- and $P$-waves:
\bea
\eqlab{pwavehigh}
&&\langle nl\vert  \nicefrac{e^{-\sqrt{t} r}}{r}  \vert nl \rangle\\
&&=\frac{4^{1+l} \,t^{n-1-l}  \left(\frac{2}{a n}+\sqrt{t}\right)^{-2 n}}{(an)^{3+2l}\,n}\frac{\Gamma[1+l+n]}{\Gamma[n-l]}\;{}_2F_1\left(1+l-n,1+l-n,2+2l,\nicefrac{4}{(an)^2t}\right),\qquad\nn\\
&&\stackrel{l=0}{=}\frac{4 \,t^{n-1} \left(\frac{2}{a n}+\sqrt{t}\right)^{-2 n}}{(an)^3}\,{}_2F_1\left(1-n,1-n,2,\nicefrac{4}{(an)^2t}\right).\nn
\eea
 From the denominator one can see that the energy correction is proportional to $a^{-2l}$, cf. also \Eqref{WFdisc}. Accordingly, the effect is dominant in $S$-states and suppressed by, e.g., $\al^2$ in $P$-states.

Now, we need to derive the off-forward TPE matrix element. We start with the non-Born VVCS amplitude. For the unpolarized part, we use a parametrization of the nuclear Lagrangian in terms of polarizabilities \cite{Krupina:2014nfa}:
\beq
\mathcal{L}_{NN\gamma\gamma}=\pi \beta_{M1}\bar{N}NF^2-\frac{2\pi(\al_{E1}+\be_{M1})}{M^2}\,(\partial_\al \bar{N})(\partial_\be N)F^{\al \mu}F^{\be \nu}g_{\mu\nu}, \eqlab{Lagrangian}
\eeq
with the e.m.\ field-strength tensor $F^{\mu \nu}$, the e.m.\ dipole polarizabilities $\al_{E1}$ and $\be_{M1}$, the Dirac spinor of the nucleus $N$, and the mass of the nucleus $M$. The nuclear side of the TPE then reads:
\bea
\frac{T^{\mu \nu}}{4\pi}&=&-\frac{\al_{E1}+\be_{M1}}{2M^2}\Big\{\half\!\left[p_t^2+q\cdot q'\right]q^\mu q^{\prime\,\nu}\nn +\left[2(P\cdot q)^2-\half (q\cdot p_t) (q'\cdot p_t)\right]g^{\mu\nu}\nn\\
&&-2\,P\cdot q\left[q^\mu P^\nu+P^\mu q^{\prime\,\nu}\right]\nn+\half\!\left[q\cdot q' \,q^{\prime\,\mu}q^\nu -q^2\,q^{\prime\,\mu} q^{\prime\,\nu} -q'^2\,q^\mu q^{\nu} \right]\\
&&+2 \,q\cdot q' P^\mu P^\nu\Big\}+\beta_{M1}\Big\{q\cdot q'\,g^{\mu\nu}-q^\mu q^{\prime\,\nu}\Big\},\eqlab{Tunpol}
\eea
with $P=\half (p+p')$ and $P\cdot p_t=0$. We can check that this tensor structure is even under photon crossing and gauge invariant.


\noindent The contribution of the lowest-order spin polarizabilities, $\gamma_{E1E1}$, $\gamma_{M1M1}$, $\gamma_{M1E2}$ and $\gamma_{E1M2}$, to the polarized VVCS is given by \cite{LenskySpinPol}:
\begin{align}
\hspace{-0.3cm}\frac{T^{\mu \nu}}{4\pi}=&\;\frac{1}{M^2}\bigg[\half\!\ga_0\,q\cdot P\,\Big\{\left[\gamma^{\al \be \mu}P^\nu+\gamma^{\al \be \nu}P^\mu\right]q_\al q'_\be-q\cdot P \,\ga^{\al \mu \nu}[q+q']_\al\Big\}\eqlab{spinpol}\\
&+\ga_{M1E2}\; \Big\{\gamma^{\al \be \sigma}q_\al q'_\be P_\sigma \left[q^\mu P^\nu +q^{\prime \nu} P^\mu\right]+q\cdot q'\left[\gamma^{\al \be \mu}P^\nu P_\be q_\al-\gamma^{\al \be \nu}P^\mu P_\be q'_\al \right]\nn\\
&-q\cdot P\left[q\cdot q'\,\gamma^{\mu \nu \si}P_\sigma+g^{\mu \nu}\gamma^{\al \be \sigma}q_\al q'_\be P_\sigma\right]\Big\}\nn\\
&+\gamma_{E1M2} \Big\{q\cdot q'\left[\gamma^{\al \be \nu}P^\mu q_\al P_\be-\gamma^{\al \be \mu}P^\nu q'_\al P_\be \right]+q\cdot P\left[\gamma^{\al \be \mu}q'^\nu q'_\al P_\be-\gamma^{\al \be \nu}q^\mu q_\al P_\be \right]\Big\}\nn\\
&+\ga_{M1M1}\; q\cdot P\Big\{\gamma^{\al \be \mu}q'^\nu q_\al P_\be -\gamma^{\al \be \nu}q^\mu q'_\al P_\be -q\cdot q'\,\gamma^{\mu \nu \si}P_\sigma+g^{\mu \nu}\gamma^{\al \be \sigma}q_\al q'_\be P_\sigma \Big\}\bigg],\nn
\end{align}
with the FSP $\ga_0$ defined in \Eqref{FSPdef} and the nuclear spinors omitted. This tensor is likewise even under photon crossing and gauge invariant. Due to its nuclear-spin dependence, \Eqref{spinpol} contributes to the HFS. As we found no $(Z\al)^6 \ln Z \al$ contribution to the HFS, we move any further discussion of it to \appref{5HFS}{6.5} and focus for now on the nuclear-spin-independent VVCS amplitude.

In the next step, we need to close the TPE box diagram, 
\beq
\mathscr{M}=\int \frac{\dd^4 q}{i(2\pi)^4}\frac{g_{\mu \al}}{q'^2}\,L^{\al \be}\,\frac{g_{\be \nu}}{q^2}\,T^{\mu \nu},\eqlab{TPEBox}
\eeq
 by including the photon interaction with the lepton:
\beq
L^{\al \be}=4\pi \al\,\bar{u}(l')\left[\ga^\al \frac{\slashed{l}-\slashed{q}+m}{(l-q)^2-m^2}\ga^\be\right]u(l), \eqlab{LeptonTensor}
\eeq
where $m$ is the mass of the lepton and $u$ is the lepton Dirac spinor. Note that for the lepton tensor, $L^{\al \be}$, it is enough to considered the $u$-channel VVCS process. Taking the full CS off the lepton, meaning the sum of $s$- and $u$-channels, would lead to exact double counting in the TPE matrix element.\footnote{The full $s$- and $u$-channel result can be deduced from \appref{chap5}{offVVCSBorn} by setting the FFs to their structureless values: $F_1=1$ and $F_2=0$.}
 
Just as in the previous derivation of the OPE Breit potential, cf.\ \Eqref{BreitPotentialScatAmp}, we will make a semi-relativistic expansion of the TPE amplitude. All relevant replacements are listed in \appref{chap5}{SemRelExpDSpin}. For the spin-independent case, the expansion is rather trivial. The appearing energy-dependent prefactor is absorbed into the Dirac spinors, cf.\ \Eqref{eprefactor}. Since the unpolarized amplitude, \Eqref{Tunpol}, is independent of the nuclear spin, i.e., not containing any Dirac matrices, we can write: $\ol{\mathpzc{N}}(p')\, T^{\mu \nu}\mathpzc{N}(p)=T^{\mu \nu}\ol{\mathpzc{N}}(p')\mathpzc{N}(p)\approx T^{\mu \nu}$, where in the last step we took the leading term in the semi-relativistic expansion of $\;\ol{\mathpzc{N}}(p')\mathpzc{N}(p)\approx1+\mathcal{O} (c^{-2})$, see \Eqref{NN1}. In this way, the nuclear tensors in Eqs.~\eref{Tunpol} and \eref{TPEBox} are taken between the spinors, just like the lepton tensor in \Eqref{LeptonTensor}. 

We are interested in the contribution of order $(Z\al)^6 \ln Z\al$, hence, everything that cancels a photon propagator can be removed from the integrand of \Eqref{TPEBox}. Therefore, we replace, e.g., $q \cdot q' \rightarrow \nicefrac{-p_t^2}{2}$ and $(q\cdot p_t)(q'\cdot p_t)\rightarrow \nicefrac{-p_t^4}{4}$. 

The integration over the loop momenta is performed by means of the Feynman-parameter method. Shifting the integration momentum as $q\to q + y l -(x-y) p_t $, 
 and subsequently scaling $y\to  xy $, the denominator is replaced by:
\bea
&& \frac{1}{q^2 (q+p_t)^2 [(l-q)^2 -m^2]} = 2 \int_0^1 \dd x \,x \int_0^1\dd y 
 \,\frac{1}{\big[ q^2 - \MM^2(x,y) \big]^3}, \eqlab{FeynmanTrick} \\
&&\hspace{1cm} \mbox{with}   \quad  \MM^2 = x\left[x y^2m^2- (1-x)(1-y) p_t^2\right] - i0^+. \nn
 \eea

The kinematics of the present bound-state problem, cf.\ Fig.~\ref{TPEFRW}, match those of elastic scattering.
In the CM frame, the four-momenta of the incoming and outgoing particles read:
\begin{subequations}
\eqlab{COM}
\bea
p=\{E_p/c,\boldsymbol{p}\}, &\qquad& l=\{E_l/c,-\boldsymbol{p}\},\\
p'=\{E_p/c,\boldsymbol{p'}\}, &\qquad& l'=\{E_l/c,-\boldsymbol{p}'\},
\eea
\end{subequations}
with $\vert \boldsymbol{p}\vert^2=\vert \boldsymbol{p}'\vert^2$ and $E_p/c=\sqrt{(Mc)^2+\vert \boldsymbol{p}\vert^2}\approx Mc$, etc.
All particles are considered to be on-shell, hence, the energies of the lepton and the nucleus are conserved, respectively. 
Accordingly, there is no retardation in the CM frame: $p_t^0=0$.

Likewise to the nuclear spinors, we expand the lepton spinors semi-relativistically: $\bar{\mathpzc{u}}(l')\mathpzc{u}(l)\approx1+\mathcal{O} (c^{-2})$ and $\bar{\mathpzc{u}}(l')\,\ga_0\,\mathpzc{u}(l)\approx1+\mathcal{O} (c^{-2})$, see \appref{chap5}{SemRelExpDSpin}. Independent of the frame, we have:
\beq
\bar{\mathpzc{u}}(l')\left[p_t\cdot \gamma \cdot l\right] \mathpzc{u}(l)=p_t^2/2.
\eeq
Furthermore,  we evaluate the following structures in the CM frame:
\begin{subequations}
\bea
&&p \cdot l\approx Mmc^2+\mathcal{O} (c^0),\\
&&\bar{\mathpzc{u}}(l')\left[p_t\cdot \gamma \cdot p\right] \mathpzc{u}(l)\approx 0+\mathcal{O} (c^0).\eqlab{10b}
\eea
\end{subequations}
The calculation simplifies slightly because the leading term in \Eqref{10b} is canceling.

Introducing the Feynman integrals:
\begin{subequations}
\bea
J_n(\mathcal M^2)&=& \int \frac{\dd ^d q}{i(2\pi)^d} \frac{1}{\left[q^2-\MM^2\right]^n},\\
&=&\frac{(-1)^n}{(4\pi)^2}\frac{\Gamma(n-2)}{\Gamma(n)}\frac{1}{\mathcal{M}^{2(n-2)}},\\
J_n^{\mu_1 \dots \mu_s}(\mathcal M^2)&=& \int \frac{\dd ^d q}{i(2\pi)^d} \frac{q^{\mu_1}\cdots q^{\mu_s}}{\left[q^2-\MM^2\right]^n},
\eea
(with $\MM^2 \neq 0, n>2$ and $d=4$), as well as,
\beq
J_n^{\mu_1 \mu_2}(\mathcal M^2)=\frac{1}{2(n-1)}\,J_{n-1}(\mathcal M^2)\,g^{\mu_1 \mu_2},
\eeq
and particularly,
\bea
J_2(\mathcal M^2)&=&-\frac{1}{(4\pi)^2}\left[L_\eps+\ln \mathcal M^2\right],\\
J_3(\mathcal M^2)&=&-\frac{1}{2(4\pi)^2}\frac{1}{\mathcal M^2},\eqlab{J3def}
\eea
\end{subequations}
with $L_\eps$ being a real constant from dimensional regularization, we integrate over the loop momentum. To leading order in the semi-relativistic expansion, the matrix element of the diagram in Fig.~\ref{TPEFRW} can be expressed as:
\newpage
\vspace*{-1.5cm}
\begin{align}
\mathscr M(p_t^2)=&\;8\pi^2 \al m\int_0^1 \dd x\, x\int_0^1 \dd y\,\Big\{(\al_{E1}+\be_{M1})\times\Big(J_3(\mathcal M^2) \Big[p_t^2 \Big(2-3 x y+(xy)^2\Big)\eqlab{MM}\\
&+4 m^2 (xy)^2 (1-2 x y)\Big]+J_2(\mathcal M^2) \left[1-6 x y\right]\Big)\nn\\
&+(\al_{E1}-\be_{M1})\times\Big(J_3(\mathcal M^2) \Big[4 m^2 (xy)^2\nn+ p_t^2 \Big(2 + x y-(xy)^2\Big)\Big]+J_2(\mathcal M^2)\Big)\Big\}.
\end{align}
In this result, we neglected recoil effects because of their additional suppression by $1/M$ or $1/M^2$.

Thanks to the DR in \Eqref{DRM}, it will be sufficient to calculate the imaginary part of $\mathscr M$. The integrations over the Feynman parameters $x$ and $y$, cf.\  \Eqref{MM}, are performed in \appref{chap5}{AI}, see \Eqref{AuxInt1}. Our final result is:
\begin{subequations}
\eqlab{ImMt}
\bea
\eqlab{test}
\im \mathscr M(t)\eqlab{ImM}&=&\frac{\pi \al m}{6(1-\tau)^{7/2}}\Big\{[\al_{E1}+\be_{M1}]\,\tau \sqrt{1-\tau}\,(10-\tau+6\tau^2)\\
&&-3\,\sqrt{\tau} \,\Big[(4-7\tau+10\tau^2-2\tau^3)\,\al_{E1}+\tau(5-2\tau+2\tau^2)\,\be_{M1}\Big]\arccos \sqrt{\tau}\,\Big\},\qquad\nn\\
&\approx& \pi \al m\left[ - \pi\alpha_{E1}  \sqrt{\tau }+\nicefrac{\tau}{3}\left(11 \alpha_{E1}+5\beta_{M1}\right) \right]+\mathcal{O}(\tau^{3/2}),\eqlab{tauExp}
\eea
\end{subequations}
with $\tau=\nicefrac{t}{4m^2}$. In the last row, we expanded for small $\tau$. To the given order, our result agrees with Refs.~\cite{Bernabeu:1976jq} and \cite[Eq.~(24)]{Holstein:2016cyq}.


In \Eqref{potentialTPEoff} and below, we discussed the perturbative TPE potential in coordinate space.
For the numerical evaluation, it will be more convenient to work with the alternative momentum-space approach. In first-order PT, the energy correction to the $2P_{1/2}-2S_{1/2}$ LS due to a momentum-space potential is given by, cf.\ \Eqref{1PTsphsym}:
\beq
 E_\text{LS} = \frac{1}{2\pi^2}
 \int_0^\infty\! \dd \vert\boldsymbol{p_t}\vert\, \vert\boldsymbol{p_t}\vert^2\, w_{2P-2S}(\vert\boldsymbol{p_t}\vert)\, 
 V (\vert\boldsymbol{p_t}\vert),\eqlab{LS}
 \eeq
where $w_{2P-2S}$ is the convolution of momentum-space Coulomb wave functions, cf.\ \Eqref{wfconvolution}:
\beq
w_\text{2P-2S}(\vert\boldsymbol{p_t}\vert)= 
\frac{2\,(a\vert\boldsymbol{p_t}\vert)^2\left[1-(a\vert\boldsymbol{p_t}\vert)^2\right]}{\left[1+(a\vert\boldsymbol{p_t}\vert)^2\right]^4},
\eeq
and the retardation-free off-forward TPE potential reads as: $V(\vert\boldsymbol{p_t}\vert)=\mathscr{M}(-\vert\boldsymbol{p_t}\vert^2)$. Plugging the off-forward TPE amplitude, meaning the dispersive integral in \Eqref{DRM}, into \Eqref{LS}, we arrive at:
\beq
 E_\text{LS} =\frac{1}{8\pi^2a}\int_0^\infty \dd t \left[\frac{1}{a^2t}-\frac{a^2t}{\left[1+a\sqrt{t}\right]^4}\right]\im \mathscr M (t).\eqlab{EDR}
\eeq
In \appref{chap5}{App1}, we list corresponding formulas for the lowest $S$-levels and the $2P$-level. For the leading off-forward TPE contribution, one can equivalently evaluate \Eqref{EDR} or \Eqref{2S}, see \Eqref{pwavehigh} and discussion below. In \appref{chap5}{AI}, cf.\ \Eqref{AInt2}, we give all necessary integrals to evaluate \Eqref{EDR}. 

In Eqs.~\eref{EDR} and \eref{ImMt}, we derived the nuclear-polarizability effect on the LS from off-forward TPE. Our numerical results are presented in \secref{5LS}{offTPE} for different light muonic atoms.

\section{Summary and Conclusion}

In this Chapter, we discussed the theory of TPE effects in hydrogen-like atoms. The well-known forward TPE formalism was presented in \secref{chap5}{TPE}, where we clarified the terminology and defined the polarizability effect. In \secref{chap5}{matchingOTPE}, we matched the  non-recoil effects from the nucleon-pole contributions of one- and two-photon exchange.

In \secref{chap5}{PolExpTPE}, we derived an expression for the TPE polarizability contribution to the HFS in terms of spin polarizabilities. We will apply this formalism in \secref{5HFS}{polexpHFS} to interpret our results for the HFS. 

In \secref{chap5}{6.3}, we derived the nuclear dipole polarizability contribution to the LS through off-forward TPE. 
In Chapters \ref{chap:5LS} and \ref{chap:5HFS}, we will numerically evaluate forward and off-forward TPE polarizability effects on the LS and HFS, respectively.

\begin{subappendices}
\section{Hyperspherical Coordinates and Wick Rotation}
\seclab{App:Wick}
In general, $n$-dimensional hyperspherical coordinates are defined as:
\beq
\begin{aligned}
x_1&=r \cos \phi_1,\\
x_2&=r \sin \phi_1\cos \phi_2,\\
x_3&=r \sin \phi_1\sin \phi_2\ \cos \phi_3,\\
\vdots\\
x_{n-1}&=r \sin \phi_1 \cdots \sin \phi_{n-2} \cos \phi_{n-1},\\
x_{n}&=r \sin \phi_1 \cdots \sin \phi_{n-2} \sin \phi_{n-1}.
\end{aligned}
\eeq
The volume in $n$-dimensions can be calculated through:
\bea
V_n=\int_{\phi_{n-1}=0}^{2\pi}\int_{\phi_{n-2}=0}^{\pi}\cdots\int_{\phi_{1}=0}^{\pi}\int_{r=0}^\infty \dd^nV,
\eea
with the volume element
\bea
\dd^nV&=&\left\vert \det \frac{\partial(x_i)}{\partial(r,\phi_j)}\right\vert \dd r \, \dd \phi_1 \, \dd \phi_2 \cdots \dd \phi_{n-1},\nn\\
&=&r^{n-1} \sin^{n-2} \phi_1 \sin^{n-3} \phi_2 \cdots \sin \phi_{n-2}\,\dd r \, \dd \phi_1 \, \dd \phi_2 \cdots \dd \phi_{n-1}.
\eea
To interpret, f.i., Eqs.~\eref{VVCS_LS}  and \eref{VVCS_HFS}, it is convenient to Wick-rotate ($q_0\rightarrow i Q_0$), and switch to Euclidean hyperspherical coordinates. In doing so, $q=(\nu,\bq)$ becomes:
\beq
\nu=iQ\cos \chi,\quad \bq=(Q \sin \chi \sin \theta \cos \varphi,Q\sin \chi \sin \theta \sin \varphi,Q\sin \chi \cos \theta),
\eeq
and the integration, $\dd^4q=\dd \nu\, \dd \bq$, changes into:\footnote{$\int \dd^4q=\int_0^\infty\dd \nu\, \int\dd \bq$}
\bea
\dd^4q&=&Q^3 \sin^2 \chi \sin \theta \,\dd Q \,\dd \chi \,\dd\theta \,\dd \varphi,
\eea
with $\phi \in (0,2\pi)$ and $\chi,\theta \in (0,\pi)$.


\section{Semi-Relativistic Expansion of Dirac Spinors} \seclab{SemRelExpDSpin}
In the present Section, we perform the semi-relativistic expansion of different Dirac spinor structures. Such expansion is needed in Sections \ref{chap:chap2}.\ref{sec:BreitDerivation}, \ref{chap:chap5}.\ref{sec:6.3} and \ref{chap:5HFS}.\ref{sec:neutralPion}.

\noindent $N_\la(p)$ be the Dirac spinor of a spin-1/2 particle with arbitrary momentum $p=(E_p/c,\boldsymbol{p})$, mass $M$ and polarization $\la$ (spin projections along the $z$-axis). We then define \cite{LandauLifshitz4}: 
\begin{subequations}
\bea
\mathpzc{N}_{\;\la}(p)&=&\left(2E_p\right)^{-1/2} N_\la(p), \\
&=& \mbox{$\sqrt{\frac{E_p +Mc^2}{2E_p} } $}\eqlab{eprefactor}
\barr  1 \\
\frac{c\,\boldsymbol{\sigma} \cdot \boldsymbol{p}}{E_p +M c^2} \earr 
\otimes \chi_\la, \\
&\simeq&
\barr  1 -\frac{\boldsymbol{p}^2}{8M^2 c^2} \\
\frac{\boldsymbol{\sigma} \cdot \boldsymbol{p}}{2Mc}\earr 
\otimes \chi_\la, \eqlab{spinorExp}
\eea
\end{subequations}
where in the last step we expanded semi-relativistically, recalling that $E_p/c=\sqrt{M^2c^2+\bp^2}$.  Note that the energy prefactor, $\left(2E_p\right)^{-1/2} $, stems from the semi-relativistic expansion of the propagators. Furthermore, we have the
Pauli matrices, $\boldsymbol{\sigma}=\{\si_1, \si_2, \si_3\}$, and the Pauli spinors:
 \beq
 \chi_{\nicefrac{1}{2}}=\barr  1 \\
0\earr,\qquad  \chi_{-\nicefrac{1}{2}}=\barr  0 \\
1\earr.
\eeq
We now give expressions for the Dirac spinor structures appearing in our calculations:
\begin{subequations}
\eqlab{NRnucl}
\begin{small}
\bea
\ol{\mathpzc{N}}(p') \,\mathpzc{N}(p) 
&\simeq&  1 - \frac{\boldsymbol{P}^2}{2M^2c^2} + \frac{i \,\bS \cdot \boldsymbol{q} \times \boldsymbol{P}}{2M^2c^2} 
+\mathcal{O}(1/c^4), \eqlab{NN1}\\
\ol{\mathpzc{N}}(p')\,\gamma_5\, \mathpzc{N}(p) 
&\simeq&  \frac{\boldsymbol{S}\cdot \boldsymbol{q}}{Mc} +\mathcal{O}(1/c^3),\eqlab{gamma5NN} \\
\ol{\mathpzc{N}}(p') \ga^0 \mathpzc{N}(p) 
&\simeq&
 1 - \frac{\boldsymbol{p_t}^2}{8M^2c^2} - \frac{i \,\bS \cdot \boldsymbol{q} \times \boldsymbol{P}}{2M^2c^2}
 +\mathcal{O}(1/c^4) , \\
\ol{\mathpzc{N}}(p') \, \gamma^i \, \mathpzc{N}(p) 
&\simeq&  \frac{1}{Mc}\left[ P^i - i \left(\bS\times \boldsymbol{q}\right)^i  \right]+\mathcal{O}(1/c^3),
\\
\ol{\mathpzc{N}}(p') \ga^0 \gamma^i \mathpzc{N}(p) 
&\simeq& \frac{1}{2Mc}\left[p_t^i -4 i \left(\bS\times \boldsymbol{P}\right)^i  \right]+\mathcal{O}(1/c^3),\qquad
\\
\ol{\mathpzc{N}}(p') \gamma^i\gamma^j \mathpzc{N}(p) 
&\simeq&-\delta^{ij}-2i\epsilon^{ijk}\bS^k+\frac{1}{8M^2c^2}\bigg\{4\left[\boldsymbol{P}^2-i \,\bS \cdot \boldsymbol{q} \times \boldsymbol{P} \right]\delta^{ij}+2\left[P^ip_t^j-P^jp_t^i\right]\\
&&+4i\,\bigg[(\boldsymbol{P}+\half\! \boldsymbol{q} )^j \left[\bS \times (\boldsymbol{P}-\half\! \boldsymbol{q} )\right]^i-(\boldsymbol{P}+\half\! \boldsymbol{q} )^i\left[\bS \times (\boldsymbol{P}-\half\! \boldsymbol{q} )\right]^j\bigg]\nn\\
&&+4i\,\eps^{ijk }(\boldsymbol{P}+\half\! \boldsymbol{q} )^k\;\bS \cdot (\boldsymbol{P}-\half\! \boldsymbol{q} )+i\left[4\boldsymbol{P}^2+\boldsymbol{q} ^2\right]\epsilon^{ijk} \bS^k\bigg\}+\mathcal{O}(1/c^4),\nn\\
\ol{\mathpzc{N}}(p') \gamma^i\gamma^j\gamma^0 \mathpzc{N}(p) 
&\simeq&-\delta^{ij}-2i\,\epsilon^{ijk}\bS^k+\frac{1}{8M^2c^2}\bigg\{\left[\boldsymbol{q}^2+4i \,\bS \cdot \boldsymbol{q} \times \boldsymbol{P} \right]\delta^{ij}-2\left[P^ip_t^j-P^jp_t^i\right]\qquad\quad\\
&&-4i\,\bigg[(\boldsymbol{P}+\half\! \boldsymbol{q} )^j \left[\bS \times (\boldsymbol{P}-\half\! \boldsymbol{q} )\right]^i-(\boldsymbol{P}+\half\! \boldsymbol{q} )^i\left[\bS \times (\boldsymbol{P}-\half\! \boldsymbol{q} )\right]^j\bigg]\nn\\
&&-4i\,\eps^{ijk }(\boldsymbol{P}+\half\! \boldsymbol{q} )^k\;\bS \cdot (\boldsymbol{P}-\half\! \boldsymbol{q} )+i\left[4\boldsymbol{P}^2+\boldsymbol{q} ^2\right]\epsilon^{ijk} \bS^k\bigg\}+\mathcal{O}(1/c^4),\nn\\
\ol{\mathpzc{N}}(p') \gamma^i\gamma^j\gamma^k \mathpzc{N}(p) &\simeq& \frac{1}{Mc}\Big[-\delta^{ij}P^k+\delta^{ki}P^j-\delta^{jk}P^i-2i\,\epsilon^{ijk} \boldsymbol{S}\cdot \boldsymbol{P}+i \delta^{ij}(\boldsymbol{S}\times \boldsymbol{q})^k\\
&&-i \,\delta^{ki}(\boldsymbol{S}\times \boldsymbol{q})^j+i \,\delta^{jk}(\boldsymbol{S}\times \boldsymbol{q})^i\Big]+\mathcal{O}(1/c^3),\nn
\eea
\end{small}
\end{subequations}
with $q=p-p'$, $P=\half(p+p')$ and the nuclear spin $\bS=\boldsymbol{\sigma}/2$. Analogously, one derives expressions for the lepton spinor, distinguishing only $\bs=\boldsymbol{\sigma}/2$ and $q=l'-l$, where $l$ ($l'$) is the incoming (outgoing) lepton momentum. The presented expressions are independent of the reference frame.

\section{Auxiliary Integrals} \seclab{AI}
Here, we provide a list of auxiliary integrals occurring in our calculation.
\begin{itemize}
\item
Feynman-parameter integrals as appearing in \Eqref{MM}: 
\begin{subequations}
\eqlab{AuxInt1}
\begin{small}
\bea
I_1(\tau)&=&\im\int_0^1\dd y\int_0^1\dd x\, \frac{x}{\mathcal{M}^2}=\frac{\pi}{m^2} \frac{\arccos \sqrt{\tau}}{2\sqrt{\tau ( 1-\tau) }},
\\
I_2(\tau)&=&\im\int_0^1\dd y\int_0^1\dd x\,
\frac{x^2y}{\mathcal{M}^2}=\frac{\pi}{m^2}
\frac{1}{2(1-\tau )}\left[1-\frac{\sqrt{\tau}\arccos \sqrt{\tau} }{\sqrt{1-\tau}}\right],\\
I_3(\tau)&=&\im\int_0^1\dd y\int_0^1\dd x\,
\frac{x^3y^2}{\mathcal{M}^2}=\frac{\pi}{m^2}\frac{1 }{4 (1-\tau)^2}\left[-3 \tau+\frac{(1+2 \tau ) \sqrt{\tau }\arccos \sqrt{\tau}}{\sqrt{1-\tau }}\right],\eqlab{I3}\qquad\quad\\
I_4(\tau)&=&\im\int_0^1\dd y\int_0^1\dd x\,\frac{x^4y^3}{\mathcal{M}^2}\\
&=&\frac{\pi}{m^2}\frac{\tau }{12 (1-\tau)^3}\left[(4+11 \tau)-\frac{ (9+6 \tau )\sqrt{\tau} \arccos \sqrt{\tau }}{ \sqrt{1-\tau }}\right]\nn,\\
I_5(\tau)&=&\im  \int_0^1\dd y \int_0^1\dd x \,x \ln \mathcal{M}^2=\frac{\pi\tau}{2 (1-\tau )}\left[1-\frac{\arccos \sqrt{\tau}}{\sqrt{\tau  (1-\tau )}}\right]\eqlab{I5},\\
I_6(\tau)&=&\im  \int_0^1\dd y \,y\int_0^1\dd x\, x^2 \ln \mathcal{M}^2=-\frac{\pi\tau}{2(1-\tau)^2}\left[\frac{\tau +2}{3}-\frac{\sqrt{\tau } \arccos \sqrt{\tau}}{\sqrt{1-\tau }}\right],
\eea
\end{small}
\end{subequations}
with $\MM^2 = x\left[x y^2m^2- (1-x)(1-y) p_t^2\right] - i0^+$ and $\tau=\nicefrac{p_t^2}{4m^2}$.

\begin{subequations}
\eqlab{AInt2}
\item $\Delta E_{nS}[\im \mathscr M (t)]$, cf.\ Eqs.~\eref{1S}-\eref{4S}, evaluated with \Eqref{AuxInt1} for $\tau=\nicefrac{t}{4m^2}$:
\bea
\Delta E_{nS}[t I_1]&\approx&-\frac{1}{8 \pi ^2 m}\frac{(Z\alpha m_r)^4}{ n^3}\,\ln \frac{Z\al m_r}{2nm}+\mathcal{O}(\al^6,\al^7),\\
\Delta E_{nS}[I_3]&\approx&-\frac{1}{64 \pi ^2 m^3}\frac{(Z\alpha m_r)^4}{ n^3}\,\ln \frac{Z\al m_r}{2nm}+\mathcal{O}(\al^6,\al^7),\\
\Delta E_{nS}[I_5]&\approx&\frac{1}{64 \pi ^2 m}\frac{(Z\alpha m_r)^4}{ n^3}\,\ln \frac{Z\al m_r}{2nm}+\mathcal{O}(\al^6,\al^7),
\eea
\beq
\Delta E_{nS}[t I_2]\approx0+\mathcal{O}(\al^6,\al^7), \quad \Delta E_{nS}[I_4]\approx0+\mathcal{O}(\al^6,\al^7),\quad\Delta E_{nS}[I_6]\approx0+\mathcal{O}(\al^6,\al^7).\nn
\eeq
\end{subequations}
\end{itemize}

\section{First-order Perturbation Theory in Momentum Space}\seclab{App1}
\begin{itemize}
\begin{subequations}
\eqlab{ConvolutionsWFMom}
\item Convolution of momentum-space Coulomb wave functions:
\begin{small}
\bea
w_{1S}(\boldsymbol{p_t}\vert)&=&\frac{16}{\left[4+(a\vert\boldsymbol{p_t}\vert)^2\right]^2} ,\eqlab{ConvolutionsWFMom1S}\\
w_{2S}(\vert\boldsymbol{p_t}\vert)&=& 
\frac{\left[1-2\,(a\vert\boldsymbol{p_t}\vert)^2\right]\left[1-(a\vert\boldsymbol{p_t}\vert)^2\right]}{\left[1+(a\vert\boldsymbol{p_t}\vert)^2\right]^4},\\
w_{3S}(\vert\boldsymbol{p_t}\vert)&=& \frac{16 \left[4-3 (a\vert\boldsymbol{p_t}\vert)^2\right] \left[4-27 (a\vert\boldsymbol{p_t}\vert)^2\right] \left[16-216 (a\vert\boldsymbol{p_t}\vert)^2+243 (a\vert\boldsymbol{p_t}\vert)^4\right]}{\left[4+9 (a\vert\boldsymbol{p_t}\vert)^2\right]^6},\qquad\quad\\
w_{4S}(\vert\boldsymbol{p_t}\vert)&=&\frac{\left[1-4 (a\vert\boldsymbol{p_t}\vert)^2\right] \left[\left(4 (a\vert\boldsymbol{p_t}\vert)^2-1\right)^2-16 (a\vert\boldsymbol{p_t}\vert)^2\right]  }{\left[1+4 a(a\vert\boldsymbol{p_t}\vert)^2\right]^8}\times\\
&&\times \left[1-48 (a\vert\boldsymbol{p_t}\vert)^2+288 (a\vert\boldsymbol{p_t}\vert)^4-256 (a\vert\boldsymbol{p_t}\vert)^6\right],\nn\\
w_{2P}(\vert\boldsymbol{p_t}\vert)&=& 
\frac{1-(a\vert\boldsymbol{p_t}\vert)^2}{\left[1+(a\vert\boldsymbol{p_t}\vert)^2\right]^4}.
\eea
\end{small}
\end{subequations}
\item Energy shifts in first-order PT:
\begin{subequations}
\bea
\Delta E_{1S} &=&\frac{1}{\pi^2a}\int_0^\infty \dd t \left[\frac{1}{\left[2+a\sqrt{t}\right]^2}-\frac{1}{a^2t}\right]\im \mathscr M (t),\eqlab{1S}
 \\
\Delta E_{2S} &=&\frac{1}{8\pi^2a}\int_0^\infty \dd t \left[\frac{1+2a^2t}{2\left[1+a\sqrt{t}\right]^4}-\frac{1}{a^2t}\right]\im \mathscr M (t),\eqlab{2S}\\
\Delta  E_{3S} &=&\frac{1}{9\pi^2a}\int_0^\infty \dd t \left[\frac{16+27a^2t \left[8+9a^2t\right]}{\left[2+3a\sqrt{t}\right]^6}-\frac{1}{3a^2t}\right]\im \mathscr M (t),\\
\Delta  E_{4S} &=&\frac{1}{64\pi^2a}\int_0^\infty \dd t \left[\frac{1+16a^2t\left[3+2a^2t\left(9+8a^2t\right)\right]}{\left[1+2a\sqrt{t}\right]^8}-\frac{1}{a^2t}\right]\im \mathscr M (t),\qquad\quad\eqlab{4S}\\
\Delta  E_{2P} &=&\frac{1}{16\pi^2a}\int_0^\infty \dd t \,\frac{1}{\left[1+a\sqrt{t}\right]^4}\im \mathscr M (t).
\eea
\end{subequations}
\end{itemize}

\section{Born and Elastic Parts of Off-Forward Doubly-Virtual Compton Scattering} \seclab{offVVCSBorn}
The off-forward VVCS can be decomposed into a set of $9$ tensors:
\begin{subequations}
\eqlab{CovarAmp}
 \beq
 \bar u^{\,\prime} (\veps' \cdot T \cdot \veps)  u  = e^2 \hat \scA^T (s,t)\, \bar u^{\,\prime} \hat \scO^{\mu \nu}  u \,  \mathcal{E}_{\mu}^{\prime }  \mathcal{E}_{\nu},
 \eeq
  with 
 \bea
 \hat \scA(s,t) &=& \big\{\scA_1, \, \cdots, \, \scA_9 \big\} (s,t), \\
\hat \scO^{\mu\nu} &=& \big\{ -g^{\mu \nu}, \; q^{\mu} q^{\prime\,\nu},\; 
-\gamma^{\mu \nu},\; g^{\mu \nu} (q' \cdot \gamma \cdot q),\; q^{\mu} q'_{\alpha} \gamma^{\alpha \nu}-\gamma^{\alpha \mu} q_{\alpha} q'^{\nu}, \\  
&& \;
q^{\mu} q_{\alpha} \gamma^{\alpha \nu}-\gamma^{\alpha \mu} q'_{\alpha} q'^{\nu},q^{\mu} q^{\prime\,\nu}(q' \cdot \gamma \cdot q) ,\; 
- i \gamma_5 \epsilon^{\mu \nu \alpha \beta} q'_{\alpha} q_{\beta},\nn\\
&&\;q^{\mu} q_{\alpha} \gamma^{\alpha \nu}+\gamma^{\alpha \mu} q'_{\alpha} q'^{\nu}\big\}, \nn\\
&&\hspace{-3cm} \mathcal{E}_{\mu}  = \veps_{\mu} - \frac{P \cdot \veps}{P \cdot q} \, q_{\mu},
\; \mathcal{E}_{\mu}'  = \veps_{\mu}' - \frac{P \cdot \veps'}{P \cdot q} \, q_{\mu}',
\; P_\mu = \half (p + p')_\mu, \; P\cdot q=P\cdot q' = M \xi.
 \eqlab{epsilon_b}
\eea 
\end{subequations}
The Born contribution to off-forward VVCS with FF dependent e.m.\ interaction,
\beq
\Ga^\mu =e \ga^\mu F_1(Q^2) -\frac{e}{2M} \ga^{\mu\nu} q_\nu
F_2(Q^2),
\eeq
reads:\footnote{We used: \bea
\bar{u}(p')\left[\gamma^{\mu \nu \al}a_\mu b_\nu c_\al\right]u(p)&=&\frac{1}{M}\bar{u}(p')\left[P\cdot a\, b\cdot\gamma\cdot c-P\cdot b\,a \cdot\gamma\cdot c+P\cdot c\,a \cdot\gamma\cdot b\right]u(p)\nn\\
&&+\frac{1}{2M}\bar{u}(p')\left[\gamma^{\mu \nu \al \be}a_\mu b_\nu c_\al(q'-q)_\be\right]u(p).\nn
\eea}
\begin{subequations}
\begin{footnotesize}
\begin{align}
\cA_1^\mathrm{Born} &=\frac{4F_1(q^2) F_1(q^{\prime\,2})\,(q\cdot P)^2+(q\cdot q')^2 \left[F_1(q^2) F_2(q^{\prime\,2})+F_2(q^2) F_1(q^{\prime\,2})+F_2(q^2) F_2(q^{\prime\,2})\right]}{ M \left[s-M^2\right]\left[u-M^2\right]},\\
\cA_2^\mathrm{Born} &=\frac{F_2(q^2) F_2(q^{\prime\,2})\,(q\cdot P)^2+M^2\,(q\cdot q')\left[F_1(q^2) F_2(q^{\prime\,2})+F_2(q^2) F_1(q^{\prime\,2})+F_2(q^2) F_2(q^{\prime\,2})\right]}{ M^3 \left[s-M^2\right]\left[u-M^2\right]},\\
\cA_3^\mathrm{Born} &=\frac{-4M^2\,(q\cdot P)\,(q\cdot q')\,G_M(q^2)\,G_M(q^{\prime\,2})+F_2(q^2) F_2(q^{\prime\,2})\,(q\cdot P)\left[4(q\cdot P)^2-(q \cdot q')^2\right]}{2M^3 \left[s-M^2\right]\left[u-M^2\right]}\\
&=-2\,(q \cdot P) \,\cA_8^\mathrm{Born} ,\nn\\
\cA_4^\mathrm{Born} &= -\cA_5^\mathrm{Born} =\frac{2\,(q\cdot P)\,G_M(q^2)\,G_M(q^{\prime\,2})}{M \left[s-M^2\right]\left[u-M^2\right]},\\
\cA_6^\mathrm{Born} &= \frac{(q\cdot P)\left[2F_1(q^2) F_1(q^{\prime\,2})+F_2(q^2) F_1(q^{\prime\,2})+F_1(q^2) F_2(q^{\prime\,2})\right]}{M \left[s-M^2\right]\left[u-M^2\right]},\\
\cA_7^\mathrm{Born} &= -\frac{(q\cdot P)\,F_2(q^2) F_2(q^{\prime\,2})}{2 M^3 \left[s-M^2\right]\left[u-M^2\right]},\\
\cA_9^\mathrm{Born} &=\frac{(q\cdot P)\left[F_2(q^2) F_1(q^{\prime\,2})-F_1(q^2) F_2(q^{\prime\,2})\right]}{M \left[s-M^2\right]\left[u-M^2\right]},
\end{align}
\end{footnotesize}
\end{subequations}
where $P=\nicefrac{1}{2}\left(p+p'\right)$, which satisfies $q'\cdot P = q\cdot P$ (since 
$q-q'=p'-p$ and $p^2=p^{\prime\,2}=M^2$).
In deriving this, we used:
\begin{subequations}
\bea 
\frac{1}{ s-M^2} &=& \frac{q\cdot q'  - 2q\cdot P }{ [s-M^2][u-M^2]},\\
\frac{1}{ u-M^2} &=& \frac{q\cdot q'  + 2q\cdot P }{ [s-M^2][u-M^2]},
\eea
\end{subequations}
or equivalently:
\begin{subequations}
\bea
4\,q\cdot P &=&\left[s-M^2\right]-\left[u-M^2\right],\\
2\,q\cdot q' &=&\left[s-M^2\right]+\left[u-M^2\right].
\eea
\end{subequations}
In the real limit, one amplitude vanishes: $\cA_9=0$. Therefore, the RCS can be described by only $8$ independent amplitudes \cite{Hagelstein:2015egb}:
\begin{subequations}
\bea
 \hat \scA(s,t) &=& \big\{\scA_1, \, \cdots, \, \scA_8 \big\} (s,t), \\
 \hat \scO^{\mu\nu} &=& \big\{ -g^{\mu \nu}, \; q^{\mu} q^{\prime\,\nu},\; 
-\gamma^{\mu \nu},\; g^{\mu \nu} (q' \cdot \gamma \cdot q),\; q^{\mu} q'_{\alpha} \gamma^{\alpha \nu}-\gamma^{\alpha \mu} q_{\alpha} q'^{\nu},\nn \\  
&& 
\;q^{\mu} q_{\alpha} \gamma^{\alpha \nu}-\gamma^{\alpha \mu} q'_{\alpha} q'^{\nu},q^{\mu} q^{\prime\,\nu}(q' \cdot \gamma \cdot q) ,\; 
- i \gamma_5 \epsilon^{\mu \nu \alpha \beta} q'_{\alpha} q_{\beta}\big\}.
\eea 
\end{subequations}
The Born part is equivalent to the elastic nucleon-pole part up to non-pole pieces, cf.\ \Eqref{BornElasticDiff}.
In the forward limit, we have the simplification that $q\cdot P\rightarrow M\nu$ and $q\cdot q' \rightarrow -Q^2$. Hence, in the following, we identify poles in $q\cdot P$.
Introducing:
\beq 
\xi = \frac{s-u}{4M} = \frac{q\cdot P}{M},
\eeq
we can see that:
\beq
4(q\cdot P)^2 \equiv 4M^2 \xi^2 =(q\cdot q')^2 -\,\left[s-M^2\right]\left[u-M^2\right].
\eeq
We then rewrite the Born part of off-forward VVCS and separate nucleon-pole and non-pole pieces:
\begin{subequations}
\begin{small}
\begin{align} 
\cA_1^\mathrm{Born} &=-\frac{F_1(q^2) F_1(q^{\prime\,2})}{M}+\frac{(q\cdot q')^2\, G_M(q^2)\,G_M(q^{\prime\,2})}{ M \left[s-M^2\right]\left[u-M^2\right]},\\
\cA_2^\mathrm{Born} &=-\frac{F_2(q^2) F_2(q^{\prime\,2})}{4M^3}\\
&\quad+\frac{(q\cdot q') \left[F_1(q^2) F_2(q^{\prime\,2})+F_2(q^2) F_1(q^{\prime\,2})+F_2(q^2) F_2(q^{\prime\,2})\left(1+\frac{(q\cdot q')}{4M^2}\right)\right]}{ M \left[s-M^2\right]\left[u-M^2\right]},\qquad\nn\\
\cA_3^\mathrm{Born} &=-\frac{(q\cdot P)\,F_2(q^2) F_2(q^{\prime\,2})}{2M^3}-\frac{2\,(q\cdot P)\, (q \cdot q')\,G_M(q^2)\,G_M(q^{\prime\,2})}{M\left[s-M^2\right]\left[u-M^2\right]},\\
\cA_4^\mathrm{Born} &= -\cA_5^\mathrm{Born} =\frac{G_M(q^2)\,G_M(q^{\prime\,2})}{2M }\left\{\frac{1}{u-M^2}-\frac{1}{s-M^2}\right\},\\
\cA_6^\mathrm{Born} &= \frac{\left[2F_1(q^2) F_1(q^{\prime\,2})+F_2(q^2) F_1(q^{\prime\,2})+F_1(q^2) F_2(q^{\prime\,2})\right]}{4M}\left\{\frac{1}{u-M^2}-\frac{1}{s-M^2}\right\},\\
\cA_7^\mathrm{Born} &= \frac{F_2(q^2) F_2(q^{\prime\,2})}{8 M^3 }\left\{\frac{1}{s-M^2}-\frac{1}{u-M^2}\right\},\\
\cA_8^\mathrm{Born} &= \frac{F_2(q^2) F_2(q^{\prime\,2})}{4M^3}+\frac{(q\cdot q')\,G_M(q^2)\,G_M(q^{\prime\,2})}{M \left[s-M^2\right]\left[u-M^2\right]},\\
\cA_9^\mathrm{Born} &=\frac{\left[F_1(q^2) F_2(q^{\prime\,2})-F_2(q^2) F_1(q^{\prime\,2})\right]}{4M}\left\{\frac{1}{s-M^2}-\frac{1}{u-M^2}\right\}.
\end{align}
\end{small}
\end{subequations}
Obviously, the amplitudes $\cA_1^\mathrm{Born}, \cA_2^\mathrm{Born}, \cA_3^\mathrm{Born}$ and $\cA_8^\mathrm{Born}$ contain non-pole pieces: 
\begin{subequations}
\begin{small}
\bea  
\cA_1^\text{non-pole} &=&-\frac{F_1(q^2) F_1(q^{\prime\,2})}{M},\\
\cA_2^\text{non-pole} &=&-\frac{F_2(q^2) F_2(q^{\prime\,2})}{4M^3},\qquad\\
\cA_3^\text{non-pole} &=&-\frac{F_2(q^2) F_2(q^{\prime\,2})\,q\cdot P}{2M^3},\\ 
\cA_8^\text{non-pole} &=& \frac{F_2(q^2) F_2(q^{\prime\,2})}{4M^3},
\eea
\end{small}
\end{subequations}
which need to be subtracted to calculate the elastic contribution to off-forward VVCS.
Using Ref.~\cite[Appendix A]{Hagelstein:2015egb}:
\begin{subequations}
\bea 
 T_1 &=&  e^2 \cA_1, \\
   T_2 &=&  \frac{e^2 Q^2}{\nu^2} \big( \cA_1
+ Q^2 \cA_2 \big), \\
 S_1 &=&  \frac{e^2 M}{\nu} \big[ \cA_3
+ Q^2 \big(\cA_5+\cA_6\big) \big] ,\quad \\
S_2&=&e^2 M^2  \big(\cA_5+\cA_6\big),
\eea 
\end{subequations}
we reproduce the well-known VVCS amplitudes in the forward limit, see \Eqref{T12Born}.

\end{subappendices}
\chapter{Lamb Shift in Chiral Perturbation Theory} \chaplab{5LS}

In the previous Chapter, we gave an introduction into the theory of TPE effects. In the next two Chapters (\ref{chap:5LS} and \ref{chap:5HFS}), we calculate nucleon- and nuclear-polarizability contributions to the LS and HFS of light muonic atoms in the framework of BChPT.

In the present Chapter, we deal with the LS. In particular, we give a prediction for the order-$\al^5$ proton-polarizability contribution to the $\mu$H LS at NLO in BChPT.  As input, we use the VVCS amplitudes and photoabsorption cross sections from \chapref{chap4}. We compare our results to HBChPT (\secref{5LS}{HBcompLS}) and dispersive calculations (\secref{5LS}{ComparisonLS}).

In \secref{5LS}{offTPE}, we evaluate the order-$(Z\al)^6$ effect of the nuclear e.m.\ dipole polarizabilities from off-forward TPE, cf.\ \secref{chap5}{6.3}, on the LSs in $\mu$H, $\mu$D, $\mu^3$H, $\mu^3\text{He}^+$ and $\mu^4\text{He}^+$.
In \secref{5LS}{LRP}, we demonstrate that the off-forward TPE represents a viable alternative approach to the 
Coulomb-distortion long-range polarization potential known from the literature.

In \secref{5LS}{chap6.3}, we update the theoretical descriptions of the LSs in $\mu$H, $\mu$D and $\mu^4\text{He}^+$ based on the results of this Chapter, and re-extract the proton and deuteron charge radii.

\section[Proton-Polarizability Contribution at Order $\al^5$]{Proton-Polarizability Contribution at Order $\boldsymbol{\al^5}$}\seclab{ExpInPolLS}

In what follows, the NLO BChPT prediction for the order-$\al^5$ proton-polarizability effect in the LS of $\mu$H is presented.  In general, ChPT predictions of the TPE polarizability effects provide a genuine alternative to the more common dispersive evaluations based on empirical information, see \secref{5LS}{ComparisonLS}. One advantage of the ChPT approach is that calculating the non-Born diagrams gives direct access to the polarizability contribution. The dispersive calculations, on the other hand, are naturally working with the separation into contributions from elastic FFs and inelastic structure functions and require a subsequent rearrangement into Born and polarizability contributions, cf.\ \secref{chap5}{1334}

As explained in \secref{chap5}{subtrfuncsec}, the polarizability contribution to the LS can not be extracted solely from experimental data. In the dispersive calculations, the contribution of the subtraction function, $\ol T_1(0,Q^2)$, has to be modeled. In the ChPT framework, however, no modeling is needed. Since all LECs appearing in the $\mathcal{O}(p^{7/2})$ calculation of the VVCS amplitudes are known from other processes, the NLO nucleon polarizabilities come out as a pure prediction of ChPT. Consequently, ChPT contains definite predictions for the proton structure effects from TPE. 

 \seclab{HBcompHFS}
 \begin{figure}[tbh]
\centering
\includegraphics[scale=0.75]{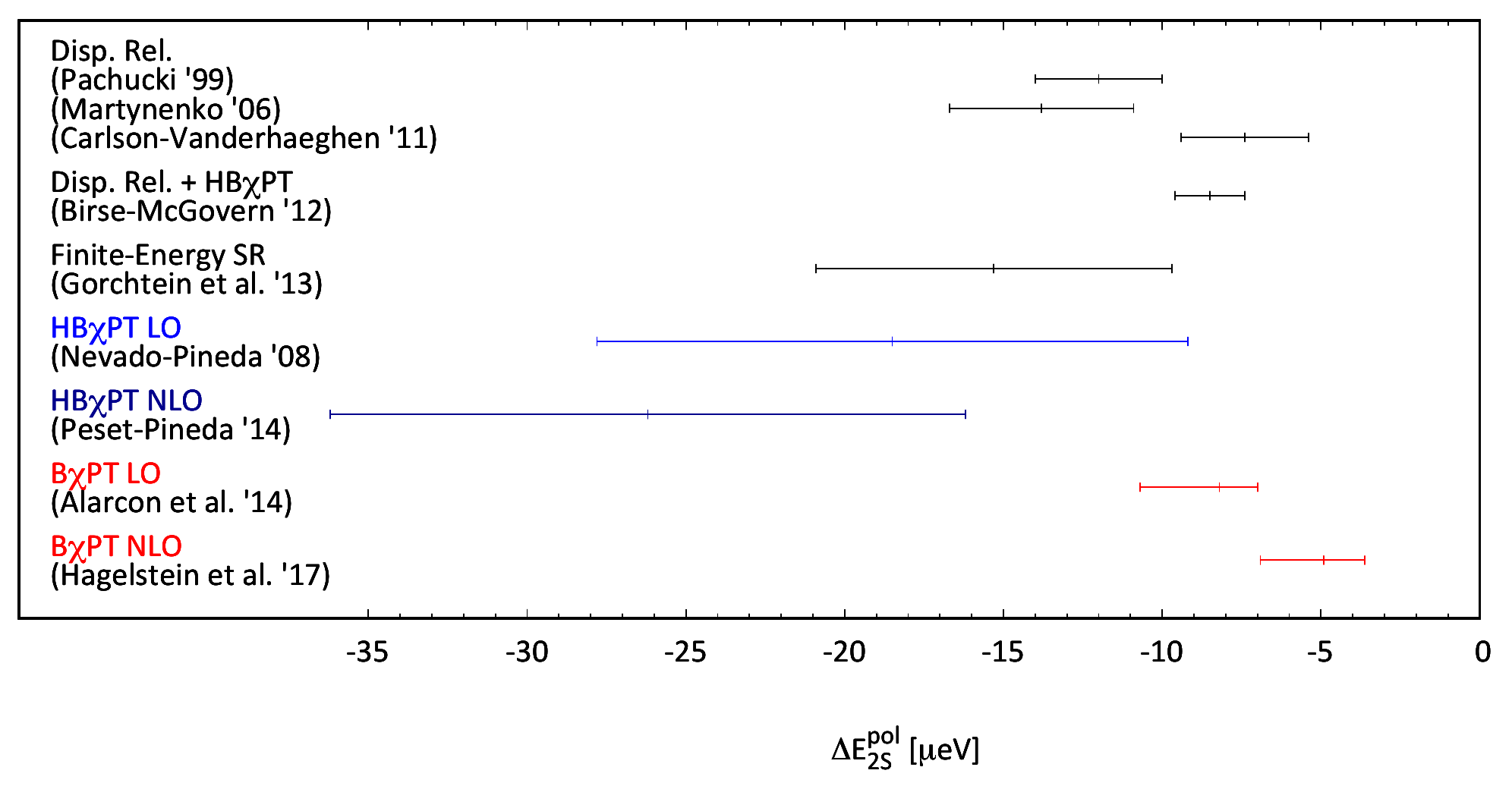}
\caption{Comparison of predictions for the polarizability contribution to the $2S$-level shift in muonic hydrogen, see also Tables \ref{Table:Summary2} and \ref{Table:Summary1}.}
\label{ComparisonLS}
\end{figure}

The energy shift of the $n$-th $S$-level due to forward TPE, \Eqref{VVCS_LS}, is dominated by low $Q$. Assuming the photon energy $\nu$ is small compared to all other scales, one finds \cite{Alarcon:2013cba}:
\beq
\Delta E^{\,\text{pol.}}(nS)=\frac{\al}{\pi}\,\phi_n^2\int_0^\infty \frac{\dd Q}{Q^2} \,w(\tau_l)\left[\ol T_1(0,Q^2)-\ol T_2(0,Q^2)\right],
\eqlab{bcancels}
\eeq
with the weighting function $w(\tau_l)=\sqrt{1+\tau_l}-\sqrt{\tau_l}$. Substituting the LEXs,
\begin{subequations}
\eqlab{bcancelsLEX}
\bea
\lim_{\nu,Q^2\rightarrow0}\frac{\ol T_1(\nu,Q^2)}{4\pi}&=&\left[\alpha_{E1}(Q^2)+\beta_{M1}(Q^2)\right]\nu^2+\beta_{M1}Q^2+\mathcal{O}(\nu^4,\nu^2Q^2,Q^4),\qquad\\
\lim_{\nu,Q^2\rightarrow0}\frac{\ol T_2(\nu,Q^2)}{4\pi}&=&\left(\alpha_{E1}+\beta_{M1}\right)Q^2+\mathcal{O}(\nu^4,\nu^2Q^2,Q^4),
\eea
\end{subequations}
one observes that the dependence on the magnetic polarizability $\beta_{M1}$ is removed. From this it follows that the polarizability part of the TPE effect in the LS is dominated by the electric dipole polarizability $\alpha_{E1}$, while the contribution of the magnetic dipole polarizability $\beta_{M1}$ is suppressed.

Another implication that the electric dipole polarizability dominates was found in Ref.~\cite{Pachucki:2011xr}, where the nuclear structure corrections in $\mu$D are calculated. Surprisingly, the total deuteron structure correction is approximately given by electric dipole polarizability contributions at various orders, while other corrections, e.g., from higher multipole polarizabilities, the magnetic dipole polarizability or relativistic nature, cancel each other out.

As the contribution of the magnetic dipole polarizability to the TPE is suppressed, likewise should be the dominant magnetic-dipole part of the e.m.\ nucleon-to-delta transition. Therefore, \citet{Alarcon:2013cba} calculate the order-$\al^5$ polarizability contribution at LO in BChPT and neglect the NLO $\Delta(1232)$-excitation. This should be a good approximation for the total polarizability effect. Nevertheless,  the separation into $E^{\mathrm{subtr.}}_\mathrm{LS}$ and $E^{\mathrm{inel.}}_\mathrm{LS}$, Eqs.~\eref{subpol} and \eref{inelasticpol}, suffers because the effect of the delta does not cancel out in the independent terms. In the following, we will include the $\Delta(1232)$-exchange to make the prediction of the subtraction function more reliable and confirm its expected insignificance in the total polarizability contribution.

\noindent Anticipating the result of this Section, the NLO BChPT prediction of the order-$\al^5$ proton-polarizability contribution to the LS in $\mu$H evaluates to:
\beq
E^\mathrm{pol.}_\mathrm{LS}(\mu\text{H})=4.9\,^{+2.0}_{-1.3}\,\upmu\text{eV},\eqlab{LSfinalvalue}
\eeq
where the contribution of the subtraction function equals: 
\begin{subequations}
\bea
E^\mathrm{subtr.}_\mathrm{LS}(\mu\text{H})&=&-5.8\pm 2.3\,\upmu\text{eV},\eqlab{totalSub}\\
E^\mathrm{inel.}_\mathrm{LS}(\mu\text{H})&=&10.7\,^{+2.3}_{-2.1}\,\upmu\text{eV}.
\eea
\end{subequations}
The latter compares best to the result of Ref.~\cite{Carlson:2011zd}. In general, the BChPT prediction compares in a satisfactory manner with the dispersive calculations, see Fig.~\ref{ComparisonLS}.

Based on the elastic FF parametrization of \citet{Bradford:2006yz}, the Born contribution of TPE amounts to:
\beq
E^\mathrm{Born}_\mathrm{LS}(\mu\text{H})=22.9\pm1.7\,\upmu\text{eV},
\eeq
where we estimated the error by taking the spread of different FF fits \cite{Kelly:2004hm,Arrington:2007ux}. Our final result for the forward TPE effect then reads:
\beq
E^\mathrm{TPE}_\mathrm{LS}(\mu\text{H})=27.8\,^{+2.6}_{-2.1}\,\upmu\text{eV}. \eqlab{LSfinalvaluePlusBorn}
\eeq

In the following, we present the individual contributions from chiral loops and the $\Delta$-exchange. Afterwards, we will compare to HBChPT and dispersive calculations. Tables \ref{Table:Summary2} and \ref{Table:Summary1} summarize relevant calculations of the TPE corrections to the $\mu$H LS performed by various authors. 

\begin{figure}[t]
\centering
\includegraphics[width=10cm]{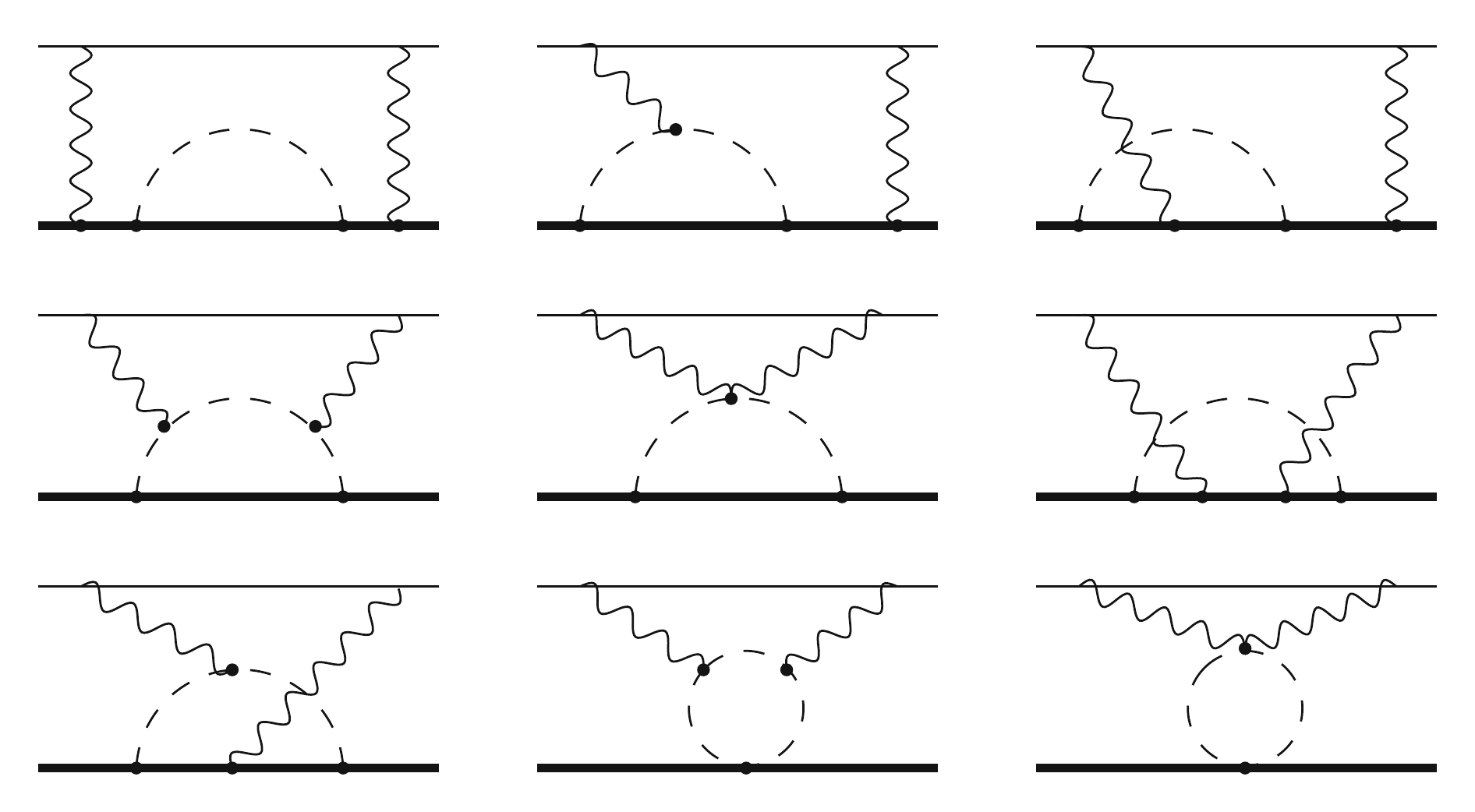}
\caption{The two-photon-exchange diagrams with chiral loops. Figure taken from Ref.~\cite{Alarcon:2013cba}.}
\label{fig1}
\end{figure}

\subsection{Chiral Loops}

In the $\delta$-expansion of ChPT, the LO polarizability contribution is given by the TPE diagrams with chiral loops, shown in Fig.~\ref{fig1}. They were calculated in Ref.~\cite{Alarcon:2013cba} with the results given in Table \ref{Table:Summary2}. Note that the VVCS structures in Figures \ref{fig:PiNLoopDiags} and \ref{fig1} differ due to a redefinition of the nucleon field,\footnote{$N\rightarrow \xi N$ with $\xi=\exp \left(\nicefrac{ig_A\pi^a \tau^a \gamma_5}{2f_\pi}\right)$} which is described in Ref.~\cite[Section 3.1]{Len10}.

\citet{Alarcon:2013cba} established the LEX in \Eqref{bcancels} as a very good approximation for the TPE polarizability effect in the LS.  The high-energy contribution  to their result was found to be small enough to not contradict the use of ChPT as a low-energy effective field theory. Nevertheless, we improve the cut-off behavior on higher momentum transfers by including the pion FF \cite{Bebek:1977pe}:
\beq
F_{\pi}(Q^2)=\left(1+\frac{Q^2}{\Lambda_\pi^2}\right)^{-1},\eqlab{PionFF}
\eeq
with $\Lambda_\pi^2=0.462\pm0.024\,\text{GeV}^2$. It enters twice in each CS diagram, cf.\ \Figref{PiNLoopDiags}, and by this reduces the contribution to the $Q$-integral in \Eqref{bcancels} from $Q>m_\rho\approx 775$ MeV to  $\sim 1\,\%$.   

Plugging the non-Born CS amplitudes into \Eqref{bcancels}, we obtain the following polarizability effect on the LS in $\mu$H:
\beq
 E^{\langle\pi N \rangle\,\mathrm{pol.}}_\mathrm{LS}(\mu\text{H})=5.86\,^{+1.76}_{-0.88}\,\upmu\text{eV}. \eqlab{individual}
\eeq
It is then easy to isolate the contribution of the $\ol T_1(0,Q^2)$ subtraction function to \Eqref{bcancels}:
\begin{subequations}
\bea
E^{\langle\pi N \rangle\,\mathrm{subtr.}}_\mathrm{LS}(\mu\text{H})&=&1.81\,^{+0.54}_{-0.27}\,\upmu\text{eV},\\
 E^{\langle\pi N \rangle\,\mathrm{inel.}}_\mathrm{LS}(\mu\text{H})&=&4.05\,^{+1.21}_{-0.61}\,\upmu\text{eV}.
\eea
\end{subequations}
As explained in Ref.~\cite{Alarcon:2013cba}, the value of the polarizability contribution is expected to increase when going to the next order (i.e., including pion-delta loops). Therefore, we assigned an uncertainty of $30\,\%$ ($\simeq \Delta/M$) towards the magnitude increase and $15\,\%$ ($\simeq m_\pi/M$) towards the magnitude decrease.

 \renewcommand{\arraystretch}{1.5}
\begin{table} [t]
\caption{$\Delta$-exchange contribution to the $2S$-level shift in muonic hydrogen. All values in $\upmu\mathrm{eV}$. \label{nonpoleLS}}
\centering
\begin{small}
\begin{tabular}{|c|c|c|c|c|}
\hline
\rowcolor[gray]{.7}
{\bf Eq.}&{\bf Input}&$\boldsymbol{\Delta E_\mathrm{2S}(T_1)}$&$\boldsymbol{\Delta E_\mathrm{2S}(T_2)}$&$\boldsymbol{\Delta E_\mathrm{2S}}$\\
\hline
\eref{LSAngular}&$T_1(0,Q^2)$ \eref{T1su}&$7.58$&/&$7.58$\\
\hdashline
\rowcolor[gray]{.95}
\eref{LSMasterSub}&$f_i$ \eref{structurefunc}&$-2.22$&$-6.01$&$-8.23$\\
\eref{LSAngular} & $T_i^{\Delta\mathrm{-pole}}$ \eref{Dpole}&$-2.22$&$-6.01$&$-8.23$\\
\hdashline
\rowcolor[gray]{.95}
\eref{LSAngular} & $\widetilde T_i$ \eref{noDpole}&$0.40$&$1.19$&$1.59$\\
\hline
\eref{LSAngular} & $T_i$ \eref{DeltaAmps}&$5.76$&$-4.82$&$0.95$\\
\hline
\end{tabular}
\end{small}
\end{table}
 \renewcommand{\arraystretch}{1.3}

\subsection[$\Delta$-Exchange]{$\boldsymbol{\Delta}$-Exchange}\seclab{deltaExLS}

  \begin{figure}[h!] 
    \centering 
       \includegraphics[width=4.5cm]{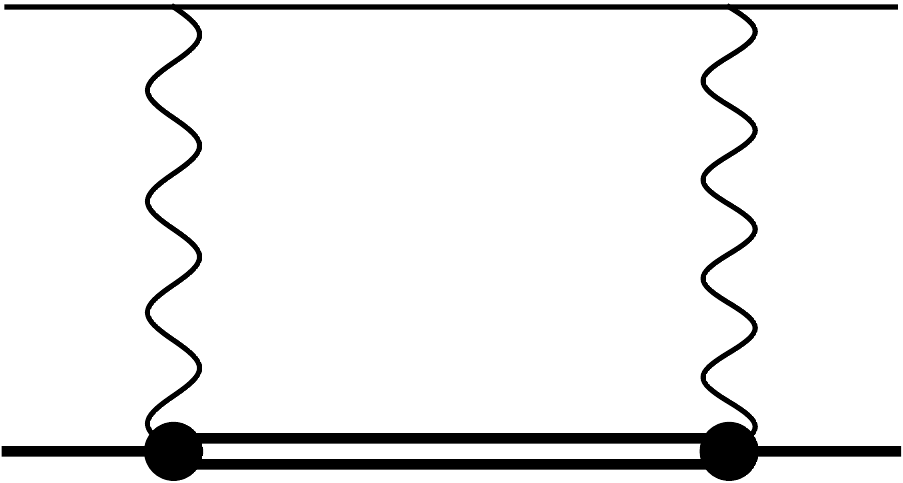}
       \caption{Two-photon-exchange diagram with intermediate $\Delta(1232)$-excitation.}
              \label{fig:TPEDelta}
\end{figure}

At NLO in the ChPT power-counting, we need to consider the $\Delta(1232)$-exchange shown in \Figref{TPEDelta}. The tree-level VVCS amplitudes with intermediate $\Delta$-excitation and the $\Delta$-production cross sections are given in \secref{chap4}{DeltaExchangeSec}. We calculated them in BChPT and afterwards related the ChPT couplings to the Jones-Scadron FFs, cf.\ \secref{chap4}{Jones}. The Jones-Scadron FFs, in turn, were replaced by the finite-momentum extension of their large-$N_c$ limit, cf.\ Eqs.~\eref{GM*old2}, \eref{GM*old3} and \eref{GM*} with $C_M^*=\frac{3.02}{\sqrt{2}\,\kappa_p}$. In this way, the $\Delta$-exchange process could be described through nucleon FFs, which are known from experiments. We chose to mainly work with the elastic FF parametrization of \citet{Bradford:2006yz}. For another thing, we use the dipole FF to study the sensitivity of our results on the FF parametrization.

The total contribution of the $\Delta(1232)$-exchange to the $2P_{1/2}-2S_{1/2}$ LS in $\mu$H amounts to:
\beq
E_\text{LS}^{\langle\Delta\rangle\,\mathrm{pol.}}(\mu \text{H})=-0.95\pm0.95\, \upmu\mathrm{eV}. \eqlab{DeExLS}
\eeq
Our results are summarized in Table \ref{nonpoleLS}, where we distinguish contributions from the subtraction function $T_1(0,Q^2)$, the $\Delta$-pole amplitudes $T_i^{\Delta\mathrm{-pole}}$ and the non-pole amplitudes $\widetilde T_i$, cf.\ decomposition in \Eqref{DeltaAmps}. The size of the individual contributions is comparable to the leading chiral loop effect, \Eqref{individual}. Combining the individual contributions, the $\Delta$-pole parts of the VVCS amplitudes largely cancel the subtraction function and the non-pole parts. Due to the large cancelations, we prefer to assign a conservative error of $100\,\%$. 

Surprisingly, the $\Delta$-exchange contribution to the subtraction function, 
\begin{subequations}
\eqlab{EqDelta11}
\bea
E_\text{LS}^{\langle\Delta\rangle\,\mathrm{subtr.}}(\mu \text{H})&=&-7.58\pm 2.27\, \upmu\mathrm{eV},\eqlab{deexsub}\\
E_\text{LS}^{\langle\Delta\rangle\,\mathrm{inel.}}(\mu \text{H})&=&6.63\pm 1.99\, \upmu\mathrm{eV},
\eea
\end{subequations}
is much larger than the LO contribution from the pion-nucleon loops. Therefore, it has a substantial effect on our BChPT prediction for the subtraction term, which is collected in \Eqref{totalSub}. Note that for the $\Delta$-exchange contribution in \Eqref{EqDelta11} we assigned a $30\,\%$ error due to higher orders in the chiral expansion.

 \renewcommand{\arraystretch}{1.5}
\begin{table} [t]
\centering
\begin{small}
\caption{Contribution of different multipole ratios to the $2S$-level shift in muonic hydrogen. All values in $\upmu\mathrm{eV}$. \label{multipolesLS}}
\centering
\begin{tabular}{|l|cc|c|c|}
\hline
\rowcolor[gray]{.7}
&\multicolumn{2}{c|}{$\boldsymbol{\Delta E_\mathrm{2S}(T_1)}$}&$\boldsymbol{\Delta  E_\mathrm{2S}(T_2)}$&$\boldsymbol{\Delta  E_\mathrm{2S}}$\\
\rowcolor[gray]{.7}
&$\boldsymbol{T_1(0,Q^2)}$&$\boldsymbol{f_1}$&$\boldsymbol{f_2}$&\\
\hline
$G_M^{*2}$&$7.960$&$-1.929$&$-5.117$&$0.915$\\
\rowcolor[gray]{.95}
$G_M^{*2}R_\mathrm{EM}$&$-0.221$&$0.041$&$0.154$&$-0.027$\\
$G_M^{*2}R_\mathrm{SM}$&$-0.146$&$0.043$&$0.080$&$-0.023$\\
\rowcolor[gray]{.95}
$G_M^{*2}R_\mathrm{EM}^2$&$-0.029$&$0.007$&$0.047$&$0.026$\\
$G_M^{*2}R_\mathrm{EM}R_\mathrm{SM}$&$0.022$&$0.011$&$-0.002$&$0.031$\\
\rowcolor[gray]{.95}
$G_M^{*2}R_\mathrm{SM}^2$&$-0.004$&$0.010$&$0.021$&$0.027$\\\hline
total&$7.581$&$-1.817$&$-4.816$&$0.948$\\
\hline
\end{tabular}
\end{small}
\end{table}
 \renewcommand{\arraystretch}{1.3}

We verified that the contribution from large momentum transfers ($Q>m_\rho$) is less than $1\,\%$. Also, it was confirmed that the dependence on the applied nucleon FF parametrization is small. Using a dipole FF for $G_{Ep}$ and $G_{Mp}$, as well as the Galster parametrization for $G_{En}$ \cite{Galster:1971kv}:\footnote{These are no parametrizations fitted to experimental data, thus, they only give a rough description of the basic $Q^2$ dependence of the nucleon FFs. There advantage, however, is that they display a physical pole structure: The dipole FF has second-order poles at $Q=\pm i \Lambda$ and the Galster parametrization has simple poles at $Q=\pm \frac{2iM}{\sqrt{\eta}}$.}
\begin{subequations}
\eqlab{Galster}
\bea
G_{En}(Q^2)&=&-\frac{\tau \mu_n} {1+\eta \,\tau}G_D(Q^2),\\
G_{D}(Q^2)&=& (1+\frac{Q^2}{\Lambda^2})^{-2},
\eea
\end{subequations}
with $\eta=5.6$ and $\Lambda^2=0.71\,\mathrm{GeV}^2$, changes \Eqref{DeExLS} by only $\sim 6\,\%$.

 In Table \ref{multipolesLS}, we break down the $\Delta$-exchange effect on the $\mu$H LS into the contributions of different multipole ratios.
The numerically small influence of the $\Delta$-resonance on the LS, \Eqref{DeExLS}, was expected because the nucleon-to-delta transition is dominantly of magnetic dipole type, while the magnetic dipole polarizability is suppressed in the LS, as can be seen from the LEX of the VVCS amplitudes \cite{Alarcon:2013cba}, cf.\ Eqs.\ \eref{bcancels} and \eref{bcancelsLEX}.

  \subsection{Comparison with Heavy Baryon Chiral Perturbation Theory} \seclab{HBcompLS}

 There are several theory based determinations of the polarizability contribution \cite{Mohr:2013axa,Hill:2011wy}, mainly performed within the chiral framework \cite{Nevado:2007dd,Alarcon:2013cba,Peset:2014yha}, see Table \ref{Table:Summary2} for a summary. Assuming ChPT is working, it should be best applicable to atomic system, where the energies are very small. In ChPT, the NLO polarizability contribution can be obtained as a model-independent prediction. Equivalently, the VVCS process at $\mathcal{O}(p^{7/2})$ in the low-energy domain of the $\delta$-expansion can be described without fitting of LECs to CS or elastic lepton-proton scattering experiments. 

\citet{Nevado:2007dd} calculate the spin-independent structure functions in HBChPT at the leading one-loop level. The study is restricted to light quarks (u,d) and the delta is neglected. The $S$-level shifts amount to:
\begin{subequations}
\bea
\Delta E^{\mathrm{pol.}}_{nS}(\text{H})&=&-\frac{87.0488}{n^3}\text{ Hz},\\
\Delta E^{\mathrm{pol.}}_{nS}(\mu \text{H})&=&-\frac{0.147614}{n^3}\text{ meV}.
\eea
\end{subequations}
In the case of H, this agrees well with the logarithmic approximation in Ref.~\cite{Pineda:2005aa}. \citet{Peset:2014jxa} improve the calculation of Ref.~\cite{Nevado:2007dd} by including the $\Delta(1232)$-resonance. They obtain (in units of $\upmu$eV):
\beq
\Delta E_{2S}^\mathrm{pol.}(\mu \text{H})=-18.51\,(\pi\text{N-loops})+1.58\,(\Delta\text{-exch.})-9.25\,(\pi\Delta\text{-loops})=-26.2 \pm 10.0\,\upmu \text{eV}.\nn
\eeq
Exploiting these results leads to a proton charge radius of $R_{Ep} = 0.8412(15) \text{ fm}$ \cite{Peset:2014yha}.

The LO BChPT calculation performed by \citet{Alarcon:2013cba}, cf.\ Fig.~\ref{fig1}, predicts the polarizability effect on the $2S$-level in $\mu$H with:
\beq
\Delta E^{\mathrm{pol.}}_{2S}(\mu \text{H})=-8.2 \,^{+1.2}_{-2.5}\, \upmu\text{eV}.
\eeq 
The quoted error limits are unsymmetrical because the NLO is expected to increase the magnitude, as is the case with $\alpha_{E1}$ \cite{Pascalutsa:2013zha}. Indeed, our calculation confirmed that the $\Delta$-exchange leads to a slight increase of the polarizability contribution to the LS. Obviously, the BChPT result is in disagreement with the HBChPT predictions. However, expanding the non-Born amplitudes in $\mu=m_\pi/M$, while keeping the ratio of light scales $\tau_\pi=Q^2/4m_\pi^2$ fixed, i.e., performing the heavy-baryon expansion, Ref.~\cite{Alarcon:2013cba} recovers the amplitudes given in References \cite{Nevado:2007dd} and \cite{Birse:2012eb}.  They find the following simple formula for the polarizability contribution 
to the $2S$-level shift from LO HBChPT
 \cite{Alarcon:2013cba}:
\beq 
\De E^{\mathrm{pol.}}_{2S} = \frac{\al^5 m_r^3 \, g_A^2}{4(4\pi f_\pi)^2} \, \frac{m}{m_\pi}
\Big( 1- 10\,G + 6 \ln 2\Big) \simeq -16.1 \,\, \upmu\mbox{eV},
\eeq 
where $G\simeq 0.9160$ is the Catalan constant.
In this way, the LO HBChPT results from References \cite{Alarcon:2013cba} and \cite{Nevado:2007dd} are in good agreement, cf.\ Table \ref{Table:Summary2}.\footnote{Note that the difference is only due to the use of a LEX, see \Eqref{bcancels}, in Ref.~\cite{Alarcon:2013cba}.} 
 
 As already observed for the dipole polarizabilities themselves, the BChPT and HBChPT results for the polarizability contribution differ substantially at ``predictive'' orders. The LO predictions of $\beta_{M1}$ even differ in the overall sign \cite{Pascalutsa:2013zha}. For various reasons, the result from BChPT seem to be more reliable. The HBChPT result for the polarizability contribution to the $\mu$H LS obtains substantial contributions (at least 25\%) from beyond the scale ($Q>m_\rho\approx 775$ MeV) at which this effective theory is safely applicable. In contrast, for the BChPT result one only finds a contribution of less than 15\%, what is not exceeding the expected uncertainty of such calculation. Another general advantage of BChPT versus HBChPT is that in the former analyticity is obeyed exactly, while in the latter it is obeyed only approximately.

 \renewcommand{\arraystretch}{1.75}
 \begin{table*}[t]
\caption{Summary of available chiral perturbation theory calculations for the two-photon-exchange corrections to the $2S$-level shift in muonic hydrogen. Energy shifts are given in $\upmu\mathrm{eV}$.}
\label{Table:Summary2}
\centering
\begin{minipage}{\linewidth}  
\centering
\begin{small}
\begin{tabular}{|c|ccccc|}            
\hline      
 \rowcolor[gray]{.7}
&{\bf Nevado \& Pineda   } &{\bf Alarc\'on et al.}&{\bf Alarc\'on et al. }  &{\bf Peset \& Pineda}&{\bf this work}\\
 \rowcolor[gray]{.7}
&{\bf HBChPT \cite{Nevado:2007dd}}&{\bf BChPT \cite{Alarcon:2013cba}}& {\bf HBChPT \cite{Alarcon:2013cba}}&{\bf HBChPT \cite{Peset:2014yha}}\footnote{prediction at LO and NLO (including pions and deltas)}&{\bf BChPT}\\
\hline
$\Delta E^{\mathrm{subtr.}}_{2S}$                                      &                                                          & $-3.0$ &$1.3$ &&$5.8(2.3)$\\
 \rowcolor[gray]{.95}
$\Delta E^{\mathrm{inel.}}_{2S}$                                      &                                                                    & $-5.2$ &$-19.1$&&$-10.7(^{+2.1}_{-2.3})$\\ 
\hline
$\Delta E^{\mathrm{pol.}}_{2S}$                                        &    $-18.5(9.3)$\footnote{error given in Ref.~\cite{Peset:2014yha}}                                                       & $-8.2(^{+1.2}_{-2.5})$ &$-17.85$&$-26.2(10.0)$&$-4.9(_{-2.0}^{+1.3})$\\
 \rowcolor[gray]{.95}
$\Delta E^{\mathrm{Born}}_{2S}$ &$-10.1(5.1)$\footnote{value from Ref.~\cite{Pineda:2005aa} as given in Ref.~\cite{Peset:2014yha}}&&&$-8.3(4.3)$&$-22.9(1.7)$\\
\hline
$\Delta E_{2S}$ &$-28.6$&&&$-34.4(12.5)$&$-27.8(^{+2.1}_{-2.6})$\\
\hline
\end{tabular}
\end{small}
\end{minipage}
\end{table*}
 \renewcommand{\arraystretch}{1.3}

 \renewcommand{\arraystretch}{1.75}
\begin{table}[h!]
\centering
\caption{Summary of available dispersive calculations for the two-photon-exchange corrections to the $2S$-level shift in muonic hydrogen. Energy shifts are given in $\upmu\mathrm{eV}$, $\beta_{M1}$ is given as $\times 10^{-4} \,\mathrm{fm}^3$.
\label{Table:Summary1}}
\begin{minipage}{\linewidth}  
\centering
\begin{tabular}{|c| ccccc|}
\hline                  
 \rowcolor[gray]{.7}
& \footnotesize {\bf Pachucki}& \footnotesize {\bf Martynenko}&  \footnotesize {\bf Carlson \&} & \footnotesize {\bf Birse \& }&   \footnotesize {\bf Gorchtein}\\
 \rowcolor[gray]{.7}
&\footnotesize {\bf \cite{Pachucki:1999zza}}&\footnotesize {\bf \cite{Martynenko:2005rc}}&\footnotesize {\bf Vanderhaeghen \cite{Carlson:2011zd}} &\footnotesize{\bf McGovern \cite{Birse:2012eb}}&\footnotesize {\bf et al.~\cite{Gorchtein:2013yga}}\footnote{Adjusted values;
the original values of Ref.~\cite{Gorchtein:2013yga},  $\Delta E^{\mathrm{subtr.}}_{2S}=3.3$ and $\Delta E^{(\mathrm{el})}_{2S}=-30.1$, are based on a different decomposition into elastic and polarizability contributions.} \\
\hline
\footnotesize$\beta_{M1}$&\footnotesize$1.56(57)$ \cite{Tonnison:1998mi}&\footnotesize$1.9(5)$ \cite{Eidelman20041}&\footnotesize$3.4(1.2)$ \cite{Beane:2002wn,Beane:2004ra}&\footnotesize$3.1(5)$ \cite{Griesshammer:2012we}&\\
\hdashline
 \rowcolor[gray]{.95}
\footnotesize$\Delta E^{\mathrm{subtr.}}_{2S}$        & \footnotesize $1.9$                                                             &    \footnotesize$ 2.3$                                                         &  \footnotesize$5.3(1.9)$                                       & \footnotesize$4.2(1.0)$                            &   \footnotesize$ -2.3 (4.6)$                                     \\
\footnotesize$\Delta E^{\mathrm{inel.}}_{2S}$            &   \footnotesize$-13.9$ \cite{Brasse:1976bf, Abramowicz:1997}                                                         &     \footnotesize$ -16.1$                                                      &  \footnotesize$-12.7(5)$ \cite{Christy:2011,Capella:1994cr}                               & \footnotesize$-12.7(5)$\footnote{Value taken from Ref.~\cite{Carlson:2011zd}.}   &\footnotesize $-13.0(6)$ \cite{Christy:2011,Capella:1994cr, Gorchtein:2011mz}                                       \\ 
\hline
 \rowcolor[gray]{.95}
\footnotesize$\Delta E^{\mathrm{pol.}}_{2S}$              & \footnotesize$-12(2)$                                                           &    \footnotesize$-13.8(2.9)$                                              &  \footnotesize$ -7.4  (2.0)$                               & \footnotesize$-8.5(1.1)$                                     & \footnotesize $ -15.3(4.6)$                                     \\
\footnotesize $\Delta E^{\mathrm{Born}}_{2S}$&\footnotesize$-23.2(1.0)$&&\multirow{2}{*}{\footnotesize$\begin{cases}
-27.8\text{ \cite{Kelly:2004hm}}\\
\bold{-29.5(1.3)}\text{ \cite{Arrington:2007ux}}\\
-30.8\text{ \cite{Bernauer:2010wm,Vanderhaeghen:2010nd}}
\end{cases}$}&\footnotesize$-24.7(1.6)$\footnote{Result taken from Ref.~\cite{Carlson:2011zd} (FF \cite{Arrington:2007ux}) with reinstated ``non-pole'' Born piece.}&\footnotesize$-24.5(1.2)$\\
&\footnotesize \cite{Simon1980381}&&&&\footnotesize \cite{Bernauer:2010wm, Kelly:2004hm, Arrington:2007ux}\\
\hline
 \rowcolor[gray]{.95}
\footnotesize $\Delta E_{2S}$&\footnotesize$-35.2(2.2)$&&\footnotesize$-36.9(2.4)$&\footnotesize$-33(2)$&\footnotesize$-39.8(4.8)$\\
\hline
\end{tabular}
\end{minipage}
\end{table}
 \renewcommand{\arraystretch}{1.3}
\newpage
\subsection{Comparison with Dispersive Calculations} \seclab{ComparisonLS}

An early study of the effect of the electric polarizability on the $S$-level shifts in electronic and muonic atoms can be found in Ref.~\cite{Bernabeu:1982qy}, where the approximation of unretarded-dipole (long-wavelength) photons is used. In Ref.~\cite{Khriplovich:1997fi,Khriplovich1998474}, the effect of both the electric and magnetic polarizability on the $1S$ ground state in H is calculated, considering the mean excitation energy of the proton. The work of \citet{Rosenfelder:1999px} builds upon the formalism introduced by \citet{Bernabeu:1982qy}. It aims at improving the previous results by accounting for retardation and estimating further contributions, e.g., through virtual transverse excitations, in the non-relativistic harmonic oscillator quark model. The reliability of the different approaches is not undisputed. For one thing, if the average excitation energy of the proton is too large ($\sim 410 \text{ MeV}$ \cite{Rosenfelder:1999px} compared to $\sim 300 \text{ MeV}$ \cite{Khriplovich1998474}), the unretarded-dipole photons are a doubtful approximation. Also, as pointed out in Ref.~\cite{Khriplovich:1999ak}, the polarization shifts can not be correctly expressed through total photoabsorption cross section by means of an unsubtracted DR, as it is done in Refs.~\cite{Bernabeu:1982qy,Rosenfelder:1999px}. Nevertheless, the first prediction of the $nS$-level shift in $\mu$H \cite{Rosenfelder:1999px}: 
\beq
\Delta E^{\mathrm{pol.}}_{nS}(\mu\text{H})=-\frac{0.136\pm0.030}{n^3}\text{ meV} \overset{n=2}{=}-0.017\pm0.004\text{ meV},
\eeq
is of similar magnitude as later dispersive and HBChPT predictions.

At the current level of precision, the LO BChPT prediction of the polarizability contribution to the $\mu$H LS \cite{Alarcon:2013cba} and our NLO update, \Eqref{LSfinalvalue}, are in good agreement with calculations based on dispersive sum rules but purely model independent, see Fig.~\ref{ComparisonLS}. In Table \ref{Table:Summary1}, we list dispersive calculations and partially requoted the empirical information which entered as input. In the following, we will give further details on Refs.~\cite{Pachucki:1999zza,Martynenko:2005rc,Carlson:2011zd,Birse:2012eb,Gorchtein:2013yga} and briefly summarize the advancement in dispersive calculations.

One of the first modern dispersive calculations of the TPE effects can be found in Ref.~\cite{Pachucki:1999zza}, see also Refs.~\cite{Pachucki:1996zza, Veitia:2004zz}. The latest dispersive approach can be found in Ref.~\cite{Gorchtein:2013yga}. The main achievement of the latter paper is to relate the $T_1(0,Q^2)$ subtraction function to the $Q^2$ dependence of the fixed $J=0$ Regge pole \cite{Creutz:1968ds} through a  finite-energy sum rule \cite[Eq.~(29)]{Gorchtein:2013yga}.
 The evaluation relies on an empirical dataset comparable to the one used in Ref.~\cite{Carlson:2011zd}. 

The work of \citet{Martynenko:2005rc} is based on the unitary isobar model and evolution equations for parton distribution functions. The total polarizability contribution to the energy shifts of $1S$ and $2S$ energy-levels in H and $\mu$H is calculated. In the resonance region, the five dominant low-lying resonances, viz.\ $P_{33}(1232)$, $S_{11}(1535)$, $D_{13}(1520)$, $P_{11}(1440)$ and $F_{15}(1680)$, and the $N\pi$, $N\eta$ and $N \pi \pi$ final-states are taken into account, as well as the contribution from $K$ mesons. The contribution of the non-resonance region is calculated from experimental data on the structure functions for deep-inelastic lepton-nucleon scattering and parton distributions.

\citet{Birse:2012eb} calculate the subtraction function at NLO in HBChPT. They include the leading contribution of the nucleon-to-delta transition FFs, while other effects of the delta are absorbed into the LECs. Strictly speaking, this ``physical cutoff'' dependent result lies outside the ChPT framework. Therefore, we list it in Table \ref{Table:Summary1} and not in Table \ref{Table:Summary2}. Their ``elastic'' and ``inelastic'' contributions are adopted from Ref.~\cite{Carlson:2011zd} with the necessary adjustments to achieve the proper separation into Born and non-Born terms.

With time, the recommended value for the magnetic dipole polarizability of the proton increased, see the first row of Table \ref{Table:Summary1}. The PDG value for instance changed from $\beta_{M1}=2.1(9)\times 10^{-4} \,\mathrm{fm}^3$ \cite[PDG '98]{Caso:1998tx} to $\beta_{M1}=2.5(4)\times 10^{-4} \,\mathrm{fm}^3$ \cite[PDG '16]{Olive:2016xmw}. Likewise, the value of $\Delta E^{\,\mathrm{subtr.}}_{2S}$ increased in recent calculations, cf.\ Refs.~\cite{Carlson:2011zd,Birse:2012eb}, by a factor of two compared to earlier calculations, cf.\ Refs.~\cite{Pachucki:1999zza,Martynenko:2005rc}. This follows from the LEX of $T_1$, \Eqref{T1betaM1}, and the customary models for the subtraction function.

Besides the dispersive and ChPT based approaches, there are other theoretical studies. They usually differ significantly from what we discussed thus far, nevertheless, we want  to mention them here for completeness.  For one thing, there is an analysis of the proton structure corrections in the framework of non-relativistic QED effective field theory \cite{Hill:2011wy}. This work requires matching to low-energy observables, such as $\beta_{M1}$ \cite{Nakamura:2010}. For another thing, Ref.~\cite{Mohr:2013axa} ultilizes a bound-state field theory approach. Here, the binding field of the proton is generated by a one-parameter static-well model (simplified MIT bag model). Despite the exceptional method, the result is in good agreement with other predictions and the radius of the well, $R=1.2 \text{ fm}$, would correspond to $R_{Ep}=0.87 \text{ fm}$ and $R_{\mathrm{Z}p}=1.3 \text{ fm}$.

\subsection{Experimental Two-Photon-Exchange Effect in Muonic Hydrogen}

In Ref.~\cite{Pohl1:2016xoo}, an experimental prediction for the TPE effect in the $\mu$D LS is given:
\beq
E^\mathrm{TPE}_\mathrm{LS}(\mu\text{D})=1.7638(68)\,\mathrm{meV}.
\eeq
This value is based on the deuterium charge radius, \Eqref{isotopicRd}, extracted from the $\mu$H result for the proton charge radius, \Eqref{remuH}, and the hydrogen-deuterium isotope shift of the $1S-2S$ transition, cf.\ \Eqref{raddif}. In the same way, one can determine the experimental value of the TPE effect in $\mu$H. The theory prediction for the $\mu$H LS is (in units of $\mathrm{meV}$) \cite{Antognini:2012ofa}:
\beq
E_{\mathrm{LS}}^{\,\text{th.}}(\mu\text{H})=206.0336(15)-5.2275(10)\,\left(\nicefrac{R_{Ep}}{\mathrm{fm}}\right)^2+E^{\mathrm{TPE}}_{\mathrm{LS}}\, .
\eeq
The proton radius, as obtained from the $\mu$D LS and the isotopic deuteron-proton charge radius difference is given in \Eqref{isotopicRp}. Comparing the theoretical expectation with the experimental value of the $\mu$H LS
\cite{Antognini:1900ns}:
\beq
E^{\,\text{exp.}}_\text{LS}(\mu\text{H})=202.3706(23) \,\text{meV},\eqlab{expmuHLS}
\eeq
and substituting the charge radius from \Eqref{isotopicRp}, we can solve for the experimental TPE effect:
\beq
E^\mathrm{TPE}_\mathrm{LS}(\mu\text{H})=-0.0130(177)\,\mathrm{meV}.
\eeq
The error bar of this value covers zero. For comparison, our theoretical prediction, \Eqref{LSfinalvaluePlusBorn}, and the one included in the summary paper \cite{Antognini:2012ofa}, $E^{\mathrm{TPE}}_{\mathrm{LS}}(\mu\text{H})=0.0332(20)$ meV, are both giving a positive value for the TPE contribution to the $\mu$H LS.

\section[Nuclear-Polarizability Contribution at Order $(Z\al)^6 \ln Z \al$]{Nuclear-Polarizability Contribution at Order $\boldsymbol{(Z\al)^6 \ln Z \al}$ } \seclab{offTPE}
 
In the following Section, we study the nuclear-polarizability contribution to the LS of light muonic atoms from off-forward TPE. The basic theory was derived in \secref{chap5}{6.3}. In comparison to the forward limit, the off-forward TPE is suppressed by an additional factor of $Z\al$. Nevertheless, it is interesting in view of the $t$-channel cut, leading to an enhancement of order $(Z\al)^6 \ln Z\al$.  Here, the LS is calculated explicitly for $\mu$H, $\mu^2$H, $\mu^3$H, $\mu^3\text{He}^+$ and $\mu^4\text{He}^+$.  In \appref{5HFS}{6.5}, we briefly outline our calculation for the HFS and explain why no logarithmic enhancement is found. 
 
The influence of the nuclear polarizabilities on the spectrum of hydrogen-like atoms has been investigated for many years. Several early works focused on the leading logarithmic contribution to the LS, which is of order $(Z\al)^5 \ln \frac{\bar E}{m}$, with $\bar E$ being the mean excitation energy and $m$ the lepton mass (cf.\ Ref.~\cite{Khriplovich:1997fi} for hydrogen and Refs.~\cite{Friar:1997aa,Martorell:1995zz, Khriplovich:1995yd} for deuterium). Also, Ref.~\cite{Friar:1978wv} found numerically important terms at order $(Z\al)^6 \ln Z\al$. Refs.~\cite{Pachucki:2015uga, Ji:2013oba, Dinur:2015vzv} recently calculated the $(Z\al)^6 \ln (Z\al)^2$
 Coulomb-distortion correction to the LS. Here and in \secref{chap5}{6.3}, we will present an alternative approach to compute the Coulomb distortion.


At the intended order, $(Z\al)^6 \ln Z\al$, the effect of the nuclear scalar polarizabilities from off-forward TPE on the LS amounts to: 
\beq
E_\text{LS}^{\left\langle(Z\al)^6 \ln Z\al\right\rangle}=\frac{4(Z\al m_r)^4\al\,\al_{E1}}{n^3}\,\ln \frac{Z\al m_r}{2nm}, \eqlab{finalresult}
\eeq
with a factor of $Z^2 \al$ embedded in the electric dipole polarizability $\al_{E1}$. The part of the potential in \Eqref{ImM}, relevant to \Eqref{finalresult}, is found as the leading term in small $p_t^2$. Therefore, the $(Z\al)^6 \ln Z\al$ effect stems from the $\mathcal{O}(\sqrt{\tau})$ piece:
\beq
\im \mathscr M(p_t^2)\approx-\frac{2\pi \al m}{(1-\tau)^{7/2}}\,\sqrt{\tau}\arccos \sqrt{\tau} \;\al_{E1}+\mathcal{O}(\tau).\eqlab{relevantIm}
\eeq
It is worth pointing out that the magnetic dipole polarizability canceled from the result. Likewise, the electric polarizability dominates the forward TPE, as it is expected from studying the LEX of the relevant VVCS amplitudes \cite{Alarcon:2013cba}, see \secref{5LS}{ExpInPolLS}.

Based on the PDG '16 \cite{Olive:2016xmw} recommended value for the proton electric dipole polarizability,
\Eqref{PDG2016}, we arrive at the off-forward TPE polarizability contribution to the LS in $\mu$H:
\beq
\eqlab{result}
E_\text{LS}^{\left\langle(Z\al)^6 \ln Z\al\right\rangle}(\mu\text{H})=-0.79\pm0.03\,\upmu\mathrm{eV},
\eeq
where the stated error is propagated from \Eqref{PDG2016}. For comparison, the leading polarizability contribution to the $\mu$H LS from forward TPE was found to be: $
E^{\langle \al^5\rangle}_\mathrm{LS}(\mu\text{H})=4.9\,^{+2.0}_{-1.3}\,\upmu\text{eV}$. It is therefore fair to say that the numerical result presented in here is larger than expected. Furthermore, its size is comparable to the effects from light-by-light scattering (Wichmann-Kroll, virtual Delbr\"uck), which all amount to about $\approx 1\,$meV \cite{Antognini:2012ofa}.

Besides $\mu$H, we want to study the effect in $\mu^2$H, $\mu^3$H, $\mu^3\text{He}^+$ and $\mu^4\text{He}^+$.\footnote{In the present Section, we switch notations and denote muonic deuterium as $\mu^2$H and the deuteron as $^2\text{H}^+$. In the rest of the thesis, we use $\mu$D and d.} Since we are studying the spin-independent LS, the formalism is the same for all nuclei. The nuclear masses are \cite{Mohr:2015ccw}:
 \begin{subequations}
\bea
M(^2\text{H}^+)&=&1.875\,612\,928(12)\,\mathrm{GeV},\\
M(^3\text{H}^+)&=&2.808\,921\,112(17)\,\mathrm{GeV},\\
M(^3\text{He}^{2+})&=&2.808\,391\,586(17)\,\mathrm{GeV},\\
M(^4\text{He}^{2+})&=&3.727\,379\, 378(23)\,\mathrm{GeV}.
\eea
\end{subequations}
The mass of the triton is bigger than the one of its mirror nuclei, the helion. This shows that the triton, $E_\text{B}(^3\text{H}^+)=8.481798(2)\,\mathrm{MeV}$ \cite{Purcell:2010hka}, is stronger bound than the helion, $E_\text{B}(^3\text{He}^{2+})=7.718043(2)\,\mathrm{MeV}$ \cite{Purcell:2010hka}. Due to the heavier mass of the ($A=2,3,4$) nuclei, the approximation $m_r\approx m$ is even better than for the proton. 

Unfortunately, there is a spread in the predictions of the nuclear electric dipole polarizabilities:\footnote{See Ref.~\cite[Table I]{Stetcu:2009py} for an overview on electric dipole polarizabilities of hydrogen and helium isotopes.}
\begin{subequations}
\eqlab{nuclearPol}
\bea
\al_{E1}(^2\text{H}^+)&=&\begin{cases}0.6314(19)\,\text{fm}^3\;\text{\cite{Phillips:1999hh}},\\
0.70(5)\,\text{fm}^3\;\text{\cite{Rodning:1982zz}}\end{cases}\\
\al_{E1}(^3\text{H}^+)&=&\begin{cases}
0.23\,\mathrm{fm}^3\; \text{\cite{Kharchenko:2010zz,Kharchenko:2010yw}},\\
0.139(2)\,\mathrm{fm}^3\;\text{\cite{Stetcu:2009py}},
\end{cases}\\
\al_{E1}(^3\text{He}^{2+})&=&\begin{cases}
0.250(40)\,\mathrm{fm}^3\;\text{\cite{Goeckner:1990nw}},\\
0.149(5)\,\mathrm{fm}^3\;\text{\cite{Stetcu:2009py}},
\end{cases}\\
\al_{E1}(^4\text{He}^{2+})&=&
0.0683(8)(14)\,\mathrm{fm}^3\;\text{\cite{Stetcu:2009py}}.
\eea
\end{subequations}
We decide to use the theoretical predictions of Refs.~\cite{Stetcu:2009py,Phillips:1999hh} for all nuclei studied herein. However, one should not forget that there are experimental extractions of the deuteron and helium-3 polarizabilities from elastic scattering of the nuclei from $^{208}\text{Pb}$ \cite{Goeckner:1990nw}, which are incompatible  with the value from Refs.~\cite{Stetcu:2009py,Phillips:1999hh}. 

Our results are summarized in Table \ref{Tab:results}. Based on the smaller polarizability values, we arrive at:
\begin{subequations}
\eqlab{result2}
\bea
E_\text{LS}^{\left\langle(Z\al)^6 \ln Z\al\right\rangle}(\mu^2\text{H})&=&-0.541\pm0.002\, \mathrm{meV},\eqlab{resultD}\\
E_\text{LS}^{\left\langle(Z\al)^6 \ln Z\al\right\rangle}(\mu^3\text{H})&=&-0.128\pm0.002\, \mathrm{meV},\eqlab{result3H}\\
E_\text{LS}^{\left\langle(Z\al)^6 \ln Z\al\right\rangle}(\mu^3\text{He}^{+})&=&-1.950\pm0.065\, \mathrm{meV},\eqlab{result3He}\\
E_\text{LS}^{\left\langle(Z\al)^6 \ln Z\al\right\rangle}(\mu^4\text{He}^{+})&=&-0.925\pm 0.022\, \mathrm{meV}\eqlab{result4He},
\eea
\end{subequations}
where the errors are propagated from the nuclear polarizabilities. The $(Z\al)^6 \ln Z\al$ polarizability effect in the light muonic atoms is considerably bigger than the one in $\mu$H. This has several reasons. First of all, nuclei are easier to polarize, as one can read off from the electric polarizabilities in Table \ref{Tab:results}. Secondly, the nuclear charge of helium is $Z=2$, which enters \Eqref{finalresult} to the fourth power (neglecting the $Z^2$ factor embedded in the nuclear polarizability). For a last thing, the reduced mass of the lepton-nucleus system is increasing with the mass of the nucleus, and approaching the lepton mass. This effect mainly influences the $m_r^4$ prefactor in \Eqref{finalresult}, but also enters the logarithmic term. To summarize, the $(Z\al)^6 \ln Z\al$ polarizability effect in light muonic atoms, i.e., deuterium, tritium and helium, is about three orders of magnitude bigger than the corresponding effect in the lightest hydrogen isotope. Especially for $\mu^2$H, one finds a large nuclear-polarizability effect. This is a result of the deuterons weak binding and the concomitant large size \cite{Friar:2005je}.

\noindent In addition to the $(Z\al)^6 \ln Z\al$ polarizability effect in Eqs.~\eref{finalresult} and \eref{result2}, we also give 
numerical results for the non-recoil effect of the dipole polarizabilities in off-forward TPE with non-vanishing photon cuts, cf.\ Eqs.~\eref{ImM} and \eref{EDR}. This is not the full order-$(Z \al)^6$ effect of the dipole polarizabilities, since we are neglecting the higher-$Q^2$ contributions by keeping only terms proportional to $1/Q^4$ and omitting recoil terms which are suppressed by $1/M$. Nevertheless, it should be a good  approximation for the subleading polarizability effects. For the LSs in muonic-hydrogen and muonic-helium isotopes, we find the electric dipole polarizability contribution as:
\begin{subequations}
\eqlab{numresult}
\bea
E_\text{LS}^{\left\langle(Z\al)^6,\;\al_{E1}\right\rangle}(\mu\text{H})&=&-0.138\pm0.005\, \upmu\mathrm{eV},\\
E_\text{LS}^{\left\langle(Z\al)^6,\;\al_{E1}\right\rangle}(\mu^2\text{H})&=&-0.403\pm0.001\, \mathrm{meV},\\
E_\text{LS}^{\left\langle(Z\al)^6,\;\al_{E1}\right\rangle}(\mu^3\text{H})&=&-0.095\pm0.001\, \mathrm{meV},\\
E_\text{LS}^{\left\langle(Z\al)^6,\;\al_{E1}\right\rangle}(\mu^3\text{He}^{+})&=&-1.403\pm0.047\, \mathrm{meV},\\
E_\text{LS}^{\left\langle(Z\al)^6,\;\al_{E1}\right\rangle}(\mu^4\text{He}^{+})&=&-0.665\pm0.016\, \mathrm{meV},
\eea
\end{subequations}
where the included contribution of $P$-levels is less than one percent of the dominant $S$-level shifts. 

The effect of the magnetic polarizability is expected to be small. As for the electric polarizabilities in \Eqref{nuclearPol}, there is a spread in the predictions for the nuclear magnetic dipole polarizabilities:
\begin{subequations}
\eqlab{nuclearPolbe}
\bea
\be_{M1}(^2\text{H}^+)&=&\begin{cases}
0.0777(3)  \,\text{fm}^3\;\text{\cite{Friar:1997aa}},\\
0.067\,\text{fm}^3\;\text{\cite{Chen:1998vi}},\\
0.072(5)\,\text{fm}^3\;\text{\cite{Carlson:2013xea}},\\
4.4\left(^{+1.6}_{-1.5}\right)(0.2)\times10^{-4}\,\mathrm{fm}^3\,\text{\cite{Chang:2015qxa}},\end{cases}\\
\be_{M1}(^3\text{H}^+)&=&2.6(1.7)(0.1)\times10^{-4}\,\mathrm{fm}^3,\text{\cite{Chang:2015qxa}},\\
\be_{M1}(^3\text{He}^{2+})&=&\begin{cases}5.7(0.5)\times10^{-3} \,\text{fm}^3\,\text{\cite{Carlson:2016cii}},\\
5.4\left(^{+2.2}_{-2.1}\right)(0.2)\times10^{-4}\,\mathrm{fm}^3\,\text{\cite{Chang:2015qxa}},\end{cases}\\
\be_{M1}(^4\text{He}^{2+})&=& 3.4\left(^{+2.0}_{-1.9}\right)(0.2)\times10^{-4}\,\mathrm{fm}^3 \;\text{\cite{Chang:2015qxa}}.
\eea
\end{subequations}
The values from Ref.~\cite{Chang:2015qxa} are  lattice predictions at a pion mass of $m_\pi \sim 806$ MeV.\footnote{In the values from Ref.~\cite{Chang:2015qxa}, the first uncertainty combines the statistical and systematic errors, and the second uncertainty estimates the effects of discretization and finite volume effects.} LQCD predicts the magnetic dipole polarizabilities of the light nuclei to be of the same magnitude as the nucleon polarizabilities. Effective field theory predictions \cite{Chen:1998vi}, however, find bigger values for the nuclear polarizabilities. For our calculation, we chose the magnetic dipole polarizabilities used by Carlson et al.\ \cite{Carlson:2013xea,Carlson:2016cii} and the lattice predictions for $\be_{M1}(^3\text{H}^+)$ and $\be_{M1}(^4\text{He}^{2+})$ \cite{Chang:2015qxa}. For the proton, we are using the PDG average \cite{Olive:2016xmw}, see \Eqref{PDG2016}.
As expected, including the magnetic dipole polarizability only leads to minimal changes in our predictions from \Eqref{numresult}:
\begin{subequations}
\eqlab{numresultsbeta}
\bea
E^{\left\langle(Z\al)^6,\;\al_{E1},\,\beta_{M1}\right\rangle}_\text{LS}(\mu\text{H})&=&-0.128\pm0.005\, \upmu\mathrm{eV},\eqlab{numresultsbetaH}\\
E^{\left\langle(Z\al)^6,\;\al_{E1},\,\beta_{M1}\right\rangle}_\text{LS}(\mu^2\text{H})&=&-0.398\pm0.001\, \mathrm{meV},\eqlab{numresultsbetaD}\\
E^{\left\langle(Z\al)^6,\;\al_{E1},\,\beta_{M1}\right\rangle}_\text{LS}(\mu^3\text{H})&=&-0.095\pm0.001\, \mathrm{meV},\eqlab{numresultsbetaH3}\\
E^{\left\langle(Z\al)^6,\;\al_{E1},\,\beta_{M1}\right\rangle}_\text{LS}(\mu^3\text{He}^{+})&=&-1.395\pm0.047\, \mathrm{meV},\eqlab{numresultsbetaHe3}\\
E^{\left\langle(Z\al)^6,\;\al_{E1},\,\beta_{M1}\right\rangle}_\text{LS}(\mu^4\text{He}^{+})&=&-0.665\pm0.016\, \mathrm{meV}.\eqlab{numresultsbetaHe4}
\eea
\end{subequations}
For the light nuclei, the contribution of the magnetic polarizability to \Eqref{numresultsbeta} is less than $1.5\,\%$. The biggest effect ($7.5\,\%$) is observed in $\mu$H, as one could expect from the ratio of dipole polarizabilities, $\al_{E1}/\be_{M1}$, which is biggest for the proton.

\subsection{Comparison to Long-Range Polarization Potentials} \seclab{LRP}
In order to avoid possible double counting, we need to identify effects in the theory of hydrogen-like atoms which already include part of the off-forward TPE polarizability contribution presented above.

Using only the leading terms of $\im \mathscr M(t)$, \Eqref{tauExp},
and the unsubtracted equivalent of \Eqref{potentialTPEoff},
\beq
V(r)=\frac{1}{4\pi^2 r}\int_0^\infty \dd t \, \im \mathscr M(t)\,e^{-r\sqrt{t}},\quad \eqlab{unsubtractedCoorPot}
\eeq
we deduce the following coordinate-space potential:\footnote{In momentum-space, this corresponds to:
\beq
\mathscr M(p_t^2)=V(\vert \boldsymbol{p_t}\vert)=\frac{\pi^2 \al \, \al_{E1}}{2} \vert \boldsymbol{p_t}\vert+\dots.
\eeq}
\beq
V_\text{l.r.}(r)=-\frac{\al\, \al_{E1}}{2r^4}\left[1-\frac{11}{2 \pi  m r}\right]+\frac{5\,\al\, \be_{M1}}{4\pi m r^5}. \eqlab{lr}
\eeq
The first term of the potential is the well-known effective long-range polarization potential \cite{Ericson:1973sz}. It is an attractive potential which falls off rapidly outside the nucleus and, hence, primarily overlaps with $S$-waves. Since the perturbation of $S$-levels due to a potential of type $V(r) \propto r^{-4}$ is divergent \cite{BetheSalpeter},
\beq
\overline{r^{-4}}=\langle nl\vert r^{-4}\vert nl\rangle=\frac{(Z\al m_r)^4 \left[3n^2-l(l+1)\right]}{2n^5(l+\nicefrac{3}{2})(l+1)(l+\nicefrac{1}{2})l(l-\nicefrac{1}{2})}\,,
\eeq
its evaluation for $l=0$ requires a cut-off \cite{Friar:1977cf}. Our treatment of the nuclear potential, however, requires no such cutoff.

In the short-range limit, the nuclear potential from Eqs.~\eref{unsubtractedCoorPot} and \eref{test} becomes:
\beq
V_\text{s.r.}=\frac{\al  m}{2\pi r^3}\left\{-2\al_{E1}+(\al_{E1}-\be_{M1})[\gamma_E+\ln mr]\right\}, \eqlab{sr}
\eeq
where $\gamma_E$ is the Euler-Mascheroni constant. In Figures \ref{fig:LRPotential} and \ref{fig:SRPotential}, we show our nuclear potential, $V_\mathrm{exact}$, as derived from \Eqref{test} (black solid line), the short-range potential from \Eqref{sr} (blue dotted line), the long-range potential from \Eqref{lr} (red short dashed and orange long dashed lines) and the Coulomb potential (green dash-dotted lines) for $\mu$D. The short- and long-range limits give a good approximation of the potential in the respective regions.

\begin{figure}[tbh]
\centering
\includegraphics[width=11.5cm]{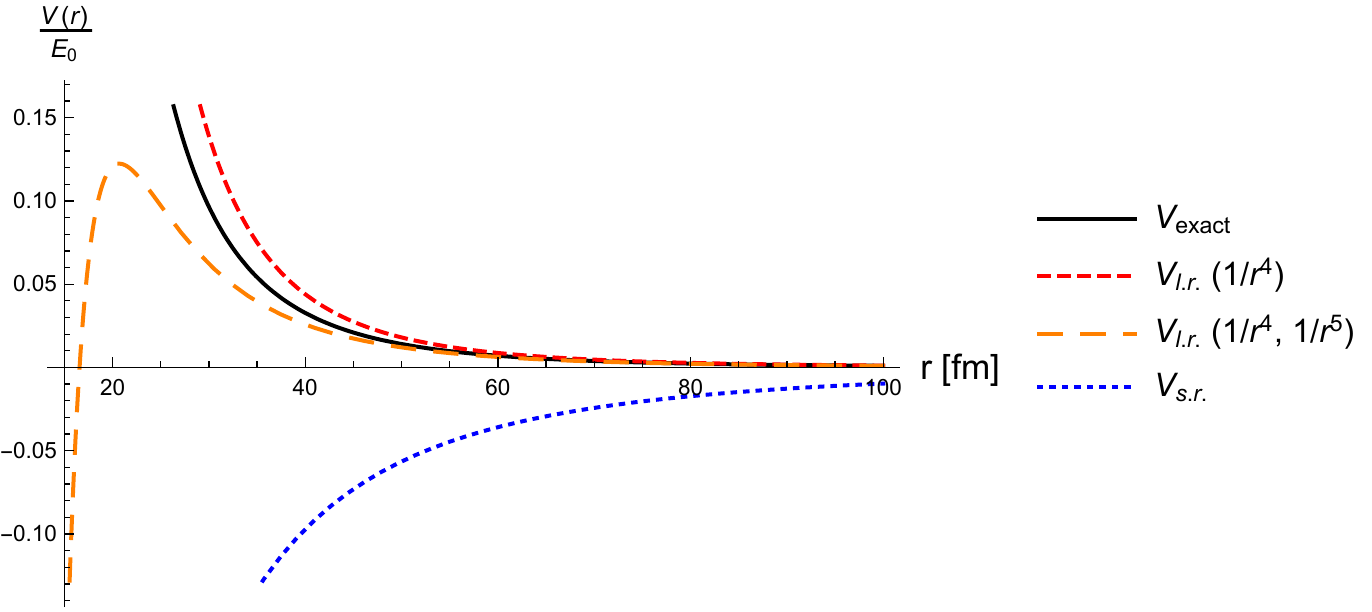}
\caption{Nuclear potential in the long-range limit normalized to the muonic deuterium ground-state energy.} 
\figlab{LRPotential}
\end{figure}

\begin{figure}[tbh]
\centering
\includegraphics[width=11.5cm]{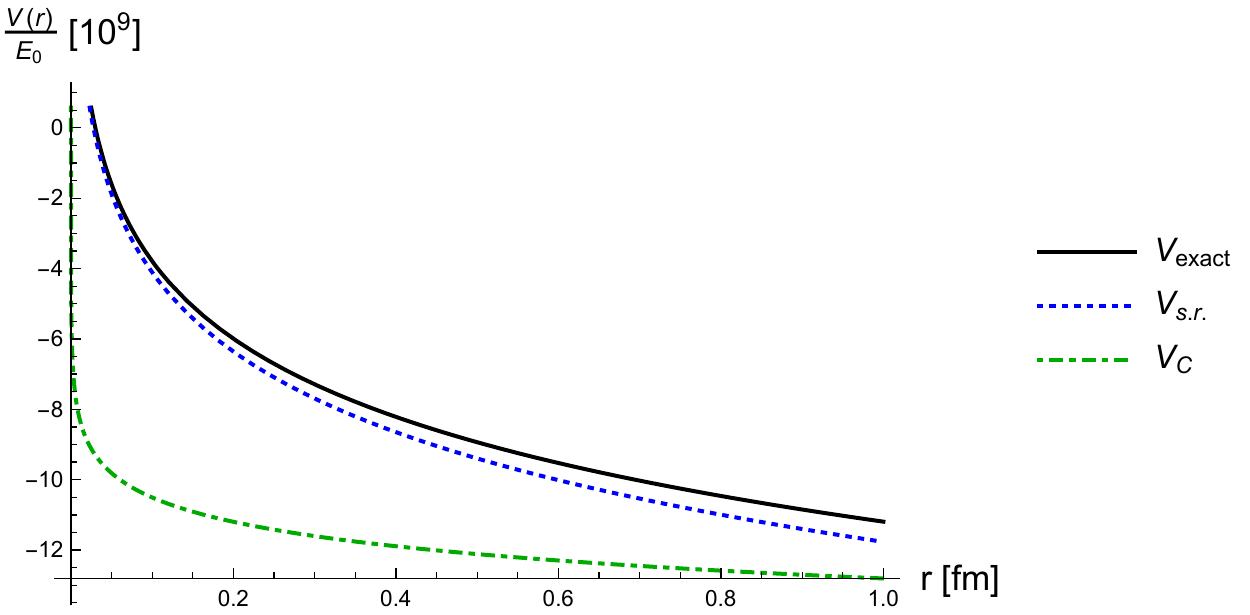}
\caption{Nuclear potential in the short-range limit  normalized to the muonic deuterium ground-state energy.} 
\figlab{SRPotential}
\end{figure}

In Ref.~\cite{Friar:1977cf}, the nuclear polarization potential is studied and the Coulomb-distortion effects are estimated in the unretarded dipole approximation, which is supposed to set an upper limit on the effects. The dominant nuclear polarization stems from virtual dipole excitations, and is given by the electric dipole polarizability:
\beq
\al_{E1} = \frac{2Z^2 \al}{3} \int_{E_\mathrm{th}}\!\frac{\dd E}{E} \,\big\vert \langle \phi_0\vert\boldsymbol{d} \vert E \rangle\big\vert^2,
\eeq
where $E$ is the nuclear excitation energy, $\boldsymbol{d}$ is the nuclear dipole operator, $\vert \phi_0 \rangle$ is the ground state and $\vert E \rangle$ is an excited state of the nucleus with energy $E$. Equivalently, the electric dipole polarizability can be expressed through a sum rule:
\beq
\al_{E1} =\frac{1}{2\pi^2} \int \dd \nu\, \frac{\sigma_\gamma^\mathrm{ud}(\nu)}{\nu^2},
\eeq
where $\sigma_\gamma^\mathrm{ud}(\nu)$ is the cross section for photoabsorption of unretarded-dipole (long-wavelength) photons by the nucleus \cite{Friar:1997jy}. The mean excitation energy, $\ol E$, is defined by a similar logarithmic sum rule \cite{Friar:1977cf}:
\beq
\al_{E1}\,\ln \ol E =\frac{2Z^2 \al}{3} \int_{E_\mathrm{th}}\!\frac{\dd E}{E} \,\big\vert \langle \phi_0\vert\boldsymbol{d} \vert E \rangle\big\vert^2 \ln E.\\
\eeq
In general, the nuclear excitation energy is much larger than a typical atomic excitation energy.\footnote{Refs.~\cite{Friar:1997jy, Friar:2005je} calculate the scalar, vector and tensor polarizabilities of the deuteron, as well as its mean excitation energy, based on different nucleon-nucleon potential models.}  Using the notation from Ref.~\cite{Pachucki:2011xr}, the LO and NLO Coulomb-distortion effects on the $2P_{1/2}-2S_{1/2}$ LS read:
\begin{subequations}
\bea
\delta_{\mathrm{C}1}&=&\frac{(Z\al)^6 m_r^4}{6} \int_{E_\mathrm{th}}\!\frac{\dd E}{E} \,\big\vert \langle \phi_0D\vert\boldsymbol{d} \vert E \rangle\big\vert^2 \left[\frac{1}{6}+\ln \frac{2m_r (Z\al)^2}{E}\right],\\
&=&\frac{\al (Z \al m_r)^4}{4}\, \al_{E1} \left[\frac{1}{6}+\ln \frac{2m_r (Z\al)^2}{\ol E}\right], \eqlab{deltaC1Coulomb}\\
\delta_{\mathrm{C}2}&=&-\frac{(Z\al)^7 m_r^{9/2}}{6\sqrt{2}}\left[\frac{19}{8}+\frac{\pi^2}{3}\right]\int_{E_\mathrm{th}}\!\frac{\dd E}{E^{3/2}} \,\big\vert \langle \phi_0\vert\boldsymbol{d} \vert E \rangle\big\vert^2.
\eea
\end{subequations}
Obviously, the Coulomb-distortion is a subleading polarizability effect, starting at order $(Z\al)^6$, and proportional to the nuclear electric dipole polarizability.
As one can see from the squared matrix elements, the Coulomb-distortion is arising at second order in PT. It is thus equivalent to our TPE effect in first-order PT. The logarithmic enhancement is similar to our expression in \Eqref{finalresult}. 

Our calculation of the off-forward TPE relies on an expansion in energies. We assume that the energies in the atomic bound state are small compared to the binding energy of the nucleus. As one can see from Table \ref{Tab:results}, the nuclear binding energies are typically of the order of MeV. The atomic binding energies are roughly proportional to $\sim Z \al m_r$. For electronic atoms, this is a keV scale. For the light muonic atoms, i.e., hydrogen and helium isotopes, it should be $\sim1$ MeV, and hence, supporting our assumption that the atomic energies are small compared to the nuclear binding energies.
Provided the LEX of the CS process is a good approximation, and we can write the off-forward VVCS amplitude as a Born part plus the contribution of dipole polarizabilities, the results presented here and in Section~\ref{chap:chap5}.\ref{sec:6.3} can be considered as an alternative to the order-$(Z\al)^6$ Coulomb-distortion effect. As compared to \Eqref{deltaC1Coulomb}, our numerical results also take the the magnetic dipole polarizability into account, cf.\ \Eqref{numresultsbeta}.

In the literature, the leading $(Z\al)^6 \ln (Z\al)$ Coulomb-distortion correction, $\delta_C^{(0)}$, to the $2P_{1/2}-2S_{1/2}$ LS is estimated as (for different nuclear potentials):\footnote{\Eqref{C0He4} also includes subleading Coulomb-distortion effects.}
\begin{subequations}
\eqlab{Cdist}
\bea
&&\hspace{-0.4cm}\delta_\mathrm{C}^{(0)}(\mu^2\text{H})=\begin{cases}-0.262\, \text{meV [AV18] \cite{Hernandez:2014pwa}},\\
-0.262\, \text{meV [N$^3$LO-EM] \cite{Hernandez:2014pwa}},\\
-(0.262,0.264)\, \text{meV [N$^3$LO-EGM] \cite{Hernandez:2014pwa}}, \end{cases}\\
&&\hspace{-0.4cm}\delta_\mathrm{C}^{(0)}(\mu^3\text{H})=\begin{cases}-0.0718(1)\, \text{meV [AV18/UIX] \cite{Dinur:2015vzv},}\\
-0.0732(0)\, \text{meV [$\chi$EFT] \cite{Dinur:2015vzv},} \end{cases}\qquad\;\\
&&\hspace{-0.4cm}\delta_\mathrm{C}^{(0)}(\mu^3\text{He}^{+})=\begin{cases}-1.000(01)\, \text{meV [AV18/UIX] \cite{Dinur:2015vzv}},\\
-1.020(3)\, \text{meV [$\chi$EFT] \cite{Dinur:2015vzv}}, \end{cases}\qquad\;\\
&&\hspace{-0.4cm}\delta_\mathrm{C}^{(0)}(\mu^4\text{He}^+)=
\begin{cases}-0.512\, \text{meV [AV18/UIX] \cite{Ji:2013oba}},\\
-0.546\, \text{meV [$\chi$EFT] \cite{Ji:2013oba}}, \end{cases}\eqlab{C0He4}\qquad\;
\eea
\end{subequations}
while their full prediction of the nuclear-polarizability contribution, $\delta_\text{pol}^\text{A}$, to the LS is \cite{Hernandez:2016jlh}:
\begin{subequations}
\bea
\delta_\text{pol}^\text{A}(\mu^2\text{H})&=&1.245(19)\, \text{meV},\\
\delta_\text{pol}^\text{A}(\mu^3\text{H})&=&0.473(17)\, \mathrm{meV},\\
\delta_\text{pol}^\text{A}(\mu^3\text{He}^{+})&=&4.17(17) \, \mathrm{meV},\\
\delta_\text{pol}^\text{A}(\mu^4\text{He}^{+})&=&2.36(14) \, \mathrm{meV}.
\eea
\end{subequations}
The $(Z\al)^6 \ln Z\al$ polarizability contributions to the LSs in muonic deuterium, tritium, helium-3 and helium-4 from off-forward TPE, see Table \ref{Tab:results}, are a factor of $1.2\div 1.5$ bigger than the presently accounted for Coulomb-distortion effects, cf.\ \Eqref{Cdist}. The differences are at the level of accuracy of the present LS theories, cf.\ \Eqref{theorypredictions}. 
We therefore conclude that the off-forward TPE is not negligible and it is worth to take it into account upon evaluating the muonic-atom experiments.  In the next Section, we will extract nuclear charge radii from the muonic spectroscopy measurements. Our extraction will be based on an updated theoretical description of the LSs, including the off-forward TPE polarizability effects from \Eqref{numresultsbeta} and the forward TPE polarizability effect in \Eqref{LSfinalvalue}.

\section{Extraction of the Nuclear Charge Radii from Spectroscopy} \seclab{chap6.3}
The theoretical predictions for the $2P_{1/2}-2S_{1/2}$ splittings in $\mu$H, $\mu$D and $\mu^4\text{He}^{+}$ have been compiled in Refs.~\cite{Antognini:2012ofa,Krauth:2015nja,Diepold:2016cxv} (in units of meV):
\begin{subequations}
\eqlab{theorypredictions}
\bea
&&\hspace{-1.1cm}E_{\mathrm{LS}}^{\mathrm{th.}}(\mu\text{H})=206.0668(25)-5.2275(10)\left(\nicefrac{R_{Ep}}{\mathrm{fm}}\right)^2\!\!,\eqlab{HydrogenTheory}\\
&&\hspace{-1.1cm}E_{\mathrm{LS}}^{\mathrm{th.}}(\mu\text{D})=230.486(20)-6.1103(3)\left(\nicefrac{R_{Ed}}{\mathrm{fm}}\right)^2\!\!,\eqlab{DeuteronTheory}\\
&&\hspace{-1.1cm}E^{\mathrm{th.}}_\text{LS}(\mu^4\text{He}^{+})=1678.544(205)-106.358(7)\left(\nicefrac{R_{E\al}}{\mathrm{fm}}\right)^2\!\!, \eqlab{He4Theory}
\eea
\end{subequations}
where $R_{Ep}$, $R_{Ed}$ and $R_{E\al}$ are the proton, deuteron and $\al$-particle charge radii, respectively. Equation \eref{HydrogenTheory} includes the order-$\al^5$ proton-polarizability effect, but no Coulomb-distortion effects. Equation~\eref{DeuteronTheory}, cf.\ Ref.~\cite[Table 3]{Krauth:2015nja}, presently accounts for a Coulomb-distortion effect of $\delta_\mathrm{C}^{(0)}(\mu \text{D})=-0.2625(15)\,\mathrm{meV}$, what also includes the next order, i.e., order $(Z\al)^7$, Coulomb-distortion effect of $\delta_{\mathrm{C}2}(\mu \text{D})=-0.006\,\mathrm{meV}$ \cite{Pachucki:2015uga,Pachucki:2011xr}. \Eqref{He4Theory} uses  the Coulomb distortion from Ref.~\cite{Ji:2013oba}, cf.\ Ref.~\cite[Table IV]{Diepold:2016cxv} and \Eqref{C0He4}. We assume the average values $\delta_\mathrm{C}^{(0)}(\mu^4\text{He}^+)=-0.529\,\mathrm{meV}$ and $\delta_{\mathrm{C}2}(\mu^4\text{He}^+)=-0.0065\,\mathrm{meV}$.

The theory budget of the $\mu$H LS, we improve by including the off-forward TPE polarizability contribution from \Eqref{numresultsbetaH}. Furthermore, we substitute the included forward TPE polarizability effect, $E^\mathrm{pol}_\mathrm{LS}=0.0085(11)$ meV \cite{Carlson:2011zd}, with our prediction, cf.\ \Eqref{LSfinalvalue}. The updated version of \Eqref{HydrogenTheory} then reads (in meV):
\beq
E_{\mathrm{LS}}^{\mathrm{th.}}(\mu\text{H})=206.0631\left(^{+30}_{-26}\right)-5.2275(10)\left(\nicefrac{R_{Ep}}{\mathrm{fm}}\right)^2\!\!.\eqlab{newHtheory}
\eeq
Together with the experimental value of the $\mu$H LS
\cite{Antognini:1900ns}, \Eqref{expmuHLS}, we extract the proton rms charge radius as:
\beq
R_{Ep}=0.84045(44)\,\text{fm}.\eqlab{Renew}
\eeq
The value quoted in \Eqref{remuH} \cite{Antognini:1900ns} is within our error. The discrepancy to the CODATA recommended proton charge radius, \Eqref{REpCodata}, increases from 5.6 to $5.7\,\sigma$.

Let us now turn to $\mu$D. Substituting the Coulomb-distortion with our estimate for the off-forward TPE polarizability effect, \Eqref{numresultsbetaD}, the theory budget of the $\mu$D LS becomes (in meV):
\beq
E_{\mathrm{LS}}^{\mathrm{th.}}(\mu\text{D})=230.344(20)-6.1103(3)\,\left(\nicefrac{R_{Ed}}{\mathrm{fm}}\right)^2\!\!.\eqlab{newDtheory}
\eeq
Comparing the theory prediction with the measured LS in $\mu$D \cite{Pohl1:2016xoo} (in meV):
\beq
E_{\mathrm{LS}}^{\mathrm{exp.}}(\mu\text{D})=202.8785(31)_\mathrm{stat}(14)_\mathrm{syst}\,\text{meV},
\eeq
we deduce the deuteron charge radius as:
\beq
R_{Ed}=2.12013(78)\,\text{fm}. \eqlab{REdnew} 
\eeq
In the case of $\mu$D, the radius puzzle between $\mu$D and D spectroscopy, Eqs.~\eref{rmuD} and \eref{Dspec}, gets slightly worse. Thus, using our input, the discrepancy amounts to $4.7\,\sigma$.

Analogously to the case of $\mu$D, we re-evaluate the $\mu^4$He$^+$-theory budget with the contribution given in Eq.~\eref{numresultsbetaHe4}. We arrive at (in meV):
\beq
E^{\mathrm{th.}}_\text{LS}(\mu^4\text{He}^{+})=1678.402(206)-106.358(7)\left(\nicefrac{R_{E\al}}{\mathrm{fm}}\right)^2\!\!.\eqlab{newHe4theory}
\eeq
A determination of the $\al$-particle charge radius will be possible once the LS is experimentally measured.

 \section{Summary and Conclusion}

In this Chapter, we have studied the influence of forward and off-forward TPE on the classic LS in hydrogen-like atoms. We have presented the BChPT prediction for the order-$\al^5$ proton-polarizability contribution to the $\mu$H LS, \Eqref{LSfinalvalue}, and compared it to the results of HBChPT and dispersive calculations.  We have extended the LO BChPT
calculation of Ref.~\cite{Alarcon:2013cba} to the next order in the $\delta$-expansion. At this order, we considered chiral loops \cite{Alarcon:2013cba} and the $\Delta(1232)$-exchange, see Figures \ref{fig1} and \ref{fig:TPEDelta}. The $Q^2$ behavior of the pion-nucleon loops was improved by including the pion FF. The $Q^2$ behavior of the $\Delta$-exchange contribution is regularized with the help of Jones-Scadron FFs and their relations to the nucleon elastic FFs in the large-$N_c$ limit. 
As one can see from Fig.~\ref{ComparisonLS}, our BChPT calculation is in good agreement with the dispersive calculations
which are currently used as input for the $\mu$H-theory budget. 

In \secref{5LS}{offTPE}, we presented the off-forward TPE polarizability effect as a natural extension of the forward TPE effects. The relevant theory was established in \secref{chap5}{6.3}. It is basically an alternative assessment of  the Coulomb-distortion effects. We focused on the polarizability contribution to the LS in hydrogen-like atoms at order $(Z\al)^6 \ln Z\al$. This logarithmic contribution is generated by the $t$-channel cut enhancement in Fig.~\ref{TPEFRW}. The contribution of the magnetic dipole polarizability is observed to vanish, whereas the electric dipole polarizability induces a downwards shift of the $n$-th $S$-level, see \Eqref{finalresult} and Table \ref{Tab:results}. Since the nuclear TPE is at present the limiting factor in the charge radius extractions from experiments with light muonic atoms, we approximated the order-$(Z\al)^6$ polarizability contributions to the LSs in $\mu$H, $\mu$D, $\mu^3$H, $\mu^3\text{He}^+$ and $\mu^4\text{He}^+$, see \Eqref{numresultsbeta}.  Despite the $Z\al$ suppression compared to the LO polarizability contribution, the off-forward TPE polarizability effect of the nuclear e.m.\ dipole polarizabilities to the spectra of light muonic atoms is found to be comparable in size to the leading nuclear-polarizability effect. Since the proton polarizability is much smaller than a typical nuclear polarizability, the $(Z\al)^6 \ln Z\al$ effect in $\mu$H is too. Choosing the CM frame, we were by default free of retardation effects. Our only assumptions in the course of the order $(Z\al)^6 \ln Z\al$ calculation were the semi-relativistic expansion and the omission of recoil effects. Furthermore, in the order-$(Z\al)^6$ calculation, we only kept terms proportional to $1/Q^4$, i.e., soft photons. In the future, it would be interesting to repeat the off-forward TPE calculation without neglecting contributions from higher $Q^2$.

\noindent In \secref{5LS}{chap6.3}, we applied the energy shifts calculated in Sections \ref{chap:5LS}.\ref{sec:ExpInPolLS} and \ref{chap:5LS}.\ref{sec:offTPE} to the nuclear charge radius extractions from muonic spectroscopy experiments.  We updated the theory predictions for the LSs in $\mu$H, $\mu$D and $\mu^4\text{He}^+$, cf.\ Eqs.\ \eref{newHtheory}, \eref{newDtheory} and \eref{newHe4theory}, and re-evaluated the proton and deuteron rms charge radii, cf.\ Eqs.~\eref{Renew} and \eref{REdnew}. While the proton charge radius remains basically unchanged, the deuteron shrinks further.
\newpage
\clearpage

\thispagestyle{empty}

\begin{sidewaystable}
\caption{Summary of our numerical results for the $(Z\al)^6 \ln (Z\al)$ polarizability contribution to the Lamb shift of light muonic atoms. \label{Tab:results}}
  \renewcommand{\arraystretch}{1.75}
\begin{scriptsize}
\begin{tabular}{|c|c|c|c|c|c|}
\hline
 \rowcolor[gray]{.7}
&{\bf $\boldsymbol{\mu}$H}&{\bf $\boldsymbol{\mu^2}\text{H}$}&{\bf $\boldsymbol{\mu^3}\text{H}$}&{\bf $\boldsymbol{\mu^3}\text{He}\boldsymbol{^+}$}&{\bf $\boldsymbol{\mu^4}\text{He}\boldsymbol{^+}$}\\
\hline
nucl.\ mass $M$ [GeV]&$0.938\,272\, 0813(58)\,$\cite{Mohr:2015ccw}&$1.875\,612\,928(12)\,$\cite{Mohr:2015ccw}&$2.808\,921\,112(17)\,$\cite{Mohr:2015ccw}&$2.808\,391\,586(17)\,$\cite{Mohr:2015ccw}&$3.727\,379\, 378(23)\,$\cite{Mohr:2015ccw}\\
 \rowcolor[gray]{.95}
nucl.\ pol.\ $\al_{E1}$ [fm$^3$]&$11.2(0.4)\times10^{-4}\,$\cite{Olive:2016xmw}&$0.6314(19)\;\text{\cite{Phillips:1999hh}}$&$0.139(2)\,$\cite{Stetcu:2009py}&$0.149(5)\,$\cite{Stetcu:2009py}&$0.0683(8)(14)\,$\cite{Stetcu:2009py}\\
nucl.\ binding en.&&&&&$28.422\,\text{\cite[AV18/UIX]{Ji:2013oba}}$\\
$E_B\,\text{[MeV]}$&\multirow{-2}{*}{$144.7\,$ [pion production \eref{PionProductionThreshold}]}&\multirow{-2}{*}{$2.224573(2)\,$\cite{Augid:1997ee}}&\multirow{-2}{*}{$8.481798(2)\,$\cite{Purcell:2010hka}}&\multirow{-2}{*}{$7.718043(2)\,$\cite{Purcell:2010hka}}&$28.343\,\text{\cite[$\chi$EFT]{Ji:2013oba}}$\\
 \rowcolor[gray]{.95}
&&&&&\\
 \rowcolor[gray]{.95}
\multirow{-2}{*}{nucl.\ pol.\ effect [meV] }&\multirow{-2}{*}{$E_\text{LS}^{\left\langle(Z\al)^5\right\rangle}=4.9^{+2.0}_{-1.3}\times10^{-3}$}&\multirow{-2}{*}{$\delta_\text{pol}^\text{A}=1.239(5)\,\text{\cite{Hernandez:2014pwa}}$}&\multirow{-2}{*}{$\delta_\text{pol}^\text{A}=0.476(10)(13)\, \text{\cite{Dinur:2015vzv}}$}&\multirow{-2}{*}{$\delta_\text{pol}^\text{A}=4.16(06)(16) \,\text{\cite{Dinur:2015vzv}}$}&\multirow{-2}{*}{$\delta_\text{pol}^\text{A}= 2.47(15) \,\text{\cite{Ji:2013oba}}$}\\
this work:&&&&&\\
$E_{LS}^{\left\langle(Z\al)^6 \ln Z\al\right\rangle}$ [meV]&\multirow{-2}{*}{$-0.79(3)\times 10^{-3}\;$ \eref{result}}&\multirow{-2}{*}{$-0.541(2)$ \eref{resultD}}&\multirow{-2}{*}{$-0.128(2)\;$\eref{result3H}}&\multirow{-2}{*}{$-1.950(65)\;$\eref{result3He}}&\multirow{-2}{*}{$-0.925(22)\;$ \eref{result4He}}\\
 \rowcolor[gray]{.95}
this work: &&&&&\\
 \rowcolor[gray]{.95}
$E^{\left\langle(Z\al)^6,\;\al_{E1},\,\beta_{M1}\right\rangle}_\text{LS}$ [MeV]&\multirow{-2}{*}{$-0.128(5)\times 10^{-3}\;$\eref{numresultsbetaH}}&\multirow{-2}{*}{$-0.398(1)\;$\eref{numresultsbetaD}}&\multirow{-2}{*}{$-0.095(1)\;$\eref{numresultsbetaH3}}&\multirow{-2}{*}{$-1.395(47)\;$\eref{numresultsbetaHe3}}&\multirow{-2}{*}{$-0.665(16)\;$\eref{numresultsbetaHe4}}\\
\hline
\end{tabular}
\end{scriptsize}
\end{sidewaystable}
  \renewcommand{\arraystretch}{1.3}
\chapter{Hyperfine Splitting in Chiral Perturbation Theory} \chaplab{5HFS}

We now turn to the TPE effects in the HFSs of hydrogen-like muonic atoms. As in the previous Chapter, we start with the NLO BChPT prediction for the proton-polarizability contribution at order $\al^5$ (\secref{5HFS}{ExpInPolHFS}). For one thing, our prediction is based on BChPT and a finite momentum transfer extension of the large-$N_c$ relations for Jones-Scadron FFs (\secref{5HFS}{deltaExHFS}). For another thing, we present a model which uses the $\Delta$-pole contribution and an elastic FF piece instead of the $\Delta$-exchange contribution (\secref{5HFS}{Depole}). We briefly compare our findings to indications from HBChPT (\secref{5HFS}{HBcompHFS}). Also, we compare to dispersive calculations based on empirical input (\secref{5HFS}{ComparisonHFS}). Here, the focus is set on the low-$Q$ region (\secref{5HFS}{lowQ}). 

We intensify the analyses of our results in \secref{5HFS}{polexpHFS}. Based on the polarizability expansion derived in \secref{chap5}{PolExpTPE}, we compare the contribution of individual polarizabilities to the HFS as implied by: 1) the Simula parametrization of spin-dependent structure functions \cite{Simula:2001iy,Simula:2002tv}, 2) the MAID model \cite{MAID}, and 3) our predictions.

For the first time, we present a model-independent calculation of the neutron-polarizability contributions to the HFSs of light muonic atoms, e.g., $\mu$D and $\mu^3$He$^+$ (\secref{5HFS}{polHFSneutron}). The size of the neutron-polarizability effect is then compared to the proton-polarizability effect.

We are also interested in contributions to the HFS from off-forward TPE. In \secref{5HFS}{neutralPion}, we derive the neutral-pion exchange and evaluate its contribution to the hydrogen HFSs \cite{Hagelstein:2015b,Hagelstein:2015egb}. In \appref{5HFS}{6.5}, we explain why there is no nuclear-polarizability contribution at order $(Z\al)^6 \ln Z \al$.

We will conclude the Chapter by extracting a new value of the proton Zemach radius based on our predictions for the polarizability effects in the $2S$ HFS of $\mu$H (\secref{5HFS}{newRZ}).

\section[Proton-Polarizability Contribution at Order $\al^5$]{Proton-Polarizability Contribution at Order $\boldsymbol{\al^5}$}\seclab{ExpInPolHFS}

\renewcommand{\arraystretch}{1.5}
\begin{table}[htb]
\caption{Summary of our numerical results for the order-$\al^5$ proton-polarizability contribution to the $2S$ hyperfine splitting in muonic hydrogen from forward two-photon exchange. \label{Tab:results1}}
\centering
\begin{small}
\begin{tabular}{|c|c|c|c|}
\hline
 \rowcolor[gray]{.7}
{\bf forward TPE}&$\boldsymbol{\Delta_1}${\bf [ppm]}&$\boldsymbol{\Delta_2}${\bf [ppm]}&{\bf $\boldsymbol{E_\text{HFS}(2S,\mu\text{H})}$ [$\boldsymbol{\upmu}$eV]}\\
\hline
Chiral loops (Fig.~\ref{fig1})&$-18$&$8$&$-0.23\,^{+1.08}_{-0.23}$\\
 \rowcolor[gray]{.95}
$\Delta$-exchange (Fig.~\ref{fig:TPEDelta})&$55$&$-106$&$-1.15\pm1.15$\\
\hline
combined result&$37$&$-97$&$-1.39^{\,+1.58}_{\,-1.17}$\\
\hline
\end{tabular}
\end{small}
\end{table}
\renewcommand{\arraystretch}{1.3}

In what follows, the NLO BChPT prediction for the order-$\al^5$ proton-polarizability contribution to the HFS in $\mu$H is presented \cite{Hagelstein:2015b,Hagelstein:2015egb}. This Section is very much analogue to \secref{5LS}{ExpInPolLS}, where we presented results for the forward TPE effects on the $\mu$H LS. The results for the proton-polarizability contribution to the $2S$ HFS are summarized in Table \ref{Tab:results1}. The combined effect of chiral loops and $\Delta(1232)$-exchange on the HFS of the $n$-th $S$-level in $\mu$H amounts to:
\begin{subequations}
 \eqlab{HFSfinalvalue}
\beq
E^\mathrm{pol.}_\mathrm{HFS}(nS, \mu\text{H})=-11.1\,^{+12.7}_{-9.4}\,\frac{\upmu\text{eV}}{n^3} \eqlab{HFSfinalvalueN}\
\eeq
For the relevant $1S$ and $2S$ HFSs, that is:
\bea
E^\mathrm{pol.}_\mathrm{HFS}(1S, \mu\text{H})&=&\left[6.71-17.79\right]\,\upmu\text{eV}=-11.1\,^{+12.7}_{-9.4}\,\upmu\text{eV},\\
E^\mathrm{pol.}_\mathrm{HFS}(2S, \mu\text{H})&=&\left[0.84-2.22\right]\,\upmu\text{eV}=-1.4\,^{+1.6}_{-1.2}\,\upmu\text{eV}. \eqlab{HFS2Sfinal}
\eea
\end{subequations}

The elastic TPE corrections, Eqs.~\eref{ZemachTerm} and \eref{recoilHFS}, are calculated as described in Ref.~\cite{Carlson:2008ke} and include the two-loop recoil correction from Refs.~\cite{Bodwin:1987mj} and the radiative corrections given in Refs.~\cite{Bodwin:1987mj,Karshenboim:1996ew}:
\begin{subequations}
\eqlab{elasticTPEnum}
\bea
\Delta_\mathrm{Z}(\mu\text{H})&=&-7628\pm 149\,\text{ppm},\\
\Delta_\mathrm{recoil}(\mu\text{H})&=&929\pm 10\,\text{ppm}.
\eea
\end{subequations}
As for the large-$N_c$ description of the Jones-Scadron FFs, we used the elastic FF parametrization of \citet{Bradford:2006yz}. In this way, we hope to minimize our error due to insufficient cancelation between the elastic and polarizability effects. The error was estimated by comparing to the selection of FF parametrizations applied in Ref.~\cite{Carlson:2008ke}. Together with the Zemach radius contribution and the recoil effects, \Eqref{elasticTPEnum}, the total TPE effect amounts to:
\begin{subequations}
\bea
E^\mathrm{TPE}_\mathrm{HFS}(1S, \mu\text{H})&=&-1.233\,^{+0.030}_{-0.029}\,\text{meV},\\
E^\mathrm{TPE}_\mathrm{HFS}(2S, \mu\text{H})&=&-0.1541\,^{+0.004}_{-0.004}\,\text{meV}.
\eea
\end{subequations}
This compares to the Fermi energy:
\begin{subequations}
\eqlab{Fermi12}
\bea
E_\mathrm{F}(1S, \mu\text{H})&=&182.4468 \text{ meV},\\
E_\mathrm{F}(2S, \mu\text{H})&=&22.8058 \text{ meV}.
\eea
\end{subequations}

\subsection{Chiral Loops} \seclab{ChiralLoopsHFSP}
\begin{figure}[htb]
\centering
\includegraphics[width=13cm]{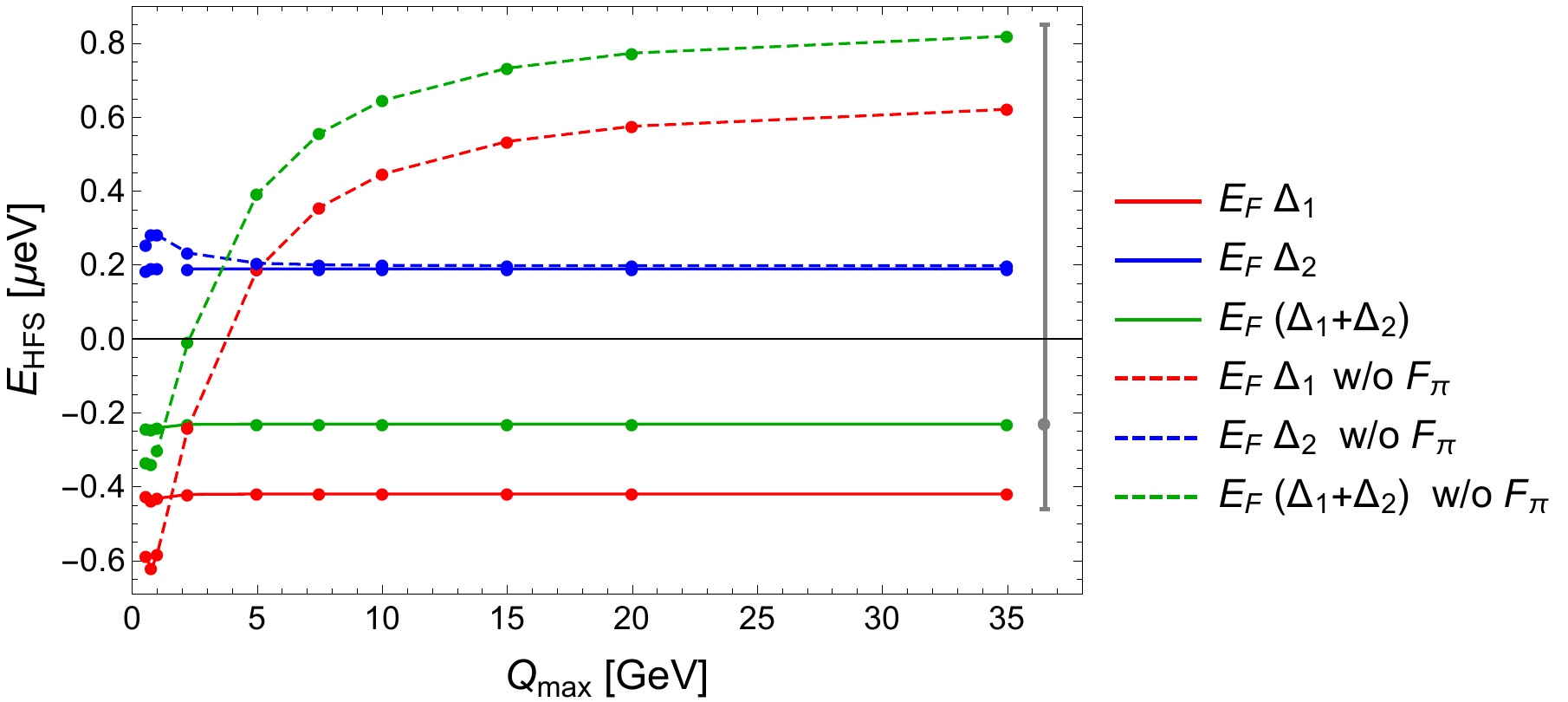}
\caption{Proton-polarizability effect in the $2S$ hyperfine splitting: Cutoff-dependence of the $\pi N$-loop contribution. Our result, \Eqref{pionresultHFS2S}, is indicated by the gray error band.}
\label{fig1piP}
\end{figure}
The contribution of the leading chiral loops can be calculated from \Eqref{POL} with the $\pi N$-production cross sections \cite{Lensky:2014dda} or from \Eqref{HFSS1S2} with the $\pi N$-loop spin-dependent VVCS amplitudes. As it turns out, the result is compatible with zero:
\begin{subequations}
\eqlab{pionresultHFS}
\bea
E_\mathrm{HFS}^{{\langle \pi N\rangle}\,\mathrm{pol.}}(1S, \mu\text{H})&=&\left[-3.36+1.51\right]\,\upmu\text{eV}=-1.84^{\,+8.65}_{\,-1.84}\,\upmu\text{eV}, \\
E_\mathrm{HFS}^{{\langle \pi N\rangle}\,\mathrm{pol.}}(2S, \mu\text{H})&=&\left[-0.42+0.19\right]\,\upmu\text{eV}=-0.23^{\,+1.08}_{\,-0.23}\,\upmu\text{eV}.\eqlab{pionresultHFS2S}
\eea
\end{subequations}
Here, the first number gives the contribution of the $S_1$ VVCS amplitude and the second number gives the contribution of the $S_2$ VVCS amplitude, see first row of Table \ref{Tab:results1}.
Since the contribution is numerically small and, hence, indicating a cancellation of LO contributions, we assign an error of $100\,\%$. In addition, we increase the upper error to incorporate the change upon including the pion FF, cf.\ \Eqref{PionFF}. 

Figure~\ref{fig1piP} shows the dependence of \Eqref{pionresultHFS2S} on the upper limit of the $Q$ integration in \Eqref{POL}. In contrast to the LS,  the HFS result is strongly cut-off dependent unless the pion FF is included. With pion FF, the contribution from beyond the scale of ChPT applicablity ($Q>m_\rho\approx 775$ MeV) is small ($\sim7\,\%$).

\subsection[$\Delta$-Exchange]{$\boldsymbol{\Delta}$-Exchange} \seclab{deltaExHFS}

While the contribution of the magnetic dipole polarizability to the LS is expected to be small, and likewise is the contribution of the $\Delta(1232)$-exchange, this is not the case for the HFS. To the contrary, we find an effect of: 
\begin{subequations}
\eqlab{DeExHFS}
\bea
E_{\text{HFS}}^{{\langle\Delta\text{-exch.}\rangle}\,\mathrm{pol.}}(1S, \mu\text{H})&=&\left[10.07-19.30\right]\,\upmu\text{eV}=-9.23\pm 9.23\, \upmu\mathrm{eV},\\
E_{\text{HFS}}^{{\langle\Delta\text{-exch.}\rangle}\,\mathrm{pol.}}(2S, \mu\text{H})&=&\left[1.26-2.41\right]\,\upmu\text{eV}=-1.15\pm 1.15\,\upmu\mathrm{eV},
\eea
\end{subequations}
which is certainly relevant in comparison to the leading chiral loops, cf.\ \Eqref{pionresultHFS}. 

Here, we used the large-$N_c$ relations given in Eqs.~\eref{GM*old2}, \eref{GM*old3} and \eref{GM*} with $C_M^*=\frac{3.02}{\sqrt{2}\,\kappa_p}$ and the nucleon FF parametrization of \citet{Bradford:2006yz}.
Individual contributions, e.g., the $\Delta$-pole and non-pole parts, are given in Table \ref{nonpoleHFS}, where we cross-checked the numerical evaluation by calculating the TPE through either the nucleon structure functions or the CS amplitudes. The non-pole contribution to $\Delta_2$ is negligible because \Eqref{S2npg} gives a purely BC-like contribution and only \Eqref{S2npgp2} remains. 

Again, we verified that the contribution from $Q<m_\rho$ is small ($5\,\%$) and that the result is not sensitive to the choice of a nucleon FF parametrization. In fact, using the dipole and Galster FFs leads to a less than $1\,\%$ change.

\renewcommand{\arraystretch}{1.5}
\begin{table} [htb]
\centering
\begin{small}
\caption{$\Delta$-exchange contribution to the $2S$ hyperfine splitting in muonic hydrogen. All values in $\upmu\mathrm{eV}$. \label{nonpoleHFS}}
\centering
\begin{tabular}{|c|c|c|c|c|}
\hline
 \rowcolor[gray]{.7}
{\bf Eq.}&{\bf Input}&$\boldsymbol{E_\mathrm{HFS}(\Delta_1)}$&$\boldsymbol{E_\mathrm{HFS}(\Delta_2)}$&$\boldsymbol{E_\mathrm{HFS}(2S,\mu\mathrm{H})}$\\
\hline
\eref{fullHFS}&$g_i$ \eref{structurefunc}&$-38.27$&$50.32$&$12.05$\\
\rowcolor[gray]{.95}
\eref{ExpBC}&BC sum rule only, $g_2$ \eref{g2Delta}&/&$52.75$&$52.75$\\
\eref{POL}&$g_i$ \eref{structurefunc}&$-38.27$&$-2.43$&$-40.69$\\
\rowcolor[gray]{.95}
\eref{HFSS1S2} & $S_i^{\Delta\mathrm{-pole}}$\eref{Dpole}&$-38.27$&$-2.43$&$-40.69$\\
\hdashline
\eref{fullHFS}&$\widetilde g_i$ \eref{npstrucfunc} \eref{S2npg} \eref{S2npgp2}&$39.53$&$-52.73$&$-13.21$\\
\rowcolor[gray]{.95}
\eref{ExpBC}&BC sum rule only, $\widetilde g_2$ \eref{S2npg}&/&$-52.75$&$-52.75$\\
\eref{POL}&$\widetilde g_i$ \eref{npstrucfunc} \eref{S2npg} \eref{S2npgp2}&$39.53$&$0.02$&$39.54$\\
\rowcolor[gray]{.95}
\eref{HFSS1S2} & $\widetilde S_i$ \eref{noDpole}&$39.53$&$0.02$&$39.54$\\
\hline
\eref{HFSS1S2} & $S_i$ \eref{DeltaAmps}&$1.26$&$-2.41$&$-1.15$\\
\hline
\end{tabular}
\end{small}
\end{table}
\renewcommand{\arraystretch}{1.3}

\renewcommand{\arraystretch}{1.5}
\begin{table} [h!]
\centering
\begin{small}
\caption{$\Delta$-exchange contribution of different multipole ratios to the $2S$ hyperfine splitting in muonic hydrogen, cf.\ \Eqref{POL}. All values in $\upmu\mathrm{eV}$. \label{multipoleHFS}}
\centering
\begin{tabular}{|l|c|c|c|}
\hline
 \rowcolor[gray]{.7}
{\bf multipoles}&$\boldsymbol{E_\mathrm{HFS}(\Delta_1)}$&$\boldsymbol{E_\mathrm{HFS}(\Delta_2)}$&$\boldsymbol{E_\mathrm{HFS}(2S,\mu\mathrm{H})}$\\
\hline
$G_M^{*2}$&$1.85$&$-2.28$&$-0.43$\\
 \rowcolor[gray]{.95}
$G_M^{*2}R_\mathrm{EM}$&$-0.12$&$-0.21$&$-0.32$\\
$G_M^{*2}R_\mathrm{SM}$&$-0.36$&$0.09$&$-0.28$\\
 \rowcolor[gray]{.95}
$G_M^{*2}R_\mathrm{EM}^2$&$-0.02$&$-0.01$&$-0.03$\\
$G_M^{*2}R_\mathrm{EM}R_\mathrm{SM}$&$-0.07$&$-0.02$&$-0.09$\\
 \rowcolor[gray]{.95}
$G_M^{*2}R_\mathrm{SM}^2$&$-0.03$&$0.02$&$-0.01$\\
\hline
total&$1.26$&$-2.41$&$-1.15$\\
\hline
\end{tabular}
\end{small}
\end{table}
\renewcommand{\arraystretch}{1.3}

\subsection[$\Delta$-Pole Model]{$\boldsymbol{\Delta}$-Pole Model}\seclab{Depole}

In the following Section, we introduce a model for the order-$\al^5$ polarizability effect to the HFS. The formalism for the polarizability contribution to the HFS from forward TPE is given with \Eqref{POL}. In \Eqref{Delta1b}, the subtraction function of the $\ol S_1$ DR, cf.\ \Eqref{S1subtrDR}, is isolated. This subtraction function is proportional to:
\beq
\ol S_1(0,Q^2)\propto \left[F_2^2(Q^2)+4I_1(Q^2)/Z^2\right], \eqlab{S1subol}
\eeq 
what vanishes in the real photon limit. Recalling \Eqref{genGDHnonpole}, we can understand that the non-polarizability part of the generalized $I_1$ integral is canceled by the Pauli FF squared. This cancelation is crucial for the dispersive calculations of TPE effects, which are based on empirical structure functions and FFs. The ChPT approach, however, can work around it. Our BChPT prediction for the polarizability effect, comprising the TPE with chiral loops (\secref{5HFS}{ChiralLoopsHFSP}) and $\Delta$-exchange (\secref{5HFS}{deltaExHFS}), only uses non-Born diagrams as input, and hence, is by definition of pure polarizability type. Therefore,  \Eqref{HFSfinalvalue} ignores the Pauli FF in \Eqref{POL}. 

As explained in Sections \ref{chap:chap4}.\ref{sec:Jones} and \ref{chap:chap4}.\ref{sec:DeltaPol}, the nucleon-to-delta transition is predominantly a magnetic-dipole transition, thus, can be in good approximation described by the magnetic Jones-Scadron FF. This attribute is reflected in Fig.~\ref{fig:I12} and Tables \ref{multipolesLS} and \ref{multipoleHFS}. According to \Eqref{GM*}, the magnetic Jones-Scadron FF is related to the nucleon Pauli FF by a large-$N_c$ relation. These observations and the fact that \Eqref{S1subol} is vanishing for $Q^2=0$ make us believe that the $\ol S_1$ subtraction function should be comparable to zero over the whole $Q^2$ range. In \secref{5HFS}{lowQ}, we study $\ol S_1(0,Q^2)$ based on empirical parametrizations for the nucleon structure functions and based on our BChPT structure functions. Confirming our presumption, the BChPT calculation yields a numerically small negative contribution from the $\ol S_1$ subtraction function to the HFS, cf.\ \Eqref{subtractionD1}. In the following, we present a model which fixes $\ol S_1(0,Q^2)=0$. It will provide an upper bound on the polarizability contribution to the HFS.

At the $\Delta$-resonance position, the spin-dependent structure functions, Eqs.~\eref{g1Delta}-\eref{g2Delta}, roughly equal
\Eqref{approxg1}, where we neglected $R_\mathrm{SM}$ and $R_\mathrm{EM}^2$, and otherwise used the static value $R_\mathrm{EM}(0)$. Together with the modified large-$N_c$ relation in \Eqref{GM*new}, we achieve a perfect cancelation of the $\ol S_1$ subtraction function, see \Figref{I1Plot}, and the polarizability contribution to the HFS reads:
\bea
\frac{E_\mathrm{HFS}^{{\langle\Delta\text{-pole.}\rangle}\,\mathrm{pol.}}(nS)}{E_\mathrm{F}(nS)}&=&-\frac{Z \al m}{\pi(1+\kappa)M} \int_0^\infty  \dd Q\,Q\,\left[\nu_\Delta(1+v_l)\left(1+\sqrt{1+Q^2 \nu_\Delta^{-2}}\right)\right]^{-2}\times\quad\eqlab{DeltaPoleModelFormula}\qquad\\
&&\times\left[1+\frac{2v_l\sqrt{1+Q^2 \nu_\Delta^{-2}}}{v_l+\sqrt{1+Q^2 \nu_\Delta^{-2}}}\right]F_2^2(Q^2).\qquad\nn
\eea
This formula is derived from \Eqref{POL} with the elastic Pauli FF of the proton and the $\Delta$-pole contribution to the spin-dependent proton structure functions, \Eqref{approxg1}.
Numerically, this amounts to:
 \begin{subequations}
\eqlab{DePoleHFS}
\bea
E_{\text{HFS}}^{{\langle\Delta\text{-pole.}\rangle}\,\mathrm{pol.}}(1S, \mu\text{H})&=&[8.42-11.93]\, \upmu\mathrm{eV}=-3.51\,^{+3.51}_{-5.72}\, \upmu\mathrm{eV},\\
E_{\text{HFS}}^{{\langle\Delta\text{-pole.}\rangle}\,\mathrm{pol.}}(2S, \mu\text{H})&=&[1.05-1.49]\, \upmu\mathrm{eV}=-0.44\,^{+0.44}_{-0.71}\, \upmu\mathrm{eV},
\eea
\end{subequations}
where the first number gives the contribution of the $g_1$ structure function and the second number gives the contribution of the $g_2$ structure function. The errors are chosen asymmetric and cover the $\Delta$-exchange contribution, \Eqref{DeExHFS}. Clearly, this is only a model. Nevertheless, these numbers can serve as a lower bound on the absolute magnitude of the  effect, cf.\ \Figref{gMdata}. Note that the region of $Q>m_\rho$ contributes with about $3\,\%$ to the result given in \Eqref{DePoleHFS}. 

In \Eqref{POL},  the contribution of the zeroth moment of $g_2$ is subtracted. Therefore, it is not crucial that our model satisfies the BC sum rule exactly. Nevertheless, we would like to point out that the $\Delta$-pole is indeed able to cancel part of the elastic FF contribution to the BC integral, as shown in \Figref{BCPlot}.

  \subsection{Comparison with Heavy Baryon Chiral Perturbation Theory}
 \seclab{HBcompHFS}
 \begin{figure}[tbh]
\centering
\includegraphics[scale=0.85]{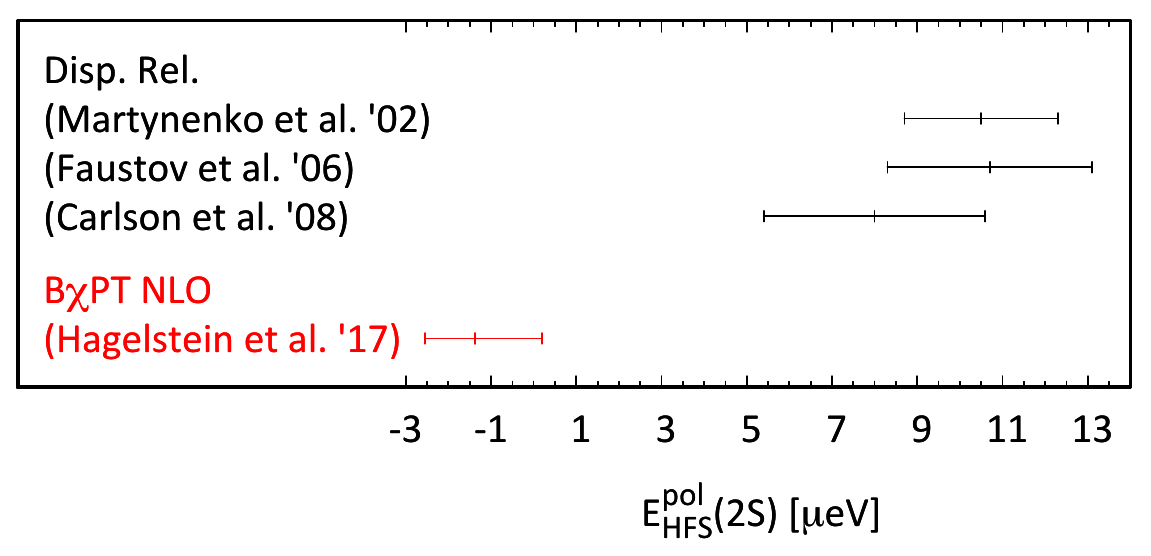}
\caption{Comparison of predictions for the order-$\al^5$ proton-polarizability contribution to the $2S$ hyperfine splitting in muonic hydrogen. The dispersive calculations are from Refs.~\cite{Carlson:2011af,Carlson:2008ke,Faustov:2006ve,Faustov:2001pn,Martynenko:2004bt}.}
\label{Comparison}
\end{figure}
 
 So far, there are only two quantitative predictions for TPE effects in the $\mu$H HFS derived within a model-independent framework, such as ChPT. One was presented in this thesis and Refs.~\cite{Hagelstein:2015b,Hagelstein:2015egb}, the other is from Pineda et al. \cite{Pineda:2003,Peset:2016wjq}. In Ref.~\cite{Pineda:2003}, the contribution of the (leading chiral) logarithms, $\mathcal{O}(m^3 \al^5/M^2\times\left[\ln m_q, \ln \Delta, \ln m\right])$, is calculated in HBChPT matched to potential NRQED. The almost analytical result given in there motivate the relative order of the Zemach and polarizability corrections. The non-Born contributions from pion-nucleon and pion-delta loops cancel each other in the large-$N_c$ limit, while the $\Delta$-exchange cancels part of the Zemach contribution \cite{Ji:1999mr,Pineda:2003}. An updated prediction for the complete TPE effect on the $\mu$H spectrum has recently been presented with Ref.~\cite{Peset:2016wjq}:
 \begin{subequations}
 \bea
   E^\mathrm{TPE}_\mathrm{HFS}(1S,\mu\text{H})&=&-1.161\pm0.020\,\text{meV},\\
 E^\mathrm{TPE}_\mathrm{HFS}(2S,\mu\text{H})&=&-0.1451\pm 0.025\,\text{meV}.
 \eea
 \end{subequations}
For the HFS, the BChPT and HBChPT predictions are closer than for the LS.
 
 \subsection{Comparison with Dispersive Calculations} \seclab{ComparisonHFS}
To judge the quality of our BChPT prediction, we compare with evaluations based on experimental data.  Dispersive calculations usually rely on empirical information for the spin structure functions, the elastic FFs and the proton polarizabilities. Some authors also work with the unitary isobar model and evolution equations for parton distribution functions \cite{Faustov:2006ve,Martynenko:2005rc}. 

Early works mainly studied the proton structure corrections to the HFS in H \cite{Iddings:1965zz, Drell:1966kk, DeRafael:1971mc,Gnaedig:1973qt}, for more recent works on the H HFS see Refs.~\cite{Faustov:2000xu,Faustov:2002yp,Martynenko:2005rc,Nazaryan:2005zc,Carlson:2006yj,Carlson:2006mw}. In Table \ref{Table:Summary3}, we summarize
the available dispersive calculations \cite{Carlson:2011af,Carlson:2008ke,Faustov:2006ve,Faustov:2001pn,Martynenko:2004bt} for the TPE corrections to the HFS in $\mu$H. $\Delta_\mathrm{Z}$, $\Delta_\mathrm{recoil}$ and $\Delta_\mathrm{pol.}$ are given in Eqs.~\eref{HFSwoEF} and \eref{FS_HFS}. 

In Fig.~\ref{Comparison}, our final number for the $\mu$H HFS of the $2S$-level, \Eqref{HFS2Sfinal}, as obtained at NLO in BChPT, is compared with the dispersive results from Table \ref{Table:Summary3}. As is apparent, the dispersive approach and the BChPT prediction  disagree by about $3.6\,\sigma$. There are a number of possible origins for the observed discrepancy between the different results for the polarizability effect. As for the prediction presented in here, we think it is fair to say that: assuming ChPT is working, it should be best applicable to atomic systems, where the energies are very small. Unfortunately, empirical information on the spin structure functions is limited (especially for $g_2$). New data from JLab should soon improve the situation in the important low-$Q$ region.

In the following Sections \ref{chap:5HFS}.\ref{sec:lowQ} and \ref{chap:5HFS}.\ref{sec:polexpHFS}, we will make more detailed comparisons based on our own dispersive analyses. 
 For one thing, we use the sum rule evaluations of MAID \cite{MAID} in combination with the FF parametrization of \citet{Bradford:2006yz}. For another thing, we use the parametrization of the proton structure functions $g_1(x,Q^2)$ and $g_2(x,Q^2)$ by Simula et al.\ \cite{Simula:2001iy}\footnote{We apply the parametrization for $x\in\{0,x_0\}$, however, it is suggested that the parametrization should be only trusted for $x \gtrsim 0.02$.} and the elastic FFs of \citet{Kelly:2004hm}, this choice agrees with Ref.~\cite{Nazaryan:2005zc}. We do not intend to derive a full dispersive analyses with error estimates, for this we trust, e.g., Ref.~\cite{Carlson:2008ke}. We merely try to get a handle on the low-$Q$ region, in particular the $\ol S_1(0,Q^2)$ subtraction function, and the contribution of individual spin polarizabilities to the HFS. The Simula parametrization was chosen because of its intuitive separation into resonance and background descriptions. Its drawback is that the latest low-$Q$ data are not included in the fit. Among other things, the MAID model provides useful output for generalized  nucleon polarizabilities below $Q^2<5$ GeV$^2$. More details of our dispersive analyses can be found in \appref{5HFS}{kappaMatching}. For now we quote the resulting 2S HFS in $\mu$H with:
\begin{subequations}
\eqlab{OurEmpRes}
\bea
E_{\mathrm{HFS}}^{\text{emp.}\,\text{pol.}}(2S, \mu \text{H})&=&\left[6.18-1.67\right]\, \upmu \text{eV}=4.51\, \upmu \text{eV  based on Simula $g_i$ \cite{Simula:2001iy}},\qquad\quad\\
E_{\mathrm{HFS}}^{\text{emp.}\,\text{pol.}}(2S, \mu \text{H})&=&\left[4.66-2.31\right]\, \upmu \text{eV}=2.35\, \upmu \text{eV based on MAID \cite{MAID}},
\eea
\end{subequations}
where the first number corresponds to the $g_1$ contribution and the second number corresponds to the $g_2$ contribution.

\renewcommand{\arraystretch}{1.5}
\begin{table}
\centering
\caption{Summary of available dispersive calculations for the two-photon-exchange corrections to the $2S$ hyperfine splitting in muonic hydrogen. 
}
\label{Table:Summary3}
\begin{minipage}{\linewidth}  
\footnotesize
\centering
\begin{tabular}{|p{0.10\linewidth}| p{.10\linewidth} |p{.07\linewidth}p{.05\linewidth}|p{.06\linewidth}| p{.07\linewidth}p{.06\linewidth}p{.06\linewidth}|p{.09\linewidth}p{.11\linewidth}|}
\hline                
 \rowcolor[gray]{.7}
\footnotesize {\bf Reference} &\footnotesize {\bf FF}&\footnotesize $\boldsymbol{R_\mathrm{Z}}$ &\footnotesize$\boldsymbol{\Delta_\mathrm{Z}}$&\footnotesize$\boldsymbol{\Delta_\mathrm{recoil}}$&\footnotesize$\boldsymbol{\Delta_\mathrm{pol.}}$&\footnotesize$\boldsymbol{\Delta_1}$&\footnotesize$\boldsymbol{\Delta_2}$&\footnotesize$\boldsymbol{\Delta_\mathrm{FSE}}$&\footnotesize$\boldsymbol{E_{\mathrm{HFS}}(2S)}$\\
 \rowcolor[gray]{.7}
&&\footnotesize{\bf [fm]}&\footnotesize{\bf [ppm]}&\footnotesize{\bf [ppm]}&\footnotesize{\bf [ppm]}&\footnotesize{\bf [ppm]}&\footnotesize{\bf [ppm]}&\footnotesize{\bf [ppm]}&\footnotesize{\bf [meV]}\\
\hline
\multirow{5}{*}{\parbox{3cm}{\footnotesize Carlson\\  et al.\ \\ \cite{Carlson:2008ke,Carlson:2011af}\footnote{QED, higher-order and other small corrections included in $E_{\mathrm{HFS}}(2S,\mu\text{H})$ are taken from Ref.~\cite{Martynenko:2004bt}. The Zemach term includes radiative corrections: $\Delta_\mathrm{Z}=-2\al m_r R_\mathrm{Z} (1+\delta^\mathrm{rad}_Z)$ with $\delta^\mathrm{rad}_\mathrm{Z}$ given in Refs.~\cite{Bodwin:1987mj,Karshenboim:1996ew}. Empirical information on structure functions and form factors are taken from Refs.~\cite{Dharmawardane:2006zd, Christy:2011, Arneodo1995107,Abe1999194, Simula:2001iy, Anthony200019}.}}}&\footnotesize AMT \cite{Arrington:2007ux}&\footnotesize$1.080$&\footnotesize$-7703$&\centering\footnotesize$931$&\footnotesize$351(114)$&\footnotesize$370(112)$&\footnotesize$-19(19)$&\footnotesize$-6421(140)$&\footnotesize$22.8123$\\
&\footnotesize AS \cite{Arrington:2006hm}&\footnotesize $1.091$&\footnotesize $-7782$&\centering\footnotesize $931$&\footnotesize$353$&&&\footnotesize$-6498$&\footnotesize$22.8105$\\
&\footnotesize Kelly \cite{Kelly:2004hm}&\footnotesize$1.069$&\footnotesize$-7622$&\centering\footnotesize$931$&\footnotesize$353$&&&\footnotesize$-6338$&\footnotesize$22.8141$\\
&\footnotesize MAMI \cite{Bernauer:2010wm, Vanderhaeghen:2010nd,Distler:2010zq}&\footnotesize$1.045$&&&&&&&\footnotesize$22.8187$\\
&\footnotesize combined\footnote{slightly moved average of the selected form factors}&&&&&&&&\footnotesize$22.8146(49)$\\
\hline
\rowcolor[gray]{.95}
&&&&&&&&&\\
 \rowcolor[gray]{.95}
\multirow{-2}{*}{\parbox{3cm}{\footnotesize Faustov\\  et al.~\cite{Faustov:2006ve}\footnote{The calculation is based on experimental data for the nucleon polarized structure functions obtained at SLAC, DESY and CERN \cite{Abe:1998wq, Abe:1996ag, Anthony:1999py, Mitchell:1999gw, Adams:1997tq, Adeva:1999pa}.}}}&&&&&\multirow{-2}{*}{\parbox{3cm}{\footnotesize$470(104)$}}&\multirow{-2}{*}{\parbox{.04\linewidth}{\footnotesize$518$}}&\multirow{-2}{*}{\parbox{.05\linewidth}{\footnotesize$-48$}}&&\\
\hline 
\multirow{2}{*}{\parbox{3cm}{\footnotesize
Martynenko \\et al.~\cite{Faustov:2001pn}\footnote{The calculation is based on experimental data for the nucleon polarized structure functions obtained at SLAC, DESY and CERN \cite{Abe:1998wq, Abe:1996ag, Anthony:1999py, Mitchell:1999gw, Adams:1997tq, Adeva:1999pa, Aidala:1999nc,Hughes:1995ef}.}}}&\multirow{2}{*}{\parbox{3cm}{\footnotesize Dipole}}&\multirow{2}{*}{\parbox{3cm}{\footnotesize$1.022$}}&\multirow{2}{*}{\parbox{3cm}{\footnotesize$-7180$}}&&\multirow{2}{*}{\parbox{3cm}{\footnotesize$460(80)$}}&\multirow{2}{*}{\parbox{3cm}{\footnotesize$514$}}&\multirow{2}{*}{\parbox{3cm}{\footnotesize$-58$}}&&\multirow{2}{*}{\parbox{3cm}{\footnotesize$22.8138(78)$\footnote{Adjusted value; as suggested in Ref.~\cite{Carlson:2011af}, the original value, $22.8148(78)\,\mathrm{meV}$, is corrected by adding $-1\,\upmu\mathrm{eV}$ because the conventions of ``elastic'' and ``inelastic'' contributions applied in Ref.~\cite{Martynenko:2004bt} are inconsistent.}}}\\
&&&&&&&&&\\
\hline
 \rowcolor[gray]{.95}
 &&&&&&&&&\\
  \rowcolor[gray]{.95}
\multirow{-2}{*}{\parbox{3cm}{\footnotesize Experiment\\ \cite{Antognini:1900ns}}}&&\multirow{-2}{*}{\parbox{3cm}{\footnotesize$1.082(37)$}}&&&&&&&\multirow{-2}{*}{\parbox{3cm}{\footnotesize$22.8089(51)$}}\\
\hline
\end{tabular}
\end{minipage}
\end{table}
\renewcommand{\arraystretch}{1.3}

\subsubsection[Importance of the low-$Q$ Region]{Importance of the low-$\boldsymbol{Q}$ Region} \seclab{lowQ}
In Table \ref{lowQcomCarlson}, we give a detailed comparison of our BChPT prediction to the dispersive calculation of Ref.~\cite[Table IV]{Carlson:2008ke}. The values quoted from Ref.~\cite{Carlson:2008ke} have statistical, systematic and modeling errors in parentheses. Values without any error specification were added by us based on the results given in Ref.~\cite{Carlson:2008ke} and the FF parametrization in \Eqref{AMT} \cite{Arrington:2007ux}. For our BChPT results, we do not assign errors to the individual $Q^2$-regions. Obviously, ChPT is supposed to work at low energies only. In Fig.~\ref{fig1piP}, one can for instance see that the $\ol S_1$ amplitude of the $\pi N$-loop CS diagram  is strongly sensitive to a cutoff at intermediate and high $Q$ if no pion FF is included.

From Table \ref{lowQcomCarlson}, it becomes obvious that BChPT and the dispersive calculation give a different weight to the $\Delta_1$ and $\Delta_2$ contributions. In BChPT, the value of $\Delta_1$ is about ten times smaller than in the dispersive calculation. Vice versa, $\Delta_2$ is about five times larger than in the dispersive calculation.
The latter is especially interesting since the available data set for $g_2$ is very small. The former is interesting because the dispersive approach shows a critical  cancelation between the $F_2$ term and the $g_1$ part.    Such cancelation was partially discussed in \secref{5HFS}{Depole} in view of the $\ol S_1$ subtraction function and we will come back to it in \secref{5HFS}{polexpHFS}. On the contrary, the ChPT approach can calculate the pure (non-Born) polarizability contribution to the TPE, in which no elastic FF appears. 

There are no CS or $ep$ data at the real photon point. Accordingly, both FF and structure function parametrizations require an interpolation to $Q^2=0$. In the very low-$Q$ region, the dispersive calculations therefore substitute empirical polarizabilities \cite{Carlson:2008ke}. For $Q \in \{0,Q_\mathrm{min}\}$, we expand the polarizabilities in Eqs.~\eref{polD1} and \eref{polD2} in $Q/M\ll 1$:
\begin{subequations}
\bea
\Delta_1^{\mathrm{low}\,Q}&\approx&\frac{Z\al m}{\pi(1+\kappa)M}\int_0^{Q_\mathrm{min}}\dd Q\,Q\,\frac{1}{(v_l+1)^2}\bigg\{2(5+4v_l)\Big[ \kappa F_2'(0)+2I_1'(0)\Big]\qquad\\
&&\qquad \qquad\qquad-\frac{11+9v_l}{1+v_l}\frac{M^2}{2\al}\gamma_0(0)\bigg\}\nn,\\
\Delta_2^{\mathrm{low}\,Q}&\approx&\frac{6Z mM}{\pi(1+\kappa)}\int_0^{Q_\mathrm{min}}\dd Q\,Q\,\frac{1}{(v_l+1)^2}\Big[\gamma_0(0)-\delta_{LT}(0)\Big].
\eea
\end{subequations}
In this way, we keep $v_l$ and the formulas are applicable for H and $\mu$H. In the same way, we can calculate the low-$Q$ effect of individual polarizabilities.

The derivatives of the generalized GDH integrals are related in the following way:
\beq
I_A'(0)=I_1'(0)+\frac{M^2}{2\al} \Big[\gamma_0(0)-\delta_{LT}(0)\Big].
\eeq
For the proton, we use $I_1'(0)=\left[7.6\pm2.5\right]\,\mathrm{GeV}^{-2}$ \cite{Prok:2008ev}, $\gamma_0(0)=\left[-0.93\pm0.06\right]\times10^{-4}\,\mathrm{fm}^4$ \cite{Gryniuk:2015aa} and $\delta_{LT}(0)=\left[1.34\pm0.02\right]\times10^{-4}\,\mathrm{fm}^4$ \cite{MAID} to obtain $I_A'(0)=\left[-1.42\pm2.5\right]\,\mathrm{GeV}^{-2}$. Similar to Ref.~\cite{Carlson:2008ke}, the upper limit is set to $Q_\mathrm{min}=0.0452\,$GeV$^2$, in the full knowledge that the Simula parametrization does not include the latest data at these low-$Q$ values. We then have:
\begin{subequations}
\bea
\Delta_1^{\mathrm{low}\,Q}(\mu\text{H})&=&\begin{cases}35.77\pm 31.48\,\text{ppm}&\text{FF \cite{Bradford:2006yz}},\\
38.07\pm 31.48 \,\text{ppm}&\text{FF \cite{Kelly:2004hm}},
\end{cases}\\
\Delta_2^{\mathrm{low}\,Q}(\mu\text{H})&=&-27.98\pm 0.78\,\text{ppm},
\eea
\end{subequations}
where the errors are propagated from the empirical polarizabilities.
Note that the low-$Q$ contribution to $\Delta_2$ given here is about five times larger than the value given by Ref.~\cite{Carlson:2008ke}, see Table \ref{lowQcomCarlson}, what brings it closer to our BChPT prediction.

In \secref{chap5}{PolExpTPE}, we presented a polarizability expansion up to and including second moments of the structure functions. In Table \ref{higherMom}, we show the contribution of higher moments to the HFS, i.e., effectively we expanded in $x/\tau$. This is a good approximation, as is confirmed by the MAID model, the Simula parametrization and the $\Delta$-exchange.

 \section{Hyperfine Splitting in Terms of Polarizabilities}\seclab{polexpHFS} 
Based on the polarizability expansion derived in \secref{chap5}{PolExpTPE}, we compare the contribution of individual polarizabilities to the HFS as implied by: 1) the Simula parametrization of spin-dependent structure functions \cite{Simula:2001iy,Simula:2002tv}, 2) the MAID model \cite{MAID}, and 3) our BChPT calculation. The results can be gathered from Table \ref{PolHFS}. In the first part, the polarizability decomposition  is performed according to Eqs.~\eref{polD1a} and \eref{polD2a}. In the second part, we use Eqs.~\eref{polD1b} and \eref{polD2b}. Note that the region of $Q\in \{0,\sqrt{0.0452}\,\text{GeV}\}$ was partially supplemented with empirical polarizabilities, see discussion in \secref{5HFS}{lowQ}. This is the case for the second moments of the predictions based on MAID or the Simula parametrization, i.e., not for the individual contributions from background and resonances, and not for higher moments. Furthermore, we modified the MAID prediction for the $I_1(Q^2)$ contribution by the procedure outlined in \appref{5HFS}{kappaMatching} to achieve agreement with the experimental value at the real photon point, see \Figref{I1Plot}.
  
 From \Eqref{polD1b}, one can isolate the contribution of the $\ol S_1(0,Q^2)$  subtraction term:
\beq
\eqlab{S1subtermHFS}
\frac{E_\mathrm{HFS}^{\langle S_1(0,Q^2)\rangle }(nS)}{E_\mathrm{F}(nS)}=
\frac{Z\al m}{\pi (1+\kappa) M}\int_0^\infty\frac{\dd Q}{Q}\frac{5+4v_l}{(v_l+1)^2}\left[4I_1(Q^2)+F_2^2(Q^2)\right].
\eeq
 It is given by the $F_2$ and $I_1$ entries in the second part of Table \ref{PolHFS}. Combining them, we have:
\beq
\eqlab{subtractionD1}
E_\mathrm{HFS}^{\langle S_1(0,Q^2)\rangle }(2S,\mu\text{H})
=\begin{cases} -0.49\,\upmu \mathrm{eV}&\text{BChPT},\\
3.35\,\upmu \mathrm{eV}&\text{MAID and FF~\cite{Bradford:2006yz}},\\
5.33\,\upmu \mathrm{eV}&\text{Simula param. \& FF~\cite{Kelly:2004hm}},\end{cases}
\eeq
where the MAID model forces an ultraviolet cutoff at $5\,$GeV$^2$ on the $Q$-integration.
From \Eqref{subtractionD1} and Table \ref{PolHFS} one can see that most of the discrepancy between the dispersive calculations and our BChPT prediction stems from the $\ol S_1(0,Q^2)$ subtraction function, while the contributions of $I_A$, $\delta_{LT}$, $\gamma_0$ and the fourth moment of $g_2$ agree very well. We therefore suspect that the dispersive calculations suffer from an inaccurate cancelation between $I_1(Q^2)$ and $F_2^2(Q^2)$.
  
 \section{Neutron-Polarizability Effect in Light Muonic Atoms} \seclab{polHFSneutron}
\begin{figure}[htb]
\centering
\includegraphics[width=13cm]{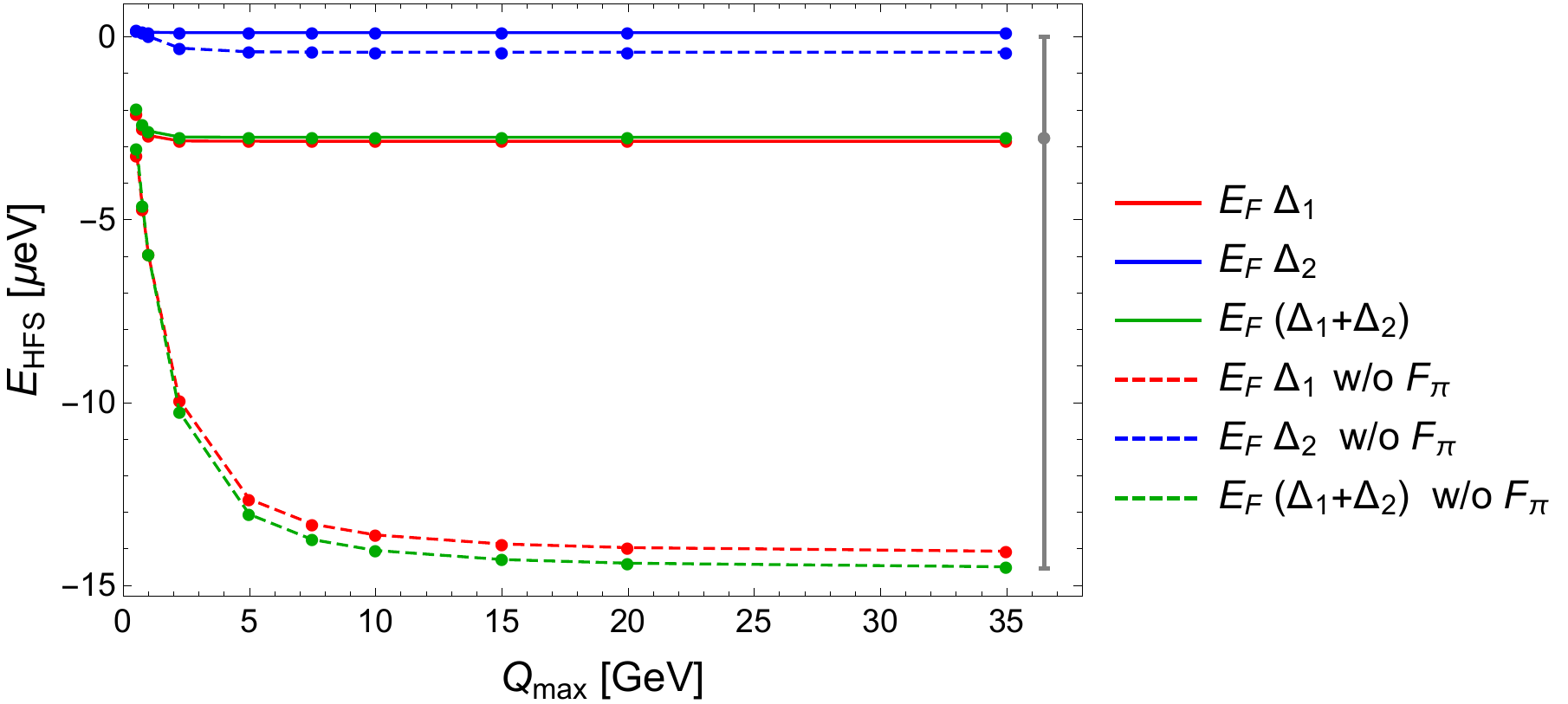}
\caption{Muon-neutron interaction in the $2S$ hyperfine splittings of muonic atoms: Cutoff-dependence of the $\pi N$-loop contribution to the $2S$ hyperfine splitting in muonic hydrogen. Our result, \Eqref{pionresultHFSn2S}, is indicated by the gray error band.}
\label{fig1pin}
\end{figure}
So far, we studied the order-$\al^5$ proton-polarizability effect, which is the main uncertainty in the theoretical description of the $\mu$H HFS. Similarly, we can calculate a neutron-polarizability effect, which is relevant for light muonic atoms.
At sufficiently large energies, cf.\ the binding energies in Table \ref{Tab:results}, a nucleus can break up into its constituents, i.e., $Z$ protons and $N$ neutrons. Accordingly, the polarizability effect in light muonic atoms comprises nuclear-polarizability, cf.\ \secref{5LS}{offTPE}, and nucleon-polarizability contributions. The intrinsic nucleon-polarizability contribution is given by the sum of polarizability contributions from each of the individual nucleons. Such effects have been calculated for the LSs in $\mu$D and $\mu^3$He$^+$ based on electron scattering data \cite{Carlson:2013xea,Carlson:2016cii}. Thus far, there has been no theoretical prediction for the neutron-polarizability effect. Theoretical studies of the LSs in $\mu$D, $\mu^3$H and $\mu^3$He$^+$ assume that the neutron-polarizability contribution can be approximated with the proton-polarizability contribution to $\mu$H \cite{Hernandez:2014pwa,Dinur:2015vzv}. In the following, we will provide a first model-independent prediction for the neutron-polarizability effect, as it enters the HFSs of light muonic atoms.

\noindent The intrinsic nucleon-polarizability contribution to the spectrum of a muonic atom with mass number $A$, also referred to as hadronic-polarizability contribution, is given by:
\beq
E^{\text{N-pol.}}(\mu\text{A})=\left[Z m_r(\mu\text{A})\right]^3 \left\{\frac{ZE^{p\text{-pol.}}(\mu\text{H})}{m_r(\mu\text{H})^3} +\frac{NE^{n\text{-pol.}}(\mu n)}{m_r(\mu n)^3}\right\},
\eeq
where $\mu n$ refers to the muon-neutron interaction within the atom and the factors of $m_r^3$ originate from the wave functions. In the following, we calculate the NLO BChPT prediction for the hadronic-polarizability contributions to the HFSs in muonic-hydrogen and muonic-helium isotopes. The NLO BChPT prediction for the order-$\al^5$ proton-polarizability contribution to the HFS in $\mu$H, denoted here as $E^{p\text{-pol.}}(\mu\text{H})$, was presented in  \secref{5HFS}{ExpInPolLS}. We are now left to study the muon-neutron interaction from forward TPE, i.e., $E^{n\text{-pol.}}(\mu n)$.

The $\Delta(1232)$-exchange mechanism is equivalent for proton and neutron, respectively, whereas the chiral loops need to be evaluated independently for the case of CS off the neutron \cite{Lensky:2014dda}.
Analogously to \secref{5HFS}{ChiralLoopsHFSP}, we obtain: 
\begin{subequations}
\eqlab{pionresultHFSn}
\bea
E_\mathrm{HFS}^{\langle \pi N\rangle}(1S,\mu n)&=&\left[-22.87+0.86\right]\upmu\text{eV}=-22.01^{\,+22.01}_{\,-94.38}\,\upmu\text{eV},\\
E_\mathrm{HFS}^{\langle \pi N\rangle}(2S,\mu n)&=&\left[-2.86+0.11\right]\upmu\text{eV}=-2.75^{\,+2.75}_{\,-11.80}\,\upmu\text{eV},\eqlab{pionresultHFSn2S}
\eea
\end{subequations}
where notations and errors are the same as in \Eqref{pionresultHFS}. The cut-off behavior is shown in Fig.~\ref{fig1pin}. Due to the inclusion of the pion FF, the contribution from beyond the ChPT scale is reasonably small ($\sim 13\,\%$). Assuming identical $\Delta$-production cross sections for photoabsorption off the proton and neutron, we rescale the proton-polarizability contribution in \Eqref{DeExHFS} by $\left[m_r(\mu n)/m_r(\mu\text{H})\right]^3 \left[M_n/M_p\right]^2$ and obtain the neutron-polarizability contribution as:
\begin{subequations}
\eqlab{DeExHFSn}
\bea
E_{\text{HFS}}^{\langle\Delta\text{-exch.}\rangle}(1S,\mu n)&=&[10.09-19.33]\, \upmu\mathrm{eV}=-9.25\pm 9.25\, \upmu\mathrm{eV},\\
E_{\text{HFS}}^{\langle\Delta\text{-exch.}\rangle}(2S,\mu n)&=&[1.26-2.42]\, \upmu\mathrm{eV}=-1.16\pm 1.16\, \upmu\mathrm{eV}.
\eea
\end{subequations}
Note that the correction factor stems from the coordinate-space wave function and the $S_i$ DRs.
In total, the neutron-polarizability contribution amounts to:
\begin{subequations}
\eqlab{neutronPolContr}
\bea
E_\mathrm{HFS}^{n\text{-pol.}}(1S,\mu n)&=&[-12.78-18.47]\,\upmu\text{eV}=-31.25\,^{+23.87}_{-94.83}\,\upmu\text{eV},\eqlab{neutronPolContr1S}\\
E_\mathrm{HFS}^{n\text{-pol.}}(2S,\mu n)&=&[-1.60-2.31]\,\upmu\text{eV}=-3.91\,^{+2.98}_{-11.86}\,\upmu\text{eV}.\eqlab{totalHFSn2S}
\eea
\end{subequations}
This is almost a factor of $3$ larger than the proton-polarizability contribution given in \Eqref{HFSfinalvalue}. We then find the first model-independent prediction for the hadronic-polarizability contributions to the $nS$ HFSs in muonic-hydrogen and muonic-helium isotopes:
\begin{subequations}
\bea
E^{\text{N-pol.}}_\mathrm{HFS}(nS, \mu\text{D})&=&-49\,^{+32}_{-111} \, \frac{\upmu\text{eV}}{n^3},\eqlab{DeuteiumNeutron}\\
E^{\text{N-pol.}}_\mathrm{HFS}(nS, \mu^3\text{H})&=&-91\,^{+61}_{-234}\, \frac{\upmu\text{eV}}{n^3},\\
E^{\text{N-pol.}}_\mathrm{HFS}(nS, \mu^3\text{He}^+)&=&-526\,^{+343}_{-953}\, \frac{\upmu\text{eV}}{n^3},\eqlab{Helium3Neutron}\\
E^{\text{N-pol.}}_\mathrm{HFS}(nS, \mu^4\text{He}^+)&=&-857\,^{+547}_{-1930}\, \frac{\upmu\text{eV}}{n^3}.
\eea
\end{subequations}

\section{Neutral-Pion-Exchange Contribution to the Hyperfine Splitting} \seclab{neutralPion}
\begin{figure}[h]
\centering
\includegraphics[scale=0.7]{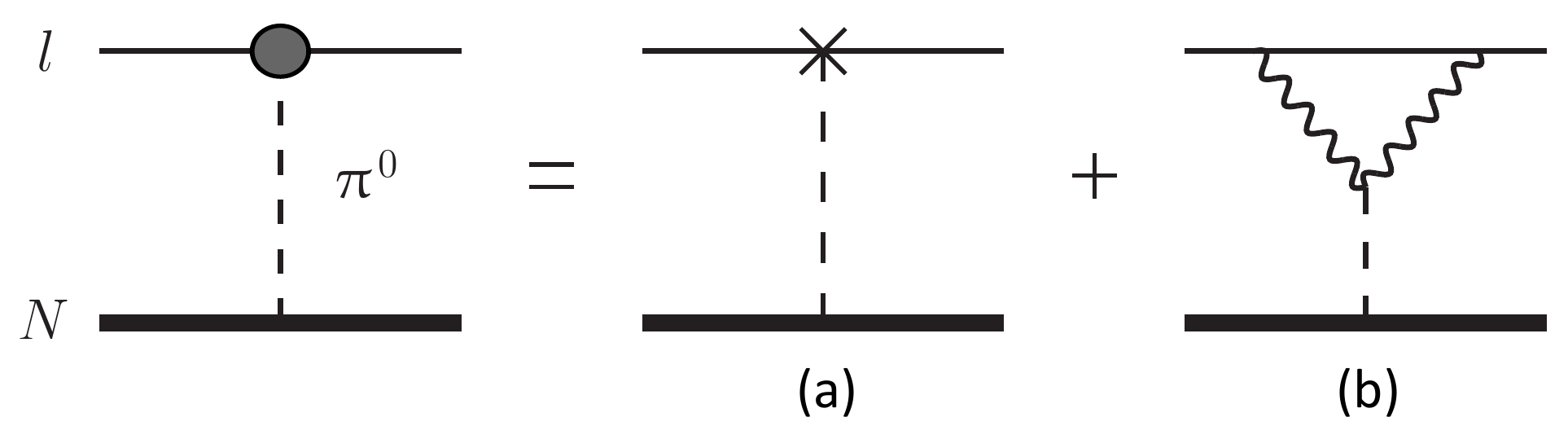}
\caption{Neutral-pion exchange in atomic bound states. }
\label{fig:PiExchange}
\end{figure}
\noindent The neutral-pion exchange between a lepton ($\ell$) and a nucleon ($N$) is shown in \Figref{PiExchange}. In Feynman diagram (a) the pion couples directly to the lepton, while in diagram (b) it couples through two photons.

We calculate the pion-exchange diagrams in the framework of ChPT. The pion coupling to the nucleon is described by the Lagrangian in \Eqref{piNN}. The coupling of the pion to the lepton is of pseudo-vector type, it can be described by the  Lagrangian:
\beq
\lag_{\pi \ell\ell} = -\frac{\al^2 g_{\pi \ell\ell} }{2m} \bar \ell \ga^\mu \ga_5 \ell \, \pa_\mu \pi^0, \eqlab{pseudovector}
\eeq
where $\ell$ is the lepton field and $m$ is the mass of the lepton.
Other relevant Feynman rules are given in \appref{chap4}{FRAppendix}.

\begin{figure}[b]
\centering
\includegraphics[width=13cm]{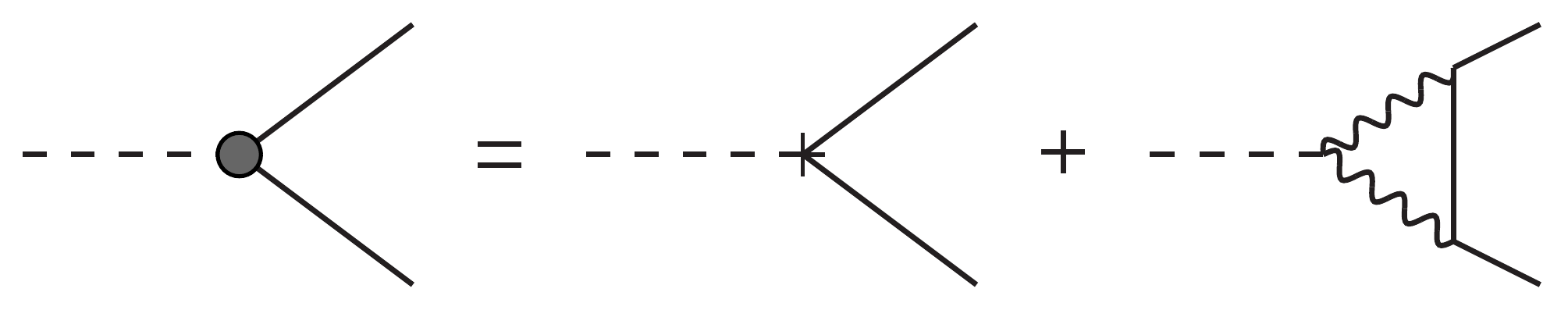}
\caption{Leading contributions to the $\pi \ell \ell$ interaction.}
\label{fig:Pi0Vertex}
\end{figure}

Let us first focus on diagram (a) and later generalize to include diagram (b).
In \chapref{chap2}, we derived the Breit potential from OPE. In the same fashion, we can get the pion-exchange potential. In momentum-space, we find:
\begin{subequations}
\eqlab{Pi0mom}
\begin{align}
 \hspace{-0.25cm}V_{\pi^0}(\bq) &=  \big(2E_k\, 2 E_{k'} \, 2E_p\,  2E_{p'}\big)^{-1/2}\;
\left[ \bar u(k') \, \Ga_{\pi\ell \ell}(q) \,u(k)\right] \frac{1}{q^2-m_\pi^2} \left[ \overline{N}(p') \,\Ga_{\pi NN}(q)\, N(p)\right]\nn,\\
&=\frac{\al^2 g_{\pi \ell \ell}\,g_A}{4mf_\pi}\left[ \bar{\mathpzc{u}}(k') \, \slashed{q}\gamma_5 \,\mathpzc{u}(k)\right]\frac{1}{q^2-m_\pi^2}  \left[ \overline{\mathpzc{N}}(p') \,\slashed{q}\gamma_5\, \mathpzc{N}(p)\right],\\
&=-\frac{\al^2 M g_{\pi \ell \ell}\,g_A}{f_\pi}\left[ \bar{\mathpzc{u}}(k') \, \gamma_5 \,\mathpzc{u}(k)\right]\frac{1}{q^2-m_\pi^2}  \left[ \overline{\mathpzc{N}}(p') \,\gamma_5\, \mathpzc{N}(p)\right],\\
&=-\frac{\al^2 g_{\pi \ell \ell}\,g_A}{m f_\pi} \frac{(\boldsymbol{S}\cdot \bq)(\boldsymbol{s}\cdot \bq)}{\boldsymbol{q}^2+m_\pi^2}  .
\end{align}
\end{subequations}
In the first step, we moved the energy prefactor into the Dirac spinors, as suggested in \appref{chap5}{SemRelExpDSpin}. In the second step, we used the Dirac equation. In the last step, we performed a semi-relativistic expansion of the Dirac spinors, substituted \Eqref{gamma5NN} and neglected retardation. The kinematics were chosen as in \appref{chap5}{SemRelExpDSpin}, with $q$ being the pion four-momentum.

The coordinate-space potential is obtained by a Fourier transformation, which we can look up in Table \ref{FTtab}. For the $S$-waves, we obtain:
\beq
V_{\pi^0}^{(l=0)}(r) =
-\frac{\al^2 g_{\pi \ell\ell} \, g_{A}}{12\pi m f_\pi} 
\Big(4\pi \de(\br) - \frac{m_\pi^2}{r} e^{-m_\pi r}\Big)
\bs\cdot \bS . \eqlab{Vpi0a}
\eeq
As one can read off, the potential has an effect on the HFS.

\begin{figure}[b]
\centering
\includegraphics[width=5.5cm]{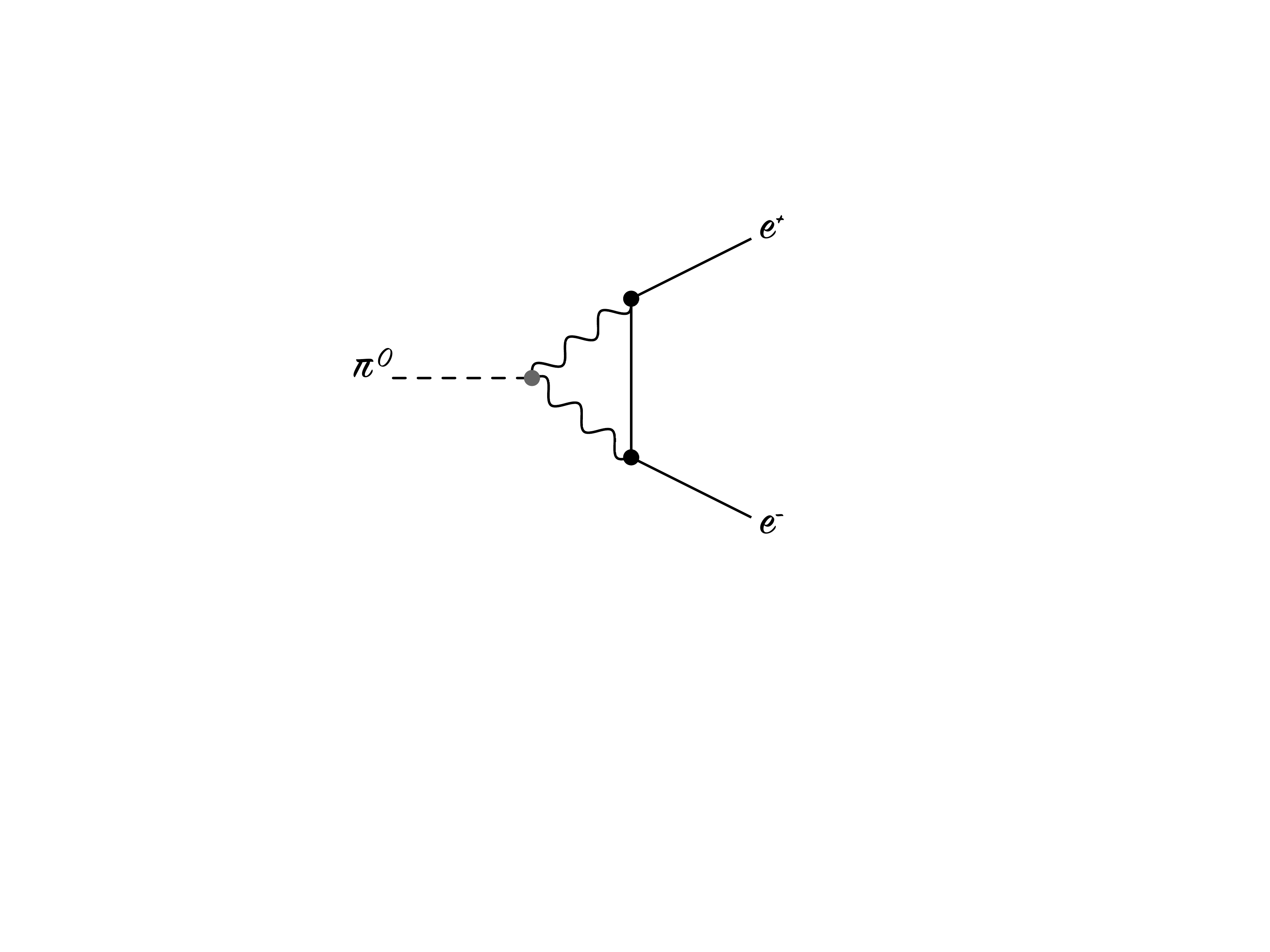}
\caption{Neutral-pion decay into an electron-positron pair. }
\label{fig:Pi0Decay}
\end{figure}

Let us now investigate the coupling of the pion to the lepton further. The leading contributions to the $\pi \ell \ell$ interaction are shown in \Figref{Pi0Vertex}. With the help of the Dirac equation, we can reduce the pseudo-vector interaction in \Eqref{pseudovector} to a pseudo-scalar interaction:
\beq
\Ga_{\pi \ell \ell}(q,p)=iF(q^2,p^2,p'^2)\gamma_5,
\eeq
where $q$ is the pion momentum, $p$ ($p'$) is the incoming (outgoing) lepton momentum and $F(0,m^2,m^2)=\al^2 g_{\pi \ell \ell}$. We want to extract the coupling strength from the experimentally measured decay of $\pi^0$ into an electron-positron pair, see \Figref{Pi0Decay}. The decay width is related to the $\pi\ell\ell$ vertex in the following way \cite{Griffiths}:
\beq
\Gamma(\pi^0\to e^+ e^-) = \frac{m_\pi}{8\pi} \sqrt{1-\frac{4m_e^2}{m_\pi^2}} \, \big| F(m_\pi^2,m_e^2,m_e^2) \big|^2.
\eeq
Calculating the diagram in \Figref{Pi0Decay} in dimensional regularization, one obtains \cite{Drell59,Ametller:1984uk}:
\begin{subequations}
\bea
 F(q^2)& \equiv & F(q^2,m^2,m^2)  = F(0) + \frac{q^2}{\pi} \int_0^\infty \!\dd s\,  \frac{\im F(s)}{(s-q^2) s}  \eqlab{Pion0FF}, \\
 && \im F(s) = -\frac{\al^2 m}{2\pi f_\pi} \frac{\arccosh(\sqrt{s}/2m)}{\sqrt{1-4m^2/s}}, \\
&& F(0)  =  \frac{\al^2 m}{2\pi^2 f_\pi} \Big[ \cA (\La) + 3\ln\frac{m}{\La} \Big],
\eea 
\end{subequations}
where $\La$ is the renormalization scale, and $\cA$ is a universal pion-lepton LEC,
related to the physical constant in an obvious way:
\beq
g_{\pi\ell\ell} = \frac{m}{2\pi^2 f_\pi} \cA (m).
\eeq

The lifetime of the neutral pion is \cite{Olive:2016xmw}:
\beq
\tau=\left[8.52\pm0.18\right]\times10^{-17} \,\mathrm{s},
\eeq
what corresponds to a decay width of:
\beq
\Gamma_\mathrm{tot}=\left[7.73\pm0.16\right]\,\mathrm{eV}.
\eeq
The dominant decay channel is $\pi^0\rightarrow \gamma \gamma$. In the literature, the fraction of decays into electron-position pairs is quoted with \cite{Olive:2016xmw}:
\beq
\frac{\Gamma(\pi^0\to e^+ e^-)}{\Gamma_\mathrm{tot}}=\left[6.46\pm0.33\right]\times 10^{-8},
\eeq
accordingly, the decay width of the leptonic channel amounts to:
\beq
\Gamma(\pi^0\to e^+ e^-)=\left[4.99\pm0.28\right]\times10^{-7}\,\mathrm{eV}.
\eeq
For the pion-lepton LEC, we then find:\footnote{In Refs.~\cite{Hagelstein:2015b,Hagelstein:2015egb}, we used an older value: $\cA(m_e)=-20(1)$.}
\begin{subequations}
\eqlab{pionleptonLEC}
\bea
\cA(m_e)&=&-22.7(6),\\
\cA(m_\mu)&=&\cA(m_e)+3 \ln \frac{m_\mu}{m_e}=-6.7(6),
\eea
\end{subequations}
and for the coupling constants:
\begin{subequations}
\eqlab{pillCouplings}
\bea
g_{\pi e e}&=&-0.00637(17),\\
g_{\pi \mu \mu}&=&\frac{m_\mu}{m_e}g_{\pi e e}+\frac{3m_\mu}{2\pi^2f_\pi} \ln \frac{m_\mu}{m_e}=-0.39(3).
\eea
\end{subequations}
Note that the coupling to the muon is stronger, due to its heavier mass.

We will now extend \Eqref{Pi0mom} by including diagram (b) of \Figref{PiExchange}, i.e., the part described by the dispersive integral in \Eqref{Pion0FF}. The momentum-space potential reads:
\beq
V_{\pi^0}(\bq) =-\frac{g_A}{m f_\pi}  
\bs\cdot \bq \,  \bS\cdot \bq  \left\{\frac{ \al^2 g_{\pi \ell\ell} }{\bq^2+m_\pi^2}-
\frac{1}{\pi} \int_0^\infty \! \frac{\dd s}{s}\,  \frac{ \im F(s)}{s-m_\pi^2} \bigg(  \frac{s}{\bq^2+s} - 
\frac{m_\pi^2}{\bq^2+m_\pi^2} \bigg)\right\},
\eeq
where we neglected retardation effects by setting $q_0=0$. After a Fourier transformation, we find the coordinate-space potential for $S$-waves:
\bea
V_{\pi^0}^{(l=0)}(r) &=&\bs\cdot \bS \,\Big\{
\frac{\al^2 g_{\pi \ell\ell} \, g_{A}}{12\pi m f_\pi} 
 \frac{m_\pi^2}{r} e^{-m_\pi r}\\
 &&+\frac{g_{A}}{12\pi^2 m f_\pi}\ \int_0^\infty \! \frac{\dd s}{s}\,  \frac{ \im F(s)}{s-m_\pi^2} \frac{1}{r}\left[m_\pi^4 e^{-m_\pi r}-s^2e^{-\sqrt{s} r}\right]\Big\}.\nn
\eea

In a last step, we calculate the effect on the $nS$ HFS from first-order PT.
For $a m_\pi \gg 1$, the $S$-wave matrix element of the Yukawa potential expands as:
\bea
\big< nS\big|  \nicefrac{ e^{-m_\pi r}}{r} \big| nS \big> &=& \frac{4\pi\phi_n^2 (0)}{m_\pi^2} 
\left[ 1 - \frac{4}{a m_\pi } + \cO\left(\nicefrac{1}{[nam_\pi]^{2}}\right) \right] ,
\eea
with $\phi_n^2 (0) = 1/(\pi a^3 n^3) $ the  wave function squared at the origin. 
In addition, we distinguishing  the cases $a \sqrt{s}\gg1$ and $s\ll m_\pi^2$, to derive an approximate formula for the HFS effect:
\beq
E_{\mathrm{HFS}}^{\langle\pi^0 \rangle}(nS)= - E_\mathrm{F}(nS)\,
\frac{g_A M m_r }{2\pi(1+\kappa) f_\pi m_{\pi}}\left[\al^2 g_{\pi \ell \ell}+\frac{\al^2 m}{2\pi^2 f_\pi}\, 
I\left(\frac{m_{\pi}}{2m}\right)\right],
\eqlab{pionHFS}
 \eeq
 where we introduce the following integral:
 \beq
 I(\gamma)\equiv 2 \int_0^\infty \!\frac{\dd \xi}{1+(\xi/\gamma)}\frac{\arccos \xi}{\sqrt{1-\xi^2}},
 \eeq
 and factored out the LO HFS, cf.\ \Eqref{FermiE}. As one can see, the neutral-pion-exchange effect is of order $\al^2(Z\al)^4$, with $(Z\al)^4$ embedded in the Fermi energy.

In the case of H, we have $\gamma\gg 1$ and
\beq
I(\gamma) = \frac{7\pi^2}{12} +\ln^2(2\gamma) -\frac{\pi}{\gamma} + \cO(1/\gamma^2),
\eeq
For the more general situation, $\gamma =\sin\th \ge 0$, we obtain:
\bea
I(\sin\th ) = \tan\th \left[ \mbox{Cl}_2(2\th) - \pi \ln \tan(\th/2) \right], 
\eea
where the Clausen integral is
\beq
\mbox{Cl}_2(\th) =- \int_0^\th \! \dd t \, \ln\big(2\sin\half t\big)
=\frac{ i}{2} \left[ \mbox{Li}_2\big(e^{-i\th}\big)-\mbox{Li}_2\big(e^{i\th}\big)\right],
\eeq
and Li$_2(x)$ is the dilogarithm. Numerical values for the electron and muon, respectively, are
\beq
 I( m_\pi/2m_e ) \simeq 36.8316, \qquad I ( m_\pi/2m_\mu ) \simeq 3.4634 .
 \eeq

In Table \ref{Tab:resultsPionEx}, we present numerical results for the $1S$ and $2S$ HFSs in H and $\mu$H, respectively. In both cases, there are a large cancellations
between the two terms in \Eqref{pionHFS}, or 
equivalently, between the two diagrams in \Figref{PiExchange}. The $F(0)$ part gives a positive contribution, while the $q^2$-dependent part gives a negative contribution. For the $2S$ HFS in $\mu$H, the individual terms in \Eqref{pionHFS} and the final HFS evaluate to:\footnote{Note that our numerical result changed due to the updated value for the pion-lepton LEC in \Eqref{pionleptonLEC}.}
\beq
E_{\mathrm{HFS}}^{\langle\pi^0 \rangle}(2S,\mu\text{H})= \left[0.245-0.126\right]\upmu\mathrm{eV}=0.119(19)\,\upmu\mathrm{eV},\eqlab{pi0finalresult}
 \eeq
where the error stems from \Eqref{pillCouplings}. What is interesting is the fact that the total effect 
in $\mu$H is positive, i.e., dominated by the direct coupling of the pion to the lepton in  \Figref{PiExchange} (a), and negative in H, i.e., dominated by the two-photon coupling in \Figref{PiExchange} (b). A similar calculation is presented in Ref.~\cite{Huong:2015naj}, the result is of the same order as \Eqref{pi0finalresult}. Another estimate can be found in Ref.~\cite{Zhou:2015bea}. The impact of the off-forward neutral-pion-exchange contribution, \Eqref{pi0finalresult}, and our updated value for the TPE polarizability contribution to the HFS, \Eqref{HFSfinalvalue}, on the extraction of the Zemach radius has already been shown in \secref{chap5}{newRZ}.

 \renewcommand{\arraystretch}{1.75}
\begin{table}
\caption{Numerical results for the neutral-pion-exchange contributions to the hyperfine splittings of electronic and muonic hydrogen. \label{Tab:resultsPionEx}}
\centering
\begin{small}
\begin{tabular}{|c|c|c|}
\hline
 \cellcolor[gray]{.7}
$\boldsymbol{E_{\mathrm{HFS}}^{\langle\pi^0 \rangle}(nS)}$&\cellcolor[gray]{.85}$\boldsymbol{1S}\;${\bf HFS}&\cellcolor[gray]{.85}$\boldsymbol{2S}\;${\bf HFS}\\
\hline
 \cellcolor[gray]{.85}{\bf H}&$-3.431(148)\,$feV&$-0.429(18)\,$feV\cellcolor[gray]{.95}\\
 \hline
\cellcolor[gray]{.85}$\boldsymbol{\mu}${\bf H}&\cellcolor[gray]{.95} $0.951(151)\, \upmu$eV&$0.119(19)\, \upmu$eV\\
\hline
\end{tabular}
\end{small}
\end{table}
 \renewcommand{\arraystretch}{1.3}

\section{Extraction of the Zemach Radius from Spectroscopy} \seclab{newRZ}
As described in \secref{5HFS}{ComparisonHFS}, our NLO BChPT prediction of the order-$\al^5$ polarizability contribution to the $\mu$H HFS does not agree with the dispersive calculations. Since the latter are used to extract the proton Zemach radius from $\mu$H spectroscopy \cite{Antognini:2012ofa}, it would be interesting to give an updated proton Zemach radius based on the model-independent predictions for the proton structure effects presented in here. 

Our prediction for the proton-polarizability contribution to the HFS from forward TPE is given in \Eqref{HFSfinalvalue}.
Combining it with the off-forward neutral-pion exchange, \Eqref{pi0finalresult}, we arrive at the following BChPT prediction for the polarizability effect in the $2S$ HFS of $\mu$H:
\begin{equation}
E_\text{HFS}^{\text{pol.}}(2S, \mu\text{H})=-1.3\,(^{+1.6}_{-1.2})\,\upmu\text{eV}. \label{resultHFS}
\end{equation}
This has to be compared to the literature value:
\begin{equation}
E_\text{HFS}^{\text{pol.}}(2S, \mu\text{H})=8.01(2.6)\,\upmu\text{eV} \text{ \cite{Carlson:2011af,Faustov:2001pn}},
\end{equation}
which is included in the theory budget of the $2S$ HFS in $\mu$H \cite{Antognini:2012ofa}, cf.\ \Eqref{muHtheoryHFS}.
Modifying \Eqref{muHtheoryHFS}, i.e., substituting our result for the polarizability effect, Eq.\ (\ref{resultHFS}), the new theory prediction reads (in units of $\mathrm{meV}$):
\beq
E_{\mathrm{HFS}}^{\text{th.}}(2S, \mu\text{H})=22.9750\,(^{+22}_{-19})-0.1621(10)\,(\nicefrac{R_{\mathrm{Z}p}}{\mathrm{fm}}). \eqlab{newHFSth}
\eeq
Combining the theory prediction, \Eqref{newHFSth}, with the hyperfine transition measured by the CREMA collaboration \cite{Antognini:1900ns}: 
\beq
E^{\,\text{exp.}}_\text{HFS}(2S, \mu\text{H})=22.8089(51) \,\text{meV},\eqlab{expmuHHFS}
\eeq
the Zemach radius reduces to: 
\begin{equation}
R_{\mathrm{Z}p}=1.025(35)\,\mathrm{fm}.\label{RZnew}
\end{equation}
This result has to be compared to the proton Zemach radius obtained previously from the $2S$ HFS in $\mu$H \cite{Antognini:1900ns}, as well as to the values extracted from the ground-state HFS in H \cite{Volotka:2004zu} and $ep$ scattering \cite{Friar:2003zg}:
\begin{subequations}
\bea
R_{\mathrm{Z}p}(\mu\text{H})&=&1.082(37)\, \text{fm},\\
R_{\mathrm{Z}p}(\text{H})&=&1.045(16)\,\text{fm},\\
R_{\mathrm{Z}p}(ep)&=&1.086(12)\,\text{fm}.
\eea
\end{subequations}
In the future, a measurement of the ground-state HFS in $\mu$H, supplemented with a  precise theory prediction, might reduce the error on the Zemach radius substantially.

 \section{Summary and Conclusion}
In this Chapter, we have studied the proton-polarizability contributions to the HFS of $\mu$H. 
The calculations are done in BChPT and within a $\Delta(1232)$-excitation model inspired by the large-$N_c$ limit of QCD. In the latter,
 the effect of the $\Delta$-resonance excitation is calculated using an extension of the large-$N_c$ relations for the magnetic $(G_M^*)$, electric $(G_E^*)$ and Coulomb $(G_C^*)$ FFs of the nucleon-to-delta transition \cite{Pascalutsa:2007wz}. In this way, the $Q^2$ behavior of the nucleon-to-delta transition is related to empirical information on the elastic nucleon FFs. In \Eqref{DeltaPoleModelFormula}, an approximate formula for the effect of the delta on the HFS was given, to which we refer to as the $\Delta$-pole model. Its main feature is that the $\ol S_1(0,Q^2)$ subtraction function was constructed to be vanishing for all $Q^2$.

Our main result is the NLO BChPT prediction for the order-$\al^5$ proton-polarizability contribution to the $\mu$H HFS, see \Eqref{HFSfinalvalue}. This model-independent prediction turned out to be significantly smaller than the results of dispersive calculations. In \secref{5HFS}{lowQ}, we tried to narrow-down the origin of the discrepancy and studied different $Q^2$ regions, as well as the contribution of the $\ol S_1(0,Q^2)$ subtraction function.  
In \secref{5HFS}{polexpHFS}, we isolated the contributions of various spin polarizabilities to the HFS effect. 

Similar to the proton-polarizability contribution, we presented a first NLO BChPT prediction for the order-$\al^5$ neutron-polarizability contribution (\secref{5HFS}{polHFSneutron}), see \Eqref{neutronPolContr}. The main difference between the proton and neutron case is due to the $\pi N$-loop diagrams, leading to a value of the neutron-polarizability contribution which is larger in magnitude than expected from the proton analogue. This first model-independent prediction of the neutron-polarizability contribution is relevant for the HFSs in light muonic atoms, and in particular for the planned measurement of the ground-state HFS in $\mu^3$He$^+$.

In \secref{5HFS}{neutralPion}, we have calculated the neutral-pion exchange in lepton-nucleus bound states. The $\pi \ell \ell$ interaction vertex can be expanded into a direct pseudo-scalar pion-lepton coupling and a coupling through two photons, see \Figref{Pi0Vertex}. The pion-lepton LEC was extracted from the experimental decay width of $\pi^0 \rightarrow e^+e^-$. The contribution of the pion-pole diagrams to the HFS was then given in \Eqref{pionHFS}, where the first term corresponds to \Figref{PiExchange} (a) and the dispersive integral corresponds to \Figref{PiExchange} (b). 
Numerical results for the $1S$ and $2S$ HFSs in H and $\mu$H are summarized in Table \ref{Tab:resultsPionEx}. The final values are the result of strong cancelations between the two Feynman diagrams in \Figref{PiExchange}. Due to the heavier muon mass, we observed different signs for the total effects in H and $\mu$H, cf.\ Table \ref{Tab:resultsPionEx}. In \appref{5HFS}{6.5}, we will explain why there is no $(Z\al)^6 \ln Z\al$ effect in the HFS generated by off-forward TPE and the lowest-order spin polarizabilities.

In \secref{5HFS}{newRZ}, we have used our results for the proton-polarizability contributions to extract the Zemach radius of the proton from the  $2S$ HFS in $\mu$H. In near future these results will become relevant for the forthcoming measurement of the ground-state HFS in $\mu$H. 

\begin{subappendices}

\section[No Nuclear Polarizability Contribution at Order $(Z\al)^6 \ln Z \al$]{Nuclear Polarizability Contribution at Order $\boldsymbol{(Z\al)^6 \ln Z \al}$ }  \seclab{6.5}
In the present Section, we discuss the polarizability contribution to the HFS from off-forward TPE.  Since no $(Z\al)^6 \ln Z\al$ effect is found, we will sketch the calculation only briefly and instead focus on the explanation why there is no logarithmic enhancement.

The master formulae for the structure effects through forward TPE are integrals over the photon 4-momentum ($\nu=q_0$, $Q^2 = \bq^2 -q_0^2$). For the $n$-th $S$-level shift and the $nS$ HFS, respectively, they are given in \chapref{chap5}, see Eqs.~\eref{VVCS_LS} and \eref{VVCS_HFS}. A comparison of Eqs.\ \eref{VVCS_LS} and \eref{VVCS_HFS} shows that the $t$-channel cut, $ 1/Q^4$, is present in the former but absent in the later. Therefore, the HFS has no TPE contribution at order $(Z\al)^5 \ln Z\al$. Similarly, one finds that there is no $(Z\al)^6 \ln Z\al$ contribution from off-forward TPE to the HFS.


The $(Z\al)^6 \ln Z\al$ polarizability contribution to the LS was discussed in \secref{5LS}{offTPE}.
In an analogous manner, we calculate the polarizability contribution to the HFS from off-forward TPE. The tensor describing VVCS off a spin-1/2 nucleus, \Eqref{spinpol}, has to be contracted with the Lepton tensor, \Eqref{LeptonTensor}. Again, terms proportional to $q^2$ or $q^{\prime\,2}$ are neglected, since they do not contribute to the $t$-channel cut enhancement. Also, the Feynman parameter trick, cf.\ \Eqref{FeynmanTrick}, stays the same.

The major complication is that we are now dealing with the spin dependence of both, the leptonic and the nuclear part. In general, we encounter the following set of gamma matrices between the spinors:
$$\ol{\mathpzc{N}}(p')\, \{\mathds{1},\gamma^\al, \gamma^\al \gamma^\be, \gamma^\al \gamma^\be \gamma^\si, \slashed{l}, \gamma^\al \slashed{l}, \gamma^\al \gamma^\be \slashed{l}\}\,\mathpzc{N}(p),$$
$$\ol{\mathpzc{u}}(l')\, \{\mathds{1},\gamma^\al, \gamma^\al \gamma^\be, \gamma^\al \gamma^\be \gamma^\si, \slashed{p}, \gamma^\al \slashed{p}, \gamma^\al \gamma^\be \slashed{l}\}\, \mathpzc{u}(l).$$

As before, it is suitable to perform the calculation in the CM frame, cf.\ \Eqref{COM}. We perform a semi-relativistic expansion of the TPE matrix element, which is especially useful in order to  simplify the appearing spinor structures and identify operators in the potential that act on the wave functions. In \appref{chap5}{SemRelExpDSpin}, we list all appearing spinor structures and their semi-relativistic expansions.

The spin operators of lepton and nucleus, $\boldsymbol{s}=\nicefrac{1}{2}\,\boldsymbol{\sigma}$ and $\boldsymbol{S}=\nicefrac{1}{2}\,\boldsymbol{\sigma}$, entering through the respective spinors, combine in different ways. Our main interest will be in the spin-spin interaction, $\boldsymbol{s} \cdot\boldsymbol{S}$, which affects both $S$- and $P$-states, cf.\ \Eqref{sSoperator}. The $\boldsymbol{s} \cdot\boldsymbol{S}$ operator is generated in the following Dirac spinor products: 
\begin{itemize}
\item  $\ol{\mathpzc{N}}(p')\,\gamma^\al \gamma^\be \gamma^\si\,\mathpzc{N}(p)\;\ol{\mathpzc{u}}(l')\,\gamma_\al \gamma_\be \gamma_\si\;\mathpzc{u}(l)$,
\item $\ol{\mathpzc{N}}(p')\,\gamma^\al \gamma^\be \slashed{l}\,\mathpzc{N}(p)\;\ol{\mathpzc{u}}(l')\,\gamma_\al \gamma_\be \slashed{p}\;\mathpzc{u}(l)$,
\item $\ol{\mathpzc{N}}(p')\,\gamma^\al \gamma^\be \slashed{l}\,\mathpzc{N}(p)\;\ol{\mathpzc{u}}(l')\,\gamma_\al \gamma_\be\;\mathpzc{u}(l)$,
\item $\ol{\mathpzc{N}}(p')\,\gamma^\al \gamma^\be \,\mathpzc{N}(p)\; \ol{\mathpzc{u}}(l')\,\gamma_\al \gamma_\be \slashed{p}\;\mathpzc{u}(l)$,
\item $\ol{\mathpzc{N}}(p')\,\gamma^\al \gamma^\be \,\mathpzc{N}(p)\;\ol{\mathpzc{u}}(l')\,\gamma_\al \gamma_\be \;\mathpzc{u}(l)$.
\end{itemize}
The CM frame allows us to rewrite, e.g.:
\beq
\ol{\mathpzc{N}}(p')\,\gamma^\al \gamma^\be \slashed{l}\,\mathpzc{N}(p)=-M\, \ol{\mathpzc{N}}(p')\,\gamma^\al \gamma^\be \,\mathpzc{N}(p)+\sqrt{s} \,\ol{\mathpzc{N}}(p')\,\gamma^\al \gamma^\be \gamma_0\,\mathpzc{N}(p),
\eeq
with the invariant mass $s=(p+l)^2$.

\noindent The leading term in the semi-relativistic expansion of the spin-dependent TPE is included in (omitting everything but $\boldsymbol{s} \cdot\boldsymbol{S}$): 
\bea
\mathscr M(p_t^2)&\sim& 32 \,\pi^2  \alpha \, p_t^2\,\int_0^1 \dd x\, x\int_0^1 \dd y\;\boldsymbol{s\cdot S}   \eqlab{spinex}\Big\{ J_2(\mathcal M^2)  \,[2\, \gamma_{E1M2}-3\, \gamma_{M1E2}-\gamma_{M1M1}]\qquad\quad\\
&&-4\,J_3(\mathcal M^2)\, m^2 (xy)^2 \,[\gamma_{M1E2}+\gamma_{M1M1}]\Big\}.\,\qquad \nn
\eea 
For the imaginary part, the integrals over the Feynman parameters can be solved by substituting Eqs.\ \eref{I3} and \eref{I5}. We recall that, following the discussion above \Eqref{relevantIm}, generating an enhancement of $(Z\al)^6 \ln Z \al$ in the energy-level shift requires $\im \mathscr M(p_t^2)$ to be of order $\mathcal{O}(\sqrt{\tau})$. Since this is not the case, \Eqref{spinex} is merely contributing to order $(Z\al)^6$.

What about the effect of other operators generated in the spin-dependent TPE, f.i., the spin-orbit interactions, $\boldsymbol{s} \cdot \boldsymbol{p_t} \times \boldsymbol{p}$ or $\boldsymbol{S} \cdot \boldsymbol{p_t} \times \boldsymbol{p}$, and the squared momentum operator? All other operators comprise two momenta, (either) the momentum operator $\boldsymbol{p}$ and (or) the momentum-transfer operator $\boldsymbol{p_t}$. Hence, the multiplying prefactor in $\im \mathscr M(p_t^2)$ has to be of order $\mathcal{O}(1/\sqrt{\tau})$ for a logarithmic enhancement. Let us illustrate this with the example of the momentum operator squared, $\boldsymbol{p}^2$, which in coordinate space translates into the Laplace operator. The $\boldsymbol{p}^2$ operator has to modify the convolution integral of the momentum-space wave functions:
 \beq
  w_{nlm}^{(\boldsymbol{p}^2)}(\vert \boldsymbol{p_t}\vert) =
  \int \dd \boldsymbol{p} \,
\vfi_{nlm}^\ast (\boldsymbol{p}+\boldsymbol{p_t}) \,\boldsymbol{p}^2 \vfi_{nlm} (\boldsymbol{p}).
\eeq
Plugging in the momentum space Coulomb wave function of the $2S$-level,
 \beq
\vfi_{200}= \frac{32 \,a^{3/2} \left[4 (a \vert \boldsymbol{p}\vert)^2-1\right]}{\sqrt{\pi } \left[1+4 (a\vert \boldsymbol{p}\vert)^2\right]^3},
 \eeq
 we obtain:
\beq
  w_{200}^{(\boldsymbol{p}^2)}(\vert \boldsymbol{p_t}\vert)=\frac{1+3(a\vert \boldsymbol{p_t}\vert)^2+4 (a\vert \boldsymbol{p_t}\vert)^6}{4 a^2 \left[1+(a\vert \boldsymbol{p_t}\vert)^2\right]^4}.
\eeq
Adopting this convoluted momentum-space wave function in first-order PT, cf.\ \Eqref{LS}, we can check that $V(\vert \boldsymbol{p_t} \vert)$ has to be of order $\mathcal{O}(1/\vert \boldsymbol{p_t} \vert)$, and in turn, $\im \mathscr M(p_t^2)$ has to be of $\mathcal{O}(1/\sqrt{\tau})$, for the $t$-channel cut enhancement to show up. The bottom line is, neither forward nor off-forward TPE display a $\ln Z \al$ enhancement in the HFS.

\section{Dispersive Calculation of the Polarizability Contribution} \seclab{kappaMatching}

In this Section, we give more details on our dispersive calculation of the proton-polarizability contribution to the $\mu$H HFS. In \secref{5HFS}{lowQ}, we described how one replaces the interpolated low-$Q$ region below $Q_\mathrm{min}$ with empirical polarizabilities. 

Another crucial point is the cancelation between $I_1(0)$ and $ F_2^2(0)$ in the $\ol S_1(0,Q^2)$ subtraction function as entering through \Eqref{S1subtermHFS}. Generally speaking, we need to match the values of the proton anomalous magnetic moment $\kappa$ used in the FF and structure function parametrizations. The parametrization of \citet{Simula:2001iy} was imposed to reproduce the GDH sum rule $I_1(0)=-\kappa^2/4$ with $\kappa\approx1.7905$, which is in good agreement with the presently recommended value $\kappa\approx1.7929$ \cite{Mohr:2012aa}. The MAID model on the other hand deviates substantially with $\kappa\approx1.6124$. Figure~\ref{fig:I1Plot} shows that the MAID model of $I_1(Q^2)$ gives a good description of the CLAS data  \cite{Prok:2008ev}, while it fails to reproduce the experimental value at $Q^2=0$. Splitting the $Q$-integration  in \Eqref{S1subtermHFS} at $\lambda$ allows for the appropriate correction at the real photon point, while keeping the good description of data at higher $Q$-values. Expanding \Eqref{S1subtermHFS} around $Q=0$ while keeping $W(Q)$ fixed, we find:
\bea
&&\int_{Q_\mathrm{min}}^\infty \dd Q \left[4I_1(Q^2)+F_2^2(Q^2)\right]W(Q)\nn\\
&=& 4 \int_{Q_\mathrm{min}}^\lambda \dd Q\left[I_1(Q^2)-I_1(0)-\frac{\kappa^2}{4}+\frac{1}{4}F_2^2(Q^2)\right]W(Q)+\underbrace{\int_\lambda^\infty \dd Q \left[4I_1(Q^2)+F_2^2(Q^2)\right]W(Q)}_{\delta_1(\lambda)}\nn
\eea
\begin{small}
\bea
&=& 4 \int_{Q_\mathrm{min}}^\lambda \dd Q\left[I_1(Q^2)-I_1(0)-\frac{\kappa^2}{4}+\frac{1}{4}F_2^2(Q^2)\right]W(Q)+\underbrace{\int_\lambda^\infty \dd Q \left[4I_1(Q^2)+F_2^2(Q^2)\right]W(Q)}_{\delta_1(\lambda)}\nn\\
&\approx&\underbrace{4 \int_{Q_\mathrm{min}}^\lambda \dd Q\, Q^2\, W(Q)}_{a(\lambda)}  \left[I_1'(0)+\frac{\kappa}{2}F_2'(0)\right]+\delta_1(\lambda)\nn\\
&=&a(\lambda)\left[I_1'(0)+\frac{\kappa}{2}F_2'(0)\right]+\delta_1(\lambda), \qquad \text{with}\quad W(Q)=\frac{Z\al m}{\pi (1+\kappa) M} \frac{1}{Q}\frac{5+4v_l}{(1+v_l)^2},\eqlab{kappaCorr}
\eea
\end{small}
where the derivatives are taken with respect to $Q^2$.  A suited estimate for the subtraction term is then found by minimising the result of \Eqref{kappaCorr} with respect to $\lambda$.

Analogously, we correct the relevant part of \Eqref{polHFSExpa}, where it should be $I_1(0)=I_A(0)=-\kappa^2/4$:
\begin{small}
\bea
&&\int_{Q_\mathrm{min}}^\infty \dd Q\, W(Q) \left[F_2^2(Q^2)+\frac{1}{(v_l+1)(5+4v_l)}\left\{(1+3v_l)\,I_A(Q^2)+(19+33v_l+16v_l^2)\,I_1(Q^2)\right\}\right]\nn\\
&=&\int_{Q_\mathrm{min}}^\lambda \dd Q\, W(Q) \left[F_2^2(Q^2)+\frac{1}{(v_l+1)(5+4v_l)}\left\{(1+3v_l)\,\left[I_A(Q^2)-I_A(0)-\frac{\kappa^2}{4}\right]\right.\right.\nn\\
&&\left.\left.\qquad\qquad\qquad+(19+33v_l+16v_l^2)\,\left[I_1(Q^2)-I_1(0)-\frac{\kappa^2}{4}\right]\right\}\right]\nn\\
&&+\underbrace{\int_\lambda^\infty \dd Q\, W(Q) \left[F_2^2(Q^2)+\frac{1}{(v_l+1)(5+4v_l)}\left\{(1+3v_l)\,I_A(Q^2)+(19+33v_l+16v_l^2)\,I_1(Q^2)\right\}\right]}_{\delta_2(\lambda)}\nn\\
&\approx&a(\lambda)\left[I_1'(0)+\frac{\kappa}{2}F_2'(0)\right]+\delta_2(\lambda). \eqlab{kappaCorr2}
\eea
\end{small}
To correct the MAID prediction in Table \ref{PolHFS}, we applied \Eqref{kappaCorr} (lower half of the Table) and \Eqref{kappaCorr2} (upper half of the Table). The results from Eqs.~\eref{kappaCorr} and \eref{kappaCorr2} are comparable and both affect the $I_1$ term in $\Delta_1$. For our prediction based on the Simula parametrization of spin structure functions \cite{Simula:2001iy}, the correction turned out to be very small, hence, we neglect it.

Final results of our dispersive calculation are given in \Eqref{OurEmpRes}. Partial results appear in Sections \ref{chap:5HFS}.\ref{sec:lowQ} and \ref{chap:5HFS}.\ref{sec:polexpHFS}, and Tables \ref{higherMom} and \ref{PolHFS} therein.

\renewcommand{\arraystretch}{1.75}
\begin{table}
\centering
\begin{scriptsize}
\centering
\caption{Contribution of different $Q^2$ regions to the hyperfine splitting in muonic hydrogen. \label{lowQcomCarlson}}
\label{Table:Summary3}
\centering
\begin{tabular}{|c|c|c|c|c|c|c|}
\hline
 \rowcolor[gray]{.7}
&&&&\multicolumn{3}{c|}{{\bf Our values [ppm]}}\\
\cline{5-7}
 \rowcolor[gray]{.7}
\multirow{-2}{*}{\bf Term}&\multirow{-2}{*}{$\boldsymbol{Q^2\,[\mathrm{GeV}^2]}$}&\multirow{-2}{*}{{\bf From}}&\multirow{-2}{*}{{\bf Values of Carlson et al. [ppm]}}&$\boldsymbol{\pi N}${\bf -loops}&$\boldsymbol{\Delta}${\bf -exch.}&{\bf Total}\\
\hline
  \rowcolor[gray]{.95} 
$\Delta_1$&$[0,0.0452]$&$F_2$&$770.39$&&&\\     
\cellcolor[gray]{.95}&$[0,0.0452]$&$g_1$&$-730.12$&$$&&\\       
 \rowcolor[gray]{.95} 
 \cellcolor[gray]{.95}&$[0,0.0452]$&$F_2$ and $g_1$&$40.27(7.96)(31.37)()$ &$-7.54$&$14.13$&$6.59$\\      
\cellcolor[gray]{.95}&$[0.0452,20]$&$F_2$&$317.03()(9.83)()$&&&\\      
 \rowcolor[gray]{.95}    
 \cellcolor[gray]{.95}&$[0.0452,20]$&$g_1$&$8.43(8.43)(74.92)(29.97)$&$$&&\\    
 \cellcolor[gray]{.95}&$[0.0452,20]$&$F_2$ and $g_1$&$325.46$&$-10.87$&$41.08$&$30.20$\\
  \rowcolor[gray]{.95} 
 \cellcolor[gray]{.95}&$[20,\infty]$&$F_2$&$0()(0)()$&&&\\     
\cellcolor[gray]{.95}&$[20,\infty]$&$g_1$&$5.15()()(0.47)$&$$&&\\        
   \rowcolor[gray]{.95} 
 \cellcolor[gray]{.95}&$[20,\infty]$&$F_2$ and $g_1$&$5.15$&$0.00$&$0.00$&$0.00$\\   
\hline
Total $\Delta_1$&&&$370.88(11.71)(107.71)(30.91)$&$-18.41$&$55.20$&$36.79$\\
\hline
 \rowcolor[gray]{.95}
$\Delta_2$&$[0,0.0452]$&$g_2$&$-5.62()()(5.62)$&$3.24$&$-23.51$&$-20.27$\\      
 \cellcolor[gray]{.95}&$[0.0452,20]$&$g_2$&$-13.58()()(13.58)$&$5.06$&$-82.28$&$-77.22$\\      
 \rowcolor[gray]{.95}
&$[20,\infty]$&$g_2$&$0()()(0)$&$0.00$&$0.00$&$0.00$\\    
\hline
Total $\Delta_2$&&&$-19.20()()(19.20)$&$8.30$&$-105.79$&$-97.48$\\  
\hline
 \rowcolor[gray]{.95}
$\Delta_\mathrm{pol}$&&&$351.68(11.71)(107.71)(36.06)$&$-10.11$&$-50.58$&$-60.69$\\
\hline
\end{tabular}
\end{scriptsize}
\end{table}
\renewcommand{\arraystretch}{1.3}

    \renewcommand{\arraystretch}{1.75}
\begin{table} [htb]
\centering
\begin{scriptsize}
\caption{Contribution of higher moments to the $2S$ hyperfine splitting in muonic hydrogen. All values in $\upmu\mathrm{eV}$. \label{higherMom}}
\begin{tabular}{|c|cc|cc|cc|cc|cc|}
\hline
 \rowcolor[gray]{.7}
{\bf Input}&\multicolumn{2}{c|}{{\bf up to }$\boldsymbol{x^2\,g_{1,2}}$}&\multicolumn{2}{c|}{$\boldsymbol{x^4\,g_{1,2}}$}&\multicolumn{2}{c|}{$\boldsymbol{x^6\,g_{1,2}}$}&\multicolumn{2}{c|}{$\boldsymbol{x^8\,g_{1,2}}$}&\multicolumn{2}{c|}{$\boldsymbol{x^{10}\,g_{1,2}}$}\\
\hline
$g_{1,2}$ \citet{Simula:2001iy}&$6.70$&&$-0.63$&&$0.46$&&$-0.43$&&$0.46$&\\
and FF \cite{Kelly:2004hm} $\kappa\approx1.7905$&$-2.14$&\multirow{-2}{*}{$4.56$}&$0.80$&\multirow{-2}{*}{$0.17$}&$-0.57$&\multirow{-2}{*}{$-0.12$}&$0.53$&\multirow{-2}{*}{$0.11$}&$-0.58$&\multirow{-2}{*}{$-0.12$}\\
\hdashline
 \rowcolor[gray]{.95}
&$44.52$&&$0.65$&&$-0.57$&&$0.70$&&$-1.04$&\\
 \rowcolor[gray]{.95}
\multirow{-2}{*}{Background and FF}&$1.33$&\multirow{-2}{*}{$45.85$}&$-0.63$&\multirow{-2}{*}{$0.02$}&$0.56$&\multirow{-2}{*}{$-0.02$}&$-0.68$&\multirow{-2}{*}{$0.02$}&$1.02$&\multirow{-2}{*}{$-0.02$}\\
&$-40.04$&&$-1.28$&&$1.03$&&$-1.13$&&$1.50$&\\
\multirow{-2}{*}{Resonances}&$-3.18$&\multirow{-2}{*}{$-43.22$}&$1.43$&\multirow{-2}{*}{$0.15$}&$-1.13$&\multirow{-2}{*}{$-0.10$}&$1.21$&\multirow{-2}{*}{$0.09$}&$-1.60$&\multirow{-2}{*}{$-0.10$}\\
 \rowcolor[gray]{.95}
&$-34.46$&&$-1.31$&&$1.05$&&$-1.14$&&$1.52$&\\
 \rowcolor[gray]{.95}
\multirow{-2}{*}{$\Delta(1232)$}&$-3.43$&\multirow{-2}{*}{$-37.89$}&$1.50$&\multirow{-2}{*}{$0.19$}&$-1.17$&\multirow{-2}{*}{$-0.12$}&$1.25$&\multirow{-2}{*}{$0.11$}&$-1.64$&\multirow{-2}{*}{$-0.12$}\\
\hline
&$1.19$&&$-0.45$&&$0.04$&&$0.31$&&$-0.82$&\\
\multirow{-2}{*}{BChPT}&$-2.83$&\multirow{-2}{*}{$-1.64$}&$0.87$&\multirow{-2}{*}{$0.42$}&$-0.38$&\multirow{-2}{*}{$-0.34$}&$0.10$&\multirow{-2}{*}{$0.40$}&$0.22$&\multirow{-2}{*}{$-0.61$}\\
\hdashline
 \rowcolor[gray]{.95}
&$-0.82$&&$0.72$&&$-0.69$&&$0.87$&&$-1.31$&\\
 \rowcolor[gray]{.95}
\multirow{-2}{*}{$\pi$-cloud}&$0.35$&\multirow{-2}{*}{$-0.47$}&$-0.31$&\multirow{-2}{*}{$0.42$}&$0.32$&\multirow{-2}{*}{$-0.36$}&$-0.43$&\multirow{-2}{*}{$0.44$}&$0.67$&\multirow{-2}{*}{$-0.65$}\\
&$2.01$&&$-1.17$&&$0.73$&&$-0.56$&&$0.49$&\\
\multirow{-2}{*}{$\Delta$-exchange}&$-3.17$&\multirow{-2}{*}{$-1.17$}&$1.17$&\multirow{-2}{*}{$0.00$}&$-0.70$&\multirow{-2}{*}{$0.03$}&$0.52$&\multirow{-2}{*}{$-0.03$}&$-0.45$&\multirow{-2}{*}{$0.04$}\\
\hline
\end{tabular}
\end{scriptsize}
\end{table}
  \renewcommand{\arraystretch}{1.3}

\newpage
\clearpage
\thispagestyle{empty}
  \renewcommand{\arraystretch}{1.5}
\begin{sidewaystable} [htb]
\begin{scriptsize}
\caption{Contribution of individual polarizabilities to the $2S$ hyperfine splitting in muonic hydrogen. All values in $\upmu\mathrm{eV}$. \label{PolHFS}}
  \scalebox{0.9}{
\begin{tabular}{|c|cc|cc|cc|cc|cc|cc|cc|cc|}
\hline
\rowcolor[gray]{.7}
{\bf Input}&\multicolumn{2}{c|}{{\bf Eq.}}&\multicolumn{2}{c|}{$\boldsymbol{F_2^2}$}&\multicolumn{2}{c|}{$\boldsymbol{I_1}$}&\multicolumn{2}{c|}{$\boldsymbol{I_A}$}&\multicolumn{2}{c|}{$\boldsymbol{\delta_{LT}}$}&\multicolumn{2}{c|}{$\boldsymbol{\gamma_0}$}&\multicolumn{2}{c|}{$\boldsymbol{x^4g_2}${\bf remainder}}&\multicolumn{2}{c|}{{\bf total}}\\
\thickhline
MAID ($\kappa\approx1.6124$) \cite{MAID}&\eref{polD1a}&&$24.56$&&$-24.34$&&$5.21$&&$-0.77$&&$$&&$$&$$&$4.66$&\\
and FF \cite{Bradford:2006yz}&\eref{polD2a}&\multirow{-2}{*}{\eref{polHFSExpa}}&$0$&\multirow{-2}{*}{$24.56$}&$4.10$&\multirow{-2}{*}{$-20.25$}&$-6.40$&\multirow{-2}{*}{$-1.20$}&$0$&\multirow{-2}{*}{$-0.77$}&&$$&&$$&$-2.31$&\multirow{-2}{*}{$2.35$}\\
\hline
 \rowcolor[gray]{.95}
Simula $g_i$ ($\kappa\approx1.7905$) \cite{Simula:2001iy} &\eref{polD1a}&&$24.83$&&$-22.29$&&$4.56$&&$-0.40$&&$$&&$$&$$&$6.70$&\\
 \rowcolor[gray]{.95}
and FF \cite{Kelly:2004hm}&\eref{polD2a}&\multirow{-2}{*}{\eref{polHFSExpa}}&$0$&\multirow{-2}{*}{$24.83$}&$3.49$&\multirow{-2}{*}{$-18.80$}&$-5.63$&\multirow{-2}{*}{$-1.07$}&$0$&\multirow{-2}{*}{$-0.40$}&&$$&&$$&$-2.14$&\multirow{-2}{*}{$4.56$}\\
\hdashline
&\eref{polD1a}&&$24.83$&&$24.88$&&$-4.92$&&$-0.27$&&$$&&$$&&$44.52$&\\
\multirow{-2}{*}{Background and FF}&\eref{polD2a}&\multirow{-2}{*}{\eref{polHFSExpa}}&$0$&\multirow{-2}{*}{$24.83$}&$-4.69$&\multirow{-2}{*}{$20.19$}&$6.02$&\multirow{-2}{*}{$1.10$}&$0$&\multirow{-2}{*}{$-0.27$}&&$$&&$$&$1.33$&\multirow{-2}{*}{$45.85$}\\
 \rowcolor[gray]{.95}
&\eref{polD1a}&&$$&&$-49.88$&&$9.68$&&$0.16$&&$$&&$$&$$&$-40.04$&\\
 \rowcolor[gray]{.95}
\multirow{-2}{*}{Resonances}&\eref{polD2a}&\multirow{-2}{*}{\eref{polHFSExpa}}&$$&\multirow{-2}{*}{$$}&$8.73$&\multirow{-2}{*}{$-41.15$}&$-11.90$&\multirow{-2}{*}{$-2.22$}&$0$&\multirow{-2}{*}{$0.16$}&&$$&&$$&$-3.18$&\multirow{-2}{*}{$-43.22$}\\
&\eref{polD1a}&&$$&&$-43.55$&&$9.09$&&$0.00$&&$$&&$$&$$&$-34.46$&\\
\multirow{-2}{*}{$\Delta(1232)$}&\eref{polD2a}&\multirow{-2}{*}{\eref{polHFSExpa}}&$$&\multirow{-2}{*}{$$}&$7.73$&\multirow{-2}{*}{$-35.82$}&$-11.16$&\multirow{-2}{*}{$-2.07$}&$0$&\multirow{-2}{*}{$0.00$}&&$$&&$$&$-3.43$&\multirow{-2}{*}{$-37.89$}\\
 \rowcolor[gray]{.95}
&\eref{polD1a}&&$$&&$-10.12$&&$1.48$&&$0.00$&&$$&&$$&&$-8.64$&\\
 \rowcolor[gray]{.95}
\multirow{-2}{*}{$D_{13}(1520)$}&\eref{polD2a}&\multirow{-2}{*}{\eref{polHFSExpa}}&$$&\multirow{-2}{*}{$$}&$1.70$&\multirow{-2}{*}{$-8.42$}&$-1.83$&\multirow{-2}{*}{$-0.35$}&$0$&\multirow{-2}{*}{$0.00$}&&$$&&$$&$-0.13$&\multirow{-2}{*}{$-8.77$}\\
\hline
&\eref{polD1a}&&$$&&$-0.61$&&$2.45$&&$-0.65$&&$$&&$$&&$1.19$&\\
\multirow{-2}{*}{BChPT}&\eref{polD2a}&\multirow{-2}{*}{\eref{polHFSExpa}}&&\multirow{-2}{*}{}&$0.15$&\multirow{-2}{*}{$-0.46$}&$-2.97$&\multirow{-2}{*}{$-0.52$}&$0$&\multirow{-2}{*}{$-0.65$}&&&&\multirow{-2}{*}{}&$-2.83$&\multirow{-2}{*}{$-1.64$}\\
\hdashline
 \rowcolor[gray]{.95}
&\eref{polD1a}&&$$&&$0.57$&&$-0.40$&&$-0.99$&&$$&&$$&&$-0.82$&\\
 \rowcolor[gray]{.95}
\multirow{-2}{*}{$\pi$-cloud}&\eref{polD2a}&\multirow{-2}{*}{\eref{polHFSExpa}}&$$&\multirow{-2}{*}{$$}&$-0.14$&\multirow{-2}{*}{$0.43$}&$0.49$&\multirow{-2}{*}{$0.09$}&$0$&\multirow{-2}{*}{$-0.99$}&&$$&&$$&$0.35$&\multirow{-2}{*}{$-0.47$}\\
&\eref{polD1a}&&$$&&$-1.18$&&$2.85$&&$0.34$&&$$&$$&&&$2.01$&\\
\multirow{-2}{*}{$\Delta$-exchange}&\eref{polD2a}&\multirow{-2}{*}{\eref{polHFSExpa}}&$$&\multirow{-2}{*}{$$}&$0.29$&\multirow{-2}{*}{$-0.90$}&$-3.46$&\multirow{-2}{*}{$-0.61$}&$0$&\multirow{-2}{*}{$0.34$}&$$&\multirow{-2}{*}{$$}&$$&\multirow{-2}{*}{$$}&$-3.17$&\multirow{-2}{*}{$-1.17$}\\
\thickhline \rowcolor[gray]{.95}
Simula $g_i$ ($\kappa\approx1.7905$) \cite{Simula:2001iy} &\eref{polD1b}&&$24.83$&&$-19.49$&&$$&&$0$&&$2.70$&&$-1.37$&$$&$6.67$&\\
\rowcolor[gray]{.95}
and FF \cite{Kelly:2004hm}&\eref{polD2b}&\multirow{-2}{*}{\eref{polHFSExpb}}&$0$&\multirow{-2}{*}{$24.83$}&$0$&\multirow{-2}{*}{$-19.49$}&$$&$$&$-0.48$&\multirow{-2}{*}{$-0.48$}&$-3.27$&\multirow{-2}{*}{$-0.57$}&$1.66$&\multirow{-2}{*}{$0.29$}&$-2.10$&\multirow{-2}{*}{$4.56$}\\
\hdashline
&\eref{polD1b}&&$24.83$&&$21.05$&&$$&&$0$&&$-2.45$&&$1.09$&$$&$44.52$&\\
\multirow{-2}{*}{Background and FF}&\eref{polD2b}&\multirow{-2}{*}{\eref{polHFSExpb}}&$0$&\multirow{-2}{*}{$24.83$}&$0$&\multirow{-2}{*}{$21.05$}&$$&\multirow{-2}{*}{$$}&$-0.33$&\multirow{-2}{*}{$-0.33$}&$2.98$&\multirow{-2}{*}{$0.53$}&$-1.32$&\multirow{-2}{*}{$-0.23$}&$1.33$&\multirow{-2}{*}{$45.85$}\\
\rowcolor[gray]{.95}
&\eref{polD1b}&&$$&&$-42.82$&&$$&&$0$&&$5.23$&&$-2.46$&&$-40.04$&\\
\rowcolor[gray]{.95}
\multirow{-2}{*}{Resonances}&\eref{polD2b}&\multirow{-2}{*}{\eref{polHFSExpb}}&$$&\multirow{-2}{*}{$$}&$0$&\multirow{-2}{*}{$-42.82$}&$$&\multirow{-2}{*}{$$}&$0.19$&\multirow{-2}{*}{$0.19$}&$-6.35$&\multirow{-2}{*}{$-1.11$}&$2.98$&\multirow{-2}{*}{$0.52$}&$-3.18$&\multirow{-2}{*}{$-43.22$}\\
&\eref{polD1b}&&$$&&$-37.28$&&$$&&$0.00$&&$5.40$&&$-2.58$&&$-34.46$&\\
\multirow{-2}{*}{$\Delta(1232)$}&\eref{polD2b}&\multirow{-2}{*}{\eref{polHFSExpb}}&$$&\multirow{-2}{*}{$$}&$0$&\multirow{-2}{*}{$-37.28$}&$$&\multirow{-2}{*}{$$}&$0$&\multirow{-2}{*}{$0.00$}&$-6.55$&\multirow{-2}{*}{$-1.15$}&$3.13$&\multirow{-2}{*}{$0.54$}&$-3.43$&\multirow{-2}{*}{$-37.89$}\\
\rowcolor[gray]{.95}
&\eref{polD1b}&&$$&&$-8.75$&&$$&&$0$&&$0.12$&&$-0.02$&&$-8.64$&\\
\rowcolor[gray]{.95}
\multirow{-2}{*}{$D_{13}(1520)$}&\eref{polD2b}&\multirow{-2}{*}{\eref{polHFSExpb}}&$$&\multirow{-2}{*}{$$}&$0$&\multirow{-2}{*}{$-8.75$}&$$&\multirow{-2}{*}{$$}&$0.00$&\multirow{-2}{*}{$0.00$}&$-0.15$&\multirow{-2}{*}{$-0.03$}&$0.02$&\multirow{-2}{*}{$0.00$}&$-0.13$&\multirow{-2}{*}{$-8.77$}\\
\hline
&\eref{polD1b}&&$$&&$-0.49$&&$$&&$0$&&$3.15$&&$-1.48$&$$&$1.19$&\\
\multirow{-2}{*}{BChPT}&\eref{polD2b}&\multirow{-2}{*}{\eref{polHFSExpb}}&&\multirow{-2}{*}{}&$0$&\multirow{-2}{*}{$-0.49$}&&&$-0.79$&\multirow{-2}{*}{$-0.79$}&$-3.81$&\multirow{-2}{*}{$-0.66$}&$1.78$&\multirow{-2}{*}{$0.31$}&$-2.83$&\multirow{-2}{*}{$-1.64$}\\
\hdashline
\rowcolor[gray]{.95}
&\eref{polD1b}&&$$&&$0.45$&&$$&&$0$&&$-1.80$&&$0.53$&&$-0.82$&\\
\rowcolor[gray]{.95}
\multirow{-2}{*}{$\pi$-cloud}&\eref{polD2b}&\multirow{-2}{*}{\eref{polHFSExpb}}&$$&\multirow{-2}{*}{$$}&$0$&\multirow{-2}{*}{$0.45$}&&$$&$-1.20$&\multirow{-2}{*}{$-1.20$}&$2.19$&\multirow{-2}{*}{$0.39$}&$-0.64$&\multirow{-2}{*}{$-0.11$}&$0.35$&\multirow{-2}{*}{$-0.47$}\\
&\eref{polD1b}&&$$&&$-0.94$&&$$&&$0$&&$4.96$&&$-2.00$&$$&$2.01$&\\
\multirow{-2}{*}{$\Delta$-exchange}&\eref{polD2b}&\multirow{-2}{*}{\eref{polHFSExpb}}&$$&\multirow{-2}{*}{$$}&$0$&\multirow{-2}{*}{$-0.94$}&&&$0.41$&\multirow{-2}{*}{$0.41$}&$-6.00$&\multirow{-2}{*}{$-1.04$}&$2.42$&\multirow{-2}{*}{$0.42$}&$-3.17$&\multirow{-2}{*}{$-1.17$}\\
\hline
\end{tabular}}
\end{scriptsize}
\end{sidewaystable}
  \renewcommand{\arraystretch}{1.3}

\end{subappendices}

\chapter{Summary, Conclusion and Outlook} \chaplab{chap7}
Protons and neutrons (collectively, {\it nucleons}) comprise the atomic nuclei and thus are present in all the visible matter around us. 
They are some of the most fundamental building blocks of matter as we know it. Yet, they are no elementary particles --- they consist of quarks and gluons.
The exact composition of the nucleon structure should be calculable from quantum chromodynamics (QCD), which is the renormalizable
quantum field theory describing the fundamental interaction among quarks and gluons.  Unfortunately, such {\it ab initio} calculations of nucleon
structure have proven to be extremely difficult due to non-perturbative QCD phenomena such as color confinement, mass-gap generation, 
spontaneous chiral-symmetry breaking.

Nonetheless, a lot of progress has been made in calculating the low-energy effects of the nucleon structure using lattice QCD on one hand 
and effective field theories on the other.
In this thesis, we have resorted to the frameworks of chiral perturbation theory (ChPT), which is a low-energy effective field theory of QCD, 
and of dispersion theory which is based  on
general principles of unitarity, causality  and low-energy theorems. These theoretical tools have allowed us to systematically assess the nucleon polarizabilities and
their effect on atomic spectroscopy.  

Our calculations are timely and relevant in the context of the {\it proton radius puzzle}.
The proton charge radius, $R_{Ep}$, has been receiving a lot of attention after its first-time extraction from $\mu$H spectroscopy in 2010 \cite{Pohl:2010zza}. The measurements of the $2P-2S$ transitions in $\mu$H and $\mu$D by the CREMA collaboration at PSI
yielded astonishing results, which prompted a significant reduction of the proton and deuteron charge radii~\cite{Pohl:2010zza,Antognini:1900ns,Pohl1:2016xoo}. The other ways to extract the charge radii --- the electron scattering and the spectroscopy of electronic atoms --- had mainly been consistent
with each other. At present, the $Z=1$ {\it (hydrogen isotope) charge radius puzzle} deals with the $5.6\,\sigma$ and $3.5\,\sigma$ discrepancies between experiments with either electrons or muons probing the proton respectively deuteron. New experiments with muonic atoms are underway and require precise theory input for their performance and interpretation. For example, the next series of CREMA experiments will be devoted to the ground-state hyperfine splittings in $\mu$H and $\mu^3$He$^+$, where a reliable theoretical evaluation of the hyperfine splittings  is simply indispensable for narrowing down the search for these transitions.

Shortly after the puzzle appeared,  it was suggested that an underestimation of the proton structure effects beyond the charge radius (e.g., Friar radius and polarizability effects) could be responsible for the discrepancy. This possibility has essentially been ruled out by dispersive \cite{Carlson:2011zd, Birse:2012eb,Gorchtein:2013yga} and ChPT calculations \cite{Alarcon:2013cba}. Nevertheless, the proton-polarizability contribution remains to be the major theoretical uncertainty in the description of $\mu$H. 
More generally, the theory of light muonic atoms, e.g., muonic-hydrogen and muonic-helium isotopes, is limited by our knowledge on nuclear (and nucleon) structure effects. This thesis is to a large extent devoted to model-independent and systematic calculations of these effects. Our results are of interest for the analyses of Lamb shift and hyperfine splitting measurements in muonic atoms, e.g., for the proposed ground-state hyperfine splitting measurements in $\mu$H \cite{Pohl:2016xsr,Bakalov:2014hda, Bakalov:2015xya, Adamczak:2016pdb,Sato:2014uza}.

To briefly summarize what we have done in this thesis, let us  recall that in \chapref{chap2} we derived the one-photon-exchange Breit potential for a lepton-nucleus bound state with nuclear form factors. We employed a dispersive approach and showed that all the finite-size and one-loop vacuum polarization effects are reproduced correctly. In \chapref{chap3}, we reported on the status of the static nucleon polarizabilities measured in real Compton scattering and formulated sum rules for the Compton contribution to photoabsorption. In \chapref{chap4}, we dealt with forward doubly-virtual Compton scattering. In particular, we calculated the tree-level contribution of the lowest nucleon-resonance --- $\Delta(1232)$ --- and studied the $\pi \Delta$-production cross sections in view of the discrepancy between baryon and heavy baryon predictions for the longitudinal-transverse polarizability of the proton, i.e., the {\it $\delta_{LT}$ puzzle}. In \chapref{chap5}, the basic theory of forward and off-forward two-photon exchange in hydrogen-like atoms was discussed. In Chapters \ref{chap:5LS} and \ref{chap:5HFS}, polarizability contributions to the Lamb shifts and hyperfine splittings in $\mu$H and other light muonic atoms have been derived from baryon ChPT.
More detailed  summaries can be  found at the end of each Chapter.

We now turn to stating our main conclusions as well as an outlook for near-future studies.  
\begin{description}
\item[$\square$]{\bf Finite-Size Effects in Hydrogen-Like Atoms (\chapref{chap2})}
\begin{itemize}
\item[\checkmark] Deriving the one-photon-exchange Breit potential within a dispersive framework is advantageous because the dispersive ansatz  can be equally applied to nuclear finite-size, electroweak and QED corrections (e.g., one-loop vacuum polarization). In Eqs.~\eref{BreitPotentialFinal} and \eref{BreitMomentum}, we give the nuclear form factor dependent coordinate-space and momentum-space potentials, describing the finite-size effects.
\item[\checkmark] Our formalism provides exact formulas for the finite-size effects, which do not rely on any expansion in moments of charge and magnetization distributions, see for instance Eqs.~\eref{rmsLSa} and \eref{wGall} for the Lamb shift, and Eqs.~\eref{HFS1PT1S} and \eref{HFS1PT} for the hyperfine splitting.
\item[\checkmark]
With the help of a toy model, we illustrated that ``soft'' contributions to the electric Sachs form factor can break down the usual accounting of finite-size effects in the Lamb shift \cite{Hagelstein:2015aa,Hagelstein:2016jgk}. It is then not enough to express the Lamb shift in terms of charge radii. Instead, the exact treatment provided by our formalism is required, leaving room for a possible explanation of the {\it proton radius puzzle}.
\item In the future, one has to find physical justifications for the presented toy models. A strong candidate is the weak correction to the lepton vertex shown in \Figref{MuonDecay}. It has to be studied whether this ``soft'' contribution to the lepton form factor is able to resolve the {\it proton radius puzzle}.
\end{itemize}
\item[$\square$]{\bf Static Nucleon Polarizabilities (\chapref{chap3})}
\begin{itemize}
\item[\checkmark] The Compton contribution to photoabsorption generates divergent pieces in the Compton scattering sum rules. We have shown that an infrared cutoff on the $\nu$-integration and infrared subtractions on the photoabsorption cross sections can remove the divergent pieces from the dispersion relations of the Compton scattering amplitudes, see \Eqref{LEXdiv}. A proper definition of the sum rules for the Compton contribution to the quasi-static
 polarizabilities was derived with \Eqref{newSRs}, and verified within quantum electrodynamics \cite{Gryniuk:2015aa,Gryniuk:2016gnm}. 
\end{itemize}
\item[$\square$]{\bf Generalized Nucleon Polarizabilities (\chapref{chap4})}
\begin{itemize}
\item[\checkmark] In ChPT, the contribution of  tree-level $\Delta(1232)$-exchange  to the process of forward doubly-virtual Compton scattering is described by three coupling constants, viz.\ the magnetic, electric and Coulomb couplings: $g_M$, $g_E$ and $g_C$. The effect of the magnetic coupling on the Compton scattering is of order $p^{7/2}$ in the low-energy domain of the $\delta$-expansion in ChPT, while terms including the electric or Coulomb couplings are attributed to higher orders.
We have shown that inclusion of the Coulomb coupling, despite its higher order in the power-counting, produces an appreciable effect in all of the generalized polarizabilities. The Coulomb coupling also influences the static limit of the longitudinal and longitudinal-transverse polarizabilities, $\al_L$ and $\delta_{LT}$.
\item[\checkmark] In view of the $\delta_{LT}$ polarizability puzzle arising within ChPT, the $\pi\Delta$-production photoabsorption cross sections were calculated.
\item[\checkmark] We have shown that the Born and polarizability contributions to the Burkhardt-Cottingham sum rule vanish independently. This is for instance the case for the sum rule contributions from $\Delta$-exchange, pion-nucleon loops and pion-delta loops, which all evaluate to zero. 
\item Since the $\pi\Delta$-production cross sections display a bad high-energy behavior, they cannot be used to reconstruct the $\pi\Delta$-loop Compton scattering amplitudes with unsubtracted dispersion relations. Therefore, at present, the $\pi\Delta$-production cross sections have no predictive power for the lower-order polarizabilities. In a future project, we plan to improve the high-energy asymptotics of the cross sections by some kind of ultraviolet completion, which will allow us to obtain a result for $\delta_{LT}$. It would also be desirable to achieve a better compliance of the baryon ChPT prediction of the nucleon structure functions with experimental data at the $\Delta$-resonance peak.

\end{itemize}
\item[$\square$]{\bf Polarizability Effects in the Lamb Shift (Chapters \ref{chap:chap5} and \ref{chap:5LS})}
\begin{itemize} 
\item[\checkmark] The next-to-leading order baryon ChPT prediction for the order-$\al^5$ proton-polarizability contribution to the Lamb shift in $\mu$H, including diagrams with pion-nucleon loops and $\Delta(1232)$-exchange, yields: 
\beq
E^\mathrm{pol.}_\mathrm{LS}(\mu\text{H})=4.9\,^{+2.0}_{-1.3}\,\upmu\text{eV}. \tag{\ref{eq:LSfinalvalue}}
\eeq 
This model-independent theory prediction is found to be in good agreement with the dispersive results, which are based on empirical information on proton form factors and structure functions.
\item[\checkmark] At next-to-leading order in baryon ChPT, the contribution of the subtraction function to the $\mu$H Lamb shift amounts to: 
\beq
E^\mathrm{subtr.}_\mathrm{LS}(\mu\text{H})=-5.8\pm 2.3\,\upmu\text{eV}. \tag{\ref{eq:totalSub}}
\eeq
This result shows that the tree-level $\Delta$-exchange gives a significant contribution to $\ol T_1(0,Q^2)$. 
\item[\checkmark] Deducing the order-$(Z\al)^6$ polarizability effects to the Lamb shift from off-forward two-photon exchange provides an alternative if not favourable approach to the Coulomb-distortion effects and the classical long-range polarization potentials. In accordance with the literature, the nuclear-polarizability contribution from off-forward two-photon exchange is found to be non-negligible in the case of light muonic atoms, 
\begin{align}
E^{\left\langle(Z\al)^6,\;\al_{E1},\,\beta_{M1}\right\rangle}_\text{LS}(\mu\text{D})&=-0.398\pm0.001\, \mathrm{meV},\tag{\ref{eq:numresultsbetaD}}\\
E^{\left\langle(Z\al)^6,\;\al_{E1},\,\beta_{M1}\right\rangle}_\text{LS}(\mu^3\text{He}^{+})&=-1.395\pm0.047\, \mathrm{meV},\tag{\ref{eq:numresultsbetaHe3}}\\
E^{\left\langle(Z\al)^6,\;\al_{E1},\,\beta_{M1}\right\rangle}_\text{LS}(\mu^4\text{He}^{+})&=-0.665\pm0.016\, \mathrm{meV}.\tag{\ref{eq:numresultsbetaHe4}}
\end{align}
 This can be mainly ascribed to the logarithmic enhancement generated by the $t$-channel cut, giving a $(Z \alpha)^6 \ln Z \alpha$ contribution proportional to the nuclear electric dipole polarizability, see \Eqref{finalresult}. 
\item For an improved precision, the calculation of the off-forward polarizability effects could be repeated, taking into account the full $Q^2$ behavior, i.e., not focusing on the $t$-channel cuts.
\item[\checkmark] Utilizing our results for the various polarizability contributions, we re-extracted the proton and deuteron charge radii from the Lamb shifts in $\mu$H and $\mu$D:
\begin{align}
R_{Ed}&=2.12013(78)\,\text{fm}, \tag{\ref{eq:REdnew}} \\
R_{Ep}&=0.84045(44)\,\text{fm}. \tag{\ref{eq:Renew}}
\end{align}
\end{itemize}
\item[$\square$]{\bf Polarizability Effects in the Hyperfine Splitting (Chapters \ref{chap:chap5} and \ref{chap:5HFS})}
\begin{itemize}
\item[\checkmark] The next-to-leading order baryon ChPT prediction for the order-$\al^5$ proton-polarizability effect in the hyperfine splitting of $\mu$H,
\beq
E^\mathrm{pol.}_\mathrm{HFS}(nS,\mu\text{H})=-11.1\,^{+12.7}_{-9.4}\,\frac{\upmu\text{eV}}{n^3}\tag{\ref{eq:HFSfinalvalueN}},
\eeq
 is considerably smaller than the results from the dispersive calculations which are presently used to extract the proton Zemach radius from $\mu$H spectroscopy. If compared to the dispersive result of Ref.~\cite{Carlson:2008ke}, the discrepancy amounts to $3.6\,\sigma$. The discrepancy can be pinned down to the region of small $Q^2$, where one observes severe cancelations between the proton structure function $g_{1}(x,Q^2)$ and the Pauli form factor $F_{2p}(Q^2)$ of the proton, which both enter $\ol S_1(0,Q^2)$.
\item In the future, further investigations are needed to better understand the discrepancy between the chiral prediction and the dispersive ansatz. At present, we suppose the empirical information in the low-$Q$ region, used as input for the dispersive approach, are not sufficient. Especially the cancelations within $\ol S_1(0,Q^2)$ might be hard to resolve based on the available empirical data. Therefore, new data from the ongoing ``spin physics program'' at the Jefferson Laboratory, which is mapping out the spin structure 
functions of the nucleon \cite{Slifer:2007bn,Chen:2008ng,Chen:2011zzp}, might lay the groundwork for a re-evaluation of the two-photon-exchange effects based on an improved empirical data set.
\item[\checkmark] The $\Delta$-pole model is described by the approximate formula in \Eqref{DePoleHFS}. Effectively, it provides a lower bound on the absolute effect of the $\Delta(1232)$-resonance excitation in the hyperfine splitting of $\mu$H.
\item[\checkmark] We have derived an expansion of the non-Born two-photon-exchange effect in hyperfine splitting in terms of the 
spin polarizabilities.
\item[\checkmark] A first prediction of the order-$\al^5$ neutron-polarizability contribution to the hyperfine splitting of
muonic atoms  in the framework of baryon ChPT, shows that the neutron-polarizability effect is more sensitive to the high-$Q^2$ region than the proton-polarizability effect and suggests that 
it might be bigger. Together with the order-$\al^5$ proton-polarizability contribution, we obtain the following hadronic-polarizability contribution to the hyperfine splitting in $\mu$D and $\mu^3$He:
\begin{align}
E^{\text{N-pol.}}_\mathrm{HFS}(nS,\mu\text{D})&=-49\,^{+32}_{-111} \, \frac{\upmu\text{eV}}{n^3},\tag{\ref{eq:DeuteiumNeutron}}\\
E^{\text{N-pol.}}_\mathrm{HFS}(nS,\mu^3\text{He}^+)&=-526\,^{+343}_{-953}\,\frac{\upmu\text{eV}}{n^3},\tag{\ref{eq:Helium3Neutron}}
\end{align}
\item[\checkmark] The neutral-pion exchange contributes to the hyperfine splitting at order $\al^6$ \cite{Hagelstein:2015b,Hagelstein:2015egb}. As a result of cancelations between the diagrams with direct lepton-pion coupling and pion coupling to the lepton through two photons, its quantitative effect is small,
\beq
E_{\mathrm{HFS}}^{\langle\pi^0 \rangle}(2S,\mu\text{H})=0.119(19)\,\upmu\mathrm{eV},\tag{\ref{eq:pi0finalresult}}
 \eeq
 and differs in sign for H and $\mu$H, respectively, see Table \ref{Tab:resultsPionEx}.
\item[\checkmark]Employing the proton-polarizability effects calculated in this work, the proton Zemach radius was extracted from the measurement of the $2S$ hyperfine splitting in $\mu$H with a shrunken value:
\begin{equation}
R_{\mathrm{Z}p}=1.025(35)\,\mathrm{fm}.\tag{\ref{RZnew}}
\end{equation}
\end{itemize}
\end{description}

We believe that many of these results will be useful for the upcoming search of the very narrow ground-state hyperfine transitions in $\mu$H and $\mu^3$He$^+$.
Once these transitions are found we will have an unprecedentedly precise measurement of the low-energy nucleon structure. It will be an exciting time
for the excited nucleon! 

\begin{small}
\bibliographystyle{model1a-num-names}
\bibliography{lowQ}
\end{small}

\appendix
\addtocontents{toc}{\protect\setcounter{tocdepth}{0}}

\mychapter{8}{List of Acronyms}
\begin{itemize}[noitemsep,leftmargin=2cm]
\item[BC]\hspace{0.25cm} Burkhardt-Cottingham 
\item[BChPT]\hspace{0.25cm} baryon chiral perturbation theory 
\item[ChPT]\hspace{0.25cm} chiral perturbation theory 
\item[CM]\hspace{0.25cm} center-of-mass
\item[CS]\hspace{0.25cm} Compton scattering 
\item [1$\Delta$I]\hspace{0.25cm} one-delta-irreducible
\item[DIS] \hspace{0.25cm} deep inelastic scattering
\item[DOF] \hspace{0.25cm} degree of freedom
\item [1$\Delta$R]\hspace{0.25cm} one-delta-reducible
\item[DR] \hspace{0.25cm} dispersion relation
\item[e.m.]\hspace{0.25cm} electromagnetic
\item[eVP]\hspace{0.25cm} electronic vacuum polarization
\item[FF] \hspace{0.25cm} form factor
\item[FS] \hspace{0.25cm} fine structure
\item[FSE] \hspace{0.25cm} finite-size effect
\item[FSP] \hspace{0.25cm} forward spin polarizability
\item[GDH] \hspace{0.25cm} Gerasimov-Drell-Hearn
\item[GP] \hspace{0.25cm} generalized polarizability 
\item[HBChPT]\hspace{0.25cm} heavy-baryon chiral perturbation theory 
\item[HFS] \hspace{0.25cm} hyperfine splitting
\item [LEC]\hspace{0.25cm} low-energy constant
\item [LET]\hspace{0.25cm} low-energy theorem
\item [LEX]\hspace{0.25cm} low-energy expansion
\item [lhs]\hspace{0.25cm} left-hand side 
\item [LO]\hspace{0.25cm} leading order
\item [LQCD]\hspace{0.25cm} lattice QCD
\item [LS]\hspace{0.25cm} Lamb shift
\item [$\mu$SE]\hspace{0.25cm} muonic self-energy
\item [$\mu$VP]\hspace{0.25cm} muonic vacuum polarization
\item [NLO]\hspace{0.25cm} next-to-leading order
\item [NNLO]\hspace{0.25cm} next-to-next-to-leading order
\item [OPE]\hspace{0.25cm} one-photon exchange
\item [PT]\hspace{0.25cm} perturbation theory 
\item[1PT/2PT]\hspace{0.25cm} first-order/second-order perturbation theory
\item [QCD]\hspace{0.25cm} quantum chromodynamics
\item [QED]\hspace{0.25cm} quantum electrodynamics
\item [QM]\hspace{0.25cm} quantum mechanics
\item[RCS]\hspace{0.25cm} real Compton scattering  
\item[SE]\hspace{0.25cm} self-energy
\item [rhs]\hspace{0.25cm} right-hand side 
\item [rms]\hspace{0.25cm} root-mean-square 
\item[TPE]\hspace{0.25cm} two-photon exchange 
\item[VCS]\hspace{0.25cm} virtual Compton scattering  
\item [VP]\hspace{0.25cm} vacuum polarization
\item[VVCS]\hspace{0.25cm} doubly-virtual Compton scattering  
\end{itemize}
\mychapter{9}{List of Collaborations and Experiments}

\begin{itemize}[noitemsep,leftmargin=2cm]
\item[ALPHA]\hspace{0.25cm} Antihydrogen Laser PHysics Apparatus
\item[BNL]\hspace{0.25cm} Brookhaven National Laboratory
\item[CEBAF]\hspace{0.25cm} Continuous Electron Beam Accelerator Facility
\item[CERN]\hspace{0.25cm} European Organization for Nuclear Research
\item[ ]\hspace{0.25cm} (Conseil Europ\'een pour la Recherche Nucl\'eaire)
\item[CLAS]\hspace{0.25cm} CEBAF Large Acceptance Spectrometer
\item[CODATA]\hspace{0.25cm} Committee on Data for Science and Technology
\item[CM]\hspace{0.25cm} center of mass 
\item[CREMA]\hspace{0.25cm} Charge Radius Experiment with Muonic Atoms
\item[DESY] \hspace{0.25cm} German Electron Synchrotron (Deutsches Elektronen-SYnchrotron)
\item[DORIS] \hspace{0.25cm} Double Ring Store (DOppel-RIng-Speicher)
\item[ELSA] \hspace{0.25cm}  Electron Stretcher and Accelerator (ELektronen-Stretcher Anlage)
\item[ETMC]\hspace{0.25cm} European Twisted Mass Collaboration
\item[FAMU] \hspace{0.25cm} Fisica degli Atomi MUonici (Physics with muonic atoms)
\item[FNAL] \hspace{0.25cm} Fermi National Accelerator Laboratory
\item[HERA] \hspace{0.25cm} Hadron-Electron Ring Accelerator facility (Hadron-Elektron-Ring-Anlage)
\item[JLab] \hspace{0.25cm} Thomas Jefferson National Accelerator Facility
\item[KEK] \hspace{0.25cm} The High Energy Accelerator Research Organization (Japanese)
\item[J-PARC] \hspace{0.25cm} Japan Proton Accelerator Research Complex
\item [LEAR]\hspace{0.25cm} Low Energy Antiproton Ring 
\item [LEGS]\hspace{0.25cm} Laser Electron Gamma Source
\item [LKB]\hspace{0.25cm} Laboratoire Kastler Brosse
\item [LHPC]\hspace{0.25cm} Lattice Hadron Physics Collaboration
\item [LPTF]\hspace{0.25cm} Laboratoire Primaire du Temps et des Fr\'equences
\item [MAMI]\hspace{0.25cm} MAinz MIcrotron
\item [MIT]\hspace{0.25cm} Massachusetts Institute of Technology
\item [MPQ]\hspace{0.25cm}  Max Planck Institut für Quantenoptik
\item [MUSE]\hspace{0.25cm} MUon proton Scattering Experiment
\item [MuSEUM]\hspace{0.25cm} Muon Spectroscopy Experiment Using Microwave
\item [MUSL]\hspace{0.25cm} Microtron Using a Superconducting Linac
\item [NIST]\hspace{0.25cm} National Institute of Standards and Technology
\item[OLYMPUS]\hspace{0.25cm} pOsitron-proton and eLectron-proton elastic scattering to test the
\item [ ]\hspace{0.25cm} hYpothesis of Multi-Photon exchange Using DoriS
\item[PACS]\hspace{0.25cm} Parallel Array Computer System
\item[PDG]\hspace{0.25cm} Particle Data Group
\item[PNDME]\hspace{0.25cm} Precision Neutron Decay Matrix Elements
\item [PRad]\hspace{0.25cm} Proton Charge Radius Experiment
\item [PSI]\hspace{0.25cm} Paul Scherrer Institute
\item[RAL]\hspace{0.25cm} Rutherford Appleton Laboratory
\item [SAL]\hspace{0.25cm} Saskatchewan Accelerator Laboratory 
\item [SLAC]\hspace{0.25cm} Stanford Linear Accelerator Center
\item[TREK]\hspace{0.25cm} Time Reversal Experiment with Kaons
\item[TRIUMF]\hspace{0.25cm} TRI University Meson Facility
\item[VEPP-3]\hspace{0.25cm} $e^+ e^-$ Collider at Budker Institute for Nuclear Physics
\end{itemize}
\mychapter{10}{Acknowledgements}
This work was supported by the Deutsche Forschungsgemeinschaft (DFG) through the Collaborative Research Center SFB 1044 [The Low-Energy Frontier of the Standard Model] and the Graduate School DFG/GRK 1581 [Symmetry Breaking in Fundamental Interactions].\\

\noindent F.~Hagelstein and V.~Pascalutsa thank K.~Griffioen for providing an interpolation table for the JLab parametrization of the spin-dependent proton structure functions, S.~Simula for providing a {\sc FORTRAN} code with his latest parametrization of the spin-dependent proton structure functions, and L.~Tiator for support with the MAID isobar model. \\

\noindent The author would like to thank V.~Pascalutsa, V.~Lensky and J. M.~Alarc\'on, and O.~Gryniuk for collaboration.\\

\noindent Additional thanks go to M.~Vanderhaeghen and the members of his research group for useful discussions. 



\end{document}